\documentclass{ucbthesis}
\usepackage[
    backend=biber,
    style=numeric-comp,
    sorting=none,
    minnames=1,
    maxnames=10,
    date=year,
    uniquename=init,
    block=space,
    giveninits,
]{biblatex}
\usepackage{hyperref}
\usepackage{rotating} 
\usepackage{amsmath}
\usepackage[mathlines]{lineno}
\usepackage{color}
\usepackage[normalem]{ulem}
\usepackage{graphicx}
\usepackage{dcolumn}
\usepackage{bm}
\usepackage{mathrsfs}
\usepackage{xfrac}
\usepackage{subcaption}
\usepackage{multirow}
\usepackage[T1]{fontenc}

\renewcommand{\thefigure}{\arabic{chapter}.\arabic{figure}}

\DeclareRobustCommand{\uvec}[1]{{%
  \ifcsname uvec#1\endcsname
     \csname uvec#1\endcsname
   \else
    \bm{\hat{\mathbf{#1}}}%
   \fi
}}

\bibliography{references}
\AtEveryBibitem{
    \clearfield{issue}
    \clearfield{issn}
    \clearfield{number}
    \clearfield{note}
    \clearfield{url}
}

\DeclareFieldFormat[article,periodical]{volume}{\mkbibbold{#1}}
\renewbibmacro{in:}{}
\newbibmacro{string+doi}[1]{%
  \iffieldundef{doi}{#1}{\href{https://doi.org/\thefield{doi}}{#1}}}
\DeclareFieldFormat{title}{\usebibmacro{string+doi}{\mkbibemph{#1}}}
\DeclareFieldFormat[article, inproceedings]{title}{\usebibmacro{string+doi}{#1\isdot}}
\DeclareFieldFormat[thesis]{title}{\usebibmacro{string+doi}{#1}}
\DeclareFieldFormat[article]{pages}{\mkfirstpage{#1}\isdot}
\DeclareFieldFormat{date}{\mkbibparens{#1}}
\DefineBibliographyStrings{english}{%
    andothers = {\em et\addabbrvspace al\adddot}
}

\DeclareSourcemap{
 \maps[datatype=bibtex,overwrite=true]{
  \map{
    \step[fieldsource=Collaboration, final=true]
    \step[fieldset=usera, origfieldval, final=true]
  }
 }
}

\renewbibmacro*{author}{%
  \iffieldundef{usera}{%
    \printnames{author}%
  }{%
    \printnames{author} (\printfield{usera}),%
  }%
}%

\renewbibmacro*{issue+date}{}

\renewbibmacro*{note+pages}{
  \printfield{note}%
  \setunit{\bibpagespunct}%
  \printfield{pages}%
  \setunit{\addspace}
  \usebibmacro{date}
  \newunit}

\DeclareFieldFormat{eprint:reportnumber}{%
  Report Number\addcolon\space\nolinkurl{#1}}

\begin{document}

\title{Athermal Phonon Sensors in Searches for Light Dark Matter}
\author{Samuel L. Watkins}
\degreesemester{Summer}
\degreeyear{2022}
\degree{Doctor of Philosophy}
\chair{Assistant Professor Matt Pyle}
\othermembers{Professor of the Graduate School Bernard Sadoulet \\
  Professor Martin White}

\numberofmembers{3}

\field{Physics}

\maketitle
\copyrightpage


\begin{abstract}


In recent years, theoretical and experimental interest in dark matter (DM) candidates have shifted focus from primarily Weakly-Interacting Massive Particles (WIMPs) to an entire suite of candidates with masses from the zeV-scale to the PeV-scale to 30 solar masses. One particular recent development has been searches for light dark matter (LDM), which is typically defined as candidates with masses in the range of keV to GeV. In searches for LDM, eV-scale and below detector thresholds are needed to detect the small amount of kinetic energy that is imparted to nuclei in a recoil. One such detector technology that can be applied to LDM searches is that of Transition-Edge Sensors (TESs). Operated at cryogenic temperatures, these sensors can achieve the required thresholds, depending on the optimization of the design. 

In this thesis, I will motivate the evidence for DM and the various DM candidates beyond the WIMP. I will then detail the basics of TES characterization, expand and apply the concepts to an athermal phonon sensor--based Cryogenic PhotoDetector (CPD), and use this detector to carry out a search for LDM at the surface. The resulting exclusion analysis provides the most stringent limits in DM-nucleon scattering cross section (comparing to contemporary searches) for a cryogenic detector for masses from 93 to 140 MeV, showing the promise of athermal phonon sensors in future LDM searches. Furthermore, unknown excess background signals are observed in this LDM search, for which I rule out various possible sources and motivate stress-related microfractures as an intriguing explanation. Finally, I will shortly discuss the outlook of future searches for LDM for various detection channels beyond nuclear recoils.

\end{abstract}

\begin{frontmatter}

\begin{dedication}
\null\vfil
\begin{center}
To My Family
\end{center}
\vfil\null
\end{dedication}

\begin{KeepFromToc}
  \tableofcontents
\end{KeepFromToc}
\clearpage

\begin{acknowledgements}
\markboth{}{}

I'm fairly certain it is tradition to write your acknowledgments in one fell swoop the day before submitting a thesis, so for anybody that I miss in the below: thank you so much for your love and support!

I want to thank a few standout teachers and mentors that helped lead me down this path of life from before I started graduate school. Thank you Mr. Goldberg for being an incredibly passionate high school teacher and providing the first jumping off point to pursue physics. I would also like to thank Prof. Eric Hudson from my time as an undergrad at UCLA. Your courses, especially the quantum optics lab, showed me how amazing physics could be, as well as how awesome a physicist could be! Although I ended up doing dark matter detection, AMO will always have a soft spot in my heart. After my undergraduate education, I worked as an algorithm engineer at PNI Sensor Corporation in Santa Rosa, where my boss Andrew Taylor saw great potential in me and supported me instantly when I told him that I wanted to pursue a Physics PhD. Thank you for seeing something in me and giving me an extremely useful experience in industry.

I am so thankful that I started graduate school at UC Berkeley as a GSI for Physics 7A with an awesome group of people: Patrick, Isaac, Neha, Energy, Reed, Liz, the other Sam, and Best. I'm not sure how I would have gotten through the first year without you all---we absolutely must attend a 7A's game again one of these days! After that crazy first year, I joined the Pyle group in its infancy, joining the motley crew of Matt, Bernard, Bruno, Caleb, and Suhas. At that time, we didn't have much of a lab, and we frequently worked with our SuperCDMS collaborators at SLAC. There I was fortunate to work with Paul, Tsuguo, and Noah. Paul and Tsuguo, thank you eternally for running the lab that provided the majority of the data for this thesis! Noah, you filled the role of our group's senior grad student, and have always provided advice and mentorship at a moment's notice (even when you're in the midst of building your own group at SLAC, you still always found time for me). Here's looking forward to many more conferences where we wander the streets of wherever we are!

For the rest of SuperCDMS, I am immensely grateful to have been able to be a part of the collaboration. For the CPD DM Search in this thesis, I want to thank Wolfgang, Ray, Steve, Scott, and all the others that helped ensure this was an excellent result. Thank you Emanuele for helping shepherd it through the various steps to publication (as well as hosting Caleb and me in Italy for the hike of a lifetime in the Dolomites, though let's bring more water next time...). Thank you Belina for your incredible support from across the globe, working with you has been some of the best collaborating of my graduate school career, and your support in my postdoc search will always immensely appreciated.

Back to the Pyle group, I, as is standard in almost every SuperCDMS thesis, must thank Bruno! I'm not sure how you manage working what seems to be the job of ten (probably more) people, but you do it incredibly well. Working with you is always a great experience, and I always appreciate your patience when I ask a question that I probably should have known the answer to years before... Bernard, although you have mostly lived on my Zoom screen for the PhD for both Hawaii and COVID-related reasons, the time spent with you has been amazing. Your depth of knowledge is awe-inspiring, as well as the perfectly timed quips here and there in meetings. Caleb (aka Dr. Best Man), you are definitely one of the main reasons I made it through graduate school. The friendship we've built over the years making it through every up and down has kept me sane. From the death metal concerts, after which neither of us can hold up our heads without pain, to the arm length burritos, CCW, and all the silly inside jokes we have, it's awesome that we're both going off to the same place! Long live Samleb.

Matt, I'm going to miss all of the random trips with the iPad to Yali's, the business school cafe, or Strada! Your willingness to always be available to talk about literally any aspect of what I was doing (or trying to do) is something that is quite rare in a PI. Of the many lessons you've taught me, I'll always remember the power of scaling laws and the power of the $\Delta \chi^2$. It's amazing what these simple tools can teach us and help guide research. I am also incredibly thankful for the work-life balance of our group---it's something that again is not a given. I'm looking forward to catching up every time we run into one another at conferences, or when I'm back in the bay area. Being part of the beginning of SPICE/HeRALD has been a fantastic experience, and I'm excited to see the awesome physics results that will come.

Liz, you have been the main source of support and encouragement in my life since we met in our first year here. You have always listened to me and helped whenever you could. I can't imagine my life without you, and I'm so glad we both taught 7A together! This last year planning our wedding, finding academic jobs at the same place, and finishing off our theses has been a whirlwind and a testament to the strength of our relationship. I can't wait for our next steps in NM!

I want to thank Jonah for being another amazing friend in graduate school. I'll miss being able to just drop by your office and get some ``tuny sammys'' for lunch or some after-work beers. And of course, you were an awesome minister at Liz's and my wedding---I'll ask you for spiritual advice for years to come! To the Neaton group (and the Neaton group adjacent) in general, thank you for making me feel welcome and a part of the social aspect of the group when I started coming by to hang with Liz. And Ben, thanks for getting Caleb and I off to our legendary gym sessions, where we probably spent most of the time just chatting... But hey, those death metal shows probably were enough of a workout anyways.

Pursuing a PhD during COVID was quite the experience. I want to thank Sami, Kayee, and Andrew for being my online buddies and playing video games into the wee hours of the night (don't worry Matt, it was only on weekends!). Having this outlet during the pandemic allowed me to let go and have fun no matter how much stress was going on in my life. Reconnecting was awesome, and I'm glad we're keeping the gaming going.

I'll finish these acknowledgments by thanking my family. Mom, Dad, Troy, Jen, you have all provided unconditional love and support throughout my academic career. I always knew I could count on one of you to listen and give advice when I needed it. And you've all supported my journey to the end of this PhD (and the beginning of what comes next). To my brother Robert, it was awesome to live close together, and I'm lucky to have an awesome bro like you. And to my Kissick, Richardson, and Watkins families, thank you! And no, I haven't found dark matter... \textit{yet}...

\end{acknowledgements}

\end{frontmatter}

\chapter{\label{chap:dm}The Motivation for Dark Matter}

\begin{figure}
    \centering
    \includegraphics[width=1.0\linewidth]{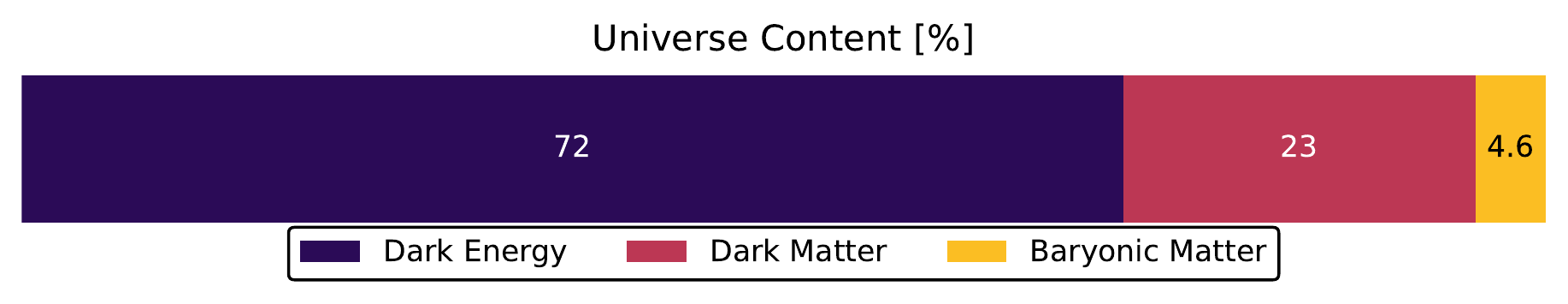}
    \caption{The percent of energy density of our universe's content for dark energy, dark matter, and baryonic matter. These data are reported in Ref.~\cite{2020A&A...641A...6P}.}
    \label{fig:uni_content}
\end{figure}

There are few unanswered questions about the universe that conjure up as much imagination as what makes up our universe beyond the baryonic matter: what is dark matter and what is dark energy? Through cosmological observations it has become clear that these two mysterious components make up the majority of the universe, as shown in Fig.~\ref{fig:uni_content}. In this chapter, I will focus on the motivation, phenomenology, and direct detection methods for dark matter, keeping ourselves working within the matter of the universe and leaving a Ph.D. on dark energy for another life.

\section{Cosmological Evidence for Dark Matter}

From various independent measurements of our universe, there is strong cosmological evidence for the existence of dark matter. They span from galaxy rotation curves to big bang nucleosynthesis to precision measurements of the cosmic microwave background.

\subsection{Galaxy Cluster Dynamics}

Evidence for dark matter is apparent at the large scale of galaxy clusters. When observing galaxy clusters, the virial theorem can be used to estimate the mass of a cluster. The general form of the virial theorem is
\begin{equation}
    \langle T \rangle = - \frac{1}{2} \sum_{k=1}^N \langle \mathbf{F}_k  \cdot \mathbf{r}_k \rangle,
\end{equation}
where $T$ is the total kinetic energy $N$ particles, $\mathbf{F}_k$ is the force on the $k$th particle, and $\mathbf{r}_k$ is the position of the $k$th particle. The brackets denote averaging of the enclosed quantities over time. For a potential that depends only on the distance between particles, i.e. $V(r) \propto r^\alpha$, this simplifies to
\begin{equation}
    \langle T \rangle = \frac{\alpha}{2} \langle V \rangle.
\end{equation}
For a galaxy cluster approximated as a sphere held together by gravity, we have that $\alpha = -1$, and this simplifies to
\begin{equation}
    \langle T \rangle = -\frac{1}{2} \langle V \rangle,
\end{equation}
where $T = \frac{1}{2} M v^2$, and $V= - \frac{3}{5} \frac{GM^2}{R}$ is the gravitational binding energy of the cluster. Plugging in these values and rearranging to solve for the mass of the cluster, we have that
\begin{equation}
    M = \frac{5 \langle v^2 \rangle}{3 G \langle 1/R \rangle},
\end{equation}
as shown and applied to galaxy clusters by Zwicky~\cite{Zwicky:1933gu, zwicky}. When Zwicky applied the virial theorem to the Coma cluster, he was able to estimate its mass-to-light ratio, i.e. the ratio of the mass of a galaxy cluster to its luminosity. For a system completely composed of visible matter, the mass-to-light ratio in units of solar mass divided by solar luminosity would be expected to be about 1. Instead, Zwicky noted that the Coma cluster had a mass-to-light ratio of a few hundred, leading to the positing of dark (nonluminous) matter making up a significant majority of the cluster.

The amount of missing luminous mass in galaxy clusters can be more accurately determined through X-ray emission and gravitational lensing techniques to study the distributions of baryonic and dark matter. At these scales, the majority of baryonic matter remains captured within the gravitational potential well of the cluster, where it forms a gas of temperatures greater than $10^6~\mathrm{K}$ and emits X-rays. From the distribution of this hot gas, and under the assumption that it is in hydrostatic equilibrium with the gravitational potential, the total mass of a cluster can be estimated~\cite{Ettori:2013tka}. For gravitational lensing, general relativity provides that the gravitational potential of the cluster's matter (or any massive object in general) will act as complex lens that distorts the images of more distant galaxies and other celestial objects, where the amount of lensing is directly related to the total mass of the cluster. 

Measurements of galaxy cluster masses using each of these techniques have been shown to be statistically consistent~\cite{Wu_1997}, confirming that there appears to be significantly more mass than what can be accounted for by the visible matter.

\subsection{Galaxy Rotation Curves}

When one is asked about the existence of dark matter, they will likely start with the common explanation based on observations of the rotational dynamics of galaxies. Using our understanding of Doppler shifts of different spectral features, astronomers can calculate the galactic rotation speed as a function of radius~\cite{Sofue_1999,doi:10.1146/annurev.astro.39.1.137}. These spectral features include the visible H spectral lines, $21~\mathrm{cm}$ H~I line, and rotational transitions in CO.

\begin{figure}
    \centering
    \includegraphics[width=0.8\linewidth]{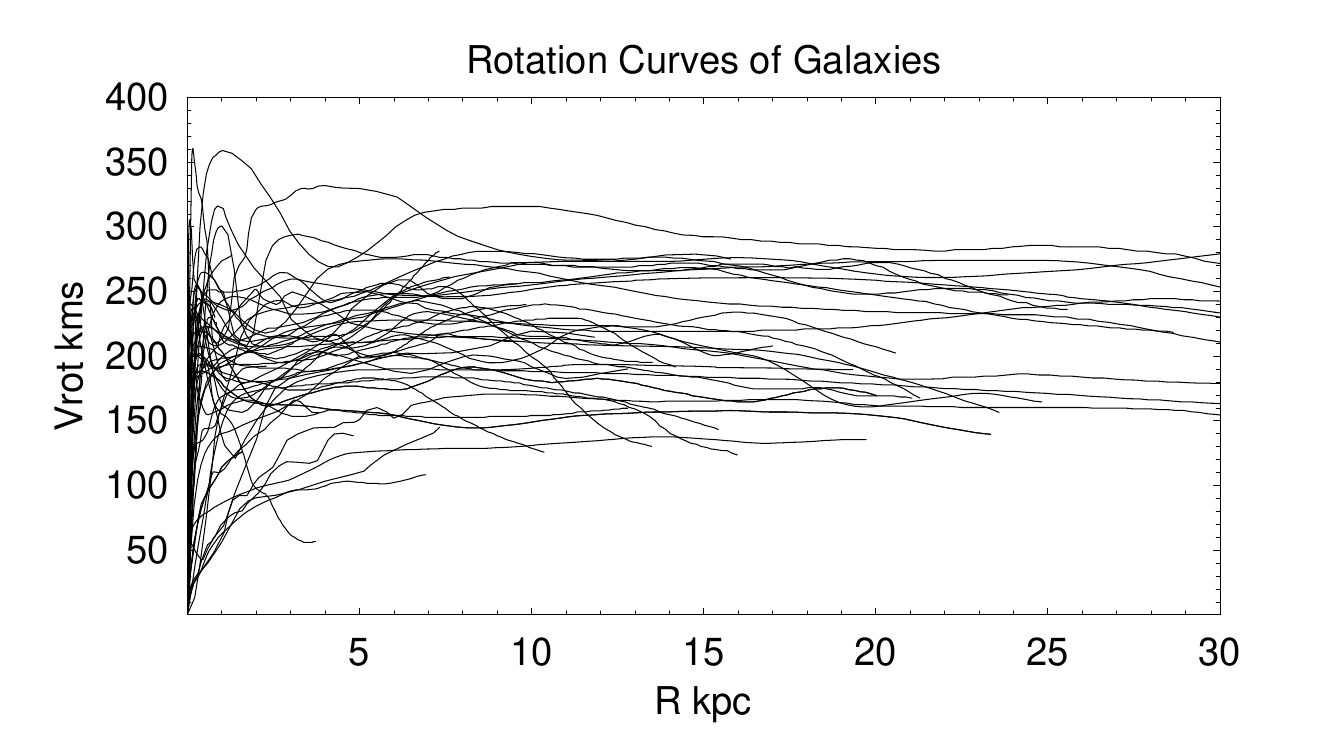}
    \caption{(Figure from Ref.~\cite{Sofue_1999}) The measured galaxy rotation curves for a variety of galaxies, showing the clear leveling off indicated that there must be more mass than is visibly apparent.}
    \label{fig:galrot}
\end{figure}

If one were to assume that visible matter represents the entirety of the mass distributions of these galaxies, then the expectation would be that the orbital velocity of galactic matter would decrease with radius $r$ as $\sim\!r^{-2}$ outside the visible disk (i.e. where the baryonic matter distribution no longer appreciably increases). However, as shown in Fig.~\ref{fig:galrot}, the observed velocity rotation curves have been observed to be constant at distances far from the center of the galaxy.

This constant velocity is consistent with a spherically symmetric distribution of mass whose enclosed mass increases linearly with radius. The suggestion becomes that there must be more mass than that which is visible, which can be explained by the existence of ``dark matter halos'' that extend much farther than the visible disk of matter. 
For non-spiral galaxies, the rotation curves may not provide insight into the amount of missing mass (e.g. an elliptical galaxy does not appreciably rotate). However, from velocity dispersion measurements and applications of the virial theorem, dark matter has been shown to be needed to match observations of dwarf and elliptical galaxies~\cite{Simon_2007,Loewenstein_1999}.

\subsection{Big Bang Nucleosynthesis}

\begin{figure}
    \centering
    \includegraphics[width=0.8\linewidth]{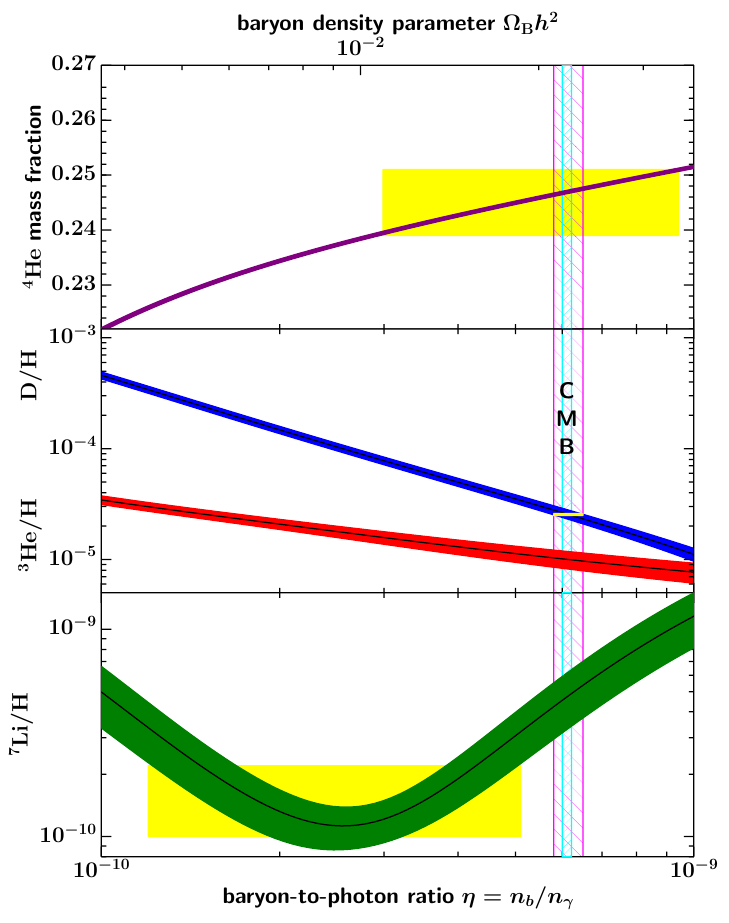}
    \caption{(Figure from Ref.~\cite{Zyla:2020zbs}) The dependence on the primordial number of various nuclei on the amount of baryonic matter in the universe. The yellow regions show the observed abundances, with the pink vertical band being the constraint set by BBN. The constraints on the baryonic matter indicate that it cannot make up the total matter content of the universe.}
    \label{fig:bbn}
\end{figure}


Through dynamics on the scales of galaxies and galaxy clusters, we have seen that dark matter makes up the majority of the mass of these celestial objects. We can turn to observations of the early universe from standard cosmology to specify how much of the matter in our universe is dark. During the cooling of the universe after the Big Bang, light elements began to form as part of Big Bang Nucleosynthesis (BBN), most notably D (deuterium), $^3$He, $^4$He, and Li. Of these elements, D is a fragile nucleus that is destroyed within stars and no longer created in the modern universe, as opposed to the latter three which are created by stars. To understand the primordial abundances of $^3$He, $^4$He, and Li is therefore quite complex, and we are fortunate to be able to precisely measure the abundance of primordial D, which pins down the baryon density extremely accurately. These measurements are based on light from distant quasars being absorbed by intervening neutral hydrogen systems at redshifts of 3--4 (at which primordial abundances have yet to be altered). The absorption spectra provide key features for extracting the ratio of D to H, which are then related to the baryon--photon ratio $\eta$ through standard BBN calculations~\cite{Burles:1997ez}. These measurements give precision results on the baryon density, as shown in Fig.~\ref{fig:bbn}. 

In this figure, the bands for each light element corresponds to predictions given various ratios of baryon--photon densities $\eta$, showing strong dependence. The vertical cyan band in this figure is the baryon density specified by primordial D measurements, corresponding to $\Omega_B h^2 = 0.019 \pm 0.0024$, or $\Omega_B \approx 0.04$ when dividing the square of the dimensionless Hubble constant $h$. When comparing the total matter density of $\Omega_M \approx 0.3$, it becomes clear that the majority of the matter content of our universe is nonbaryonic.

\subsection{Colliding Galaxy Clusters}

\begin{figure}
    \centering
    \includegraphics[width=1.0\linewidth]{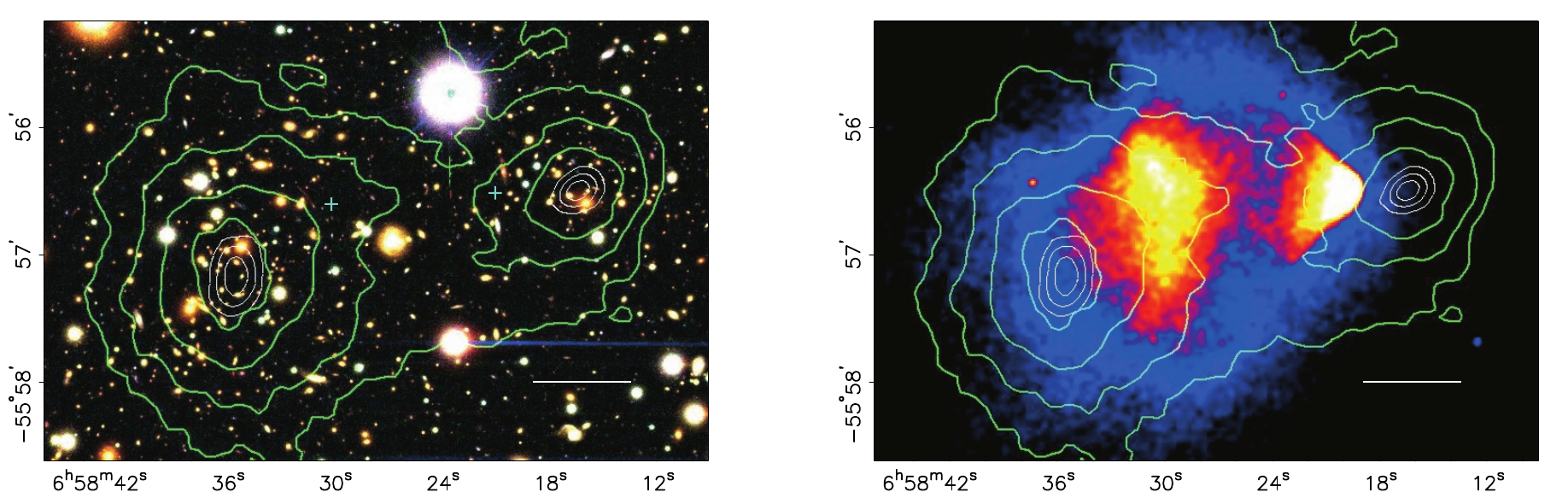}
    \caption{(Figure from Ref.~\cite{Clowe_2006}) (Left) Image of the bullet cluster, where the green contours indicate measurements via gravitational lensing. (Right) Comparison of the mass distribution measurements via hot gas (baryonic matter) and via gravitational lensing (total matter). The decoupling of the two indicate the existence of nonbaryonic matter.}
    \label{fig:cluster}
\end{figure}

In the previous sections, it has been shown that the missing matter has quite different properties than baryonic matter: it is nonluminous, does not form compact structures (i.e. does not lose energy easily), does not emit X-rays, and is not made of baryons. From the baryonic density from Big Bang Nucleosynthesis and comparing to total mass estimates, we know that the amount of baryons in the universe is not enough to explain the mass of galaxy clusters. For a dramatic display of the different properties of baryonic matter and dark matter, one of the most well-known examples is that of the Bullet Cluster (1E 0657-56), which consists of two colliding clusters of galaxies. In Ref.~\cite{Clowe_2006}, the distributions of hot gas and total mass have been reconstructed using the techniques of X-ray emission and gravitational lensing (see Fig.~\ref{fig:cluster}).

The results provide a clear decoupling between the baryonic matter and the dark matter in this cluster: the dark matter halos of each cluster have passed through one another with very little interactions (the green contours) as compared to the baryonic matter distributions that lag behind due to interactions between the hot gas of each cluster. The self-interaction cross section of dark matter can be quantitatively estimated from this cluster, giving that the cross section per mass is less than $1.25\, \mathrm{cm}^2/\mathrm{g}$~\cite{Randall_2008}. The majority of each cluster's mass is not the baryonic matter, but some other form of matter with very different properties (dark matter). This phenomenon has also been observed in the merging cluster MACS J0025.4-122~\cite{Brada__2008}, providing more strong evidence of the differences between baryonic and dark matter. These colliding clusters furthermore give strong evidence against modified Newtonian dynamics theories (i.e. that we do not understand gravity at large scale) due to the large spatial separation between the centers of the total mass and the baryonic mass ($8\sigma$ in the case of the Bullet Cluster).

\subsection{Cosmic Microwave Background and Structure Formation}

\begin{figure}
    \centering
    \includegraphics[width=0.8\linewidth]{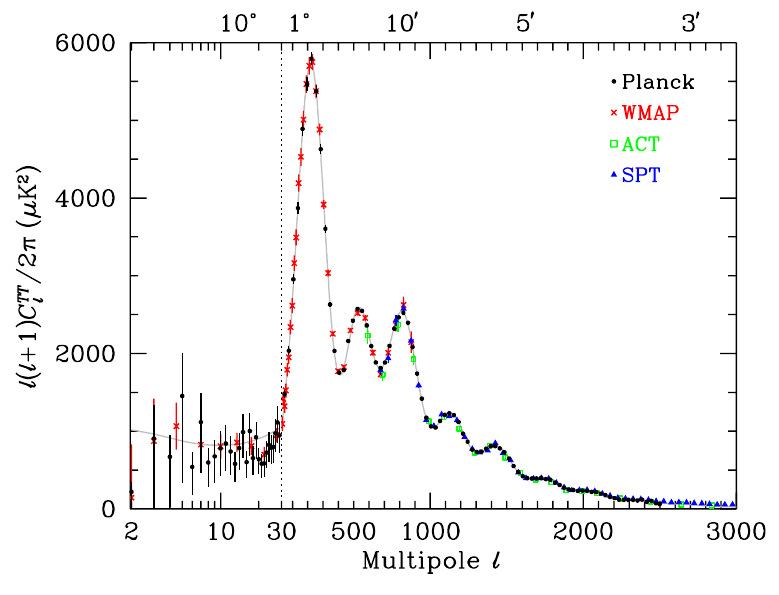}
    \caption{(Figure from Ref.~\cite{Zyla:2020zbs}) Measurements of the temperature anisotropies in the CMB expanded in multipole moment. The location and magnitude of these peaks place strong limits on the energy contents of the universe.}
    \label{fig:cmb}
\end{figure}


Through observations of the cosmic microwave background (CMB), we can find more evidence of the existence of dark matter. The CMB is the primordial blackbody radiation of the universe, which last scattered off electrons when the universe was about 400,000 years old. After this point, these photons have traveled freely through space and are observable today with a thermal blackbody spectrum at a temperature of $2.7~\mathrm{K}$. For decades, it was observed that the CMB was completely isotropic, until anisotropies were discovered in its temperature and polarization~\cite{doi:10.1146/annurev.astro.40.060401.093926,HU1997323}. These anisotropies follow a characteristic pattern that is predicted by inflation, as shown in Fig.~\ref{fig:cmb}

These temperature fluctuations of the CMB are highly dependent on the baryon density, as they come from inhomogeneities in the photon-baryon fluid before the photons decoupled to become the CMB. Thus, the density of the baryonic matter will have a substantial effect on these fluctuations, and the results give a sensitive probe on the cosmological parameters in the theory of inflation. The results shown in the figure correspond to a baryonic matter density of $\Omega_B \approx 0.04$, and a majority of the matter in the universe must be nonbaryonic. Note that this value is consistent with the value from Big Bang Nucleosynthesis with a substantially different physics derivation.

These small anisotropies in the CMB have not had enough to time to coalesce into the structures seen in today's universe. That is, if there were only baryons, these fluctuations are too small to have taken the early homogeneous universe and resulted in the observed structure of the universe today. Thus, there must be some matter that does not couple to photons, i.e. did not couple to the CMB. The fluctuations in this matter's density could then grow to make the structure we see without affecting the scale of the anisotropies of the CMB. Once radiation and baryonic matter decouple, then the baryonic matter will fall into the gravitational wells of this nonbaryonic matter, giving the structure of the universe. One important characteristic of this nonbaryonic matter is that it must be nonrelativistic (``cold''), giving us the theory of cold dark matter. If dark matter had relativistic velocities, then this would result in structure formation that does not fit our cosmological observations, as the formation paradigm would be for large-scale structure to fragment into smaller structures. Instead, our observations are consistent of a hierarchical formation of large-scale structure originating from the coalescing of smaller structures~\cite{Kolb:1990vq,2005Natur.435..629S}.

\subsection{Beyond Dark Matter: Dark Energy}


\begin{figure}
    \centering
    \includegraphics[width=0.8\linewidth]{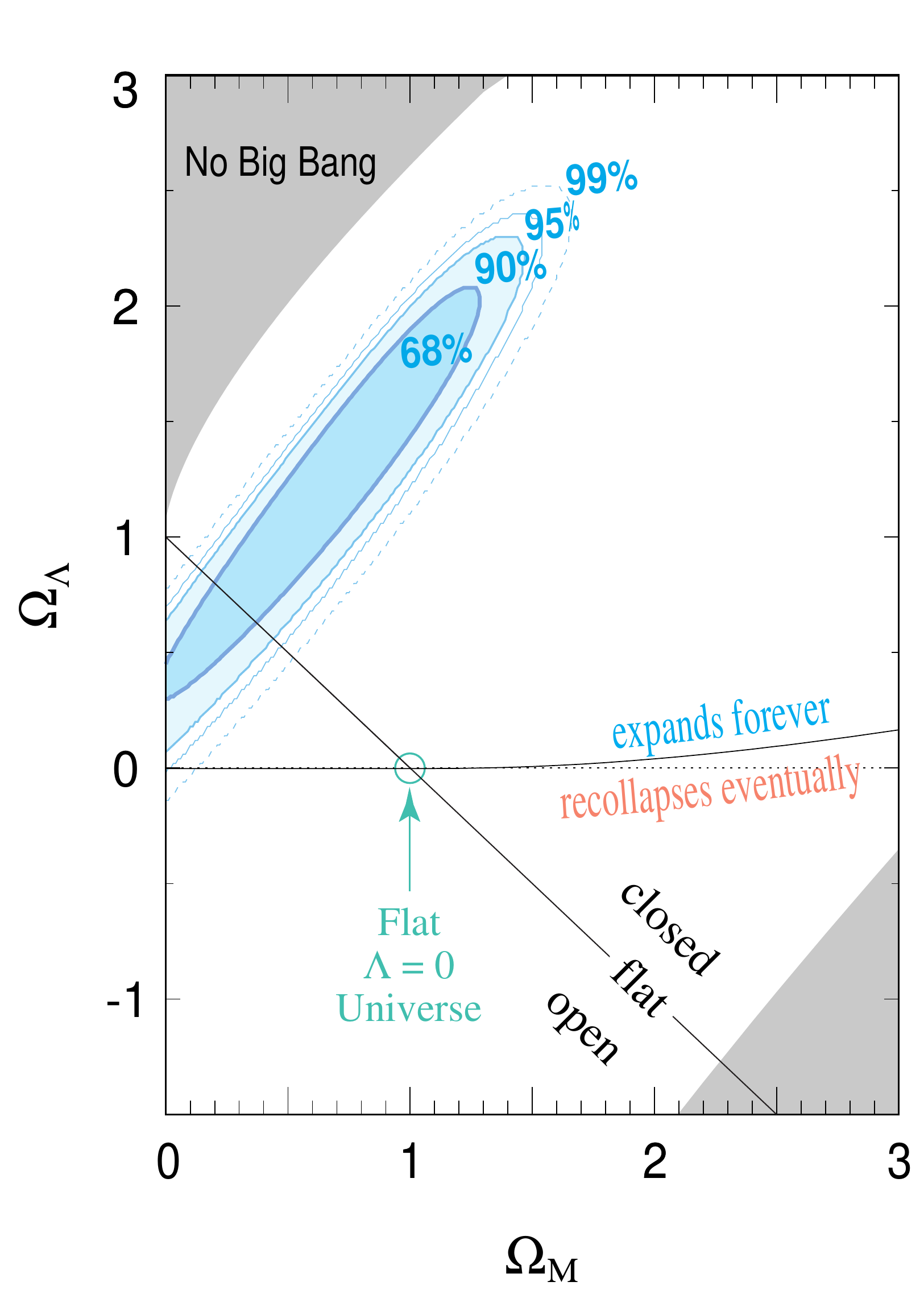}
    \caption{(Figure from Ref.~\cite{Perlmutter_1999}) Constraints set on $\Omega_\Lambda$ by The Supernova Cosmology Project in Ref.~\cite{Perlmutter_1999}. While the constraints on $\Omega_\Lambda$ have become more precise since this paper, these initial results provided direct evidence of a nonzero $\Omega_\Lambda$ and the existence of dark energy.}
    \label{fig:lambda}
\end{figure}

Although we have spent a few pages discussing the evidence for dark matter, the various types of matter in our universe cannot explain its full content (recall Fig.~\ref{fig:uni_content}). Perhaps the most convincing first argument for this is that the universe is flat from CMB anisotropy measurements. If the universe were open, then the paths of photons from the surface of last scattering (i.e. the CMB photons) starting out parallel would slowly diverge, and the physical scale with maximal anistropy (the first peak of the CMB anisotropies) would be shifted to smaller angular scales (higher $l$ in the spherical harmonics, as the angular scale is $\theta \approx \pi / l$). A closed universe would similarly shift this first peak to larger angular scales. Precisely measuring the location of this first peak then gives a precise value of the curvature of the universe---this peak has been measured to occur at around $l=200$~\cite{Boomerang:2000efg, Balbi:2000tg, Miller:1999qz}. In models where the universe is nearly flat (where the curvature is described by the total density of the universe $\Omega_0$, and a flat universe has $\Omega_0=1$), the location of this peak is $l\approx 200/\Omega_0^{1/2}$. Thus, these measurements have concluded that $\Omega_0=1$ to high precision, giving convincing evidence that the universe is flat.

To demonstrate the importance of this, we turn to the Friedmann equations, which describe the expansion of space under the assumption of a homogeneous and isotropic universe. The first Friedmann equation can be written in terms of present day values as
\begin{equation}
    \frac{H^2}{H_0^2} = \Omega_\mathrm{rad}a^{-4} + \Omega_M a^{-3} + \Omega_k a^{-2}+ \Omega_\Lambda a^{-3 (1 + w)},
\end{equation}
where $a$ is the time-dependent scale factor of the universe, $H \equiv \dot{a}/a$ is the Hubble parameter, $H_0$ is the Hubble parameter today, $\Omega_\mathrm{rad}$ is the radiation density today (about $10^{-4}$ and can be neglected), $\Omega_M$ is the total matter density today, $\Omega_k = 1 - \Omega_0$ is the ``spatial curvature density'' today, $\Omega_\Lambda$ is the cosmological constant today, and $w$ is the equation of state parameter of dark energy (the ratio of the pressure to the energy density of a perfect fluid). For a universe with flat curvature, this would mean that $\Omega_k = 0$ and (neglecting $\Omega_\mathrm{rad}$) $\Omega_M + \Omega_\Lambda = 1$. Because $\Omega_M \approx 0.3$, the value of $\Omega_\Lambda$ must be nonzero---requiring the existence of some other phenomenon. Though this is not direct evidence of dark energy, it becomes quite difficult to reconcile the observed curvature of space without including a nonzero $\Omega_\Lambda$.

The first direct evidence of dark energy was published in 1998 by Perlmutter \textit{et al.}~\cite{Perlmutter_1999} and Riess \textit{et al.}~\cite{Riess_1998}, where each analysis used high-redshift Type Ia supernovae observations to measure the expansion rate of the universe and show that this expansion is accelerating. This acceleration is well explained by the existence of dark energy with an equation of state parameter of $w=-1$. In Fig.~\ref{fig:lambda}, we reproduce the results from Perlmutter \textit{et al.}, showing that these measurements directly gave $\Omega_\Lambda>0$.

Further evidence for dark energy comes from measurements of baryon acoustic oscillations~\cite{PhysRevD.92.123516} and structure formation measurements combined with the CMB anisotropies~\cite{10.1046/j.1365-8711.2002.05215.x}. These observations have led to the $\Lambda$CDM model being the predominant description of the universe. Although the details of the $\Lambda$CDM model can be argued (e.g. perhaps $w\neq-1$, but is time varying, as proposed by Caldwell \textit{et al.} as quintessence~\cite{PhysRevLett.80.1582}), the overwhelming evidence for dark matter and dark energy means that we should figure out what they are. As mentioned in the first paragraph in this chapter, this thesis will focus on the dark matter side of these (as of 2022) unanswered questions.

\section{Phenomenology of Dark Matter}

With the strong evidence of dark matter from our cosmological observations, we have high confidence  that dark matter must be nonbaryonic, less self-interacting than baryons, stable, and cold. At its core, these are remarks that dark matter interacts gravitationally, and any theory that fits these four criteria could be a plausible explanation of the nature of dark matter. In fact, there are theories of dark matter that cover the mass range of $\mathcal{O}(\mathrm{zeV}/c^2)$ to $\mathcal{O}(\mathrm{PeV}/c^2)$ up to 30 solar masses. Of these theories, a handful have received the most attention up until roughly the 2010s, with the most given to weakly interacting massive particles (WIMPs).

\subsection{The Classic WIMP}

\begin{figure}
    \centering
    \includegraphics[width=0.5\linewidth]{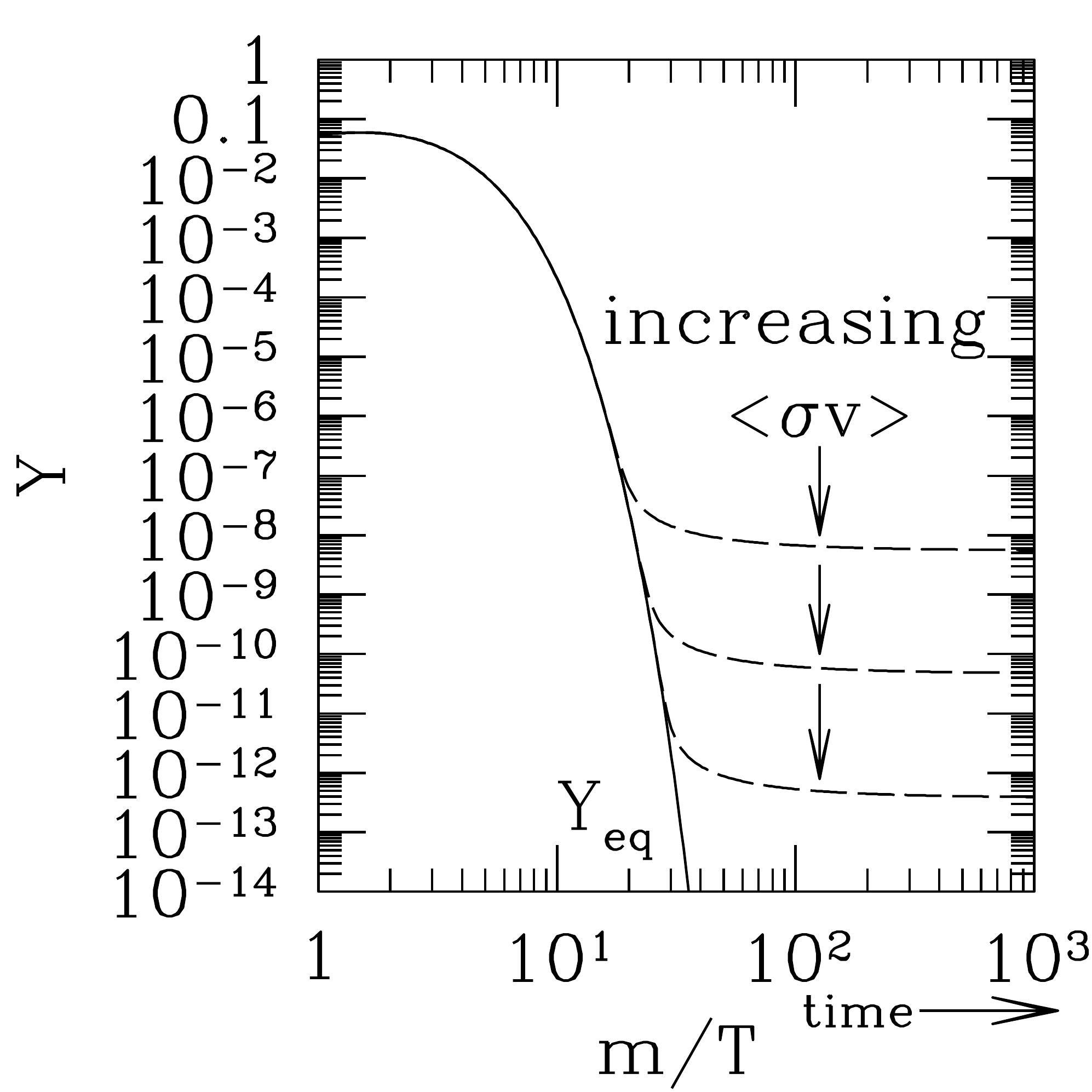}
    \caption{(Figure from Ref.~\cite{Bertone:2010zza}) The mechanism of the thermal freeze-out for WIMPs for achieving the DM relic density.}
    \label{fig:wimp_annih}
\end{figure}

Historically, in theories of dark matter, the WIMP seemed to be the most promising, due in part to the argument's relative simplicity. A WIMP $\chi$ is generally proposed at some massive nonbaryonic particle that was in thermal equilibrium with the universe at early times. At these times, when temperatures were much higher than the mass of the WIMP ($T \gg m_\chi$), the colliding particles of the thermal plasma had enough energy to efficiently create or annihilate WIMPs. As the universe expanded, the temperature of this plasma decreased until, at some point, the WIMP annihilation rate dropped below than the expansion rate of the universe $H$, a process frequently described as ``thermal freeze-out.'' After freeze-out, the number of WIMPs in the universe remained approximately constant, creating the relic cosmological abundance that we see today.

To show this quantitatively, the derivation (e.g. see Ref.~\cite{JUNGMAN1996195, Bertone:2010zza}) starts with the Boltzmann equation for the time evolution of the WIMP number density $n_\chi (t)$
\begin{equation}
    \frac{\mathrm{d}n_\chi}{\mathrm{d}t} + 3 H n_\chi = -\langle \sigma_A v \rangle \left[ \left( n_\chi \right)^2 - \left( n_\chi^\mathrm{eq} \right)^2 \right],
    \label{eq:boltzmann}
\end{equation}
where $H$ is the Hubble expansion rate. The left hand side of Eq.~(\ref{eq:boltzmann}) accounts for the expansion of the universe, while the right hand side accounts for the annihilation and creation of WIMPs. If we approximate that $\langle \sigma_A v \rangle$ is energy-independent, then we can combine our understanding of the Hubble expansion rate during the radiation-dominated universe ($H(T) \propto T^2$) and the freeze-out condition ($\Gamma_A = n_\chi\langle \sigma_A v \rangle$) to approximately solve this equation. After using the present values of today's entropy density and critical density, one finds that the present DM mass density is
\begin{equation}
    \Omega_\chi h^2 \simeq \frac{3 \times 10^{-27} \, \mathrm{cm}^3 / \mathrm{s}}{\langle \sigma_A v \rangle}.
    \label{eq:dmdensity}
\end{equation}

The result in Eq.~(\ref{eq:dmdensity}) is largely independent of the mass of the WIMP (there are small higher order corrections) and inversely proportional to its annihilation cross section. If we were to solve the Boltzmann equation numerically, then we similarly find that, as the annihilation cross section increases, the relic comoving number density decreases, as shown in Fig.~\ref{fig:wimp_annih}.

To motivate the WIMP being a natural candidate, one can quickly estimate the expected annihilation cross section for some new particle with weak-scale interactions as
\begin{equation}
\begin{split}
    \langle \sigma_A v \rangle &\sim \alpha^2 (100 \, \mathrm{GeV})^{-2} \\
    &\sim 10^{-25}\, \mathrm{cm}^3 / \mathrm{s},
    \label{eq:approxweak}
\end{split}
\end{equation}
where $\alpha \sim 10^{-2}$. This value is coincidentally very close to that which is needed to achieve the relic DM density, which provided a strong simple argument for DM to be a WIMP. Furthermore, this is the same scale at which new physics was expected from supersymmetry, which was proposed in part as a solution to the hierarchy problem (that radiative corrections would drive the Higgs mass much higher unless there are new particles at the weak-scale)~\cite{HABER198575}.

\begin{figure}
    \centering
    \includegraphics[width=1.0\linewidth]{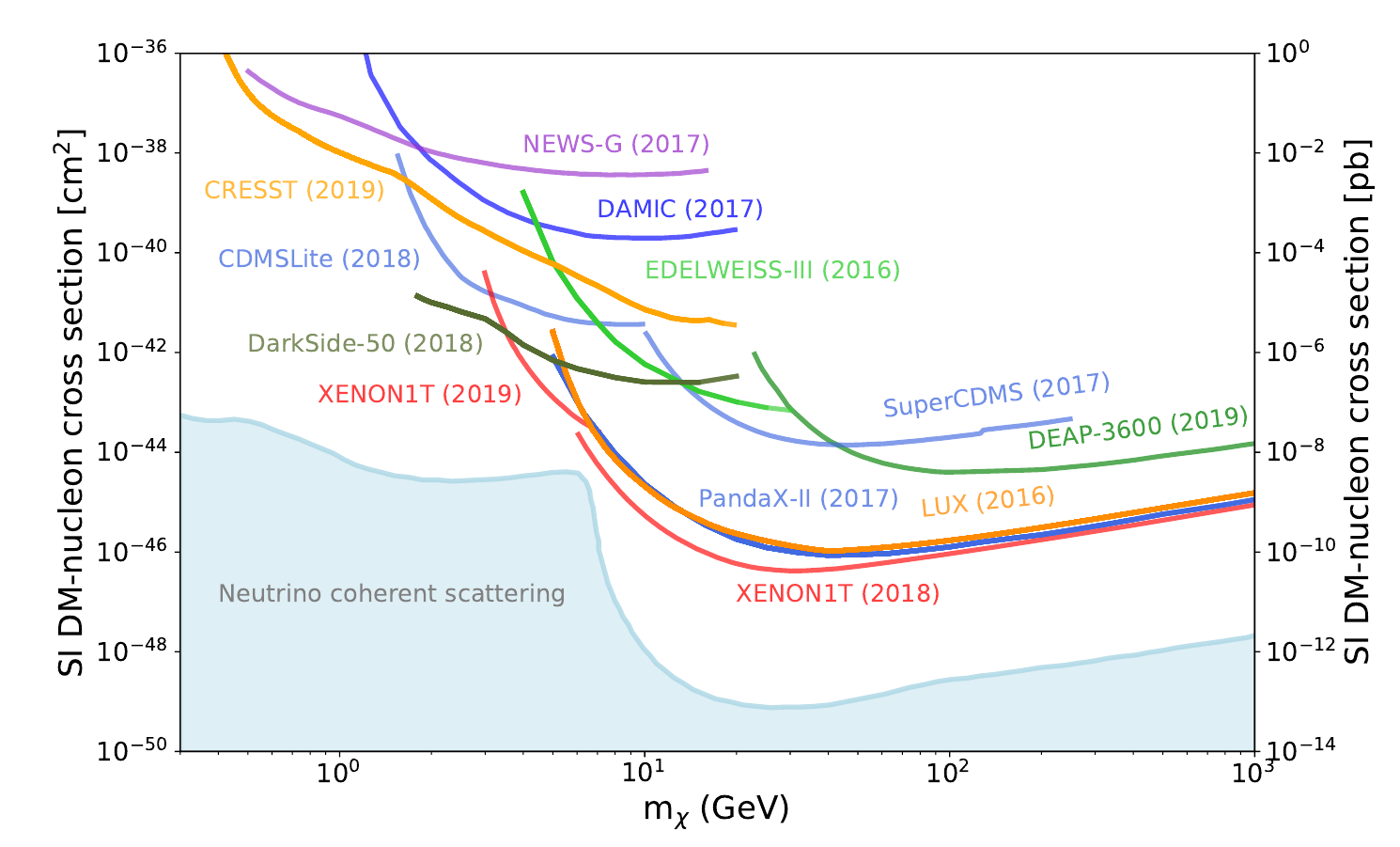}
    \caption{(Figure from Ref.~\cite{Zyla:2020zbs}) The landscape of constraints on WIMPs as of 2021.}
    \label{fig:ddplot}
\end{figure}

With the above argument in mind, experimentalists have searching for WIMP dark matter for decades to no avail. In Fig.~\ref{fig:ddplot}, various DM searches over the expected WIMP mass scale have increasingly ruled out DM-nucleon cross sections down to $\sim\!10^{-46} \, \mathrm{cm}^2$. In WIMP mass, Lee and Weinberg showed that WIMP DM should have a mass greater than $\sim\!2 \, \mathrm{GeV}$ (the Lee-Weinberg bound), otherwise the DM would have frozen out much earlier in the universe's history, leading to an overabundance of DM and the overclosing of the universe~\cite{PhysRevLett.39.165}. As these WIMP searches have breached this bound, a huge amount of the WIMP parameter space has been excluded, and expectations from supersymmetric theories have been dashed. In Fig.~\ref{fig:ddplot}, there remain theoretical supersymmetric models that predict unexplored DM-nucleon cross sections (many of which will be probed in coming experiments). However, the decades of null results on discovering supersymmetric particles at the Large Hadron Collider and WIMP dark matter particles in the expected mass range have brought recent interest to alternative dark matter theories~\cite{Battaglieri:2017aum,Alexander:2016aln,Kahn:2021ttr}.

\subsection{Light Dark Matter}

Light dark matter (LDM) is based on the idea that these DM particles are instead neutral under standard model forces, but charged under undiscovered forces (frequently referred to as hidden or dark sectors). These hidden-sector interactions provide a possible origin for DM, but without the mass restrictions provided by the WIMP hypothesis (i.e. these non-WIMP theories circumvent the Lee-Weinberg bound).

\begin{figure}
    \centering
    \includegraphics[width=1.0\linewidth]{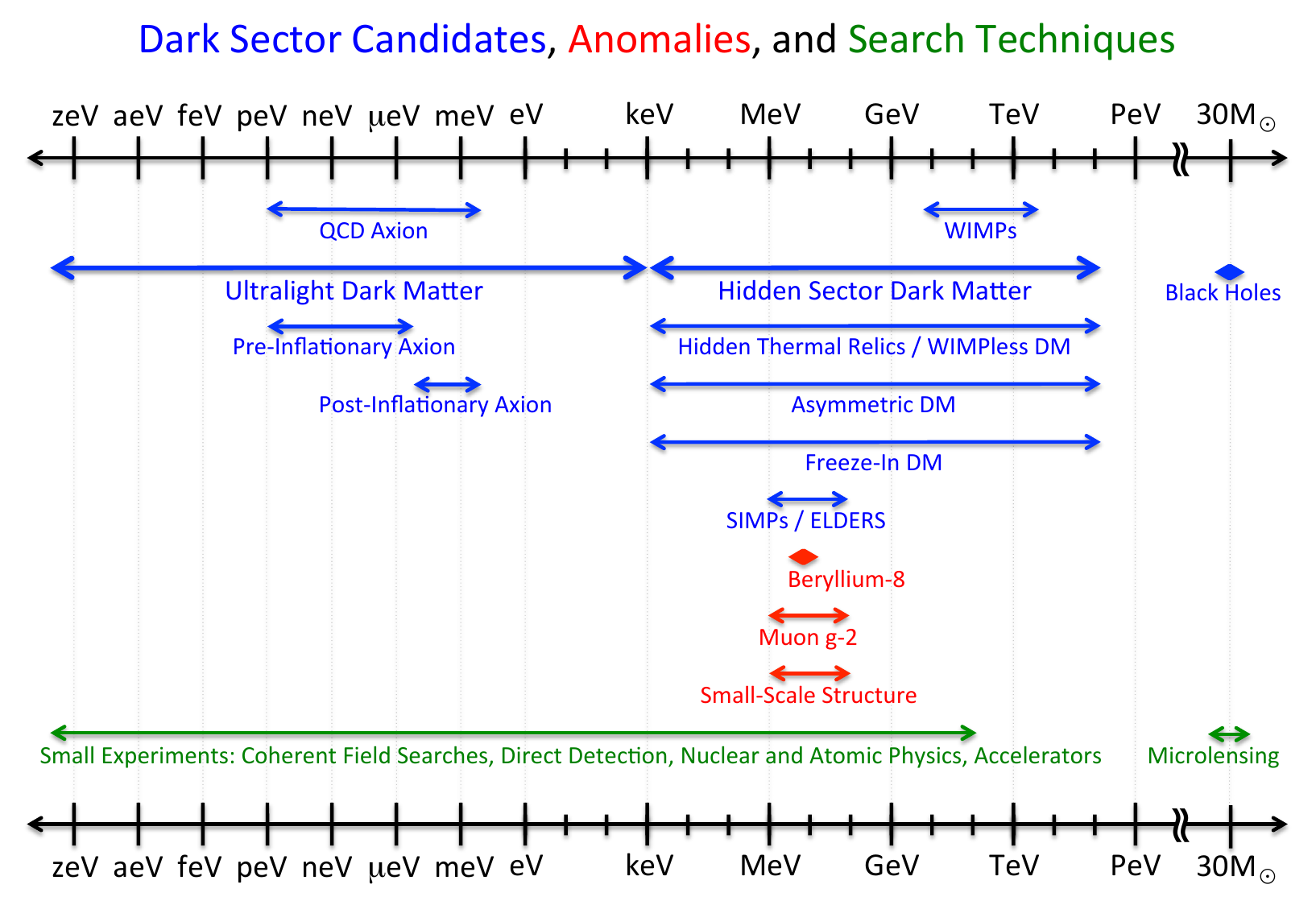}
    \caption{(Figure from Ref.~\cite{Battaglieri:2017aum}) A diagram of the masses that are spanned by all dark sector candidates. There are many orders of magnitude in mass that are populated by non-WIMP candidates.}
    \label{fig:sectors}
\end{figure}

In Fig.~\ref{fig:sectors}, these dark sector candidates cover a much wider range of DM masses, where LDM is usually thought of as the mass range between $\mathcal{O}(\mathrm{keV})$ and $\mathcal{O}(\mathrm{GeV})$. We note that the lower bound of $\mathcal{O}(\mathrm{keV})$ also corresponds to the lower bound of warm dark matter mass~\cite{PhysRevD.88.043502,Lopez-Honorez:2017csg}. Although it is entirely possible that there are no interactions between this hidden sector and the Standard Model, general symmetry arguments allow for interactions between some generic hidden sector and the Standard model, where these ``portal'' interactions are generated by radiative corrections. These couplings also can explain the mechanism by which the universe achieved its dark matter abundance via thermal contact.

When discussing hidden-sector DM, attention is generally given to simplified models of a new force that is mediated by a vector or scalar boson~\cite{Battaglieri:2017aum,PhysRevLett.115.251301}
\begin{align}
    \mathcal{L}_V &\supset V_\mu \bar{f} \left(g_f^V \gamma^\mu + a_f^V \gamma^\mu \gamma^ 5 \right) f \label{eq:vectorlag} \\
    \mathcal{L}_S &\supset \bar{f} \left(g_f^S + \gamma_5 a_f^S \right) f \phi , \label{eq:scalarlag}
\end{align}
where $V_\mu$ is a vector mediator, and $\phi$ is a scalar mediator. In these models, the structure of the $g_f$ and $a_f$ couplings will depend on how the mediator couples to ordinary matter. Two well-studied cases are the vector portal and the Higgs portal, for which there are unique renormalizable interactions of a Standard Model neutral boson that have all Standard Model symmetries~\cite{Galison:1983pa,HOLDOM1986196,Patt:2006fw}:
\begin{align}
    \mathcal{L} &\supset - \frac{\epsilon}{2 \cos \theta_W} B_{\mu \nu} F'^{\mu \nu} \ \mathrm{(vector \ portal)} \label{eq:vectorportal} \\
    \mathcal{L} &\supset \left(  \mu \phi + \lambda \phi^2 \right) H^\dagger H \ \ \ \mathrm{(Higgs \ portal)}, \label{eq:scalarportal}
\end{align}
where $B_{\mu\nu}$ is the hypercharge $U(1)_D$ vector boson field strength, $F'^{\mu \nu}$ is the dark $U(1)_D$ vector boson field strength, and $H$ is the Higgs doublet. 

Focusing on the vector portal interactions, a popular model is that of the minimal kinetically mixed dark photon. The dark photon vector field $A'$ has the Lagrangian
\begin{equation}
    \mathcal{L}_{A'} = -\frac{1}{4} F'^{\mu \nu} F'_{\mu \nu} + \frac{\epsilon}{2 \cos \theta_W} B_{\mu \nu} F'^{\mu \nu} - \frac{1}{2} m^2_{A'} A'^\mu A'_\mu,
\end{equation}
where $\epsilon$ is some kinetic mixing parameter and $m_{A'}$. This model is one of the simplest dark sectors and could also represent the mediator of a larger dark sector. That is, this model can be extended to include a DM candidate that is not the mediator itself. The DM candidate could be a fermion or a scalar boson with coupling to the dark photon through dark-sector gauge interactions. The dark photon would be the mediator for interactions between DM and the Standard Model particles.

If the DM in this dark sector achieved its current abundance through the process of thermal freeze-out, then there are two distinct annihilation processes that it could follow, depending on the hierarchy of the DM and dark photon masses.  If the DM candidate $\chi$ were heavier than the mediator ($m_\chi > m_{A'}$), then the DM would follow ``secluded'' annihilation, where it would annihilate into a pair of mediators, and the mediators would decay to Standard Model particles~\cite{POSPELOV200853}. In this case, the annihilation rate scales as
\begin{equation}
    \langle \sigma v \rangle \propto \frac{g_D^4}{m_\chi^2},
\end{equation}
 where $g_D$ is the coupling between the DM and the dark photon mediator. Note the lack of coupling between the mediator and the Standard Model $g_\mathrm{SM}=\epsilon e$. This coupling could be vanishingly small and still be consistent with thermal freeze-out of DM. Alternatively, if the DM candidate were lighter than the mediator ($m_\chi < m_{A'}$), then the DM would follow ``direct'' annihilation, where it would annihilate to Standard Model particles directly via exchange of a virtual mediator. In this case, the annihilation rate scales as
\begin{equation}
    \langle \sigma v \rangle \propto \frac{g_D^2 g_\mathrm{SM}^2 m_\chi^2}{m_{A'}^4}.
\end{equation}
With the explicit dependence on the coupling between the mediator and the Standard Model, the direct annihilation channel leads to a predictive target for discovery or falsifiability. The dark coupling and the mass ratio $m_\chi / m_{A'}$ are at most $\mathcal{O}(1)$, giving a minimum value of the coupling between the mediator and the Standard Model. For the vector portal, this is at the level of $\epsilon \sim 10^{-7} m^2_{A'}/(m_\chi \mathrm{MeV} \sqrt{\alpha_D})$, where $\alpha_D = g_D^2 / 4 \pi$, providing a benchmark for mediator and DM searches.

It is also possible that the DM relic abundance was not set by a thermal freeze-out process. One such candidate is asymmetric dark matter (ADM), where the abundance is set by a primordial asymmetry, analogous to that of the baryon asymmetry leading to more matter than anti-matter. In a hidden-sector model of ADM where its asymmetry comes from the same mechanism that causes the baryon asymmetry, the abundances of the baryons and the DM are then naturally related by $\Omega_\chi \sim r\Omega_b m_\chi / m_p $, where $r$ is an $\mathcal{O}(1)$ number that depends on the operator that maintains the chemical equilibrium between the dark matter sector and the Standard Model early in the universe~\cite{PhysRevD.79.115016,ZUREK201491}. With the observed abundances, this motivates GeV-scale ADM. On the other hand, if the mechanism of asymmetry of DM is unrelated to the baryons, then $r$ could have any value, and there would be no natural expectation of the ADM mass. Models with $r \ll 1$ would correspond to DM that is much lighter in mass than that of the proton, and those with $r \gg 1$ giving DM with very large masses.

Another possibility of DM relic abundance coming from a process that was not thermal freeze-out is for DM with masses near the QCD confinement scale of about $100 \, \mathrm{MeV}$, giving a hidden sector version of QCD~\cite{Strassler:2006im,PhysRevD.62.063506}. A number-changing process between three and two DM particles can deplete the DM abundance and achieve the correct relic density~\cite{PhysRevLett.113.171301,PhysRevLett.115.021301,Hochberg:2015vrg}. These Strongly Interacting Massive Particles (SIMPs) would require keeping the hidden sector and Standard Model particles in kinetic equilibrium via elastic scattering until the number-changing process freezes out. Instead of the number-changing process, the DM relic abundance could instead be determined by the cross-section of elastic scattering on Standard Model particles~\cite{PhysRevLett.116.221302,Kuflik:2017iqs}, in this case the DM particles are Elastically Decoupling Relics (ELDERs). For either mechanism of determining the relic abundance, both ELDERs and SIMPs have predicted masses of few to a few hundred MeV, well within the LDM range.

The DM could also not be in thermal equilibrium, but still have a small interaction with ordinary matter, leading to a ``freeze-in'' mechanism where Standard Model particles slowly annihilate or decay into DM~\cite{Hall:2009bx}. For hidden-sector DM with very weak mixing, this process could consist of the Standard Model particles freezing out into dark sector mediators, which then decay into DM. These couplings would be quite small with a light vector-portal mediator, and these freeze-in models via the hidden sector predict DM candidates with masses in the LDM scale~\cite{Essig:2015cda,Chu:2011be,PhysRevD.85.076007}.

All of these LDM candidates represent the vibrant interest in this mass scale, giving experimentalists plenty of theoretical motivation to perform LDM searches. As the mass scales are much less than the WIMP mass scales, to search for these candidates requires the development of a different class of detectors optimized instead for LDM.

\subsection{Other Candidates}

In Fig.~\ref{fig:sectors}, there is also a whole regime of ultralight dark matter (ULDM) for DM that is bosonic and has a sub-keV mass. The DM for these masses cannot be fermionic, as the Fermi degeneracy pressure would prevent the galactic substructure formation at the scale of dwarf galaxies from clumped fermionic DM~\cite{Battaglieri:2017aum}. If the DM had a mass lower than $10^{-22} \, \mathrm{eV}$ then the Compton wavelength would be larger than observed dwarf galaxies and would not be meaningfully bound to these galaxies. The most famous of these bosonic ULDM candidates is the QCD Axion, which was proposed as a solution to the strong CP problem~\cite{PhysRevLett.38.1440, PhysRevD.16.1791, PhysRevLett.40.223, PhysRevLett.40.279}. These bosonic candidates could potentially be detected through absorption in some semiconducting material (such as the detectors that will be discussed in this thesis), where, e.g., the DM would excite an electron to the conduction band and the emitted photons or phonons could be detected after relaxation to the ground state.

At the very high mass end of DM candidates, there exist the possibility of primordial black holes (PBHs) as a DM candidate. These black holes could have formed in the early universe as some regions became so compressed that they experienced gravitational collapse to form black holes~\cite{Carr:1974nx}. PBHs could have a very large range of masses, and the observation of merging black holes with masses of $30 M_\odot$ by LIGO~\cite{PhysRevLett.116.061102} has directed interest to this order of magnitude for PBHs as DM~\cite{PhysRevD.94.083504, PhysRevLett.116.201301}. To search for PBHs, a commonly used method is gravitational microlensing, where a lensing object crosses the line of sight of a background star and creates time-varying magnification the star~\cite{1986ApJ...304....1P,1991ApJ...372L..79G,Niikura:2017zjd}. Furthermore, this method has been used to look for Massive Compact Halo Objects (MACHOs) as a DM candidate, for which microlensing searches for MACHOs set an upper limit of 8\% of the DM abundance at lower masses~\cite{Lasserre:2000xw, EROS-2:2006ryy}. At the mass scale of the LIGO black holes, studies of dwarf galaxies~\cite{Brandt:2016aco,DES:2016vji} and the CMB~\cite{Ali-Haimoud:2016mbv} have ruled these out as DM candidates. However, there were complex assumptions and astrophysics involved in these constraints, whereas microlensing searches are largely independent of these, providing some motivation to continue searching for PBHs/MACHOs at these mass scales.

There is always the possibility that DM is solely gravitationally interacting. In this sometimes called ``nightmare scenario,'' there do exist novel proposals for directly detecting DM by probing its gravitational signatures, from pulsar-timing array probes~\cite{Porayko:2018sfa} to an array of quantum-limited mechanical impulse sensors for Planck-scale DM~\cite{Carney:2019pza} to accelerometer networks~\cite{Figueroa_2021}. In many theoretical models of purely gravitationally-interacting DM, a common problem becomes that there is an overabundance of DM due to a lack of a mechanism for depleting the primordial DM~\cite{Kahn:2021ttr,MOROI1993289}, reiterating that having some coupling to Standard Model particles that is non-gravitational is well-motivated to achieve our current abundances.

\section{Direct Detection of Dark Matter}

When searching for DM, the different experimental detection methods are usually split up into three categories: direct detection, indirect detection, and DM production. While this section will focus on the principles of direct detection, the importance of indirect detection and DM production should not be dismissed. While direct detection experiments aim to detect the DM itself, indirect detection experiments attempt to detect Standard Model particles that are the products of the annihilations or decays of DM, such as cosmic rays, neutrinos, or photons. For DM production, the goal is use a high-energy particle collider, such as the LHC, to produce DM particles. These particles would escape the detector, which would leave a signature that there is an excess of events that have missing energy or momentum. These different methods, alongside astrophysical probes of the universe, work in tandem to search for DM, providing clear complementarity between them~\cite{BAUER201516}.


In direct detection, we have some arbitrary detector of volume $V$ and mass density $\rho_T$, which we place in some environment to hopefully interact with DM. By using Fermi's Golden Rule, we can calculate the generic scattering rate for DM per unit target mass for this arbitrary detector~\cite{Kahn:2021ttr, Trickle_2020} 
\begin{equation}
    R_\chi =  \frac{1}{\rho_T}\frac{\rho_\chi}{m_\chi} \int  \, d^3 \mathbf{v}  f_\chi(\mathbf{v}) \frac{V\, d^3 \mathbf{p}'_\chi}{(2\pi)^3} \sum_f |\langle f, \mathbf{p}_\chi' | \Delta H_{\chi T} | i, \mathbf{p}_\chi \rangle|^2 2 \pi \delta\left(E_f - E_i + E'_{\chi} - E_\chi\right),
    \label{eq:Rgeneral}
\end{equation}
where $\rho_\chi$ is the DM energy density, $m_\chi$ is the DM mass, $f_\chi(\mathbf{v})$ is the DM velocity distribution at the detector, $| i \rangle$ is the initial detector state with energy $E_i$, $| f \rangle$ is the final detector state with energy $E_f$, and $\Delta H_{\chi T}$ is the Hamiltonian describing DM-target interactions. We will assume that this interaction Hamiltonian is nonrelativistic, which follows from the astrophysical observations that DM is cold.

In order to simplify Eq.~(\ref{eq:Rgeneral}), we follow the formalism in Ref.~\cite{Kahn:2021ttr}. We make the following assumptions: the interaction Hamiltonian can be treated as a perturbation of the free-particle DM Hamiltonian (the unperturbed eigenstates are plane waves $| \mathbf{p} \rangle$), there is no entanglement between the target and the DM ($|a, \mathbf{p}_\chi\rangle \equiv |a\rangle \otimes |\mathbf{p}_\chi \rangle$, where $a$ can be $i$ or $f$), and that there is a single operator $\mathcal{O}$ that dominates the interaction Hamiltonian. With the assumption of no entanglement, we have that $\mathcal{O} = \mathcal{O}_\chi \otimes \mathcal{O}_T$, where $\mathcal{O}_\chi$ is the operator acting on the DM state and  $\mathcal{O}_T$ is the operator acting on the target state. Thus, we can factorize the matrix element into Fourier components $\mathbf{q}$, such that
\begin{align}
    \langle f, \mathbf{p}_\chi' | \Delta H_{\chi T} | i, \mathbf{p}_\chi \rangle &\equiv \! \int  \! \frac{d^3 \mathbf{q}}{(2\pi)^3} \, \langle \mathbf{p}_\chi' | \mathcal{O}_{\chi}(\mathbf{q}) | \mathbf{p}_\chi \rangle  \times \langle f | \mathcal{O}_{T}(\mathbf{q}) |i\rangle \\
	& = \frac{1}{V} \sqrt{ \frac{\pi \bar \sigma(q)}{\mu_\chi^2} } \langle f | \mathcal{O}_{T}(\mathbf{q}) |i\rangle,
	\label{eq:factorized_matrixelement}
\end{align}
where we have used our DM plane wave states and redefined the interaction potential in terms of a interaction potential strength $\bar \sigma (\mathbf{q})$ and a mass parameter $\mu_\chi$. Note that $\mathcal{O}_T$ could still depend on a DM model, e.g. the DM coupling strengths to Standard Model particles in the target system.

We can modify Eq.~(\ref{eq:Rgeneral}) further by using a dummy variable $\omega$, such that
\begin{equation}
    \delta \left(E_f - E_i + E'_{\chi} - E_\chi \right) = \int d\omega \, \delta\left(\omega + E'_{\chi} - E_\chi\right) \delta\left(E_f - E_i + \omega\right).
    \label{eq:deltarelat}
\end{equation}
Thus, we apply our relations in Eqs.~(\ref{eq:factorized_matrixelement}) and (\ref{eq:deltarelat}) to Eq.~(\ref{eq:Rgeneral}) to find a scattering rate of
\begin{equation}
    R_\chi =  \frac{1}{\rho_T}\frac{\rho_\chi}{m_\chi} \int  \, d^3 \mathbf{v}  f_\chi(\mathbf{v}) \int \frac{d^3 \mathbf{q}}{(2\pi)^3} \,  d\omega\, \delta\left( \omega + E'_{\chi} - E_\chi\right) \, \frac{\pi \bar \sigma(q)}{\mu_\chi^2}    \times S(\mathbf{q}, \omega),
    \label{eq:Rfactorized}
\end{equation}
where we define the dynamic structure factor
\begin{equation}
    S(\mathbf{q}, \omega) \equiv \frac{2\pi}{V} \sum_f |\langle f | \mathcal{O}_{T}(\mathbf{q}) |i\rangle|^2 \delta\left(E_f - E_i -\omega\right).
\end{equation}
The dynamic structure factor can be thought of as the response of the target material to the DM interaction, as it contains all of the dynamics of the target system (e.g. the electron or phonon excitations of the crystal).

Turning our attention to the kinematics of the DM scattering, we say there is incoming DM with momentum $\mathbf{p}_\chi = m_\chi \mathbf{v}$ that scatters off a detector target, exiting with momentum $\mathbf{p'}_\chi$. For nonrelativistic DM, the energy deposited in the target is
\begin{align}
    \omega_\mathbf{q} &= E_\chi - E'_\chi \nonumber \\
    &= \frac{1}{2} m_\chi v^2 - \frac{\left( m_\chi \mathbf{v} -\mathbf{q} \right)^2}{2 m_\chi} \label{eq:kinem} \\
    &= \mathbf{q} \cdot \mathbf{v} - \frac{q^2}{2 m_\chi}, \nonumber
\end{align}
where $E_\chi = p_\chi^2 / 2 m_\chi$ and $E'_\chi = {p'}_\chi^2 / 2 m_\chi$. Thus, Eq.~(\ref{eq:kinem}) defines the kinematically-allowed region of DM scattering. For given $\omega$ and $\mathbf{q}$, the minimum DM initial velocity required for scattering is then
\begin{equation}
    v_\mathrm{min} (q, \omega) = \frac{\omega_\mathbf{q}}{q} + \frac{q}{2 m_\chi}.
    \label{eq:vmin}
\end{equation}
We can take the isotropic approximation of Eq.~(\ref{eq:Rfactorized}) in order to see this minimum velocity explicitly. Carrying out the $\omega$ integral, the isotropic rate is
\begin{equation}
	R_{\chi}^\mathrm{iso} = \frac{1}{\rho_T}\frac{\rho_\chi}{m_\chi} \int \frac{q\, dq}{(2\pi)^{2}} \,  d\omega \, \eta(v_\mathrm{min}(q,\omega)) \times \,\frac{\pi \bar{\sigma}(q)}{\mu_\chi^2} \times S(q,\omega),
	\label{eq:riso}
\end{equation}
where $\eta(v_\mathrm{min}(q,\omega))$ is defined as
\begin{equation}
    \eta(v_\mathrm{min}) \equiv  \int_{v_\mathrm{min}}^\infty d^3 \mathbf{v} \, \frac{f_{\chi}(\mathbf{v})}{v}.
\end{equation}
In DM searches via annual modulation of the DM flux or using an anisotropic target system, the isotropic approximation will not hold, and proper care of these systems is required.

For the DM velocity distribution, it is common to use the Standard Halo Model (SHM), which describes the distribution as a truncated Maxwellian
\begin{equation}
    f_\mathrm{R}(\mathbf{v}) = \frac{1}{(2\pi \sigma_v^2)^{3/2} N_\mathrm{R, esc}}\exp{\left(-\frac{|\mathbf{v}|^2}{2\sigma_v^2}\right)} \Theta(v_\mathrm{esc} - |\mathbf{v}|),
\end{equation}
where $N_\mathrm{R, esc}$ normalizes this distribution to 1:
\begin{equation}
    N_\mathrm{R, esc} = \mathrm{erf}\left(\frac{v_\mathrm{esc}}{\sqrt{2} \sigma_v}\right)- \sqrt{\frac{2}{\pi}} \frac{v_\mathrm{esc}}{\sigma_v} {-\frac{v_\mathrm{esc}^2}{2\sigma_v^2}}.
\end{equation}
In the SHM, the values used are the galactic escape velocity $v_\mathrm{esc}$, the mean DM speed $v_0 = \sqrt{2} \sigma_v$, the local DM density $\rho_0$. As discussed in Ref.~\cite{Schnee:2011ooa}, it is also common to include a small correction taking the Earth's velocity in the galaxy $v_E$ into account, which will affect the tail end of the distribution. There are systematic errors associated with all of these values (most notably in $\rho_0$)~\cite{PhysRevD.99.023012,baxter2021recommended}, such that the recommended values have changed over time. In Table~\ref{tab:shmvalues}, we include values that are historically used (and have been used in the DM search that will be discussed in this thesis).

\begin{table}
    \centering
    \caption{Example of some galactic constants used in DM searches.}
    \begin{tabular}{lrr}
    \hline \hline
    Parameter & Value & Ref. \\ \hline
    $\rho_0 \, [\mathrm{GeV}/\mathrm{cm}^3]$ & $0.3$ & \cite{Gates__1995} \\ 
    $v_0\, [\mathrm{km}/\mathrm{s}]$ & $220$ & \cite{kerr} \\ 
    $v_E\, [\mathrm{km}/\mathrm{s}]$ & $232$ & \cite{Schnee:2011ooa} \\ 
    $v_\mathrm{esc}\, [\mathrm{km}/\mathrm{s}]$ & $544$ & \cite{rave} \\ 
    \hline \hline
    \end{tabular}
    \label{tab:shmvalues}
\end{table}

For DM-nucleon scattering, we can assume that the DM mass is large enough that the kinetic energy is greater than any displacement energy of the target, and its momentum is greater than any zero-point lattice momenta of the target. This is not always true, and as searches probe lower DM masses in the LDM range, much more care will need to be taken to ensure the expected differential rate is correct. With this assumption, we can treat the nuclear target as a free particle that is initially at rest. Using Eq.~(\ref{eq:kinem}) and the energy deposited being $\omega = E_R = q^2 / (2 m_N)$ for a nucleus $N$, we find that
\begin{equation}
    \mathbf{q} \cdot \mathbf{v} = \frac{q^2}{2 \mu_{\chi N}},
    \label{eq:nrelastic}
\end{equation}
where $\mu_{\chi N}$ is the DM-nucleus reduced mass. From Eq.~(\ref{eq:nrelastic}), we have that the maximum momentum transfer is $q_\mathrm{max} = 2 \mu_{\chi N} v$, giving a maximum nuclear recoil energy of $E_{R, \mathrm{max}} = q_\mathrm{max}^2/ (2m_N) = 2 \mu_{\chi N}^2 v^2 /m_N$. The best kinematic match occurs when $m_\chi \approx m_N$, and this energy transfer becomes quite inefficient when $m_\chi \ll m_N$. From Eq.~(\ref{eq:vmin}), we can calculate the minimum DM velocity needed for a nuclear recoil of energy $E_R$
\begin{equation}
    v_\mathrm{min} = \sqrt{\frac{m_N E_R}{2\mu_{\chi N}^2}}.
    \label{eq:vminelastic}
\end{equation}

\begin{figure}
    \centering
    \includegraphics[width=0.8\linewidth]{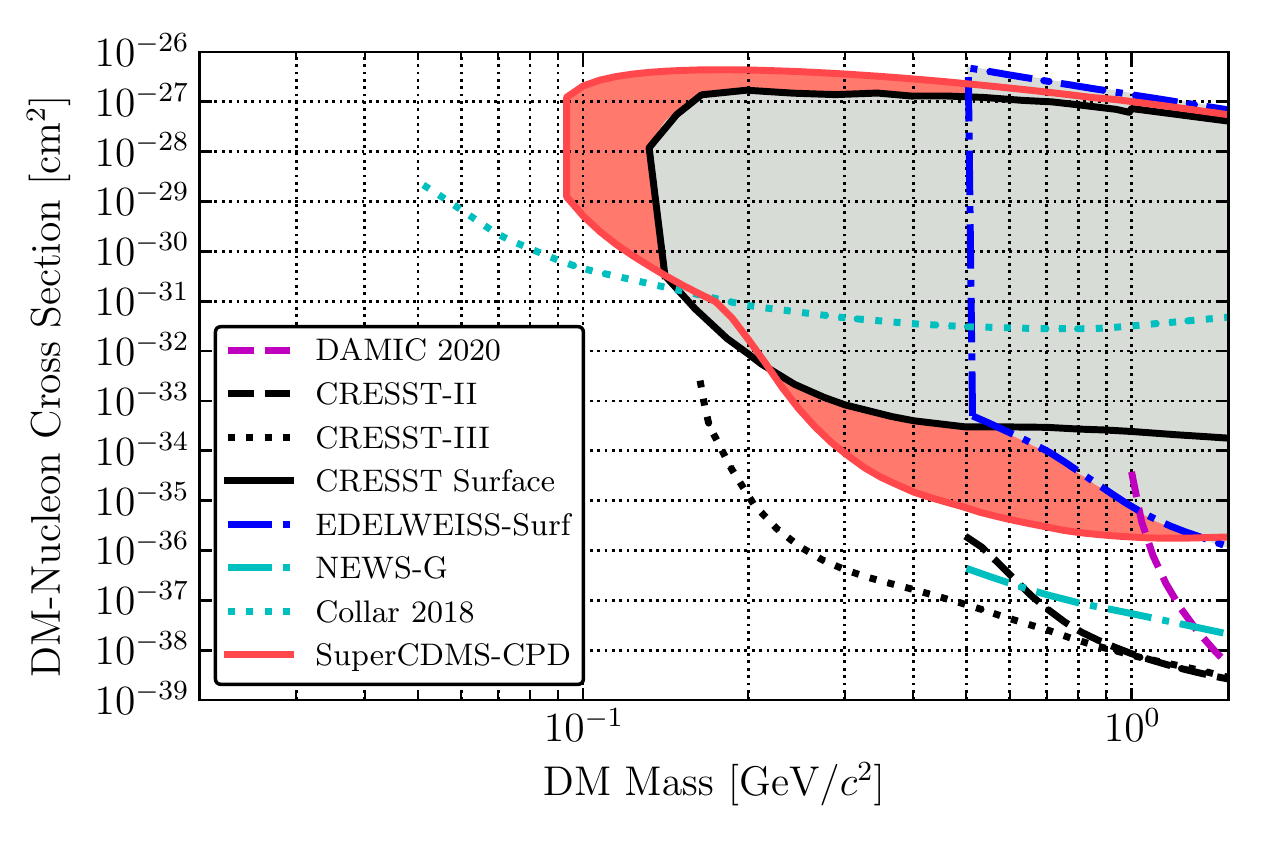}
    \caption{(Figure from Ref.~\cite{PhysRevLett.127.061801}) The landscape of constraints on sub-GeV scale LDM through the DM-nucleon scattering channel.}
    \label{fig:dmnucleonlimits}
\end{figure}

For elastic nuclear scattering, we can assume a contact potential between the DM and nucleon, such that the interaction Hamiltionian is
\begin{equation}
    \Delta H_{\chi T} =  \sqrt{ \frac{\pi \bar{\sigma}_{n}}{\mu^2_{\chi n} }} \int \frac{ d^{3} \mathbf{q}}{(2\pi)^{3}} \ e^{i \mathbf{q} \cdot (\mathbf{r}_{n} - \mathbf{r}_{\chi})},
    \label{eq:Hnuc_contact}
\end{equation}
where $\bar{\sigma}_n$ is the DM-nucleon scattering cross section and $\mu_{\chi n}$ is the DM-nucleon reduced mass (as opposed to the entire nucleus), assuming the same coupling to protons and neutrons. Summing over the target nuclei $N_\mathrm{nuc}$, the dynamic structure factor is
\begin{equation}
    S(\mathbf{q}, \omega) = \frac{ 2 \pi N_\mathrm{nuc}}{V} A^{2} |F_{N}(q)|^{2} \, \delta \! \left(\omega - \frac{q^2}{2 m_{N}} \right),
	\label{eq:Sqw_elasticNR}
\end{equation}
where $A$ is the mass number and $F_N$ is a nuclear form factor
\begin{equation}
	F_{N}(q) = \langle N | \frac{1}{A}  \sum_{\alpha = 1}^A e^{i \mathbf{q} \cdot \mathbf{r}_{\alpha}} | N \rangle.
	\label{eq:Fnuclear}
\end{equation}
For these elastic nuclear recoils, this nuclear form factor is commonly taken to be the Helm form factor~\cite{PhysRev.104.1466}. This form factor is relevant for WIMP DM, but is effectively unity for LDM. Using Eq.~(\ref{eq:riso}), we can calculate the isotropic rate and integrate over $\omega$ to arrive at the elastic nuclear recoil rate
\begin{equation}
    R_\chi = N_{T} \frac{\rho_\chi}{m_\chi} \frac{A^2\bar{\sigma}_n}{2\mu_{\chi n}^2} \int q \, dq \, \eta(v_\mathrm{min}(E_{R})) F^2_N(q),
    \label{eq:dRdERElastic}
\end{equation}
where $N_T = N_\mathrm{nuc}/(\rho_T V)$ is the number of target nuclei per detector mass. By making the substitution $q dq \longrightarrow dE_R m_N$, we arrive at the differential rate for DM-nucleon scattering
\begin{equation}
    \frac{\partial R_\chi}{\partial E_R} = m_{T} \frac{\rho_\chi}{m_\chi} \frac{A^2\bar{\sigma}_n}{2\mu_{\chi n}^2} \eta(v_\mathrm{min}(E_{R})) F^2_N(q),
    \label{eq:drde_derived}
\end{equation}
where $m_T$ is the mass of the target material. The $\eta(v_\mathrm{min})$ integral can be analytically solved, as done in Ref.~\cite{LEWIN199687}. We note that this function is exponential in nature (with some corrections from the cutoffs in the DM velocity distribution), with a characteristic form of
\begin{equation}
    \eta(v_\mathrm{min}) \sim \exp \left(-\frac{m_N E_R}{2\mu_{\chi N}^2 \sigma_v^2}\right).
\end{equation}
This differential rate spectrum has been used in many DM experiments to set limits on the DM-nucleon scattering cross section, with the ``current'' limits on LDM shown in Fig.~\ref{fig:dmnucleonlimits}.

\begin{figure}
    \begin{subfigure}{.5\textwidth}
        \centering
        \includegraphics[width=1\linewidth]{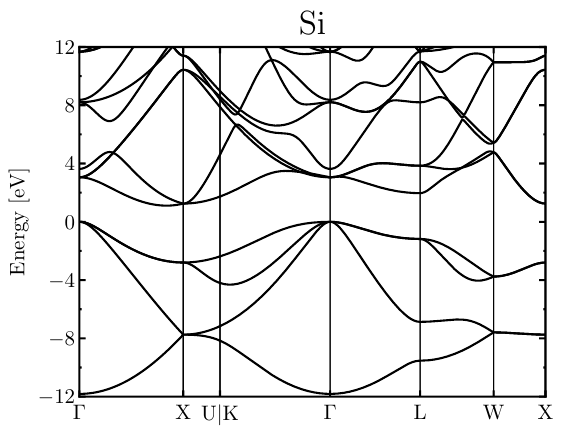}
    \end{subfigure}%
    \begin{subfigure}{.5\textwidth}
        \centering
        \includegraphics[width=1\linewidth]{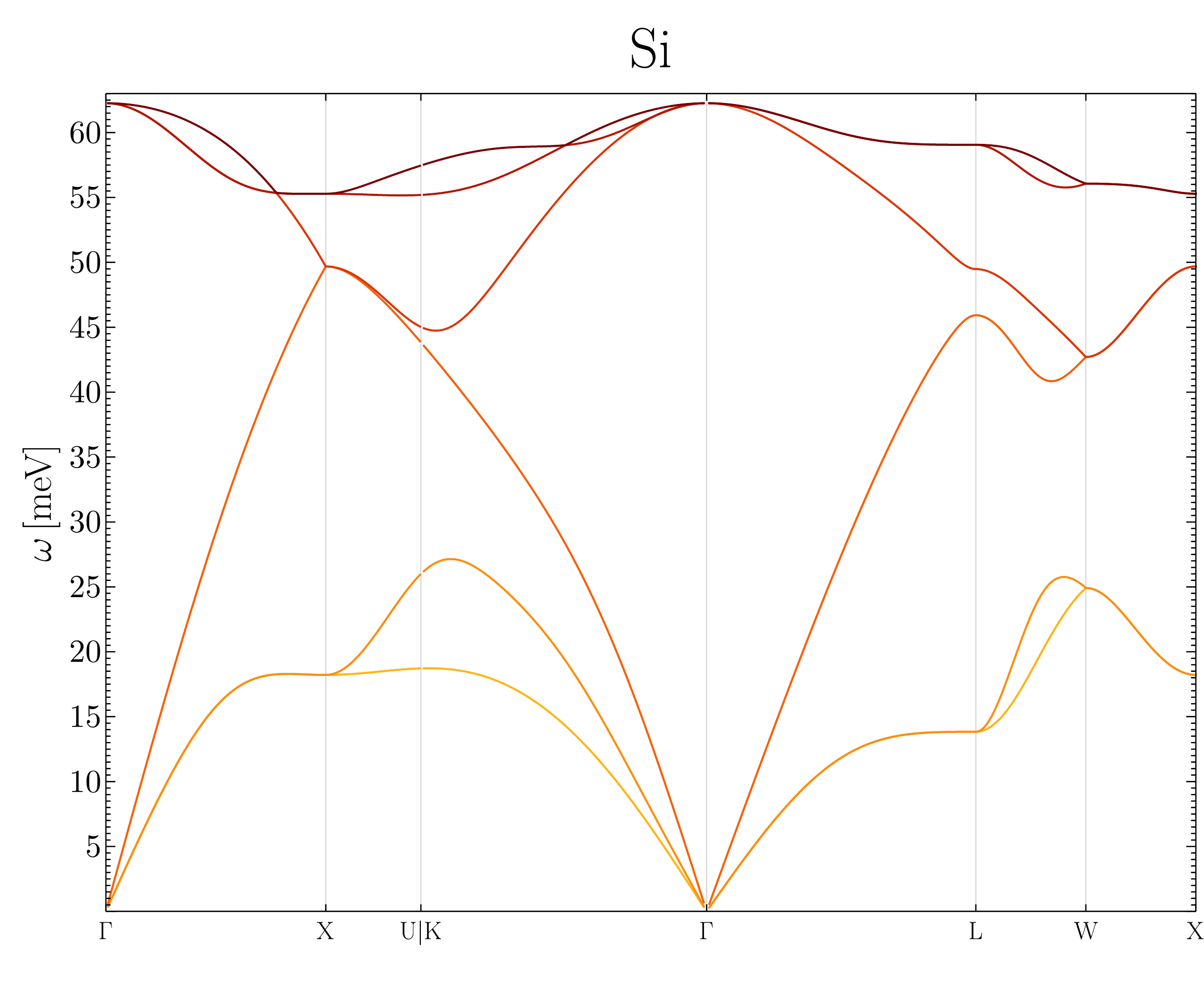}
    \end{subfigure}
    \caption{(Figure from Ref.~\cite{PhysRevD.101.055004}) (Left) Electronic band structure for Si. (Right) Phonon band structure for Si.}
    \label{fig:bandstructures}
\end{figure}

For DM-electron scattering and DM-nucleon scattering with phonon-scale recoil energies, the dynamic structure factor is no longer that of simple elastic nuclear scattering, as seen above. Instead, one must take into account the crystal structure and its wavefunctions in order to correctly take into account the expected response of the target material. In Fig.~\ref{fig:bandstructures}, we show example electron and phonon band structures for silicon. These band structures can be calculated from first principles via density functional theory (DFT). Common software to carry out these complex calculations are \textsc{VASP}~\cite{Hafner2008AbinitioSO}, \textsc{Quantum ESPRESSO}~\cite{Giannozzi_2009}, and \textsc{Phonopy}~\cite{TOGO20151}. The outputs from the DFT calculations must then be post-processed in order to calculate the dynamic structure factor for various DM models, which can be done using existing software such as \textsc{QEdark}~\cite{Essig:2015cda}, \textsc{QEdark-EFT}~\cite{PhysRevResearch.3.033149}, \textsc{EXCEED-DM}~\cite{Griffin:2021znd}, or \textsc{DarkELF}~\cite{Knapen:2021bwg,Campbell-Deem:2022fqm}.

\chapter{\label{chap:two}Athermal Phonon Sensors}

In this chapter, I will start with a brief discussion on the need for athermal phonon sensors, derive the small signal model frequently used when characterizing Transition-Edge Sensors (TESs), discuss the intrinsic noise sources of the system, and expand these concepts to Quasiparticle-trap-assisted Electrothermal-feedback Transition-edge-sensors (QETs) and the collection of athermal phonons.

\section{Athermal Calorimeters}

\begin{figure}
    \centering
    \includegraphics{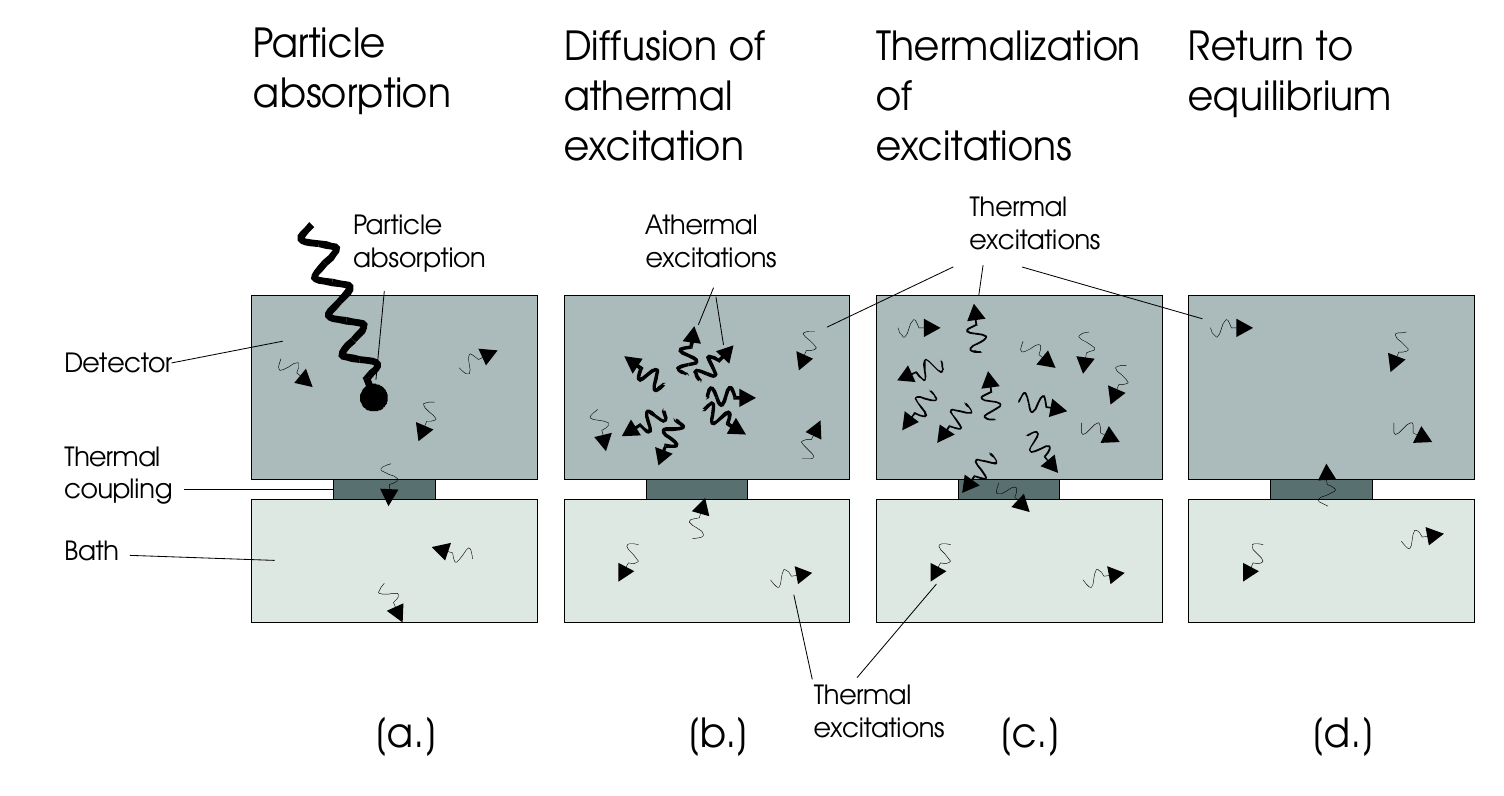}
    \caption{(Figure from Ref.~\cite{Lindeman_thesis}) (a) A particle is absorbed in a detector. (b) The absorption creates athermal excitations. (c) These athermal excitations begin to thermalize. (d) The detector returns to equilibrium.}
    \label{fig:thermcal}
\end{figure}

When a particle interacts with some detector, the standard steps that follow are that the detector absorbs some energy from the particle, this absorption creates athermal excitations in the detector (e.g. prompt phonons, ionization, photons), inelastic scattering thermalizes these excitations, and the detector cools back to equilibrium. These steps are shown diagrammatically in Fig.~\ref{fig:thermcal}, where the internal components of the detector are not shown (but would include some sensor and absorber).

\begin{figure}
    \centering
    \includegraphics{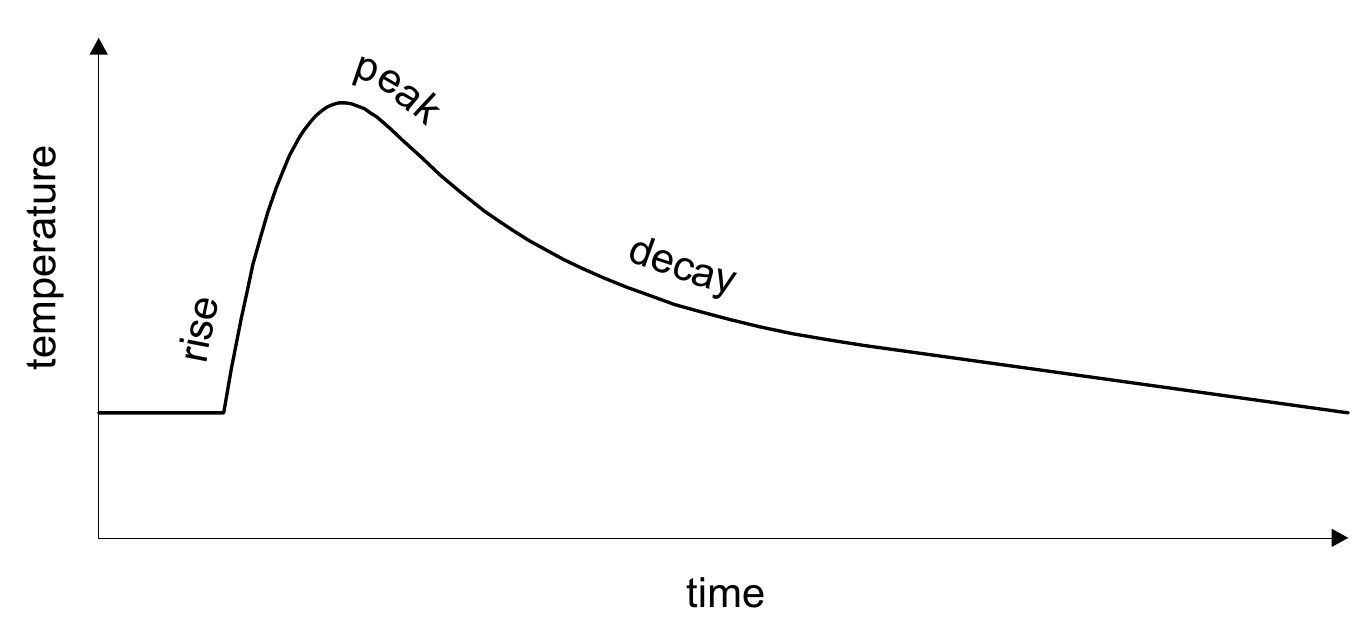}
    \caption{(Figure from Ref.~\cite{Lindeman_thesis}) Pulse behavior experienced by a sensor (or thermometer), where these a rise in temperature, a peak, and then a fall in temperature as it returns to equilibrium.}
    \label{fig:thermpulse}
\end{figure}

In this process, the detector target (absorber) has some heat capacity $C_{abs}$ which is connected to the thermal bath with a thermal conductance $G_{ab}$. When measuring thermal excitations, the sensor and the absorber should ideally have a strong thermal coupling, such that they act as a single thermal element. In Fig.~\ref{fig:thermpulse}, we show the pulse behavior that is experienced by the sensor (or thermometer). In this diagram, the temperature rises with some characteristic time constant and then falls with a different characteristic time constant. These time constants are determined by the thermalization time and $C_{abs}/G_{ab}$, where the rise is corresponds to the faster of the two, and the decay to the slower. In this system, the ideal energy sensitivity (or baseline energy resolution) is limited by thermal fluctuations across the thermal link to the bath, giving $\sigma_E^2 \approx K_BT^2C_{abs}$, implying that operating at low (cryogenic) temperatures will improve energy sensitivity.

The thermal conductance between the thermometer (sensor) and the absorber is driven in thin metal films by the electrons in the sensor being pushed far out of thermal equilibrium (the hot-electron effect~\cite{PhysRevB.49.5942}) and will scale as $T^4$. On the other hand, the thermal conductance between the absorber and the bath scales as $T^3$ from the elastic constant mismatch between dissimilar semiconducting substrates (known as phonon mismatch or Kapitza coupling)~\cite{doi:10.1139/p59-037}. Thus, we have the caveat that, as we decrease our temperature to improve energy sensitivity, it becomes difficult to thermally couple the thermometer and the absorber.

To get around this, one can use athermal calorimeters. That is, we put a sensitive sensor with fast response times that can measure the athermal excitations before they thermalize, such as a superconducting tunnel junction or a TES. In the case of TESs, this makes the limiting heat capacity that of the sensor and changes the pulse shape, where the time constants are determined by the athermal collection time (significantly faster than the thermalization time) and the sensor's thermal time constant (rather than being defined by the thermal time constant between the absorber and bath). TESs can successfully act as these sensors, especially when we expand the design to QETs, as will be discussed in the next sections.

\section{Transition-Edge Sensors}

\begin{figure}
    \centering
    \includegraphics[width=0.6\linewidth]{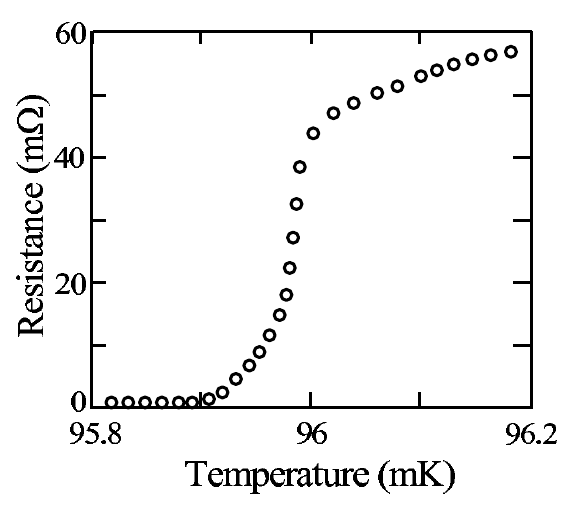}
    \caption{(Figure from Ref.~\cite{irwin}) A characteristic plot of the transition of a TES from its normal state to its superconducting state with a change in operating temperature.}
    \label{fig:sctransition}
\end{figure}

At its core, the idea behind a TES is quite simple. A TES is some metal (commonly W in SuperCDMS) which has been heated by a suitable current to operate within its superconducting transition. This superconducting transition is very sharp with widths of the order of $\mu \mathrm{K}$, with an example shown in Fig.~\ref{fig:sctransition}. Thus, we have some temperature-dependent resistor (or thermistor), which is highly sensitive to temperature perturbations. Furthermore, this thermistor will be sensitive as well to current perturbations (current supplies Joule heating), and its dynamics will be described by two nonlinear coupled differential equations. When operating these sensors in the voltage-biased limit, they become self-regulating through the idea of negative electrothermal feedback. When heat (e.g. some event hits the TES) increases the resistance of the TES, the Joule heating ($P_J = V_{TES}^2 / R_{TES})$ decreases---this decreases the temperature (or heat) of the TES, which then increases Joule heating, and we have that the TES self-regulates its temperature and returns to its steady-state. The alternative operating method of current-biasing leads instead to positive electrothermal feedback (where an increase in resistance instead increases the Joule heating, giving rise to thermal runaway), which made it historically difficult to keep TESs in this mode stable. First suggested by Irwin \textit{et al.}~\cite{QET}, TESs have since been exclusively run in the voltage-biased mode.

\begin{figure}
    \centering
    \includegraphics[width=0.6\linewidth]{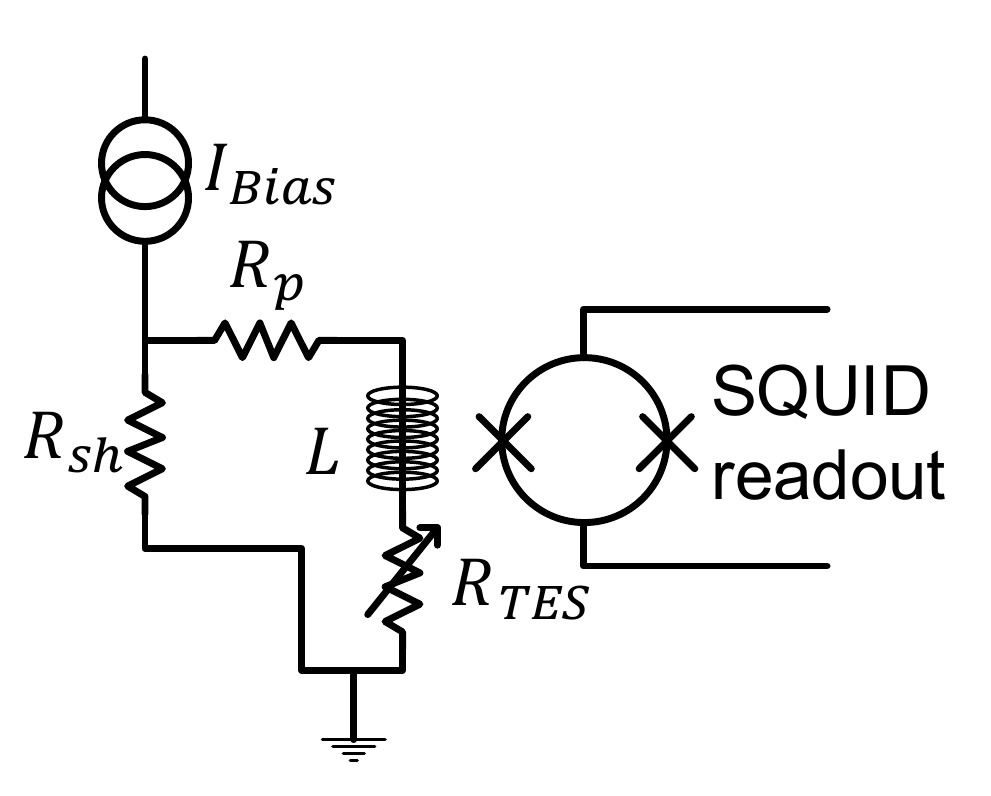}
    \caption{(Figure from Ref.~\cite{fink2020characterizing}) A schematic of the usual circuit for reading out changes in current of the TES.}
    \label{fig:tescircuit}
\end{figure}

\begin{figure}
    \centering
    \includegraphics[width=0.8\linewidth]{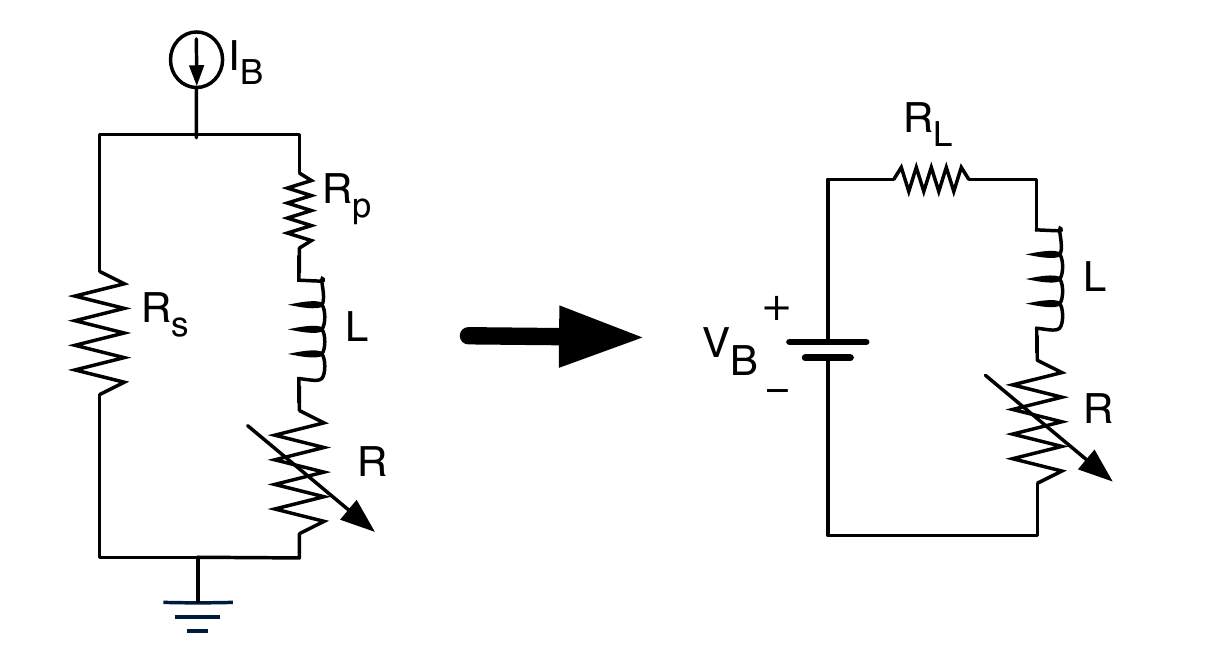}
    \caption{(Figure from Ref.~\cite{Matt_thesis}) Conversion of the circuit schematic (left) to the Th\'evenin equivalent circuit (right).}
    \label{fig:thevcirc}
\end{figure}

Because the transition is sharp, this means that any small temperature change will result in a large change in resistance. Thus, by reading out the current through the TES via a circuit, one can measure a current change, which we will see can be related to the energy of some event. Generally, the actual circuit used is of the form shown in Fig.~\ref{fig:tescircuit}, where the readout circuit is based on a Superconducting Quantum Interference Device (SQUID). To define a few terms in Fig.~\ref{fig:tescircuit}, we have some applied bias current $I_{Bias}$, a shunt resistor $R_{sh}$, some parasitic resistance $R_{p}$, an inductor with inductance $L$, and our TES with a variable resistance $R_{TES}$. With this circuit, it is common to approximate it as the Th\'evenin equivalent voltage-biased circuit, as shown in Fig.~\ref{fig:thevcirc}, as we are generally operating in the voltage-biased mode to take advantage of negative electrothermal feedback.

In the Th\'evenin equivalent circuit, we define the voltage bias $V_b = I_{Bias} R_{sh}$ and the load resistance $R_\ell \equiv R_p + R_{sh}$. The resulting equations are mathematically equivalent to those of the circuit with the bias current, but redefining the circuit in this way simplifies some of the algebra with these new definitions. From this circuit, we can use Kirckhoff's laws to come to a differential equation describing this circuit
\begin{equation}
    L \frac{\mathrm{d}I}{\mathrm{d}t} = V_b - I R_\ell - I R(T,I) + \delta V,
    \label{eq:circdiffeq}
\end{equation}
where we have explicitly shown that the TES resistance is a function of both TES current and temperature, i.e. $R_{TES}\equiv R(T,I)$, and $\delta V$ represents any change (assuming a first order expansion)  in bias voltage from internal (e.g. noise fluctuations) or external sources (e.g. a change in TES resistance due to an upstream-related change in the voltage bias).

The other differential equation comes from thermal considerations of our system. Our TES will be in contact with some thermal bath (or reservoir), which we take to have a heat capacity of effectively infinity. In other words, the heat capacity is so large that its temperature does not change with a reasonable amount of heat (i.e. an amount that we would expect some particle to impart given some recoil). Given this, we can represent the thermal interactions between the TES and the bath through its heat capacity:
\begin{equation}
    C \frac{\mathrm{d}T}{\mathrm{d}t} = P_J - P_{bath} + \delta P,
\end{equation}
where $P_J = I^2R(T,I)$ is the Joule heating, $P_{bath}$ is the power flowing from the TES to the thermal bath, and $\delta P$ represents any change (again assuming a first order expansion) in power due to internal (e.g. noise fluctuations) or external sources (e.g. energy imparted into the TES). Generally, the $P_{bath}$ term will follow a power law that is related to the temperature of the TES and the temperature of the bath. For metals with small volume and high power densities, the power flow $P_{bath}$ from the electrons to the phonons will dominate with the form $P_{bath} = K (T_c^n - T_{bath}^n)$, where $K=\Sigma_{ep} V_{TES}$ and $n=5$~\cite{irwin}. The coefficient $K$ is proportional to the volume of the TES by some material dependent constant which is usually on the order of $10^9\ \mathrm{W} \cdot \mathrm{m}^{-3} \cdot \mathrm{K}^{-5}$. Note that this exponent of $n = 5$ is specific to our assumptions on TES volume and power density. In practice, this exponent should be verified through a $G_{TA}$ measurement, as discussed in Ref.~\cite{fink2020characterizing} and further detailed in Appendix~\ref{chap:appa} of this thesis. Thus, our thermal nonlinear differential equation becomes
\begin{equation}
    C \frac{\mathrm{d}T}{\mathrm{d}t} = I^2R(T,I) - K (T^n - T_{bath}^n) + \delta P.
    \label{eq:thermaldiffeq}
\end{equation}

At this point, we have arrived at the two nonlinear differential equations that govern the simplest TES systems. Before investigating the response times of these systems, it is worth gaining intuition on the zero-frequency components, or DC characteristics.

\subsection{DC Characteristics}

When discussing DC characteristics of TESs, we return to Kirckhoff's laws as they apply to the circuit in Fig.~\ref{fig:thevcirc}. Taking the zero-frequency component of Eq.~(\ref{eq:circdiffeq}) (i.e. setting the time derivatives and the $\delta V$ terms to zero), the equation simplifies to
\begin{equation}
    V_b = I(R_\ell + R(T,I)),
    \label{eq:ohmslaw}
\end{equation}
which returns us to our well-known Ohm's law of $V=IR$. There are two natural extremes to take this equation to: the limit of superconducting resistance and the limit of normal resistance. Both of these limits provide (while seemingly basic) important insight on the DC characteristics of the TES.

Starting with the superconducting resistance regime (i.e. when the temperature of the TES is well-below its superconducting transition temperature and has zero resistance), it is important to choose (as will be apparent later when we derive the baseline energy resolution equations) metals with low superconducting transition temperatures, and conventional BCS (Bardeen-Cooper-Schrieffer)~\cite{PhysRev.108.1175} superconductors are commonly-used and preferred. In the superconducting regime of the TES circuit, the relation between the load resistance $R_\ell$ and the shunt resistance $R_{sh}$ is
\begin{equation}
    V_b = I_{Bias} R_{sh} = I R_\ell. 
\end{equation}
From an experimentalist's point of view, we should know $I_{Bias}$ (our applied current bias to the circuit) and $R_{sh}$ (the shunt resistor which we have installed on the circuit), and we can use this information to extract the parasitic resistance $R_p$ with a strong degree of numerical certainty.

Continuing to the normal resistance regime (i.e. when the temperature of the TES is well-above its superconducting transition temperature), we have that the TES can be approximated to first order as having a constant resistance $R_N$. Generally, it is not true that normal resistance is constant for a metal (or other materials) with temperature from $T_c$ to $\infty$. However, it is a very good approximation at cryogenic temperatures for metals. When operating a TES above its transition at, e.g., $50 \ \mathrm{mK}$ the change in resistance is quite negligible at $100 \ \mathrm{mK}$. For the normal regime, we have that Ohm's law gives
\begin{equation}
    V_b = I (R_\ell + R_N),
\end{equation}
where $R_N$ is the normal resistance of the TES near its superconducting transition. This is again a linear relationship between the voltage bias and the current, both of which one will know in their experimental setup. Thus, the two limits give a strong reason to vary the voltage bias and observe the change in current, as this will provide direct insight on the normal resistance and the parasitic resistance.

For the approximation of TES dynamics, it is the transition region that is of upmost importance for science results because of its sensitivity to temperature fluctuations. Historically, one could voltage-bias a TES in transition and achieve a competitive result due to the novelty of the technology. Given the technology's maturity at the time of this thesis, one must further understand the mathematical model of the TES to begin to optimize its performance, and this begins with an understanding of the $IV$ curve. In experimentally-specific terms, this is a measurement of how the variation of the bias current $I_{Bias}$ relates to the measured current through the TES $I_{TES} \equiv I$ via, e.g., the SQUID array. Importantly, our understanding within the superconducting transition comes from both $P_{bath}$ and the minuscule $\mathcal{O}(\mu \mathrm{K})$ width of the superconducting transition. In other words, we can see that this small width allows us to approximately treat the equilibrium bias power ($P_0=P_{bath}$) as a constant. Thus, within the transition with our constant bias power, we can use the voltage-current product law to give
\begin{equation}
    V_b = P_{0} / I + I R_\ell,
\end{equation}
which effectively follows a $1/x$ curve (in the usual case of $R_\ell \ll R_{TES}$).

\subsubsection{Measuring TES DC Characteristics}

\begin{figure}
    \centering
    \includegraphics[width=0.7\linewidth]{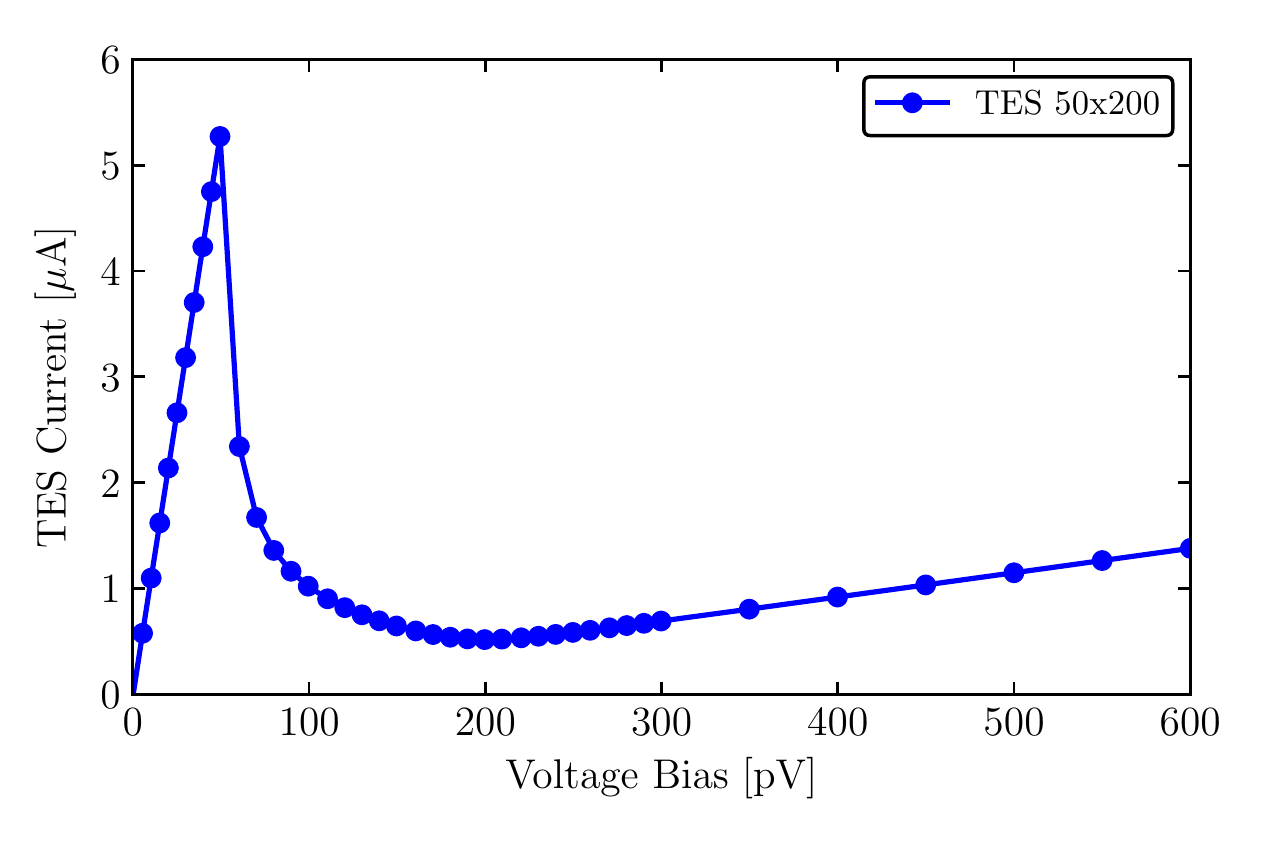}
    \caption{IV curve for a rectangular W TES of dimensions $200 \ \mu \mathrm{m} \times 50 \ \mu \mathrm{m} \times 40 \ \mathrm{nm}$.}
    \label{fig:ivcurve}
\end{figure}

Going through the above equations, we now have an expectation of how the current through the TES would change as we change the voltage bias: linearly in the normal region, approximately as $1/V_b$ within the transition to superconducting, and linearly in the superconducting region. In Fig.~\ref{fig:ivcurve}, we show this relation for a rectangular W TES of dimensions $200 \ \mu \mathrm{m} \times 50 \ \mu \mathrm{m} \times 40 \ \mathrm{nm}$. Each marker in the figure denotes a measurement of the TES current, while the connecting lines are simple linear interpolations between neighboring points. In practice, when one is measuring the current via a SQUID array, the measured current is only relative, not absolute (i.e. we can only measure changes in current). Because of this, one has to correct for the overall current offset by taking advantage of the linear normal and superconducting regions. In particular, one can linearly fit these regions, find the $y$-intercepts, and correct the IV curve such that both intercepts are zero. It is useful to do this linear fit for both of these regions, as this can correct for the possibility of some unknown offset in voltage bias (e.g. if there a voltage jitter sent down the TES bias line that has a small DC offset which adds to the current bias $I_{Bias}$).

\begin{figure}
    \centering
    \includegraphics[width=0.7\linewidth]{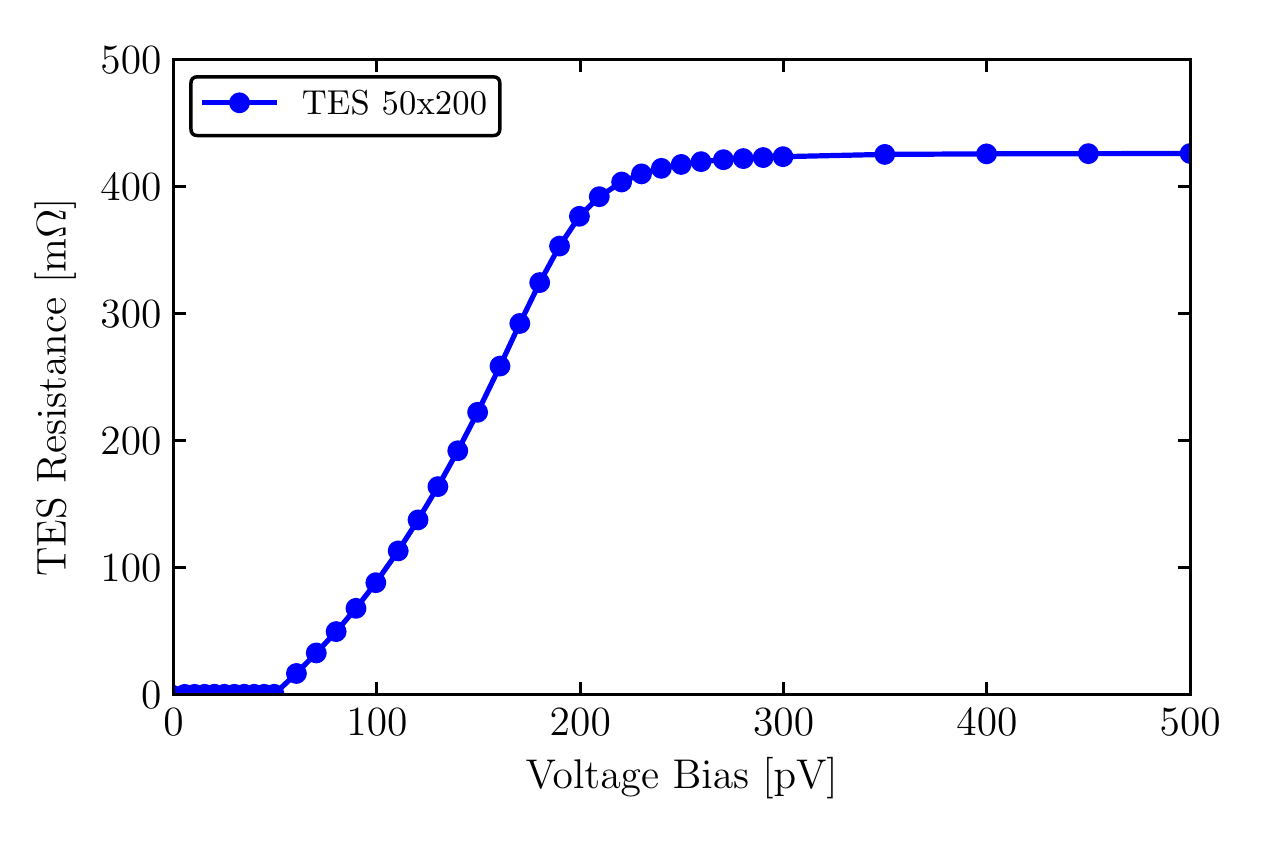}
    \caption{Resistance as a function of voltage bias for a rectangular W TES of dimensions $200 \ \mu \mathrm{m} \times 50 \ \mu \mathrm{m} \times 40 \ \mathrm{nm}$.}
    \label{fig:rvcurve}
\end{figure}

\begin{figure}
    \centering
    \includegraphics[width=0.7\linewidth]{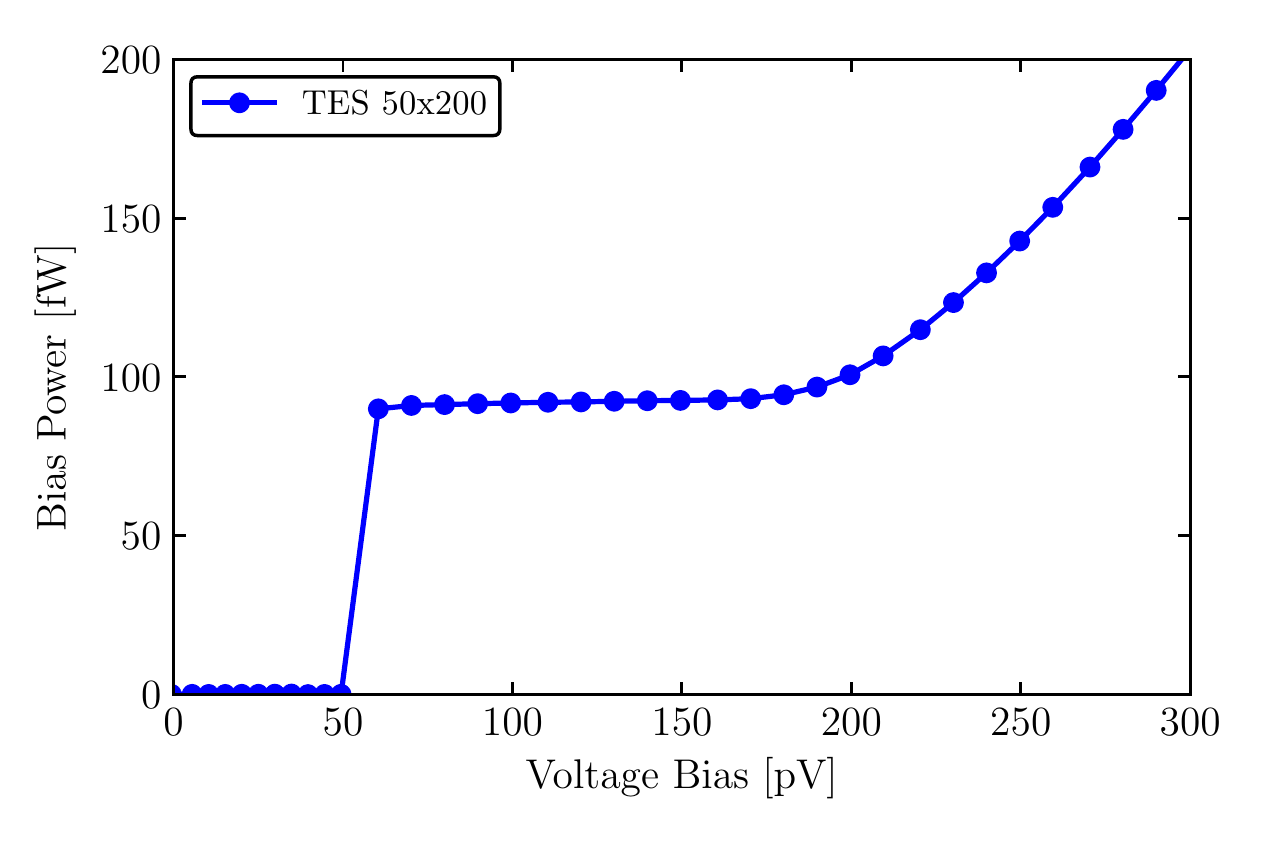}
    \caption{Bias power as a function of voltage bias for a rectangular W TES of dimensions $200 \ \mu \mathrm{m} \times 50 \ \mu \mathrm{m} \times 40 \ \mathrm{nm}$. In transition, the bias power is flat due to the sharpness of the superconducting transition in temperature and stabilized through electrothermal feedback.}
    \label{fig:pvcurve}
\end{figure}

Returning to Eq.~(\ref{eq:ohmslaw}), we can (given that we have already measured the load resistance from the superconducting region) rearrange the equation to give us both the TES resistance and the bias power as a function of voltage bias:
\begin{align}
    R_{TES} &= \frac{V_b}{I} - R_\ell, \\
    P_0 &= I V_b - I^2 R_\ell .
\end{align}
Thus, it is simple to convert our $IV$ curve in Fig.~\ref{fig:ivcurve} to curves of resistance or bias power as a function of voltage bias, as shown in Figs.~\ref{fig:rvcurve} and \ref{fig:pvcurve}. We can then use our knowledge of our DC characteristics at various bias points as part of the goal of understanding the expected performance of the TES. However, to fully understand the expected performance, one must also understand the expected time-varying response of the TES.

\subsection{\label{sec:tesresponse}TES Response}

To understand the expected response of the TES to some Dirac delta deposit of energy, we need to solve our nonlinear coupled differential equations in Eqs.~(\ref{eq:circdiffeq}) and (\ref{eq:thermaldiffeq}), which are not analytically solvable in their nonlinear form. However, we can Taylor expand about the equilibrium point assuming that the changes in current and temperature are small, i.e. we are making a small signal approximation. In making this approximation, we will need to make a series of parameter definitions, as done by Irwin and Hilton~\cite{irwin}.

Starting with the power flowing to the bath, we have that $P_{bath} = K (T^n - T_{bath}^n)$. Expanding this about some equilibrium temperature $T_0$, we have that
\begin{equation}
    P_{bath} \approx P_{bath_0} + G \delta T,
\end{equation}
where $G\equiv \mathrm{d}P_{bath} / \mathrm{d}T = nKT^{n-1}$ is the thermal conductance from TES to bath and $\delta T = T-T_0$. Next, we expand the TES resistance about its steady state values in resistance, temperature, and current
\begin{equation}
    R(T,I) \approx R_0 + \frac{\partial R}{\partial T} \bigg|_{I_0} \delta T + \frac{\partial R}{\partial I}\bigg|_{T_0} \delta I.
\end{equation}
The above equation is a somewhat messy to use in a derivation due to the partial derivatives, and it is convention to define the unitless temperature sensitivity
\begin{equation}
    \alpha \equiv \frac{\partial \log R}{\partial \log T} \bigg|_{I_0} = \frac{T_0}{R_0} \frac{\partial R}{\partial T}\bigg|_{I_0},
\end{equation}
and the unitless current sensitivity
\begin{equation}
    \beta \equiv \frac{\partial \log R}{\partial \log I} \bigg|_{T_0} = \frac{I_0}{R_0} \frac{\partial R}{\partial I}\bigg|_{T_0},
\end{equation}
which rearranges the TES resistance equation to
\begin{equation}
    R(T,I) \approx R_0 + \alpha \frac{R_0}{T_0} \delta T + \beta \frac{R_0}{I_0} \delta I.
\end{equation}
Next, we expand our Joule power to first order about the steady state values, such that
\begin{equation}
    P_J = I^2 R(T,I) \approx P_{J_0} + 2 I_0 R_0 \delta I + \alpha \frac{P_{J_0}}{T_0} \delta T + \beta \frac{P_{J_0}}{I_0} \delta I.
\end{equation}
We also define the (unitless) loop gain
\begin{equation}
    \mathscr{L} \equiv \frac{P_{J_0} \alpha}{G T_0},
\end{equation}
and the natural thermal time constant
\begin{equation}
    \tau_0 \equiv \frac{C}{G}.
\end{equation}
Thus, we can substitute all the above expansions and definitions into Eqs.~(\ref{eq:circdiffeq}) and (\ref{eq:thermaldiffeq}), giving us the linearized versions
\begin{align}
    \frac{\mathrm{d} \delta I}{\mathrm{d}t} &= -\frac{R_\ell + R_0 (1 + \beta)}{L} \delta I - \frac{\mathscr{L}G}{I_0 L} \delta T + \frac{\delta V}{L}, \\
    \frac{\mathrm{d} \delta T}{\mathrm{d}t} &= \frac{I_0 R_0 (2 + \beta)}{C} \delta I - \frac{1-\mathscr{L}}{\tau_0} \delta T + \frac{\delta P}{C}.
\end{align}
One method to solve these two differential equations is to switch to frequency space through a Fourier transform. As convention is always up in the air, we define transform pair (for some function $g$) for the continuous case as
\begin{align}
    \tilde{g}(f) &= \int_{-\infty}^\infty \mathop{dt} g(t) \mathrm{e}^{-i \omega t}, \\
    g(t) &= \int_{-\infty}^\infty \mathop{df} \tilde{g}(f) \mathrm{e}^{i \omega t},
\end{align}
where $\omega \equiv 2 \pi f$ for convenience, and for the discrete case as
\begin{align}
    \tilde{g}_n &= \frac{1}{N} \sum_{k = -N/2}^{N/2 - 1}g_k  \mathrm{e}^{-i \omega_n t_k}, \\
    g_k  &= \sum_{n = -N/2}^{N/2 - 1} \tilde{g}_n \mathrm{e}^{-i \omega_n t_k}.
\end{align}
In the context of our differential equations (using the continuous case), we have that $\mathrm{d}/\mathrm{d}t \rightarrow i \omega$ (i.e. the Fourier transform of the derivative of some function $f$ is equal to $i\omega$ times the Fourier transform of $f$). Thus, these equations become
\begin{align}
    \mathop{i \omega} \delta I &= -\frac{R_\ell + R_0 (1 + \beta)}{L} \delta I - \frac{\mathscr{L}G}{I_0 L} \delta T + \frac{\delta V}{L}, \\
    \mathop{i \omega} \delta T &= \frac{I_0 R_0 (2 + \beta)}{C} \delta I - \frac{1-\mathscr{L}}{\tau_0} \delta T + \frac{\delta P}{C}.
\end{align}
At this point, this system of equations can be put into matrix form, which we have done below after rearranging some of the terms
\begin{equation}
    \begin{bmatrix}
        i\omega + \frac{R_\ell + R_0 (1 + \beta)}{L} & \frac{\mathscr{L}G}{I_0 L}\\[6pt]
         - \frac{I_0 R_0 (2 + \beta)}{C} & i\omega + \frac{1-\mathscr{L}}{\tau_0}
    \end{bmatrix}
    \begin{bmatrix}
        \delta I\\[6pt]
        \delta T
    \end{bmatrix}
    =
    \begin{bmatrix}
        \frac{\delta V}{L} \\[6pt]
        \frac{\delta P}{C}
    \end{bmatrix}
    ,
\end{equation}
and we define the matrix at this point as $M$. Multiplying by the inverse of $M$, we can then come to our expected current and temperature responses to small power or voltage excitations
\begin{equation}
    \begin{bmatrix}
        \delta I\\[6pt]
        \delta T
    \end{bmatrix}
    =
    M^{-1}
    \begin{bmatrix}
        \frac{\delta V}{L} \\[6pt]
        \frac{\delta P}{C}
    \end{bmatrix}
    ,
    \label{eq:tesmatrix}
\end{equation}
where the inverse of the matrix is
\begin{equation}
    M^{-1}
    =
    \frac{1}{ \begin{vmatrix}
       M
    \end{vmatrix}}
    \begin{bmatrix}
         i\omega + \frac{1-\mathscr{L}}{\tau_0} & -\frac{\mathscr{L}G}{I_0 L}\\[6pt]
         \frac{I_0 R_0 (2 + \beta)}{C} & i\omega + \frac{R_\ell + R_0 (1 + \beta)}{L}
    \end{bmatrix}
\end{equation}
and
\begin{equation}
    \begin{vmatrix}
       M
    \end{vmatrix}
    =
    \left[i\omega + \frac{R_\ell + R_0 (1 + \beta)}{L}\right] \left[i\omega + \frac{(1-\mathscr{L})}{\tau_0}\right] + \frac{\mathscr{L}G}{L}\frac{R_0 (2 + \beta)}{C}.
\end{equation}

After going through the algebra, we have finally arrived at our desired form. From Eq.~(\ref{eq:tesmatrix}), we can extract two important relations for characterizing TESs: the complex admittance of the TES circuit $\partial I / \partial V$ and the power-to-current transfer function $\partial I / \partial P$. Starting with the complex admittance, we will assume that there is only a change in voltage $\delta V$ (i.e. we are setting external signal power $\delta P=0$). Thus, we can easily find the complex admittance, as it is related by a single matrix element
\begin{equation}
    \frac{\partial I}{\partial V} \equiv \frac{\delta I}{\delta V} = \frac{\left(M^{-1}\right)_{11}}{L}.
\end{equation}
It can be shown that this form simplifies to
\begin{equation}
    \frac{\partial I}{\partial V} (\omega) = \frac{1}{R_\ell + i\omega L + \underbrace{R_0 (1 + \beta) + \frac{R_0 \mathscr{L}}{1 - \mathscr{L}} \frac{2 + \beta}{1 + i \omega \tau_0 / (1 - \mathscr{L})}}_{Z_{TES}(\omega)}} = \frac{1}{Z_{circ}(\omega)}
\end{equation}
where we have defined the circuit complex impedance as
\begin{equation}
    Z_{circ}(\omega) \equiv R_\ell + i\omega L + Z_{TES}(\omega),
    \label{eq:zcirc}
\end{equation}
and the TES complex impedance as
\begin{equation}
    Z_{TES}(\omega) \equiv R_0 (1 + \beta) + \frac{R_0 \mathscr{L}}{1 - \mathscr{L}} \frac{2 + \beta}{1 + i \omega \frac{\tau_0}{1 - \mathscr{L}}}.
    \label{eq:ztes}
\end{equation}
Note that the circuit complex impedance is what we work with, as in practice we are reading out all of the load resistor, inductor, and TES in series, rather than solely the TES.

For the power-to-current transfer function (also frequently referred to as the responsivity), the steps are nearly identical, but we set $\delta V = 0$ and keep $\delta P$ nonzero. Thus, we find that the power-to-current transfer function becomes
\begin{equation}
    \frac{\partial I}{\partial P} \equiv \frac{\delta I}{\delta P} = \frac{\left(M^{-1}\right)_{12}}{C}.
\end{equation}
Here we note that the denominator will be exactly the same as it was for the complex admittance, as it is set by the determinant of our original matrix $M$. In this way, it is clear that there is a strong relationship between the complex admittance and the power-to-current transfer function. In fact, once we know the complex admittance, then we can fully define the power-to-current transfer function. A useful form of the power-to-current transfer function is
\begin{equation}
    \frac{\partial I}{\partial P}(\omega) = - \frac{1}{Z_{circ}(\omega) I_0} \frac{Z_{TES}(\omega) - R_0 (1 + \beta)}{R_0(2 + \beta)},
    \label{eq:didp}
\end{equation}
from which we see the ease of which we can calculate the power-to-current transfer function from the complex impedance (the reciprocal of the complex admittance). This functional form applies to more complex TES models with multiple thermal bodies, as used by Maasilta in Ref.~\cite{maasilta} and generally derived by Lindeman \textit{et al.} in Ref~\cite{lindeman_nep} (in which it is shown that the $\partial I / \partial P$ is determined by the $\partial I / \partial V$ without an assumption on the complexity of the thermal modes).

With the knowledge that the complex admittance and power-to-current transfer function have the same frequency dependence (i.e. have the same poles), we can solve for these poles, which will give us the characteristic time constants of the system. One method to do this is to use that the determinant of a matrix is equal to the product of its eigenvalues, giving us a simple method to rearranging the denominator of these functions. Doing this, we have that
\begin{equation}
    \frac{\partial I}{\partial V} (\omega) \propto \frac{1}{(i \omega + \lambda_+)(i \omega + \lambda_-)},
\end{equation}
where the eigenvalues are
\begin{equation}
    \lambda_\pm \equiv \frac{1}{\tau_\pm} = \frac{1}{2}\left[ \frac{1}{\tau_{el}} + \frac{1}{\tau_I} \pm \sqrt{\left( \frac{1}{\tau_{el}} - \frac{1}{\tau_I} \right)^2 - 4 \frac{\mathscr{L} R_0 (2+\beta)}{L\tau_0}}\right],
    \label{eq:eigenvals}
\end{equation}
and we have defined~\cite{irwin}
\begin{align}
    \tau_{el} &\equiv \frac{L}{R_\ell + R_0 (1 + \beta)}, \\
    \tau_I &\equiv  \frac{\tau_0}{1 - \mathscr{L}}.
\end{align}
In these definitions, $\tau_{el}$ serves as the electronic time constant of the system, while $\tau_I$ is usually thought of as the constant-current time constant. When voltage-biasing, $\tau_I$ is not an intrinsic time constant as we are not operating with constant current, and we do not worry about negative values when $\mathscr{L}>1$. If we were current-biased, then this would be an intrinsic time constant, and $\mathscr{L}>1$ (leading to a negative $\tau_I$) represents the thermal runaway experienced from positive electrothermal feedback, severely limiting the allowed values of $\mathscr{L}$ to less than 1. On the other hand, if a sensor's detection sensitivity is based on, e.g., the hopping conduction mechanism, such as neutron transmutation doped (NTD) Ge, then the temperature sensitivity $\alpha$ is negative, and current-biasing is the stable operating mode~\cite{PhysRevB.41.3761}.

We have our relationship between the TES small-signal parameters and the physical time constants $\tau_\pm$, again noting that they are the same time constants for both the complex admittance and the power-to-current transfer function. These time constants simplify greatly when we take the limit as the inductance $L$ approaches zero (the low-inductance limit, i.e. $\tau_+ \ll \tau_-$), which allows us to have a little more intuition on their dependence on the various TES parameters than if we simply look at Eq.~(\ref{eq:eigenvals}). Taking this limit for both time constants, we have the following definitions
\begin{align}
    \tau_{el} &\equiv \lim_{L \to 0} \tau_+ = \frac{L}{R_\ell + R_0 (1 + \beta)}, \\
    \tau_{eff} &\equiv \lim_{L \to 0} \tau_- = \tau_0 \frac{1 + \beta + R_\ell / R_0}{1 + \beta + R_\ell / R_0 + (1 - R_\ell / R_0) \mathscr{L}},
\end{align}
where $\tau_{el}$ is the electrical time constant and $\tau_{eff}$ is the zero-inductance effective thermal time constant. The effective time constant can be taken to another limit
\begin{equation}
    \lim \limits_{\substack{\beta \to 0 \\ R_\ell \ll R_0}} \tau_{eff} = \frac{\tau_0}{1 + \mathscr{L}},
\end{equation}
which implies that to achieve fast sensor time constants, large loop gains $\mathscr{L}$ are needed (which equivalently means large temperature sensitivity $\alpha$). Comparing to current-biased TESs with the constraint of $\mathscr{L} < 1$, their time constants are significantly slower than the voltage-biased counterparts, leading to a worse baseline energy resolution (this relationship is shown at the end of Section~\ref{sec:noisemodeling}).

\subsubsection{Measuring TES Response}

With our expected TES response understood, the next step is to experimentally measure and fit the TES response for characterization. This can been done by injecting white noise and convolving the input and response~\cite{doi:10.1063/1.1711144}. However, most TES groups have since switched over to either sine wave sweeps or measuring square wave responses. The idea here is that we can bias the TES with some voltage bias and then add on top of the voltage bias some repeating signal (sine wave or square wave) and then deconvolve the measured response to extract the measured complex admittance. This is generally done in Fourier space, as the Fourier transform of a convolution becomes a simple multiplication of the two functions' Fourier transforms. Let's define some repeating signal $V_{s} (t)$ with Fourier transform $\tilde{V}_s (f)$ which is convolved with the complex admittance to return the TES response $I_{TES} (t)$ (or $\tilde{I}_{TES} (f)$ in frequency domain. We can relate each of these values through the convolution theorem
\begin{equation}
    \tilde{V}_s (f) \frac{\partial I}{\partial V} (\omega) = \tilde{I}_{TES} (f),
\end{equation}
remembering that $\omega = 2 \pi f$. We divide by the Fourier transform of the signal, and we have that
\begin{equation}
    \frac{\partial I}{\partial V} (\omega) = \frac{\tilde{I}_{TES} (f)}{\tilde{V}_s (f)}.
    \label{eq:convthm}
\end{equation}
Thus, we can directly measure the complex admittance by measuring its time-domain response to some known repeating signal, which we can then fit to the small-signal model to extract all of our TES parameters. Practically, it is much faster to use a single square wave as far as data taking, as opposed to a single sine wave. This is because the Fourier transform of a sine wave returns Dirac delta functions at the sine wave frequency (and its negative frequency), whereas the Fourier transform of a square wave is defined across many discrete frequencies. For a single sine wave, we would have to take data over a sweep of frequencies, which can be quite slow when going from, e.g, 1 Hz to 10 kHz. To remedy this, one could use multiple overlapping sine waves, but calibrating the phase becomes difficult as compared to the well-defined start time from a single square wave.

To extract the measured complex admittance when using a square wave, we can start with the Fourier series of a square wave with frequency $f_{sg}$ and peak-to-peak amplitude in voltage bias $A_{sg}$
\begin{equation}
    V_s (t) = A_{sg} \frac{2}{\pi} \sum_{n = 1, 3, 5,...}^\infty \frac{\sin\left( 2 \pi n f_{sg} t \right)}{n}.
\end{equation}
Taking the (discrete) Fourier transform and using Eq.~(\ref{eq:convthm}), we have that, for each frequency that is an odd integer multiple of the square wave frequency, the measured complex admittance is
\begin{equation}
    \frac{\partial I}{\partial V} (\omega) = \begin{cases}
    \frac{\tilde{I}_{TES} (f)}{A_{sg} \frac{2}{n \pi}} & \omega=2\pi n f_{sg} \ \mathrm{for} \ n = 1, 3, 5,...\\
    0 & \mathrm{otherwise}
    \end{cases},
\end{equation}
and we can measure the complex admittance over a large range of frequencies in a single measurement. In practice, we will measure the response of the TES in time domain many times, select data that does not have external backgrounds (e.g. pulses, electronic glitches), Fourier transform and deconvolve the data, and then average the measure complex admittances (keeping track of the standard error of the mean for error propagation). This workflow has been written in \texttt{Python} as one of the many features of the \textsc{QETpy} package, co-created and co-maintained by C. W. Fink and myself~\cite{qetpy}.

\begin{figure}
    \centering
    \includegraphics[width=0.7\linewidth]{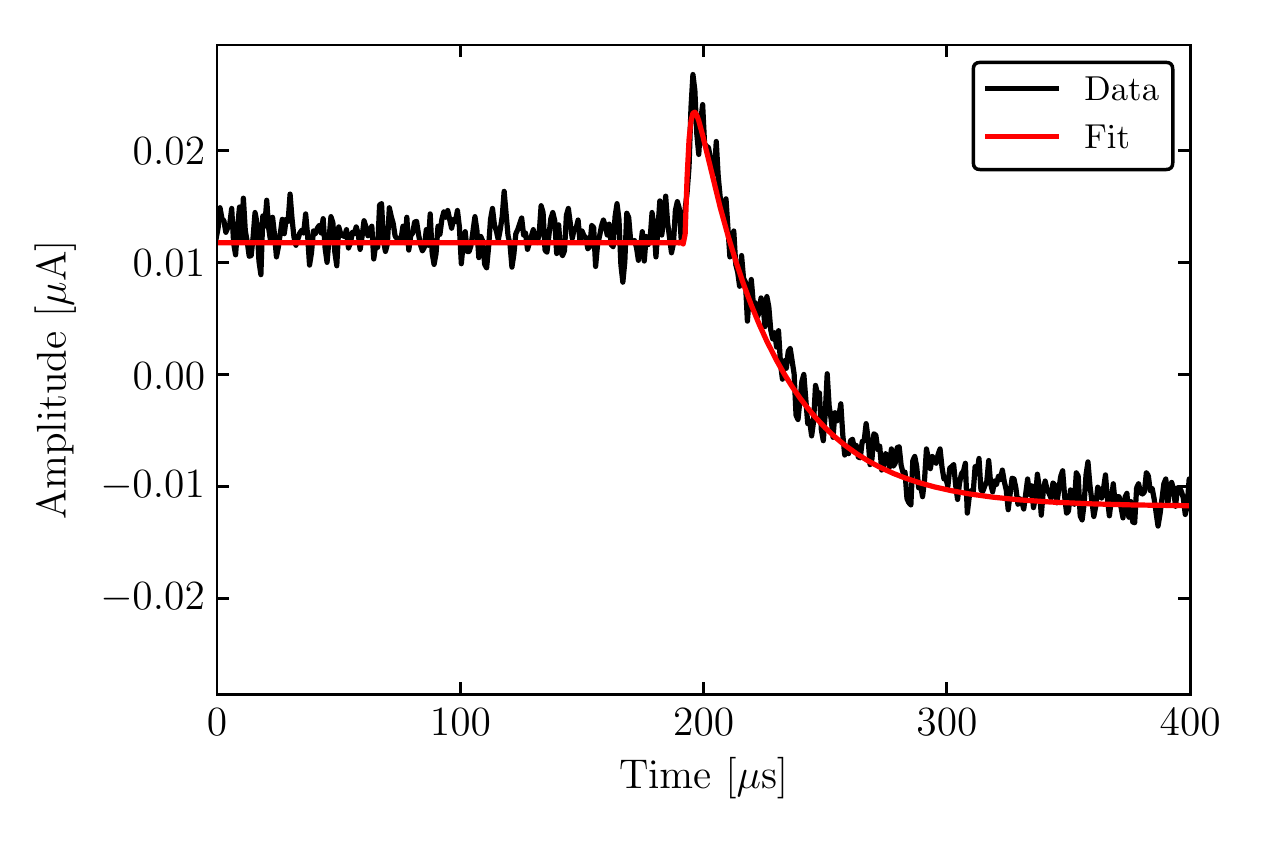}
    \caption{Example complex admittance fit in time domain for a rectangular W TES of dimensions $200 \ \mu \mathrm{m} \times 50 \ \mu \mathrm{m} \times 40 \ \mathrm{nm}$.}
    \label{fig:didv_timedomain}
\end{figure}

\begin{figure}
    \begin{subfigure}{.5\textwidth}
        \centering
        \includegraphics[width=\linewidth]{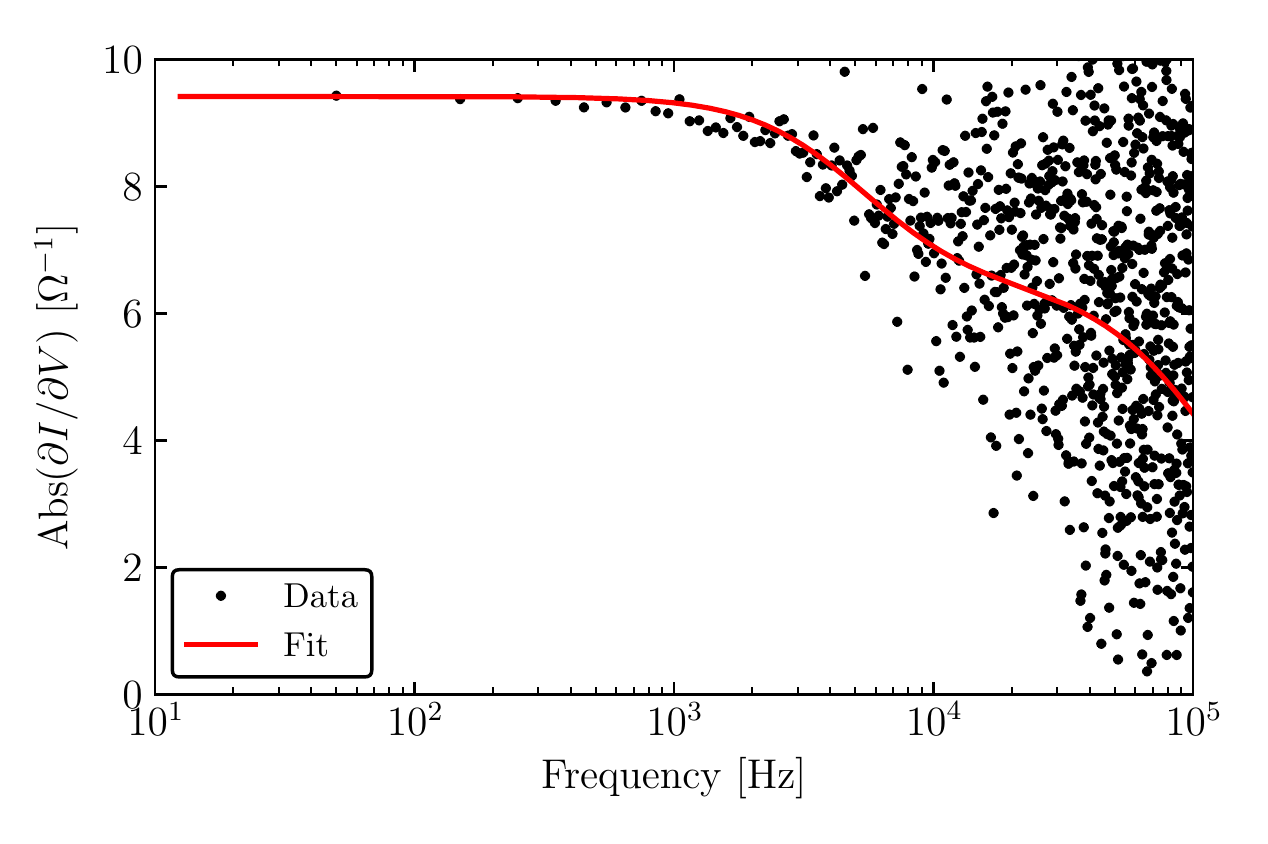}
    \end{subfigure}%
    \begin{subfigure}{.5\textwidth}
        \centering
        \includegraphics[width=\linewidth]{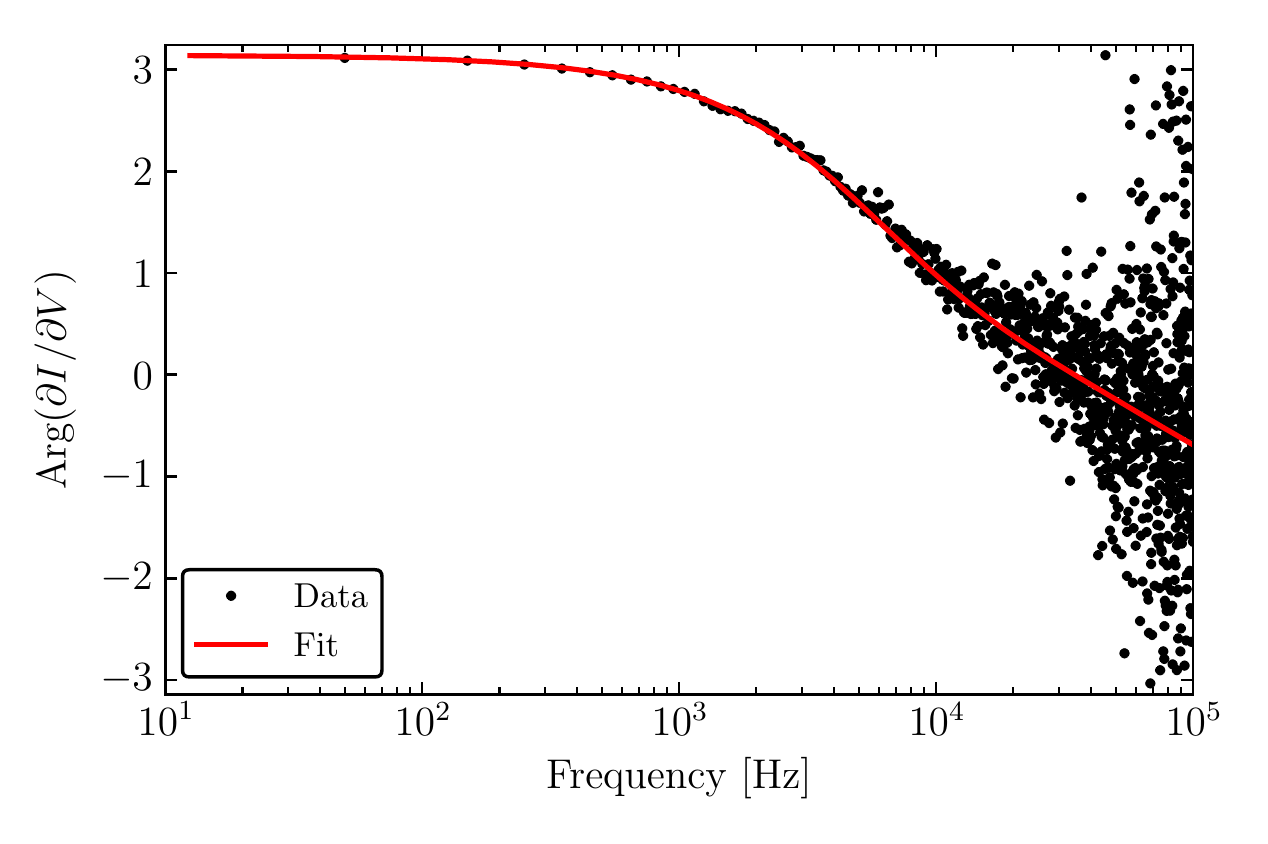}
    \end{subfigure}
    \caption{(Left) Example absolute value of complex admittance fit in frequency domain for a rectangular W TES of dimensions $200 \ \mu \mathrm{m} \times 50 \ \mu \mathrm{m} \times 40 \ \mathrm{nm}$. (Right) Example phase of complex admittance fit in frequency domain for a rectangular W TES of dimensions $200 \ \mu \mathrm{m} \times 50 \ \mu \mathrm{m} \times 40 \ \mathrm{nm}$.}
    \label{fig:didv_freqdomain}
\end{figure}

\begin{figure}
    \begin{subfigure}{.5\textwidth}
        \centering
        \includegraphics[width=\linewidth]{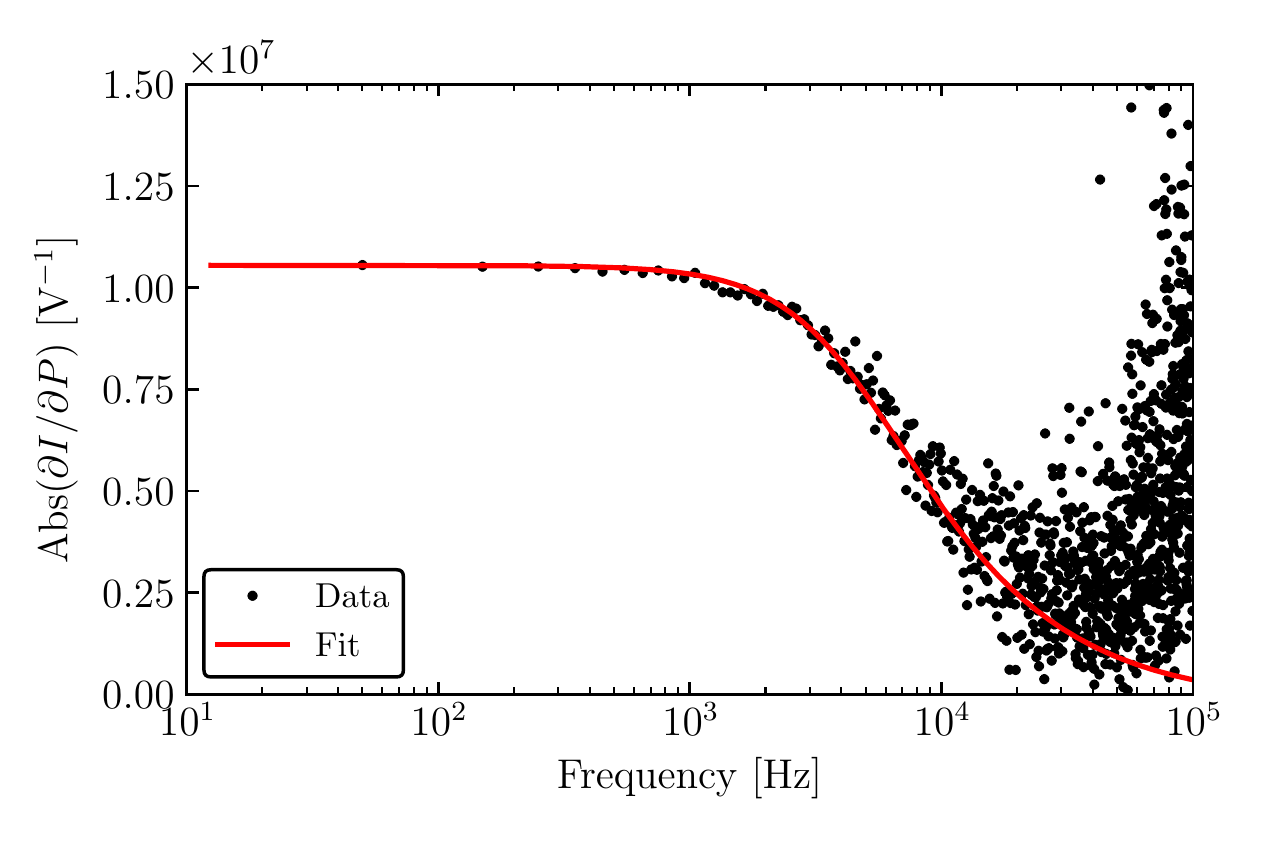}
    \end{subfigure}%
    \begin{subfigure}{.5\textwidth}
        \centering
        \includegraphics[width=\linewidth]{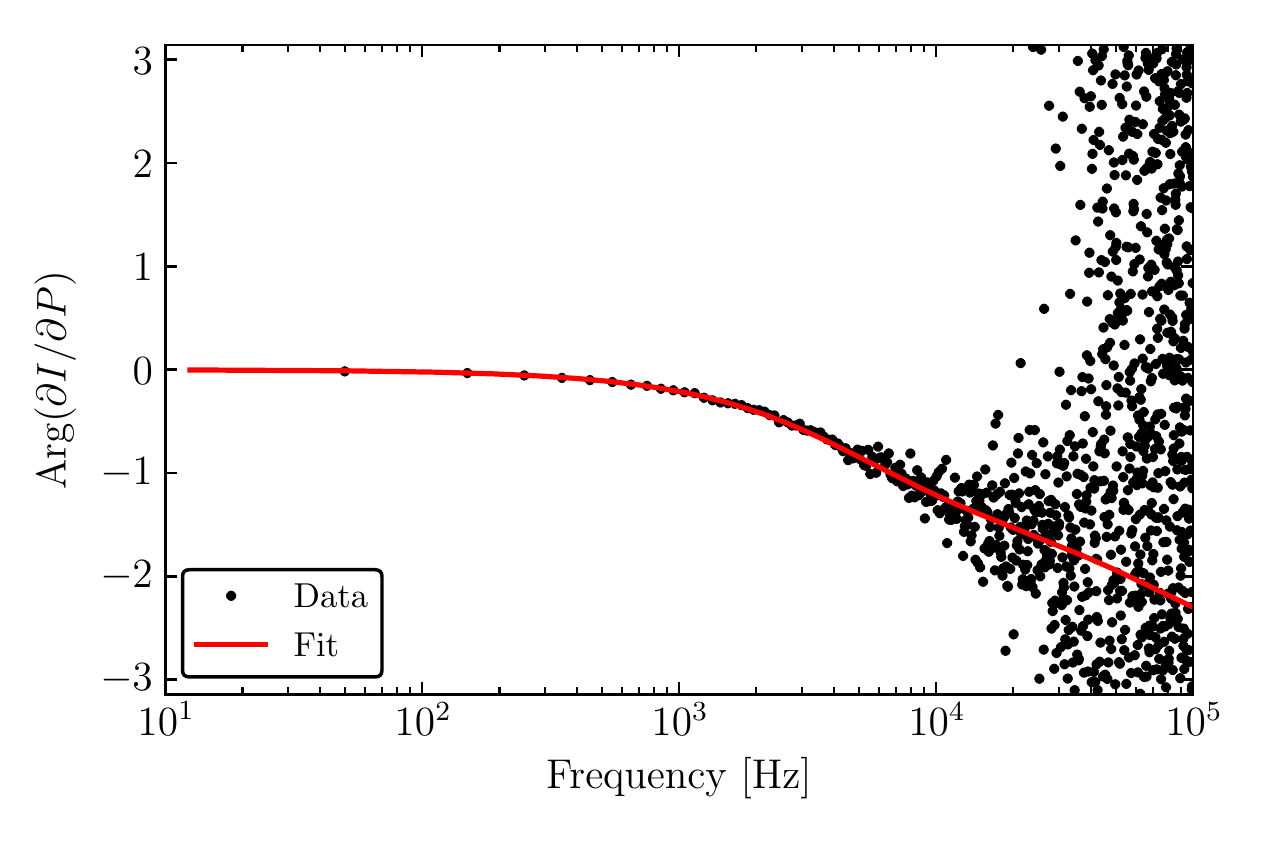}
    \end{subfigure}
    \caption{(Left) Example absolute value of $\partial I/ \partial P$ using values from the complex admittance fit in frequency domain for a rectangular W TES of dimensions $200 \ \mu \mathrm{m} \times 50 \ \mu \mathrm{m} \times 40 \ \mathrm{nm}$. (Right) Example phase of $\partial I/ \partial P$ using values from the complex admittance fit in frequency domain for a rectangular W TES of dimensions $200 \ \mu \mathrm{m} \times 50 \ \mu \mathrm{m} \times 40 \ \mathrm{nm}$.}
    \label{fig:didp_freqdomain}
\end{figure}

In Fig.~\ref{fig:didv_timedomain}, we show an example fit to a TES response to a square wave, where the rise time of the exponential is $\tau_+=1.7\ \mu \mathrm{s}$ and the fall time is $\tau_- = 32.3 \ \mu \mathrm{s}$, showing that this measurement truly gives us insight in these physical time constants. We can also show this fit in frequency domain, as done in Fig.~\ref{fig:didv_freqdomain}. We can see that the roll-off from $\tau_-$ corresponds to the bandwidth $1 / (2 \pi \tau_-) \approx 5 \ \mathrm{kHz}$. We can also use the fitted parameters ($R_0$ and $\beta$) and Eq.~(\ref{eq:didp}) to give corresponding example $\partial I/ \partial P$ curves, as shown in Fig.~\ref{fig:didp_freqdomain}. The $\partial I/ \partial P$ curve has the same poles as the complex admittance.

With these fits, there is an important difference in the number of TES parameters and the number of degrees of freedom in the complex admittance. Specifically, we have six TES parameters as defined in Eqs.~(\ref{eq:zcirc}) and (\ref{eq:ztes}): $R_\ell$, $L$, $R_0$, $\beta$, $\mathscr{L}$, and $\tau_0$. However, we only have four effective degrees of freedom. To explicitly show this, we reparameterize the complex impedance as
\begin{equation}
    Z_{circ}(\omega) = A (1 + i \omega \tau_2) + \frac{B}{1 + i \omega \tau_1},
\end{equation}
where we define
\begin{align}
    A &\equiv R_\ell + R_0 (1 + \beta) , \\
    B &\equiv \frac{R_0 \mathscr{L} (2 + \beta)}{1 - \mathscr{L}} , \\
    \tau_1 &\equiv \frac{\tau_0}{1 - \mathscr{L}}, \\
    \tau_2 &\equiv \frac{L}{R_\ell + R_0 (1 + \beta)}.
\end{align}
Thus, we cannot actually fit all six TES parameters individually, as we have a degeneracy of parameters. To solve this problem, we go back to our $IV$ curve analysis, where we solved for the TES resistance $R_0$ at different bias voltages, as well as the load resistance $R_\ell$. We must measure these two quantities before finding the rest of TES parameters from a $\partial I/ \partial V$ measurement in order to break these degeneracies. The optimal workflow for characterizing a TES is then carrying out an $IV$ sweep as a function of bias voltage and taking $\partial I / \partial V$ data at each of these bias points, which allows us to fully characterize our TES. For completion's sake, we include the values of the our TES parameters and time constants in Table~\ref{tab:tes_params} for the TES used as our example.

\begin{table}
    \centering
    \caption{Fitted and calculated TES parameters for the $200 \ \mu \mathrm{m} \times 50 \ \mu \mathrm{m} \times 40 \ \mathrm{nm}$ W TES at $R_0 = 113.5\,\mathrm{m}\Omega$, noting that the shunt resistance $R_{sh}$ is assumed to be $5 \ \mathrm{m}\Omega$. These measurements were carried out in the CryoConcept dilution refrigerator (with $T_{bath} \approx 8 \, \mathrm{mK}$) located at UC Berkeley in the Pyle lab. Values are grouped by the measurement method used to extract them.}
    \begin{tabular}{lrc}
    \hline \hline
    Parameter & Value & Measurement Method \\ \hline
    $R_\ell \ [\mathrm{m}\Omega]$ & 9.35 & \multirow{4}{*}{$IV$ curve}  \\ 
    $R_0 \ [\mathrm{m}\Omega]$ & 113.5  & \\ 
    $I_0 \ [\mu\mathrm{A}]$ & -0.901  &  \\ 
    $P_0 \ [\mathrm{fW}]$ & 92.1  &  \\  \hline
    $L\ [\mathrm{nH}]$ & 261  & \multirow{6}{*}{Transition $\partial I / \partial V$ curve}\\ 
    $\beta$ & 0.470 & \\ 
    $\mathscr{L}$ & 137 &   \\ 
    $\tau_0 \ [\mathrm{ms}]$ & 2.98  & \\ 
    $\tau_+ \ [\mu\mathrm{s}]$ & 1.76  & \\ 
    $\tau_- \ [\mu\mathrm{s}]$ & 31.3 & \\ 
    \hline \hline
    \end{tabular}
    \label{tab:tes_params}
\end{table}

As shown by I. Maasilta~\cite{maasilta}, our small signal model can become more complex if there are extra thermal bodies coupling to the TES and the thermal bath. However, the workflow proposed above will still be optimal, as this only means that we must work with a more complicated complex admittance from the same information gained from the $IV$ and $\partial I / \partial V$ sweeps. It is also worth noting that these concepts can be applied to the normal and superconducting regions of the TES. That is, we can calculate the expected complex impedance for both these regions, which suffer from less degeneracies. The complex impedances for these two regions are
\begin{align}
    Z_{sc} (\omega) &\equiv R_\ell + i \omega L \label{eq:zsc} \\
    Z_N (\omega) &\equiv R_\ell + i \omega L + R_N. \label{eq:znormal}
\end{align}
Thus, each of these equations have a single time constant: the electrical time constant. For each region, the time constant is $L/ R_\ell$ when superconducting and $L/(R_\ell + R_N)$ when normal. After fitting these two regions' responses to a square wave, a natural cross-check is provided for the corresponding values obtained by the $IV$ curve, as well as a quick way of measuring them before taking a full $IV$ sweep. Though, the $R_\ell$ and $R_N$ measurements generally have much less uncertainty in the $IV$ sweep than from measurements to the complex admittance in these regions.

\subsubsection{Infinite Loop Gain Limit and Energy Absorbed by a TES}

When characterizing TESs that are operating with large loop gains (i.e. low in transition), the infinite loop gain limit can be useful both for quick estimation of parameters, as well as sanity checks of values. To show this, we will take the limit as $\mathscr{L} \to \infty$ for both the complex admittance and the power-to-current transfer function, and apply them to a few useful applications.

Starting with the complex admittance, the zero-frequency term in the infinite loop gain limit becomes
\begin{equation}
    \lim_{\mathscr{L} \to \infty} \frac{\partial I }{\partial V}(0) = \frac{1}{R_\ell - R_0},
\end{equation}
where this is always negative for a voltage-biased TES (i.e. $R_0 \gg R_\ell$). We did not have to make any assumptions on the value of $\beta$, as it fortuitously cancels out in this limit. From this limit, we can measure the zero-frequency of our complex admittance and immediately have an estimate of $R_0$ (assuming we know $R_\ell$), without needing an $IV$ curve. This of course will just be an estimate, and the values from the $IV$ curve will be much more accurate in cases of $\mathscr{L}$ is $\mathcal{O}(1$--$10)$.

For our example TES we have been studying, our fit in terms of the $A$, $B$, $\tau_1$, and $\tau_2$ parameters gives $A=0.176 \,  \mathrm{m}\Omega$, $B= -0.282 \, \mathrm{m}\Omega$, $\tau_1 = -2.20 \, \mu \mathrm{s}$, and $\tau_2 = 1.48 \, \mu \mathrm{s}$. Thus, the zero frequency component in this form is
\begin{align}
    \frac{\partial I }{ \partial V} (0) &= \frac{1}{A + B}, \\
    &= -9.41 \, \Omega^{-1}.
\end{align}
Solving for $R_0$ in the infinite $\mathscr{L}$ limit (using $R_\ell = 9.35 \, \mathrm{m}\Omega$ from, e.g., the superconducting complex admittance), we find that $R_0 \approx 115.5 \mathrm{m} \Omega$. Comparing to the value in Table~\ref{tab:tes_params} of $R_0 = 113.5 \, \mathrm{m} \Omega$, this is only $2 \,\mathrm{m}\Omega$ (about 2\% error), which is quite close due to $\mathscr{L}$ being $\mathcal{O}(100)$. We can go one step further and estimate the bias power
\begin{equation}
    P_0 = \left( \frac{V_b}{R_0 + R_\ell} \right)^2 R_0
\end{equation}
using that $I_0 = V_b / (R_0 + R_\ell)$. As we should know the bias voltage ($V_b = -110.5 \, \mathrm{nV}$ for our example TES), we have estimates of each quantity, and we have that $P_0 \approx 90.5 \, \mathrm{fW}$, again only about $2\%$ error off of the value in Table~\ref{tab:tes_params} of $92.1 \, \mathrm{fW}$. Thus, simply knowing the zero-frequency of the complex admittance is quite powerful in the estimates that we can make when in the case of infinite $\mathscr{L}$.

For a very quick back-of-the-envelope calculation of $R_0$, one can estimate off of a time domain complex admittance plot. In Fig.~\ref{fig:didv_timedomain}, for example, the zero-frequency component of the complex admittance is the change in TES current divided by the change in voltage bias. In this data, we have a $2.5 \, \mathrm{nV}$ peak-to-peak square jitter in bias voltage, and a $23.5 \, \mathrm{nA}$ change in TES current (comparing the current change before and after the TES responds), such that we can estimate the zero-frequency value as $-9.4 \, \Omega^{-1}$ (recalling the minus sign due to $\partial I / \partial V(0)<0$ when $\mathscr{L}>1$) and follow the same steps to estimate various parameters. Thus, even if we were only looking at the TES response on, e.g., an oscilloscope, we could still quickly estimate $R_0$ and $P_0$ just from this change in TES current.

Next, we take the same infinite $\mathscr{L}$ limit for the power-to-current transfer function defined in Eq.~(\ref{eq:didp}) at zero-frequency, which gives \begin{equation}
    \lim_{\mathscr{L} \to \infty} \frac{\partial I }{\partial P}(0) = \frac{1}{I_0 (R_\ell - R_0)},
    \label{eq:didp0_infloop}
\end{equation}
and can be positive or negative, depending on the current $I_0$ through the TES. In other words, this becomes the zero-frequency complex admittance divided by the TES current, and we can estimate the $\partial I/ \partial P (0)$ quickly as well.

We can rearrange Eq.~(\ref{eq:didp0_infloop}) such that
\begin{align}
    \mathop{dP} &= I (R_\ell - R_0) \mathop{dI}, \\
    &= - (V_b - 2 I R_\ell) \mathop{dI},
\end{align}
where we have substituted $R_0 = V_b/I - R_\ell$. To calculate the energy absorbed by the TES, we note that the change in Joule heating power is equal and opposite in sign to the power absorbed by the TES, i.e. $dP_J = - dP_{abs}$, by energy conservation. Thus, we have that
\begin{equation}
    \mathop{dP_{abs}} = (V_b - 2 I R_\ell) \mathop{dI},
\end{equation}
and we can integrate both sides
\begin{align}
    \Delta P_{abs} &= \int_{I_0}^{I_0 + \Delta I(t)} \mathop{dI} (V_b - 2I R_\ell), \\
    &= (V_b - 2I_0) \Delta I(t) - \Delta I(t)^2 R_\ell,
\end{align}
Finally, we integrate this change in power over time to calculate the energy absorbed by the TES, also known as the energy removed by electrothermal feedback in the large $\mathscr{L}$ limit
\begin{equation}
    E_\mathrm{ETF} = \int_0^\infty \mathop{dt} \left[ (V_b - 2I_0) \Delta I(t) - \Delta I(t)^2 R_\ell \right].
\end{equation}
In practice, we are looking at a finite chunk of data, such that we cannot actually integrate to infinity. Thus, we must choose some time cutoff to the integral $T_{trunc}$, such that
\begin{equation}
    E_\mathrm{ETF} = \int_0^{T_{trunc}} \mathop{dt} \left[ (V_b - 2I_0) \Delta I(t) - \Delta I(t)^2 R_\ell \right].
\end{equation}
where this cutoff is generally far enough from the change in current from injected energy to capture the entire change (e.g. if the change in the current has a characteristic fall time $\tau_-$, then integrate up to $T_{trunc} = 7\tau_-$). Note that the direction of pulses have no effect on the integral, as a negative bias would give a negative $V_b$, a negative $I_0$, and a negative $\Delta I (t)$. Whereas a positive bias would result in each of these quantities being positive. Thus, in either case, each term in the integral would end up with the same sign.

\section{TES Noise and Energy Sensitivity}

Understanding our TES parameters is only part of the battle. As we will be using TESs to set DM limits, we need to understand the baseline energy resolution of these devices in order to know what energy thresholds are feasible. This will be done through the modeling and calculation of the power spectral densities (PSD) for a description of the power of the noise at various frequencies.

When working with these PSDs, it must be remembered that there are two common versions: one-sided and two-sided. The two-sided PSD is defined over positive and negative frequencies, while the one-sided PSD is only defined for positive frequencies. Because data being taken are real-valued, the two-sided PSD is an even function (i.e. the two-sided PSD has the same value for some frequency $f$ and its negative counterpart). Thus, it is common practice to ``fold over'' the PSD, where we effectively dispose of the negative frequencies and multiply the positive frequencies by a factor of two, giving the one-sided PSD. In this way, the PSD becomes easier to plot on a log-log scale, and we have not lost the total power of the system. We note that all of the PSD definitions that will be made in Section~\ref{sec:noisederiv} are using one-sided PSDs, and we have properly taken care of our factors of two. Furthermore, it is also common practice (as the reader will see in Section~\ref{sec:noisemodeling}) to take the square root of the PSD when plotting, creating strange units such as $\mathrm{A}/\sqrt{\mathrm{Hz}}$, as the corresponding PSD would have units of $\mathrm{A}^2/\mathrm{Hz}$. We will see that there is convenience to doing this when we finally calculate the expected baseline energy resolution of a TES.

\subsection{\label{sec:noisederiv}Intrinsic Noise Sources}

With a TES, we have a handful of intrinsic noise sources that are directly related to the TES characteristics. We will begin with the Johnson noise, discovered by J. B. Johnson~\cite{PhysRev.32.97} and subsequently explained by H. Nyquist via a calculation of the electromotive force due to thermal agitation of electric charge carriers~\cite{PhysRev.32.110}.

In our TES circuit, we have that any resistive element will have Johnson noise. For our purposes, it is easiest to split the Johnson noise into the load Johnson noise and the TES Johnson noise. The load Johnson noise will include the Johnson noise of the shunt resistor and any parasitic resistances. As Johnson voltage noise is flat in frequency space, we have that the load Johnson noise is represented by the one-sided PSD
\begin{equation}
    S_{V_{load}} = 4 k_B \left(\sum_i T_{sh_i} R_{sh_i} + \sum_j T_{p_j} R_{p_j} \right),
\end{equation}
where we generalize that there could be multiple shunt resistances and parasitic resistances all at different temperatures. In practice, we do not have a way of splitting all of the degeneracies between these different resistors, and it is useful to reparameterize this in terms of an effective load temperature
\begin{equation}
    T_\ell \equiv \frac{\sum_i T_{sh_i} R_{sh_i} + \sum_j T_{p_j} R_{p_j}}{R_\ell},
\end{equation}
which allows us to simplify the load Johnson noise in voltage to
\begin{equation}
    S_{V_{load}} = 4 k_B T_\ell R_\ell.
    \label{eq:svload}
\end{equation}
Since we know $R_\ell$ from our $IV$ analysis, we only need to measure the effective load temperature $T_\ell$ to model this noise component, as will be discussed in Section~\ref{sec:noisemodeling}. When actually measuring the noise of our system, we will be measuring the current noise (as we are reading out the current through the TES circuit). We can simply convert this voltage noise to current noise by using our complex admittance (i.e. our voltage-to-current transfer function)
\begin{equation}
    S_{I_{load}}(\omega) = 4 k_B T_\ell R_\ell \left| \frac{\partial I}{\partial V} (\omega) \right|^2,
\end{equation}
showing the frequency dependence of the load Johnson noise in current space.

The TES Johnson noise is more complex due to the thermal and current dependencies of the TES creating nonlinear effects. Using nonlinear Markov fluctuation-dissipation techniques for nonlinear resistors developed by R. L. Stratonovich~\cite{stratonovich1970thermal, stratonovich2012nonlinear}, Irwin and Hilton show that the TES Johnson noise in voltage can be approximated as~\cite{irwin}
\begin{equation}
    S_{V_{TES}} = 4 k_B T_0 R_0 (1+\beta)^2,
\end{equation}
showing that the voltage noise is elevated by our current sensitivity $\beta$. Converting this to current noise is more complex than the using the complex admittance due to the current and temperature dependence of the TES. In Ref.~\cite{Matt_thesis}, M. Pyle shows that the conversion to current space follows the compact form
\begin{equation}
    S_{I_{TES}}(\omega) = 4 k_B T_0 R_0 (1+\beta)^2 \left|\frac{\partial I}{\partial V}(\omega) - I_0 \frac{\partial I }{\partial P}(\omega) \right|^2.
\end{equation}

We now move to the next intrinsic noise source: thermal fluctuation noise (TFN), also referred to as phonon noise. Because our TES is by design sensitive to temperature changes in transition, this brings a sensitivity to thermal fluctuations between the TES and the thermal bath (each having different temperatures). As shown by R. F. Voss and J. Clarke~\cite{PhysRevB.13.556}, and applied to TESs in, e.g., M. Pyle's thesis~\cite{Matt_thesis}, these fluctuations across the thermal link will give a noise in power space that is flat across all frequencies:
\begin{equation}
    S_{P_{TFN}} = 4 k_B T_0^2 G F_{TFN}(T_0, T_{bath}),
    \label{eq:tfn}
\end{equation}
where the unitless $F_{TFN}$ factor~\cite{Boyle:59,McCammon2005,Mather:82}
\begin{equation}
    F_{TFN}(T_0, T_{bath}) = \begin{cases}
    \frac{\left(T_{bath} / T_0 \right)^{n+1} + 1}{2} & \mathrm{ballistic \ phonons}\\
    \frac{n}{2 n + 1}\frac{\left(T_{bath} / T_0 \right)^{2n+1} - 1}{\left(T_{bath} / T_0 \right)^{n} - 1} & \mathrm{diffuse \ phonons}\\
    \end{cases},
\end{equation}
takes a value from 0.5 to 1. Usually, we are in the case of ballistic phonons. As we want the TFN contribution in current space, we easily do this via our power-to-current transfer function
\begin{equation}
    S_{I_{TFN}}(\omega) = 4 k_B T_0^2 G F_{TFN}(T_0, T_{bath}) \left| \frac{\partial I}{\partial P}(\omega) \right|^2.
\end{equation}

Lastly, as we are measuring the current through the TES via a SQUID amplifier, this will give us a SQUID and downstream electronics noise component. The treatment of these noise sources can be quite complex, but we generally use a simplified model of $1/f$ noise~\cite{Noah_thesis}
\begin{equation}
    S_{I_{SQUID}}(\omega) = \left\{DC_{SQUID} \left[1 + \left( \frac{\omega_{SQUID} }{\omega} \right)^{n_{SQUID}} \right] \right\}^2,
\end{equation}
where $DC_{SQUID}$ is some constant value (usually on the order of a few pA/$\sqrt{\mathrm{Hz}}$), $\omega_{SQUID}$ determines the characteristic frequency of the $1/f$ behavior, and $n_{SQUID}$ is some exponent for generalization of the $1/f$ behavior. In practice, each of these values will be fit from noise measurements in the normal regime of the TES, as explained in Section~\ref{sec:noisemodeling}.

We can combine all of our noise sources to find the total expected noise spectrum in current for our TES in transition
\begin{equation}
    S_{I_{tot}}(\omega) = S_{I_{load}}(\omega) + S_{I_{TES}}(\omega) + S_{I_{TFN}}(\omega) + S_{I_{SQUID}}(\omega).
\end{equation}

For the normal and superconducting regimes, the derivations are much simpler, as the resistors act as normal resistors (i.e. there is no TFN). For the normal regime, the load Johnson noise has the same voltage form as Eq.~(\ref{eq:svload}), but we use our normal resistance complex impedance from Eq.~(\ref{eq:znormal}) to convert to current noise
\begin{equation}
    S_{I_{load}}^{N}(\omega) = \frac{4 k_B T_\ell R_\ell} {\left| Z_N (\omega) \right|^2}.
\end{equation}
The TES Johnson noise simplifies greatly, where we assume that the temperature of the TES does not change appreciably from its equilibrium (transition) temperature
\begin{equation}
    S_{V_{TES}}^{N}(\omega) = 4 k_B T_0 R_N,
\end{equation}
with the corresponding current noise being
\begin{equation}
    S_{I_{TES}}^{N}(\omega) = \frac{4 k_B T_0 R_N} {\left| Z_N (\omega) \right|^2}.
\end{equation}
Thus, the total expected normal state current noise becomes
\begin{equation}
    S_{I_{tot}}^{N}(\omega) = S_{I_{load}}^{N}(\omega) + S_{I_{TES}}^{N}(\omega) + S_{I_{SQUID}}(\omega),
\end{equation}
where the SQUID current noise term is unchanged, as it has no dependence on the status of the TES. In a well-designed system, the SQUID noise term will dominate when a TES is normal and can be easily measured.

When superconducting, the TES Johnson noise will go to zero (the TES resistance is zero), and we have only the load Johnson noise and the SQUID noise. The load Johnson current noise when superconducting is determined by Eq.~(\ref{eq:zsc}), such that
\begin{equation}
    S_{I_{load}}^{sc}(\omega) = \frac{4 k_B T_\ell R_\ell} {\left| Z_{sc} (\omega) \right|^2},
\end{equation}
and the total superconducting current noise is
\begin{equation}
    S_{I_{tot}}^{sc}(\omega) = S_{I_{load}}^{sc}(\omega) + S_{I_{SQUID}}(\omega).
\end{equation}

With the expected current noise and our transfer functions known for the normal, superconducting, and in transition regimes of the TES, this allows us to model the expected noise performance of our system, given that we have knowledge of all of our TES parameters.

\subsection{\label{sec:noisemodeling}Noise Modeling}

As we have been using an example W TES of dimensions $200 \ \mu \mathrm{m} \times 50 \ \mu \mathrm{m} \times 40 \ \mathrm{nm}$ throughout this chapter, we will use what we learned about the intrinsic TES noise to allow us to model it, via the small-signal approximation. In order to measure our TES current noise, we generally record many random time domain traces of data at a single bias point. We then select traces that should describe the intrinsic noise well (i.e. remove traces that have noise from external sources such as pulses, vibrations, electronic glitches, etc.), Fourier transform these traces, take the modulus squared of the Fourier transforms, properly normalize (divide by the digitization rate and the number of time bins) to calculate the power spectral density (PSD), and take the mean at each frequency. This is programmatically done in the \texttt{calc\_psd} function in \textsc{QETpy}~\cite{qetpy}.

\begin{figure}
    \centering
    \includegraphics{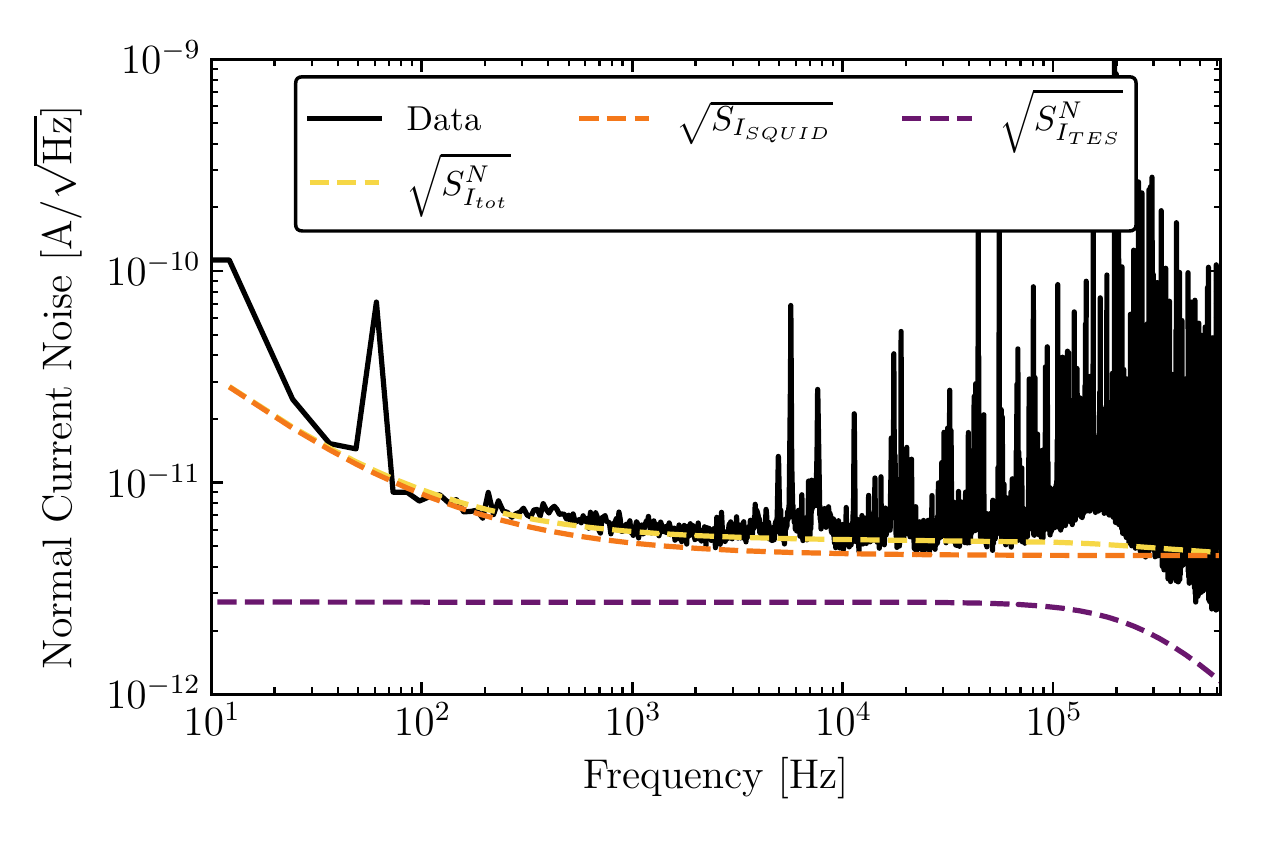}
    \caption{Modeling of normal current noise for the $200 \ \mu \mathrm{m} \times 50 \ \mu \mathrm{m} \times 40 \ \mathrm{nm}$ W TES. The $1/f$ noise is attributed to the SQUID, with a small discrepancy between the model and the data at $\sim\!10\,\mathrm{Hz}$ due to the fit range being restricted to 0.1--1$\,$kHz to be above the $60\, \mathrm{Hz}$ spike. The many peaks at $10\, \mathrm{kHz}$ and above can be attributed to RF pickup in the electronics.}
    \label{fig:normalnoisemodel}
\end{figure}

At this point, we have all of our TES parameters, but we will need to use our noise measurements to help us understand our SQUID noise and our load Johnson noise when modeling all of our noise components. To do this, we first start with our normal noise. In this regime, the SQUID noise is dominant, with a small correction from the Johnson noise of the normal TES. Thus, we can model our SQUID noise by fitting the data to our SQUID noise model (including the correction from the TES Johnson noise). From this fit, we have the parameters to model our SQUID noise in the other regimes, where it may not be (and usually is not) the dominant noise source. We show an example of this in Fig.~\ref{fig:normalnoisemodel}. Note that, while we have defined a load Johnson noise component when normal in our equations, it is so small (less than $1 \ \mathrm{pA}/ \sqrt{\mathrm{Hz}}$ over all frequencies) in the normal regime that we neglect it. We have give our fit values in Table~\ref{tab:noise_params}, which also contains the superconducting fit values and will be discussed in the next paragraph.

\begin{figure}
    \centering
    \includegraphics{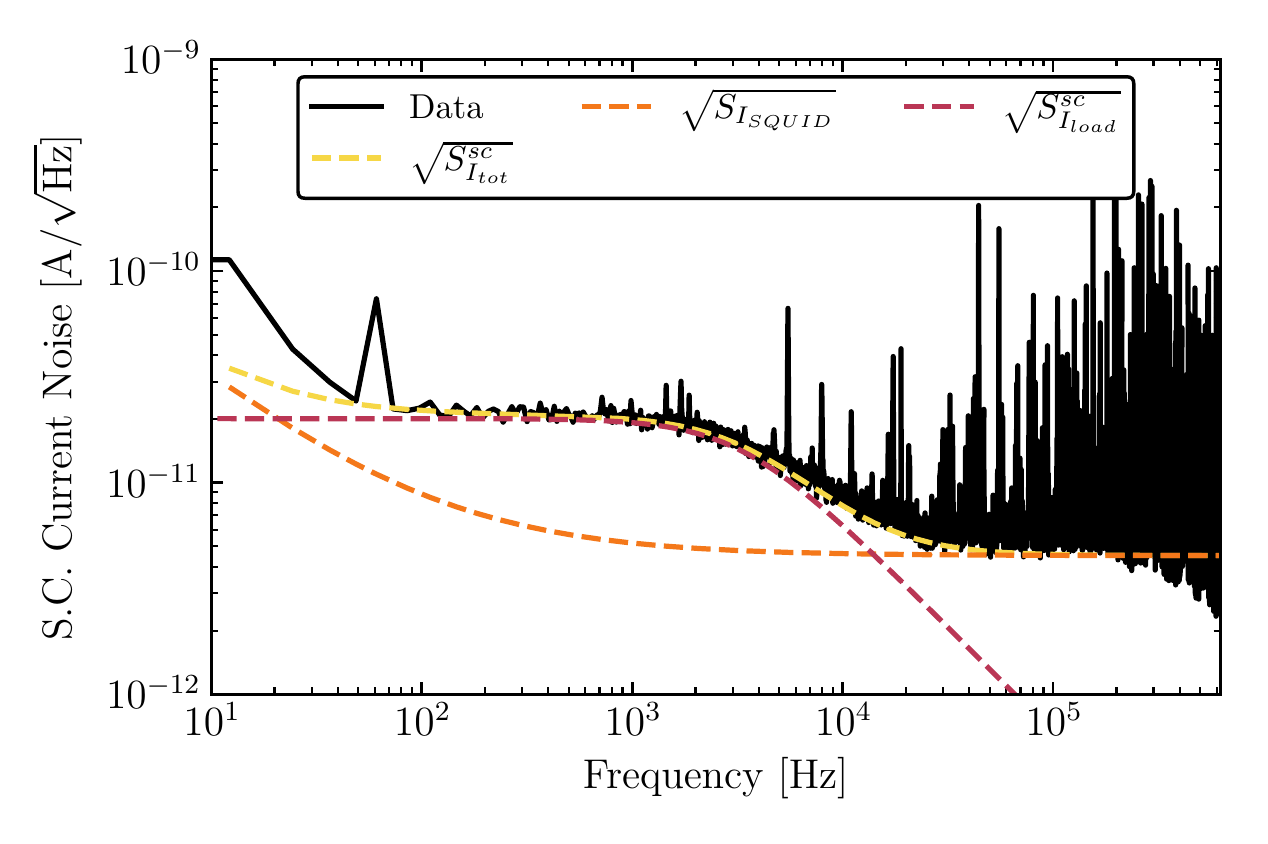}
    \caption{Modeling of superconducting (S.C.) current noise for the $200 \ \mu \mathrm{m} \times 50 \ \mu \mathrm{m} \times 40 \ \mathrm{nm}$ W TES.}
    \label{fig:scnoisemodel}
\end{figure}

After modeling our SQUID noise, we return to the effective load temperature $T_\ell$. When defining $T_\ell$ in Section~\ref{sec:noisederiv}, we noted that it must be measured for proper noise modeling. Fortunately, it can be measured from the superconducting noise. As we know the rest of the parameters in the load Johnson noise component, we can use our superconducting PSD data to estimate (i.e. fit for) $T_\ell$, where we use our total noise model and our previous SQUID noise fit. In Fig.~\ref{fig:scnoisemodel}, we show both of these components and the total noise, where we have fit the superconducting noise model to find our effective load temperature, which was found to be $T_\ell = 68.5 \ \mathrm{mK}$, as reported in Table~\ref{tab:noise_params}.

\begin{figure}
    \centering
    \includegraphics{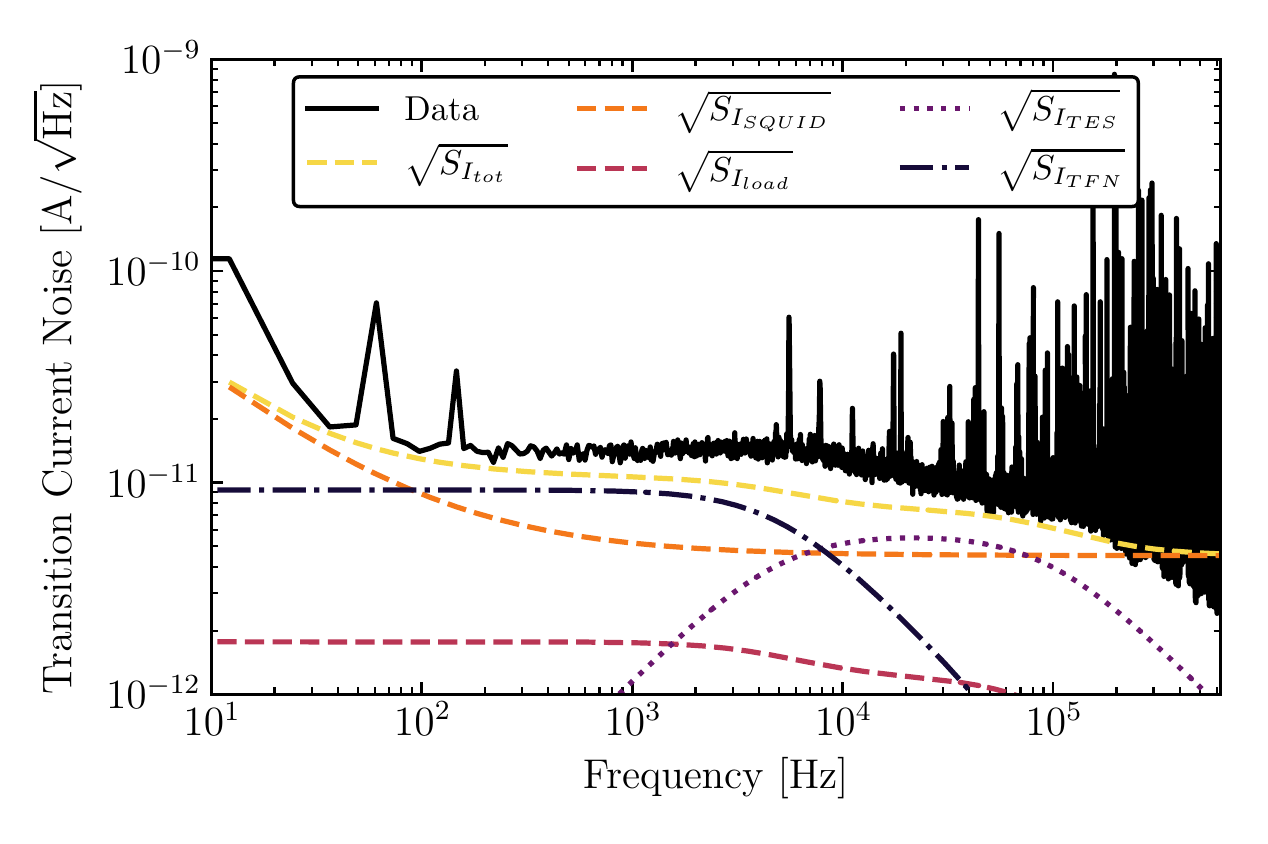}
    \caption{Modeling of transition current noise for the $200 \ \mu \mathrm{m} \times 50 \ \mu \mathrm{m} \times 40 \ \mathrm{nm}$ W TES. There is a small amount of excess noise above the total modeled noise, which is of unknown origin. It is seldom seen that the noise model from the small signal approximation perfectly matches the measured noise, and most groups see similar excesses~\cite{fink2020characterizing,cpdcollaboration2020performance}.}
    \label{fig:trannoisemodel}
\end{figure}

With a TES, the most interesting region is the transition region. Our work on the normal and superconducting regions was simply to allow us to fully model the noise of a TES in transition. There is one more step that we have neglected so far, which is calculating the thermal conductance $G$. If we are in the limit that $T_0 \gg T_{bath}$ and small parasitic power (which is usually the case), then we can estimate the thermal conductance as
\begin{equation}
    G \approx \frac{nP_0}{T_0},
\end{equation}
giving $G \approx 7.75 \ \mathrm{pW}/\mathrm{K}$. In Appendix~\ref{chap:appa}, we explain methods to directly measure the thermal conductances of the system, but we will use this approximation for our noise modeling purposes. With all of our parameters now known and in Table~\ref{tab:noise_params}, we can use our derived equations from Section~\ref{sec:noisederiv} to calculate our expected total current noise model and compare to the measured current noise, as compared in Fig.~\ref{fig:trannoisemodel}.

\begin{table}
    \centering
    \caption{TES noise parameters for the $200 \ \mu \mathrm{m} \times 50 \ \mu \mathrm{m} \times 40 \ \mathrm{nm}$ W TES.}
    \begin{tabular}{lr}
    \hline \hline
    Parameter & Value  \\ \hline
    $DC_{SQUID} \ [\mathrm{pA}/ \sqrt{\mathrm{Hz}}]$ & 4.52  \\ 
    $\omega_{SQUID} \ [s^{-1}]$ & 594  \\ 
    $n_{SQUID}$ & 0.812  \\ 
    $T_\ell \ [\mathrm{mK}]$ & 68.5 \\ 
    $G \ [\mathrm{pW}/\mathrm{K}]$ & 7.75  \\ 
    $\sigma_E \ [\mathrm{meV}]$ & 45.9  \\ 
    \hline \hline
    \end{tabular}
    \label{tab:noise_params}
\end{table}

In our comparison, we see that the model does not perfectly overlap with what we measured. This is common in the TES field. There can be external noise sources that increase our noise, such as infrared photons or electromagnetic interference. In general, excess noise is frequently experienced, and there have been many explanations put forward~\cite{maasilta, fink2020characterizing, doi:10.1063/1.5086045}, but it is always a difficult problem to understand due to the nature of TES sensitivity.

\begin{figure}
    \centering
    \includegraphics{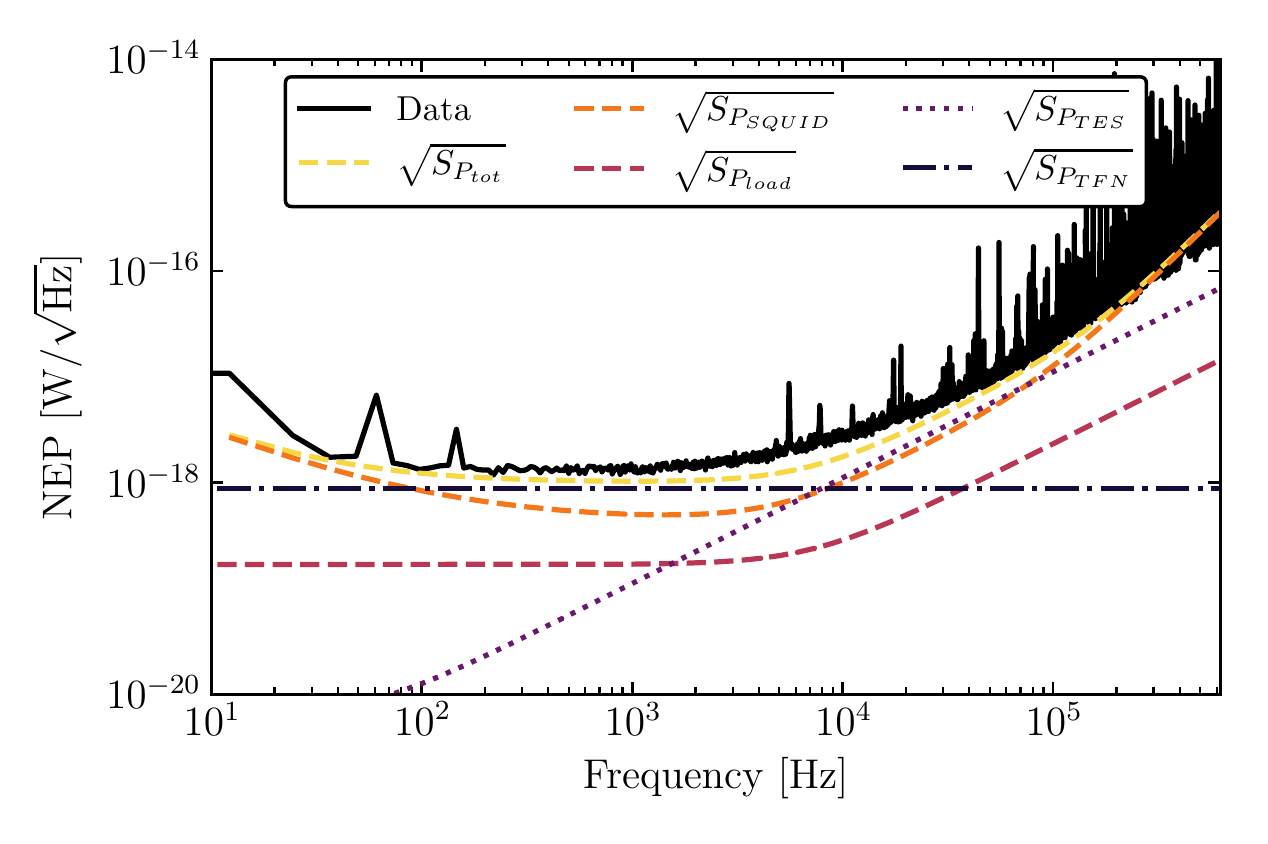}
    \caption{Modeling of transition power noise, or Noise-Equivalent Power (NEP), for the $200 \ \mu \mathrm{m} \times 50 \ \mu \mathrm{m} \times 40 \ \mathrm{nm}$ W TES.}
    \label{fig:nepmodel}
\end{figure}

At this point, we have only discussed current noise, due to the easy comparison between the different regions from normal to transition to superconducting. However, we can calculate the expected energy sensitivity of a TES via an understanding of its power noise. Since we already know the power-to-current transfer function, we can convert from TES current noise to TES power noise in one step
\begin{equation}
    S_{P_{tot}}(\omega) = \frac{S_{I_{tot}(\omega)}}{\left| \frac{\partial I}{\partial P}(\omega) \right|^2}.
    \label{eq:totalpowernoise}
\end{equation}
As discussed earlier, we will work instead with the square root of the power noise, which we will define as the Noise-Equivalent Power (NEP)
\begin{equation}
    \mathrm{NEP}(\omega) \equiv \sqrt{S_{P_{tot}}(\omega)}.
\end{equation}
In Fig.~\ref{fig:nepmodel}, we show this conversion, where each component of the total modeled power noise has also been converted from current noise to power noise by taking advantage of our (at this point) well-known power-to-current transfer function.

\subsection{TES Baseline Energy Resolution}

The importance of calculating the NEP is best shown when using all of our knowledge gained to calculate the expected baseline energy resolution (the expected standard deviation of zero-energy fluctuations) of a TES to Dirac delta deposits. This is usually done via the optimal filter (OF) formalism, as discussed in detail in, e.g., Refs.~\cite{irwin, Matt_thesis, McCammon2005, fink2020characterizing, golwala}. The common derived result is that this baseline energy resolution can be expressed as (in our Dirac delta deposit approximation)
\begin{equation}
    \sigma_E = \left[ \int_0^\infty \frac{\mathrm{d}\omega}{2 \pi} \frac{4}{S_{P}(\omega)}\right]^{-1/2}.
    \label{eq:dirac_eres}
\end{equation}
From our measured (one-sided) $S_P(\omega)$, we can numerically integrate this baseline energy resolution, returning the reported value in Table~\ref{tab:noise_params} of $\sigma_E = 45.9 \ \mathrm{meV}$.

In practice, we can find an approximate analytic solution to this integral if we have an estimate of the NEP. The idea is that $S_P(\omega)$ can be approximated as
\begin{equation}
    S_P(\omega) \approx NEP(0)^2 (1 + i \omega \tau_-).
\end{equation}
Using this form, we can analytically solve Eq.~(\ref{eq:dirac_eres}), which gives us an expected baseline energy resolution to Dirac delta deposits of
\begin{equation}
    \sigma_E \approx NEP(0)\sqrt{\tau_-},
    \label{eq:approxdirac}
\end{equation}
and an estimated baseline energy resolution of $\sigma_E \approx 47.8 \ \mathrm{meV}$. Comparing to the the value obtained in Table~\ref{tab:noise_params}, this is quite close, which allows us to achieve a simple way of estimating the expected baseline energy resolution without needing to carry out numerical integration.

\subsection{\label{sec:mfactor}Parameterizing Excess Noise}

In Fig.~\ref{fig:trannoisemodel}, we noted the existence of excess noise. It can be useful to parameterize this noise into two different categories: excess TFN--like noise and excess Johnson--like noise. To describe excess TFN--like noise, we define the parameter $\xi$ as related to the previously derived $S_{P_{TFN}}$ in Eq.~(\ref{eq:tfn}) by
\begin{equation}
    S^\mathrm{*}_{P_{TFN}} = (1+\xi) S_{P_{TFN}},
\end{equation}
where we have added an asterisk to denote the inclusion of an excess noise factor. We assume $\xi$ is frequency-independent, such that this same relation holds in current space. Similarly, the excess Johnson--like noise can be described using the $M$ factor~\cite{doi:10.1063/1.3292343}, defined in relation to the modeled Johnson voltage noise by
\begin{equation}
    S^\mathrm{*}_{V_{TES}} = S_{V_{TES}} (1 + M^2),
\end{equation}
where again have added an asterisk. Assuming that $M$ is frequency-independent, this same relation holds for the TES Johnson noise in current space. In order to actually determine $\xi$ and $M$, the predicted TES noise must be compared to the measured noise. One method is to carry out a least squares fit to the data using the total modeled noise including the excess, defined in current space as
\begin{equation}
    S^\mathrm{*}_{I_{tot}}(\omega) = S_{I_{load}}(\omega) + S_{I_{TES}}(\omega) (1 + M^2) +  S_{I_{TFN}}(\omega)(1+\xi) + S_{I_{SQUID}}(\omega).
    \label{eq:excess_itot}
\end{equation}

\begin{figure}
    \centering
    \includegraphics{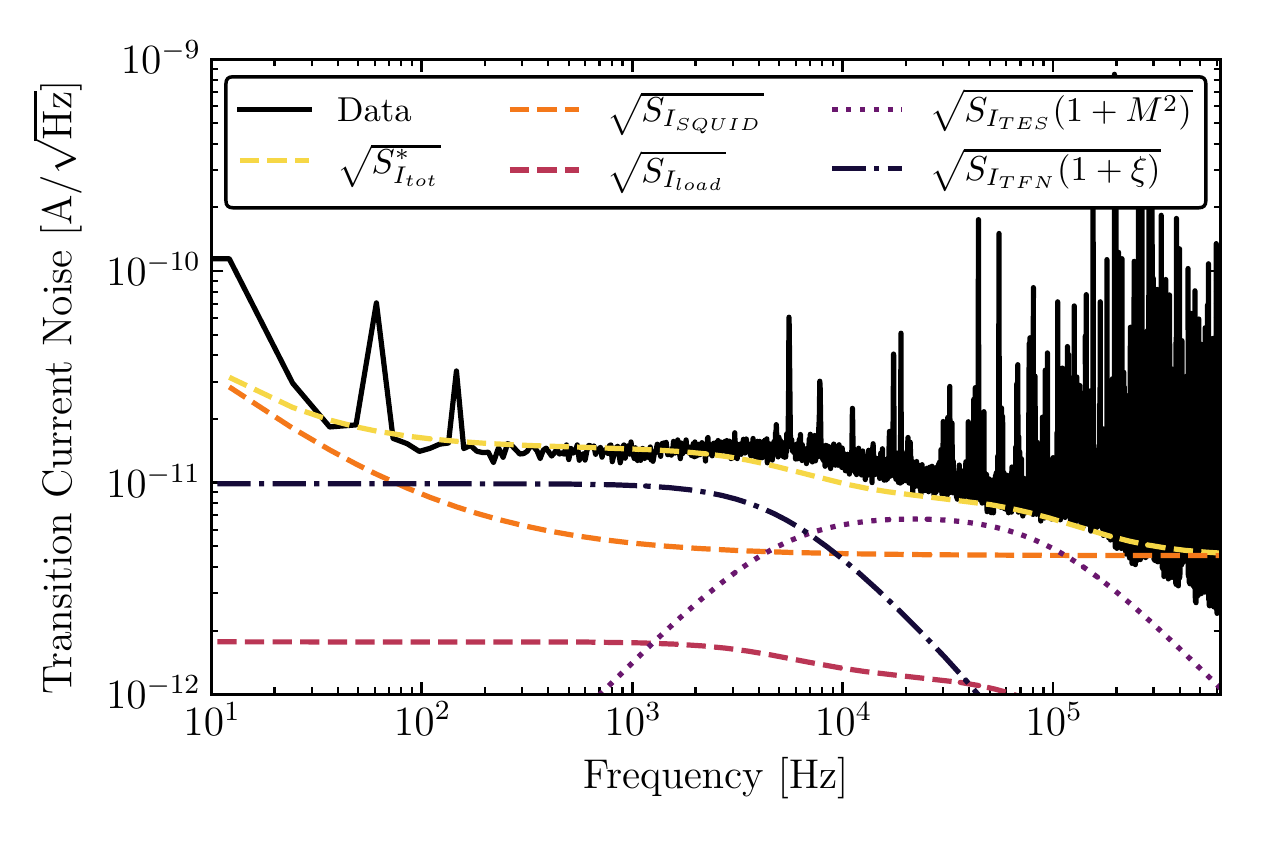}
    \caption{Parameterized excess noise using the $\xi$ and $M$ parameters for the $200 \ \mu \mathrm{m} \times 50 \ \mu \mathrm{m} \times 40 \ \mathrm{nm}$ W TES in current space.}
    \label{fig:excessnoise}
\end{figure}

For the TES we have been studying in this section, we can use the current noise in Fig.~\ref{fig:trannoisemodel} and parameterize the excess noise with $\xi$ and $M$. Fitting Eq.~(\ref{eq:excess_itot}) with $\xi$ and $M$ varying to the data, we find that $\xi=1.15$ and $M = 0.71$ provide a good parameterization of the excess noise, the result of which is shown in Fig.~\ref{fig:excessnoise}. There does still remain a small amount of excess current noise at around $10 \, \mathrm{kHz}$ that this parameterization was not able to fully explain, suggesting that there may be another source of excess noise which does not behave as TFN or Johnson noise.

\section{\label{sec:qet_res}QETs and Athermal Phonon Collection}

Now that we understand the basic concepts of a TES, we can further extend this technology to the QET (Quasiparticle-trap-assisted Electrothermal-feedback Transition-edge-sensor)~\cite{QET}. A QET is a TES that has been fabricated with a small region that overlaps with superconducting fin structures (typically Al). These fins will increase the total sensor surface area without increasing the sensor heat capacity. 

\begin{figure}
    \centering
    \includegraphics[width=\linewidth]{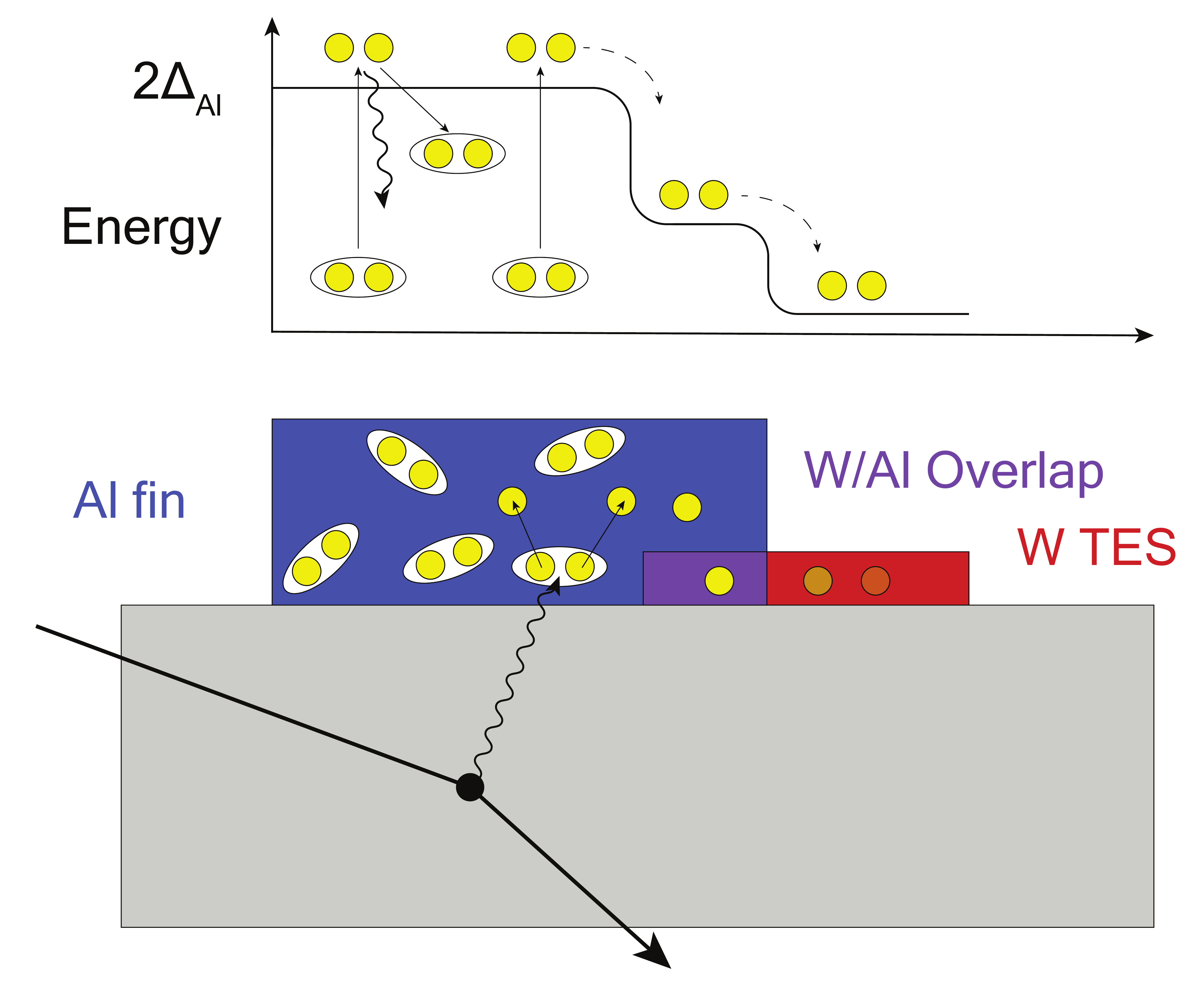}
    \caption{(Top) The spatial $T_c$ gradient of the superconducting bandgap in a QET. (Bottom) A cross section of a typical QET, showing quasiparticles being collected in the the W TES.}
    \label{fig:qetdiagram}
\end{figure}

With these fin structures, athermal excitations will interact with the superconducting Al fins and break Cooper pairs, creating quasiparticles. These quasiparticles will randomly travel throughout the Al fin, eventually reaching and becoming trapped in the overlap region, which will have a suppressed $T_c$ due to the W. These quasiparticles then travel into the W TES due to the spatial $T_c$ gradient, as depicted in Fig.~\ref{fig:qetdiagram} with a cross section of a typical QET shown. Because of the sharp $T_c$ variation, the majority of the potential energy of the Al quasiparticles will convert into kinetic energy and potential energy of the W quasiparticles, which then thermalize in the TES. The temperature of the TES will increase and then respond via the equations we derived in Section~\ref{sec:noisederiv}.

These athermal excitations come in the form of athermal phonons that are the product of some particle interaction in our substrate. The particle interaction will create electron-hole pairs and optical phonons, with the electron-hole pairs recombining into optical phonons. These optical phonons very quickly (on the order of fs to ps) downconvert to acoustic phonons, which then travel ballistically throughout the substrate that our QETs have been deposited on. As they travel throughout the substrate, they will eventually interact with a superconducting Al fin, breaking Cooper pairs, and starting the process discussed in the previous paragraph. In this process, there are efficiency losses at each interface (e.g. when phonons reflect of surfaces, when they enter into the Al fins, when quasiparticles travel between different regions of the QET), as well as efficiency losses due to downconversion of the phonons to energies below the Al superconducting bandgap, but the increased sensor coverage is worth these losses (which in total are on the order of $\varepsilon_{ph} \approx 10$--$20\%$). There is also a characteristic time constant $\tau_{ph}$ for collecting these athermal phonons, and we are no longer in the Dirac delta deposit regime, as discussed in Section~\ref{sec:noisemodeling}.

Thus, we must change our expected baseline energy resolution for collecting athermal phonons. Instead of the Dirac delta deposit, we will have a single-exponential (assuming one phonon collection time), which in Fourier space has the form
\begin{equation}
    p(\omega) = \frac{\varepsilon_{ph}}{1 + i \omega \tau_{ph}}.
\end{equation}
The expected baseline energy resolution equation from Eq.~(\ref{eq:dirac_eres}) becomes
\begin{equation}
    \sigma_E = \left[ \int_0^\infty \frac{\mathrm{d}\omega}{2 \pi} \frac{4\left| p(\omega)^2 \right|}{S_{P}(\omega)}\right]^{-1/2}.
    \label{eq:phonon_eres}
\end{equation}
Taking the same approximation that $S_p(\omega) \approx NEP(0)(1+ i \omega \tau_-)$, this equation can be analytically solved, giving
\begin{equation}
    \sigma_E \approx \frac{1}{\varepsilon_{ph}} NEP(0)\sqrt{\tau_{ph} + \tau_-},
\end{equation}
which is quite similar to Eq.~(\ref{eq:approxdirac}), but with $\tau_{ph}$ in the square root. In the case of the QET deposited on a substrate, this tells us that we must understand the phonon collection time $\tau_{ph}$ and the collection efficiency $\varepsilon_{ph}$ when calculating baseline energy resolutions.

In the next chapter, we will apply all of these concepts to a QET-based detector in order to fully characterize it.

\chapter{\label{chap:perf}Performance of a Cryogenic PhotoDetector}

In this chapter, I will present the motivation, design, and characterization of a large-area Cryogenic PhotoDetector (CPD). The analysis herein builds upon the TES characterization concepts discussed in Chapter~\ref{chap:two}, applying them to a fully-realized detector concept for active particle identification in rare event searches. An abbreviated version of this chapter was originally published in \textit{Applied Physics Letters} as Ref.~\cite{cpdcollaboration2020performance}.

\section{Motivation of a Large Area Photon Detector}

In rare event searches, experimental sensitivity is often limited by background signals~\cite{PhysRevLett.120.132501, PhysRevLett.124.122501, Andreotti:2010vj, EdelweissWIMP, Angloher_2017, edelweissHV, Agnese_2018, Abramoff_2019, PhysRevLett.123.181802, Abdelhameed_2019}. Developing precision detectors to veto background and noise signals has been a high priority in these fields.

\subsection{Neutrinoless Double Beta Decay Experiments}

\begin{figure}
    \centering
    \includegraphics[width=0.8\linewidth]{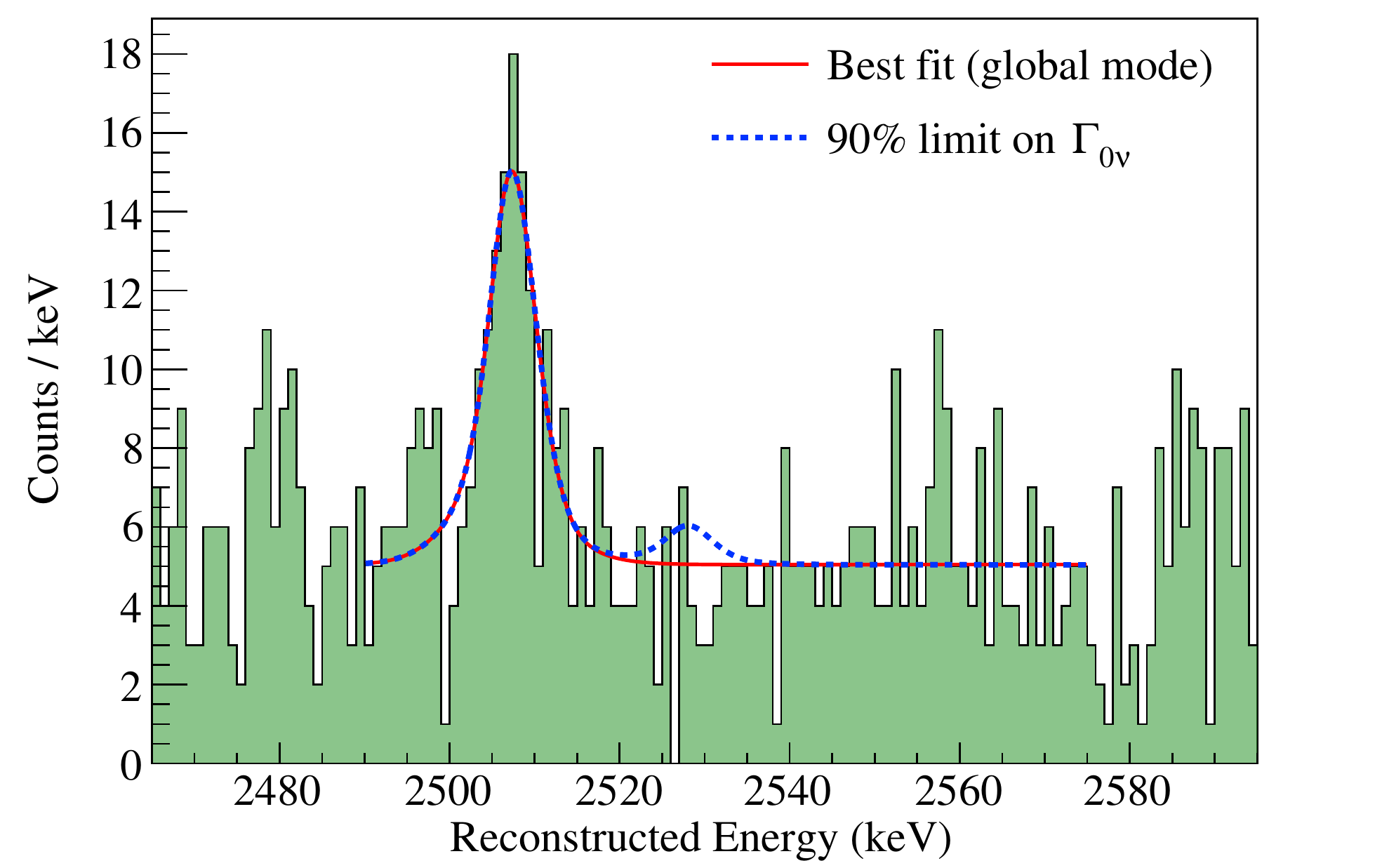}
    \caption{(Figure from Ref.~\cite{PhysRevLett.124.122501}) Limits from 2020 on $0\nu\beta\beta$ decay set by CUORE, showing the spectrum used for setting the limit. The dominant background is from $\alpha$ decays of the surrounding experimental setup.}
    \label{fig:0nbb_result}
\end{figure}

Much interest in low temperature cryogenic detector technology has been shown by groups carrying out searches for neutrinoless double beta decay~\cite{RevModPhys.80.481} ($0\nu\beta\beta$), such as the CUORE \cite{PhysRevLett.120.132501, PhysRevLett.124.122501}, CUPID~\cite{group2019cupid}, and AMoRE~\cite{Alenkov_2019} experiments. In these low-temperature calorimeters, the dominant source of background events consists of $\alpha$ decays from surface contaminants on the crystal and the surrounding environment~\cite{PhysRevLett.120.132501, PhysRevLett.124.122501, Azzolini:2019nmi}. In Fig.~\ref{fig:0nbb_result}, we show a recent result from CUORE which is indeed limited by these background signals~\cite{PhysRevLett.124.122501}.

It has been shown that Cherenkov emission or scintillation light can be used to positively identify the signal $\beta$s, allowing for background discrimination~\cite{beta_tag}. In order for these experiments to achieve a high level of rejection for these $\alpha$ backgrounds, photon detectors with large surface areas and baseline energy resolutions below $20\,\mathrm{eV}$ (RMS) for Cherenkov signals~\cite{beta_tag1}, or of $\mathcal{O}(100)\,\mathrm{eV}$ for scintillation signals~\cite{group2019cupid}, are required.

To reject the pileup background from multiple ordinary (two neutrino) double beta decay ($2\nu\beta\beta$) events, experiments need timing resolutions down to $10\,\mu\mathrm{s}$ (for the $^{100}$Mo isotope)~\cite{group2019cupid}.

\subsection{Other Applications}

There is also the theoretical and experimental motivation to search for dark matter in the mass range of $\mathrm{keV}/c^2$ to $\mathrm{GeV}/c^2\,$~\cite{Battaglieri:2017aum, PhysRevD.85.076007,dark2013,Alexander:2016aln}. However, current experiments have been limited by unknown background signals in the energy range of $\mathcal{O}$(1-100)$\,\mathrm{eV}$~\cite{EdelweissWIMP, Angloher_2017, edelweissHV, Agnese_2018, Abramoff_2019, PhysRevLett.123.181802, Abdelhameed_2019, PhysRevD.102.015017}. If the source of such backgrounds are high energy photons that deposit only an extremely small fraction of their energy in the target~\cite{PhysRevD.95.021301}, then a nearly $4\pi$ active shield composed of high-$Z$ scintillating crystals read out by these large area photon detectors could be highly efficient at suppressing these backgrounds. Additionally, a sensitive large area cryogenic detector could be useful for discriminating small energy depositions due to radiogenic surface backgrounds.

\begin{figure}
    \centering
    \includegraphics[width=0.6\linewidth]{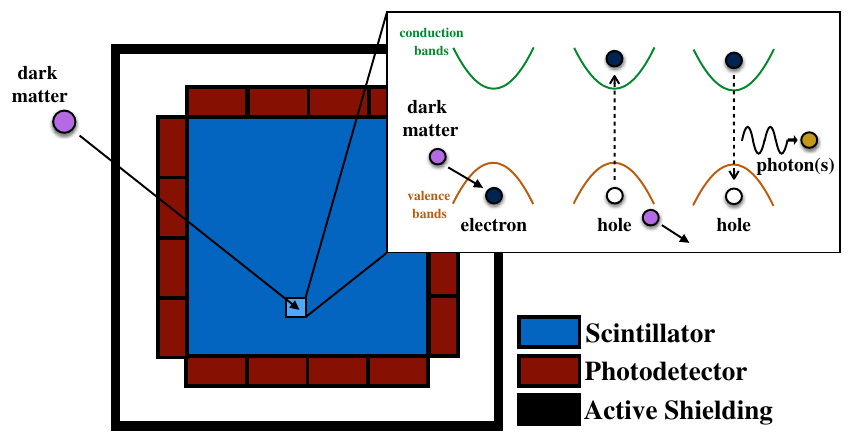}
    \caption[Scintillation diagram]{(Figure from Ref.~\cite{PhysRevD.96.016026}) A diagram of an DM-electron recoil with a scintillating crystal. The photodetectors would surround the scintillation material and read out the energy of emitted photon from the recoil.}
    \label{fig:scintillation}
\end{figure}

\begin{figure}
    \centering
    \includegraphics[width=0.5\linewidth]{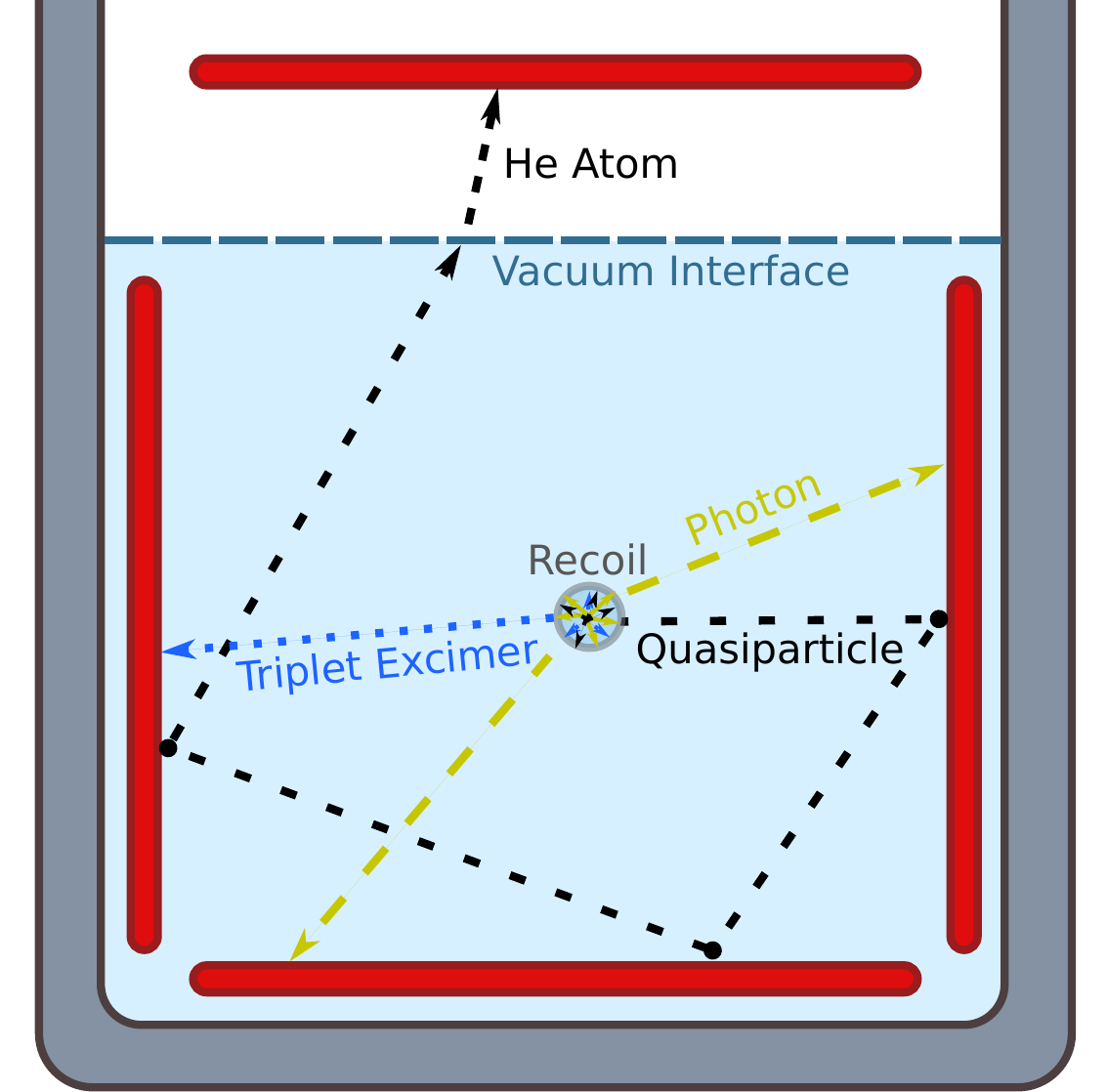}
    \caption[Superfluid $^4$He calorimetry]{(Figure from Ref.~\cite{PhysRevD.100.092007}): A diagram of the various byproducts of a recoil with a superfluid $^4$He (light blue). The photodetectors in this concept at the red rectangles, which could read out the various signals associated with some DM recoil.}
    \label{fig:he4}
\end{figure}

Other potential DM applications for this detector technology include searches for inelastic electronic recoils off scintillating crystals~\cite{PhysRevD.96.016026, KNAPEN2018386} and searches for interactions with superfluid $^4$He~\cite{PhysRevD.100.092007}, shown schematically in Figs.~\ref{fig:scintillation} and \ref{fig:he4}, respectively.

\section{QET-based Large Area Photon Detector}

\begin{figure}
    \centering
    \includegraphics[width=0.6\linewidth]{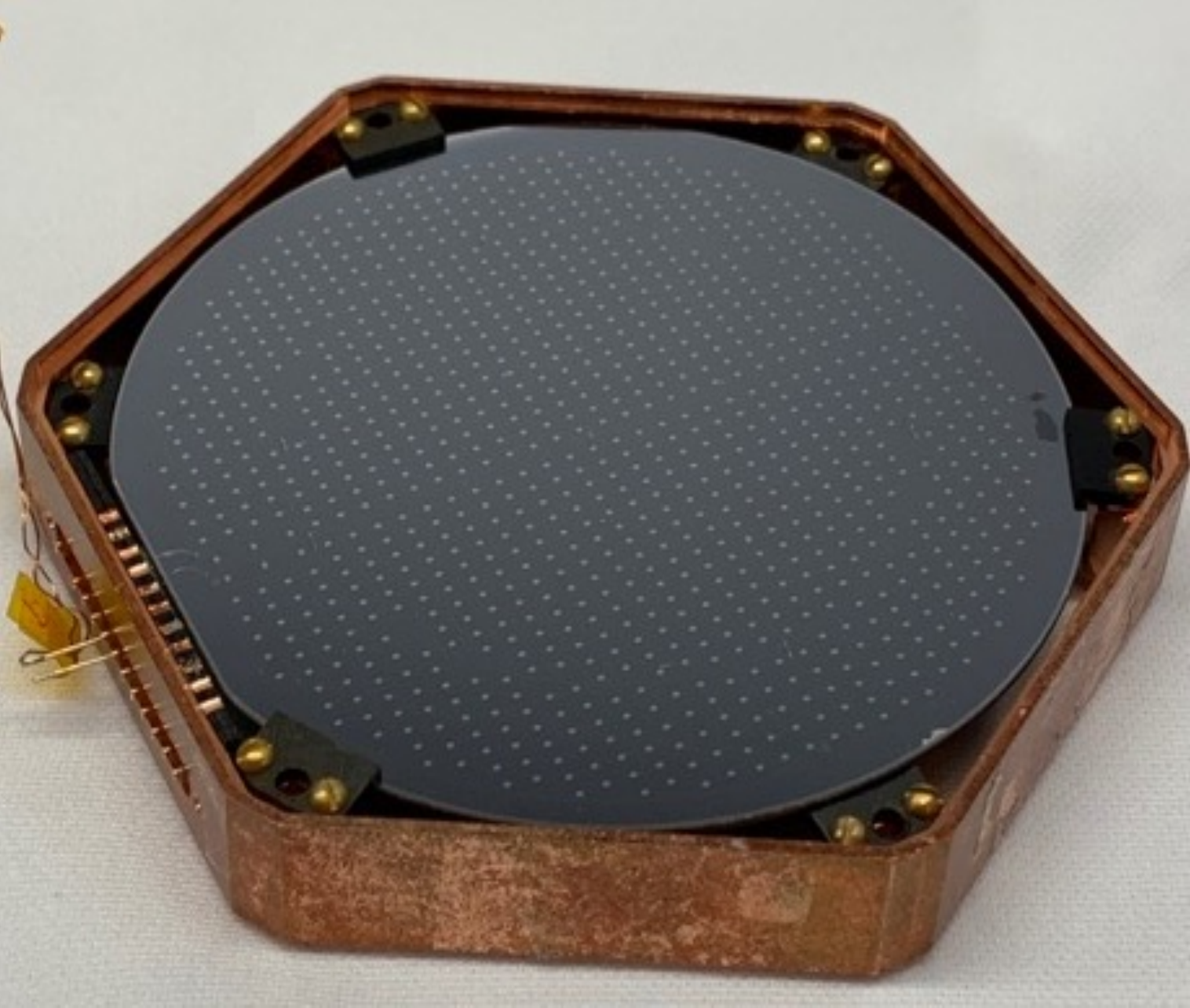}
    \caption[Photo of CPD]{A picture of the CPD installed in a copper housing. The instrumented side is shown facing up.}
    \label{fig:det}
\end{figure}

\begin{figure}
\begin{center}
(a)\includegraphics[width=0.5\linewidth]{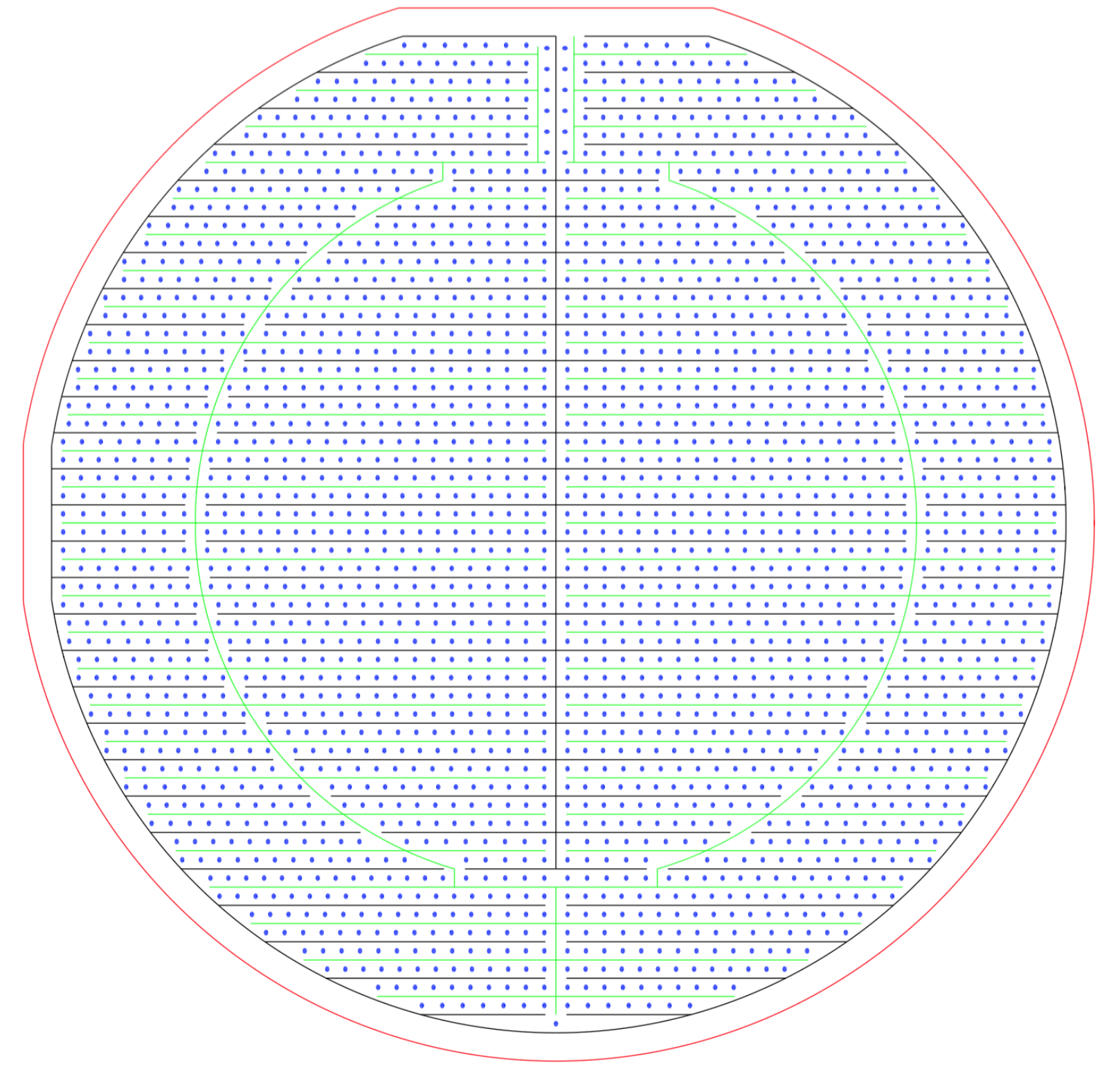}
(b)\includegraphics[width=0.35\linewidth]{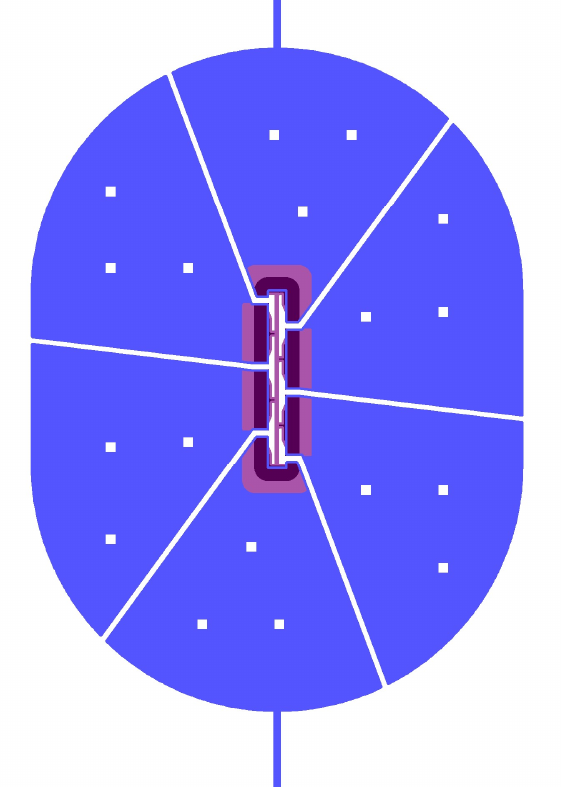} \\
\caption{Left: Mask design of the CPD. Right: The design of the QETs used for the detector. (Blue: Al fins, Purple: W TES.)}
\label{fig:det_qet}
\end{center}
\end{figure}

The (100)-oriented substrate of the CPD is a $10.6\,\mathrm{g}$ Si wafer of thickness $1\,\mathrm{mm}$ and a surface area of $45.6\,\mathrm{cm}^2$. A parallel network of 1031 QETs~\cite{irwin,QET} with $T_c=41.5\,\mathrm{mK}$ was deposited on one side of the wafer. The QETs are uniformly distributed over the wafer's surface and connected to a single readout channel. The uniform and distributed nature of the channel allows for the fast collection of athermal phonons with minimal positional dependence, reducing efficiency penalties from effects such as athermal phonon down-conversion~\cite{knaak, downconversion}. The opposite side of the Si wafer is unpolished and noninstrumented. The detector and QET mask design can be seen in Figs.~\ref{fig:det} and \ref{fig:det_qet}. In Table~\ref{tab:specs}, the QET design specifications for the CPD are listed.

\begin{table}
    \centering
    \caption[QET design specifications for the CPD]{QET design specifications for the CPD describing the W TESs and the Al fins that each QET consists of. The active surface area refers to the amount of substrate that is covered by the Al fins of the QETs, while the passive surface area is that which is not covered by the Al fins, but by the Al bias rails, bonding pads, and other structures that absorb athermal phonons, but do not add to the signal.}
    \begin{tabular}{lc}
    \hline \hline
    Specification & Value \\ \hline
    TES Length $[\mu\mathrm{m}]$ & 140 \\
    TES Thickness $[\mathrm{nm}]$ & 40 \\
    TES Width $[\mu\mathrm{m}]$ & 3.5 \\
    Number of Al Fins & 6 \\
    Al Fin Length $[\mu\mathrm{m}]$ & 200 \\
    Al Fin Thickness $[\mathrm{nm}]$ & 600 \\
    Al-W Overlap $[\mu\mathrm{m}]$ & 10 \\
    Number of QETs & 1031 \\
    Active Surface Area $[\%]$ & 1.9 \\
    Passive Surface Area $[\%]$ & 0.2 \\ \hline \hline
    \end{tabular}
    \label{tab:specs}
\end{table}

\section{Experimental Setup}

The detector was studied mostly by our Berkeley group at the SLAC National Accelerator Laboratory in a cryogen-free dilution refrigerator at a bath temperature ($T_B$) of $8\,\mathrm{mK}$. The detector was placed in a copper housing and was held mechanically with the use of six cirlex clamps. The cirlex clamps also provided the thermal link between the detector and the copper housing. The QET arrays were voltage biased and the current through the TES was measured with a DC superconducting quantum interference device (SQUID) array with a measured noise floor of ${\sim\!4~\mathrm{pA}/\sqrt{\mathrm{Hz}}}$, similar in design to the one described in Ref.~\cite{revc}.

\begin{figure}
    \centering
    \includegraphics[width=0.8\linewidth]{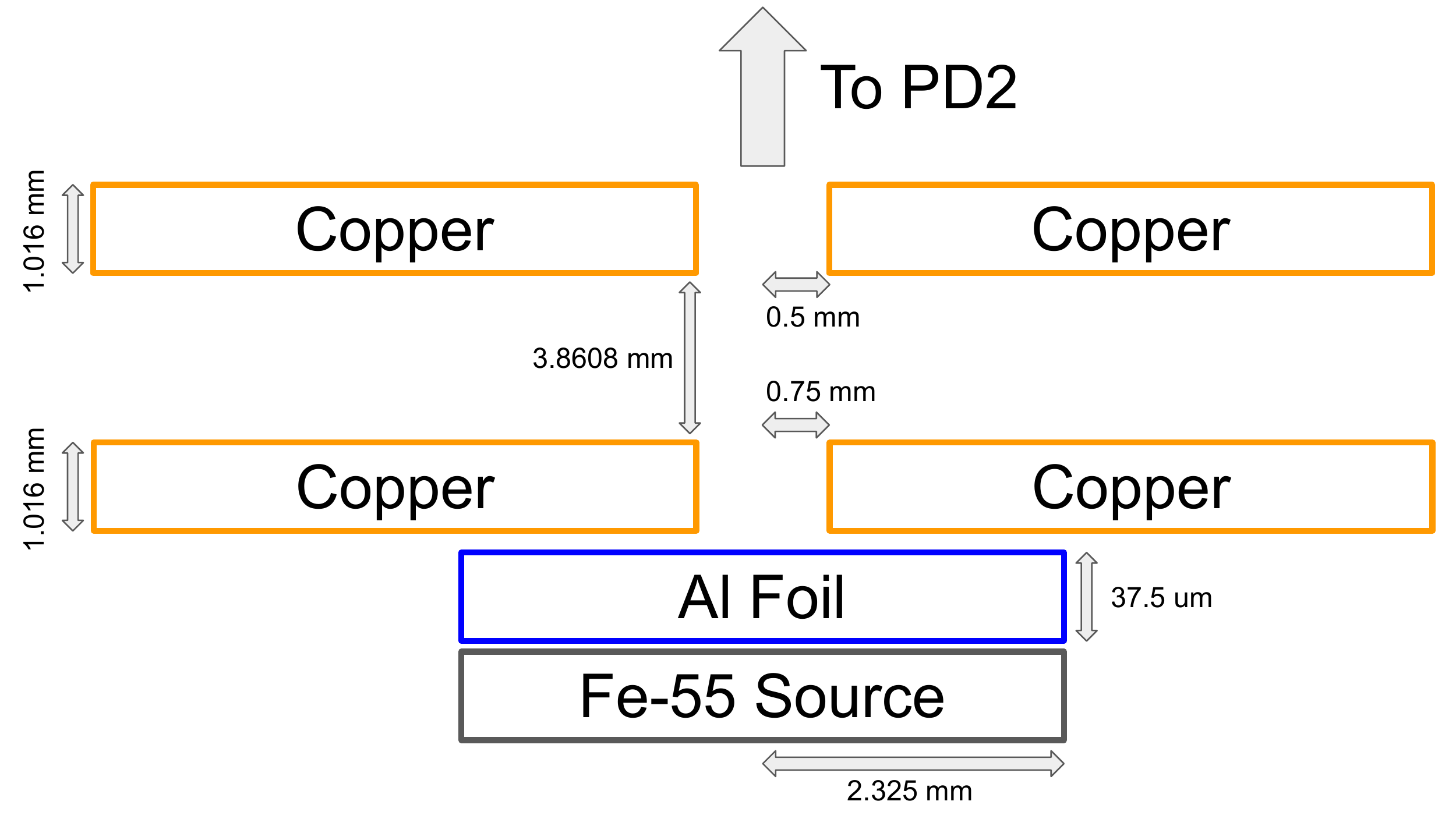}
    \caption[Collimator Design]{Design of collimator used in calibrating the CPD. Note that the diagram is not to scale. We have also included the various dimensions of the design, where horizontal measurements refer to radii.}
    \label{fig:collimator}
\end{figure}

A collimated $^{55}\mathrm{Fe}$ X-ray source was placed inside the cryostat and was incident upon the noninstrumented side of the CPD in the center of the detector. A layer of Al foil was placed inside the collimator to provide a calibration line from fluorescence at $1.5\,\mathrm{keV}$~\cite{alfluor, fe55}. The collimator was tuned such that there was $\sim\! 5\,\mathrm{Hz}$ of the $^{55}\mathrm{Fe}$ K$_\alpha$ and K$_\beta$ decays incident on the detector. A diagram of the collimator is shown in Fig.~\ref{fig:collimator}. The detector was held at a bath temperature ${T_B\ll T_c}$ for approximately two weeks to allow any parasitic heat added by the cirlex clamps to dissipate. During this time, we attempted to neutralize potential charged impurities within the Si wafer as much as possible with ionization produced by a $9.13\,\mu\mathrm{Ci}$ $^{137}\mathrm{Cs}$ source placed outside of the cryostat.

\section{Characterization of the CPD}

Following the characterization steps that have been outlined in Chapter~\ref{chap:two}, we can understand the DC characteristics, TES response, and noise of this device. Furthermore, we will discuss the phonon pulse shape for this athermal phonon sensor.

\subsection{DC Characteristics and TES Response}

\begin{figure}
    \centering
    \includegraphics{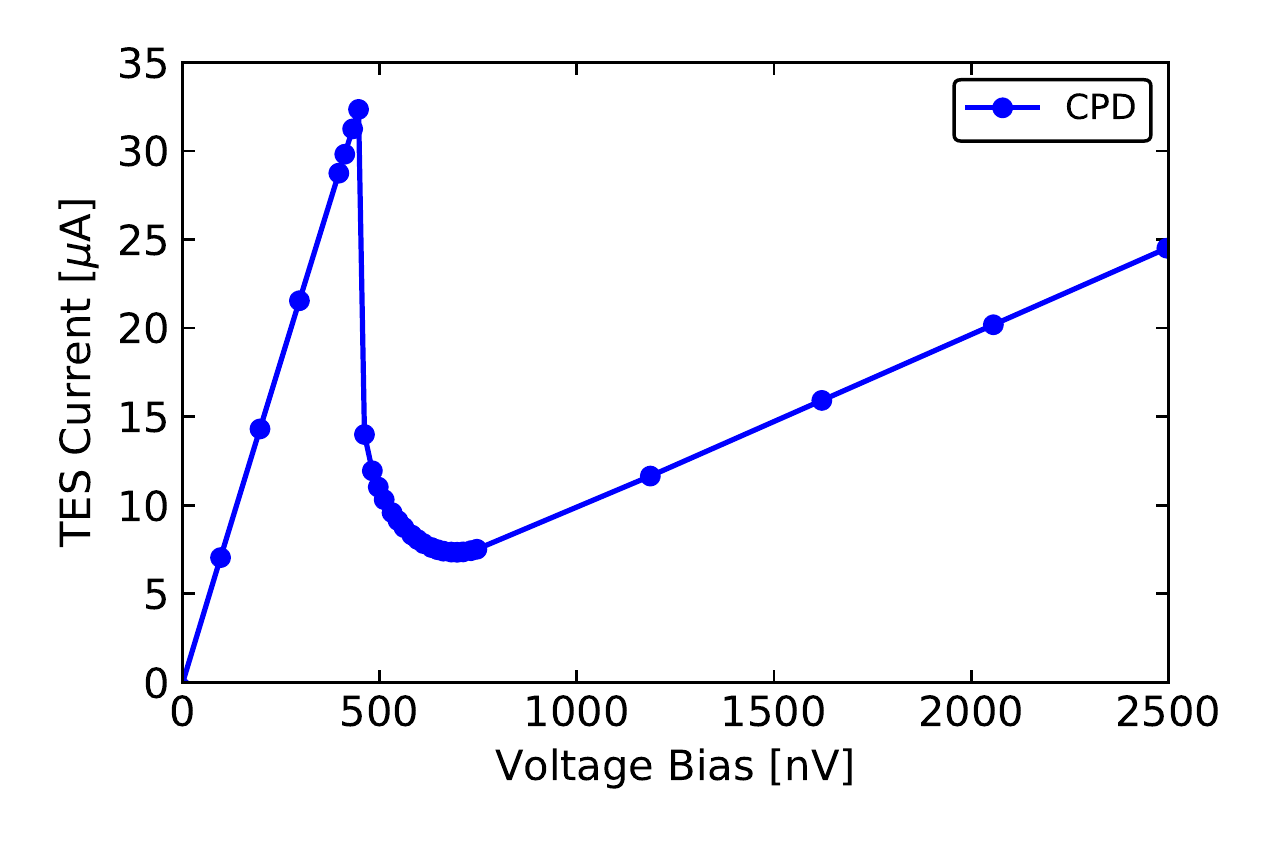}
    \caption{TES current as a function of voltage bias for the CPD at base temperature.}
    \label{fig:cpd_iv}
\end{figure}

\begin{figure}
    \centering
    \includegraphics{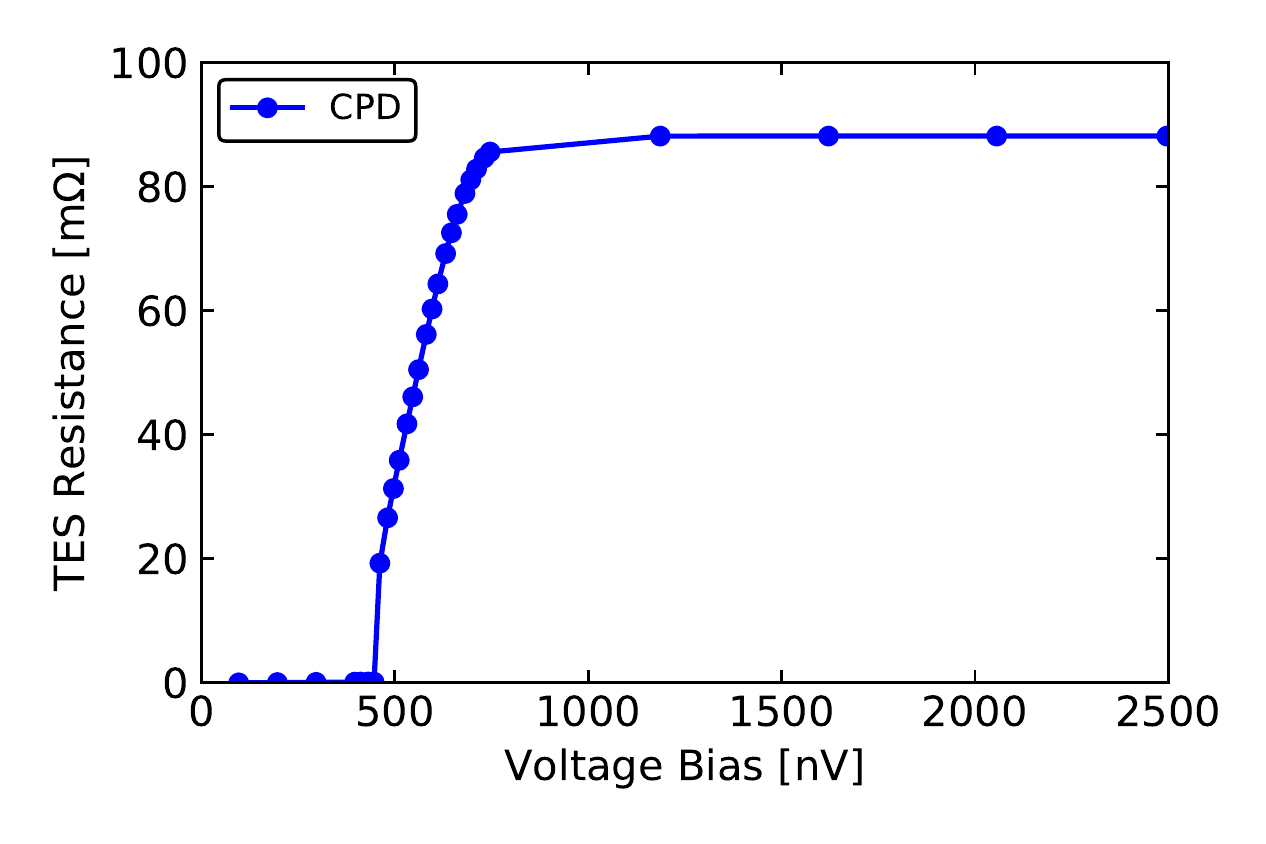}
    \caption{TES resistance as a function of voltage bias for the CPD at base temperature.}
    \label{fig:cpd_rv}
\end{figure}

\begin{figure}
    \centering
    \includegraphics{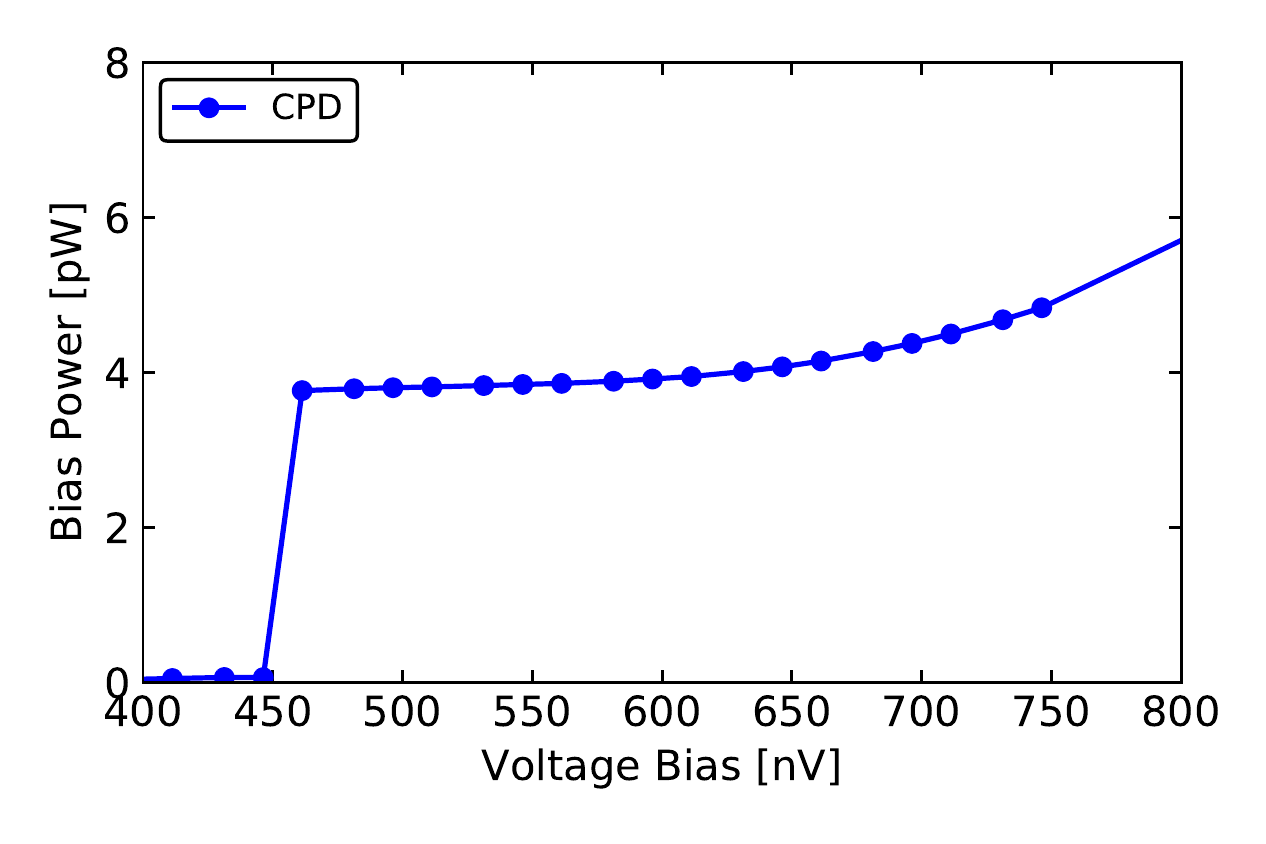}
    \caption{TES bias power as a function of voltage bias for the CPD at base temperature.}
    \label{fig:cpd_pv}
\end{figure}

To characterize the QETs, $IV$ sweeps were taken at various bath temperatures by measuring TES quiescent current as a function of bias current, with superimposed small square pulses providing complex admittance~\cite{irwin} at each point in the $IV$ curve~\cite{fink2020characterizing, Matt_thesis, Noah_thesis}.  Since all the QETs are connected in parallel in a single channel, the channel was treated as if it were a single QET, describing the average characteristics of the total array. The $IV$ data allowed for the estimation of the parasitic resistance in the TES line ($R_p$), the normal state resistance ($R_N$), and the nominal bias power ($P_0$). In Figs.~\ref{fig:cpd_iv}--\ref{fig:cpd_pv}, we show the TES current, resistance, and bias power as a function of voltage bias ($V_b = I_b R_{sh}$) at base temperature. The effective thermal conductance between the QETs to the Si wafer ($G_{TA}$) and $T_c$ were measured by fitting a power law to the measured bias power as a function of bath temperature~\cite{fink2020characterizing}, also discussed further in Appendix~\ref{chap:appa}. This measurement is a lower bound of both these values, as it assumes no parasitic bias power in the system. If there were parasitic power, then this would cause the measured $P_0$ curve as a function of temperature to be systematically shifted down as compared to the intrinsic curve, such that the measured values for $G_{TA}$ and $T_c$ are lower (see Appendix~\ref{chap:appa} for examples of these curves). We summarize the DC characteristics of the detector in Table~\ref{tab:rp}.

\begin{table}
    \centering
    \caption[DC characteristics of the TES]{DC characteristics of the TES from $IV$ curves, where a specific $P_0$ is reported at the bias point given by $R_0$. The systematic errors on $G_{TA}$ and $T_c$ represent the upper bounds on these values, using the hypothesis that the observed excess noise in the sensor bandwidth is entirely due to parasitic bias power.}
    \begin{tabular}{ll}
    \hline \hline
    Parameter & Value \\ \hline
    $R_{sh}\, [\mathrm{m}\Omega]$ & $5\pm 0.5$ \\
    $R_p\, [\mathrm{m}\Omega]$ & $8.7\pm 0.8$ \\
    $R_N\, [\mathrm{m}\Omega]$ & $88 \pm 10$ \\
    $G_{TA}\, [\mathrm{nW}/\mathrm{K}]$ & $0.48 \pm 0.04\,(\mathrm{stat.})^{+0.49}_{-0.00}\,(\mathrm{syst.})$ \\
    $T_c\, [\mathrm{mK}]$ & $41.5\pm 1.0\,(\mathrm{stat.})^{+10}_{-0}\,(\mathrm{syst.})$ \\
    $R_0\,[\mathrm{m}\Omega]$ & $31\pm 3$ \\
    $P_0\, [\mathrm{pW}]$ & $3.85 \pm 0.45$ \\
    \hline \hline
    \end{tabular}
    \label{tab:rp}
\end{table}

\begin{figure}
    \centering
    \includegraphics{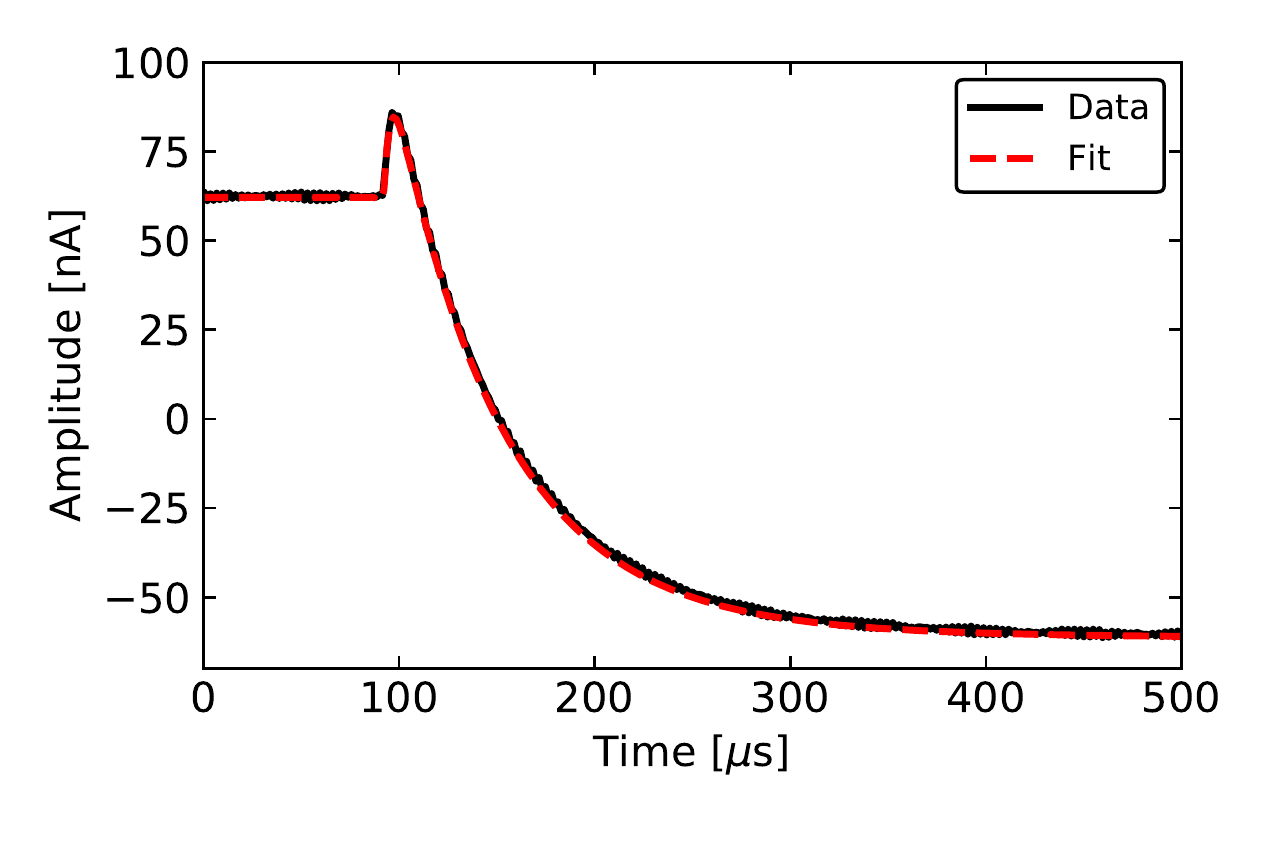}
    \caption[Complex impedance of the CPD in time domain]{The TES response of the CPD to a square wave jitter in time domain (black). The model for the fit (red) includes the extra time constant beyond the usual TES thermal time constant.}
    \label{fig:cpd_didv_time}
\end{figure}

\begin{figure}
    \includegraphics[width=\linewidth]{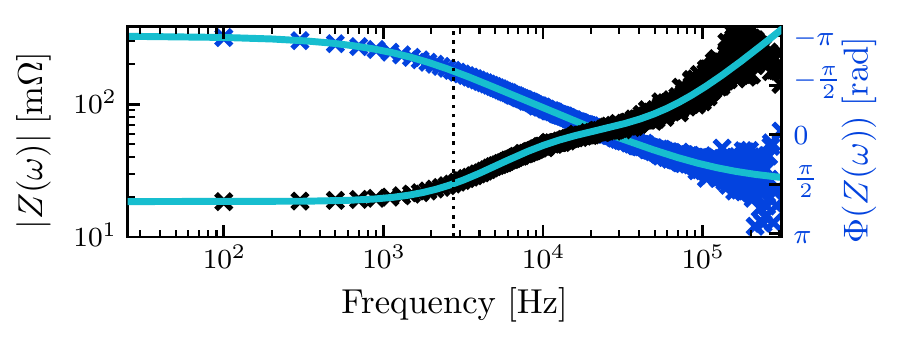}
    \caption[Complex impedance of the CPD]{The magnitude and phase of the measured complex impedance are shown as the black and blue markers, respectively. The modeled complex impedance is shown as the cyan solid line. The black dotted line denotes the corresponding bandwidth of ${2.7\,\mathrm{kHz}}$ for the thermal time constant ${\tau_-=58\,\mu\mathrm{s}}$.}
    \label{fig:dvdi}
\end{figure}

The complex admittance data allows us to estimate the dynamic properties of the sensors. Throughout the superconducting transition, primary and secondary thermal fall times were observed, e.g. $58\,\mu\mathrm{s}$ and $370\,\mu\mathrm{s}$, respectively, at $R_0\approx 35\%\,R_N$. The origin of this additional time constant could be due to a more complex thermal or electrical system, e.g. phase separation~\cite{2008JLTP..151...82C, Matt_thesis} or an extra heat capacity connected to the TES heat capacity~\cite{maasilta}. The electronic rise time of the TES at at $R_0\approx 35\%\,R_N$ was also measured to be $3 \, \mu\mathrm{s}$. Characteristic plots of the complex impedance of the TES circuit in time domain and frequency domain can be seen in Figs.~\ref{fig:cpd_didv_time} and \ref{fig:dvdi}, respectively. Note the excellent agreement between the fits and the model, showing that we are modeling are TES response well. We have summarized the TES parameters at this bias point in Table~\ref{tab:cpdresponse}, which will be used for noise modeling.

\begin{table}
    \centering
    \caption[TES response parameters]{TES response parameters from the complex impedance data for the bias point of $R_0\approx 35\%\, R_N$.}
    \begin{tabular}{ll}
    \hline \hline
    Parameter & Value \\ \hline
    $\tau_0 \,[\mu\mathrm{s}]$ & $1700\pm 200 $ \\
    $L\, [n\mathrm{H}]$ & $190 \pm 10$\\
    $\beta$ & $1.1\pm 0.1$ \\
    $\mathscr{L} $ & $80\pm 15$ \\
    \hline \hline
    \end{tabular}
    \label{tab:cpdresponse}
\end{table}

\subsection{Noise Modeling}

\begin{figure}
    \centering
    \includegraphics[width=0.8\linewidth]{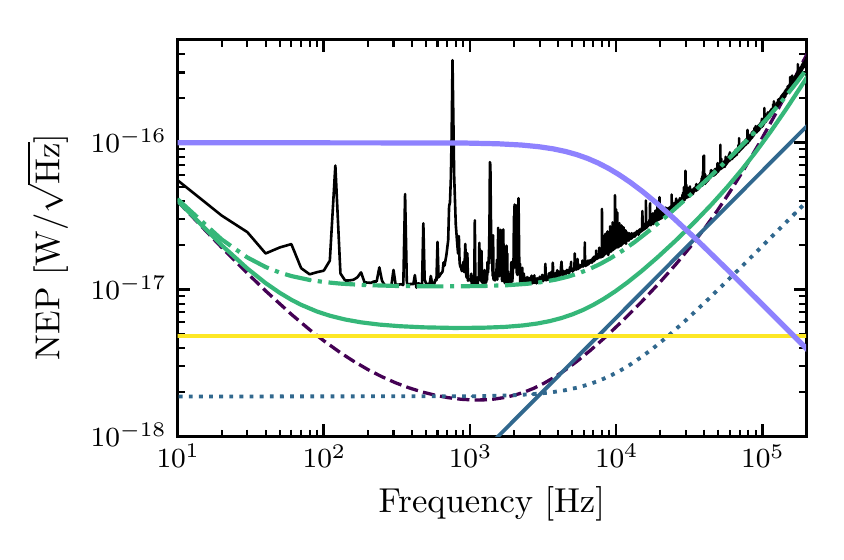}
    \caption[Noise equivalent power of the CPD]{Modeled noise components: TES Johnson noise (blue solid), load resistor Johnson noise (blue dots), electronics noise (purple dashed), thermal fluctuation noise (TFN) between the TES and the bath (yellow solid), and total modeled noise (green solid) compared with the measured NEP (black solid) for $R_0 \approx 35\% R_N$. We additionally show the total noise model (green alternating dashes and dots), which includes a hypothetical environmental noise source of ${8\times 10^{-18}\, \mathrm{W}/\sqrt{\mathrm{Hz}}}$ and excess TES Johnson noise with $M=1.8$. The light-purple line in upper portion of the figure denotes the power-pulse shape (arbitrarily scaled), which consists of a single pole at the observed rise time of $1 / \left(2 \pi \tau_\mathrm{ph}\right)=8\,\mathrm{kHz}$.}
    \label{fig:transition_noise}
\end{figure}

Knowledge of the TES parameters, given in Tables~\ref{tab:rp} and \ref{tab:cpdresponse}, allowed for the calculation of the power-to-current responsivity, which was used to convert the measured current-referred power spectral density (PSD) to the noise equivalent power (NEP). These parameters were used to predict the expected noise spectrum using the single-heat-capacity thermal model. A comparison of the NEP to the model at $R_0\approx 35\%R_N$ can be seen in Fig~\ref{fig:transition_noise}. The excess noise spikes above approximately $500\, \mathrm{Hz}$ have been experimentally confirmed to be largely caused by vibrations from the operation of the pulse tube cryocooler. The observed noise is also elevated above our model at frequencies in the effective sensor bandwidth interval (approximately the inverse of the thermal time constant~$\tau_-$~\cite{irwin}) by a factor of $\sim\!2$, as compared to the prediction. This ``in-band'' excess noise is consistent with two different hypotheses: a white power noise spectrum incident on the detector of $8\times 10^{-18}\, \mathrm{W}/\sqrt{\mathrm{Hz}}$ (e.g. a light leak) or a parasitic DC power in the bias circuit of approximately $6\,\mathrm{pW}$. If we assume the latter is the source, this allows us to calculate the upper bounds on our estimates of $G_{TA}$ and $T_c$, as reported in Table~\ref{tab:rp}. There remains bias-dependent excess noise above the sensor bandwidth. We parameterize the excess TES Johnson--like noise with the $M$ factor, as discussed in Section~\ref{sec:mfactor}. Using values of $M$ up to 1.8, depending on bias point, can account for the discrepancy between observation and prediction at these frequencies. We note that this excess noise could possibly also be explained with a more complex thermal model.

The lowest integrated NEP was achieved at an optimum bias point of $R_0=31\,\mathrm{m}\Omega\approx 35\%R_N$. In addition to the characterization data, approximately 500,000 threshold triggered events and 80,000 randomly triggered events were recorded at this bias. These data will be used to understand the phonon pulse shape of events, energy calibration, and baseline energy resolution.

\subsection{\label{sec:phononpulseshape}Phonon Pulse Shape}

For the measured phonon-pulse shape, there are multiple characteristic time constants. The pulse rise time was measured as ${\tau_{ph}=20\,\mu\mathrm{s}}$, which is the expected characteristic time scale for athermal phonons being absorbed by the Al collection fins of the QETs for this design. The dominant pulse fall time is consistent with the expectation from the complex impedance as we approach zero-energy, where we confirmed the expected thermal time constant ${\tau_-=58\,\mu\mathrm{s}}$ via a fit of the rise and fall times of the pulses. The secondary time constant from the complex impedance of $370\,\mu\mathrm{s}$ was also seen in these low-energy pulses. The secondary time constant from the complex impedance of $370\,\mu\mathrm{s}$ was also seen in these low-energy pulses, with an amplitude ratio of less than $2\%$ to the dominant decay exponential.

We observed an additional long-lived behavior in the pulses, which can be estimated as a low-amplitude $\sim\!3\,\mathrm{ms}$ exponential tail whose magnitude scales linearly with the event energy. This time constant is not seen in the complex impedance data, which implies that it is not due to intrinsic TES response. Due to phonon down-conversion, there will be a phonons with energies less than the superconducting Al bandgap. In this case, these phonons still have a chance of being directly absorbed by the W TES (as opposed to being caught by the Al fins). Because the W TES coverage is significantly smaller than the Al coverage, we would expect the corresponding time constant to be much slower. The ratio of Al coverage to W coverage is about a factor of 340, while the ratio of Al coverage to W coverage, including the Al/W overlap, is about a factor of 33. Estimating the subgap phonon collection time as this factor times the collection time of the above-gap phonons $\tau_{ph}=20\, \mu\mathrm{s}$ (the phonon collection time is a geometric quantity that is inversely proportional to sensor coverage, as shown, e.g., in C. W. Fink's thesis~\cite{finkthesis}), we find that the subgap phonon collection time could be in the range of $0.9$--$6.9 \, \mathrm{ms}$, depending on how much of the Al/W overlap is used, where $3\,\mathrm{ms}$ sits firmly in this range. Thus, this long-lived behavior might be due to direct absorption of phonons with energy smaller than the Al superconducting band gap into the TES, as previously suggested by K. D. Irwin \emph{et al.} in Ref.~\cite{QET}.

\begin{figure}
    \centering
    \includegraphics[width=0.8\linewidth]{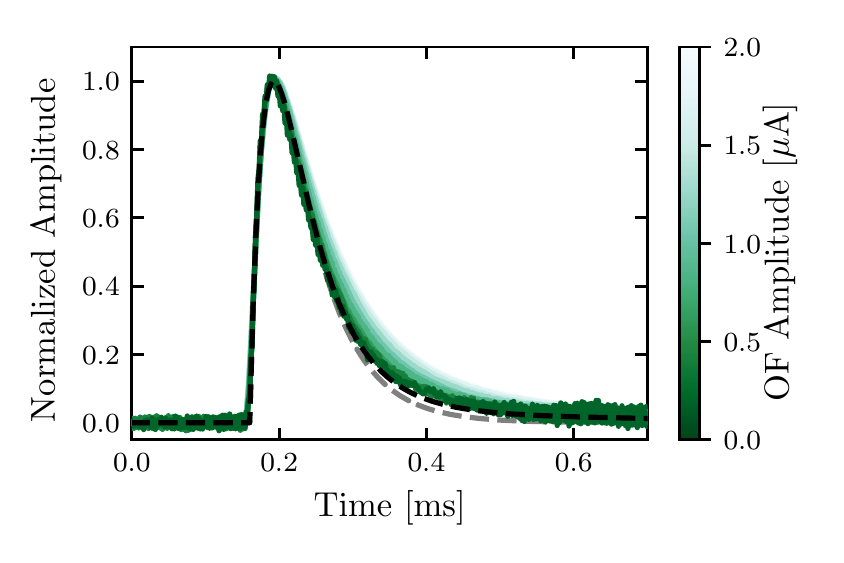}
    \caption[CPD phonon pulse shapes as a function of OF amplitude]{We show averaged pulse shapes (green solid) normalized by the fitted OF amplitude (which explains the small amount of dispersion of the maximum values at the peak current), for which the shade of green lightens with increased OF amplitude. For reference, $0.1\,\mu\mathrm{A}$ corresponds to about $0.1\,\mathrm{keV}$, $1.1\,\mu\mathrm{A}$ corresponds to about $1.5\,\mathrm{keV}$, and $2.0\,\mu\mathrm{A}$ corresponds to about $3.4\,\mathrm{keV}$. Each averaged pulse consists of about $100$ events averaged in $0.04\,\mu\mathrm{A}$ bin-widths. The lengthened fall time of the averaged pulse with increased OF amplitude (an energy estimator) is evident. The phonon-pulse template used in this analysis (black dashed) shows good agreement with the low energy (dark green) pulses. We also show an analytic phonon-pulse with only the first sensor fall time (gray dashed). Comparing to the phonon-pulse template, we see that the second sensor fall time has a small effect in this limited time interval.}
    \label{fig:pulse}
\end{figure}

For energies above $300\,\mathrm{eV}$, we observed a local saturation effect that manifests as the dominant fall time lengthening with increased energy. In Fig.~\ref{fig:pulse}, we show averaged pulses for various event amplitudes, showing the dependence of the pulse fall time on energy. We associate this effect with high-energy, single-particle events pushing nearby QETs into the normal resistance regime, slowing down the response of the total single-channel device. We also note that there is a position-dependent effect for a subset of high-energy events, notable by a varying fall time for events with the same amplitude. Our hypothesis for this phenomenon is that events close to the edge of the detector have less solid angle to deposit the energy, which leads to longer recovery times as opposed to events in the center of the detector (e.g. the calibration events). These effects are specific to the single-particle nature of the measured events. For scintillation events, the isotropic nature of the photons would spread out the event energy across the entire detector channel, avoiding these local saturation and position-dependent effects.

\section{\label{sec:ofcalibration}Energy Calibration and Resolution}

To reconstruct event energies, two energy estimators were used in this analysis: the optimal filter (OF) amplitude (see Appendix~\ref{chap:of} and Refs.~\cite{OF52,Gatti:1986cw,golwala}) and the energy removed by electrothermal feedback ($E_{\mathrm{ETF}}$)~\cite{irwin}. For the OF, we used an offline algorithm to reconstruct energies. A single noise spectrum was used, which was computed from the randomly triggered events. The phonon-pulse template used was an analytic template that matches the measured low-energy pulse shape, neglecting the $3\,\mathrm{ms}$ low-amplitude tail. Because we could not directly measure the low-energy phonon-pulse shape with high statistics, we used a template without the long-lived behavior.

The integral estimator $E_{\mathrm{ETF}}$ was calculated for each triggered event by measuring the decrease in Joule heating via
\begin{equation}
    E_{\mathrm{ETF}} = \int_0^{T_{trunc}} \left[(V_b-2I_0R_\ell)\Delta I(t) - \Delta I(t)^2R_\ell\right]\mathrm{d}t,
    \label{eq:eabs}
\end{equation}
where $T_{trunc}$ is the time at which the integral is truncated, $\Delta I(t)$ is the baseline-subtracted pulse in current, $I_0$ is the quiescent current through the TES, $R_\ell$ is the load resistance, and $V_b$ is the voltage bias of the TES circuit, as discussed at the end of Section~\ref{sec:tesresponse}.

In comparison to the OF amplitude, this integral estimator was less sensitive to saturation effects, but had a worse baseline energy resolution. When characterizing this device, we used the integral truncation of $T_{trunc}\approx7\tau_-$ for $E_{\mathrm{ETF}}$. This was done to preserve good baseline energy sensitivity in this integral estimator when calibrating the OF amplitude energy estimator at low energies.

\begin{figure}
    \centering
    \includegraphics[width=0.8\linewidth]{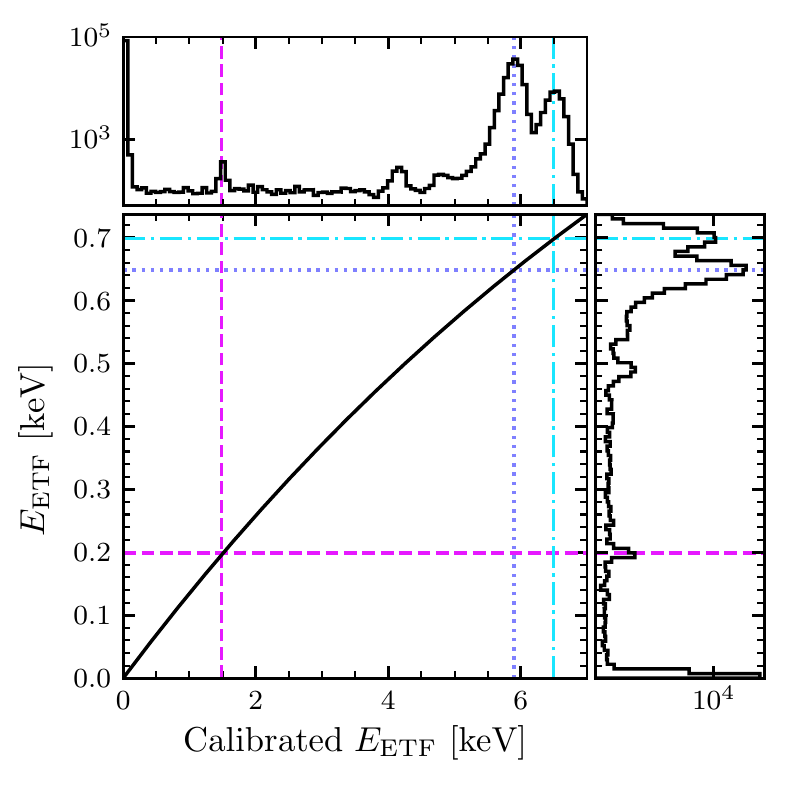}
    \caption[Calibration of $E_\mathrm{ETF}$]{Upper: The calibrated $E_\mathrm{ETF}$ (which estimates $E_\mathrm{true}$) spectrum for the CPD (solid black) grouped in bins of width $70\,\mathrm{eV}$. Right: The energy spectrum in $E_\mathrm{ETF}$ (solid black) grouped in bins of width $7.4\,\mathrm{eV}$. Lower left: The fitted saturation model using Eq.~(\ref{eq:sat}) (solid black). In each of these panels, we have shown, for both the calibrated and uncalibrated $E_\mathrm{ETF}$ energy scales, the location of the K$_\alpha$, K$_\beta$, and Al fluorescence calibration peaks (pink dashed, blue dotted, and cyan alternating dashes and dots, respectively). In the lower left panel, the intersections of the lines corresponding to each spectral peak represent the points used for calibration of $E_\mathrm{ETF}$ via Eq.~(\ref{eq:sat}). The unmarked peaks at $4.2\,\mathrm{keV}$ and $4.8\,\mathrm{keV}$ in calibrated $E_\mathrm{ETF}$ are the Si escape peaks~\cite{Reed_1972}.}
    \label{fig:spectrum}
\end{figure}

For pulse-shape saturation at high energies, we use the following empirical model:
\begin{equation}
    E_{\mathrm{ETF}} = a \left(1 - \mathrm{exp}\left( - \frac{E_\mathrm{true}}{b}\right)\right).
    \label{eq:sat}
\end{equation}
This functional form has the expected behavior: it intercepts zero, approaches an asymptotic value at high energies, and becomes linear for small values of $E_{\mathrm{true}}$. In Fig.~\ref{fig:spectrum}, the fitted saturation model, as well as the calibrated and uncalibrated $E_{\mathrm{ETF}}$ spectra, are shown, as compared to the energies of various spectral peaks in both energy scales. For the event spectra, we observed an unknown background at low energies, which we will discuss in detail in Chapter~\ref{chap:excess}.

\begin{figure}
    \centering
    \includegraphics{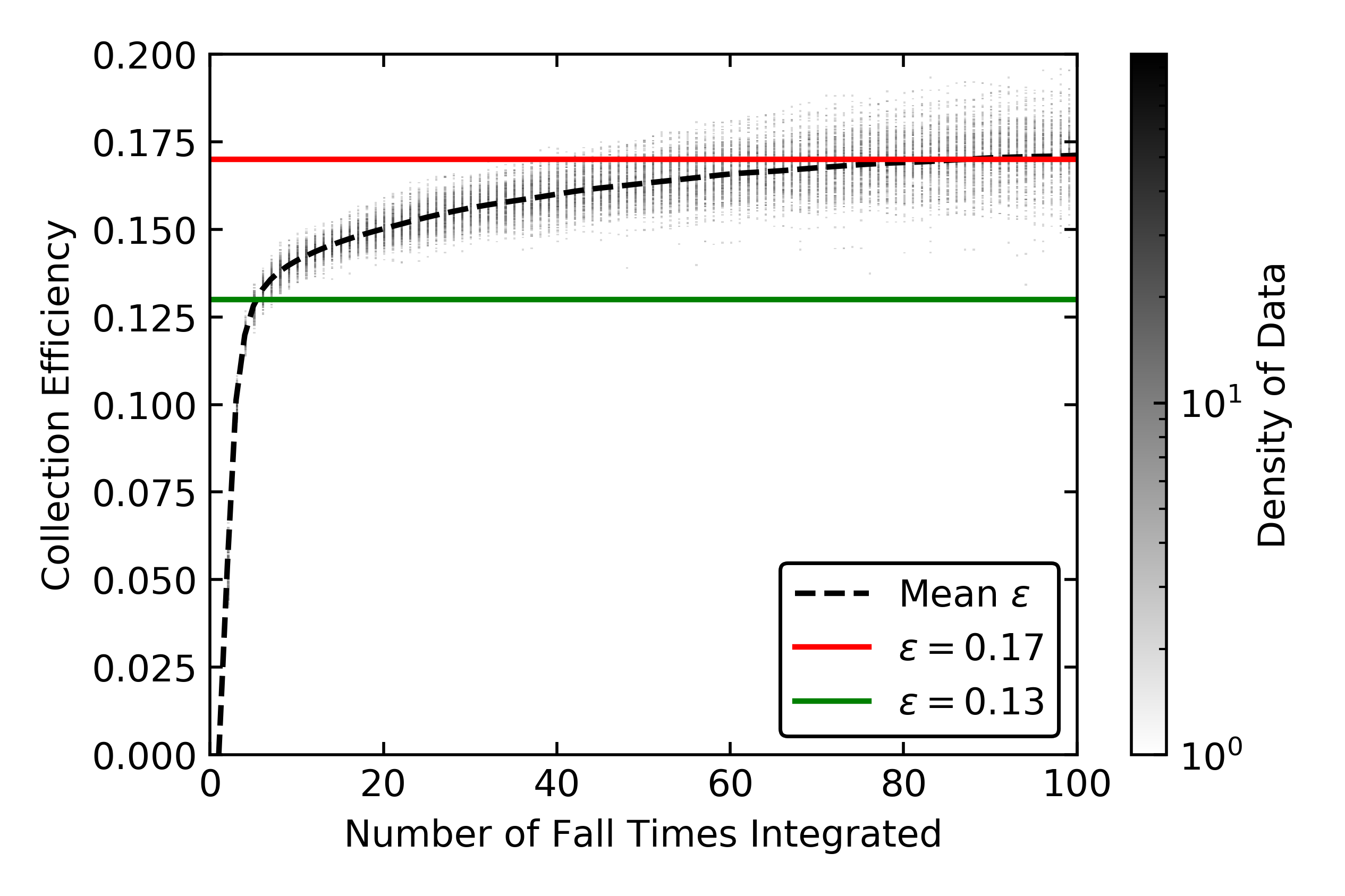}
    \caption{How the collection efficiency of the Al line changes with amount of integration of $E_\mathrm{ETF}$.}
    \label{fig:ceff_dependence}
\end{figure}

The absolute phonon collection efficiency ($\varepsilon_{ph}$) of the detector was estimated by measuring $E_{\mathrm{ETF}}$ at the lowest energy calibration line (Al fluorescence) and dividing by the known energy of that line. Because of the long-lived behavior in the phonon-pulse shapes, the measured collection efficiency of this detector depends on the integration truncation time $T_{trunc}$. If it is chosen to only include energy collected over a few times the first sensor fall time $\tau_-$ (e.g. ${T_{trunc}\approx7\tau_-}$), then we find that ${\varepsilon_{ph}=13\pm1\%}$. Alternatively, if we integrate to effectively infinity, this includes the low-amplitude long-lived behavior of the phonon pulses. In this case, the collection efficiency increases to ${\varepsilon_{ph}^\infty=17\pm1\%}$, which implies that about ${30\%}$ of the collected energy for a given event is associated with the low-amplitude tail of the phonon-pulse shape (about 8\% and 22\% from the $370\,\mu\mathrm{s}$ and $3\,\mathrm{ms}$ components, respectively). This is shown in Fig.~\ref{fig:ceff_dependence}, where we are plotting the collection efficiency calculated at the Al fluorescence line as a function of number of fall times $\tau_-$ integrated. Without the long-lived behavior of the phonon pulses, we would have expected the collection efficiency to asymptote at 13\%. Because we see the collection efficiency continue to increase long after the first sensor fall time, this shows the importance of the long-lived behavior in determining the amount of energy collected by the TES for an event.

\begin{figure}
    \centering
    \includegraphics{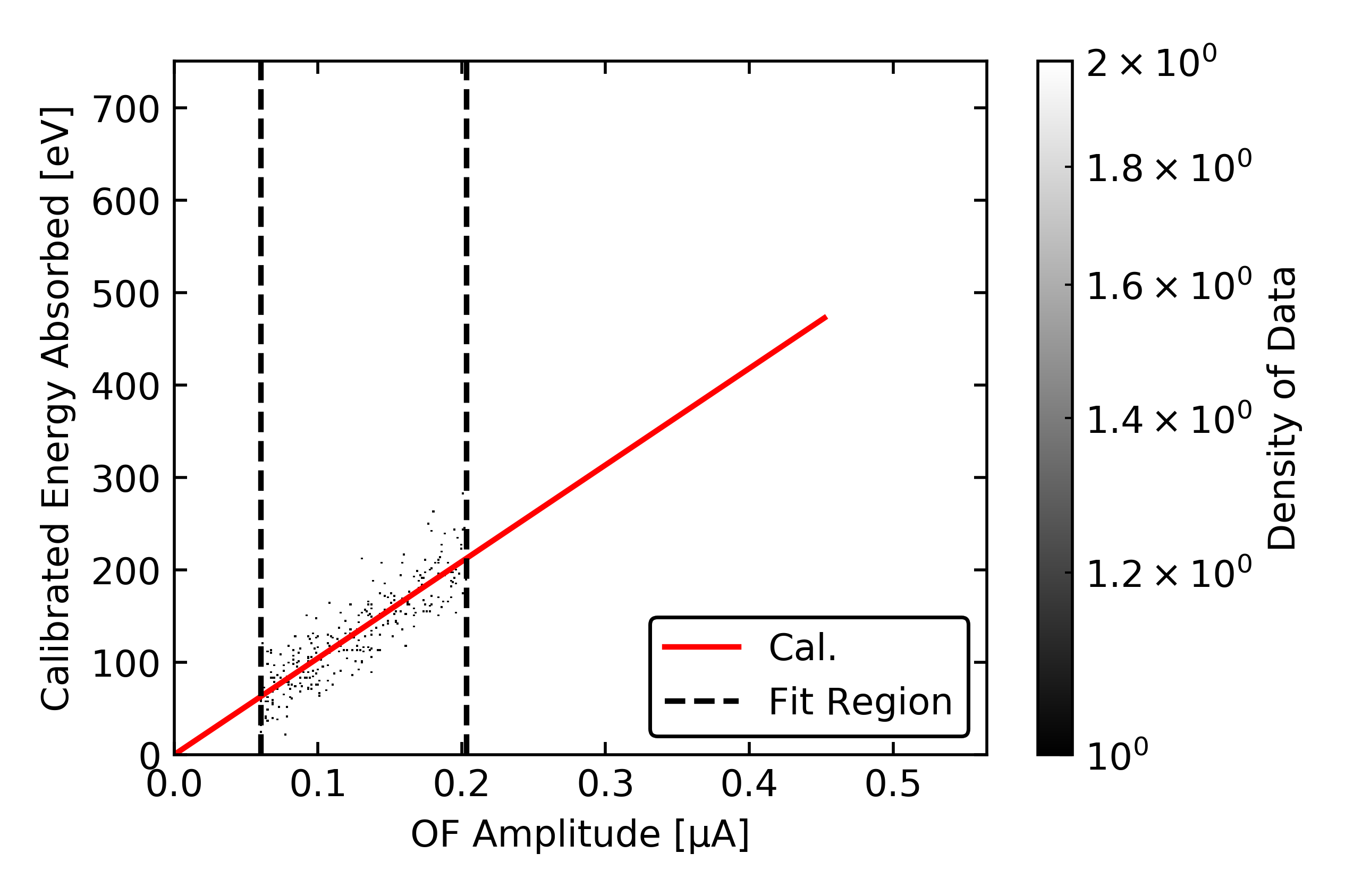}
    \caption{Correction of OF amplitude to energy, using linear region. Note that the variation of the black dashed lines that define the fit region result in negligible change in the calibration.}
    \label{fig:slope_fit}
\end{figure}

To calibrate the OF amplitude to units of energy, we fit the relationship between the calibrated $E_{\mathrm{ETF}}$ and the OF amplitude to a linear slope at low energies (below approximately $300\,\mathrm{eV}$), as shown in Fig.~\ref{fig:slope_fit}. The value of the slope of the red line in the plot gives a calibration constant of $1.045 \pm 0.009 \ \mathrm{keV}/\mu\mathrm{A}$. This method does not provide a calibration of the OF amplitude at high energies, but allows for the calculation of the baseline energy resolution.

\begin{figure}
    \centering
    \includegraphics{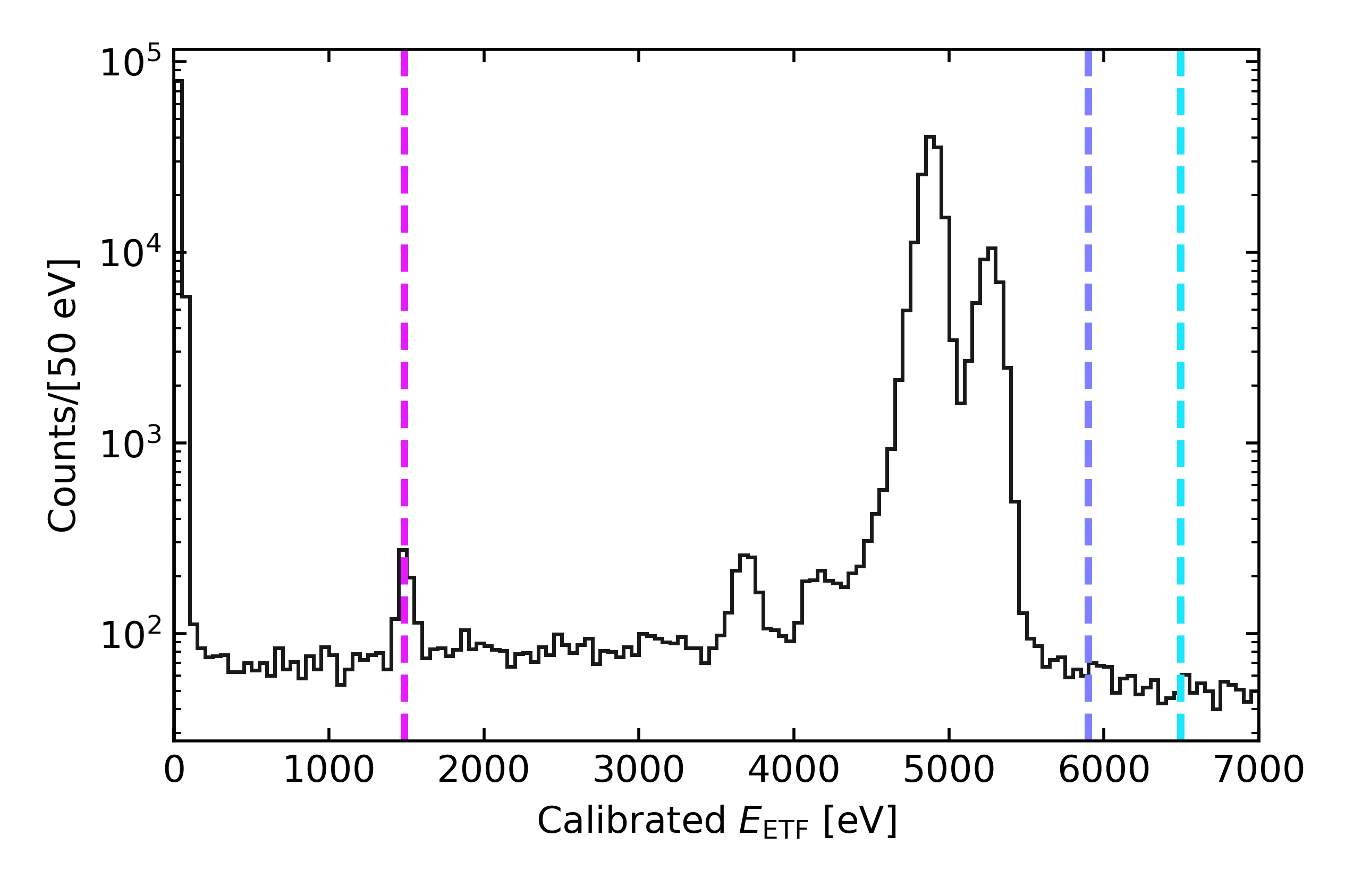}
    \caption{Calibrated $E_\mathrm{ETF}$ energy spectrum if calibrating directly to the Al line via Eq.~(\ref{eq:eetf_syst}). This alternative calibration determines the systematic upper bound on the baseline energy resolution reported for the CPD.}
    \label{fig:syst_spectrum}
\end{figure}

For the calibration method used, the main source of systematic error is the saturation model in Eq.~(\ref{eq:sat}). Since it is empirical, its use introduces uncertainty in its applicability. We can estimate a rough upper bound of the effect of this systematic on the baseline energy resolution as the value that would be reached if we instead calibrated $E_\mathrm{ETF}$ linearly using the Al fluorescence line. That is, the formula for calibration in this case would be
\begin{equation}
    E_\mathrm{ETF} = a \cdot E_\mathrm{true}|_{\mathrm{Al \ line}}.
    \label{eq:eetf_syst}
\end{equation}
In this case, this worsens the baseline energy resolution, as we are not taking into account the expected response that is seen in Fig.~\ref{fig:spectrum}, and we know that there is saturation from our observations (however, we will use this is a conservative upper bound nonetheless). The full calibrated $E_\mathrm{ETF}$ spectrum would then become what is shown in Fig.~\ref{fig:syst_spectrum}, where the true locations of the $^{55}$Fe K$_\alpha$ and K$_\beta$ lines are clearly no longer reconstructed well. At low energies, this also has the effect of increasing the conversion factor from OF amplitude to units of energy to $1.098 \pm 0.009 \ \mathrm{keV}/\mu\mathrm{A}$. Because the baseline energy resolution is simply the OF amplitude baseline energy resolution in units of Amperes multiplied by this conversion factor, this shows that this calibration method returns a worsened baseline energy resolution when not treating saturation effects with our model in Eq.~(\ref{eq:sat}). This worsened baseline energy resolution will henceforth be reported as the upper bound of the systematic error on this value.

The baseline energy resolution was calculated as the RMS of 46,000 randomly triggered events, after removing data contaminated by pileup events, electronic glitches, or thermal tails. This gave a resolution of ${\sigma_E=3.86\pm0.04\,(\mathrm{stat.})^{+0.19}_{-0.00}\,(\mathrm{syst.})\,\mathrm{eV}}$ (RMS) for the OF energy estimator, where these data are consistent with a normal distribution. This is in agreement with our estimation from the observed NEP and the power-referred phonon-pulse shape (a single-exponential with fall time $\tau_{ph}$ and collection efficiency $\varepsilon_{ph}$), which gave an expected baseline energy resolution of ${\sigma_E^{th}=3.9\pm 0.4\,\mathrm{eV}}$ (RMS), as discussed in Section~\ref{sec:qet_res} and was similarly done in Ref.~\cite{fink2020characterizing}.

Using the OF formalism, we can also calculate the expected timing resolution $\sigma_t$ (see Section~\ref{sec:timingres}) of the CPD, which provides an estimate of the minimum resolving time for two pileup events (i.e. we would expect to be able to determine that two pulses spaced a few $\sigma_t$ apart to be detectable as two pulses). For a $5\sigma_E$ event, the corresponding timing resolution of this detector is $2.3\, \mu\mathrm{s}$, and this timing resolution improves for higher energies, as it is inversely proportional to energy. For many $0\nu\beta\beta$ experiments, the minimum resolving time requirement to make pileup of multiple $2\nu\beta\beta$ events a negligible background is on the order of $1\,\mathrm{ms}$~\cite{Chernyak, artusa, pyle_CRESST, Casali_2019}. For the CUPID and CUPID-1T experiments, this requirement is about $300\,\mu\mathrm{s}$ and $10\,\mu\mathrm{s}$, respectively~\cite{group2019cupid}. In the future, one should study the minimum resolving time using an OF-based pileup detection algorithm with a simulation based on the expected $2\nu\beta\beta$ spectrum for $^{100}$Mo, where one could compare the $\chi^2$ goodness-of-fit metric from the OF fits for one pulse and two pulses (i.e. calculate the $\Delta \chi^2$) and find the minimum separation time of two pulses for which this comparison breaks down.

\section{Discussion}

When comparing the baseline energy resolution of the CPD to the requirements of the CUPID experiment, the value surpasses the requirement of less than ${20\,\mathrm{eV}}$ (RMS) by a factor of five. While the CPD is a TES-based detector, it has been shown that Microwave Kinetic Inductance Detectors (MKIDs) and Neutron-Transmutation-Doped (NTD) Ge detectors are also promising avenues for achieving the sub-${20\,\mathrm{eV}}$ baseline goal. In Table~\ref{tab:comp}, we report this result alongside those of other detectors for this application. In comparison to the devices that have met or exceeded the requirement, the CPD does not require Neganov-Trofimov-Luke (NTL) amplification~\cite{Neganov:1985khw, doi:10.1063/1.341976} (which often results in excess dark counts due to current injection at large electric field) and has the best baseline energy sensitivity for its size.

\begin{table}
    \centering
    \caption[Comparison of the CPD to various other devices]{Comparison of this work to various state-of-the-art devices for degraded $\alpha$ rejection in $0\nu\beta\beta$ experiments. The table is sorted by decreasing $\frac{\sigma_E}{\sqrt{\mathrm{Area}}}$, a common figure-of-merit of devices for this application, for which smaller is better. The column labeled ``NTL?'' denotes whether or not each detector relies on NTL amplification to achieve the corresponding result.}
    \begin{tabular}{lcccc}
    \hline \hline
    \rule{0pt}{10pt} Device & Area $\left[\mathrm{cm}^2\right]$ & $\sigma_E$ $\left[\mathrm{eV}\right]$ & $\frac{\sigma_E}{\sqrt{\mathrm{Area}}}$ $\left[\frac{\mathrm{eV}}{\mathrm{cm}}\right]$ & NTL? \\ \hline
    MKID~\cite{Cardani_2018} & 4.0 & 26 & 13 & No \\
    W-TES~\cite{SCHAFFNER201530} & 12.6 & 23 & 6.5 & No \\
    Ge-NTD~\cite{BARUCCI2019150} & 15.6 & 20 & 5.1 & No \\
    Ge-NTD~\cite{Pattavina_2015} & 19.6 & 19 & 4.3 & Yes \\
    IrAu-TES~\cite{Willers_2015} & 4.0 & 7.8 & 3.9 & Yes \\
    Ge-NTD~\cite{Armengaud_2017}& 4.9 & 7.6 & 3.5 & Yes \\
    Ge-NTD~\cite{PhysRevC.97.032501} & 15.2 & 10 & 2.6 & Yes \\
    Ge-NTD~\cite{Novati_2019} & 15.2 & 8 & 2.1 & Yes \\
    W-TES~\cite{2018JLTP..193.1160R} & 12.6 & 4.1 & 1.2 & No \\
    W-TES (this) & 45.6 & 3.9 & 0.6 & No \\ \hline \hline
    \end{tabular}
    \label{tab:comp}
\end{table}

The measured baseline energy resolution of ${3.86\pm0.04\,(\mathrm{stat.})^{+0.19}_{-0.00}\,(\mathrm{syst.})\,\mathrm{eV}}$ and the expected timing resolution of ${2.3\, \mu\mathrm{s}}$ (at $5\sigma_E$), combined with its large surface area, makes this detector an excellent candidate for background rejection in both $0\nu\beta\beta$ and DM experiments. Because of the energy sensitivity, this device can be used as a dark matter detector itself, as we have done in collaboration with SuperCDMS to set limits on spin-independent dark matter-nucleon interactions for sub-$\mathrm{GeV}/c^2$ dark matter particle masses~\cite{PhysRevLett.127.061801}. Similarly, this 10-gram--scale device could be applied to coherent elastic neutrino-nucleus scattering experiments~\cite{coh}. The performance of the CPD can be further optimized through adjustment of characteristics such as the Al-W overlap and overall Al coverage. From these considerations, we anticipate up to a factor of two improvement in baseline energy resolution for a future iteration of the CPD, which has been designed by S. Zuber and C. W. Fink, and is currently being tested (as of 2022).

\chapter{A Light Dark Matter Search Operated Above Ground}

In this chapter, we use the CPD to carry out an exclusion analysis that set the most stringent (comparing to contemporary searches) dark matter-nucleon scattering cross-section limits achieved by a cryogenic detector for dark matter particle masses from $93$ to $140\,\mathrm{MeV}/c^2$, with a raw exposure of $9.9\,\mathrm{g\,d}$ acquired at an above-ground facility. This work illustrates the scientific potential of detectors with athermal phonon sensors with eV-scale energy resolution for future dark matter searches. An abbreviated version of this chapter was originally published in \textit{Physical Review Letters} as Ref.~\cite{PhysRevLett.127.061801}.

\section{Introduction}

As detailed in Chapter~\ref{chap:dm}, numerous observations have shown that the majority of the Universe is composed of dark matter~\cite{pdg, planck,Clowe_2006}, for which the WIMP~\cite{wimp} has long been a favored candidate for this dark matter. However, direct detection experiments have ruled out a significant portion of the most compelling WIMP parameter space~\cite{xenon1t_wimp,PhysRevLett.118.021303,PhysRevLett.119.181302}, which has provided the motivation for both theoretical and experimental exploration of alternative DM models~\cite{Battaglieri:2017aum}. In this chapter, we will detail a search for light dark matter (LDM) with a mass in the $\mathrm{keV}/c^2$ to $\mathrm{GeV}/c^2$ range and coupling to Standard Model particles via a new force mediator, which provides a well-motivated alternative to the WIMP hypothesis~\cite{PhysRevD.85.076007,dark2013,Alexander:2016aln}. While many LDM searches focus on DM-electron interactions~\cite{Agnese_2018,Abramoff_2019,PhysRevLett.123.181802,xenon10,xenon1t_erdm}, detectors with eV-scale energy thresholds can also be used to study LDM via DM-nucleon interactions.

Though the CPD was designed for active particle identification in rare event searches, we will use its excellent baseline energy resolution to probe LDM parameter space. As a combined effort of the SuperCDMS and CPD collaborations, a DM search was carried out with $9.9\,\mathrm{g\,d}$ of raw exposure from Sept. $9^\mathrm{th}$ to $10^\mathrm{th}$ 2018. The data were acquired at the SLAC National Accelerator Laboratory in a surface facility of $\sim\!100\,\mathrm{m}$ in elevation. In the following sections, we will discuss data acquisition techniques, device performance, and the results of an exclusion analysis for spin-independent DM-nucleon interactions.

\section{Experimental Setup and Data Acquisition}

The experimental setup for the DM search is unchanged as compared to that described in Chapter~\ref{chap:perf}, as the DM search dataset was taken as part of the same analysis run as the characterization of the detector. Because of project time constraints and large cosmogenic backgrounds, the DM search was limited to $22\,\mathrm{hrs}$. Data were acquired over this period using a field-programmable gate array (FPGA) triggering algorithm based on the optimal filter (OF) formalism (see Appendix~\ref{chap:of} and Refs.~\cite{OF52,Gatti:1986cw,golwala}). Throughout the exposure, randomly triggered samples of the baseline noise were acquired (``in-run random triggers''), which allowed us to observe any changes in the noise over the course of the search and to calculate and monitor the baseline energy resolution.

Using the OF formalism (see Appendix~\ref{chap:of}), the trigger threshold was set at $4.2\,\sigma_E$ above the normally-distributed baseline noise level (i.e. the baseline energy resolution of randomly triggered data using the OF with no delay), corresponding to $16.3\,\mathrm{eV}$ after calibration. Any pulse with an OF amplitude above this threshold was tagged as an event and saved. The phonon-pulse template used for the FPGA triggering algorithm was a double-exponential pulse with a rise time of $\tau_r = 20 \, \mu\mathrm{s}$ and a fall time of $\tau_f = 58\, \mu\mathrm{s}$. The rise time was taken from the expected collection time of athermal phonons, and the fall time was taken from the thermal response time of the QET estimated from a measurement of the complex admittance~\cite{irwin}. Each of these time constants was confirmed by a nonlinear least squares fit to nonsaturated pulses. Although Chapter~\ref{chap:perf} discusses the existence of extra fall times, their effect on the OF amplitude measured for each event is negligible. Before starting the DM search, a separate, small subset of random triggers was collected. After removing data contaminated by effects such as elevated baselines and phonon pulses, the noise spectrum used by the FPGA algorithm was generated from these random triggers.

For overlapping triggered pulse traces, the triggering algorithm was set to save a trace centered on the pulse with the largest OF amplitude. We note that the FPGA triggering algorithm acted on a trace that was downsampled by a factor of 16, from the digitization rate of $625\,\mathrm{kHz}$ to $39\,\mathrm{kHz}$. Additionally, the FPGA triggering algorithm considered only $26.2\,\mathrm{ms}$ of the total $52\,\mathrm{ms}$ long time trace saved for each triggered pulse trace (``event''). Because of these factors, the energy resolution of the FPGA triggering algorithm is not as good as can be achieved by reconstructing event energies using an offline OF, as described in the following sections.

\section{Data Reconstruction and Calibration}

\begin{figure}
    \centering
    \includegraphics[]{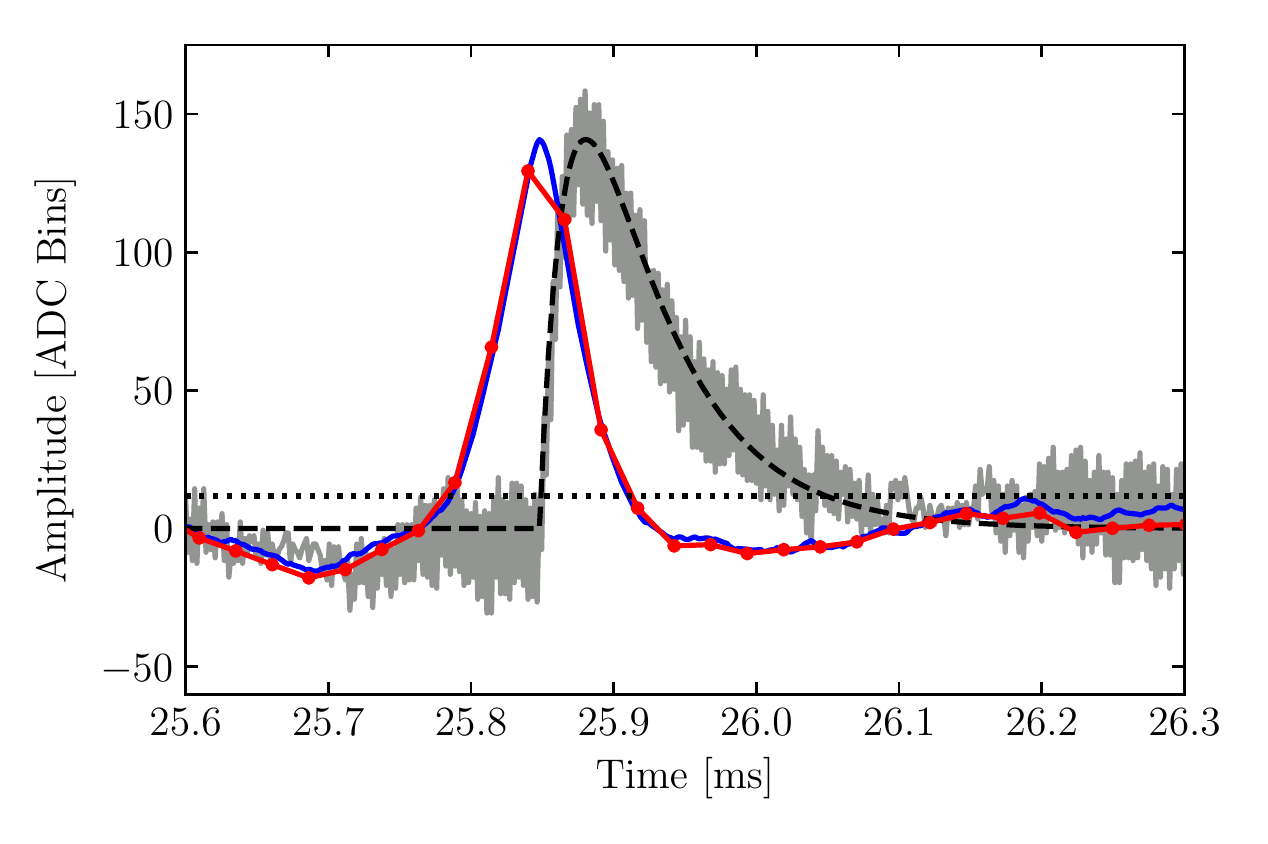}
    \caption[Example event within the dark matter search region-of-interest]{\label{fig:event}A zoomed-in portion of an example event within the analysis ROI. The raw pulse (gray) is compared to the offline optimal filter result (blue), the pulse template scaled by the fit result (black dashed), the FPGA filter result (red with dots), and the FPGA trigger threshold (black dotted). The noticeable spacing between the FPGA optimal filter data points is due to its downsampling by a factor of 16 as compared to the offline optimal filter. The offline and FPGA optimal filters are highly correlated, but not exactly the same, with corresponding energy estimates for this event of $187\,\mathrm{eV}$ and $179\,\mathrm{eV}$, respectively. The offset between the optimal filters and the raw pulse is by design of the filters, as they were set up to determine the time of the beginning of the pulse, as opposed to the maximum of the pulse.}
\end{figure}

While the FPGA-based OF was used to trigger the experiment in real time, we ultimately used our offline algorithm to reconstruct event energies also based on the OF formalism. For this offline OF, we were able to use a single noise spectrum computed from the in-run random triggers to represent the entire data set because there was negligible time variation of the noise over the course of the full exposure. Pulse amplitudes and start times were reconstructed using the same phonon-pulse template as in the FPGA triggering algorithm. Thus, there are two different pulse amplitudes for each event---one from the FPGA triggering algorithm and one from the offline OF. In Fig.~\ref{fig:event}, we compare the different energy estimators for a representative event.

\begin{figure}
    \centering
    \includegraphics[]{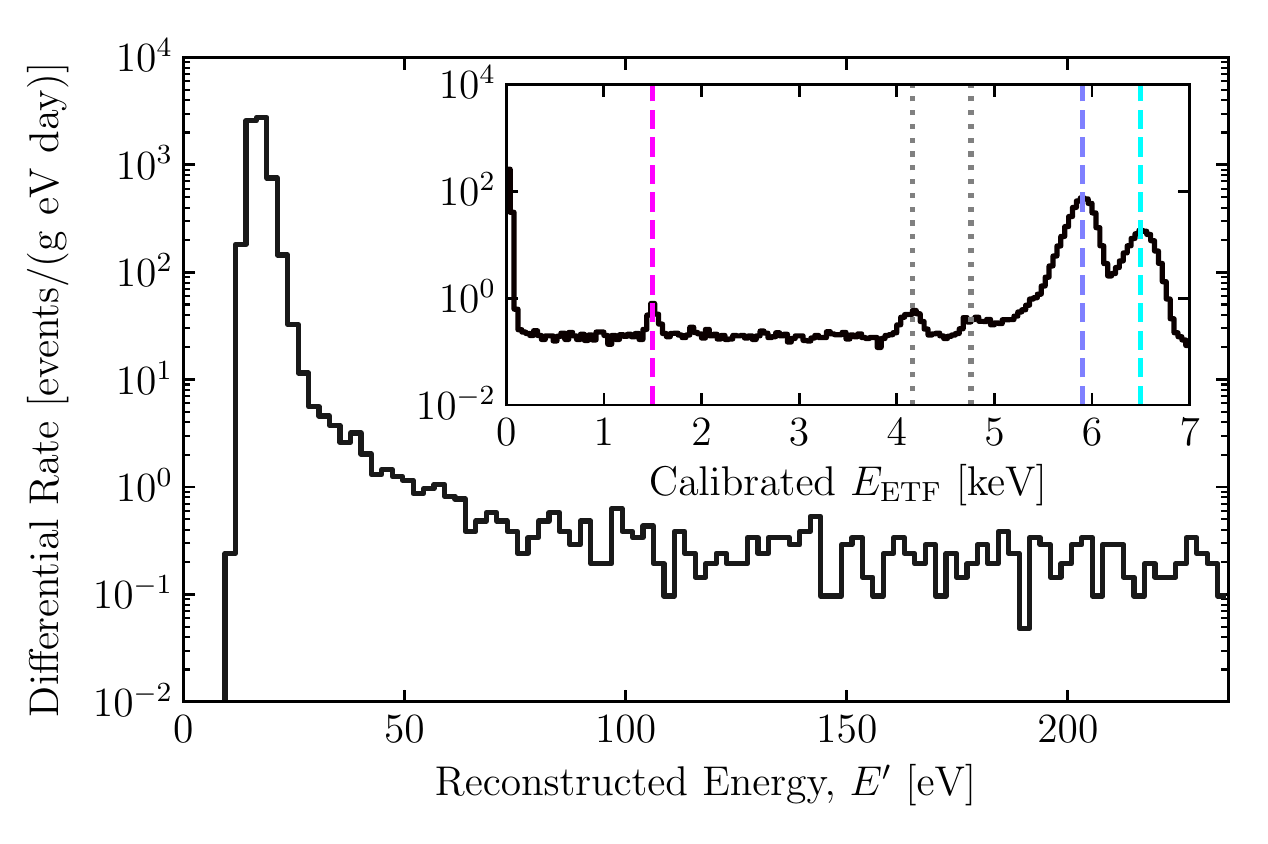}
    \caption[Energy spectrum for the dark matter search region-of-interest]{\label{fig:spectrum_full}Measured energy spectrum in the DM-search ROI for the full exposure after application of the quality cuts. The data have been normalized to events per gram per day per eV and have been corrected for the event-selection efficiency, but not the trigger efficiency. The inset shows the calibrated $E_\mathrm{ETF}$ spectrum up to $7\,\mathrm{keV}$, noting the locations of the different spectral peaks. The known values of the dashed lines are $1.5$, $5.9$, and $6.5\,\mathrm{keV}$ for the Al fluorescence (pink), $^{55}\mathrm{Fe}$ $\mathrm{K}_\alpha$ (blue), and $^{55}\mathrm{Fe}$ $\mathrm{K}_\beta$ (cyan) lines, respectively. The two dotted gray lines between 4 and $5\,\mathrm{keV}$ in calibrated $E_\mathrm{ETF}$ are the Si escape peaks~\cite{Reed_1972}.}
\end{figure}

This detector was optimized for maximum energy sensitivity at low energies and does not have a large enough dynamic range to observe the calibration lines without nonlinear effects from saturation of the QETs. The nonlinearity is minimal within our region-of-interest (ROI), which is below $240 \, \mathrm{eV}$. Above the ROI, the fall time of the pulses increases monotonically with energy, which can be explained by effects of local saturation. Localized events can saturate nearby QETs to above the superconducting transition, while QETs far from the event stay within the superconducting transition. Because this is a single-channel device, the saturated and unsaturated QETs are read out in parallel and thus effectively combine into a single phonon pulse with an increased fall time. The calibration method as outlined in Section~\ref{sec:ofcalibration} is used to correct out these effects. In Fig.~\ref{fig:spectrum_full}, we show the differential rate spectrum of the calibrated offline OF amplitude wihtin the ROI, with the inset showing the differential rate spectrum for the calibrated $E_\mathrm{ETF}$, which matches the spectrum shown in Fig.~\ref{fig:spectrum}.

For the calibration of the FPGA trigger energy estimator, we know the conversion of the arbitrary FPGA units to Amperes. In this analysis, we had a 16-bit FPGA trigger, which meant that the largest FPGA amplitude that could be saved was 32767 in arbitrary FPGA units. This corresponds to 0.47 uA after converting (about $500\,\mathrm{eV}$), which is far below the Al fluorescence line. Any events that had an amplitude greater than $0.47 \, \mu \mathrm{A}$ effectively have no FPGA trigger energy estimator associated with it. This means that we are unable to carry out an independent calibration of the FPGA trigger energy estimator, where we would apply the same method as the offline energy estimator. One possible method would be to apply the same calibration of the offline energy estimator $E'$ to the FPGA trigger energy estimator $E_T$. However, there could be an error in the factors we are using to convert from the arbitrary units of the FPGA to Amperes or some other effect such that it is not guaranteed that we can use the same calibration as we did for the offline reconstructed energy. 

\begin{figure}
    \centering
    \includegraphics{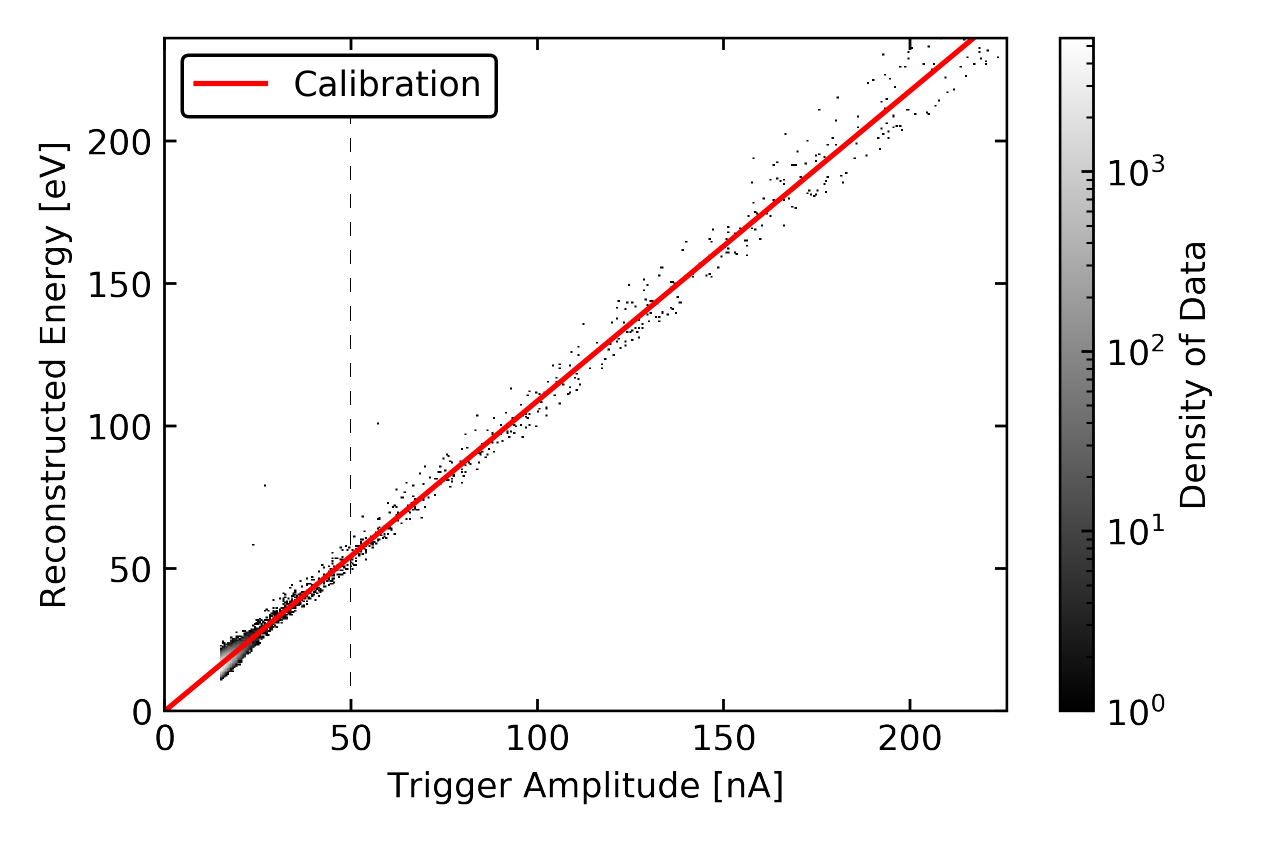}
    \caption{Calibration of the trigger amplitude to energy using the reconstructed energy. Data to the left of the dashed black line were not included in the calibration.}
    \label{fig:fpgacalibration}
\end{figure}

Rather than using the same calibration as the reconstructed energy, an alternative calibration will be done where we calibrate the trigger amplitude to the reconstructed energy using a relative normalization, defined by $E_T = m \cdot A_T$. This normalization ensures that zero energy events have zero energy. As we do not know the true energy of the events in our spectrum, we will use the reconstructed energy instead as an estimate. This should be valid, as we expect that the offline reconstructed energy should have an average that is equal to the true energy. We will also not include the low energy events, as these events are at low enough energies where the energy estimators become significantly biased (where the nature of the bias may be slightly different in the two estimators) due to an OF being biased towards large noise fluctuations. For this fit, as shown in Fig.~\ref{fig:fpgacalibration}, we did not include any data points with a trigger amplitude below $50 \, \mathrm{nA}$. Note that that figure shows values below this cutoff, but they were not included in the calibration.

\begin{figure}
    \centering
    \includegraphics{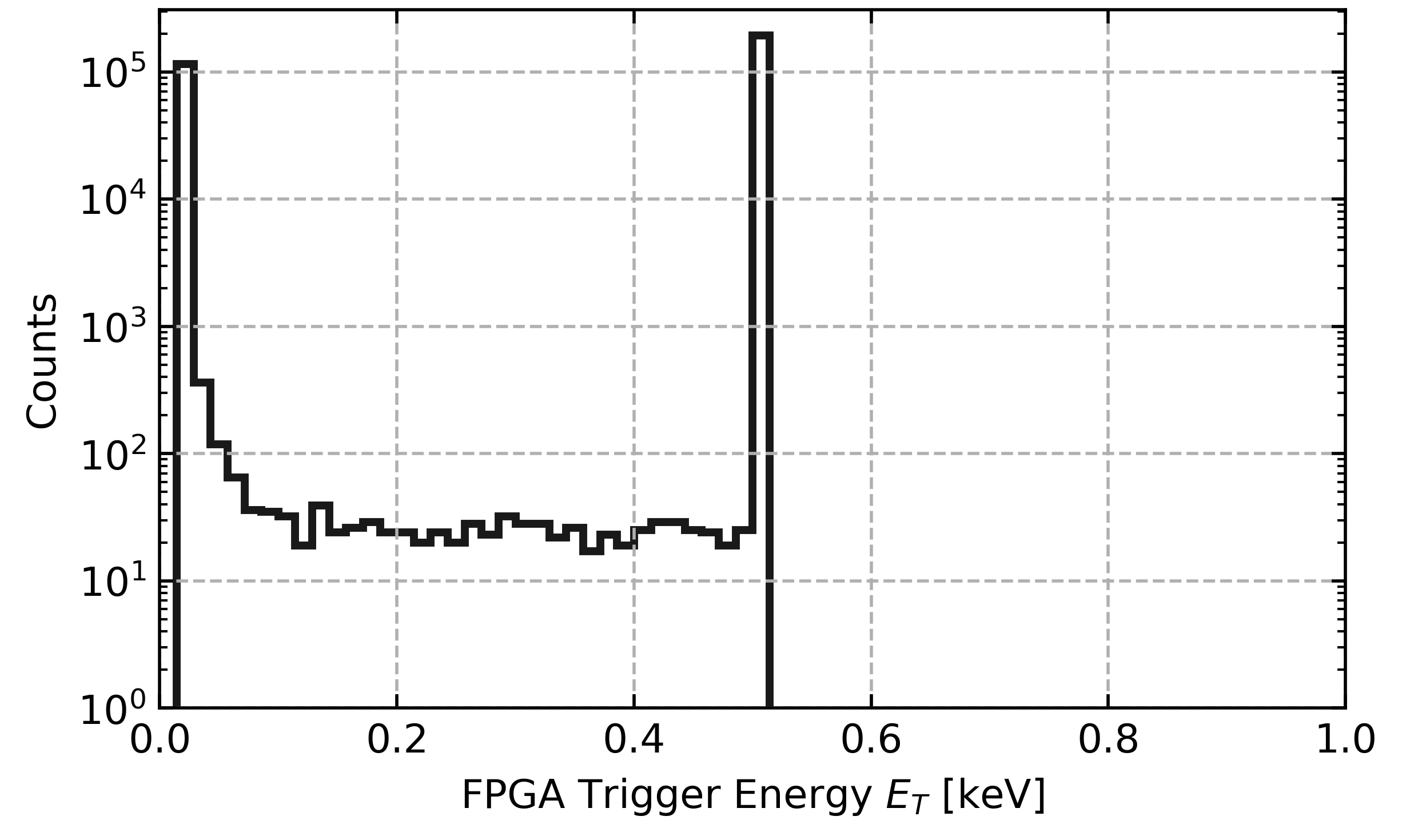}
    \caption{FPGA trigger energy estimator spectrum in units keV. The large population of events at about $0.5 \, \mathrm{keV}$ is due to saturation of the 16 bit FPGA trigger amplitude.}
    \label{fig:fpgaspectrum}
\end{figure}

The resulting spectrum for the FPGA trigger amplitude is shown in Fig.~\ref{fig:fpgaspectrum}. We can see that, in the figure, we only have FPGA trigger energy estimator information for event up to about $0.5 \, \mathrm{keV}$, which is above our ROI. At this point, we see a large population of data, which is all events with an FPGA trigger amplitude of 32767 (i.e. all of these events do not have a meaningful value for the FPGA trigger amplitude). This does not change anything for the DM search, as we only need FPGA trigger amplitudes for events within our ROI.

\begin{figure}
    \centering
    \includegraphics{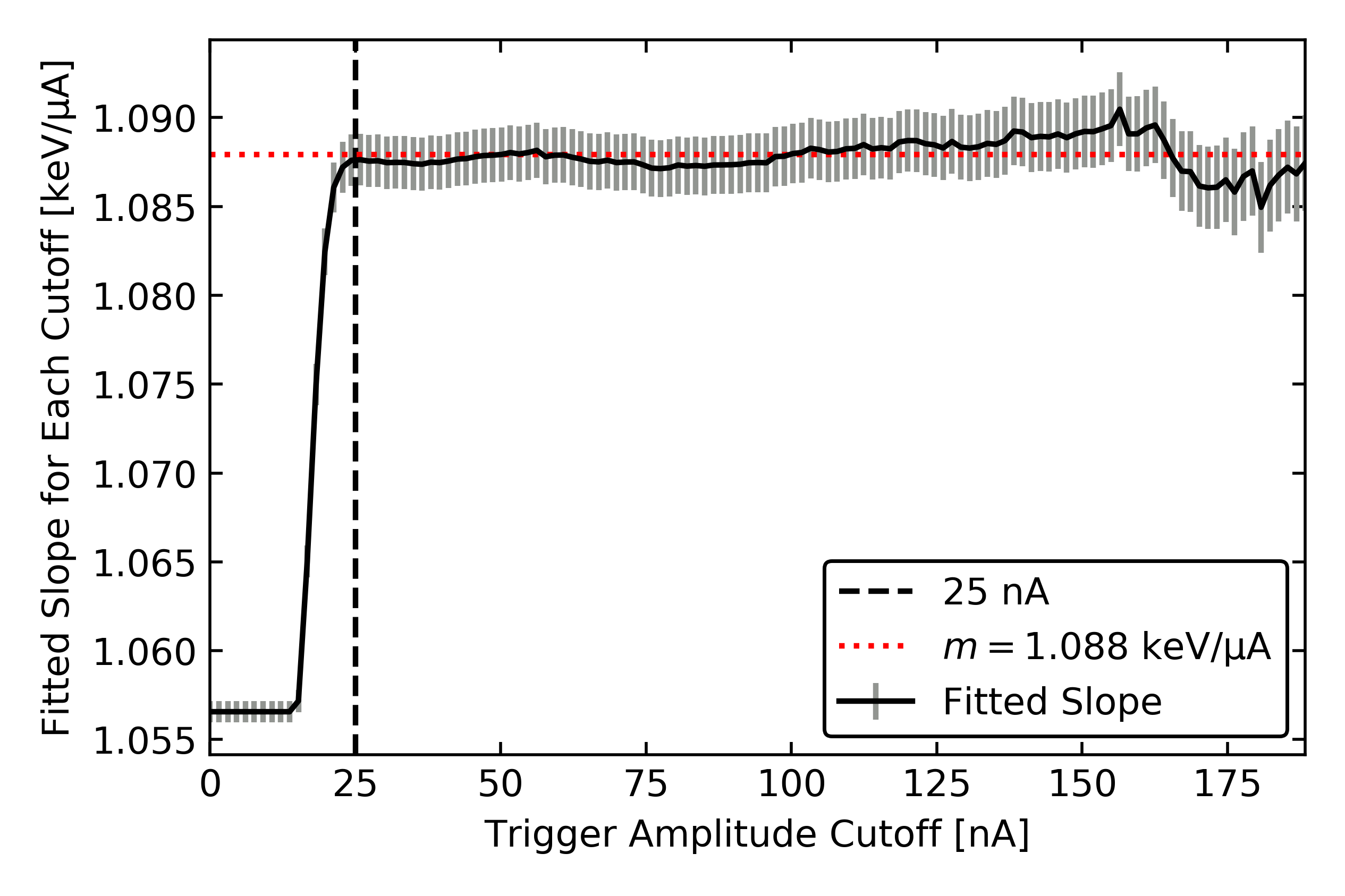}
    \caption[FPGA trigger cutoff systematics]{Dependence of the relative normalization of the trigger energy on the trigger amplitude cutoff. The red, dotted line is the fitted value with cutoff of $50\, \mathrm{nA}$, and the black, dashed line shows the $25 \, \mathrm{nA}$ cutoff. Note that the slope is very consistent above $25 \, \mathrm{nA}$, and then changes drastically below this point. This is due to the effect of the low energy events with a biased amplitude towards positive fluctuations.}
    \label{fig:fpgasyst}
\end{figure}

As previously noted, we did not include any data points with an FPGA trigger amplitude below $50 \, \mathrm{nA}$ in its calibration. To show the effect of different lower energy cutoffs, we can plot the fitted slope as a function of cutoff, as done in Fig.~\ref{fig:fpgasyst}. We see that our fitted value is within the statistical deviation of the slope for fits with cutoff above $25 \, \mathrm{nA}$ in trigger amplitude. Below this point, the low energy fluctuations bias the fit, showing that we should not include them. The chosen cutoff of $50 \, \mathrm{nA}$ (rather than $25 \, \mathrm{nA}$) ensures that we are not too close to where the bias begins. Because the error bars are consistent with the chosen value throughout, the effect of a different cutoff would be negligible. We also note that, since the plotted fit is within the statistical deviation of the fitted slope for all other cutoffs, this implies that the cutoff itself does not inject systematic error into our analysis. The statistical error in the fit ($\sim\!0.1\%$ error) is very small, which tells us that this statistical deviation is negligible.

\section{Data Selection and Efficiency}

We make our final event selection with a minimal number of selection criteria (``cuts'') to remove poorly reconstructed events without introducing energy dependence into the selection efficiency. This approach helps to reduce the complexity of the analysis and thus avoid introduction of systematic uncertainties. We apply two data-quality cuts: a prepulse baseline cut and a chi-square cut.

\subsection{Baseline Cut}

\begin{figure}
    \centering
    \includegraphics[width=\linewidth]{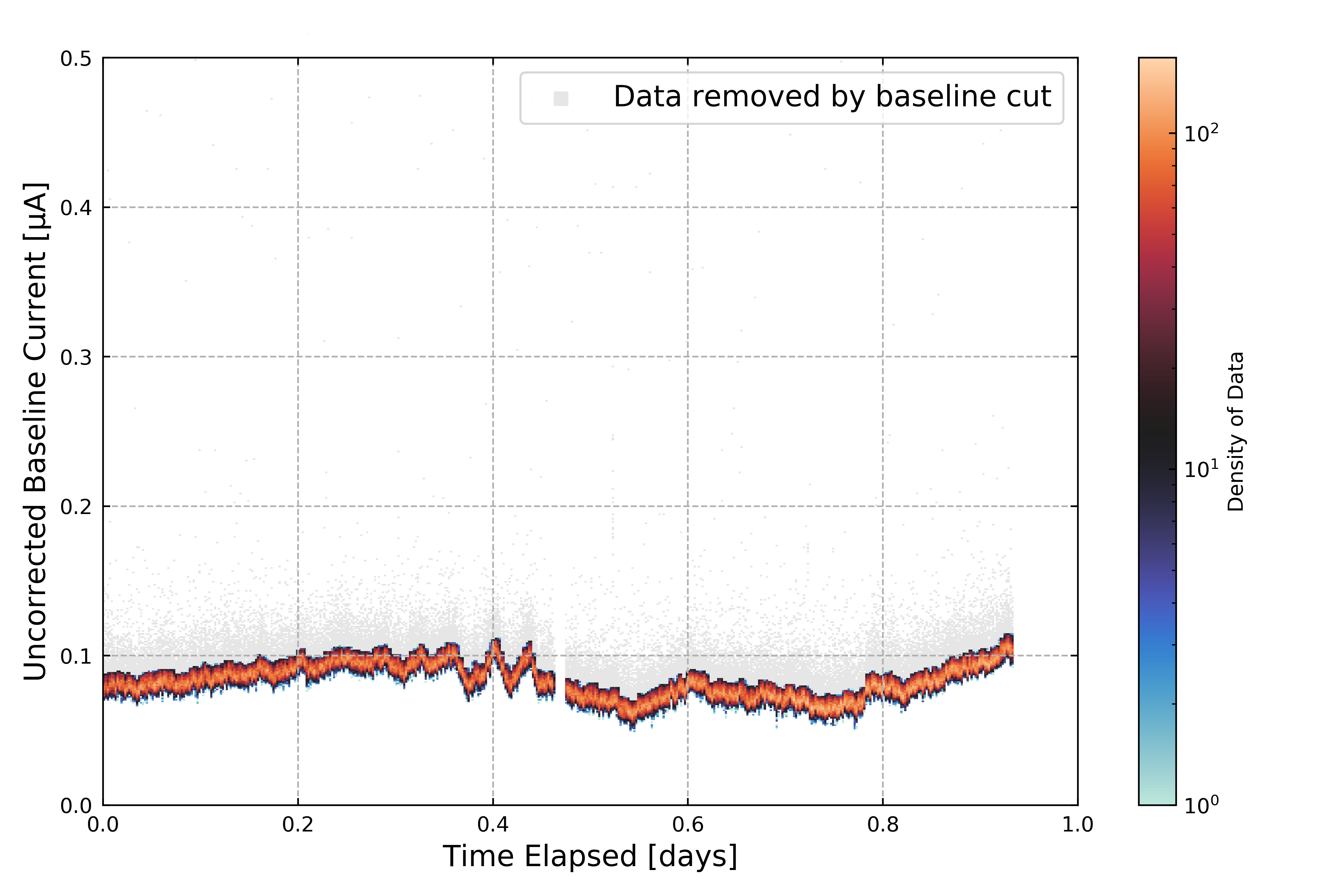}
    \caption{The data passing the baseline cut over time, to show the relatively constant baseline. The greyed out data are those that were deemed to have high baselines and were removed from the DM search dataset.}
    \label{fig:baselinecut}
\end{figure}

We define the event baseline as the average output in the prepulse section of each event, which is the first $25.6\,\mathrm{ms}$ of each trace. Large energy depositions have a long recovery time, which may manifest itself as a sloped baseline for subsequent events. Our trigger has reduced efficiency for any low-energy events occurring on such a baseline. We expected roughly 10\% of the events to sit on the tail of a high energy event in part because of the high muon flux at the surface of  $\sim\!1\,\mathrm{muon}/\mathrm{cm}^2/\mathrm{min}$~\cite{muonflux}. The baseline cut is performed by binning the data across the search in $400\,\mathrm{s}$ long bins and removing from each bin the 10\% of events that have the highest baseline, as shown in Fig.~\ref{fig:baselinecut}.

\begin{figure}
    \centering
    \includegraphics[width=0.8\linewidth]{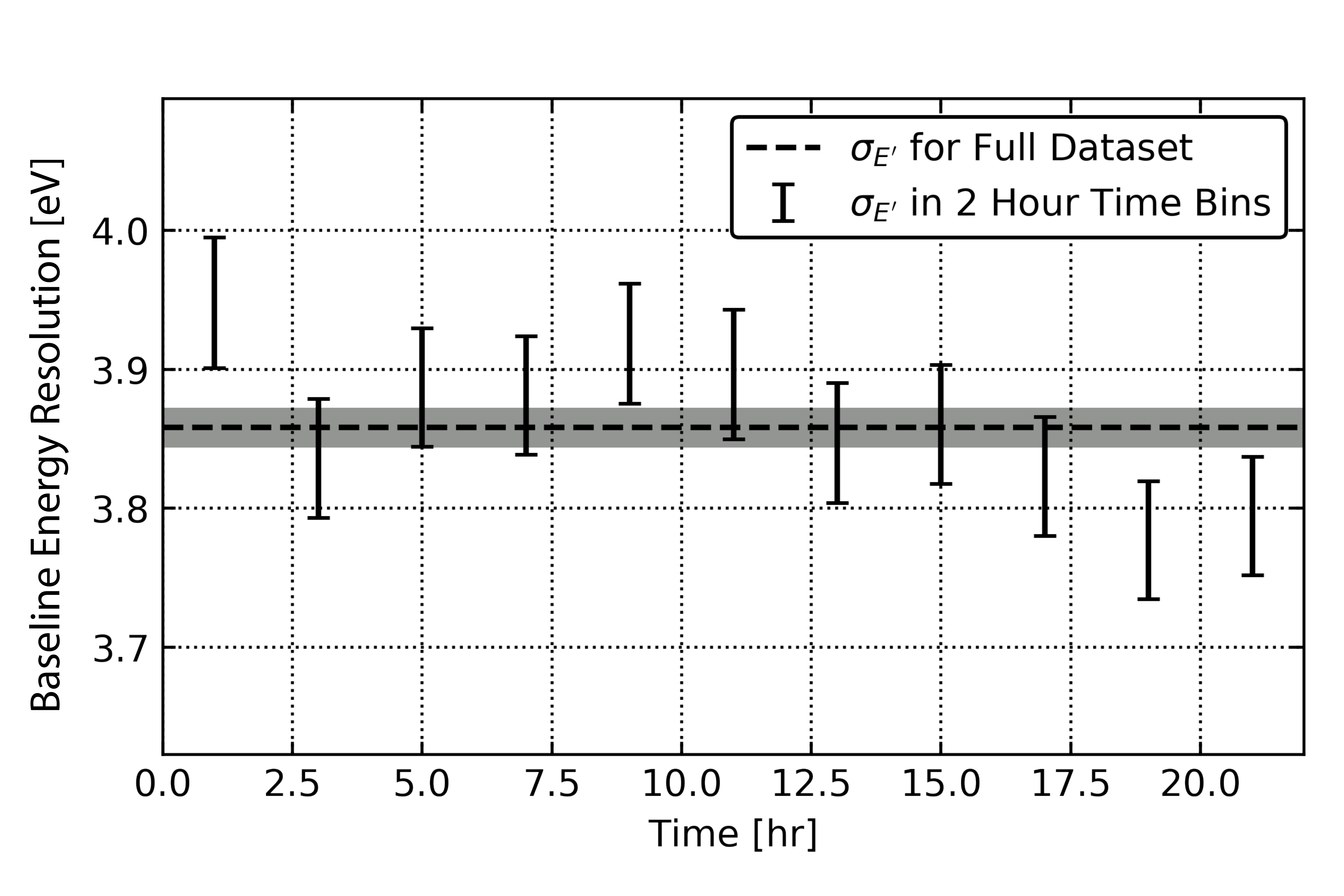}
    \caption{Using the in-run randomly triggered events, we have calculated the baseline energy resolution (RMS) of the reconstructed energy (i.e. calibrated OF amplitude) for two-hour sections of the data, which are shown in the plot as black error-bar  markers (where the error is of the $1\sigma$ level). We have also plotted the energy resolution for the entire run as the black dashed line with 68\% confidence level bands. Their agreement over time represents the stability of the system.}
    \label{fig:resolution_time}
\end{figure}

We see that there is a small amount of variation in the baseline over time, meaning that we should check if the TES bias point is changing appreciably. From Fig.~\ref{fig:baselinecut}, we can see that the largest change in current is approximately 30 nA over the course of the experiment. To estimate the corresponding change in resistance, we can estimate the bias voltage as constant with some change in current $\Delta I$ and related change in TES resistance $\Delta R$
\begin{equation}
    V_b = I_0 (R_0 + R_\ell) = (I_0 + \Delta I) (R_0 + \Delta R + R_\ell).
\end{equation}
Thus, rearranging, we have the the change in resistance given some change in current (i.e. baseline) is
\begin{equation}
    \Delta R = \frac{-\Delta I}{I_0 + \Delta I} (R_0 + R_\ell).
\end{equation}
Using the values in Table~\ref{tab:rp}, a change in current at this bias point of $\Delta I = 30 \, \mathrm{nA}$ corresponds to a change in resistance of $\Delta R = 0.12 \, \mathrm{m}\Omega$. Since the operating resistance of the TES at this bias point is $R_0 = 31 \, \mathrm{mOhms}$, this is a less than 0.4\% effect, and we can consider our baseline to be stable. We can further measure the baseline energy resolution throughout the DM search exposure, as we have done in Fig.~\ref{fig:resolution_time}. We see from above that the measured energy resolution for any two hour section is consistent with the value for the entire run, as they are all within $3\sigma$ of the value for the entire run. This is expected from our discussion of the stable baseline, where we argued that we should have a stable energy resolution. We are confident that our energy resolution is stable and represents the system well.

\subsection{Low Frequency Chi-Square Cut}

The chi-square cut is a general cut on our goodness-of-fit metric, for which we use the low-frequency chi-square $\chi^2_\mathrm{LF}$ calculated from the offline OF fit~\cite{golwala}. This metric is similar to the $\chi^2$ from the offline OF fit, but we exclude frequencies over $f_\mathrm{cutoff}$ from the sum. That is, in the OF formalism, we have the following quantities: the Fourier transform of our signal $\tilde{v}(f)$, the Fourier transform of our pulse template $\tilde{s}(f)$, the fitted OF amplitude $\hat{A}$, and the two-sided noise PSD $J(f)$. To calculate the chi-square for a given fit, as discussed in Appendix~\ref{chap:of}, it is given by
\begin{equation}
    \chi^2 = \int_{-\infty}^\infty \mathop{df} \frac{\left|\tilde{v}(f) - \hat{A} \tilde{s}(f) \right|^2}{J(f)},
\end{equation}
where we use the continuous Fourier transform for convenience (we convert to discrete when calculating these values on data). However, this equation includes high frequency values outside of our signal band. Because the mean of a chi-square distribution is $n_{dof}$ and its variance is $2 n_{dof}$, where $n_{dof}$ is the number of degrees of freedom is the number of frequencies that we are integrating over, the inclusion of the frequencies above our signal band serve to reduce the sensitivity of the chi-square metric. Thus, to improve this sensitivity, we truncate the integral at some specified $f_\mathrm{cutoff}$, giving
\begin{equation}
    \chi^2_\mathrm{LF} = \int_{-f_\mathrm{cutoff}}^{f_\mathrm{cutoff}} \mathop{df} \frac{\left|\tilde{v}(f) - \hat{A} \tilde{s}(f) \right|^2}{J(f)}.
\end{equation}

\begin{figure}
    \centering
    \includegraphics[width=\linewidth]{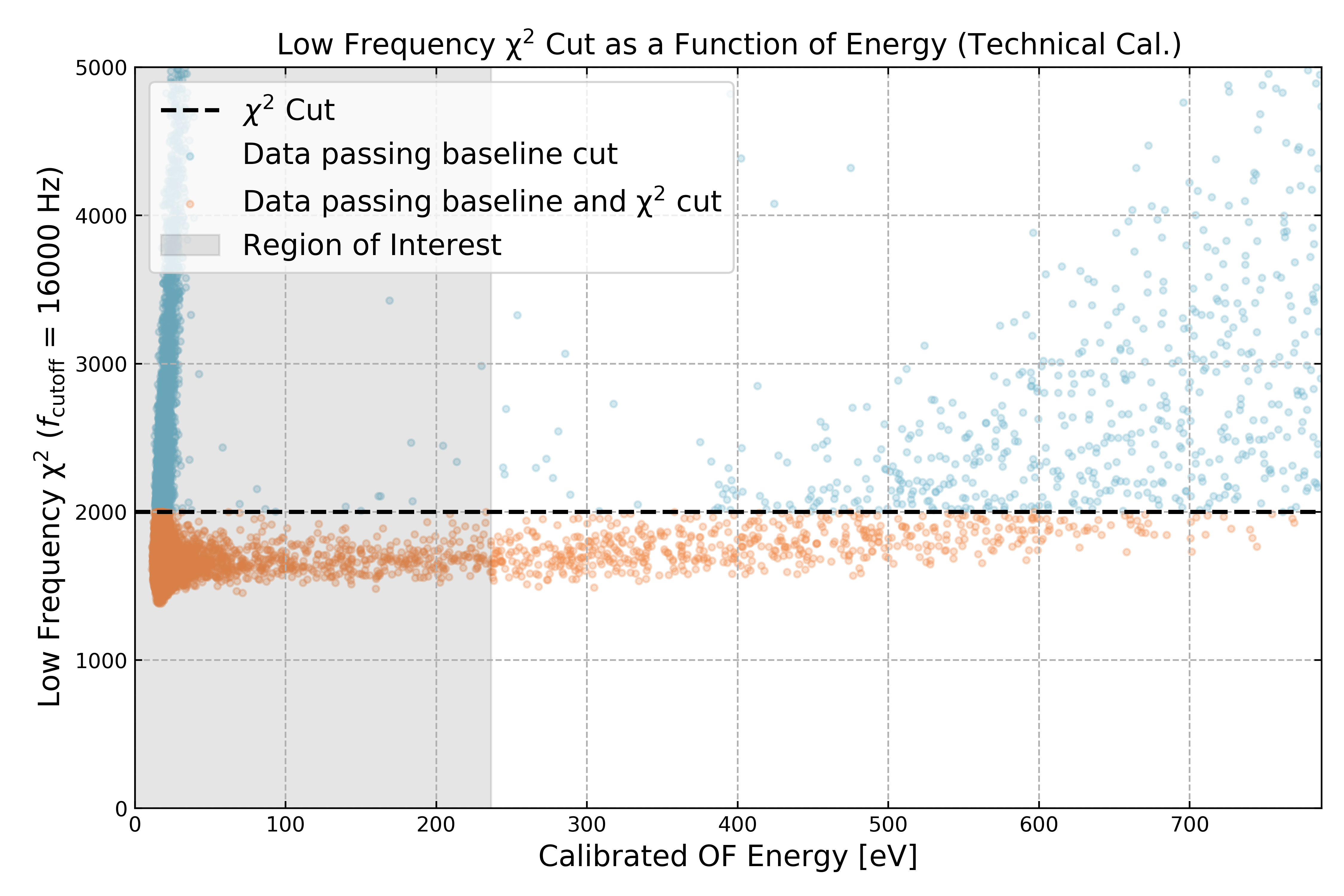}
    \caption{The data passing the low frequency chi-square cut at $\chi^2_\mathrm{LF} = 2000$. The grey region represents the DM search ROI.}
    \label{fig:chi2cut}
\end{figure}

This truncation allows us to remove sensitivity to superfluous degrees of freedom outside of our signal band from the chi-square, thereby reducing both the expected mean and the expected variance of the chi-square distribution. In this analysis, we used $f_\mathrm{cutoff} = 16\,\mathrm{kHz}$ because the rise and fall times of our expected pulse shape correspond to frequencies of $8.0\,\mathrm{kHz}$ and $2.7\,\mathrm{kHz}$ respectively. The inclusion of this cutoff reduces $n_{dof}$ from 32500 to 1663, improving the chi-square variance substantially (changing its standard deviation from 255 to 58). The pulse-shape variation within the DM-search ROI is minimal; this leads to a chi-square distribution that is largely independent of reconstructed event energy within this range. This in turn allows us to set an energy-independent cutoff value for this metric at $\chi^2_\mathrm{LF}=2000$, as applied to the data in Fig.~\ref{fig:chi2cut}. We note that this cutoff value is nearly $6$ standard deviations away from the mean, suggesting that this cut should have a high efficiency for accepting events.

\begin{figure}
    \centering
    \includegraphics[width=\linewidth]{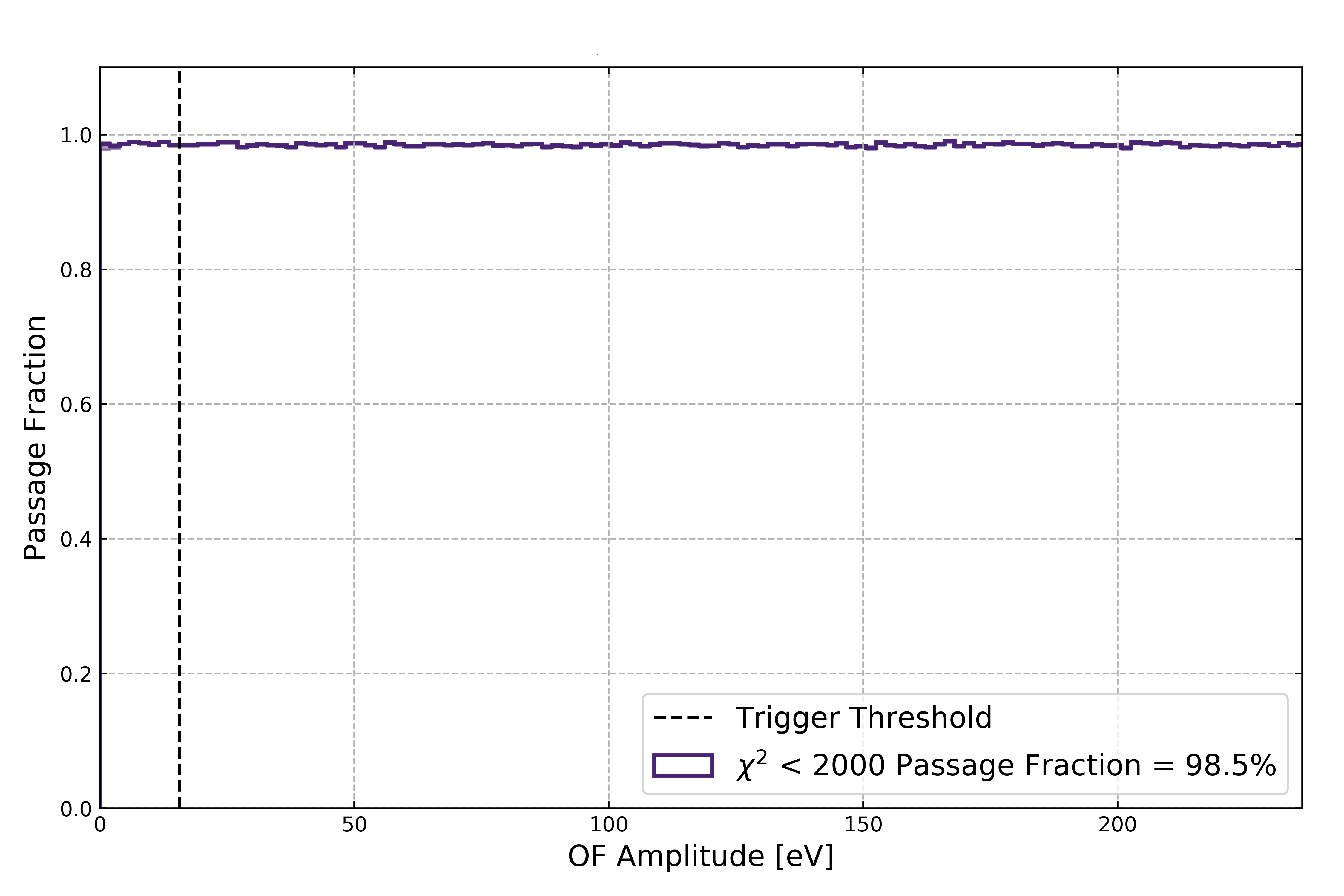}
    \caption{The simulated passage fraction as a function of energy for the low frequency chi-square cut.}
    \label{fig:chi2sim}
\end{figure}

Our measured events cannot be used directly to measure the signal efficiency of the chi-square cut because they include some that are not representative of the expected DM signal, e.g. vibrationally-induced events, electronic glitches, pileup events, etc. Therefore, we created a pulse simulation by adding noise from the in-run random triggers to the pulse template, systematically scaling the latter over the range of energies corresponding to the DM-search ROI. We then process and analyze the simulated data in the same way as the DM-search data. This simulation workflow has been included in the now-archived GitHub repository \textsc{RQpy}, co-created by C. W. Fink and myself~\cite{rqpy}. For the chi-square cut, we use randomly triggered data that have passed the baseline cut, and have no (real) pulses in them that have an energy outside of the ROI. The passage fraction of this chi-square cut, which has an energy-independent value of $98.53\pm0.01\%$, represents the cut's efficiency, as shown in Fig.~\ref{fig:chi2sim}. While this value does not match the expectation of 6 standard deviations (which would be 99.99\ldots\%), this is is not unexpected: the PSD we use to calculate the chi-square assumes that the noise is stationary (we know there is some time-variation in the noise, but it is small from Fig.~\ref{fig:resolution_time}) and uncorrelated (of which we cannot make a claim due to the single-channel nature of the device). Furthermore, the randomly triggered events can include real pulses, such that we could have both a simulated and a real pulse in the same event. This scenario would be excluded and is not included in the expected efficiency when assuming uncorrelated and stationary noise. Thus, it is reasonable to expect the observed deviation.

We can also compare the value of the simulated passage fraction due to the chi-square cut to our passage on in-run randoms, which are equivalent to zero-energy dark matter events. We would expect to see that the two are approximately equal, as our chi-square cut is seen to be energy independent in our region of interest. The passage fraction for our in-run randoms is $98.6 \pm 0.6\%$, showing that our result matches expectations.

\begin{table}
    \centering
    \caption{The distribution of fall times and the estimated standard deviation due to (presumably) positional dependence at different pulse energies within the ROI.}
    \begin{tabular}{rrr}
    \hline \hline
    Pulse Energy [eV] & $\tau_f$ $\left[\mu \mathrm{s}\right]$  & $\sigma_\mathrm{positional}$ $\left[\mu \mathrm{s}\right]$  \\ \hline
    16 & 58 & 1  \\
    80 & 62 & 1.5 \\
    240 & 63 & 5.8 \\ \hline \hline
    \end{tabular}
    \label{tab:falltimes}
\end{table}

As previously mentioned, the pulse-shape variation within the DM-search is minimal. To show that this possible source of systematic is negligible, we will show that inclusion of pulse-shape variation does not change the chi-square cut's efficiency. Within the ROI, there are two possible sources of pulse-shape variation: local saturation and positional dependence. The main observable of local saturation is an increasing fall time of the pulses as energy increases. From Fig.~\ref{fig:chi2cut}, we see that chi-square values are roughly flat within the ROI, suggesting that we are likely unaffected by local saturation within this range. The main observable feature of positional dependence is, even with no local saturation, there being a variance of measured fall times that is greater than the expected statistical variance, regardless of energy. To estimate the statistical variance in a given population of the data, we average the variances for each data point taken from a nonlinear fit to the event fall times. Then, we estimate the total variance of the measured fall times from its distribution, which should be a value that is greater than or equal to the statistical variance. Given the single-channel nature of the CPD, we cannot prove its existence, but it is a reasonable explanation for variance of pulse shapes large than that expected from statistical variance (within CDMS, positional dependence has been frequently observed in similarly designed multi-channel devices). In order to extract the estimated variance due to (presumably) positional dependence, we subtract the statistical variance from the total variance
\begin{equation}
    \sigma^2_\mathrm{positional} \approx \sigma^2_\mathrm{total} - \sigma^2_\mathrm{statistical}.
\end{equation}
In order to study the effect of the chi-square cut value, we have measured the distribution of the fall times for pulses of energies $16 \, \mathrm{eV}$, $80 \, \mathrm{eV}$, and $240\, \mathrm{eV}$, where the various distributions of the fall times $\tau_f$ due to this assumption of positional dependence are reported in Table~\ref{tab:falltimes}.

\begin{figure}
    \centering
    \includegraphics[width=\linewidth]{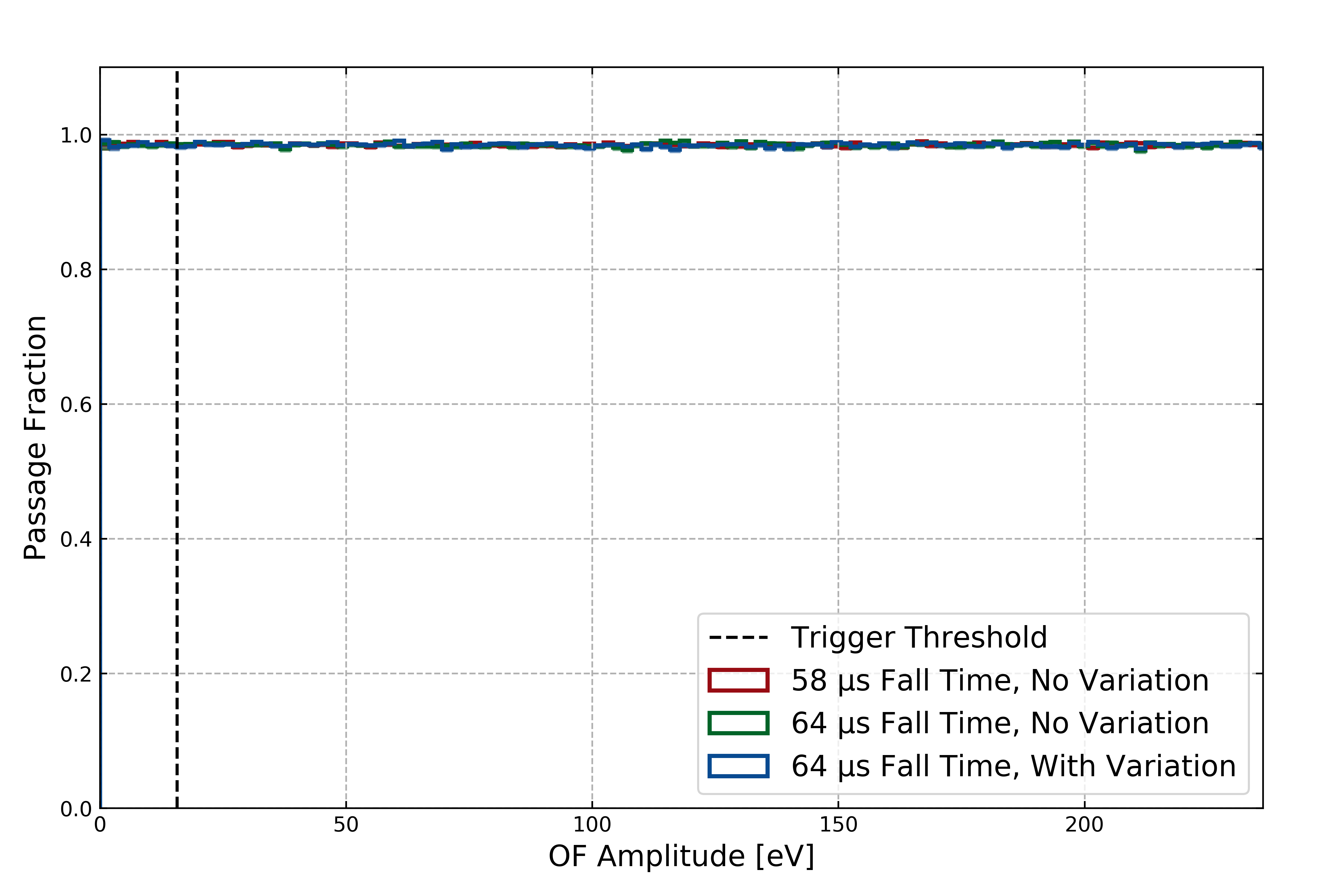}
    \caption{The calculated passage fraction of the chi-square cut for different simulated datasets corresponding to different pulse shapes used, showing that the pulse shape variation does not have an effect on this efficiency.}
    \label{fig:chi2syst}
\end{figure}

To study the effect of the pulse shape variation on the chi-square, we can use the pulse shapes at $240\, \mathrm{eV}$ for the three cases of pulse shapes, as this is where the effect should be most apparent. The three cases are:  pulses have $58 \, \mu \mathrm{s}$ fall times with no variation, pulses have $64 \, \mu \mathrm{s}$ fall times with no variation, and  pulses have $64 \, \mu \mathrm{s}$ fall times with an estimated $5.8 \, \mu \mathrm{s}$ standard deviation in fall time variation. The assumption that we used when measuring the chi-square passage fraction is the first case. With these three cases, we will see what the maximum effect of the different pulse shapes is. The expectation would be that the last case would have the largest effect, due to there being pulses with fall times far from the template's fall time of $58 \, \mu \mathrm{s}$. To test this idea, we have simulated data from recoil energies 0 to $240 \, \mathrm{eV}$ for the three cases above, run the data through the same processing routines as the DM search, and calculated the passage fraction as a function of energy for the chi-square cut in each case. In Fig.~\ref{fig:chi2syst}, we show the chi-square cut efficiency for these three cases, showing that the pulse shape variation within the ROI is indeed minimal and can be neglected.

We do not apply any other cuts to the DM-search data. The total signal efficiency is thus 88.7\%  and is independent of energy. A variation of the cut values within reasonable bounds was found to have no significant impact on the experimental sensitivity.

\section{Signal Model and Results}

In our DM signal model for spin-independent nuclear-recoil interactions~\cite{LEWIN199687}, we use the standard astrophysical parameters for the dark matter velocity distribution~\cite{kerr,rave,schonrich}: a velocity of the Sun about the galactic center of $v_0 = 220\,\mathrm{km}/\mathrm{s}$, a mean orbital velocity of the Earth of $v_E = 232\,\mathrm{km}/\mathrm{s}$, a galactic escape velocity of $v_\mathrm{esc} = 544\,\mathrm{km}/\mathrm{s}$, and a local DM density of $\rho_0 = 0.3 \,\mathrm{GeV}/\mathrm{cm}^3$. To relate this to the reconstructed energy, there are a multitude of possible techniques one might try. In the following paragraphs, we discuss the different possibilities and motivate the one used in this analysis.

\subsection{Discussion of Flawed Reconstructed Signal Models}

The first reconstructed signal model option would be a method where we simply fit the trigger efficiency in reconstructed energy and multiplied that curve by the DM differential rate model
\begin{equation}
    {\frac{\partial R}{\partial E'}(E')} = \varepsilon(E') {\frac{\partial R}{\partial E_0}(E')}.
\end{equation}
In this method, we are modelling the trigger efficiency as an error function, where this is the efficiency curve of the data after applying a cut on the trigger energy $E_T$. Note that we have written the differential rate as a function of the reconstructed energy. This is equivalent to assuming that the reconstructed energy has zero baseline energy resolution (RMS), which is a false statement. This immediately tells us that this is a flawed model, and that the nonzero baseline energy resolution must be correctly taken into account.

One possible method of taking into account the nonzero baseline energy resolution of the reconstructed energy $E'$ is to convolve the signal model in true energy with a Gaussian that describes the resolution of our reconstructed energy
\begin{equation}
    {\frac{\partial R}{\partial E'}(E')} = \varepsilon(E') \int_0^\infty \mathop{dE_0} \mathcal{N}(E'|E_0, \sigma_{E'}) {\frac{\partial R}{\partial E_0}(E_0)}.
\end{equation}
However, this model is also flawed. This would only be applicable if we were setting our trigger threshold on $E'$ (in which cased $\varepsilon(E')$ would be a step function centered on the threshold). For the CPD dataset, we have set our threshold instead on the FPGA amplitude $E_T$, while using a separate energy estimator $E'$ for our reconstructed energy. We must take into account the relation between these two (highly correlated) energy estimators when modeling the signal in reconstructed energy. We do note that if this model were used, then the Gaussian used would have to be truncated at some value of $E'$, as we have finite statistics in our sample. There is some range where we would not have enough statistics to be able to confidently say that we have a Gaussian to some level. In this case, we would have to decide on the extent of the smearing based on our confidence in the Gaussianity of the distribution far from the mean value.

The next model we consider takes into account the relationship of the two highly correlated energy estimators $E'$ and $E_T$ by modelling their relationship as a bivariate normal distribution (with the baseline energy resolutions of each estimator being related through some covariance matrix $\Sigma$) and convolving the signal with this distribution
\begin{equation}
    {\frac{\partial R}{\partial E'}(E')}=\int_0^\infty \mathop{dE_T} \int_0^\infty  \mathop{dE_0} \Theta(E_T - \delta)\mathcal{N}(E',E_T|E_0, \Sigma)\frac{\partial R}{\partial E_0}(E_0) .
\end{equation}
In this signal model, we note the following assumptions that have been made: the trigger efficiency function the FPGA energy estimator $E_T$ is a step function (as shown by the Heaviside step function, where the step occurs at the trigger threshold $\delta$), the relationship between the energy estimators $E'$ and $E_T$ is described by a multivariate normal distribution, and the covariance matrix describing the multivariate normal distribution is independent of the true energy $E_0$. However, the bivariate normal distribution does not describe the relationship between the two energy estimators near threshold. Because of the time-shifting aspects of both the FPGA trigger and the offline OF, the algorithms will latch onto the largest noise fluctuation in their search windows (as discussed in Appendix \ref{chap:of}, the largest OF amplitude corresponds to the smallest chi-square), which are $26.2144\, \mathrm{ms}$ and $160\,  \mu\mathrm{s}$ respectively. Due to latching onto the largest fluctuation, the distribution that relates the two energy estimators is non-normal below approximately $30\, \mathrm{eV}$.

\subsection{Reconstructed Signal Model for DM Search}

To correctly take into account the trigger efficiency, we convolve the differential rate with the joint probability density function relating our two energy estimators, including the effects of the applied cuts. The signal model, which includes the estimated trigger efficiency, is
\begin{equation}
    {\frac{\partial R}{\partial E'}(E')}=\int_0^\infty \mathop{dE_T} \int_0^\infty  \mathop{dE_0} \Theta(E_T - \delta)\varepsilon(E', E_T, E_0) P(E',E_T|E_0)\frac{\partial R}{\partial E_0}(E_0),
    \label{eq:signal}
\end{equation}
where $E_0$ is the true recoil energy, $E'$ is the recoil energy measured by the offline OF, $E_T$ is the recoil energy measured by the FPGA triggering algorithm, $\delta$ is the trigger threshold set on the FPGA triggering algorithm, $\varepsilon$ is the efficiency of the two quality cuts and two cuts that are applied to simulated data (as described in the following paragraphs), $\Theta$ represents the trigger threshold cut (a Heaviside function), and $P(E',E_T|E_0)$ is the probability to extract $E'$ and $E_T$ using the two energy reconstruction algorithms given the true recoil energy $E_0$. For the efficiency $\varepsilon$, we have generalized its form to be a function of energy, knowing that the baseline and chi-square cuts themselves are energy independent. The yield factor of heat signals (i.e. the ratio of heat signals produced by nuclear and electron recoils of the same energy that accounts for effects such as displacement damage) has been assumed to be unity for this work. Though measurements of the heat yield factor have not been made for Si, similar work has been undertaken for Ge, where the heat quenching factor was shown to be very close to unity~\cite{BENOIT2007558,206Pb}.

The model in Eq.~(\ref{eq:signal}) was evaluated numerically, taking advantage of our pulse simulation. The pulse simulation includes a software simulation of the FPGA triggering algorithm, which had the same output as the hardware version when run on the DM-search data. With this simulation of the FPGA triggering algorithm, we can use the pulse simulation to determine $P(E',E_T|E_0)$ directly. Low pulse height events may have their OF energy estimate affected by a shift of the start time estimate, but the simulation automatically takes this effect into account.

We also added two cuts to the simulated data only: a confidence ellipse cut and a trigger time cut. The confidence ellipse cut removed any events with an energy estimator value outside of the 99.7\% confidence ellipse, which is defined by the covariance matrix of our two energy estimators for zero-energy events. This cut was implemented to exclude the possible scenario of calculating a finite sensitivity to zero-energy DM, which would be a nonphysical result. The trigger time cut removed events that were not within $29\,\mu \mathrm{s}$---half of a fall time of a pulse---of the true event time, as determined by the energy-scaled pulse template. This cut ensured that the triggering algorithm was able to detect the signal added, as opposed to a large noise fluctuation elsewhere in the trace. These two cuts required knowledge of the true energy of the pulse---they cannot be applied to the data, but can be applied to the signal model---and helped to ensure that our signal modeling was conservative. In adding each of these cuts, we reduced our signal efficiency estimate, which necessarily biased the results in the conservative direction.

\begin{figure}
    \begin{center}
    (a)\includegraphics[width=0.8\linewidth]{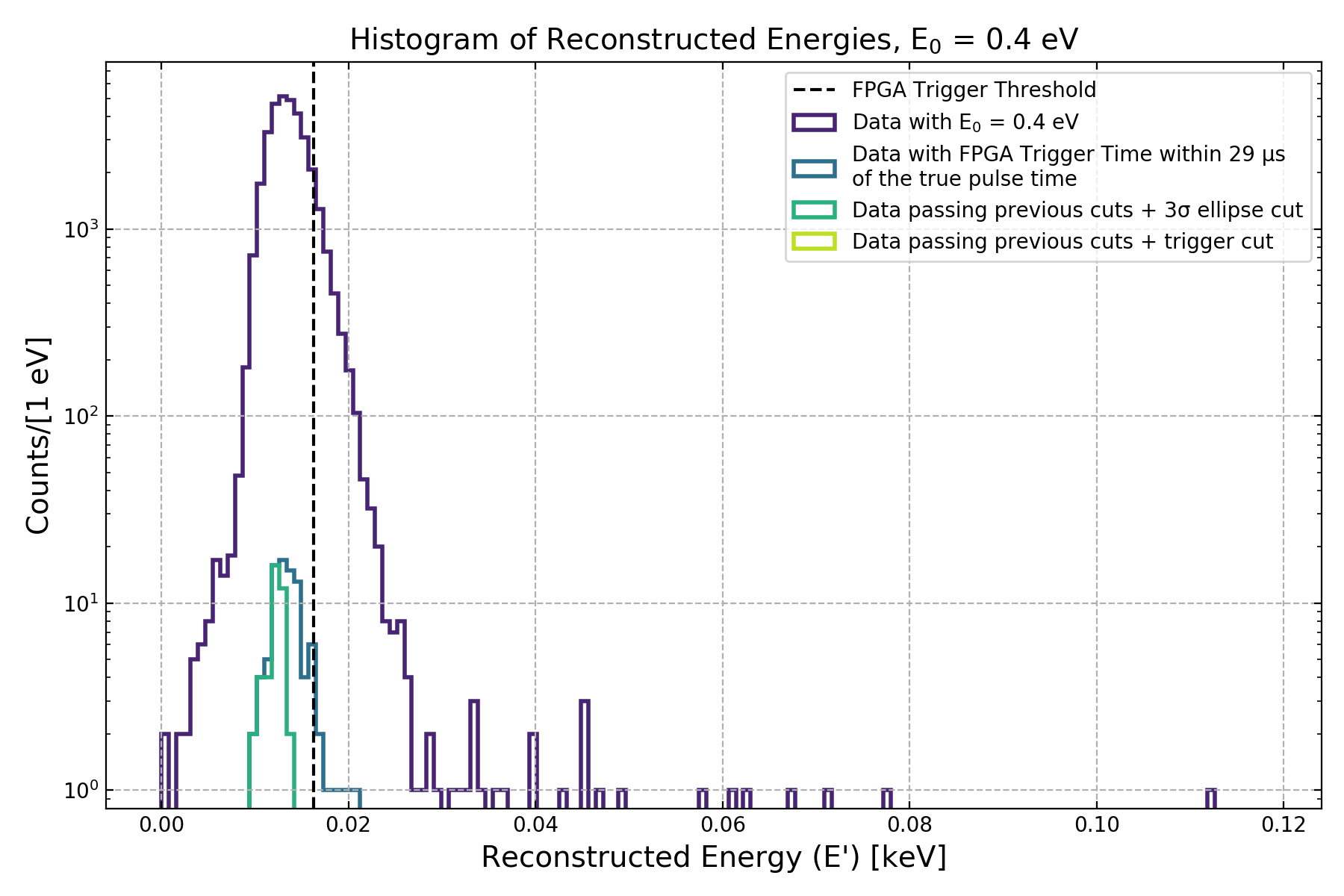} \\
    (b)\includegraphics[width=0.8\linewidth]{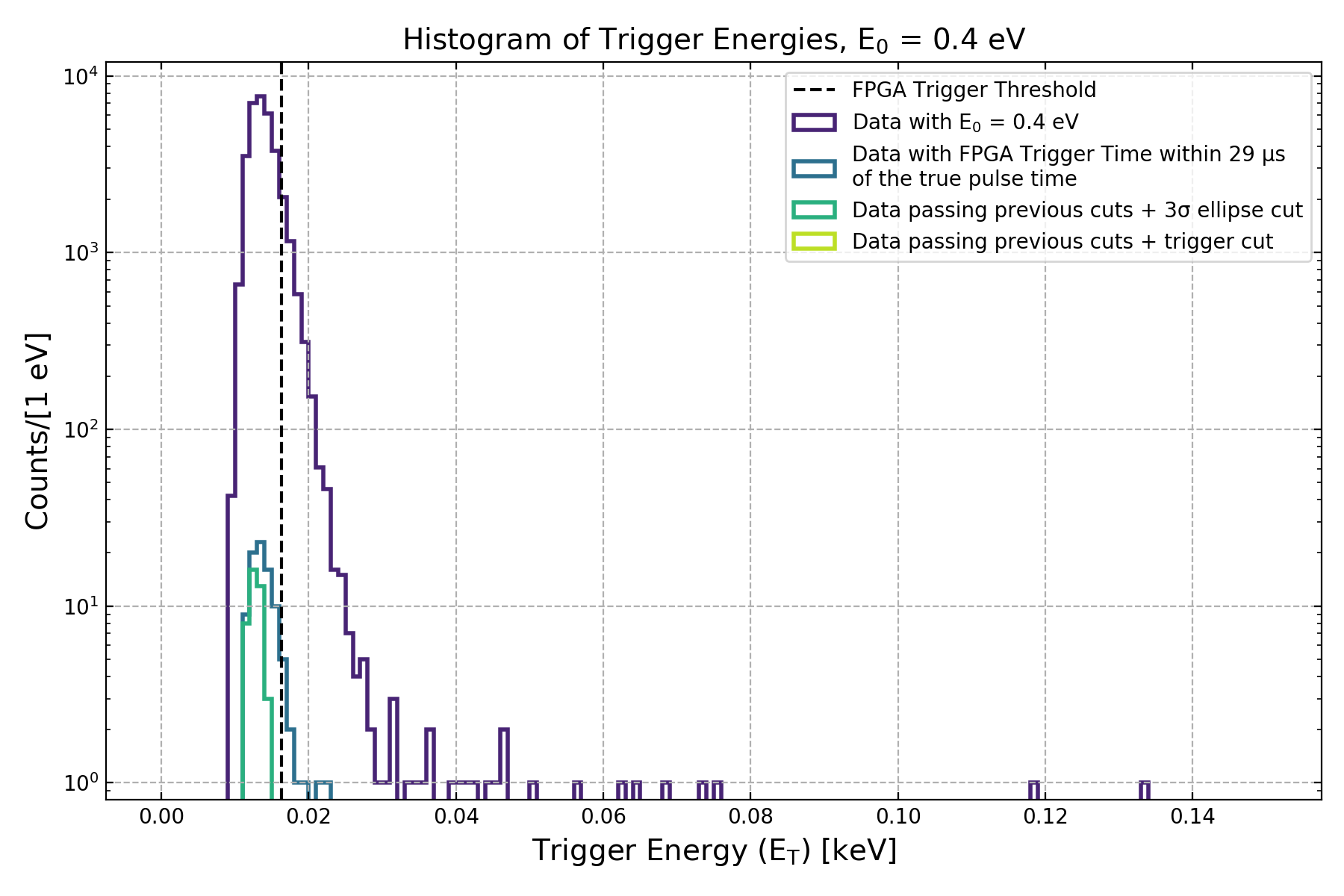} \\
    \caption{Dependence of the reconstructed energy (a) and the FPGA trigger energy (b) on the true energy inputted, including cuts on simulated data. The data consist of simulated events all with injected energies of $0.4 \, \mathrm{eV}$, and no events pass all cuts (as shown by the lack of the lime-green line). We do not have sensitivity to events with $0.4\,\mathrm{eV}$ energies or less.}
    \label{fig:pdf00.4}
    \end{center}
\end{figure}

\begin{figure}
    \begin{center}
    (a)\includegraphics[width=0.8\linewidth]{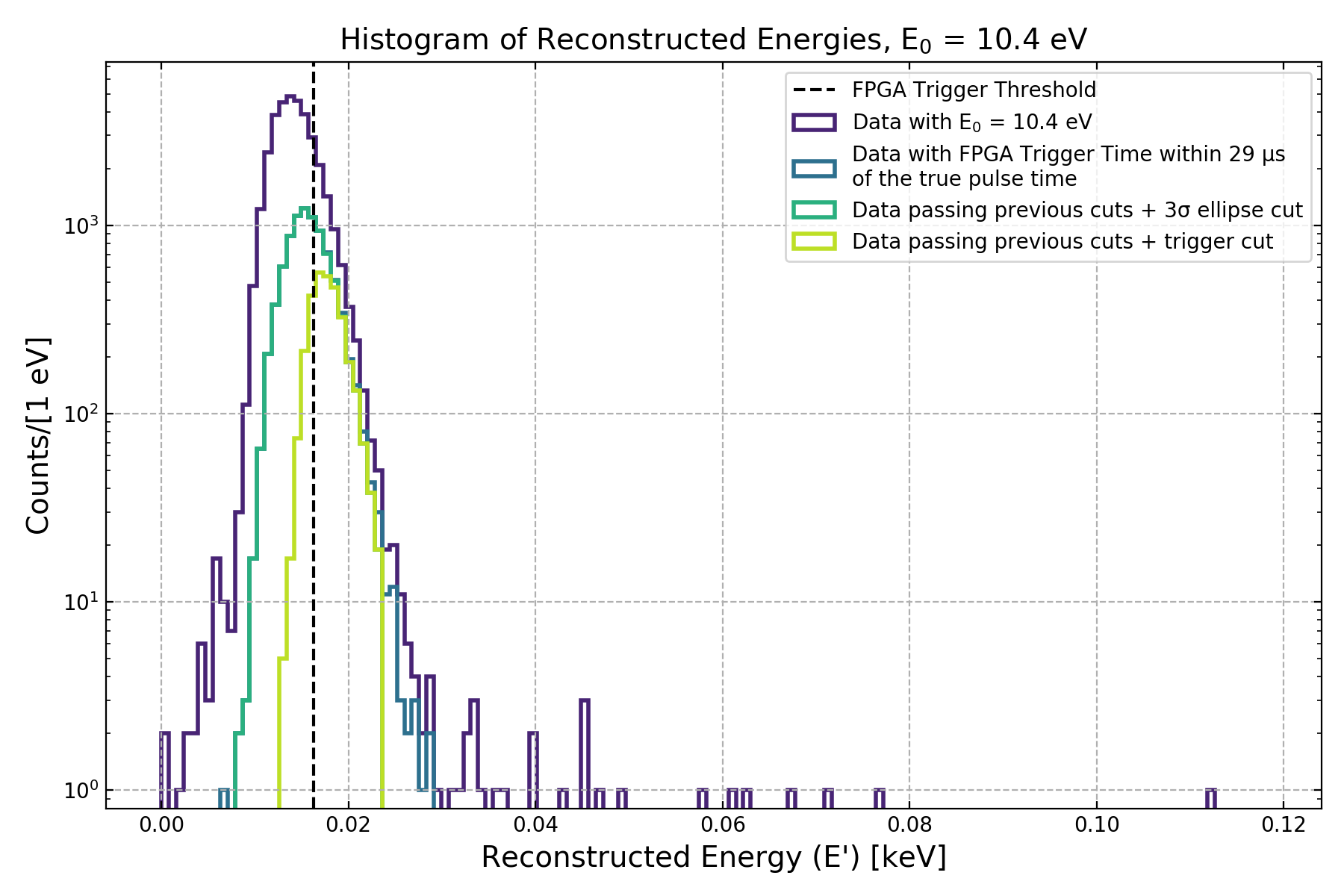} \\
    (b)\includegraphics[width=0.8\linewidth]{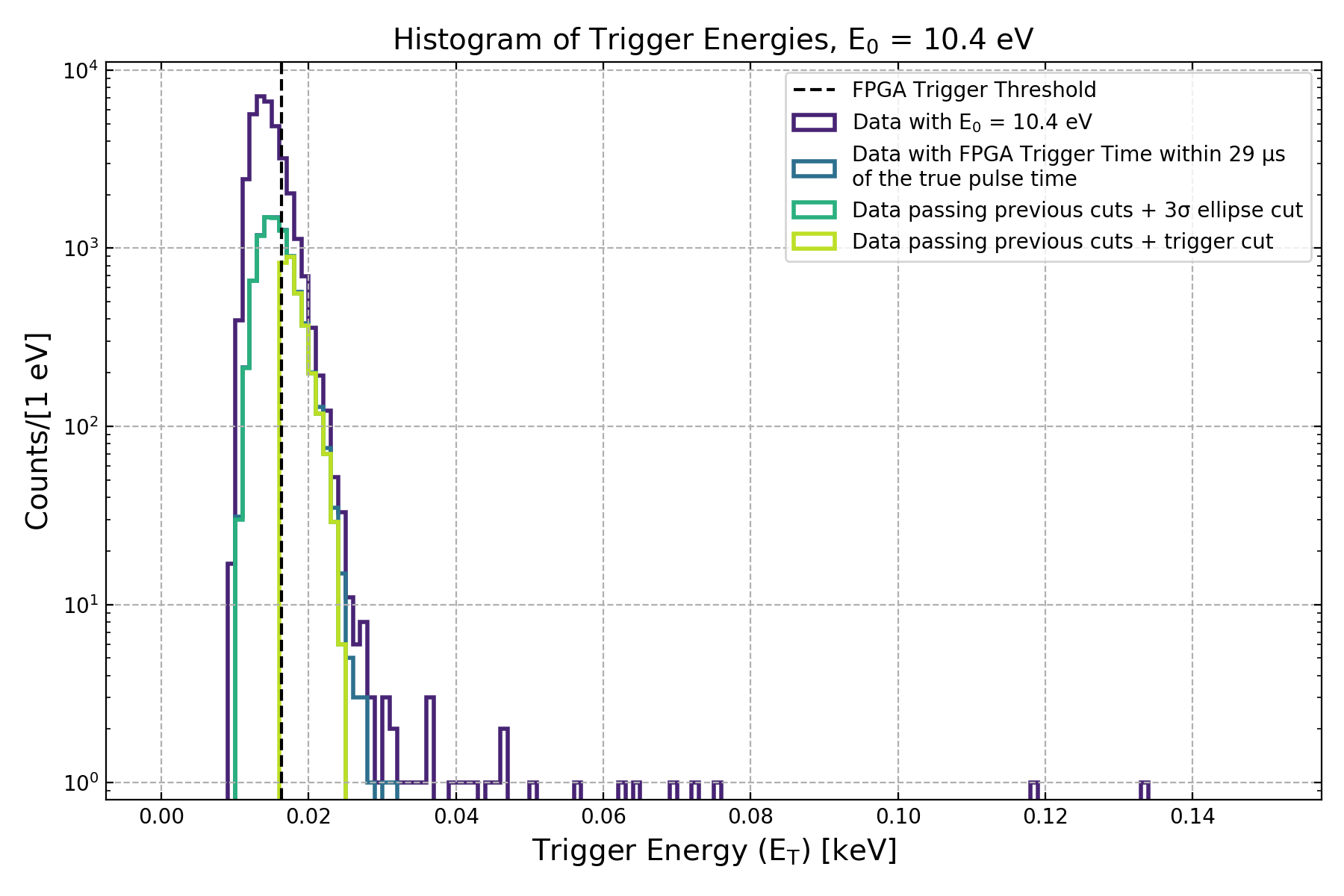} \\
    \caption{Dependence of the reconstructed energy (a) and the FPGA trigger energy (b) on the true energy inputted ($10.4 \, \mathrm{eV}$), including cuts. The distributions are non-Gaussian, and the trigger threshold cut is a step function in trigger energy.}
    \label{fig:pdf10.4}
    \end{center}
\end{figure}

\begin{figure}
    \begin{center}
    (a)\includegraphics[width=0.8\linewidth]{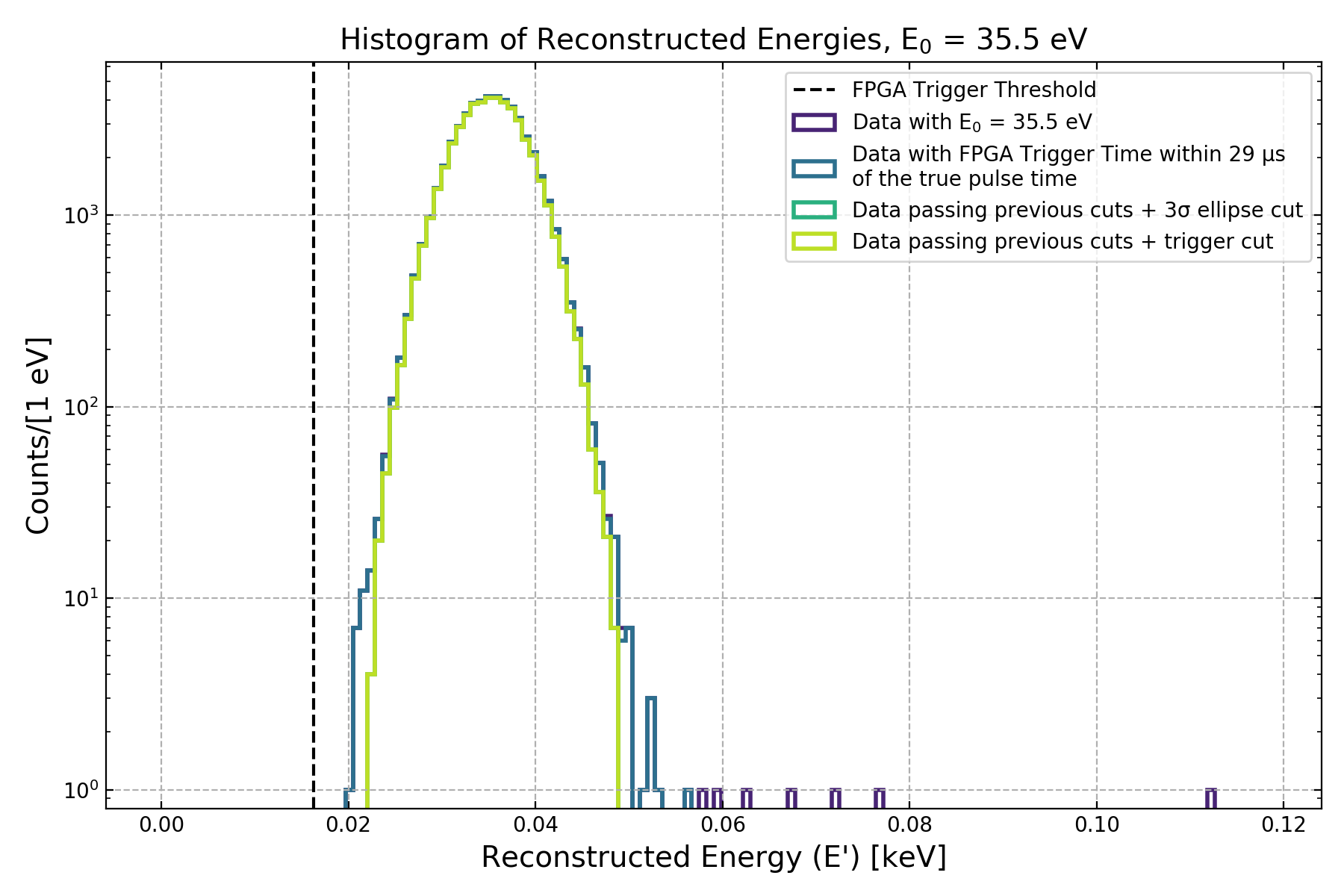} \\
    (b)\includegraphics[width=0.8\linewidth]{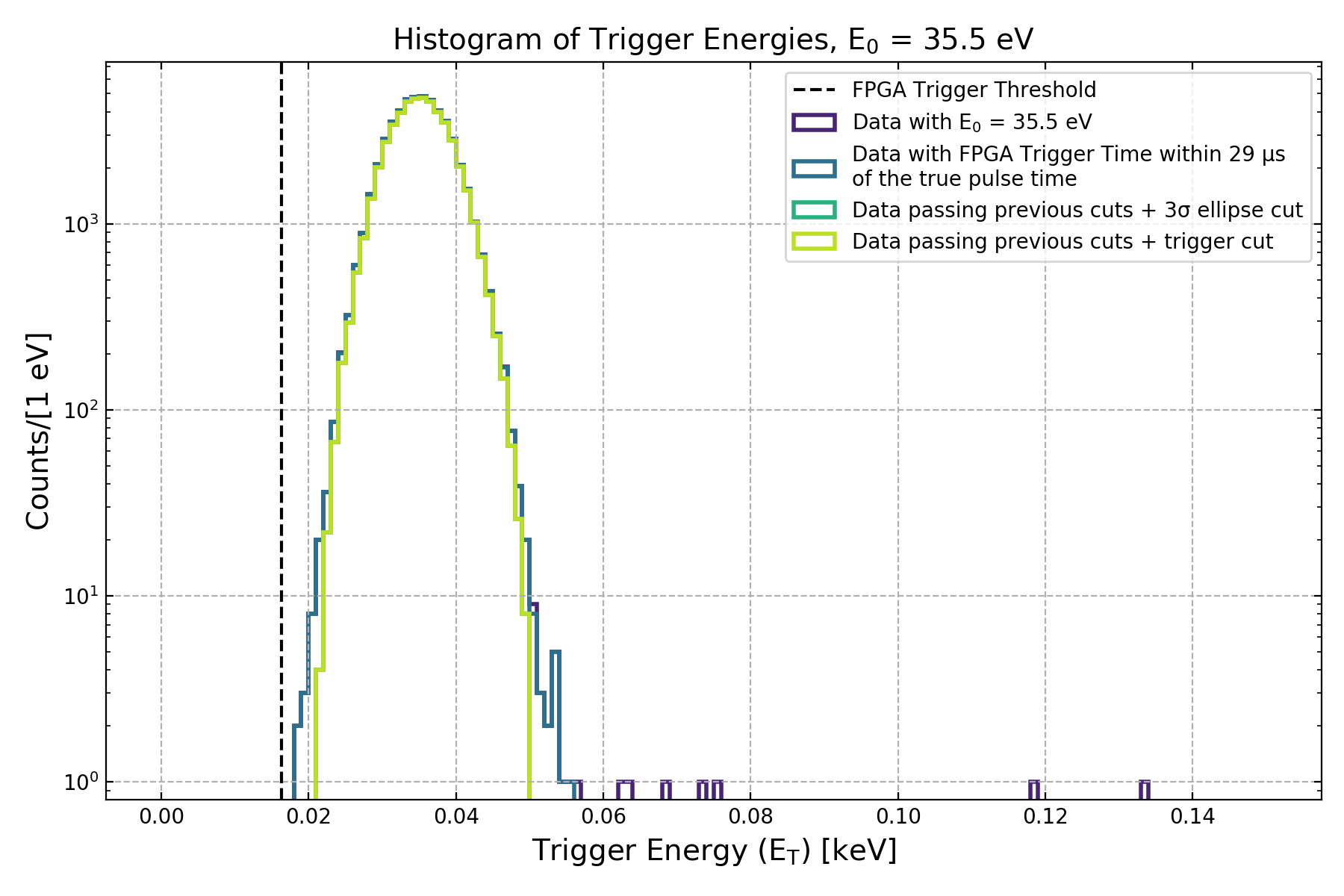} \\
    \caption{Dependence of the reconstructed energy (a) and the FPGA trigger energy (b) on the true energy inputted, including cuts on simulated data. The data consist of simulated events all with injected energies of $35.5 \, \mathrm{eV}$, and the distributions are now Gaussian, as we are far from threshold.}
    \label{fig:pdf35.5}
    \end{center}
\end{figure}

Carrying out the pulse simulation by injecting events to the in-run randoms with energies ranging from 0 to $40 \, \mathrm{eV}$ in $0.4 \, \mathrm{eV}$ steps, we can take the energy estimators, relate them to the known injected energies, and extract the probability density function. With our pulse simulation and cuts, we have shown three characteristic estimations of the probability density function in Eq.~(\ref{eq:signal}) in Figs.~\ref{fig:pdf00.4}--\ref{fig:pdf35.5} with three cuts being applied to the simulated dataset (purple): trigger time cut (blue), 99.7\% confidence ellipse cut (turquoise), and the trigger threshold cut (lime green). Each figure shows the probability density function projected to the reconstructed energy axis and to the FPGA trigger energy axis (for easier demonstration of the effects of the various cuts). In Fig.~\ref{fig:pdf00.4}, we see that no events pass all of the cuts, as desired from the inclusion of the 99.7\% confidence ellipse cut (i.e. that we do not have sensitivity to zero-energy events). In Fig.~\ref{fig:pdf10.4}, we see that there is a nonzero population of events that have been reconstructed with energies above our trigger threshold. As these events all have a true energy of $10.4 \, \mathrm{eV}$, this demonstrates the concept of being sensitive to noise-boosted events, as well as the distribution being non-normal. In Fig.~\ref{fig:pdf35.5}, the injected energy is now far above threshold, and the distributions now appear to be normal. In fact, the width of the distribution matches the baseline energy resolution in both energy estimators, as we would expect.

We can have further conceptual understanding of our trigger model by looking at the probability that an event with true energy $E_0$ is triggered (this can be thought of as the trigger efficiency in true energy), which we define as
\begin{equation}
    P_\mathrm{trig} = \int_0^\infty \mathop{dE_T} \int_0^\infty  \mathop{dE'} \Theta(E_T - \delta)\varepsilon(E', E_T, E_0) P(E',E_T|E_0),
    \label{eq:trigeff}
\end{equation}
This equation is quite similar to Eq.~(\ref{eq:signal}), with the difference being that the true differential rate $\partial R/ \partial E_0$ is replaced by 1, and we integrate over both of the energy estimators. The expectation would be that the shape is similar to an error function, but not exactly that same due to the non-Gaussianity of our joint probability density function at and below our trigger threshold.

\begin{figure}
    \begin{center}
    (a)\includegraphics[width=0.8\linewidth]{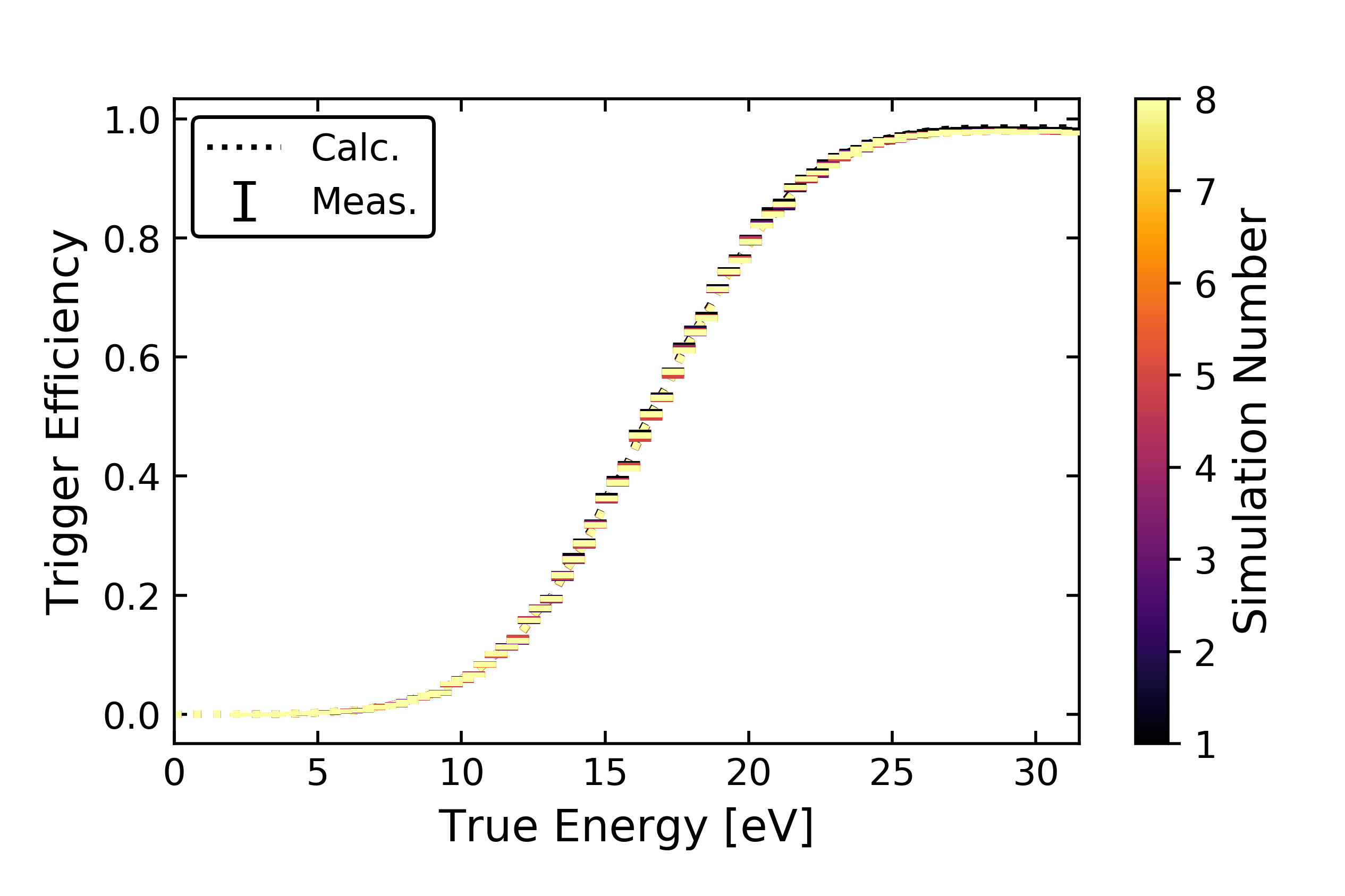} \\
    (b)\includegraphics[width=0.8\linewidth]{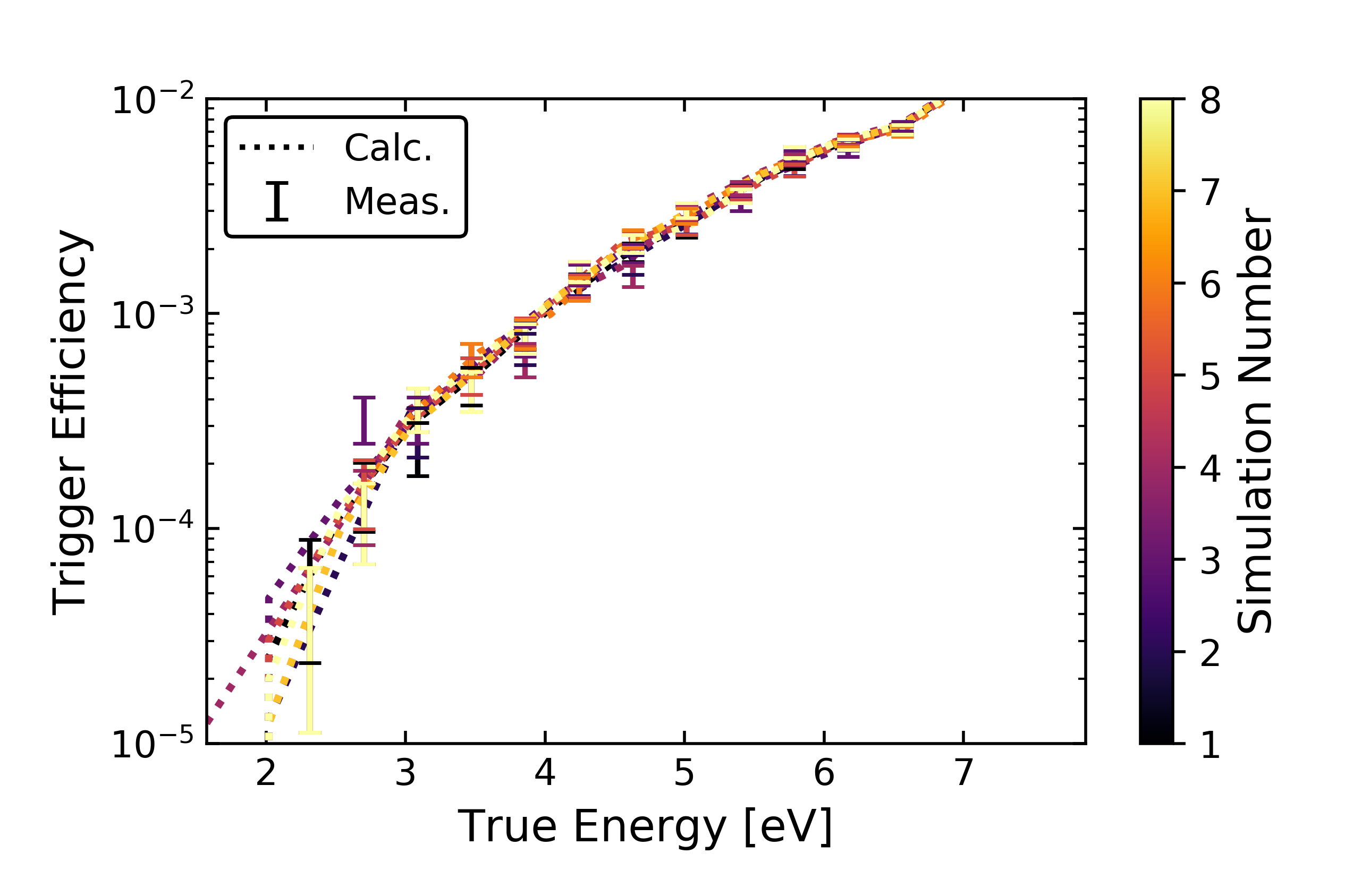} \\
    \caption{(a) The measured and calculated trigger efficiency with the FPGA trigger time and 99.7\% confidence ellipse cuts on a linear scale. (b) Zoomed in to low energies and plotted on a log-scale.}
    \label{fig:trig_eff}
    \end{center}
\end{figure}

At this point it is worth noting that plots of our probability density function were shown for just one simulation. In order to have an understanding of the variation of the trigger efficiency and our eventual limits, we have redone the simulations an additional seven times. Each simulation ensures that the random seed is uniquely set and the pulses are allowed to be injected over a $10 \, \mathrm{ms}$ window in the center of each random trace (to ensure edge effects are negligible in reconstructing the energies). Thus, we carry out the same workflow as far as measuring the probability density function, as well as calculating the trigger efficiency. There are two ways to calculate the trigger efficiency for the simulated data: a direct measure of the passage fraction as a function of true energy and calculating it from our interpolated probability density. The expectation is that these values should agree well for all energies. For the direct measure of the passage fraction as a function of true energy (``measured''), this is the passage fraction of the 99.7\% confidence ellipse cut, FPGA trigger time cut, and the FPGA trigger threshold cut. To calculate the uncertainty, we use the formalism described by Paterno in Ref.~\cite{Paterno:2004cb}, such that we avoid nonphysical values for efficiencies near 0\% or 100\%. For the calculation for our interpolated PDF (``calculated''), we again use Eq.~(\ref{eq:trigeff}). For each simulation and each calibration, we have estimated the trigger efficiency using both methods, as shown in Fig.~\ref{fig:trig_eff}.

\begin{figure}
    \centering
    \includegraphics[width=0.8\linewidth]{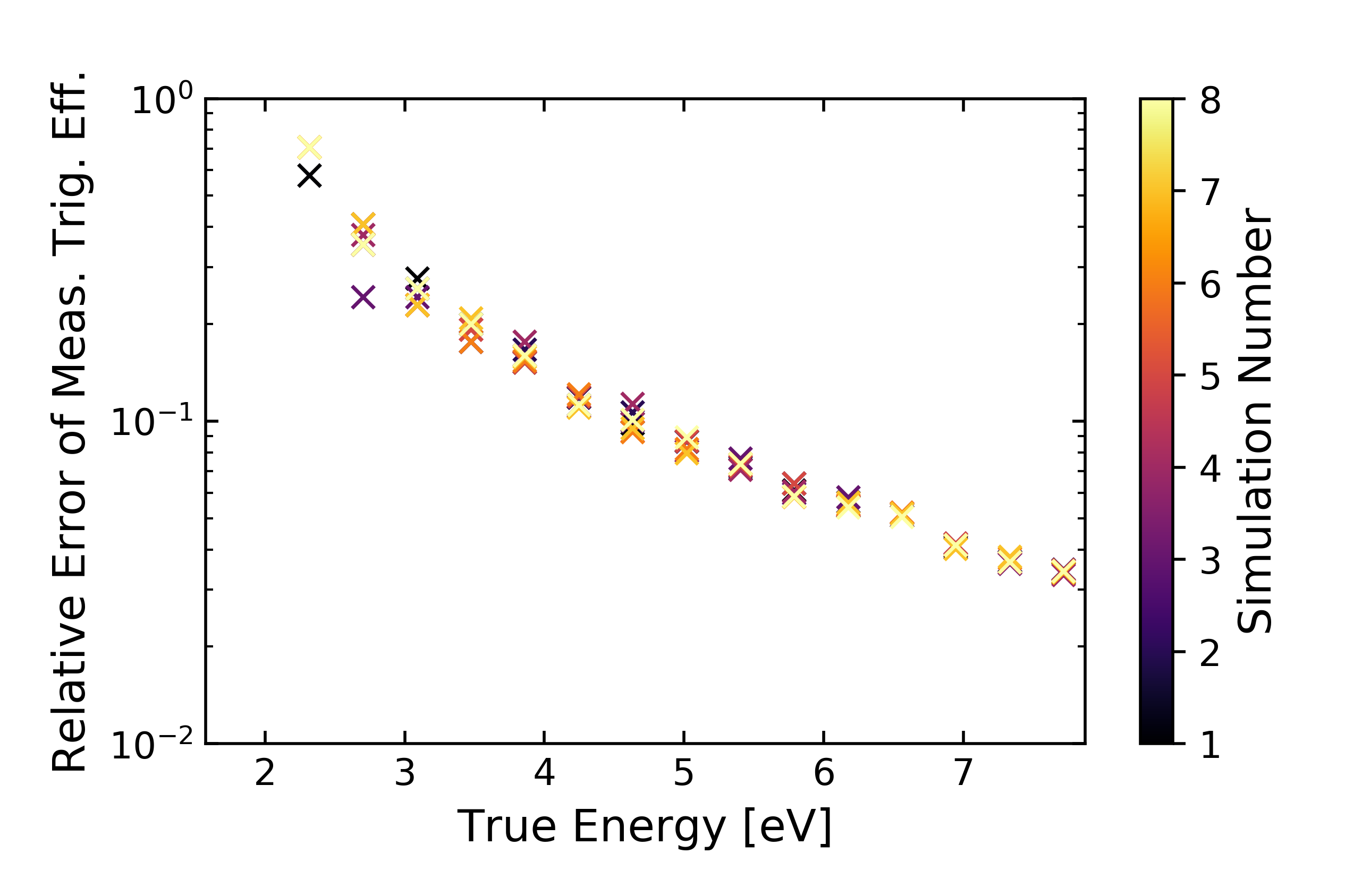}
    \caption{Relative error in the measured trigger efficiency.}
    \label{fig:trig_eff_error}
\end{figure}

We see that the measured and calculated trigger efficiency for each curve agree well with each other. More importantly, we see that there is some variation between the curves (both calculated and measured) for values of true energy below $5\, \mathrm{eV}$. We also see that the calculated curve finds nonzero values of the trigger efficiency below the lowest nonzero value of the measured trigger efficiency. This is due to the binning between energies when interpolating between energies in the determination of the PDF. Thus, we need to add a cut on the true energy to ensure we are not allowing nonzero trigger efficiencies where it should be zero. In this case, we have added a cut on the values where, below this number ($2.7 \, \mathrm{eV}$), the measured trigger efficiency is zero for at least one simulation. To get a more intuitive idea of the uncertainties at these low energies, we have shown the relative error in the measured trigger efficiency in Fig.~\ref{fig:trig_eff_error}.

We see that the relative error decreases with increasing energy, as expected (we have more passing events, which implies better statistics). We also see that below about 5 eV, we have that the relative error increases above 10\%. This tells us to expect a similar behavior for the limits set at low masses, as many of these are being set by our sensitivity to events with recoil energies this low.

\subsection{Limit Setting and Results}

The objective of this DM search was to set conservative limits on the spin-independent interaction of dark matter particles with masses below $1.5\,\mathrm{GeV}/c^2$. For the lower edge of the limit contour, we use the optimum interval (OI) method with unknown background, which is a method for selecting an interval within the ROI that contains few events as compared to what would have been expected from (in this case) a DM signal.\cite{yellin_upper, yellin_extended}. Practically, this is done by scaling the DM-nucleon scattering cross section of the expected signal for a given DM mass, and finding the cross section above which (at some specified confidence level) we would have expected to see a DM signal in some range of values above the measured spectrum. For the upper edge of the limit contour, we use a modified version of the publicly available \textsc{verne} code~\cite{verne}, which uses a Poisson counting method to calculate the effects of overburden~\cite{starkman1990, zaharijas2005, overburden} on the DM signal, as similarly used in Refs.~\cite{PhysRevD.97.123013, PhysRevLett.123.241803, EdelweissWIMP}. Both limit-setting methods assume that the full measured event rate could be due to a DM signal and set the limits at the 90\% confidence level (C.L.). 

Because the velocity distribution for DM with large interacting cross sections has a dependence on the interaction cross section itself, this means that the calculated DM spectra for the overburden are not simply proportional to the scattering cross section (which prevents use of the OI method). To set this Poisson limit for the overburden, a grid search is performed in both DM mass and scattering cross section. For the ROI in the DM search, the code compares the number of events observed with the number of expected events for each DM mass and interaction cross section when including the overburden effects. The 90\% Poisson limit is calculated by interpolating the expected event numbers at each DM mass as a function of cross section, finding the cross section that corresponds to the 90\% Poisson limit on the total observed number of events. This code does make the assumption that DM particles scatter continuous and travel along straight line trajectories, which is a better approximation for heavy DM. However, it has been shown that, for $\mathcal{O}(1) \, \mathrm{GeV}$ masses and below, the \textsc{verne} code and Monte Carlo techniques without this assumption (e.g. \textsc{DaMaSCUS-CRUST}) achieve similar results~\cite{overburden}. For the overburden assumption, we include the $5\,\mathrm{cm}$ thick, five-walled Cu ``box'' surrounding the section of the dilution refrigerator that contains the detector (which dominates over the shielding of the building), the shielding from the atmosphere, and the shielding from the Earth.

\begin{figure}
    \centering
    \includegraphics[width=0.8\linewidth]{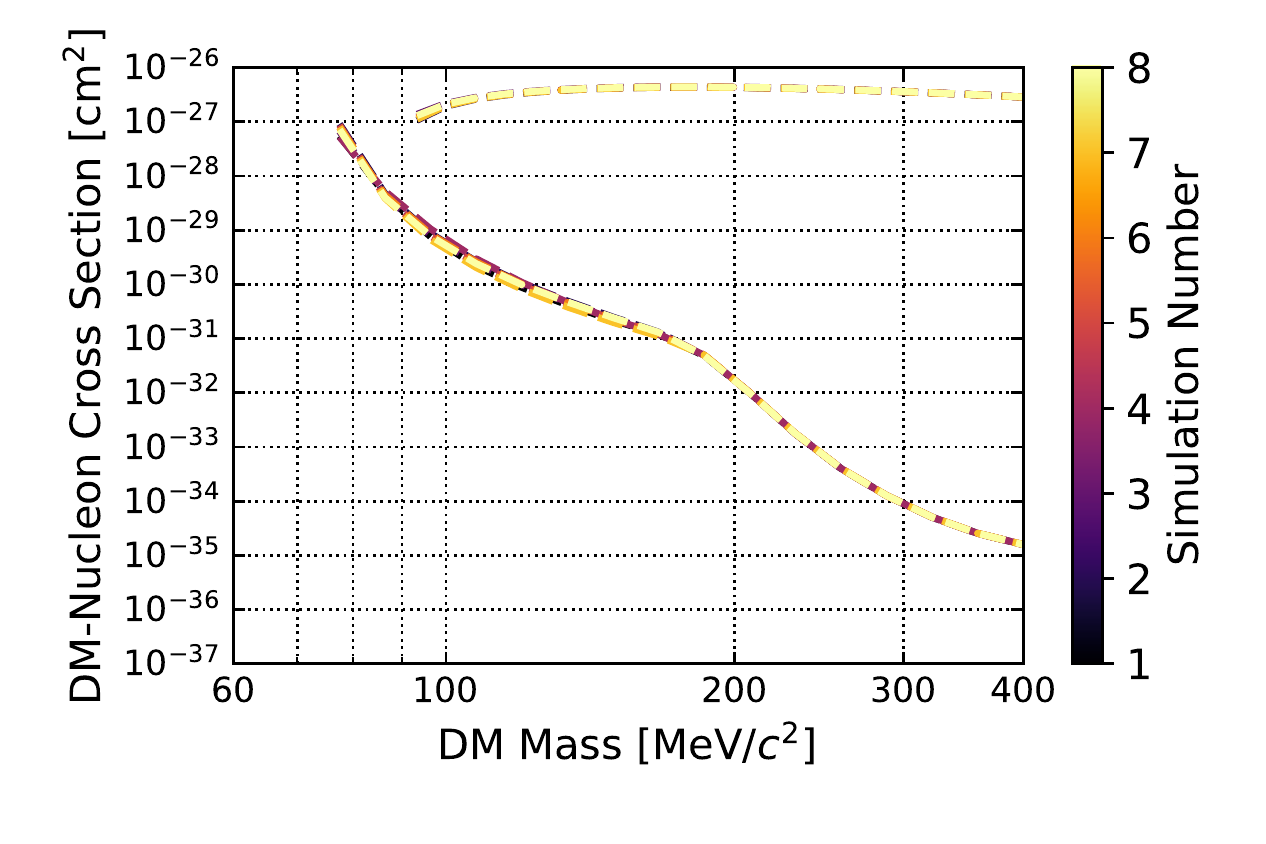}
    \caption{The OI and overburden limit for each simulation. The overburden calculation rules out sensitivity to DM masses below $93\, \mathrm{MeV}/c^2$. Though the two curves do not connect, this is because the overburden drops effectively vertically in cross section below this, and a very fine grid search would be required to capture the behavior. Because this would have been computationally expensive, and initial studies showed that the reach would have changed by less than $1\, \mathrm{MeV}/c^2$, we will simply cut the exclusion limit off at a DM mass of $93\, \mathrm{MeV}/c^2$.}
    \label{fig:full_limit_error}
\end{figure}

\begin{figure}
    \centering
    \includegraphics[width=0.8\linewidth]{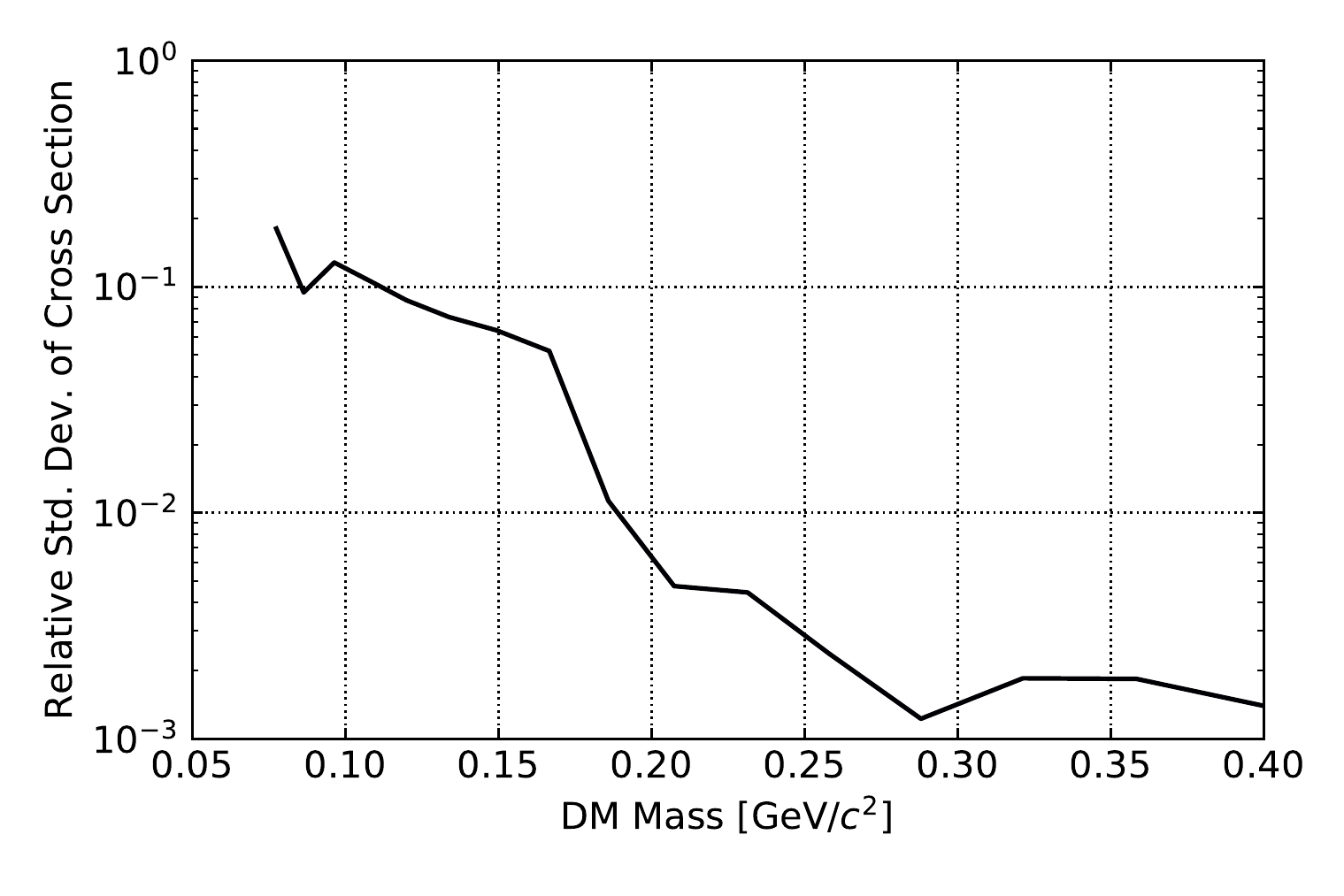}
    \caption{The relative error of the OI limit cross sections at each DM mass}
    \label{fig:rel_limit_error}
\end{figure}

\begin{figure}
    \centering
    \includegraphics[width=0.8\linewidth]{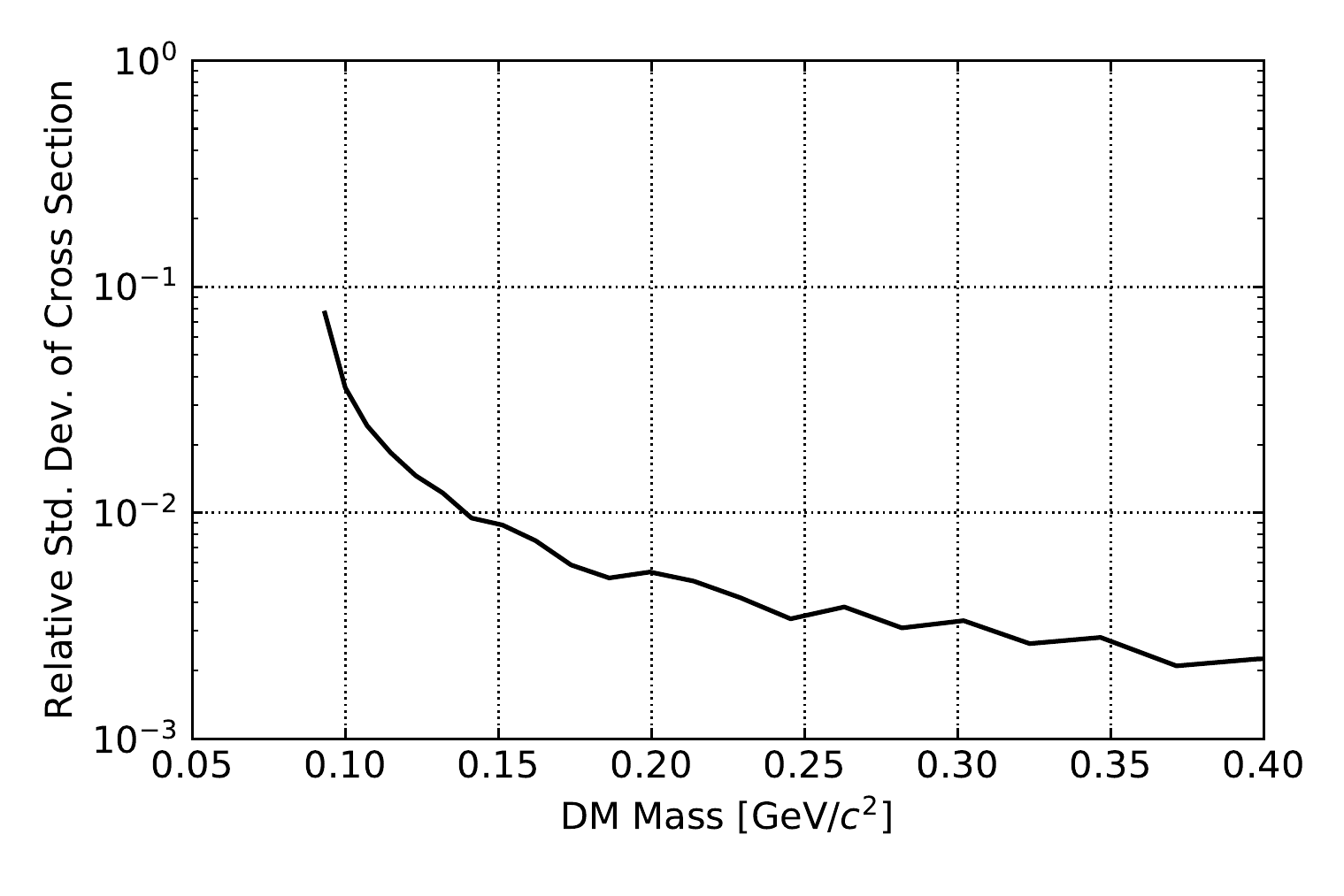}
    \caption{The relative error of the cross sections in the overburden limit at each DM mass.}
    \label{fig:rel_overburden_error}
\end{figure}

To estimate the statistical error in the limit contour, we compared the results obtained by calculating the signal model using the eight different sets of pulse simulations. Showing this in Fig.~\ref{fig:full_limit_error}, there is visual noticeable variation in the lower edge of the limits for DM masses below $200\, \mathrm{MeV}/c^2$. The corresponding plot of the relative standard deviation (Fig.~\ref{fig:rel_limit_error}) of the cross section for each DM mass shows that, above $200\, \mathrm{MeV}/c^2$, the error is less than 1\%. However, below this mass, the error increases to $\mathcal{O}(10\%)$. This matches our expectations looking at where the OI limits appear to ``split'' in Fig.~\ref{fig:full_limit_error}, as well as confirms the idea that the lowest masses in the limit will have more variation due to the higher uncertainty in the trigger efficiency. For the overburden limits, we see that the errors on the overburden cross section (Fig.~\ref{fig:rel_overburden_error}) are also of $\mathcal{O}(10\%)$ as compared to the errors on the DM OI limit. The error on the overburden limit does increase as we go lower in DM mass, but it does not increase much more than 10\%, and it only does so at the very lowest masses below $100\, \mathrm{MeV}/c^2$.

\begin{figure}
    \centering
    \includegraphics[width=0.8\linewidth]{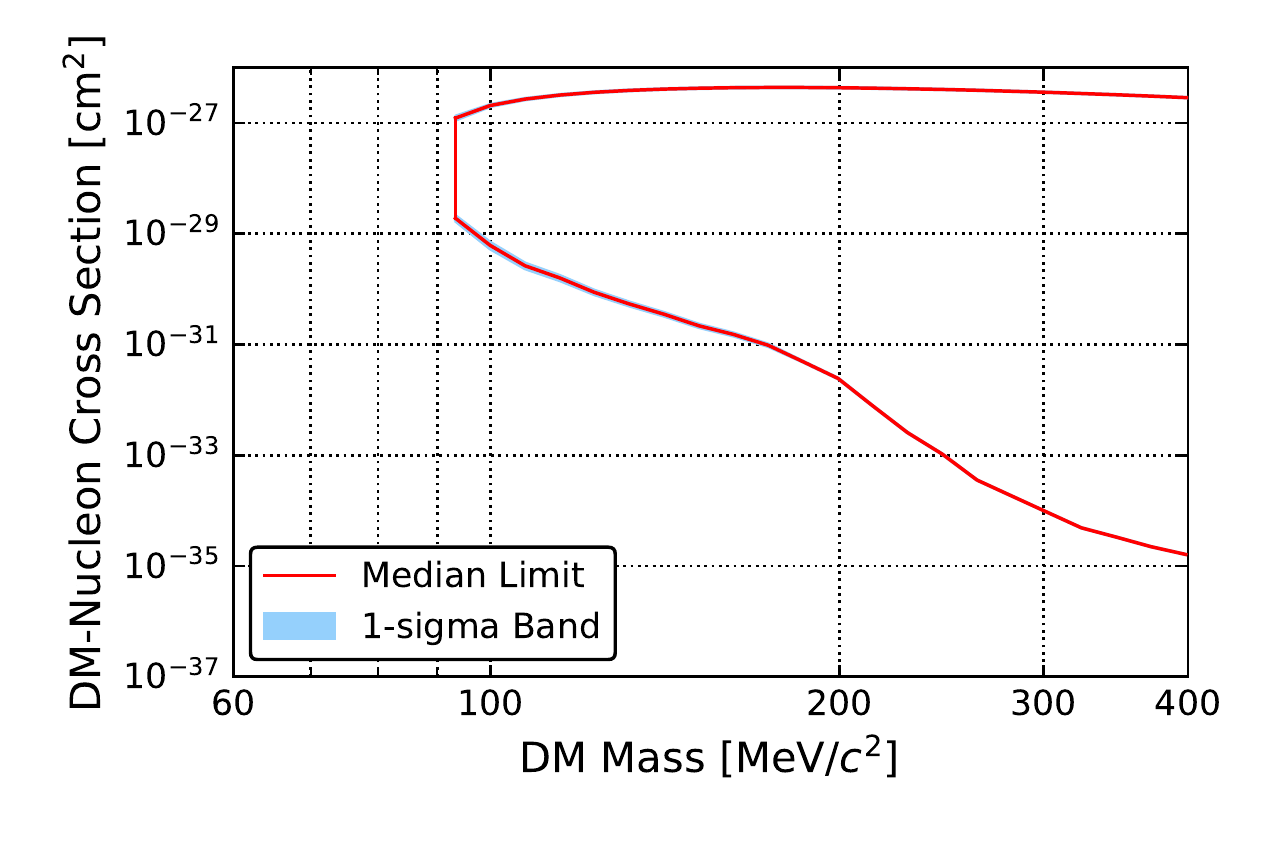}
    \caption{The DM exclusion contour set by the upper limit and the overburden calculation, including the $1\sigma$ error band in the limit set from the eight pulse simulations.}
    \label{fig:limit_error_band}
\end{figure}

With variation in both the OI limit and the overburden, we should understand how this variation propagates to the final limit. To do this, we take the central value (median) of the limits for each DM mass, and estimate its error as the standard deviation of the limits---as opposed to taking the mean value and the standard error of the mean, which would reduce the error by a factor of $\sqrt{8}$. Compared to the latter, the central value method is more conservative in its error estimation (i.e. it will give larger errors), chosen to ensure that we are at worst overestimating the uncertainty. We show the effect of this central value method for the calculation of the $1\sigma$ band in Fig.~\ref{fig:limit_error_band}.

We see that the $1\sigma$ band for the OI limit is quite small, which is as expected when plotting 10\% errors on a log-scale over many orders of magnitude. For the overburden limit, the $1\sigma$ band is effectively not visible on this scale, which is as expected due to its smaller error. We also see that the overburden calculation removed our ability to set limits on some of the lowest masses in the upper limit calculation, as shown by the vertical line connecting the two limit techniques. One might wonder if there should be a smoother connection between the two, as the overburden limit should continue to decrease in cross section for lower DM masses. In practice, this is true, but it would require a very fine grid-search to find this exact path due to how quickly the limit changes after this lowest mass. Due to computational limitations and the negligible improvement in the limit, this was not pursued.

\begin{figure}
    \centering
    \includegraphics[width=0.8\linewidth]{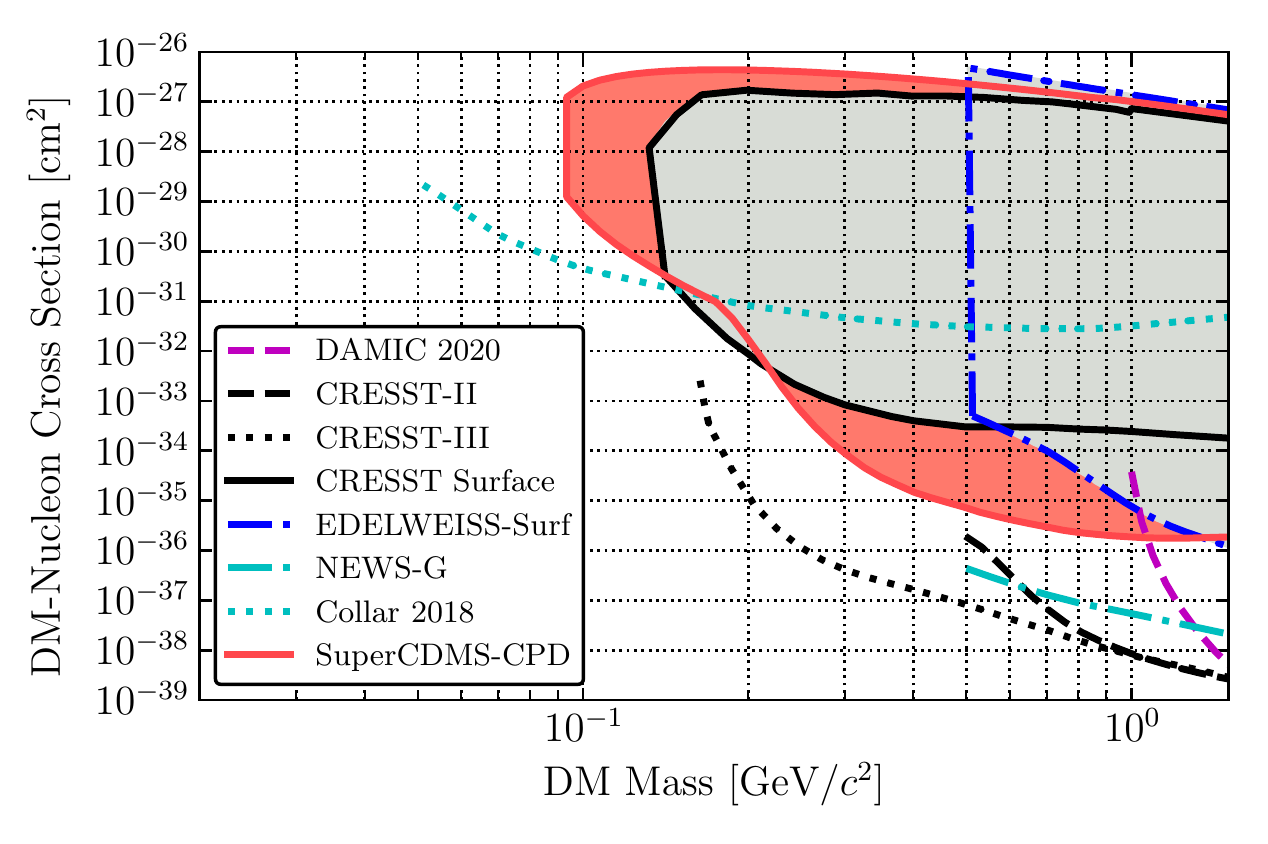}
    \caption[Dark matter exclusion limit set with the CPD]{\label{fig:limit}The 90\% C.L. limits on the spin-independent DM-nucleon cross section as a function of DM mass for this work (solid red line), compared to results from other experiments~\cite{EdelweissWIMP,Abdelhameed_2019,Angloher_2017,PhysRevLett.125.241803,cresst2,newsg,PhysRevD.98.023005}. For above-ground experiments with overburden calculations, the previously ruled out parameter space is shown as the gray shaded region, and the new parameter space ruled out from this search is shown as the red shaded region. For the Collar 2018 surface result, which uses a liquid scintillator cell operated at $1^{\circ} \, \mathrm{C}$, an overburden calculation would be useful for comparison to the upper edges of the various contours for the surface searches. We note that the systematic error in the baseline energy resolution changes the result within the error of the limit's line width, thus we only include the result from the $3.86\,\mathrm{eV}$ calibration.}
\end{figure}

The results of the dark matter search are shown in Fig.~\ref{fig:limit}, where we have taken the median of the limits calculated for the eight simulations at each DM mass, and compared to other pertinent DM searches in the same parameter space~\cite{PhysRevLett.125.241803,cresst2,Abdelhameed_2019,Angloher_2017,EdelweissWIMP,newsg,PhysRevD.98.023005}. The 10\% variation is not plotted, as it would not be visible in the figure. For DM masses between $93$ and $140\,\mathrm{MeV}/c^2$, these results provide the most stringent limits for nuclear-recoil DM signals using a cryogenic detector. For DM masses between $220\,\mathrm{MeV}/c^2$ and $1.35\,\mathrm{GeV}/c^2$, they are the most stringent limits achieved in an above-ground facility. For these low DM masses, the large cross sections approach the level at which the Born approximation used in the standard DM signal model begins to fail~\cite{PhysRevD.100.063013}. However, in the absence of a generally accepted alternative model and to be comparable to other experiments (all of which also use the Born approximation in this regime), we decided to keep it in our signal model as well.

\begin{figure}
    \centering
    \includegraphics[width=0.8\linewidth]{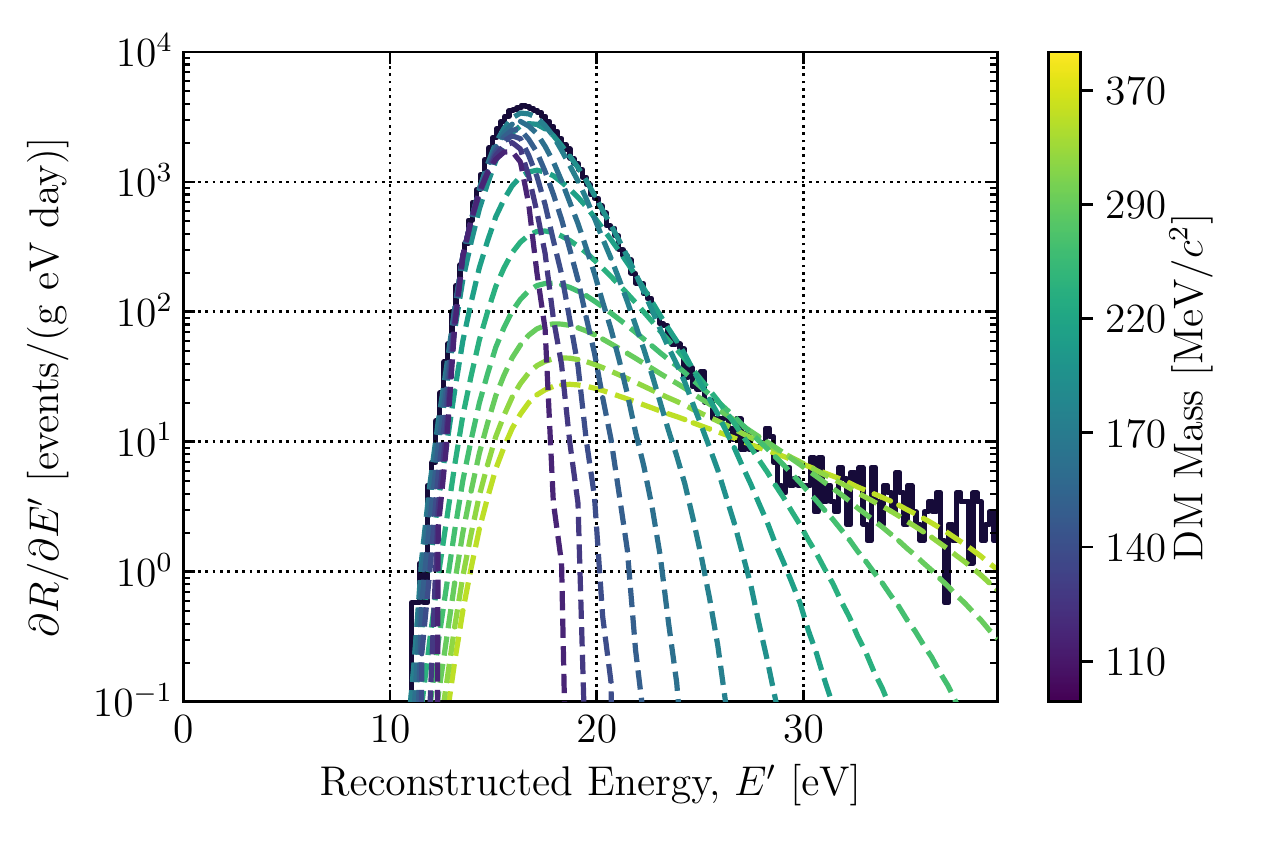}
    \caption[Expected event rates for various dark matter cross sections and masses from the exclusion limit]{\label{fig:drde}The event spectrum for the DM search data below $40\,\mathrm{eV}$ in reconstructed energy. The data have been normalized to events per gram per day per eV and have been corrected for the signal efficiency of the data-quality cuts, but not the confidence ellipse and trigger time cuts. The colored dashed lines represent the calculated event rates for selected DM cross sections and masses from the 90\% C.L. OI limit for a single pulse simulation, where the optimum intervals in recoil energy are below $40\,\mathrm{eV}$. Sensitivity to DM masses below $400\,\mathrm{MeV}/c^2$ corresponds to recoil energies below $40\,\mathrm{eV}$, with the lowest masses requiring energy sensitivity down to about $15\,\mathrm{eV}$.}
\end{figure}

In Fig.~\ref{fig:drde}, we show the data spectrum for reconstructed energies below $40\,\mathrm{eV}$ and DM signal curves for various DM masses for a single pulse simulation, where the cross sections from the OI limit are used. The approximate location of the optimum interval is apparent for each dark matter mass. We see that the sensitivity to DM masses below $400\,\mathrm{MeV}/c^2$ corresponds to recoil energies below $50\, \mathrm{eV}$, with the lowest masses requiring energy sensitivity down to $15 \, \mathrm{eV}$. In future experiments, we will need to continue to improve our baseline  energy resolution in order to be able to set thresholds at lower values.

In this search, we see an excess of events for recoil energies below about $100\,\mathrm{eV}$, emerging above the roughly flat rate from Compton scattering of the gamma-ray background. This excess of events could be from an unknown external background or due to detector effects such as stress-induced microfractures~\cite{ASTROM2006262}. As other experiments have observed excess events in searches for low-mass nuclear-recoiling DM~\cite{Abdelhameed_2019, Angloher_2017, EdelweissWIMP, PhysRevD.102.015017}, understanding this background is of pivotal importance. In Chapter~\ref{chap:excess}, we will discuss further the possible various backgrounds, as well as studies that have been done to rule some out for the CPD.

\section{Conclusion}

Using a detector with $\sigma_E = 3.86 \pm 0.04 \, (\mathrm{stat.})^{+0.19}_{-0.00} \, (\mathrm{syst.}) \, \mathrm{eV}$ baseline energy resolution operated in an above-ground facility with an exposure of $9.9\,\mathrm{g\,d}$, we probe parameter space for spin-independent interactions of DM with nucleons for dark matter particles with masses above $93\,\mathrm{MeV}/c^2$. The range from $93$ to $140\,\mathrm{MeV}/c^2$ was previously not accessible to cryogenic detectors. These results also set the most stringent limits for above-ground nuclear-recoil signals from dark matter for masses between $220\,\mathrm{MeV}/c^2$ and $1.35\,\mathrm{GeV}/c^2$. This was achieved using a single readout channel composed of QETs distributed on a Si substrate, with a recoil energy threshold set at $16.3\,\mathrm{eV}$.

The results of this work were accomplished despite the high background rates in our surface facility because of the excellent baseline energy resolution of the detector. There are plans to operate this detector in an underground laboratory, where we expect to have a significantly lower Compton scattering background rate. This will allow further study of the excess events observed in the ROI, hopefully providing insight into the origin of the event rate that is limiting the results reported here, as discussed in Chapter~\ref{chap:excess}.

These results also demonstrate the potential of athermal phonon sensors with eV-scale baseline energy resolution for future dark matter searches via DM-nucleon interactions. Because this detector has a large surface area relative to its small volume, it is not optimal for a DM search. The baseline energy resolution of such devices scales with the square root of the number of QETs, which itself is proportional to the instrumented area (assuming the same QET design used by the CPD)~\cite{Hochberg_2016,KNAPEN2018386}. Thus, a decrease in the instrumented area, with an increase in volume, should lead to improvements in baseline energy resolution. Future work is planned to design detectors of volume $\sim\!1\,\mathrm{cm}^3$ (detailed in C. W. Fink's thesis~\cite{finkthesis}), for which it is reasonable to expect roughly an order of magnitude improvement in baseline energy resolution through these geometric considerations alone. With improved baseline energy resolution comes a lower energy threshold, allowing a search for spin-independent DM-nucleon interactions for even lower DM masses and a clear path to surpassing the existing noncryogenic detector constraints on sub-$100\, \mathrm{MeV}/c^2$ DM interacting with nucleons.

\chapter{\label{chap:excess}Investigation of Excess Signals}

Many LDM experiments have observed various excess signals above the expected background, as detailed in Ref.~\cite{adari2022excess}. As noted in the previous chapter, the SuperCDMS-CPD DM search too has its own excess signals. In this chapter, we will motivate our hypothesis of stress-induced microfracture backgrounds via the ruling out of other backgrounds, as well as showing its likely existence in other devices run by our research group. We also propose a solution to removing this background from our detectors via stress-free holding schemes and passive vibration isolation.

\section{Excess Signals in CPD}

\begin{figure}
    \centering
    \includegraphics{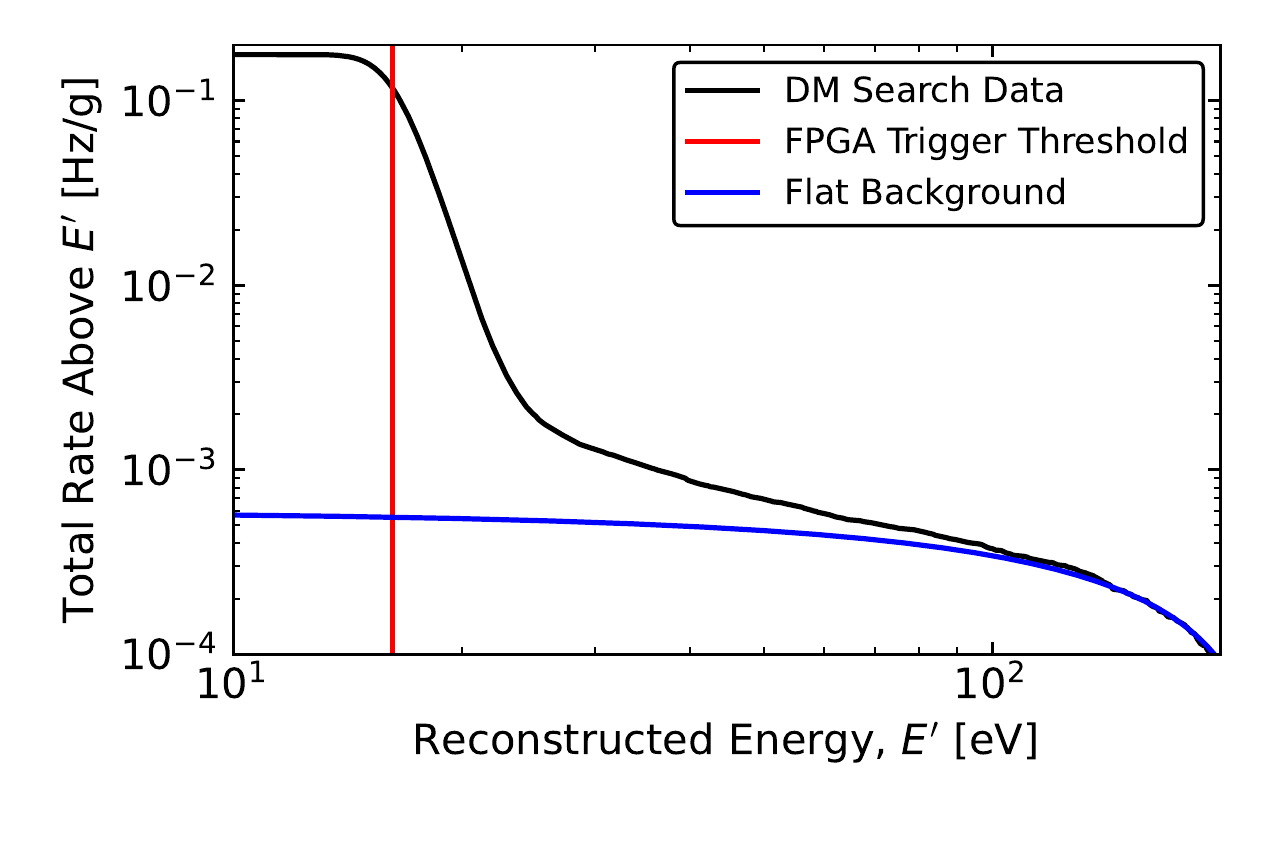}
    \caption{The mass-normalized integrated rate of the SuperCDMS-CPD DM search spectrum within the ROI, with the location of the FPGA trigger threshold noted. Because we are plotting as a function of the offline reconstructed energy, the smearing of the energies shows events below threshold---these events are above threshold in FPGA trigger energy.}
    \label{fig:cpd_total_rate}
\end{figure}

There are a multitude of possible sources of the excess signals observed in the CPD analysis that rise over the roughly flat expected differential rate of Compton scattering of the gamma ray background. Before going through them, we first return to the spectrum from the DM search. For convenience in studying the excess, we will integrate the observed reconstructed energy spectrum (i.e. the main plot of Fig.~\ref{fig:spectrum_full}) in terms of the reconstructed energy via
\begin{equation}
    R(E') = \int_x^{E_{ROI}} \mathop{dE'} \frac{\partial R}{\partial E'}(E'), 
    \label{eq:integ_rate}
\end{equation}
where $E_{ROI}= 240 \, \mathrm{eV}$ is the maximum energy within the DM search ROI and $\partial R / \partial E'$ is defined by 
\begin{equation}
    {\frac{\partial R}{\partial E'}(E')}=\int_0^\infty \mathop{dE_T} \int_0^\infty  \mathop{dE_0} \Theta(E_T - \delta)\varepsilon(E', E_T, E_0) P(E',E_T|E_0)\frac{\partial R}{\partial E_0}(E_0),
\end{equation}
which has been reproduced from Eq.~(\ref{eq:signal}). We can count up the total number of events within the ROI and divide by the 22 hours of live time, which returns the total rate within the ROI of $1.9 \, \mathrm{Hz}$. If we normalize this to the CPD mass of $10.6 \, \mathrm{g}$, this is a rate of $0.18 \, \mathrm{Hz}/\mathrm{g}$ (i.e. the rate of events that were tagged with FPGA trigger energies above threshold). Calculating $R(E')$ over the entire ROI, we have the mass-normalized rate above some offline reconstructed energy $E'$, as plotted in Fig.~\ref{fig:cpd_total_rate}. In this figure, we also compare to the rate expected from the flat background at $2 \times 10^5 \, \mathrm{events}/(\mathrm{kg} \, \mathrm{keV} \, \mathrm{day})$, noting that the functional form is simply a constant multiplied by $(E_{ROI} - E')$. In this case, that constant is $2.5 \, \mathrm{mHz}/(\mathrm{g} \, \mathrm{keV})$, taking care to convert units correctly. In other words, the $y$-intercept of the flat background is $0.6 \, \mathrm{mHz}/\mathrm{g}$, as seen in the figure.

We see that there are two characteristic exponential excess backgrounds from that expected from the flat background. There is one that is dominant from the energy threshold to about $25 \, \mathrm{eV}$ and another that is dominant from $25 \, \mathrm{eV}$ to about $100 \, \mathrm{eV}$. To make an educated guess at the origin of these backgrounds, we must go through all of the possibilities, which are: random fluctuations, Cherenkov interactions, transition radiation, luminescence, other low energy interactions with high energy particles, neutrons, radon, electromagnetic interference (EMI) signals, and stress-induced microfractures.

\subsection{Expectation from Random Fluctuations}

\begin{figure}
    \centering
    \includegraphics[width=\linewidth]{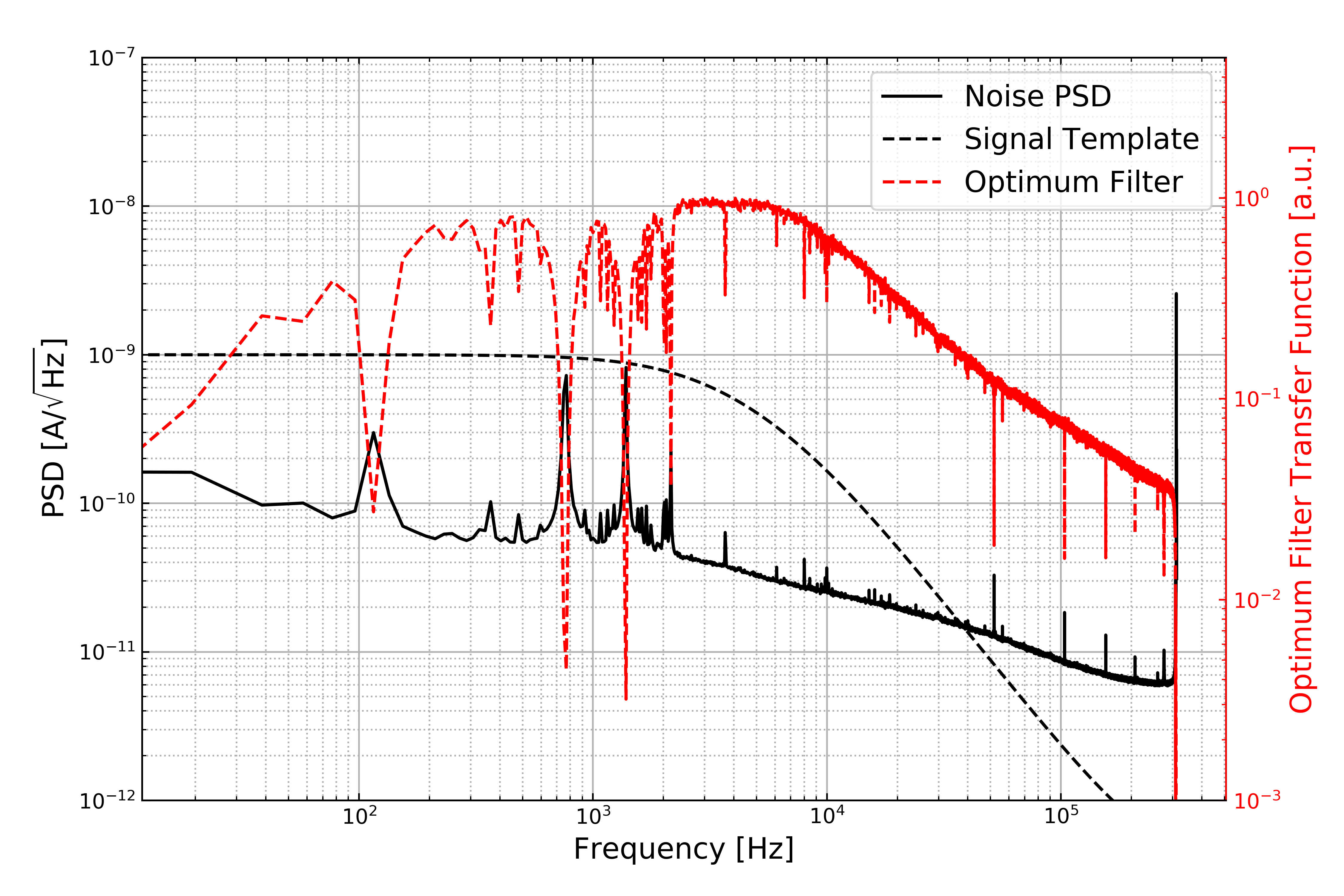}
    \caption{Comparison of the PSD, template, and optimal filter (see Appendix~\ref{chap:of} for the definition) for offline processing of both the DM search data and any simulated data.}
    \label{fig:psd_templates}
\end{figure}

To understand what is expected from random fluctuations above threshold, we can generate events from our understanding of the noise of the CPD. To do this, we take our noise PSD, as depicted in Fig.~\ref{fig:psd_templates}, and generate simulated randoms from it via the function \texttt{gen\_noise} in \textsc{QETpy}~\cite{qetpy}. The workflow of this function is to sample from a standard normal distribution to create white noise, take the Fourier transform of the sampled data, multiply by the square root of the PSD, and then inverse Fourier transform back to time domain. This is equivalent to applying a transfer function to white noise to have the desired frequency response (our noise spectrum). Note that this randoms generation assumes that the noise is uncorrelated and stationary, which is likely not a true representation of our noise environment, but is the best that we can do with our data from the single-channel CPD.

For this study, we created 50,000 simulated randoms using the above noise PSD and ran an unconstrained time degree-of-freedom OF to roughly simulate the FPGA triggering algorithm, as the FPGA triggering algorithm is effectively an unconstrained OF due to its convolution over traces with a $25 \, \mathrm{ms}$ long window and choosing the maximum amplitude. For the generated randoms, this means that we do not put the data fully through the same analysis pipeline as the DM search dataset, but we expect that this should give a good rough estimate.

\begin{figure}
    \centering
    \includegraphics{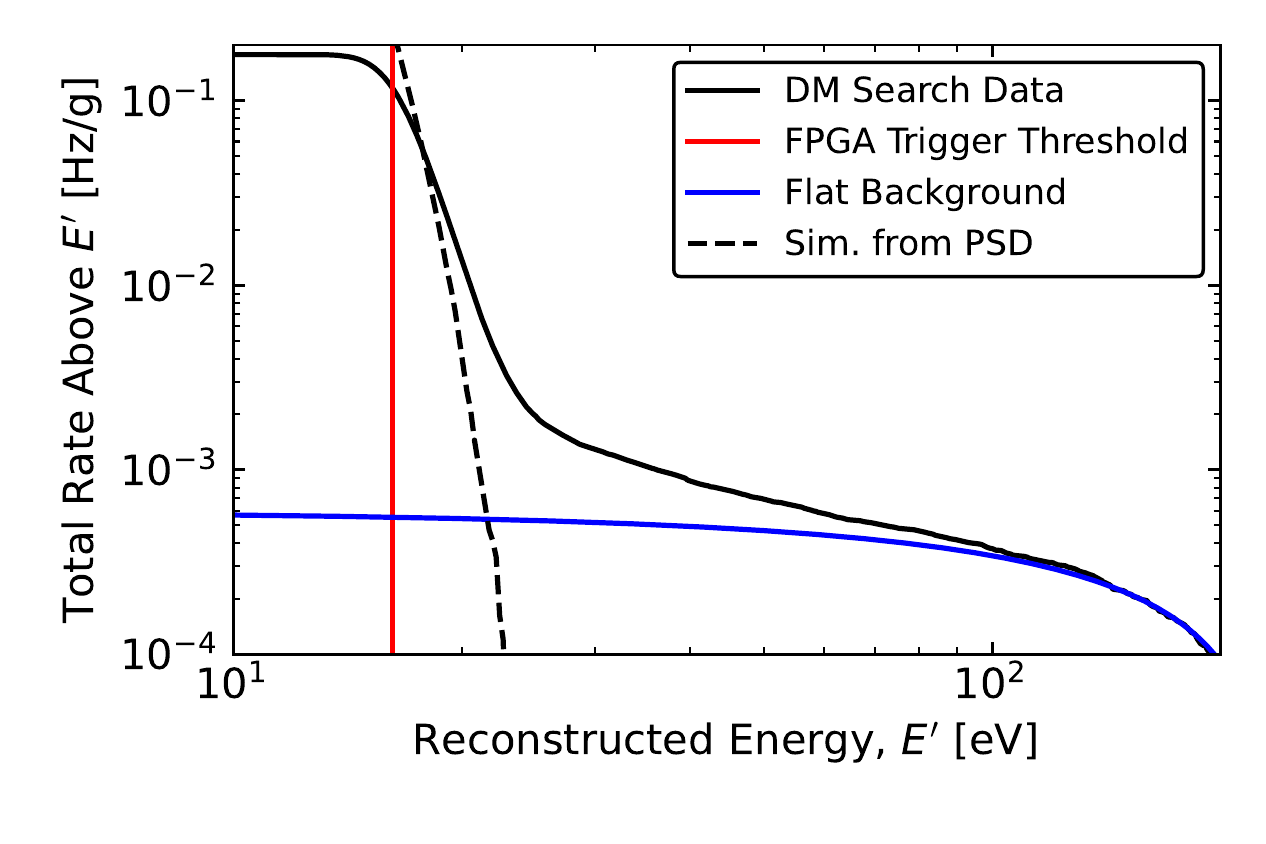}
    \caption{Comparison of integrated spectrum of DM Search to the randoms spectrum calculated from the noise generation routine, where these data have been generated from the experimentally measured noise PSD. The black dashed line cuts off at the FPGA trigger threshold with a rate of $0.2 \, \mathrm{Hz}/ \mathrm{g}$, as we set the energy threshold on the generated randoms at the FPGA threshold.}
    \label{fig:cpd_rate_rands}
\end{figure}

Taking the OF amplitude spectrum from these generated randoms, we can calculate the integrated rate via Eq.~(\ref{eq:integ_rate}) and compare it to the DM search spectrum, as well as the expectation from the flat background. The result of this is shown in Fig.~\ref{fig:cpd_rate_rands}.

We see that the randoms spectrum is roughly within a factor of two of the sub-$30 \, \mathrm{eV}$ region of the DM search spectrum. Because the simulated data from the experimental PSD assumes a normal distribution, nonstationary noise, and no correlations between frequencies, the factor of two difference is not unexpected, as each of these phenomena are likely existent in the DM search data to some degree. With this small difference, it is reasonable that the sub-$30 \, \mathrm{eV}$ background could indeed be due to randoms and does not require explanation via some physical background. With this in mind, pushing towards lower threshold DM searches will also mean pushing of this randoms background to lower energies, and the main background to be worried about is the apparent sub-$100 \,\mathrm{eV}$ excess background.

To understand the remaining component of the excess background, we carry out a least squares fit assuming an exponential background to the residual of its shape (after subtracting the expectation from the flat background). Doing this for the $30$ to $100\,\mathrm{eV}$ region, we find that the exponential background has a characteristic energy scale of $E_{char.} = 16 \, \mathrm{eV}$ (that is, it goes as $\exp{\left(-E' / E_{char.}\right)}$) and a $y$-intercept of $4.4 \, \mathrm{mHz} / \mathrm{g}$. It remains to be seen if this background is well-modeled as a single exponential below threshold, which can be confirmed in future low-threshold searches.

\begin{figure}
    \centering
    \includegraphics{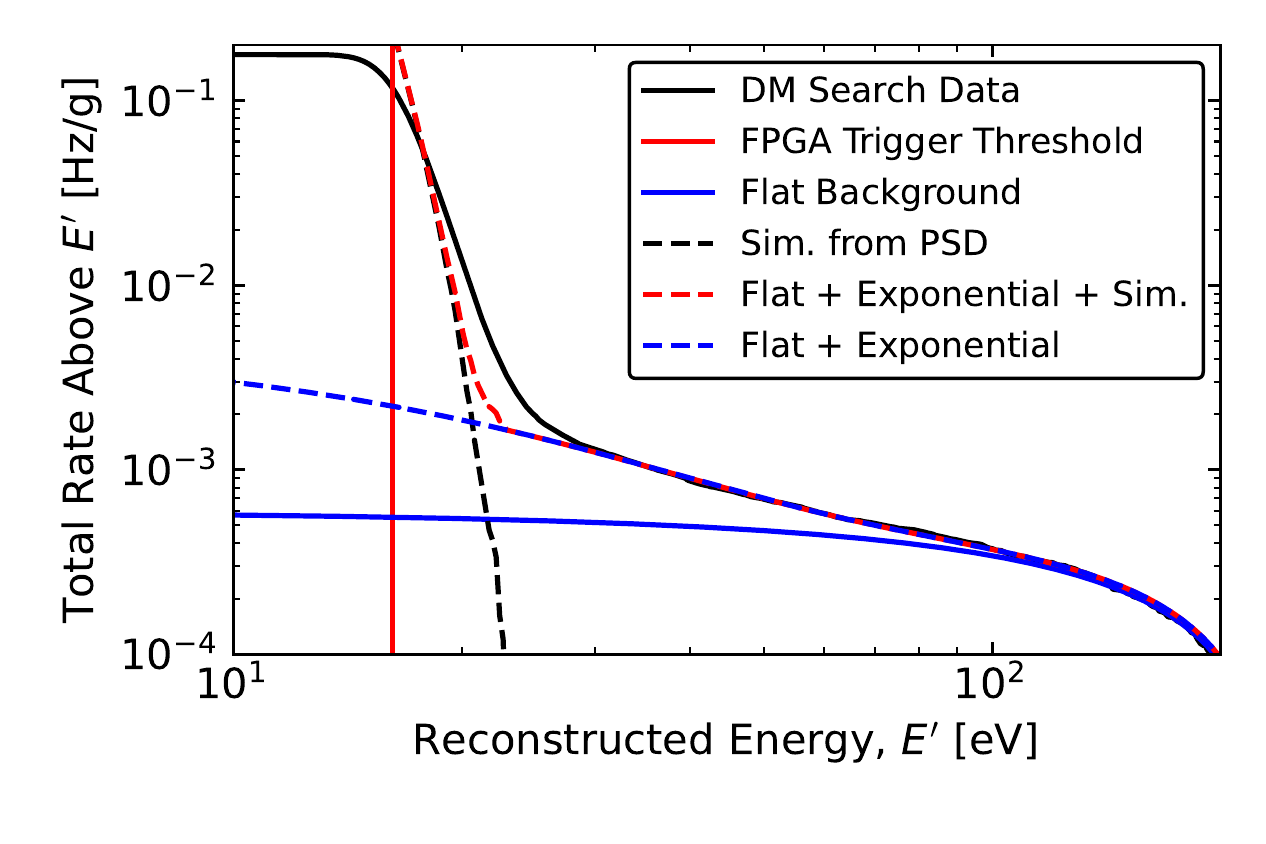}
    \caption{Comparison of the summed expected spectra from randoms, the flat background, and the fitted exponential excess to the observed DM search spectrum.}
    \label{fig:cpd_rate_fits}
\end{figure}

In Fig.~\ref{fig:cpd_rate_fits}, we compare the randoms spectrum, the DM search spectrum, the flat background, and the exponential background. Together, the flat and exponential backgrounds provide excellent agreement with the observed spectrum above $30 \, \mathrm{eV}$, with a small amount of discrepancy below this point likely due to the generated randoms not being truly representative of random events in the DM search. Thus, we have a reasonable explanation of the sub-$30 \, \mathrm{eV}$ background, while we still lack an explanation of the remaining exponential background in the 30 to $100 \, \mathrm{eV}$ region. This requires us to continue ruling out other possible sources of such a background.

\subsection{Other Possible Sources}

To recap, the remaining other sources of this exponentially increasing excess background could be one of the following: Cherenkov interactions, transition radiation, luminescence, other low energy interactions with high energy particles, neutrons, radon, EMI signals, or stress-induced microfractures. In the next few paragraphs, we will discuss whether or not any of these are likely explanations.

\subsubsection{Cherenkov Interactions, Transition Radiation, or Luminescence}

\begin{figure}
    \begin{subfigure}{.5\textwidth}
        \centering
        \includegraphics[width=1\linewidth]{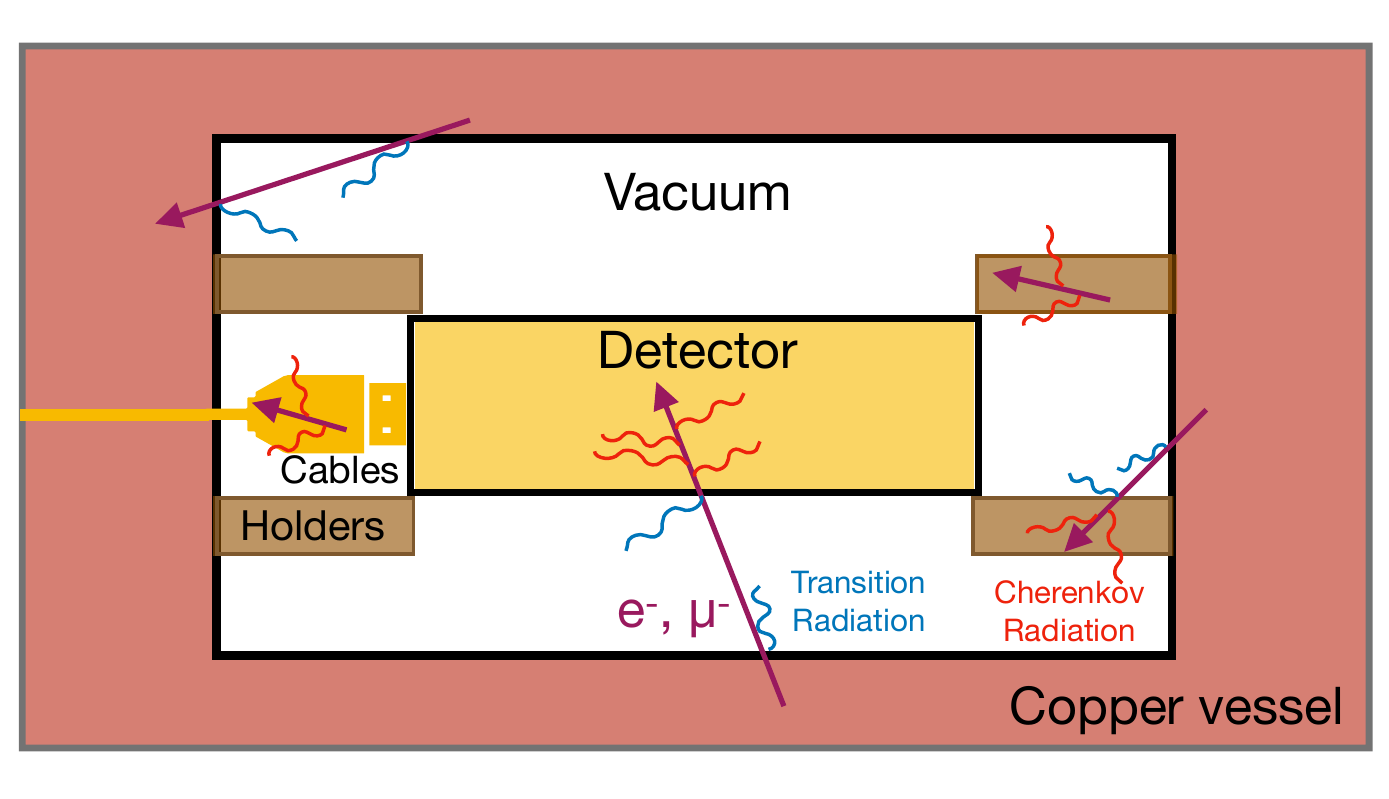}
    \end{subfigure}%
    \begin{subfigure}{.5\textwidth}
        \centering
        \includegraphics[width=1\linewidth]{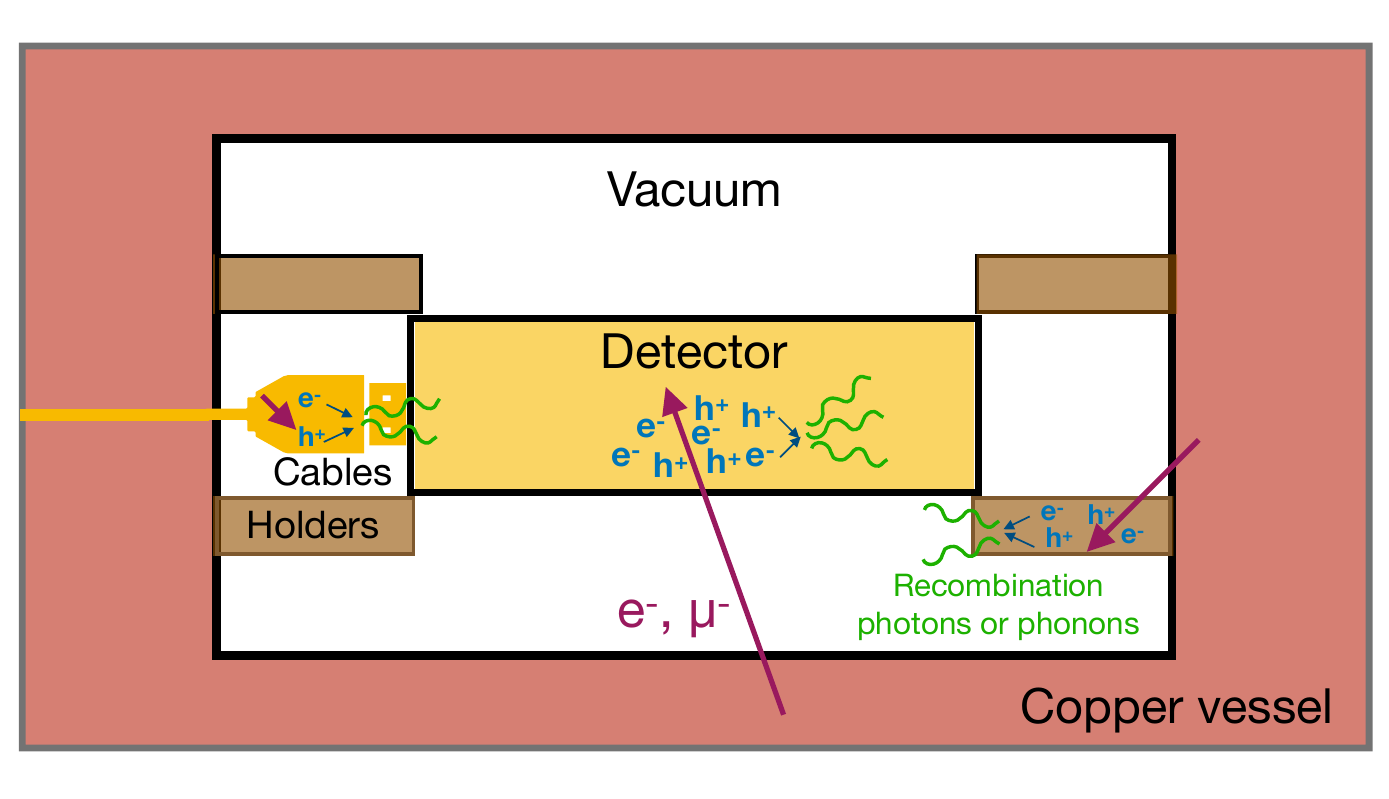}
    \end{subfigure}
    \caption{(Figure from Ref.~\cite{dupeizhi2022}) (Left) An arbitrary detector being held in place and its signals read out by some type of cabling system. Generally, these detectors are held inside a Cu housing. The purple arrows represent high-energy charged particles (electrons or muons) interacting with the various components of the system. The red lines denote photons emitted from Cherenkov interactions, and the blue lines denote those emitted from transition radiation. (Right) For the same arbitrary setup, a diagram of possible interactions of of a high energy particle exciting electrons and holes, which then relax and recombine, releasing energy in the form of photons (luminescence) or phonons (the green lines).}
    \label{fig:chertran}
\end{figure}

Cherenkov interactions, transition radiation, and luminescence have the potential to create low-energy events that could be detected by in low-threshold dark matter searches, such as in the CPD DM search. Cherenkov interactions are related to when charged particles pass through dielectric materials with a velocity greater than the in-medium speed of light. When this occurs, photons with eV-scale energies can be emitted via these high-energy particle interactions. Similarly, transition radiation occurs at the interface between two regions with different dielectric properties, where the passage of high-energy charged particles can too emit photons of with eV-scale energies. Luminescence (or scintillation) occurs when some particle (e.g. a photon, electron, etc.) scatters off a material, excites an electron into the conduction band, which then releases energy after returning to the valence band in the form of photons (or possibly phonons if an indirect bandgap or there are in-gap states from impurities). These processes are diagrammatically shown in Fig.~\ref{fig:chertran} for an arbitrary detector setup.

In the case of the CPD DM search, the CPD is a Si substrate held by cirlex clamps and read out by our PCB electronics that are within the hexagonal Cu housing. Thus, there are multiple nonconductive materials with various dielectric properties which all can emit low-energy photons via Cherenkov interactions, transition radiation, or luminescence. In Ref.~\cite{dupeizhi2022}, P. Du \textit{et al.} have theoretically calculated the expected rates of each of these backgrounds and compared to the experimentally observed excess background in the SuperCDMS-CPD DM search. They find that these backgrounds can only account for up to 10\% of the observed excess exponential background, and thus cannot be the dominant source.

\subsubsection{Other Low Energy Interactions with High Energy Particles, Neutrons}

There does remain backgrounds from other low-energy interactions with high-energy particles, as well as interactions with neutrons. From only the CPD DM dataset at SLAC, we cannot claim anything about these backgrounds. Fortunately, we were able to have our SuperCDMS collaborators run the CPD at their underground facility CUTE (Cryogenic Underground TEst facility), located at SNOLAB in Sudbury, Ontario~\cite{Camus:2018fab,Rau:2020ujh}. In particular, we were able to measure event spectra in their low-background cryostat, though without a calibration source. However, we can achieve a rough energy calibration through our estimate of the energy resolution via the NEP, as discussed in Chapter~\ref{chap:two}. In this scenario, we can take the measured OF amplitudes in current, and convert to energy scales via the ratio of the measured OF amplitude resolution and the expected baseline energy resolution of the CPD. Note that the expected baseline energy resolution has a dependence on the collection efficiency of the detector, which was measured to be $\varepsilon_{ph}=13\%$ (recall Section~\ref{sec:phononpulseshape}). In the case of running this detector at CUTE, we would expect this efficiency to be roughly the same, where changes could be due to an increase in charged impurities as compared to the state of the detector when ran at SLAC, or perhaps oxygen and other gases remaining on the surface of the detector due to not pumping to vacuum efficiently. Thus, we can only use this as a ``rough calibration'' (that becomes nonlinear at a few hundred eV), but it should be roughly correct at low energies.

\begin{figure}
    \centering
    \includegraphics{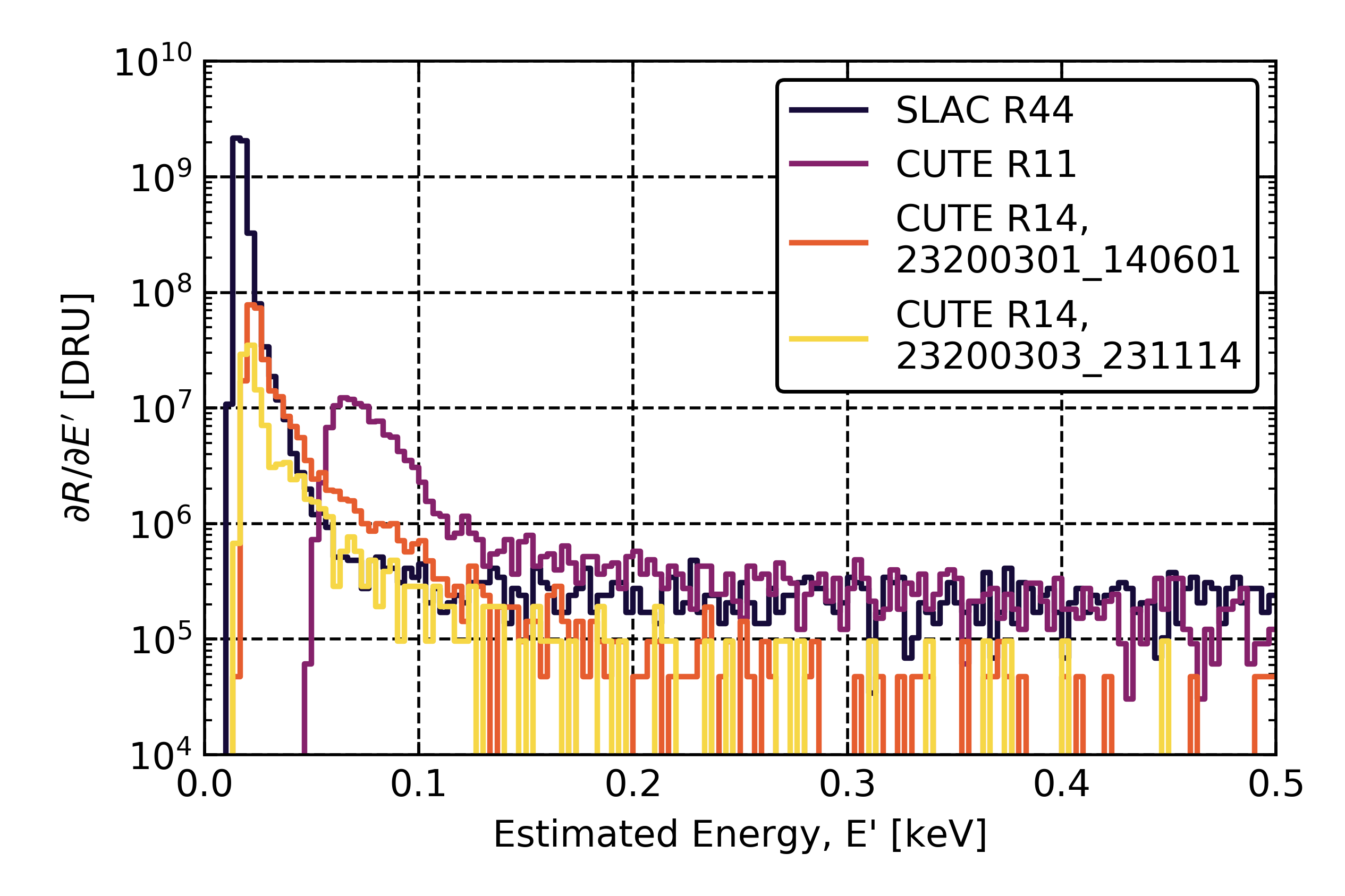}
    \caption{Comparison of the various datasets with the rough calibration. The SLAC R44 line corresponds to a rough calibration of the CPD DM search at SLAC spectrum. The $y$ values are the differential rate spectra for each dataset in units of DRU (Differential Rate Unit), a common abbreviation used to denote $\mathrm{events}/(\mathrm{kg} \, \mathrm{keV} \, \mathrm{day})$. Comparing the two CUTE R14 datasets, we see that there is a decrease in the rate after only two days, providing another reason that this background is likely not from low-energy interactions with high-energy particles (we would not expect a decreasing time-dependence from this type of background). This figure was also reproduced in Matt Pyle's talk at the EXCESS2022 Workshop~\cite{excess22}.}
    \label{fig:cutecomp}
\end{figure}

The CPD was operated at CUTE in a few runs, and we will compare the spectra from the CPD DM search at SLAC to those from Runs 11 and 14 at CUTE. In Run 11, there was a single 24-hour dataset that we will compare to from mid-January, 2020. In Run 14, there were two datasets that we will compare to: \texttt{23200301\_140601} (22 hours starting on March 1, 2020) and \texttt{232200303\_231114} (8 hours starting on March 3, 2020). These numeric sequences denote the date and time of the beginning of data taking---both in early March 2020, but two days apart. Extracting offline OF amplitudes in the same way as in the DM search at SLAC, we can compare the spectra of all of these datasets, as done in Fig.~\ref{fig:cutecomp}.

Between the CUTE runs and SLAC, there are many similarities. Comparing SLAC and CUTE Run 11, we see that the spectra are actually quite similar above about $150 \, \mathrm{eV}$. However, due to an unknown change between CUTE Runs 11 and 14, this background disappears for the Run 14 datasets. However, between all four datasets, we see a very similar exponential background with nearly the same characteristic energy scale. The continued existence of this background and similar order of magnitude suggests that going underground did not reduce the background. Because of the excellent shielding at CUTE from both the Earth and the facility's water shield, the lack of change means that low-energy interactions of high-particles (beyond Cherenkov interactions, transition radiation or luminescence), as well as neutron backgrounds, are likely not the source of this background. The figure also shows that the rate has decreased after two days between the two Run 14 datasets, suggesting a time-dependence that we would not expect for these types of backgrounds.

\subsubsection{Radon Contamination}

Furthermore, at keV-scale energies in the CUTE Run 14 datasets, the differential rate is on the order of 1000 $\mathrm{events}/(\mathrm{kg} \, \mathrm{keV} \, \mathrm{day})$. One source that could be creating these events is $^{210}$Pb surface contamination of the detector or its housing (resulting from radon exposure), as the detector had not been subject to radon control for up to two years of unmitigated exposure. In order to estimate the expected rate of events due to from $^{210}$Pb surface contamination, we start with the national average of radon concentration indoors being about $50 \, \mathrm{Bq}/\mathrm{m}^3$. This concentration will affect our detector through radon progeny depositing onto the material surfaces, a process called ``plate-out''. For a surface with some area, the amount of radon progeny deposition on that surface is related to an effective plate-out height, such that all the progeny of the radon above that area up to this height will be deposited. For a typical cleanroom, this value is about $h\approx 15 \, \mathrm{cm}$~\cite{doi:10.1063/1.5019012}. This detector was enclosed in a housing while in a cleanroom, and thus is more likely that the effective plate-out height is about $1\, \mathrm{cm}$. In either case, we will estimate the expected rate due to this contamination for both scenarios for a duration two years.

\begin{figure}
    \centering
    \includegraphics[width=0.8\linewidth]{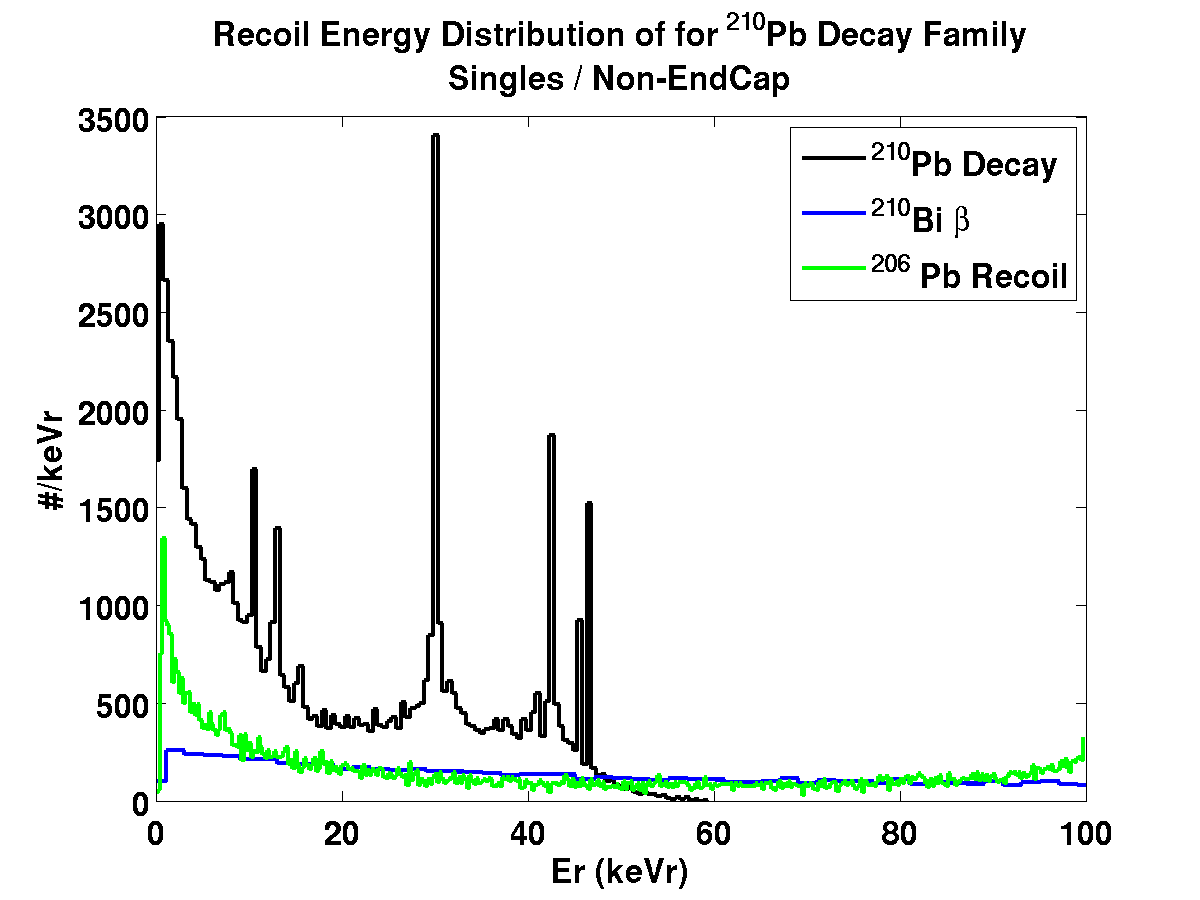}
    \caption{Energy spectra for $^{210}$Pb daughter decay (plot by Peter Redl), simulated as part of the SuperCDMS SNOLAB projected sensitivity~\cite{PhysRevD.95.082002}.}
    \label{fig:pb210}
\end{figure}

To estimate the $^{210}$Pb activity, we multiply the concentration of radon indoors by the plate-out height, the surface area of the CPD ($93.6 \, \mathrm{cm}^2$), and the ratio of contamination time to the mean lifetime of $^{210}$Pb, which gives a total $^{210}$Pb activity of $4.5 \, \mathrm{mBq}$ and $30 \, \mu \mathrm{Bq}$ for plate-out heights of $15 \, \mathrm{cm}$ and $1 \, \mathrm{cm}$, respectively. To convert these values to an expected differential rate on the CPD, we refer to the energy recoil distributions for the various progeny, as shown in Fig.~\ref{fig:pb210}, focusing on the dominant $^{210}$Pb spectrum. We can estimate the expected DRU within the energies that have been measured in the CPD (up to $6\, \mathrm{keV}$) by taking the fractional integral of the $^{210}$Pb spectrum up to $6\, \mathrm{keV}$ (about 0.27), multiplying it by the $^{210}$Pb activity, and dividing by the CPD mass, which gives an expected differential rate from $^{210}$Pb of $1760 \, \mathrm{events}/(\mathrm{kg} \, \mathrm{keV} \, \mathrm{day})$ and $117 \, \mathrm{events}/(\mathrm{kg} \, \mathrm{keV} \, \mathrm{day})$ for plate-out heights of $15 \, \mathrm{cm}$ and $1 \, \mathrm{cm}$, respectively, where the maximal value is on the order of the observed $1000 \, \mathrm{events}/(\mathrm{kg} \, \mathrm{keV} \, \mathrm{day})$ rate. However, the effective plate-out height for this scenario is quite conservative because the detector was stored in a covered housing, as opposed to being left out uncovered in a cleanroom. Thus, we do not expect that nearly as much radon had plated out onto the substrate, and the rate corresponding to effective plate-out height of $1 \, \mathrm{cm}$ is a much better estimate (an order of magnitude below what was observed). Though there are other progeny ($^{210}$Bi, $^{210}$Po) which could increase these rates, their differential rates are expected to be significantly lower than that from the $^{210}$Pb, based on takeaways from the SuperCDMS SNOLAB sensitivity projections~\cite{PhysRevD.95.082002}.

Surface contamination from $^{210}$Pb could, in principle, have been the origin in the absolute worst case scenario, but it is highly unlikely given that the expectation for a CPD stored in a housing can only explain up to roughly 10--15\% of these excess signals. If one were to consider $^{210}$Pb surface and bulk contamination in the Cu housing, this may increase the contribution to as much 50\% of the excess signals, but still too low to fully explain our observations.

\subsubsection{EMI Signals}

Signals originating from electromagnetic interference (EMI) are common sources of excess backgrounds in rare event searches. For the CPD, these signals would couple into the TES electronically, rather than coupling into the substrate. Because EMI signals are extremely fast as compared to the time constants of a TES, they can be treated as a dirac Delta pulse. In this scenario, a TES would respond to an EMI pulse with an electronic rise time and the TES thermal fall time, similarly to what is seen in response to a square wave jitter down the bias line (e.g. Fig.~\ref{fig:didv_timedomain} in Chapter~\ref{chap:two}). For the CPD, the electronic rise time was measured to be $3 \, \mu \mathrm{s}$ and the primary TES fall time was measured to be $58 \, \mu \mathrm{s}$. However, as noted in Chapter~\ref{chap:perf}, the characteristic pulse shape at low energies was seen to have a $20 \, \mu \mathrm{s}$ rise time, which is attributed to the collection of athermal phonons. As the excess exponential background becomes dominant at low energies, then we would expect a large population of events $3 \, \mu \mathrm{s}$ rise time, of which none were observed in the CPD datasets, ruling out EMI signals as the source of the excess exponential background.

Of our proposed sources of these excess signals, we have yet to rule out stress-induced microfractures. In the following sections, we will further motivate this background through other tests and discuss methods to mitigate it.

\section{The Case for Stress-Induced Microfractures}

When the CPD is installed in its housing with the cirlex clamps, the force between the clamps and the wafer provide the thermal connection for heat to flow from the Si substrate to the bath. As the system is cooled from $300 \, \mathrm{K}$ to $8 \, \mathrm{mK}$, both the clamps and the Si will undergo thermal contraction each at a rate defined by their coefficient of thermal expansion (CTE). If the CTEs are different enough, then this can create a stress at the interface after cooling down due to the difference in how much the two materials have contracted. In this metastable state, stress can be released through energy fluctuations or athermal events (e.g. recoils), dropping the interface to lower energy state while releasing energy over time into the Si. These events are what can be thought of as microfractures, i.e. some micro-dislocation of the materials at their interface.

A common alternative holding scheme to clamps for TESs on $1 \, \mathrm{cm}^2$ chips is to glue them with some type of adhesive, such as GE-7031 varnish, which should be subject to the same concept of stress induced by CTE mismatch. While cirlex has been reported to have CTEs of $30\times 10^{-6}$ and $118 \times 10^{-6} \, \mathrm{K}^{-1}$ depending on orientation~\cite{Daal_2019}, GE-7031 varnish has been estimated based on other polymers to have a CTE of about $50 \times 10^{-6} \, \mathrm{K}^{-1}$~\cite{Morelock:ks5362}. When comparing to crystalline Si, the CTE is on the order of $1 \times 10^{-6} \, \mathrm{K}^{-1}$~\cite{properties-crystalline}, showing that a CTE mismatch is reasonable to expect in these detector holding schemes. Given this, we would expect to see an excess amount of events on a TES-based sensor whose substrate has been glued with GE-7031 varnish as opposed to an unglued (or resting) version, if stress-induced microfractures are a significant background. In the next section, we will do this comparison with simple TES rectangles.

\subsection{Glued TES Rectangles}

\begin{figure}
    \centering
    \includegraphics[width=0.8\linewidth]{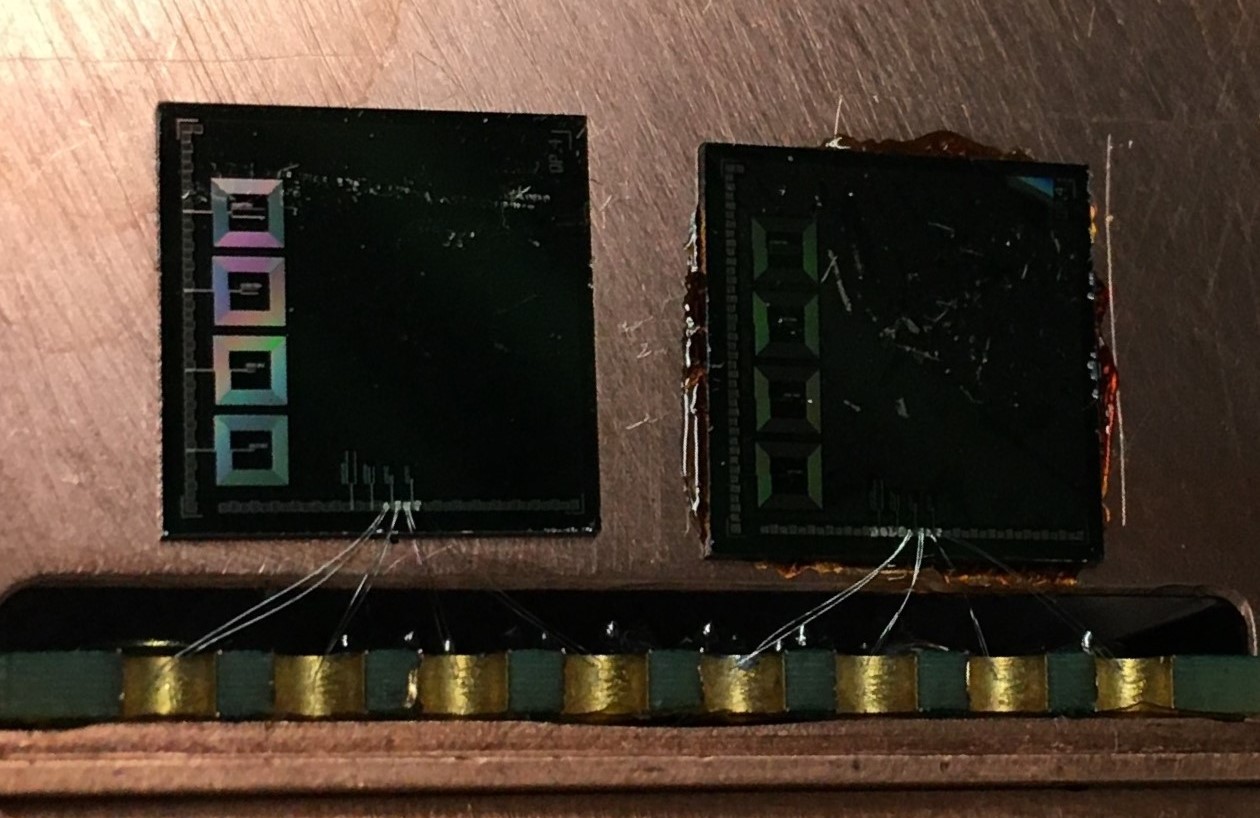}
    \caption{(Left) The QP.4 chip with the $25 \, \mu \mathrm{m} \times 100 \mu \mathrm{m} \times 40 \, \mathrm{nm}$ and $50 \, \mu \mathrm{m} \times 200 \mu \mathrm{m} \times 40 \, \mathrm{nm}$ rectangles being read out, which is simply resting on the Cu housing. (Right) The QP.4 chip with the $25 \, \mu \mathrm{m} \times 100 \mu \mathrm{m} \times 40 \, \mathrm{nm}$ and $50 \, \mu \mathrm{m} \times 200 \mu \mathrm{m} \times 40 \, \mathrm{nm}$ rectangles being read out, which is glued to the Cu housing with GE-7031 varnish. Photo by William Page.}
    \label{fig:qp4chips}
\end{figure}

\begin{figure}
    \centering
    \includegraphics[width=0.8\linewidth, angle=90]{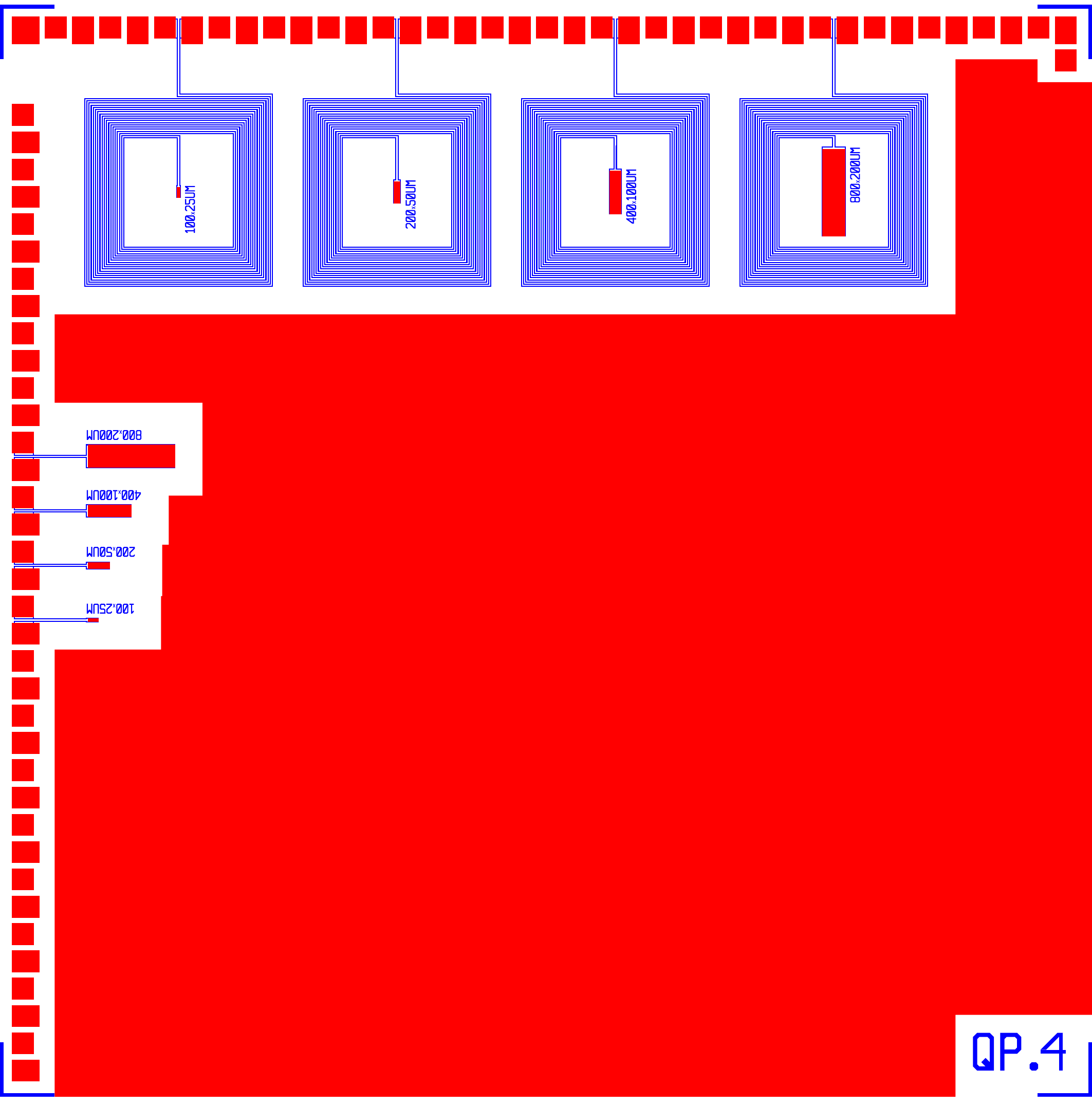}
    \caption{Full mask design of the QP.4 chip with TES rectangles of various areas, with and without antenna structures. Red is W, and blue is Al. Note that the surface of this chip is almost entirely covered with passive W (i.e. W that is not read out by our electronics). The other side of each of these $1\, \mathrm{cm}^2$ chips is bare Si (i.e. there is no W or Al added to the Si on that side).}
    \label{fig:qp4passive}
\end{figure}

\begin{figure}
    \centering
    \includegraphics{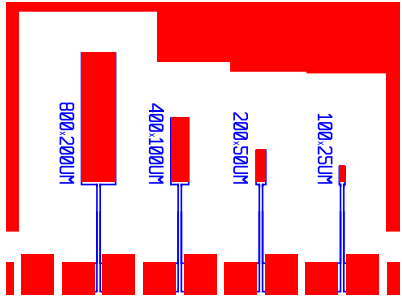}
    \caption{Zoomed in on the TES rectangles without antenna structures, these TESs scale by factors of 4 in area, while keeping the aspect ratio the same. Thus, it is expected that the normal resistances stay the same, while the in-transition bias powers will scale with area.}
    \label{fig:qp4rectangles}
\end{figure}

In our cryogen-free dilution refrigerator at UC Berkeley, we installed two $1\, \mathrm{cm}^2$ Si substrates, as pictured in Fig.~\ref{fig:qp4chips}, for which each had the mask design shown in Fig.~\ref{fig:qp4passive}. These $T_c \approx 55 \, \mathrm{mK}$ chips are from the QP.4 detector mask, designed by Suhas Ganjam in 2017. The original purpose of these simple rectangular TESs was to test power-scaling laws via the rectangles changing area with the same aspect ratios, where we have zoomed in to show more detail in Fig.~\ref{fig:qp4rectangles}. The same rectangles were also included on the chip with antenna-like structures for EMI tests (coupling with RF), but these TESs were not read out in this study.

For each chip in Fig.~~\ref{fig:qp4chips}, we are reading out the TES rectangles with dimensions $25 \, \mu \mathrm{m} \times 100 \mu \mathrm{m} \times 40 \, \mathrm{nm}$ and $50 \, \mu \mathrm{m} \times 200 \mu \mathrm{m} \times 40 \, \mathrm{nm}$, where the left chip is simply resting on the Cu housing and the right chip has been glued to the housing with GE-7031 varnish. In this configuration, resting chip thermalizes through the superconducting Al wire bonds that are used for reading out the current change in the TES, while the glued chip thermalizes mainly through the GE-7031 varnish connection to the Cu plate.

\begin{table}
    \centering
    \caption{Various parameters for the different TES rectangles on the QP.4 chips. Each dataset was taken with the cryocooler pulse-tube on, as phonons from vibrationally induced events are collected in the large amount of passive W, rather than the small TES rectangles.}
    \begin{tabular}{lrrrrrr}
    \hline \hline
    \rule{0pt}{10pt} Device            & $P_0$ $[\mathrm{fW}]$ & $\tau_+$ $[\mu \mathrm{s}]$ & $\tau_-$ $[\mu \mathrm{s}]$ & $\sigma_A$ $[\mathrm{nA}]$ & $\frac{\partial I}{ \partial P}(0)$ $[\frac{\mathrm{A}}{\mu\mathrm{W}}]$ & $\sigma_E$ $[\mathrm{meV}]$ \\ \hline
    TES25x100 Glued   &   17.9         &   1.4        &  94.6         &     1.07            &  21.6        &   31.0               \\
    TES50x200 Glued   &   73.2         &     1.6      &    38.9       &      1.91           &   10.1       &    52.4              \\
    TES25x100 Resting &   18.0         &     1.6      &     60.3      &     1.34            &   22.1       &    24.8              \\
    TES50x200 Resting &    82.6        &    1.6       &     43.5      &      1.50           &   9.7       &  49.6  \\ \hline \hline  
    \end{tabular}
    \label{tab:qp4chars}
\end{table}

With this setup, we can follow the TES characterization steps as outlined in Chapter~\ref{chap:two}, where we analyze $IV$ and $\partial I /\partial V$ sweep data for understanding our TESs. In this case, we have reproduced the pertinent parameters for studying stress-induced microfractures in Table~\ref{tab:qp4chars}, where $P_0$ is the TES bias power, $\tau_+$ is the electronic TES time constant, $\tau_-$ is the thermal TES time constant, $\sigma_A$ is the OF amplitude baseline resolution (RMS), $\frac{\partial I}{ \partial P}(0)$ is the zero-frequency value of the power-to-current transfer function, and $\sigma_E$ is the OF energy baseline energy resolution (RMS). In order to calculate $\sigma_E$, we convert OF amplitudes from current to energy for an event via
\begin{equation}
    E_{OF} = A_{OF} \frac{\int \mathop{dt} s(t)}{\frac{\partial I}{\partial P} (0)}, 
    \label{eq:ofenergy_conv}
\end{equation}
where we further define $E_{OF}$ as the OF energy of this event, $A_{OF}$ as the OF amplitude (in current) of this event, and $s(t)$ as the amplitude-normalized pulse template. For $s(t)$, we use a template with a rise time of $\tau_+$ and a fall time of $\tau_-$ for each device, as this is what would be expected for an event that is directly incident on the TES (as opposed to phonon collection via superconducting Al fins in QETs). The energy estimate provided by $E_{OF}$ only provides a rough energy estimate for single events on a TES (as it neglects phenomena such as collection efficiency or pulse shape saturation). Because substrate events would have widely varying phonon collection times due to the large amount of passive W in this QP.4 chip (see Fig.~\ref{fig:qp4passive}), any events that are coincident on multiple TESs on the same chip would not have an accurate energy estimate.

\begin{figure}
    \begin{subfigure}{.5\textwidth}
        \centering
        \includegraphics[width=1\linewidth]{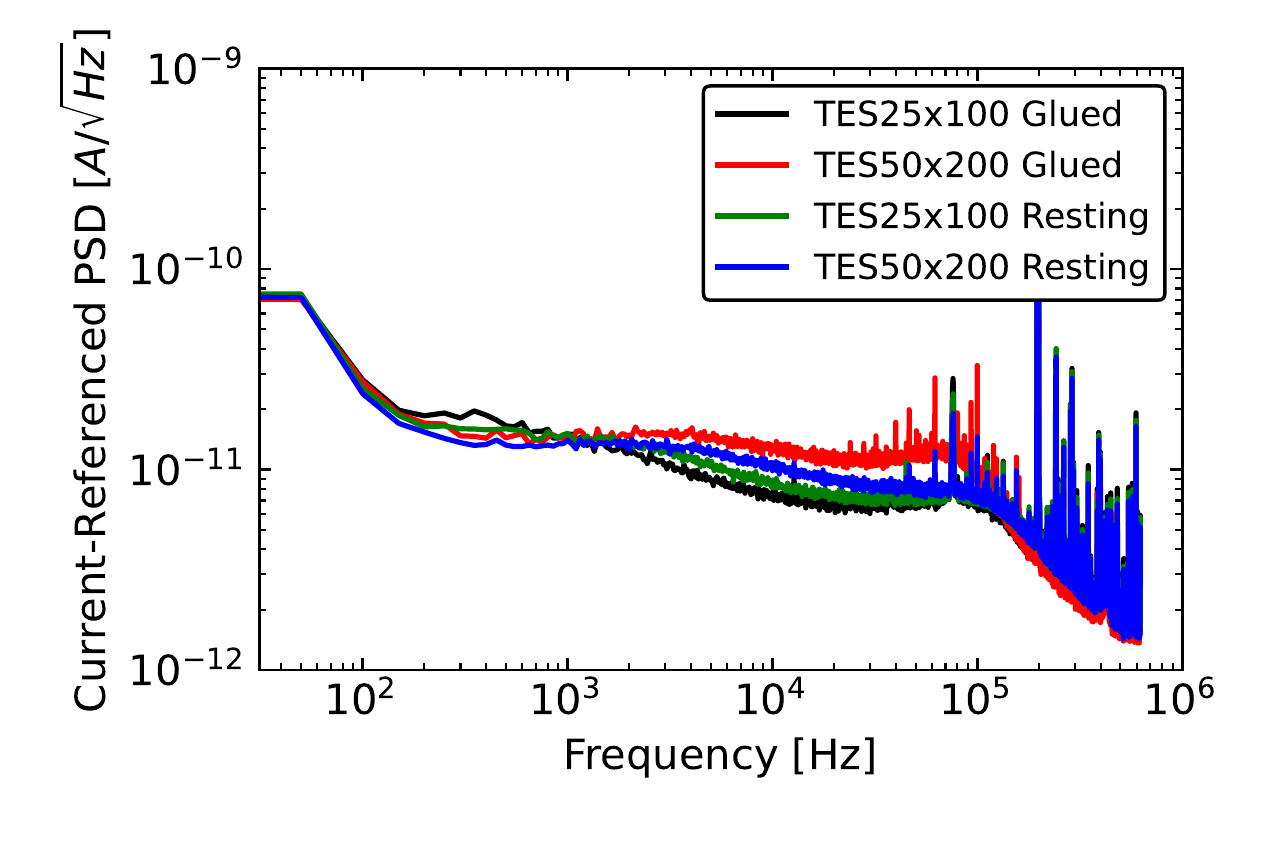}
    \end{subfigure}%
    \begin{subfigure}{.5\textwidth}
        \centering
        \includegraphics[width=1\linewidth]{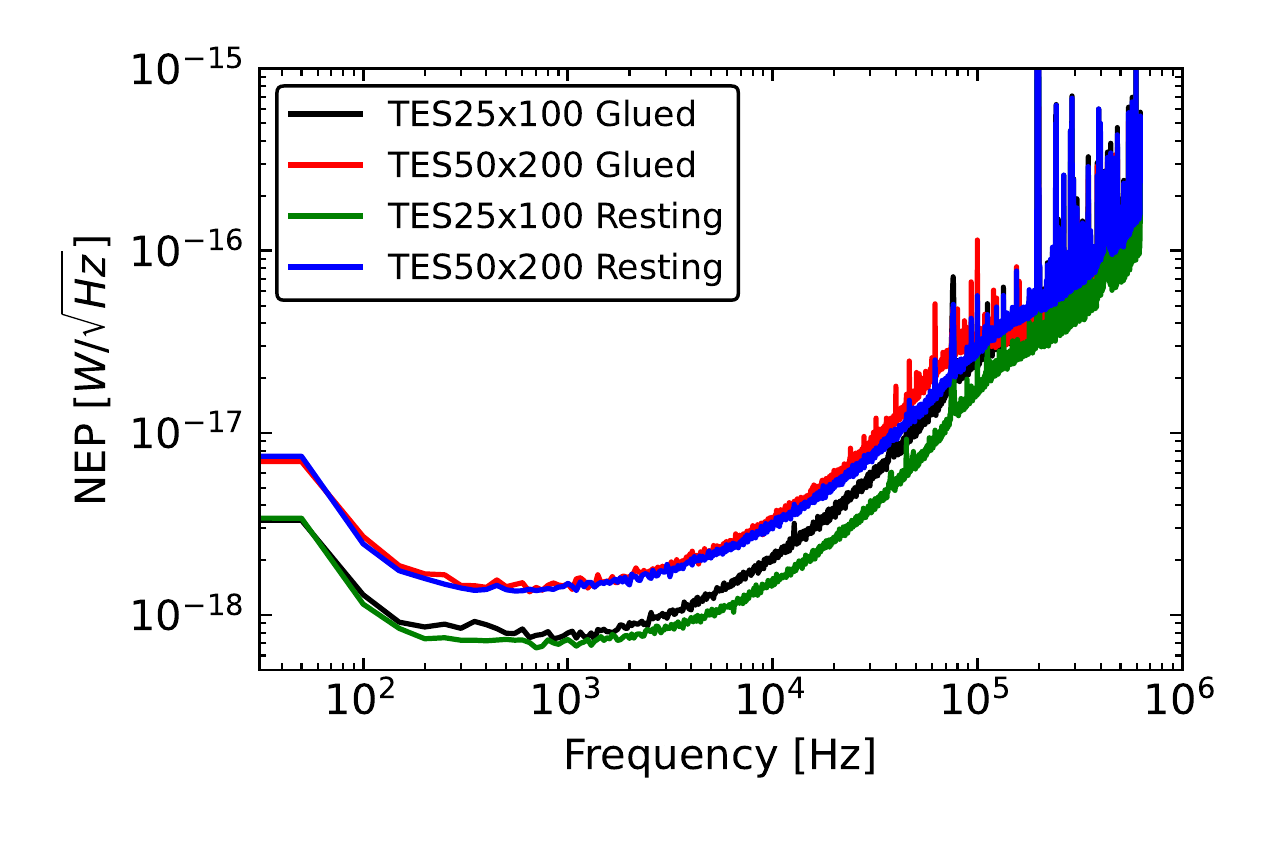}
    \end{subfigure}
    \caption{(Left) Current-referenced PSDs for each of the TESs on the two QP.4 chips. (Right) NEPs for each of the TESs on the two QP.4 chips. The smaller two TESs on each chip have slightly different high frequency responses in power space, which could be due to slightly different properties of the TESs originating from deviations in the fabrication process. The low frequency (flat) NEP is about the same for TESs of the same dimensions.}
    \label{fig:qp4noise}
\end{figure}

To study these four TES rectangles, we took continuous data for four hours. We used the triggering algorithm in \textsc{pytesdaq}~\cite{pytesdaq} for extraction of OF amplitudes from the continuous data and saving events with amplitudes that were greater than a threshold of $6 \sigma_A$. In \textsc{pytesdaq}, there is support for flagging when events occur at the same time (coincident events) or if an event was only on one TES (singles), which we used when running the event triggering algorithm. For extracting OF amplitudes, we have our previously defined pulse template $s(t)$, and we use the measured current-referenced PSDs as shown in Fig.~\ref{fig:qp4noise}. Because we have fully characterized our TESs, we can convert these current-referenced PSDs to NEPs, also shown in Fig.~\ref{fig:qp4noise}, where we see that the NEPs scale with a factor of two between the two TES geometries, which is expected for TESs dominated by phonon noise.

Because the triggering algorithm flags events as coincidences or singles, we can select data that are coincidences or singles, respectively defined as when two TESs on the same chip both have reconstructed OF amplitudes greater than $6\sigma_A$, or only a single TES does. Furthermore, we selected the region of data without large baseline fluctuations and added a loose ($5\sigma$) chi-square goodness-of-fit cut to all four TESs individually, each of which had a passage fraction of $\sim\!90\%$. With our energies estimated and our data quality cuts applied, we can measure the event spectra seen on each of the TESs. We split this up into comparing the spectra of the coincidences and the spectra of the singles.

\begin{figure}
    \begin{subfigure}{.5\textwidth}
        \centering
        \includegraphics[width=1\linewidth]{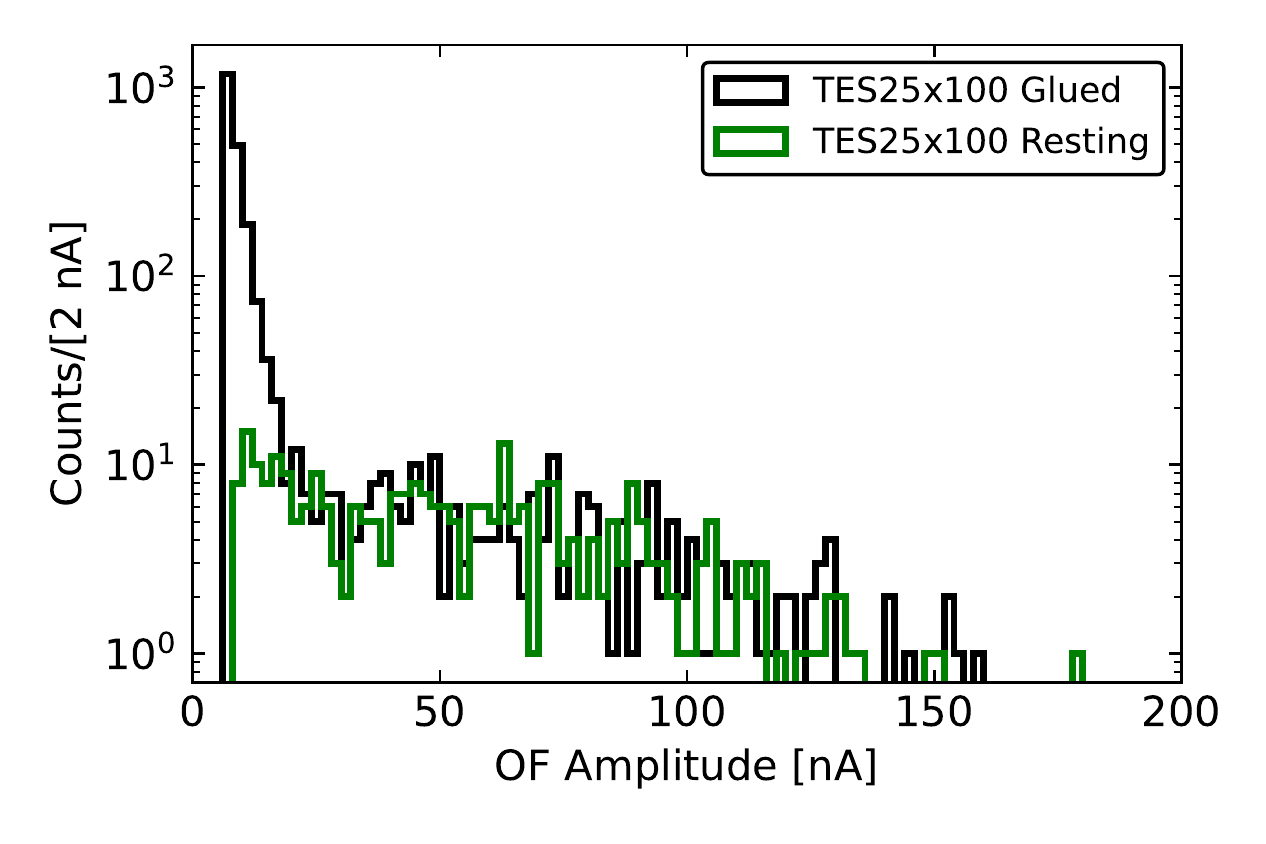}
    \end{subfigure}%
    \begin{subfigure}{.5\textwidth}
        \centering
        \includegraphics[width=1\linewidth]{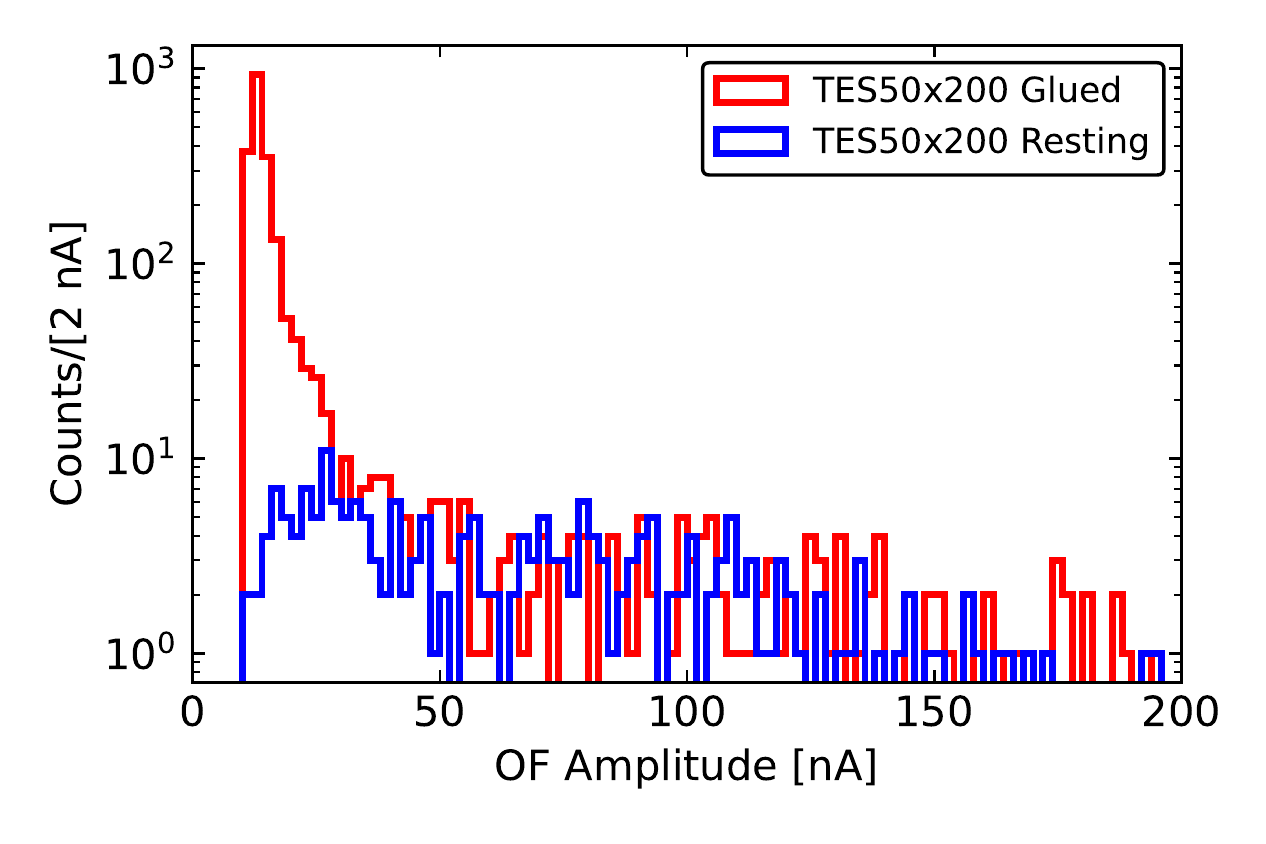}
    \end{subfigure}
    \caption{(Left) Comparison of the spectra of the TES25x100 rectangle on the glued and resting chips, where each of the events were marked as coincident with the TES50x200 rectangle on the same chip. (Right) Comparison of the spectra of the TES50x200 rectangle on the glued and resting chips, where each of the events were marked as coincident with the TES25x100 rectangle on the same chip. In each plot, we keep the OF amplitudes in units of current because the energies of these shared events are not well defined, as a significant (and variable) fraction of energy is collected by the passive W on the chip.}
    \label{fig:qp4coincs}
\end{figure}

We first select events that are coincident between TES25x100 and TES50x200 on each of the chips. To understand if there is a change in the coincident event spectrum with and without the chip being glued, we will compare the TES25x100 glued spectrum to the TES25x100 resting spectrum and the TES50x200 glued spectrum to the TES50x200 resting spectrum, as shown in Fig.~\ref{fig:qp4coincs}. Note that these spectra are shown in units of current due the large amount of passive W in the mask design, as previously discussed. The difference between the glued and resting spectra is striking: there is a large increase in the near-threshold event rate for the TESs on the glued chip, which is completely absent in the spectra of TESs on the resting chip. Because the only significant difference between the two chips are that one is glued and one is not, this provides strong evidence that the presence of the glue has increased the low-energy event rate on the TESs. Furthermore, the higher amplitude event rates are the same, meaning that the glue did not have an effect at these higher energies.

\begin{figure}
    \begin{subfigure}{.5\textwidth}
        \centering
        \includegraphics[width=1\linewidth]{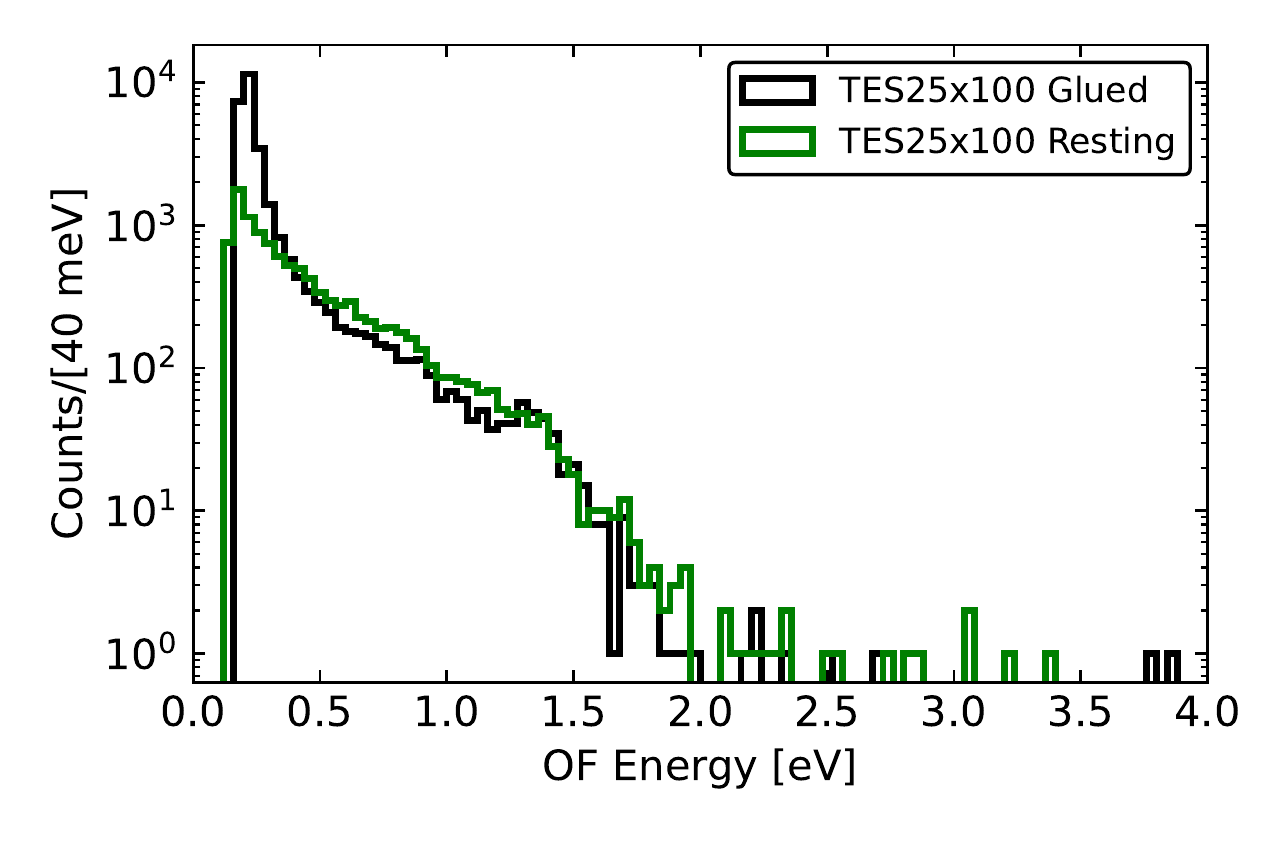}
    \end{subfigure}%
    \begin{subfigure}{.5\textwidth}
        \centering
        \includegraphics[width=1\linewidth]{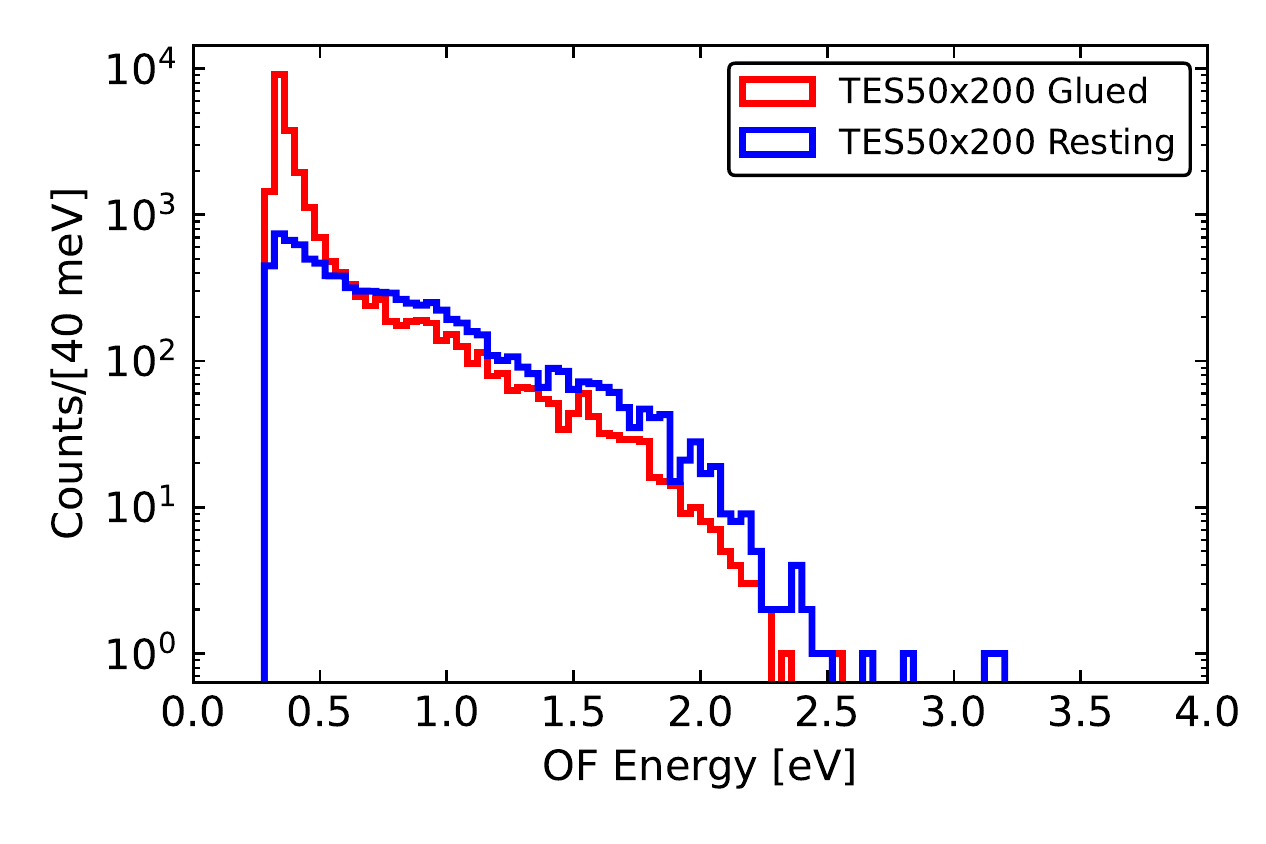}
    \end{subfigure}
    \caption{(Left) Comparison of the spectra of the TES25x100 rectangle on the glued and resting chips, where each of the events were marked as occurring independently of the TES25x100 rectangle on the same chip (i.e. was a single). (Right) Comparison of the spectra of the TES50x200 rectangle on the glued and resting chips, where each of the events were marked as occurring independently of the TES50x200 rectangle on the same chip (i.e. was a single). As opposed to Fig.~\ref{fig:qp4coincs}, events collected into a single TES are better defined in energy, and we can roughly estimate the collected energy using Eq.~(\ref{eq:ofenergy_conv}), remembering that this is only a rough estimate and that the main takeaway from this figure is that the single rates with and without glue are about the same except very near the energy threshold.}
    \label{fig:qp4singles}
\end{figure}

We next can look at the singles event rate on each of the TESs and make the same comparison. That is, we would like to know if there a change in the singles spectra with or without the presence of glue. In Fig.~\ref{fig:qp4singles}, we compare these spectra and again see that there is an excess of events for TESs on the glued chip near threshold, which is not seen for the TESs on the resting chip. Though this is seen in the singles spectrum, it is natural to explain this excess as substrate events that were not appreciably detected by the other TES on the same chip. Because of the large amount of passive W, it is not unlikely that the phonons produced by some substrate event would only be seen by one of the TESs depending on the location of the event (i.e. the energy was collected by the passive W or another TES not being read out).

The significant difference in the low-energy coincident spectra between TESs on glued and resting substrates is consistent with the idea of stress-induced microfractures. The GE-7031 varnish that provides the thermal link for the glued chip can have different thermal contraction properties than the Si substrate, such that the interfaces contracting at different scales can cause deviations of nuclei at the surface of the Si, releasing phonons into the substrate. In this study of the TES rectangles, we have been limited in understanding the energy scale of these possible microfracture events due to the passive W. In order to estimate the energy scale, we would need a device with very low amounts of passive structures, such that we have a high and position-independent phonon collection efficiency, as well as an optical calibration source. This study will be done in future work in the Pyle lab.

\section{Mitigation of Stress-Induced Events}

With stress-induced events motivated as an origin to these excess events, we must consider how to change our experimental setup for the CPD such that these types of events are mitigated. To do this, we will propose a stress-free holding scheme, thermalization through a gold bonding pad, and reduction of vibrational sensitivity via a passive vibration isolation system.

\subsection{Stress-Free Holding Schemes}

\begin{figure}
    \centering
    \includegraphics[width=0.6\linewidth]{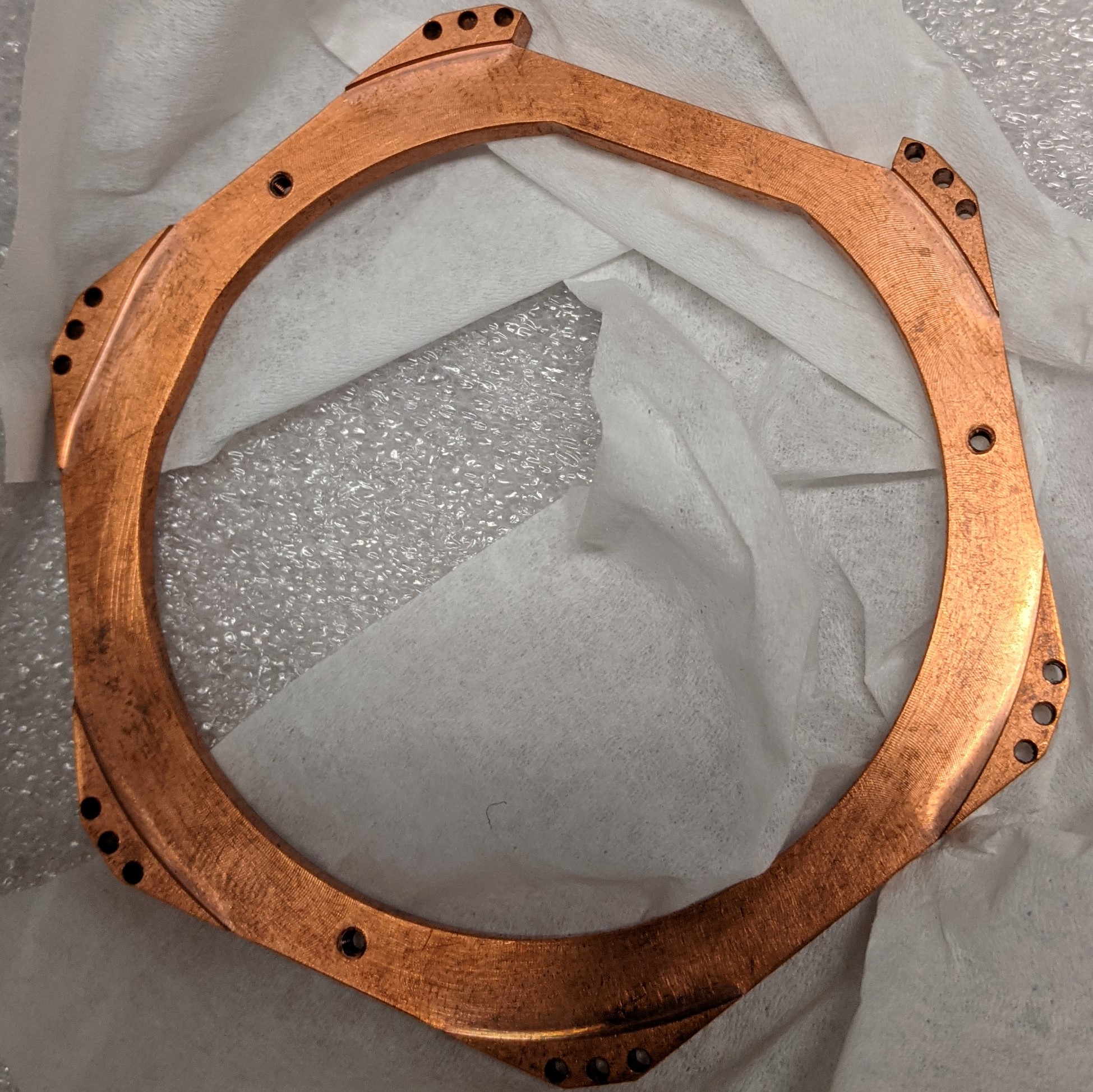}
    \caption{Stress free holder for the CPD for removal of stress-induced events, designed by Samuel Watkins. This holder will be used in a future run, and has not been run yet as of June 2022.}
    \label{fig:stressfree}
\end{figure}

To hold a CPD with minimal stress on the substrate, we know that we need to abandon the cirlex clamps, and we also should not use some type of glue given our the studies of the Si chips from the previous section. The natural solution is to simply have the CPD resting directly on Cu, such as with the resting Si chip as well as other cryogenic light detectors~\cite{BARUCCI2019150}. However, we note that the surface area of contact will be much higher with the 3-inch diameter CPD than for a $1\, \mathrm{cm}^2$ chip. Because of this large surface area, the resting CPD will have many points of contact for frictional rubbing. To reduce these points of contact, we can adopt the design pictured in Fig.~\ref{fig:stressfree}.

In this design, the points of contact between the Cu holder and a CPD would be only near the perimeter of the wafer. We also note that this region of contact is inset from the connection of the holder to the housing, in order to restrict movement of the CPD when installing the detector. This design also includes three symmetrically placed tapped holes, which allow for the installation of screws to raise the CPD from this region of contact. In this case, the three screws would be the only point of contact, which may further reduce possible frictional rubbing events. Note that, because a standard 3-inch wafer generally has primary and secondary flats for alignment, we have also included these in the cutout geometry of the holder. In Table~\ref{tab:semi}, we have placed the relevant SEMI dimensions for a standard 3-inch wafer, which are from the SEMI M1-0302 specifications.

\begin{table}
    \centering
    \caption{The SEMI specifications for a standard 3-inch wafer.}
    \begin{tabular}{lc}
    \hline \hline
    Dimension & Length [$\mu \mathrm{m}$] \\ \hline
    Diameter & $76200\pm630$ \\
    Primary Flat Length & $22220\pm3170$ \\ 
    Secondary Flat Length & $11180\pm1520$ \\ \hline \hline
    \end{tabular}
    \label{tab:semi}
\end{table}

\subsection{Thermalization through Gold Pads}

With this holder, however, we would not be able to thermalize a CPD without a thermal connection beyond that given by the superconducting Al wire bonds. In order to provide an excellent thermal connection, we can add a gold wire-bonding pad to a CPD, from which we can use gold wire to connect to the stress-free holder. In practice, the Cu holder should be gold-plated, as our group has found it difficult to achieve a good bonding connection with gold wire onto bare Cu. To ensure that the proposed gold wire thermalization scheme can handle the $3.86 \, \mathrm{pW}$ bias power, we must ensure that both the wire itself has a good thermal conductance, as well as the gold pad.

\begin{table}
    \centering
    \caption{The thermal conductive of thin gold wire for different levels of impurity, values from Ref.~\cite{NOVOTNY1977451}.}
    \begin{tabular}{cc}
    \hline \hline
    \rule{0pt}{10pt} Purity [$\%$] & $\lambda(T)$ $\left[ \frac{\mathrm{W}}{\mathrm{K}\, \mathrm{cm}}\right]$ \\ \hline
    99.7 & $0.115 T + 7 \times 10^{-4} T^2$ \\
    99.9 & $0.325 T$ \\ 
    99.999 & $4.35T $ \\ \hline \hline
    \end{tabular}
    \label{tab:goldwire}
\end{table}

\begin{figure}
    \centering
    \includegraphics{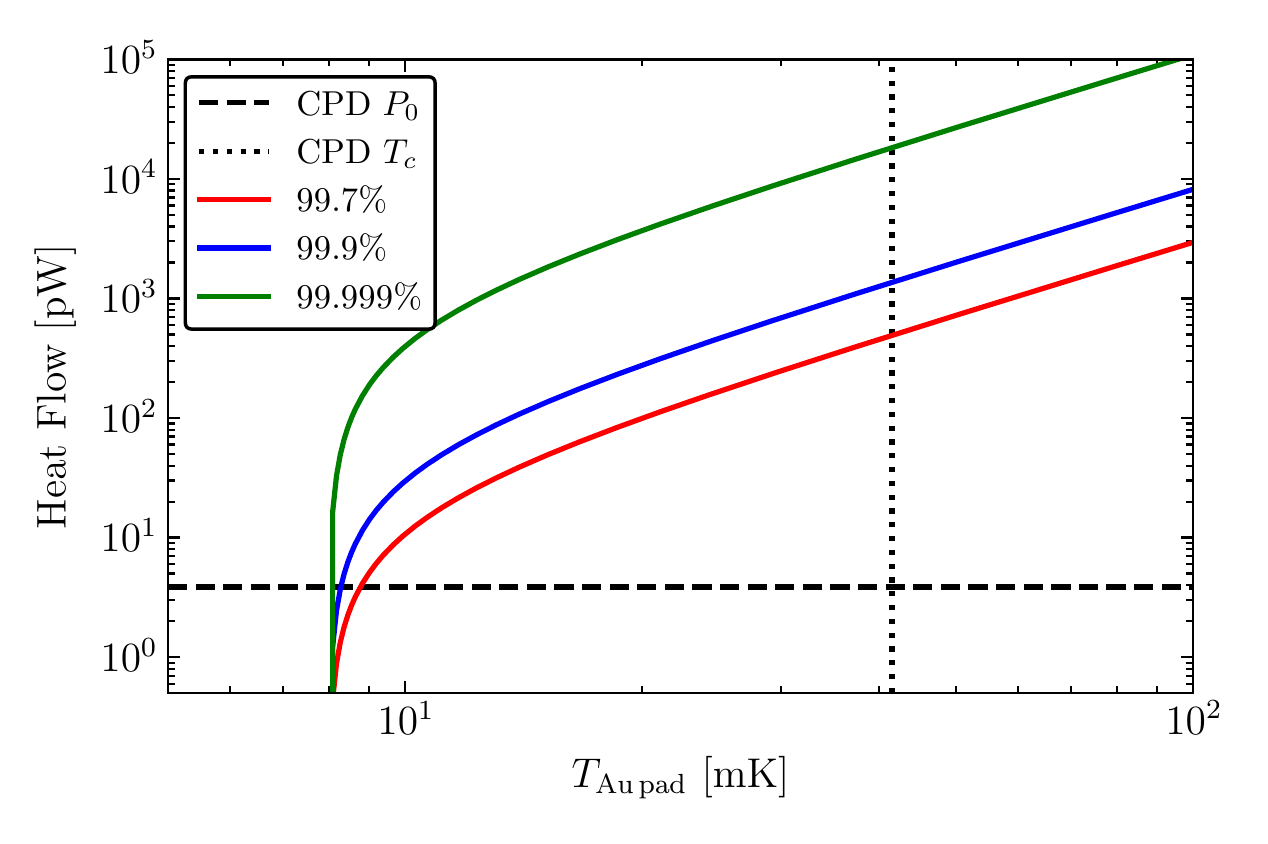}
    \caption{The heat flow through a single gold wire of diameter of $25.4 \, \mu \mathrm{m}$ and a length of $1 \, \mathrm{cm}$ as a function $T_\mathrm{Au \ pad}$, comparing to the bias power of the CPD ($P_0 = 3.86 \, \mathrm{pW}$) and its $T_c = 41.5 \, \mathrm{mK}$.}
    \label{fig:auwire}
\end{figure}

For the conductance through the gold wire, the process that limits the conductivity is scattering of electrons by impurities. In Table~\ref{tab:goldwire}, we give the thermal conductivity values that have been reported in Ref.~\cite{NOVOTNY1977451}, which we can use to check how well $3.86 \, \mathrm{pW}$ of power flows through a single gold wire bond of varying purity. We can relate the thermal conductivity $\lambda(T)$ of a wire to the power flow through it via
\begin{equation}
    P_\mathrm{wire} =\frac{A_\mathrm{wire}}{\ell_\mathrm{wire}} \int_{T_b}^{T_\mathrm{Au \, pad}} \mathop{dT} \lambda(T),
\end{equation}
where $A_\mathrm{wire}$ is the cross sectional area of the wire, $\ell_\mathrm{wire}$ is the length of the wire, $T_\mathrm{Au \, pad}$ is the temperature of the Au pad, and $T_b$ is the bath temperature. For our purposes, we will assume that we have bonded a single gold wire with a diameter of $25.4 \, \mu \mathrm{m}$ and a length of $1 \, \mathrm{cm}$, and that the bath temperature is $T_b = 8 \, \mathrm{mK}$. With these values, we can plot $P_\mathrm{wire}$ as a function of $T_\mathrm{Au \ pad}$ and compare to the bias power of the CPD, as done in Fig.~\ref{fig:auwire}.

For the lowest purity gold wire, the power load of the CPD does not increase the temperature of the Au pad to more than $9 \, \mathrm{mK}$, which is far from the $T_c$ of the device. Because commercial-grade gold wire can be found with purities of 99.99\% (or 4N in the industry nomenclature), this means that one gold wire should easily be able to hand the CPD heat load. In practice, it would be advisable to add more wires in the event of failure, which would also serve only to improve the thermal conductance.

With the gold wire known to be able to handle the CPD bias power, the next step is to ensure that the thermal conductance between the Si absorber and the gold pad is large enough, such that the Si absorber of the CPD does not equilibriate at a high temperature (i.e. significantly close to $T_c$). As measured in Appendix~\ref{chap:appa}, the heat flow between a Si absorber and a gold pad is
\begin{equation}
    P_\mathrm{pad} = \Sigma_{ep,Au} V_\mathrm{pad} (T_\mathrm{abs}^n - T_\mathrm{Au \, pad}^n),
\end{equation}
where $\Sigma_{ep,Au}$ is the gold electron-phonon coupling constant, $V_\mathrm{pad}$ is the gold bonding pad volume, $T_\mathrm{abs}$ is the temperature of the Si absorber, and $n=5$ is the power-law exponent. In Appendix~\ref{chap:appa}, we noted that our fabrication facilities at Texas A\&M University are currently creating gold pads with $750 \, \mathrm{nm}$ thickness. Thus, we will assume this thickness for a prospective gold pad on the CPD and calculate what area of pad will be needed for its $3.86 \, \mathrm{pW}$ bias power.

\begin{figure}
    \centering
    \includegraphics{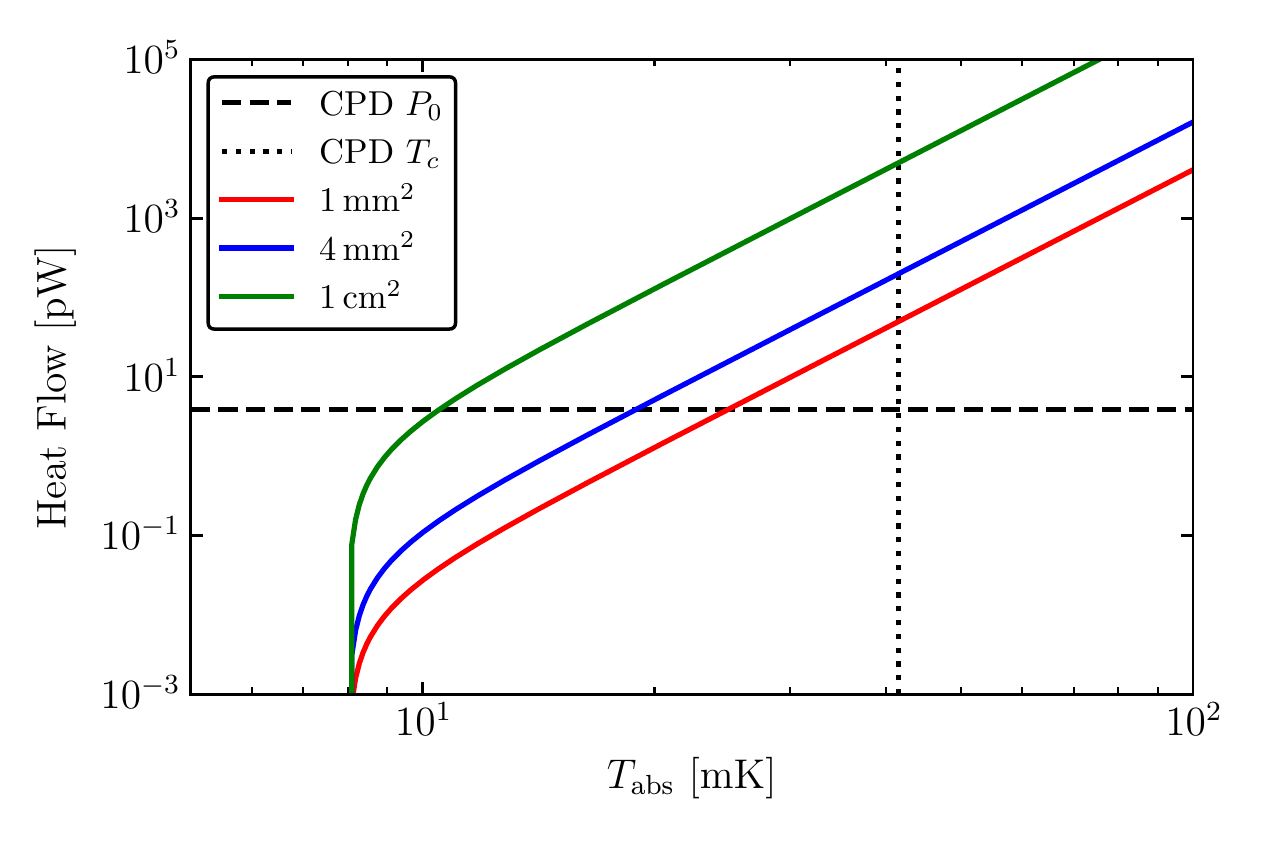}
    \caption{The heat flow from absorber to a gold pad of thickness $750 \, \mathrm{nm}$ and varying area as a function $T_\mathrm{abs}$, comparing to the bias power of the CPD ($P_0 = 3.86 \, \mathrm{pW}$) and its $T_c = 41.5 \, \mathrm{mK}$.}
    \label{fig:aupad}
\end{figure}

In Fig.~\ref{fig:aupad}, we show the heat flow from the absorber to the gold pad as a function of absorber temperature for different size gold pads, where we assume that the gold wire itself is pure enough such that the pad has a temperature of $T_\mathrm{Au \, pad}=8 \, \mathrm{mK}$. At this interface, we clearly must be careful about the size of the pad, as a volume of $V_\mathrm{pad} = 1\,\mathrm{mm}^2 \times 750 \, \mathrm{nm}$ would allow the absorber to cool down to about $T_\mathrm{abs} = 25 \, \mathrm{mK}$. The thermal flow between the TES and the absorber similarly goes as
\begin{equation}
    P_\mathrm{TA} = \Sigma_{ep,W} V_\mathrm{TES} (T_c^n - T_\mathrm{abs}^n),
\end{equation}
where $\Sigma_{ep,W}$ is the W electron-phonon coupling constant, $V_\mathrm{TES}$ is the total TES volume, and $T_c$ is the TES superconducting critical temperature. Thus, the fractional change in bias power for a device with an absorber at a temperature at $T_\mathrm{abs}$ rather than effectively at the bath temperature $T_b$ is given by
\begin{equation}
    \varepsilon_\mathrm{abs} = \frac{T_c^n - T_\mathrm{abs}^n}{T_c^n - T_b^n}.
    \label{eq:fractional_power}
\end{equation}
For $T_\mathrm{abs} = 25 \, \mathrm{mK}$ and $T_b = 8 \, \mathrm{mK}$, this would give a fractional change in bias power of $\varepsilon_\mathrm{abs} = 0.92$, meaning that the expected measured bias power of the CPD would by lowered by 8\% to $3.55 \, \mathrm{pW}$ due to the introduction of this parasitic power, significantly degrading the detector's performance. For the $4 \, \mathrm{mm}^2$ and $1\, \mathrm{cm}^2$ gold pad areas, we can follow the same calculation, and we have summarized the results in Table~\ref{tab:aupad}.

\begin{table}
    \centering
    \caption{The expected Si absorber temperature and fractional change in bias power $\varepsilon_\mathrm{abs}$ of the CPD for different pad areas, where $\varepsilon_\mathrm{abs}$ is defined in Eq.~(\ref{eq:fractional_power}). Each of these values assume a bath temperature of $T_b = 8 \, \mathrm{mK}$.}
    \begin{tabular}{crr}
    \hline \hline
    Pad Area &  $T_\mathrm{abs}$ $[\mathrm{mK}]$ & $\varepsilon_\mathrm{abs}$ \\ \hline
     $1 \, \mathrm{mm}^2$  & 25 & 0.921 \\
     $4 \, \mathrm{mm}^2$  & 18 & 0.985 \\ 
     $1 \, \mathrm{cm}^2$  & 10 & 0.999 \\ \hline \hline
    \end{tabular}
    \label{tab:aupad}
\end{table}

The effect on the CPD becomes smaller as we increase the gold pad size. However, we must remember that we do not want the pad to be so large that its passive area begins to appreciably collect the signal. From Table~\ref{tab:specs} when we characterized the CPD, we have that it has an active surface area of 1.9\% and passive surface area of 0.2\%, where the total area of the CPD wafer is about $93.6 \, \mathrm{cm}^2$. Thus, the $1 \, \mathrm{mm}^2$, $4 \, \mathrm{mm}^2$, and $1\, \mathrm{cm}^2$ gold pads would increase the passive surface area to 0.21\%, 0.24\%, and 1.27\%, respectively. Although the $1\, \mathrm{cm}^2$ pad would have excellent heat flow, the passive area would be on the order of the active area, and this pad would likely act as a significant area for collection of phonons that we would not read out, thus degrading the phonon collection efficiency and the detector resolution. Instead, a pad on the order of $4\, \mathrm{mm}^2$ would likely have a small effect on the phonon signal, while ensuring that the Si absorber can cool down to a low enough temperature to have a negligible effect on the bias power.

With care taken to ensure that the thermal connections of the gold pad are designed well, we should be able to thermalize a stress-free CPD, where we have added a gold pad of area $4\, \mathrm{mm}^2$. The next-generation CPD will have a mask design such that the equilibrium bias power will be lower due the optimization of the QETs for lower surface coverage, which should allow thermalization with a smaller gold pad than would be needed for this CPD with a  $3.86 \, \mathrm{pW}$ bias power.

\subsection{Vibrational Sensitivity}

\begin{figure}
    \begin{subfigure}{.5\textwidth}
        \centering
        \includegraphics[width=1\linewidth]{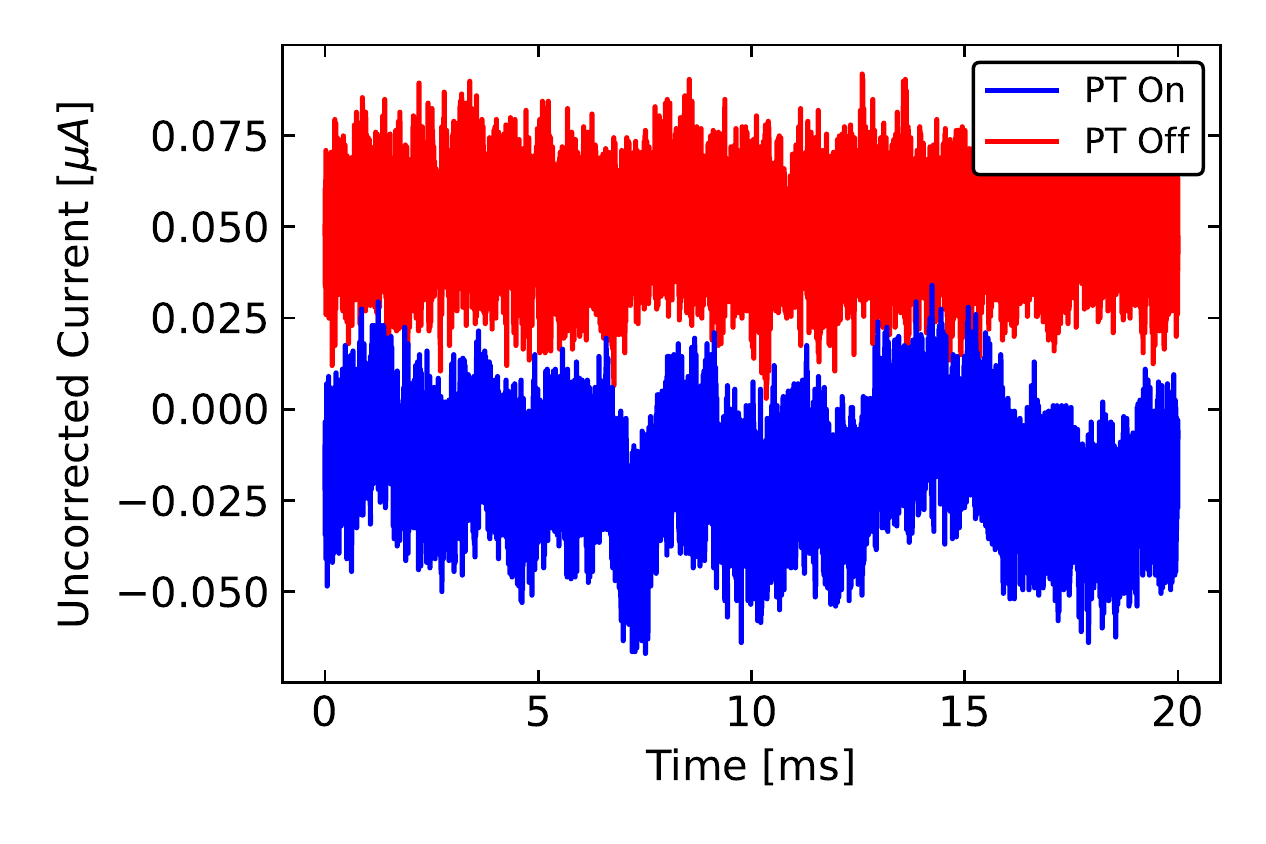}
    \end{subfigure}%
    \begin{subfigure}{.5\textwidth}
        \centering
        \includegraphics[width=1\linewidth]{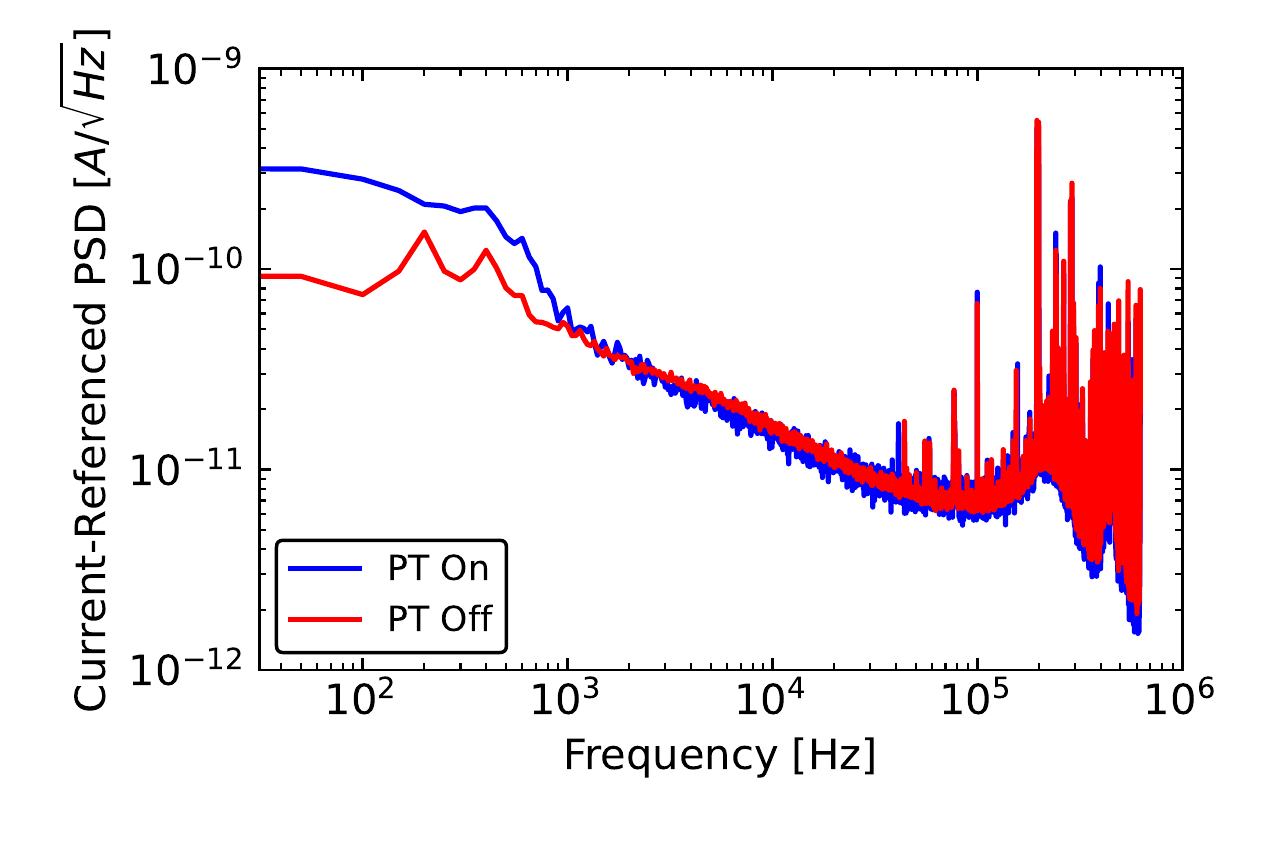}
    \end{subfigure}
    \caption{(Left) Comparison of characteristic time-domain traces for the 0.25\% coverage SPICE Melange detector with PT on and off. (Right) Comparison of current-referenced PSD for the 0.25\% coverage SPICE Melange detector with PT on and off.}
    \label{fig:melangeptonoff}
\end{figure}

Although we have proposed a scheme for holding a CPD without creating stress-induced events and its thermalization, we must also take into account that these sensitive devices will be run in a cryogen-free dilution refrigerator. In these systems, a common source of environmental noise can be attributed to the pulse-tube (PT) cryocooler, which is critical for achieving the sub-$10 \, \mathrm{mK}$ base temperatures. The oscillation of the PT at $1.4 \, \mathrm{Hz}$ will create a source of vibrational noise within the dilution refrigerator, which can then be seen by sensitive detectors. In the case of resting detectors, the effects of these vibrations can be stark, as shown for one of our devices in Fig.~\ref{fig:melangeptonoff}. The device shown is the 0.25\% coverage SPICE Melange detector, designed by C. W. Fink~\cite{finkthesis}. Here, we use this detector to show that the effect of PT vibrations is significant for a resting device, in that it is visibly apparent in time domain, and is seen as excess low frequency noise in the PSD. Thus, we would expect these events to degrade our baseline energy resolution as well as create excess vibrationally-induced events.

The decoupling of sensitive detectors or instruments from environmental vibrations has been pursued by many cryogenic experiments, usually using a three-dimensional elastic pendulum~\cite{LYNCH2002345} for passive vibration isolation~\cite{Maisonobe_2018, doi:10.1063/1.5088364, doi:10.1063/1.1149839, Caparrelli2006, PIRRO2000331, PIRRO2006672, doi:10.1063/1.4794767}. These systems have all had success in reducing vibrationally-induced backgrounds, and such a system will be necessary to run a detector in a stress-free holding scheme. In Appendix~\ref{chap:vibrations}, we propose such a system for the Pyle group dilution refrigerator, which has been designed to achieve excellent attenuation of vibrations with frequencies above about $10 \, \mathrm{Hz}$.

\section{Summary}

In this chapter, we have motivated the hypothesis of stress-induced events as a likely source for the excess signals observed in the DM search dataset through ruling out many other backgrounds and a comparison of TESs on a glued and resting substrate. With the stress-free strategies put forward in the preceding sections, it should be possible to make such a background negligible. In this scenario, we would expect a CPD with improved baseline energy resolution to significantly improve upon the low-mass reach of the SuperCDMS-CPD DM search.

\chapter{Conclusion and Outlook}

In this thesis, we started with basic TES concepts, applied them to a QET-based detector, carried out an LDM search that achieved world-leading limits for a cryogenic device, investigated stress-induced events as a source of excess signals in that search, and proposed their mitigation  through stress-free holding schemes. In the near future, our lab at UC Berkeley will have an installed passive vibration isolation system, and then stress-free CPDs can be run to improve the reach of the SuperCDMS-CPD DM search. A next generation CPD energy sensitivity would be improved upon through a lowering of $T_c$ and a lowering of the surface coverage of the sensors, and such a device has been designed by S. Zuber and C. W. Fink with an expectation of $1\, \mathrm{eV}$ baseline energy resolution. This CPD would improve upon the results in this thesis, especially if ran in a low-background cryostat. However, orders of magnitude improve to lower DM masses through the nuclear recoil channel will require orders of magnitude of improvement in baseline energy resolution.

\begin{figure}
    \centering
    \includegraphics{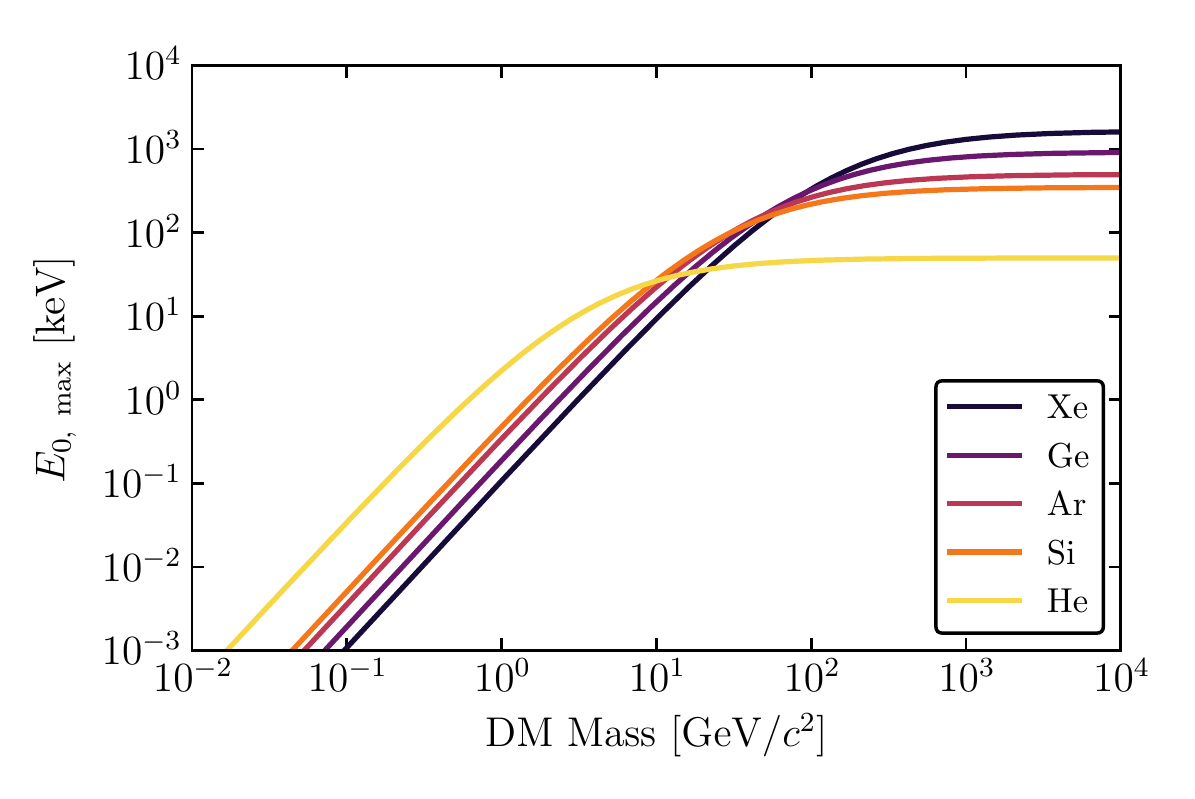}
    \caption{Maximum recoil energy of the differential rate spectra for various target materials. As DM mass drops below the mass of the target nucleus, the imparted energy to the target becomes vanishingly small.}
    \label{fig:maxrecoil}
\end{figure}

Ultimately, these detectors will be limited by kinematic mismatching of sub-$\mathrm{GeV}/c^2$ LDM with masses less than the detector nucleus when searching for LDM through elastic nuclear recoils. This concept is demonstrated in Fig.~\ref{fig:maxrecoil}, which shows the maximum recoil energies of the expected spectrum of DM-nucleon interactions for various detector materials. Significantly improving the current LDM mass reach (from $100\,\mathrm{MeV}/c^2$ to $1\,\mathrm{MeV}/c^2$) through this interaction channel with, e.g., Si-based detectors will require athermal sensors with meV-scale energy thresholds. Given that state-of-the-art TESs have been shown to be capable of energy thresholds on the order of hundreds of meV~\cite{fink2020characterizing}, TES and QET detector R\&D will need to produce detectors with improvements of energy sensitivities by a few orders of magnitude to vastly improve the DM search results through nuclear recoils.

While R\&D continues on lowering energy thresholds, DM detection schemes via inelastic recoils have been proposed to circumvent the limitations brought forward by kinematic matching. For DM-nucleon interactions, limits can be set using DM signal models based on Bremsstrahlung~\cite{PhysRevLett.118.031803} or the Migdal effect~\cite{Ibe:2017yqa, PhysRevLett.127.081805}. Each of these processes are based on secondary radiation that can accompany a nuclear recoil. For Bremsstrahlung, there is some probability of photon emission when a dark matter particle recoils with a nucleus, where detection of this photon energy with a calorimeter can allow probes of much lower LDM masses. In the case of the Migdal effect, the process is quite analogous, where an electron is ionized from a nucleus rather than a photon. In, e.g., limits set by XENON1T~\cite{PhysRevLett.123.241803}, the Migdal effect outperforms the low-mass reach of Bremsstrahlung. In fact, it has been shown that the Migdal effect dominates over Bremsstrahlung, essentially making Bremsstrahlung irrelevant for LDM direct detection~\cite{PhysRevD.101.015012}. However, the Migdal effect has not been confirmed experimentally, thus the limits set by them are mostly an expectation rather than an exclusion.

For DM-electron interactions, interactions with polar materials~\cite{KNAPEN2018386}, superfluid $^4$He~\cite{PhysRevD.100.092007}, scintillating crystals~\cite{PhysRevD.96.016026}, and excitation and readout of electron-hole pairs with narrow bandgap materials via cryogenic charge amplifiers~\cite{osti_1764965,Juillard_2019,PHIPPS2019181,Essig:2022dfa} have all been pursued as promising experimental avenues for probing of lower mass LDM. For scintillating crystals such as GaAs, a CPD with sub-eV resolution would prove to be an excellent match for the $1.33\, \mathrm{eV}$ scintillation photons. The same holds true for the scintillation signals expected from  superfluid $^4$He, which are of order $1\, \mathrm{eV}$, as well as potentially the triplet excimer or He adsorption signals in such an experiment. All of these experiments are quite novel and require R\&D for background reduction and low-energy calibration, but all offer intriguing and promising directions for future LDM searches.

The recent experimental push towards LDM searches has created an exciting new direction within the world of DM detection, where small-scale labs have the potential to make large-scale discoveries. With many different experimental directions and methods that can be pursued, the future is bright for pushing to lower DM masses and potentially making a discovery---answering the fundamental question I started with in this thesis: what makes up our universe?

\printbibliography

\appendix
\renewcommand{\thefigure}{\Alph{chapter}.\arabic{figure}}
\chapter{\label{chap:appa}Thermal Conductance Measurements}

In this appendix, we will discuss the principles of how we can measure thermal conductances with TESs and measure the conductances for various systems.

\section{Thermal Conductance Measurement Principles with TESs}

\begin{figure}
    \centering
    \includegraphics[width=0.6\linewidth]{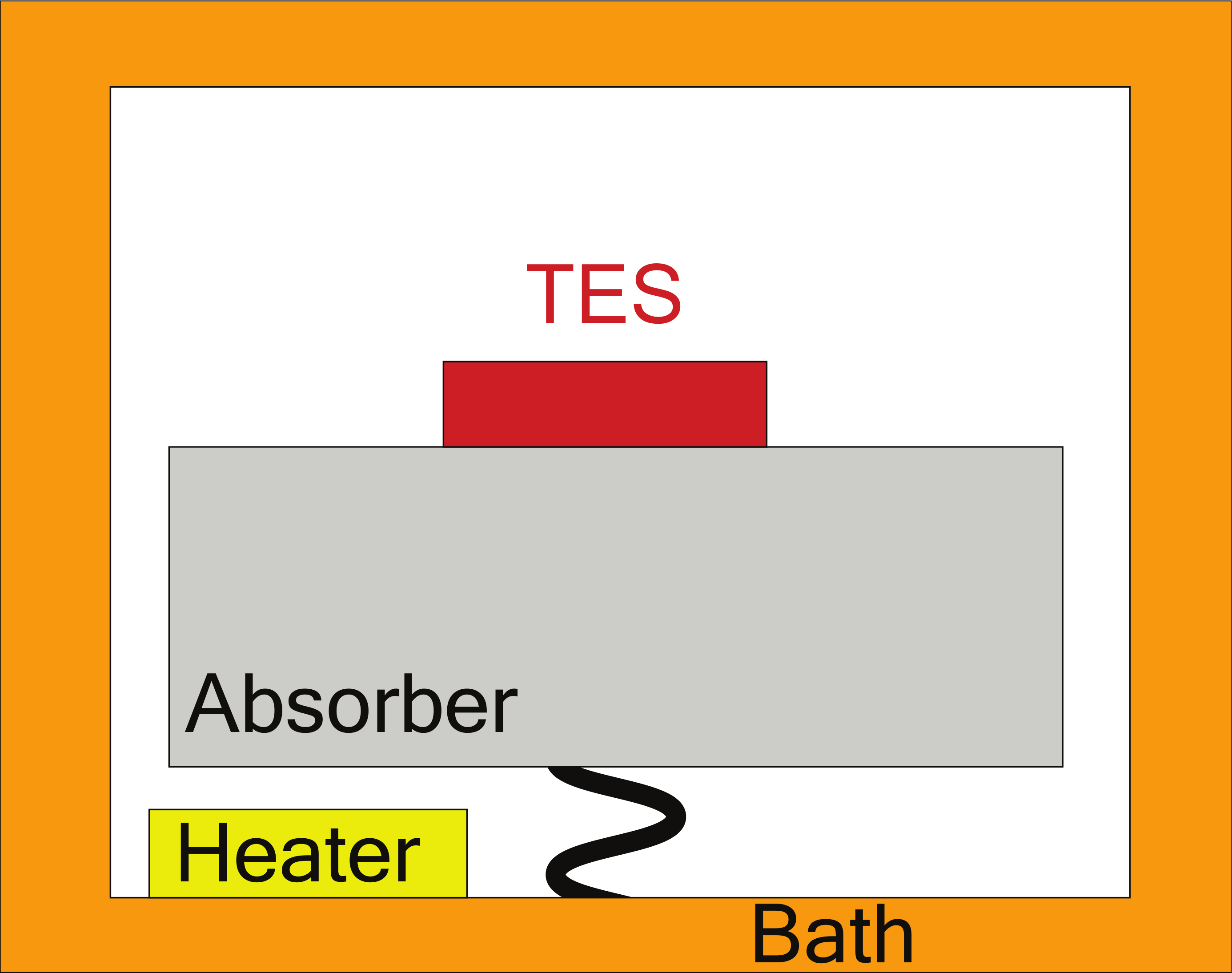}
    \caption{The setup of an arbitrary TES. The TES has been fabricated on some type of absorber, which itself is thermally connected to the thermal bath (i.e. the Cu housing in this diagram). The squiggly black line represents some type of thermal connection, which will depend on the detector holding scheme.}
    \label{fig:arbtes}
\end{figure}

The general setup for an arbitrary TES in a dilution refrigerator is pictured in Fig.~\ref{fig:arbtes}. There is a TES on an absorber, which itself is connected to the thermal bath (e.g. through detector holders or through a thermalizing Au wire bond). There is also some type of heater that can change the temperature of the thermal bath. Since the TES is in thermal contact with an absorber, which is in thermal contact with the bath, we must measure two thermal conductances, $G_{TA}$ and $G_{AB}$, to understand the thermal properties of our system. In the following sections, we explain the principles behind the measurements of these two values.

\subsection{\label{sec:gab}Measuring Thermal Conductance between Absorber and Bath}
We begin with measuring $G_{AB}$, which requires two TESs on the same absorber, one which acts as an absorber thermometer and the other as an absorber heater. The underlying equation that we want to fit is
\begin{equation}
    P_{AB} = K_{AB} (T_a^n - T_b^n),
    \label{eq:Pab}
\end{equation}
where $P_{AB}$ is the power flowing from absorber to bath, $T_a$ is the temperature of the absorber, $T_b$ is the temperature of the bath, $n$ is a power-law exponent, and $K_{AB}$ is related to $G_{AB}$ via
\begin{equation}
    G_{AB} = \frac{d P_{AB}}{dT_a} = nK_{AB}T_a^{n-1}.
    \label{eq:Gab}
\end{equation}
To extract $G_{AB}$, we must fit Eq.~(\ref{eq:Pab}), which requires measuring $P_{AB}$ as a function of $T_b$, while keeping $T_a$ constant. Because these measurements will be made when the system is in thermal equilibrium, the power flowing from the TESs to the absorber will be equal to the power flowing from the absorber to the bath. Thus, we can bias a TES in the normal regime with a Joule heating of
\begin{equation}
    P_{TES} = I_{TES}^2 R_N,
    \label{eq:Ptes1}
\end{equation}
where $I_{TES}$ is the current flowing through the TES and $R_N$ is the TES normal resistance, both of which can be known through the characterization steps outlined in Chapter~\ref{chap:two}. The TES current can be related back to the bias current $I_{bias}$ and various resistances of the system via
\begin{equation}
    I_{TES} = I_{bias} \frac{R_{sh}}{R_\ell + R_N},
    \label{eq:Ites1}
\end{equation}
where $R_{sh}$ is the shunt resistance, $R_\ell$ is the load resistance, and $R_N$ is the normal resistance.

\begin{figure}
    \centering
    \includegraphics[width=0.75\textwidth]{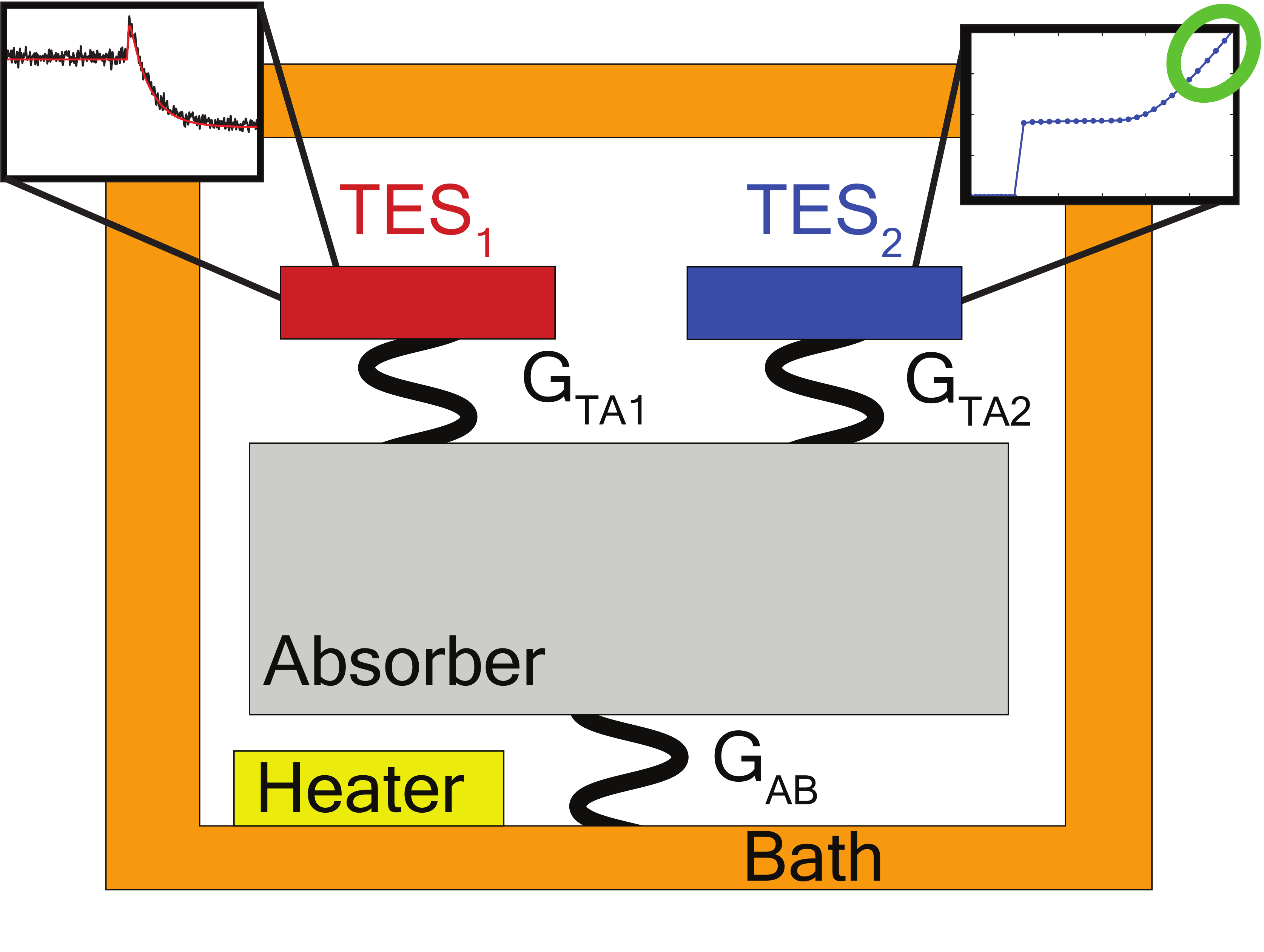}
    \caption{Diagram of the $G_{AB}$ measurement setup, where TES$_1$ is the thermometer, and TES$_2$ is the heater. For TES$_1$, the inset plot shows a characteristic in-transition TES response in time domain to a square wave. For TES$_2$, the inset plot shows the bias power as a function of bias current, where the green circle shows the approximate starting area in the normal regime for the $G_{AB}$ measurement.}
    \label{fig:Gab}
\end{figure}

To measure $P_{AB}$ as a function of $T_b$, we can use a heater placed in the bath (usually the mixing chamber in a dilution refrigerator) and measure the change in power as we change the heat load on the bath. To keep $T_a$ constant, we use our two TESs that have been fabricated on the same absorber. The thermometer TES is biased at some point in its superconducting transition, such that any change in temperature in the absorber will change the TES bias point. The heater TES is biased such that it is normal, and its Joule heating will heat up the absorber. This setup is diagrammatically shown in Fig.~\ref{fig:Gab}.

The measurement carries out as follows: we bias TES$_1$ low in transition and heat the bath up to a few mK below the $T_c$ of TES$_1$ (e.g. start at $42 \, \mathrm{mK}$ if $T_c=45 \, \mathrm{mK}$), such that biasing TES$_2$ normal will keep TES$_1$ in transition. TES$_2$ is then biased normal with a power that should be much greater than the bias power of TES$_1$, such that the latter can be neglected (this should be simple if TES$_1$ is the more sensitive device). This first point gives $P_{AB}$ at the starting $T_b$, as well as our starting bias point for TES$_1$. We then decrease the heat load on the bath, which decreases $T_a$ and equivalently changes the bias point of TES$_1$, as the power flowing in equilibrium has not yet changed (it is still equal to the Joule heating of TES$_2$). To remedy this, we increase the bias current on TES$_2$ until TES$_1$ has returned to its initial bias point and thus so has $T_a$. This gives us our second data point for $P_{AB}$ at a new $T_b$. This procedure is then continued until either TES$_2$ can no longer heat the absorber up to the same $T_a$ (due to some limitation with the amount of bias current that can be supplied) or we have reached the base temperature of the dilution refrigerator.

In Fig.~\ref{fig:Gab}, the bias point of TES$_1$ is represented by its response to a square wave. In this case, we would know that, when changing the Joule heating of TES$_2$, TES$_1$ has returned to its initial bias point when zero frequency change of its response (i.e. $\partial I / \partial V(0)$) is the same. Alternatively, an equivalent method is to not send a square wave jitter down the bias line, but instead follow the DC value of the TES current. After the DC value changes as the bath is heated, then TES$_1$ will have returned to its initial point by changing the Joule heating of TES$_2$ until TES$_1$ has the same TES current. It is up to the person doing the measurement which they prefer, the former gives more TES information, while the latter may be faster in terms of data taking time (no need to fit the TES response, but simply track the DC value of the current).

At the conclusion of the measurement, we will have $P_{AB}$ as function of $T_{b}$ at hopefully many bias points, and we can simply fit Eq.~(\ref{eq:Pab}) via a least squares routine to extract $G_{AB}$, $T_a$, and $n$. It may also be useful to calculate the natural time constant $\tau = C/G_{AB}$ where $C$ is the heat capacity of the absorber, which is the characteristic time for the absorber to cool back down to base temperature after being warmed up to about $T_c$ (e.g. a muon hits the detector and pushes the TESs normal). If the absorber is a semiconductor or insulator (usually the case), then we can estimate its heat capacity in the low-temperature Debye limit
\begin{equation}
C^{th} = \frac{12 \pi^4 k_b}{5 A} \rho V N_A \left(\frac{T_a}{\Theta_D}\right)^3,
\label{eq:Cv}
\end{equation} 
where $k_b=1.38\times 10^{-23} \, \mathrm{J}/\mathrm{K}$ is the Boltzmann constant, $\rho$ is the density of the absorber, $V$ is the volume of the absorber, $A$ is the atomic mass of the absorber, $N_A = 6.022  \times 10^{23} \, \mathrm{mol}^{-1}$ is the Avogadro constant, and $\Theta_D$ is the Debye temperature of for the absorber.


\subsection{Measuring Thermal Conductance between TES and Absorber}

To measure $G_{TA}$, we only need a single TES, but it is dependent on already having done the $G_{AB}$. The underlying equation that we want to fit is
\begin{equation}
    P_{TA} = K_{TA} (T_c^n - T_a^n),
    \label{eq:Pta}
\end{equation}
where $T_c$ is the superconducting critical temperature of the TES, $n$ is the power law exponent, and $K_{TA}$ is related to $G_{TA}$ via
\begin{equation}
    G_{TA} = \frac{d P_{TA}}{dT_c} = nK_{TA}T_c^{n-1}.
    \label{eq:Gta}
\end{equation}
As discussed in Section 3.5 of Irwin and Hilton~\cite{irwin}, the thermal impedance between the electrons and phonons in a metal with small volume and high power density (e.g. a W TES) dominates. In this case, the power flow is from the electrons to the phonons via $P = \Sigma_{ep} V_{TES}\left(T_{el}^5 - T_{ph}^5 \right)$, where $V_{TES}$ is the TES volume and $\Sigma_{ep}$ is a material-dependent electron-phonon coupling constant (usually on the order of $10^9 \, \mathrm{W} \, \mathrm{m}^{-3} \, \mathrm{K}^{-5}$). Thus, it is expected that $n=5$, $K_{TA} = \Sigma_{ep} V_{TES}$, and $G_{TA} = 5\Sigma_{ep}V_{TES} T_c^4$ for a W TES.

Note that Eq.~(\ref{eq:Pta}) is dependent on $T_a$, but the experimental setup generally only has the ability to measure $T_b$, showing why it is important to measure $G_{AB}$ first in order to accurately calculate $T_a$. Although, in practice, a good thermal connection between absorber to bath will mean that $G_{AB} \gg G_{TA}$ and $T_a \approx T_b$. Thus, in single-channel devices (i.e. only one TES channel on an absorber), it is common to make the assumption that $T_a \approx T_b$ if the thermalization scheme is known to have a large $G_{AB}$ from historical measurements and continue to do a $G_{TA}$ measurement.

For the $G_{TA}$ measurement, the setup is simple: we have a single TES on a substrate, which is thermally connected to the absorber, and the absorber is thermally connected to the bath. For this measurement, we keep the TES at the same transition point while changing the bath temperature via the heater, and we calculate the TES bias power as a function of bath temperature. Since we wait for the system to reach equilibrium, the bias power of the TES $P_0$ will be equal to the power from the TES to the absorber $P_{TA}$, which in turn is equal to the power from the absorber to the bath $P_{AB}$. $P_0$ is easily defined via
\begin{equation}
    P_0 = I_0^2 R_0,
    \label{eq:P0}
\end{equation}
where $I_0$ is the current through the TES, and $R_0$ is TES resistance. There are two common ways to calculate $P_{0}$ as a function of $T_b$ for the $G_{TA}$ measurement: the ``quick'' way and the ``clean'' way.

\subsubsection{The Quick $G_{TA}$ Measurement}

The quick measurement is to bias the TES in transition and measure the TES response. If the TES is low enough in transition, then we can use that, in the limit of large loop gain $\mathcal{L}$, the TES complex impedance (Eq.~(\ref{eq:zcirc})) becomes
\begin{equation}
    \lim_{\mathcal{L} \to \infty} Z_\mathrm{circ}(\omega) = R_\ell + i\omega L - R_0,
\end{equation}
and the zero-frequency value is $Z_\mathrm{circ}(0) = R_\ell - R_0$. Thus, we can solve for $R_0 = \left| Z_\mathrm{circ}(0) \right| + R_\ell$ and $I_0 = I_{bias} R_{sh} / (R_0 + R_\ell)$, and we have an estimate of $P_0$.

\begin{figure}
\centering
\includegraphics[width=0.6\textwidth]{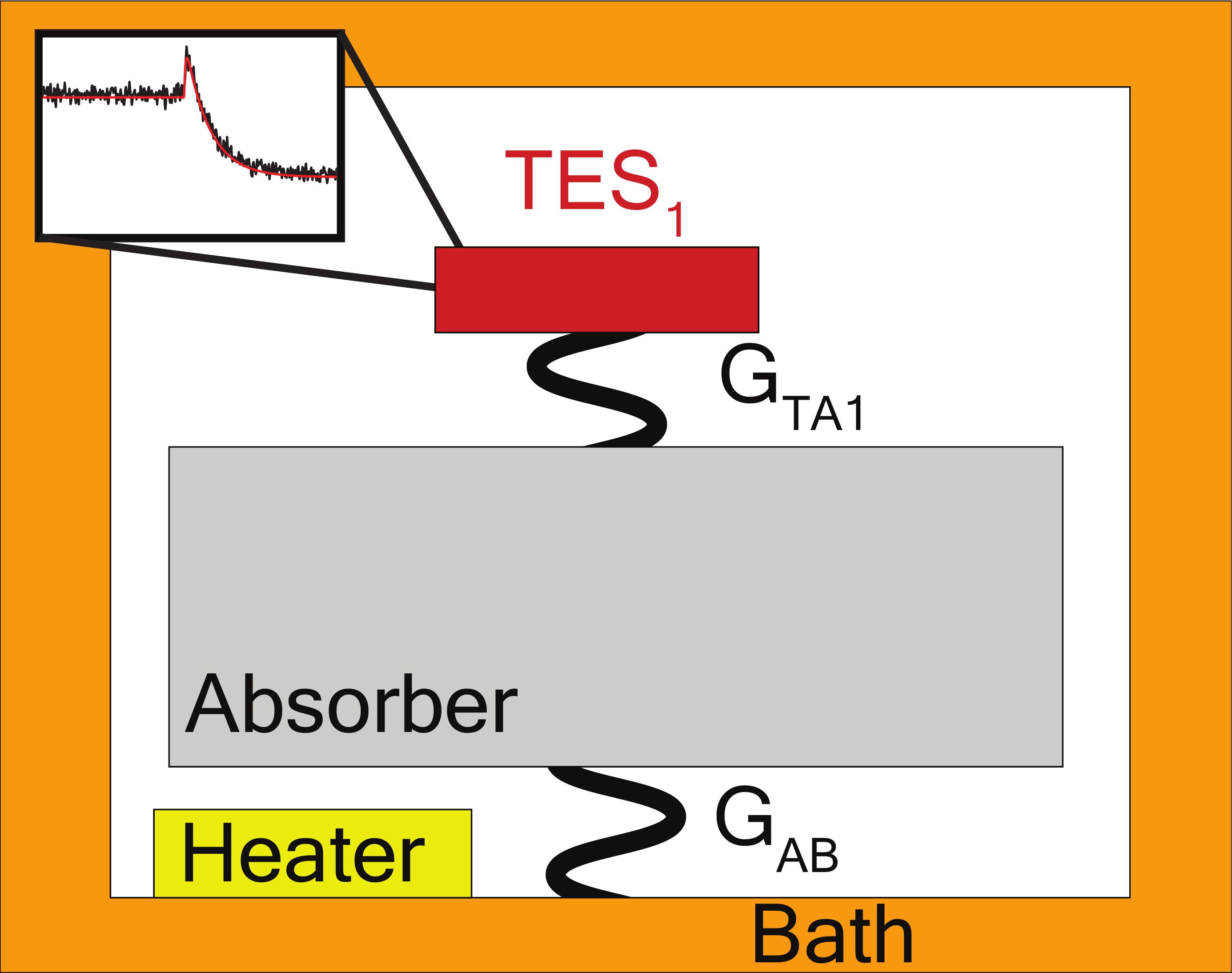}
\caption{Diagram of the $G_{TA}$ measurement setup, where TES$_1$ is kept at the same point in transition as the bath temperature changes. For TES$_1$, the inset plot shows a characteristic in-transition TES response in time domain to a square wave.}
\label{fig:Gta}
\end{figure}

The quick $G_{TA}$ measurement procedure is then as follows (and pictured in Fig.~\ref{fig:Gta}): we bias the TES low in transition and heat the bath up to a few mK below the $T_c$ of TES (e.g. start at $42 \, \mathrm{mK}$ if $T_c=45 \, \mathrm{mK}$). We measure the zero-frequency value of the circuit complex impedance, and we immediately have an estimate of $P_0$ and, equivalently, $P_{TA}$ for at this $T_b$. In order to convert this $T_b$ to $T_a$ (if we have a $G_{AB}$ measurement), we use Eq.~(\ref{eq:Gab}), giving the first data point for $P_{TA}$ as a function of $T_a$. We then, as we did with $G_{AB}$, decrease the heat load on the bath, which decreases $T_a$ and equivalently changes the bias point of the TES, as the power flowing in equilibrium has not yet changed (it is equal to the Joule heating of the TES). Note that the TES in transition is also supplying the Joule heating, in contrast to the $G_{AB}$ measurement system (in which the in-transition TES bias power was negligible). When we increase the bias current to the TES so that it returns to the same bias point (i.e. the same $R_0$) the bias power will increase, and $T_a$ will decrease. Doing this, we have a second data point for the bias power of the TES as a function of $T_a$. This procedure continues until we have reached the base temperature of the dilution refrigerator (that is, the heater on the mixing chamber has zero heat load).

This measurement can go quickly, but the approximation of $\mathcal{L} \to \infty$ may introduce a systematic on $R_0$, and we may actually have a different true bias power. To avoid this possible systematic, the $G_{TA}$ measurement can be done using clean way, at the expense of longer data taking.

\subsubsection{The Clean $G_{TA}$ Measurement}

In essence, the clean $G_{TA}$ measurement is done by taking $IV$ sweeps at many different bath temperatures. These $IV$ sweeps can then be interpolated such that we can choose the same TES resistance for each bath temperature from the sweep, immediately extracting the measured $I_0$ and $R_0$ and avoiding possible systematic error. The drawback is that these $IV$ sweeps take a fair amount of time (perhaps an hour per sweep), depending on how many points are desired. When doing these for many bath temperatures, the data taking time can quickly become nearly the entire day. Because of this, usually the quick way to do the measurement is done first, with the clean way done in the case of analysis for publication. We show an example of the clean way in Section~\ref{sec:cleangta}.


With our methods of measuring $G_{AB}$ and $G_{TA}$ explained, we will next show a few example measurements, where all nonlinear least squares fits are done using \texttt{optimize.curve\_fit} in \textsc{SciPy}~\cite{2020SciPy-NMeth}.

\section{\label{sec:g115}Thermal Conductance Measurements with G115}

\begin{figure}
    \centering
    \includegraphics[width=\linewidth]{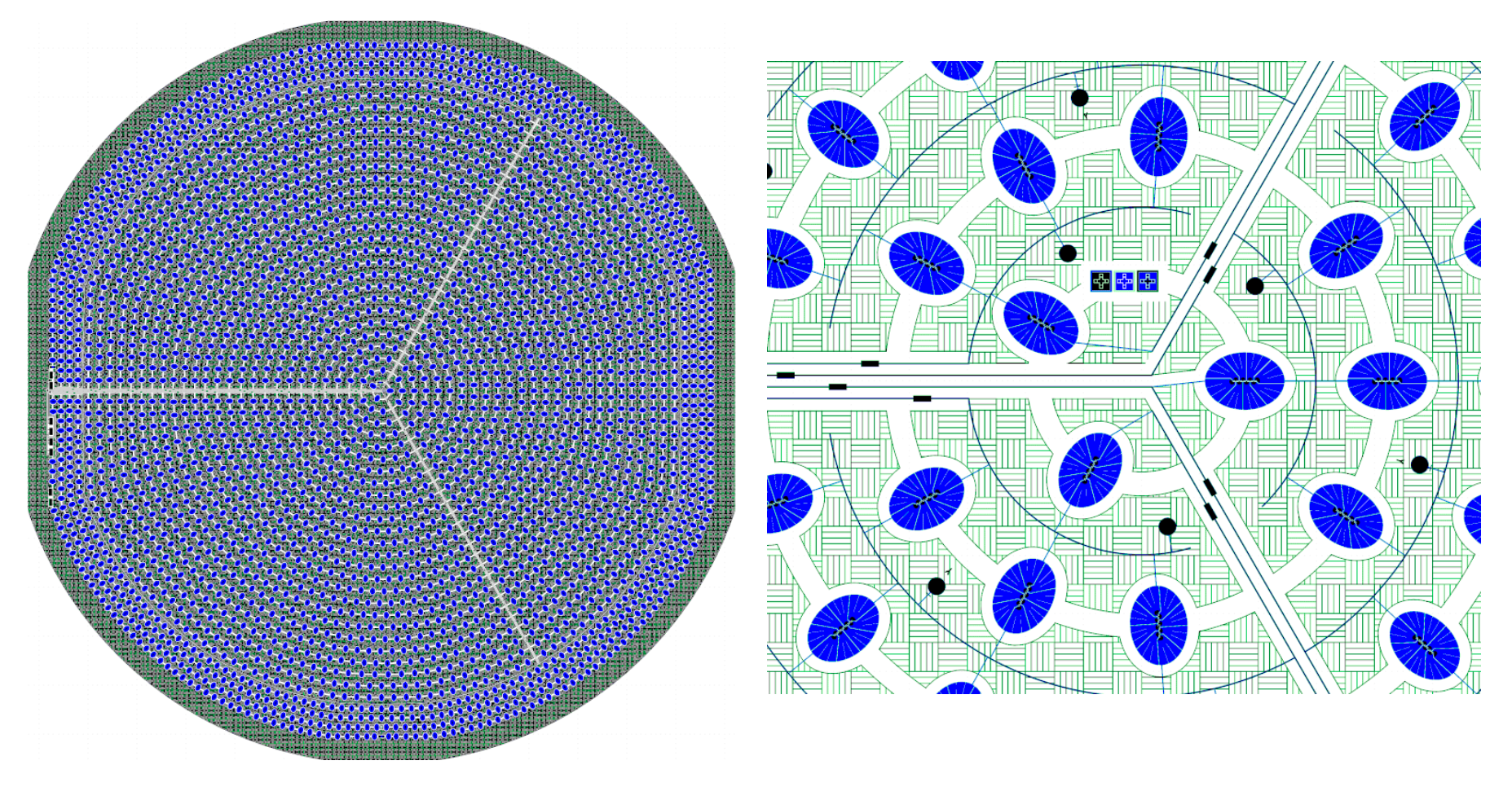}
    \caption{(Figure from Ref.~\cite{Noah_thesis}) Left: Full image of the G115 mask design, showing the Al (blue), amorphous Si (green), and W (black). There are four channels, where three are inner (note the three line denoting the split) and one is on the outer ring. Right: Zoom in of the center of the detector, showing the parquet pattern of the QETs.}
    \label{fig:g115mask}
\end{figure}

Developed by Matt Pyle and Suhas Ganjam, G115 is a prototype detector, consisting of a $3\, \mathrm{inch}$ diameter and $1\,\mathrm{mm}$ thick Ge wafer, with a four-channel QET mask deposited on one side (mask design shown in Fig.~\ref{fig:g115mask}). This detector was made to study QET performance, as well as for understanding the effect high energy events (muons and photons) on high voltage noise, as discussed in detail in Noah Kurinsky's thesis~\cite{Noah_thesis}. In this section, we will bypass that discussion and instead summarize the thermal conductance measurements made on this device in a cryogen-free dilution refrigerator as SLAC, in which the detector was thermally connected to bath via cirlex clamps.

For these measurements, we will use two of the inner channels of G115, henceforth known as channels A and B in this section. The measured normal resistances of the channels A and B are $200 \, \mathrm{m}\Omega$ and $182 \, \mathrm{m}\Omega$, respectively. For each channel, the parasitic resistance is $R_p = 12\, \mathrm{m}\Omega$.

\subsection{Cirlex Clamps}

The cirlex clamps holding scheme will have a interfacial thermal resistance due to the elastic constant mismatch between the dissimilar semiconducting substrate and the clamps themselves~\cite{doi:10.1139/p59-037}, and the thermal conductance is expected to go as $G \propto T^3$.

\subsubsection{$G_{AB}$ Measurement}

\begin{table}
    \centering
    \caption{Values measured and calculated for channel B, calculating the power flowing from absorber to bath.}
    \begin{tabular}{rrr}
    \hline \hline
     $T_b$ [mK]     &  $I_\mathrm{bias}$ $[\mu \mathrm{A}]$    & $P_{AB}$ [pW]  \\ \hline
     42.15            & 500.00                   & 32.5     \\
     41.75            & 585.00                   & 44.5         \\ 
     41.45            & 663.00                   & 57.2           \\ 
     41.05            & 733.00                   & 69.9           \\ 
     40.75            & 797.00                   & 82.6           \\ 
     40.40            & 853.00                   & 94.7           \\ 
     40.09            & 909.75                   & 107.7         \\ 
     39.05            & 1049.00                  & 143.2           \\
     37.95            & 1161.00                  & 175.4         \\ 
     37.16            & 1258.00                  & 205.9         \\ 
     35.87            & 1399.00                  & 254.6         \\ \hline \hline
    \end{tabular}
    \label{tab:tesc}
\end{table}

To measure $G_{AB}$, we follow the method as outlined in Section~\ref{sec:gab}, where channel A acted as a thermometer, and channel B acted as a heater. The data from this measurement are reported in Table~\ref{tab:tesc}, wherehe error in $P_{AB}$ is set to be 10\% of the measured power, based on an estimated 10\% uncertainty in $R_{sh}$. The dominant systematic when taking these measurements was that the fridge had a stabilization time of $\mathcal{O}(20\, \mathrm{min})$, meaning that it was difficult to tell if we had fully stabilized at each point. Furthermore, the size of the detector meant that it had a substantial rate of high energy events (muons and gammas) at the surface, which would heat the substrate. Thus, there may be some bias in the measurements, which is difficult to quantify.

\begin{figure}
\centering
\includegraphics{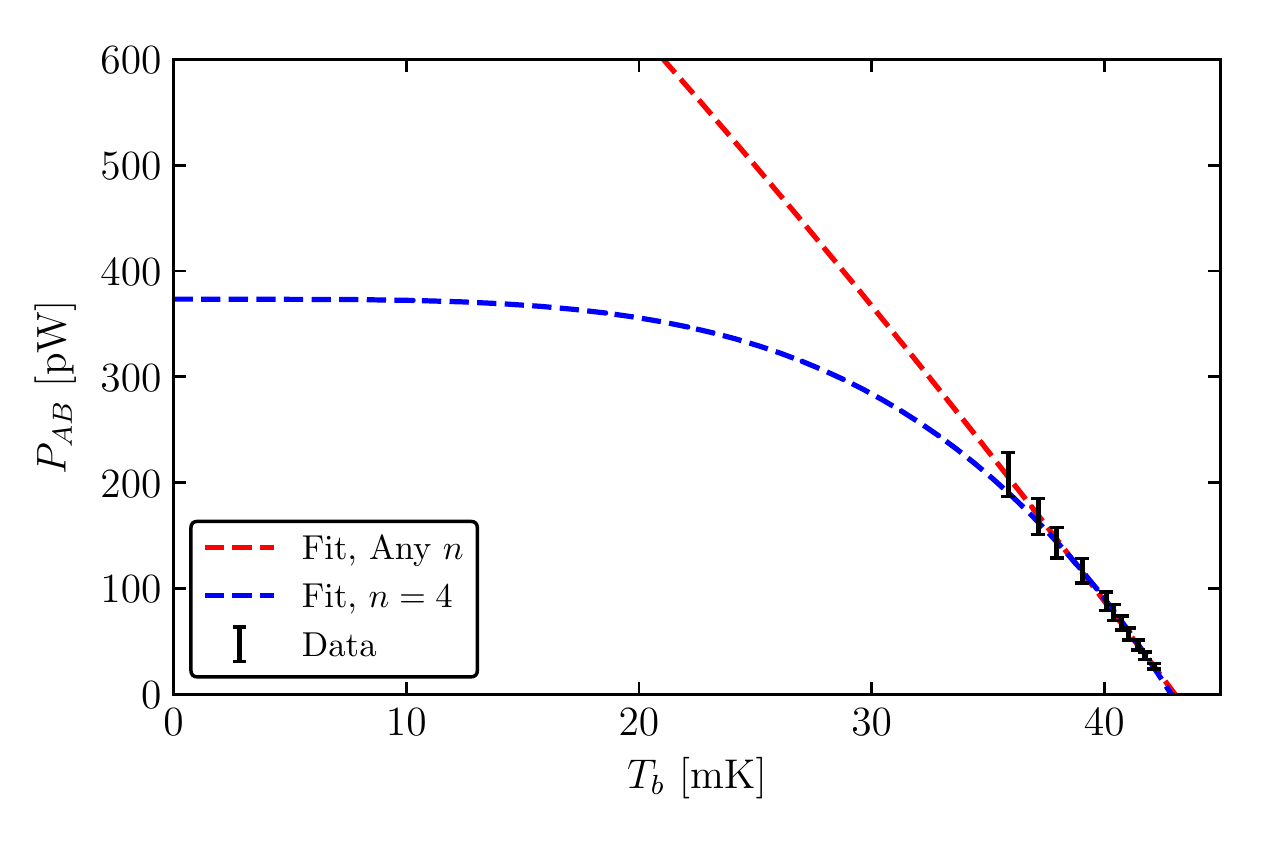}
\caption{The power flowing from absorber to bath through the cirlex clamps at the various bath temperatures taken for G115. Also shown are the least squares fits with $n$ varying and $n=4$ fixed.}
\label{fig:Gabfit}
\end{figure}

\begin{table}
    \centering
    \caption{Values from the least-squares fits to the power flowing from absorber to bath the cirlex clamps. Included in the table are the estimated heat capacity via the low-temperature Debye model using Eq.~(\ref{eq:Cv}), as well as the related thermal time constant $\tau$.}
    \begin{tabular}{crrrrr}
    \hline \hline
    \rule{0pt}{10pt} Fit & $G_{AB}$ $\left[\frac{\mathrm{nW}}{\mathrm{K}}\right]$ & $T_a$ [$\mathrm{mK}$] & $n$ & $C^{th}$ $\left[\frac{\mathrm{nJ}}{\mathrm{K}}\right]$ & $\tau$ $[\mathrm{ms}]$ \\ \hline
    Any $n$ & $29.2\pm5.0$ & $43.04\pm0.19$  & $1.6\pm2.3$ & $3.96\pm0.05$  & $136\pm 25$ \\
    $n=4$   & $34.8\pm1.8$ & $42.88\pm0.09$  & 4 & $3.92\pm 0.02$  & $112\pm 6$   \\ \hline \hline
    \end{tabular}
    \label{tab:Gabfit_g115}
\end{table}

We next fit Eq.~(\ref{eq:Gab}) with $n$ allowed to be any value, and $n=4$ fixed as expected from the phonon mismatch between the absorber and the clamps. In Fig.~\ref{fig:Gabfit}, we show the results of these two fits and compare to the measured data, with the various fit parameters reported in Table~\ref{tab:Gabfit_g115}. It is clear from the fits that we did not take enough data (i.e. go to a low enough bath temperature) to be able to describe the power law behavior with any confidence. However, the thermal conductance itself is not very sensitive to the power law exponent, nor is the absorber temperature, so we were still able to estimate the absorber heat capacity and the thermal time constant to a reasonable degree.

\subsubsection{$G_{TA}$ Measurement}

\begin{table}
    \centering
    \caption{Values measured and calculated for channel A of G115, showing that we are approximately at the same operating resistance as the bath temperature changes. Note that $T_a$ is calculated using the $G_{AB}$ fit.}
    \begin{tabular}{rrrrr}
    \hline \hline
     $T_b$ [mK]     &  $T_a$ [mK]  & $I_\mathrm{bias}$ $[\mu \mathrm{A}]$ & $R_0$ $[\mathrm{m}\Omega]$ &  $P_{TA}$ [pW]  \\ \hline
      10.90    & 11.02      & 170.0    & 134.79    & 4.23  \\ 
      13.40    & 13.52      & 168.0    &  132.20   & 4.19   \\ 
      17.80    & 17.91      & 164.5    & 132.43    & 4.01  \\ 
      26.30    & 26.40      & 151.0    & 132.43    & 3.38    \\ 
      31.80    & 31.87      & 129.5    &  132.20   & 2.49  \\ 
      36.25    & 36.29      & 102.0    & 131.51    & 1.55   \\ 
      40.80    & 40.81      & 50.0     & 132.42  &  0.37  \\ \hline \hline
    \end{tabular}
    \label{tab:tesb2}
\end{table}

To measure $G_{TA}$, we used channel A and the quick $G_{TA}$ measurement method (i.e. estimating the TES resistance from the zero-frequency value of the circuit impedance). The data from this measurement are reported in Table~\ref{tab:tesb2}, and the error in $P_{TA}$ is estimated to be 10\% of the measured power. The sources of systematic error are the same as the discussion for the $G_{AB}$ measurement, such that there may be a bias in these measurements due to the long thermalization times of the fridge and the high rate of high-energy events.

\begin{figure}
\centering
\includegraphics{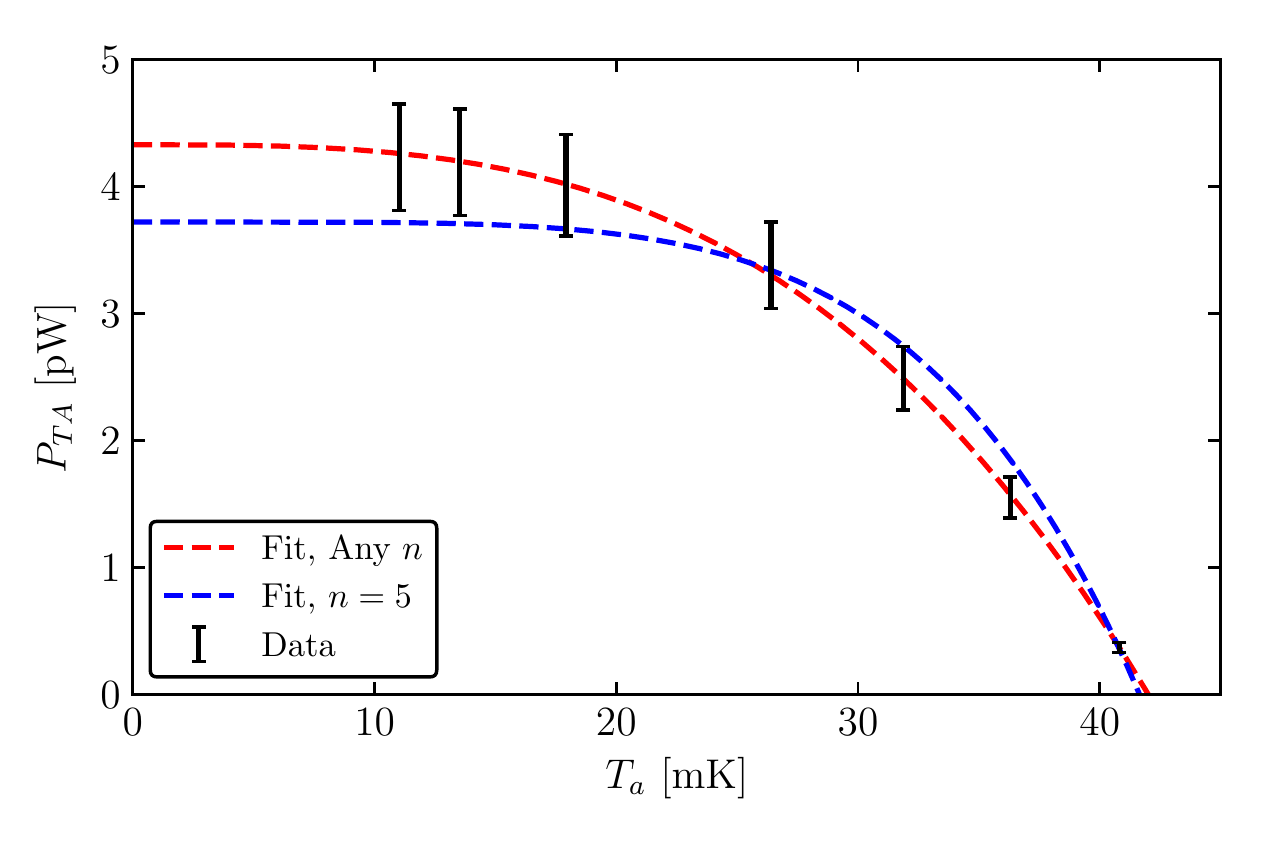}
\caption{The power flowing from TES to absorber at various absorber temperatures taken for G115. Also shown are the least squares fits with $n$ varying and $n=5$ fixed.}
\label{fig:Gtafit}
\end{figure}

\begin{table}
    \centering
    \caption{The corresponding fitted values from the $G_{TA}$ measurement for channel A of G115, where the errors come from propagating the 10\% error in the power measurement. Included is estimated electron-phonon coupling constant $\Sigma_{ep}$ for this array of W TESs.}
    \begin{tabular}{crrrr}
    \hline \hline
    \rule{0pt}{10pt} Fit & $G_{TA}$ $\left[\frac{\mathrm{pW}}{\mathrm{K}}\right]$ & $T_c$ [$\mathrm{mK}$] & $n$ & $\Sigma_{ep}$ $\left[\frac{\mathrm{GW}}{\mathrm{K}^n \mathrm{m}^3}\right]$\\ \hline
    Any $n$ & $318\pm41$ & $42.00\pm0.21$  & $3.1\pm0.5$ &\multicolumn{1}{c}{---}\\
    $n=5$ & $447\pm20$ & $41.64\pm0.11$  &  5  & $1.24\pm 0.06$\\ \hline \hline
    \end{tabular}
    \label{tab:Gtafit_g115}
\end{table}

Using Eq.~(\ref{eq:Gta}), we fit the power law with $n$ varying and $n=5$ fixed, the latter as expected for a W TES. In Fig.~\ref{fig:Gtafit}, we show the results of the two fits as compared to the measured data, and report the various fit parameters in Table~\ref{tab:Gtafit_g115}. With this data, we took enough data points, but we instead have an anomalous power exponent of $n=3.1\pm0.5$ when letting $n$ vary. As this is far from expectation, and we know that we were having trouble with the long fridge thermalization times, we can explain this value as due to the systematics from the measurement.

\section{Thermal Conductance Measurements with QP.4}

\begin{figure}
    \centering
    \includegraphics{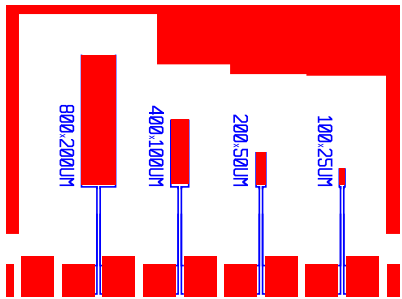}
    \caption{The TES rectangles to be used for thermal conductance measurements, which are on a $1 \, \mathrm{cm}^2 \times 1 \, \mathrm{mm}$ Si substrate. These TESs scale by factors of 4 in area, while keeping the aspect ratio the same.}
    \label{fig:qp4_rectangles_app}
\end{figure}

The entire QP.4 chip we will use for these measurements is shown in Fig.~\ref{fig:qp4passive}, where we use the TES rectangles, as pictured in Fig.~\ref{fig:qp4_rectangles_app} for convenience. We will measure thermal conductances of QP.4 in the following configurations: glued with GE-7031 varnish, resting on Cu and thermalized through the electronic wire bonds, and glued with rubber cement. We will also show the clean method of measuring $G_{TA}$ via $IV$ curves at different bath temperatures. These TES rectangles have dimensions $800 \, \mu\mathrm{m} \times 200 \, \mu\mathrm{m} \times 40 \, \mathrm{nm}$, $400 \, \mu\mathrm{m} \times 100 \, \mu\mathrm{m} \times 40 \, \mathrm{nm}$, $200 \, \mu\mathrm{m} \times 50 \, \mu\mathrm{m} \times 40 \, \mathrm{nm}$ and $100 \, \mu\mathrm{m} \times 25 \, \mu\mathrm{m} \times 40 \, \mathrm{nm}$, and will henceforth be referred to as TES800x200, TES400x100, TES200x50, and TES100x25, respectively. Each of the following measurements were done on a QP.4 chip from the same wafer, but each are different chips, thus giving variation on the various values of $G_{TA}$ that we will measure for different TESs with the same rectangular dimensions.

\subsection{\label{sec:gevarn}GE-7031 Varnish}

GE-7031 varnish is a common polymer-based adhesive for cryogenic applications. In our lab is has been frequently used to glue $1\, \mathrm{cm}^2$ substrates to our Cu housing. Because the GE-7031 varnish is dissimilar to the semiconductor substrate, just as the case with the cirlex clamps, the thermal conductance at the interface will be due to the mismatch of the elastic constants~\cite{doi:10.1139/p59-037}, giving an expected thermal conductance temperature dependence of $G \propto T^3$ (i.e. the power law exponent of the heat flow is expected to be $n=4$). Practically, we have had some trouble in the past with the GE-7031 varnish thermal contraction being large enough that our $1\, \mathrm{cm}^2$ chips would frequently pop off of the Cu housing, and, depending on the device, the substrate may no longer have a good thermal connection to bath.

\subsubsection{$G_{AB}$ Measurement}

\begin{table}
    \centering
    \caption{Values used to calculate the change in power flowing from the absorber to bath via the GE-7031 varnish at different bath temperatures, where TES200x50 was the absorber heater.}
    \begin{tabular}{rrr}
    \hline \hline
     $T_b$ [mK]     &  $I_\mathrm{bias}$ $[\mu \mathrm{A}]$    & $P_{AB}$ [pW]  \\ \hline
     56            & 44.8                   & 0.12    \\
     55.5            &  161.05                  & 1.52         \\ 
     55            & 223.875                  & 2.93          \\ 
     54.5            & 270.5                  & 4.27          \\ 
     53            & 368                   & 7.91          \\ 
     52            & 417.5                   & 10.18           \\ 
     51            & 457.75                   & 12.24        \\ 
     50            & 492.25                  & 14.16          \\ \hline \hline
    \end{tabular}
    \label{tab:pab_data_gevarnish}
\end{table}

\begin{figure}
    \centering
    \includegraphics{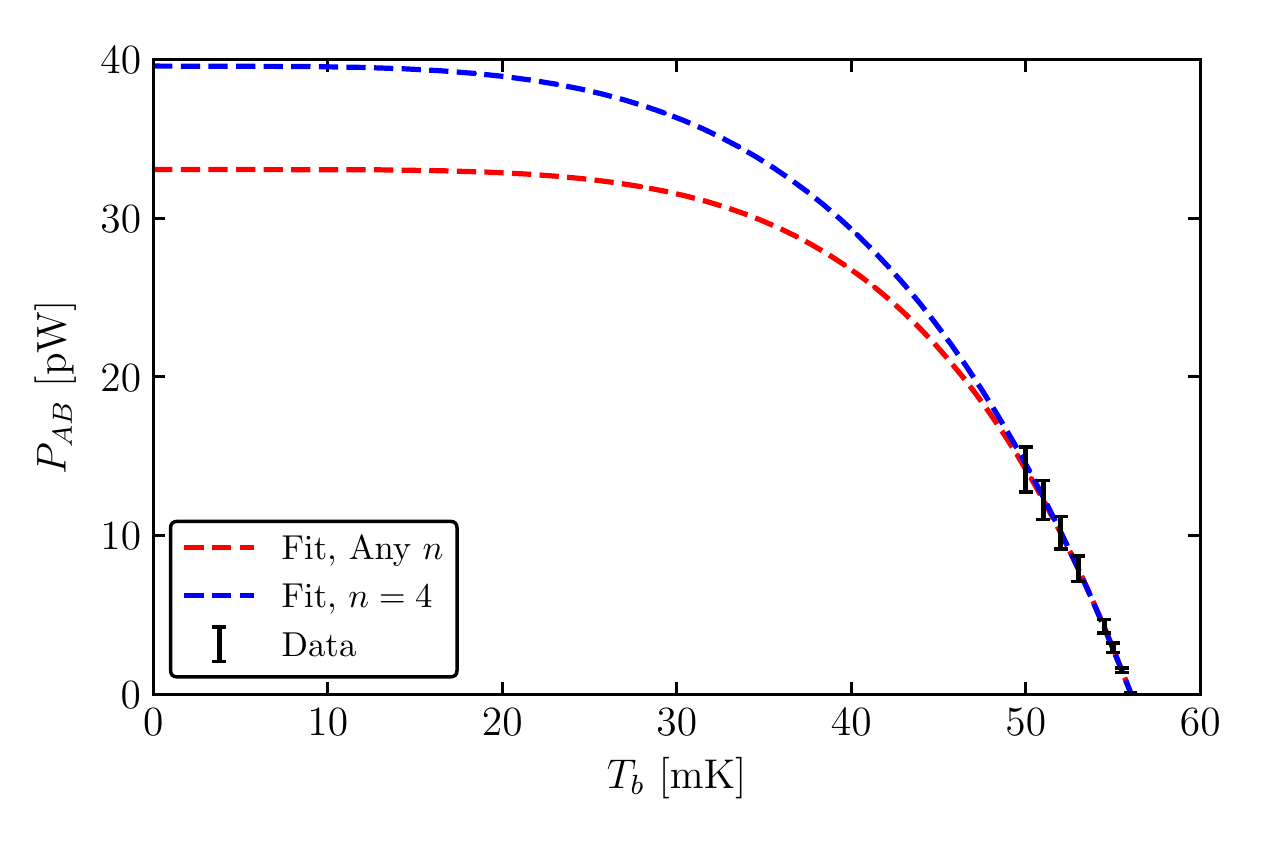}
    \caption{The power flowing from absorber to bath through the GE-7031 varnish at various bath temperatures taken for TES200x50. Also shown are the least squares fits with $n$ varying and $n=4$ fixed.}
    \label{fig:gab_glue_r6}
\end{figure}

\begin{table}
    \centering
    \caption{Values from the least-squares fits to the power flowing from absorber to bath through GE-7031 varnish. Included in the table are the estimated heat capacity via the low-temperature Debye model using Eq.~(\ref{eq:Cv}), as well as the related thermal time constant $\tau$.}
    \begin{tabular}{crrrrr}
    \hline \hline
    \rule{0pt}{10pt} Fit & $G_{AB}$ $\left[\frac{\mathrm{nW}}{\mathrm{K}}\right]$ & $T_a$ [$\mathrm{mK}$] & $n$ & $C^{th}$ $\left[\frac{\mathrm{pJ}}{\mathrm{K}}\right]$ & $\tau$ $[\mathrm{ms}]$ \\ \hline
    Any $n$ & $2.89\pm0.22$ & $56.04\pm0.01$  & $4.9\pm2.3$ & $10.572 \pm 0.003$ & $3.65 \pm 0.27$ \\
    $n=4$ & $2.83\pm0.11$ & $56.04\pm0.01$  & 4 & $10.572\pm0.003$ & $3.74\pm0.15$\\ \hline \hline
    \end{tabular}
    \label{tab:gab_glue_r6}
\end{table}

To measure $G_{AB}$, we use TES200x50 as the heater and TES100x25 as the thermometer. In addition to the $5\, \mathrm{m}\Omega$ shunt resistor, the normal resistance of TES200x50 in this measurement is $413 \, \mathrm{m}\Omega$, with a parasitic resistance of $2.4\, \mathrm{m}\Omega$. The data from this measurement are reported in Table~\ref{tab:pab_data_gevarnish}. The error in $P_{AB}$ is set to be 10\% of the measured power, based on an estimated 10\% uncertainty in $R_{sh}$. We fit Eq.~(\ref{eq:Gab}) with $n$ allowed to be any value, and $n=4$ fixed as expected from the phonon mismatch between the absorber and the GE-7031 varnish. In Fig.~\ref{fig:gab_glue_r6}, we show the results of these two fits and compare to the measured data, with the various fit parameters reported in Table~\ref{tab:gab_glue_r6}.

\subsubsection{$G_{TA}$ Measurement}

\begin{figure}
    \begin{subfigure}{.5\textwidth}
        \centering
        \includegraphics[width=1\linewidth]{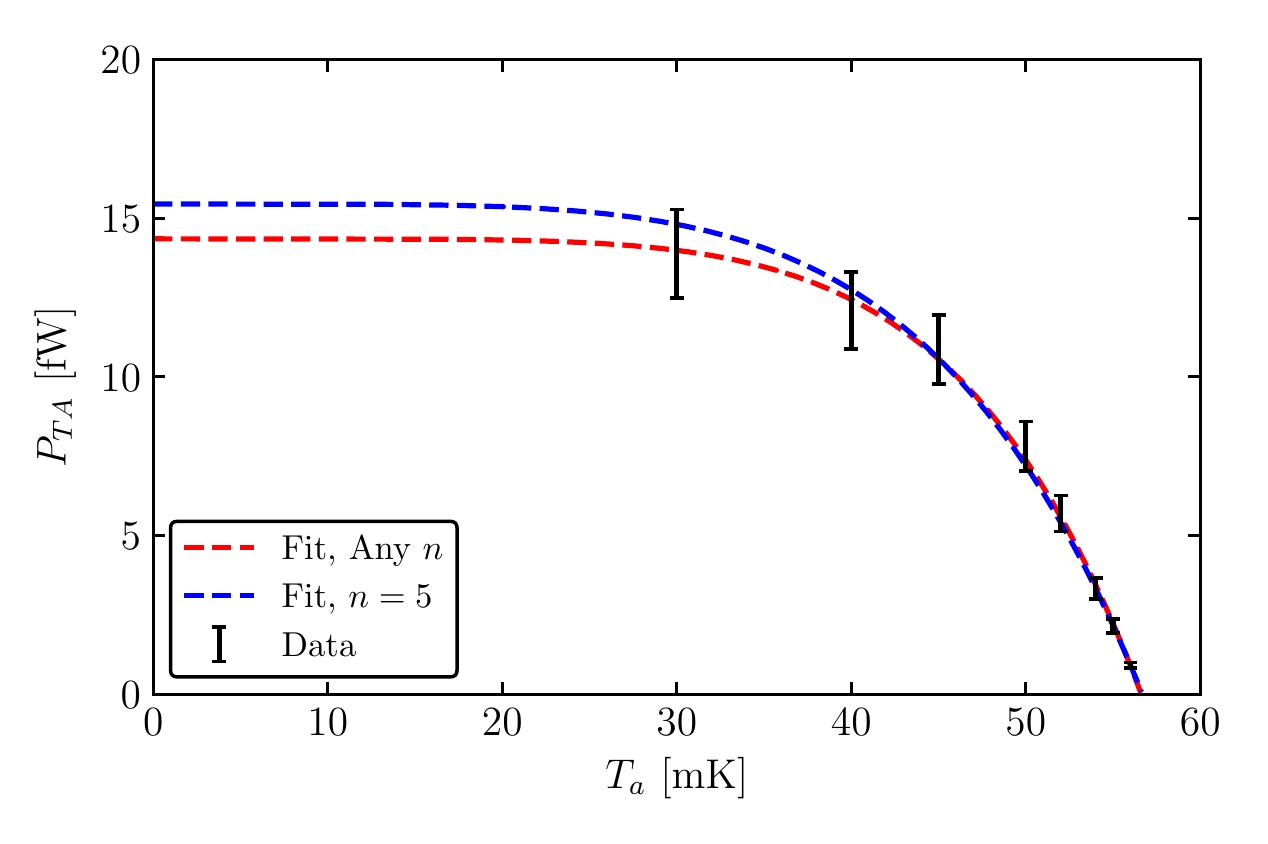}
    \end{subfigure}%
    \begin{subfigure}{.5\textwidth}
        \centering
        \includegraphics[width=1\linewidth]{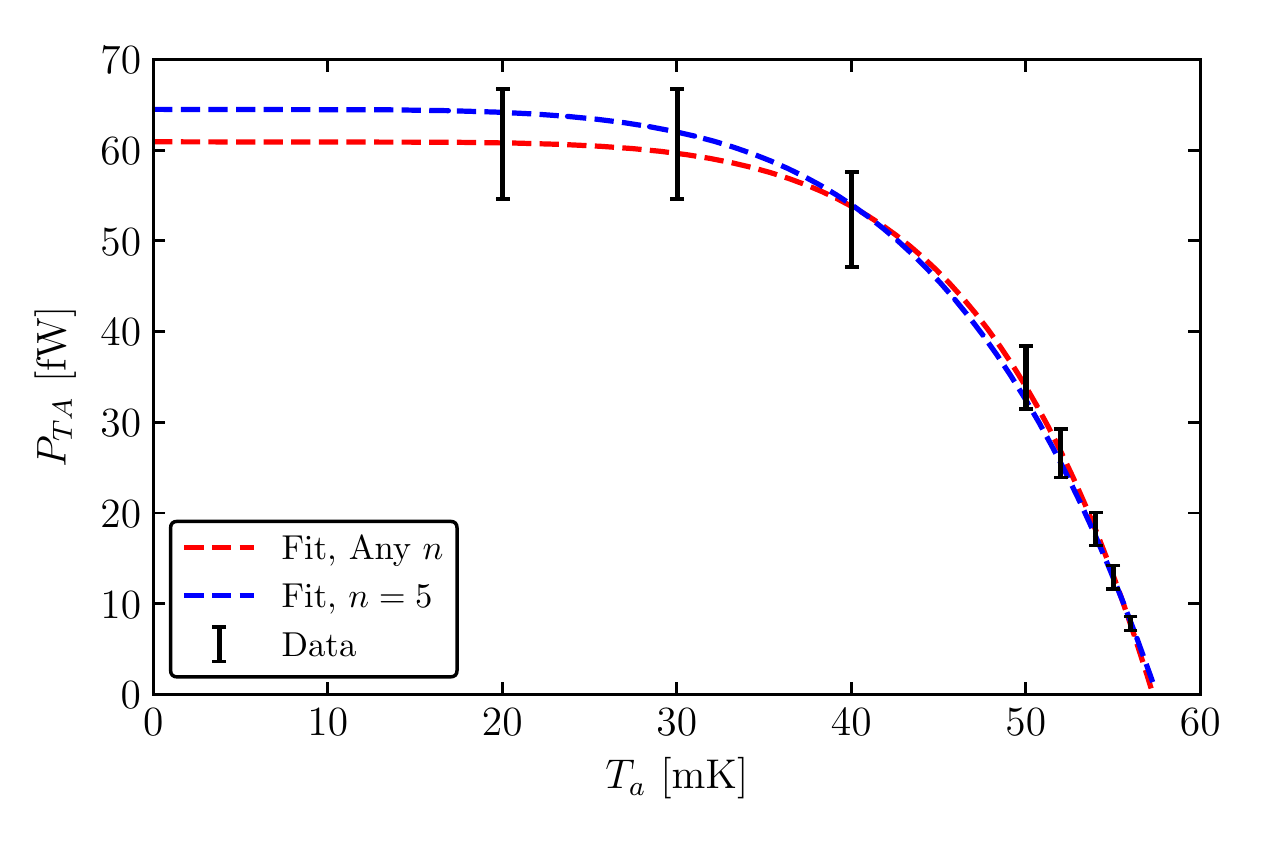}
    \end{subfigure}
    \caption{For the TESs on the chip thermalized through GE-7031 varnish: (Left) The power flowing from TES to absorber for TES100x25 as a function of absorber temperature. (Right) The power flowing from TES to absorber for TES200x50 as a function of absorber temperature. Both plots show the least squares fits with $n=5$ fixed and $n$ allowed to vary.}
    \label{fig:gta_glue_r6}
\end{figure}

\begin{table}
    \centering
    \caption{Values measured and calculated for TES100x25 on the GE-7031 varnished chip, showing that we are approximately at the same operating resistance as the bath temperature changes during the $G_{TA}$ measurement.}
    \begin{tabular}{rrrrr}
    \hline \hline
     $T_b$ [mK]     &  $T_a$ [mK]  & $I_\mathrm{bias}$ $[\mu \mathrm{A}]$ & $R_0$ $[\mathrm{m}\Omega]$ &  $P_{TA}$ [fW]  \\ \hline
      30  & 30  & 12.45  & 264    & 13.88  \\ 
      40  & 40 & 11.725 &  269   & 12.09   \\ 
      45  & 45 & 10.975 & 262    & 10.86  \\ 
      50  & 50 & 9.275  & 260    & 7.81    \\ 
      52  & 52 & 7.8     &  252   & 5.69  \\ 
      54  & 54 & 6       & 255    & 3.33   \\ 
      55  & 55  & 4.875  & 261    & 2.15   \\ 
      56  & 56 & 3.175  & 261  &  0.91  \\ \hline \hline
    \end{tabular}
    \label{tab:pta_gevarnish_tes100x25}
\end{table}

\begin{table}
    \centering
    \caption{Values measured and calculated for TES200x50 on the GE-7031 varnished chip, showing that we are approximately at the same operating resistance as the bath temperature changes during the $G_{TA}$ measurement.}
    \begin{tabular}{rrrrr}
    \hline \hline
     $T_b$ [mK]     &  $T_a$ [mK]  & $I_\mathrm{bias}$ $[\mu \mathrm{A}]$ & $R_0$ $[\mathrm{m}\Omega]$ &  $P_{TA}$ [fW]  \\ \hline
      20  & 30  & 27.35  & 293    & 60.68  \\ 
      30  & 40 & 27.35 &  293   & 60.68   \\ 
      40  & 45 & 25.4 & 293    & 52.33  \\ 
      50  & 50 & 20.75  & 293    & 34.93    \\ 
      52  & 52 & 18.075     &  292   & 26.59  \\ 
      54  & 54 & 15.125   & 299    & 18.20   \\ 
      55  & 55  & 12.45  & 285    & 12.91   \\ 
      56  & 56 & 9.75  & 290  &  7.79  \\ \hline \hline
    \end{tabular}
    \label{tab:pta_gevarnish_tes200x50}
\end{table}

\begin{table}
    \centering
    \caption{The corresponding fitted values from the $G_{TA}$ measurement for TES100x25 and TES200x50 on the GE-7031 varnished chip. Included are estimated electron-phonon coupling constants $\Sigma_{ep}$ for these W TESs.}
    \begin{tabular}{ccrrrr}
    \hline \hline
      \rule{0pt}{10pt} & Fit & $G_{TA}$ $\left[\frac{\mathrm{pW}}{\mathrm{K}}\right]$ & $T_c$ [$\mathrm{mK}$] & $n$ &  $\Sigma_{ep}$ $\left[\frac{\mathrm{GW}}{\mathrm{K}^n \mathrm{m}^3}\right]$  \\ \hline
  \multirow{2}{*}{TES100x25} & Any $n$ & $1.47\pm0.14$ & $56.63\pm0.10$ & $5.8 \pm 0.9$ & \multicolumn{1}{c}{---} \\
    & $n=5$ & $1.36\pm0.06$ & $56.69\pm0.08$  & 5 & $0.264\pm0.013$ \\ \hline 
   \multirow{2}{*}{TES200x50} & Any $n$ & $6.37\pm0.77$ & $57.29\pm0.24$  & $6.0\pm1.0$ & \multicolumn{1}{c}{---} \\
    & $n=5$ & $5.61\pm0.28$ & $57.51\pm0.0.17$  & 5 & $0.256\pm0.015$ \\ \hline \hline
    \end{tabular}
    \label{tab:gta_glue_r6}
\end{table}

For the $G_{TA}$ measurements with GE-7031 varnish, we used the quick measurement method with TES100x25 and TES200x50. The data from this measurement are reported in Tables~\ref{tab:pta_gevarnish_tes100x25} and \ref{tab:pta_gevarnish_tes200x50}, where the error in $P_{TA}$ is estimated to be 10\% of the measured power. Using Eq.~(\ref{eq:Gta}), we can fit the power law with $n$ varying and $n=5$ fixed, the latter as expected for a W TES. In Fig.~\ref{fig:gta_glue_r6}, we show the results of the fits as compared to the measured data, and report the various fit parameters in Table~\ref{tab:gta_glue_r6} for the two TESs. 

\subsection{\label{sec:albonds}Al Wire Bonds}

To study a configuration without glue, we ran a set of rectangles where the thermalization of the chip was achieved through the superconducting Al wire bonds that are used for electrical readout. In this scenario, we do not expect to be the thermal conductance to be from an interface, but instead dominated by the lattice wave (phonon) contribution. Because Al has a critical temperature $T_c = 1.2 \,\mathrm{K}$, and we are operating these TESs far below this value (tens of mK), the electrons have no contribution to the thermal transport due to being bound in Cooper pairs, and the thermal conductance is expected to be $G\propto T^3$ from the phonon contribution to the thermal transport (i.e. we expect $n=4$ for the $P_{AB}$ power law).

\subsubsection{$G_{AB}$ Measurement}

\begin{table}
    \centering
    \caption{Values used to calculate the change in power flowing from the absorber to bath via the superconducting Al wire bonds at different bath temperatures, where TES200x50 was the absorber heater.}
    \begin{tabular}{rrr}
    \hline \hline
     $T_b$ [mK]     &  $I_\mathrm{bias}$ $[\mu \mathrm{A}]$    & $P_{AB}$ [fW]  \\ \hline
     50            & 44.8                   & 113.7    \\
     49            &  53.7                  & 163.4         \\ 
     48            & 60.05                  & 204.4          \\ 
     45            & 75.25                  & 320.9          \\ 
     42            & 85                   & 409.5          \\ 
     38            & 94                   & 500.8           \\ 
     32            & 103                   & 601.2        \\ 
     26            & 107.5                   & 654.9        \\ 
     20            & 110.5                   & 692.0        \\ 
     15            & 114                  & 736.5          \\ \hline \hline
    \end{tabular}
    \label{tab:pab_data_al}
\end{table}

\begin{figure}
    \centering
    \includegraphics{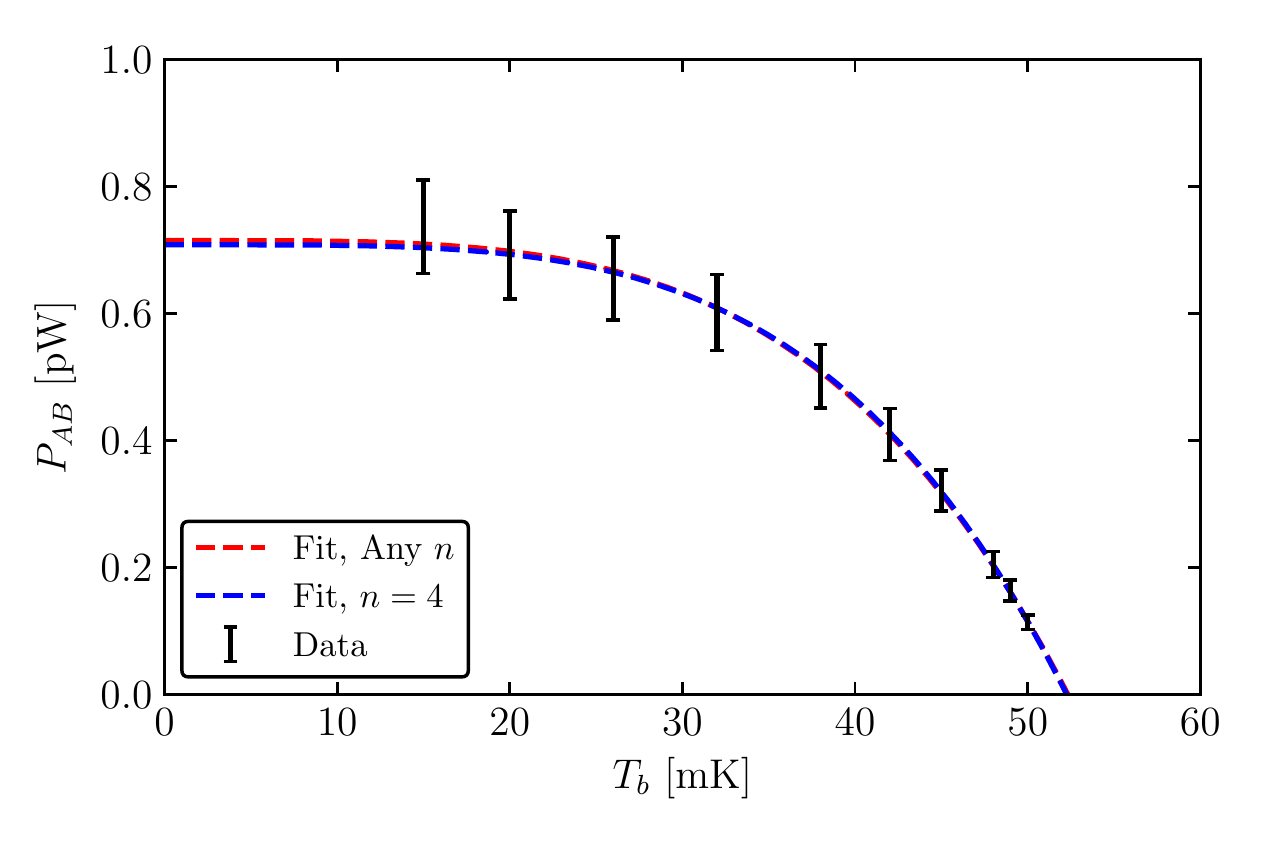}
    \caption{The power flowing from absorber to bath through the superconducting Al wire bonds at various bath temperatures taken for TES200x50. Also shown are the least squares fits with $n$ varying and $n=4$ fixed.}
    \label{fig:gab_unglue_r6}
\end{figure}

\begin{table}
    \centering
    \caption{Values from the least-squares fits to the power flowing from absorber to bath through the superconducting Al wire bonds. Included in the table are the estimated heat capacity via the low-temperature Debye model using Eq.~(\ref{eq:Cv}), as well as the related thermal time constant $\tau$.}
    \begin{tabular}{crrrrr}
    \hline \hline
    \rule{0pt}{10pt} Fit & $G_{AB}$ $\left[\frac{\mathrm{pW}}{\mathrm{K}}\right]$ & $T_a$ [$\mathrm{mK}$] & $n$ & $C^{th}$ $\left[\frac{\mathrm{pJ}}{\mathrm{K}}\right]$ & $\tau$ $[\mathrm{ms}]$ \\ \hline
    Any $n$ & $53.0\pm7.4$ & $52.32\pm0.45$  & $3.9\pm0.7$ & $8.60 \pm 0.22$ & $162\pm 26$ \\
    $n=4$ & $54.2\pm2.4$ & $52.26\pm0.25$  & 4 & $8.57 \pm 0.12$ & $156 \pm 9$  \\ \hline \hline
    \end{tabular}
    \label{tab:gab_unglue_r6}
\end{table}

To measure $G_{AB}$, we use TES200x50 as the heater and TES100x25 as the thermometer. In addition to the $5\, \mathrm{m}\Omega$ shunt resistor, the normal resistance of TES200x50 in this measurement is $426 \, \mathrm{m}\Omega$, with a parasitic resistance of $2.5\, \mathrm{m}\Omega$. The data from this measurement are reported in Table~\ref{tab:pab_data_al}, where error in $P_{AB}$ is set to be 10\% of the measured power, based on an estimated 10\% uncertainty in $R_{sh}$. We then fit Eq.~(\ref{eq:Gab}) with $n$ allowed to be any value, and $n=4$ fixed as expected from the phonon contribution to the thermal conductance. In Fig.~\ref{fig:gab_unglue_r6}, we show the results of these two fits and compare to the measured data, with the various fit parameters reported in Table~\ref{tab:gab_unglue_r6}. 

\subsubsection{$G_{TA}$ Measurement}

\begin{figure}
    \begin{subfigure}{.5\textwidth}
        \centering
        \includegraphics[width=1\linewidth]{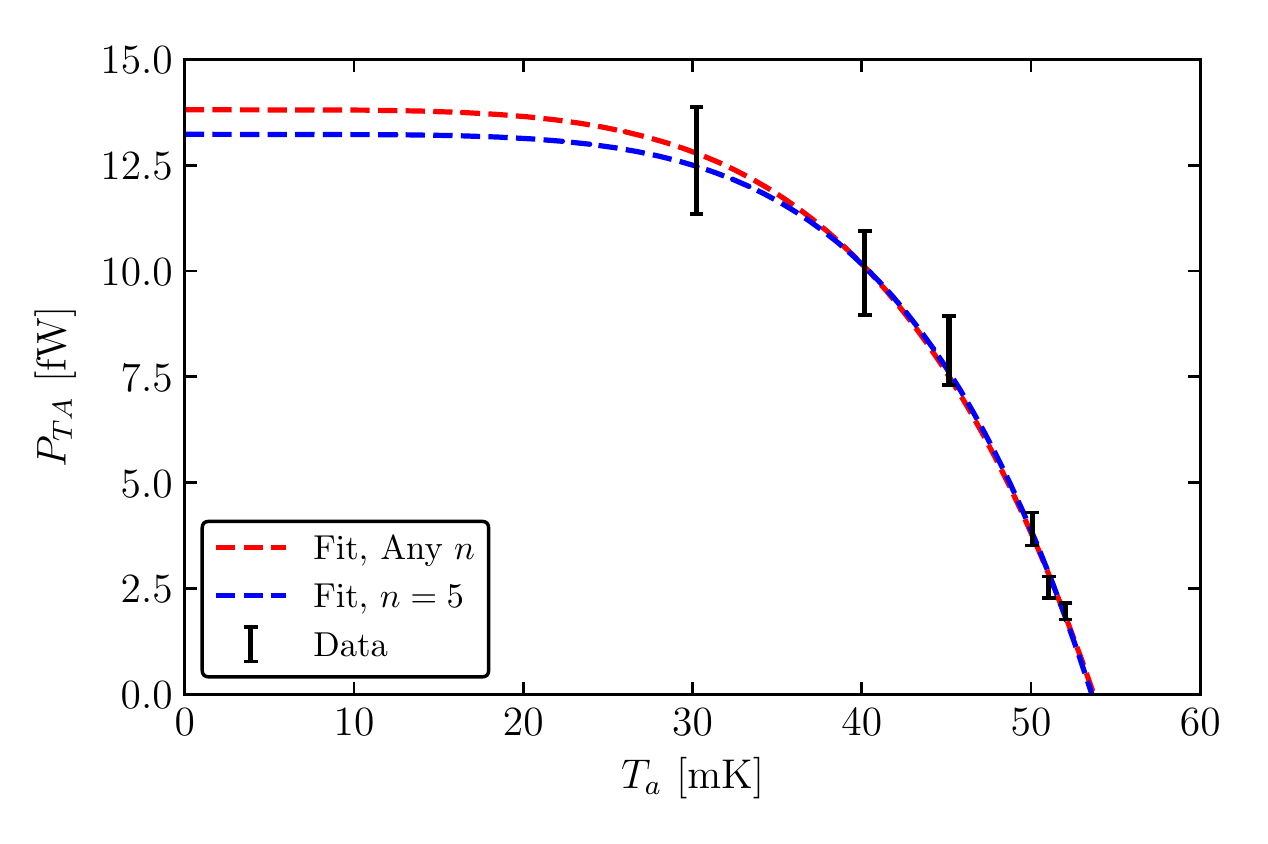}
    \end{subfigure}%
    \begin{subfigure}{.5\textwidth}
        \centering
        \includegraphics[width=1\linewidth]{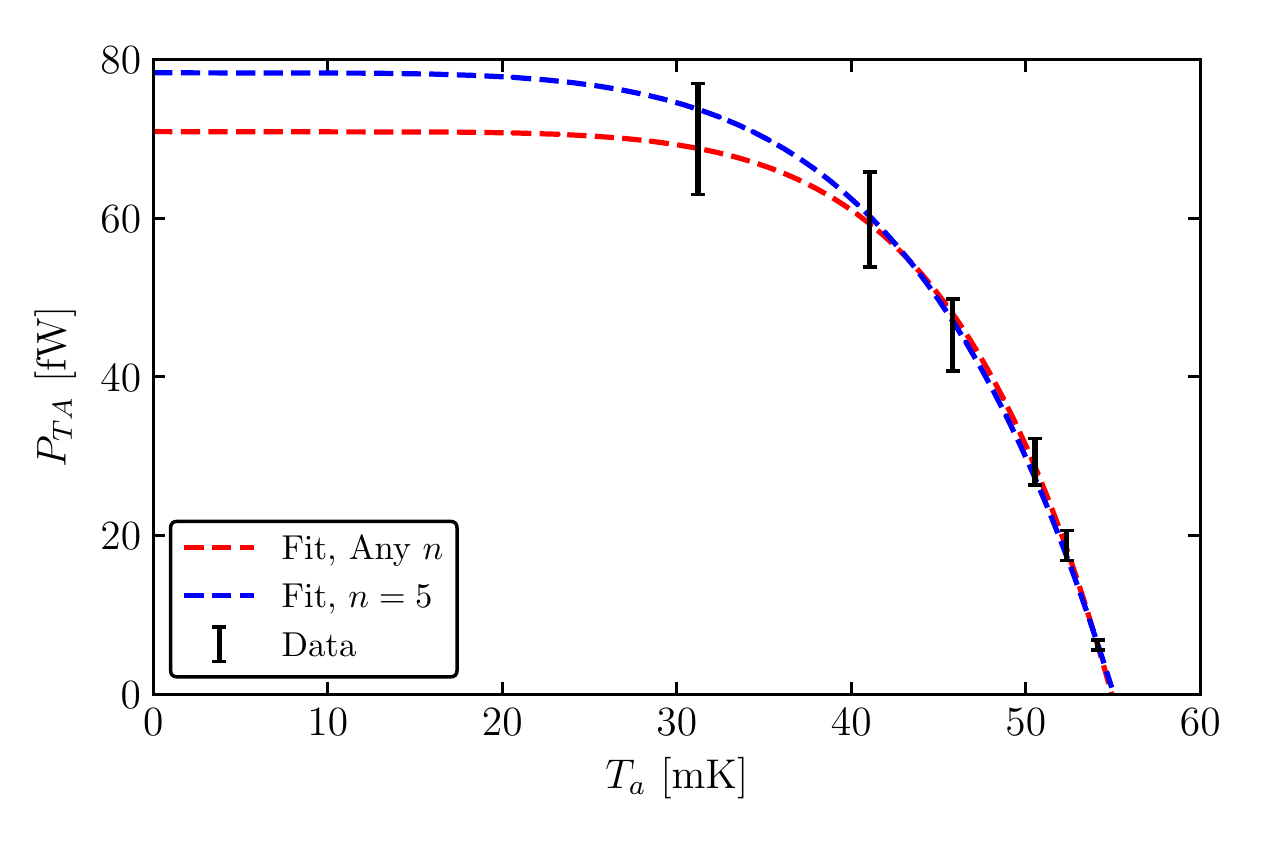}
    \end{subfigure}
    \caption{For the TESs on the chip thermalized through the superconducting Al wire bonds: (Left) The power flowing from TES to absorber for TES100x25 as a function of absorber temperature. (Right) The power flowing from TES to absorber for TES200x50 as a function of absorber temperature. Both plots show the least squares fits with $n=5$ fixed and $n$ allowed to vary.}
    \label{fig:gta_unglue_r6}
\end{figure}

\begin{table}
    \centering
    \caption{Values measured and calculated for TES100x25, showing that we are approximately at the same operating resistance as the bath temperature changes.}
    \begin{tabular}{rrrrr}
    \hline \hline
     $T_b$ [mK]     &  $T_a$ [mK]  & $I_\mathrm{bias}$ $[\mu \mathrm{A}]$ & $R_0$ $[\mathrm{m}\Omega]$ &  $P_{TA}$ [fW]  \\ \hline
      30  & 30.2  & 11.975  & 269    & 12.61  \\ 
      40  & 40.2 & 10.75 &  275   & 9.96   \\ 
      45  & 45.1 & 9.625 & 270    & 8.12  \\ 
      50  & 50.0 & 6.6  & 264    & 3.90    \\ 
      51  & 52.0 & 5.375     &  270   & 2.53  \\ 
      52  & 56.0 & 4.475  & 240  &  1.96  \\ \hline \hline
    \end{tabular}
    \label{tab:pta_al_tes100x25}
\end{table}

\begin{table}
    \centering
    \caption{Values measured and calculated for TES100x25, showing that we are approximately at the same operating resistance as the bath temperature changes.}
    \begin{tabular}{rrrrr}
    \hline \hline
     $T_b$ [mK]     &  $T_a$ [mK]  & $I_\mathrm{bias}$ $[\mu \mathrm{A}]$ & $R_0$ $[\mathrm{m}\Omega]$ &  $P_{TA}$ [fW]  \\ \hline
      30  & 30.2  & 27.35  & 252    & 12.61  \\ 
      40  & 40.2 & 25.15 &  249   & 9.96   \\ 
      45  & 45.1 & 22 & 252    & 8.12  \\ 
      50  & 50.0 & 17.825  & 256    & 3.90    \\ 
      52  & 52.0 & 14.4     &  261   & 2.53  \\ 
      54  & 56.0 & 8.075  & 247  &  1.96  \\ \hline \hline
    \end{tabular}
    \label{tab:pta_al_tes200x50}
\end{table}

\begin{table}
    \centering
    \caption{The corresponding fitted values from the $G_{TA}$ measurement for TES100x25 and TES200x50 on the chip thermalizing through the Al wire bonds. Included are estimated electron-phonon coupling constants $\Sigma_{ep}$ for these W TESs.}
    \begin{tabular}{ccrrrr}
        \hline \hline
           \rule{0pt}{10pt} & Fit & $G_{TA}$ $\left[\frac{\mathrm{pW}}{\mathrm{K}}\right]$ & $T_c$ [$\mathrm{mK}$] & $n$ &  $\Sigma_{ep}$ $\left[\frac{\mathrm{GW}}{\mathrm{K}^n \mathrm{m}^3}\right]$ \\ \hline
       \multirow{2}{*}{TES100x25} & Any $n$ & $1.16\pm0.21$ & $53.71\pm0.40$  & $4.5\pm1.3$  & \multicolumn{1}{c}{---} \\
          & $n=5$ & $1.23\pm0.08$ & $53.59\pm0.21$  & 5 &  $0.299\pm0.022$ \\ \hline
        \multirow{2}{*}{TES200x50} & Any $n$ & $8.03\pm1.00$ & $54.92\pm0.14$  & $6.2\pm1.3$ & \multicolumn{1}{c}{---} \\
          & $n=5$ & $7.11\pm0.37$ & $55.03\pm0.11$  & 5 & $0.387 \pm 0.022$ \\ \hline \hline
    \end{tabular}
    \label{tab:gta_unglue_r6}
\end{table}

For the $G_{TA}$ measurement, we use the quick measurement method for TES100x25 and TES200x50. The data from these measurements are reported in Tables~\ref{tab:pta_al_tes100x25} and \ref{tab:pta_al_tes200x50}, and the error in $P_{TA}$ is estimated to be 10\% of the measured power. Using Eq.~(\ref{eq:Gta}), we fit the power law with $n$ varying and $n=5$ fixed, the latter as expected for a W TES. In Fig.~\ref{fig:gta_unglue_r6}, we show the results of the two fits as compared to the measured data, and report the various fit parameters in Table~\ref{tab:gta_unglue_r6}.

\subsection{Rubber Cement}

As the GE-7031 varnish had problems with glued chips popping off after cooling, it was recommended by Aritoki Suzuki that we try rubber cement (Elmer's) as an alternative adhesive, as the Suzuki group had never experienced this in their lab. As with GE-7031 varnish and the cirlex clamps, rubber cement is dissimilar to the semiconductor substrate, and the thermal conductance at the interface will be due to the mismatch of the elastic constants~\cite{doi:10.1139/p59-037}, giving an expected thermal conductance temperature dependence of $G \propto T^3$.

\subsubsection{$G_{AB}$ Measurement}

\begin{table}
    \centering
    \caption{Values used to calculate the change in power flowing from the absorber to bath via rubber cement at different bath temperatures, where TES200x50 was the absorber heater.}
    \begin{tabular}{rrr}
    \hline \hline
     $T_b$ [mK]     &  $I_\mathrm{bias}$ $[\mu \mathrm{A}]$    & $P_{AB}$ [pW]  \\ \hline
     52            & 501.2                   & 14.2    \\
     51.5            &  734                  & 30.5         \\ 
     51            & 894                  &    45.2      \\ 
     50.5            & 1030               &    60.0          \\ 
     50            & 1140                   &  73.5          \\ 
     49            & 1310                   &  97.0           \\ 
     48            & 1440                   & 117.2        \\ 
     46            & 1640                   & 152.1        \\ 
     43            & 1830                   & 189.4        \\ 
     40            & 1960                  &  217.2    \\ \hline \hline
    \end{tabular}
    \label{tab:pab_data_rubber}
\end{table}

\begin{figure}
    \centering
    \includegraphics{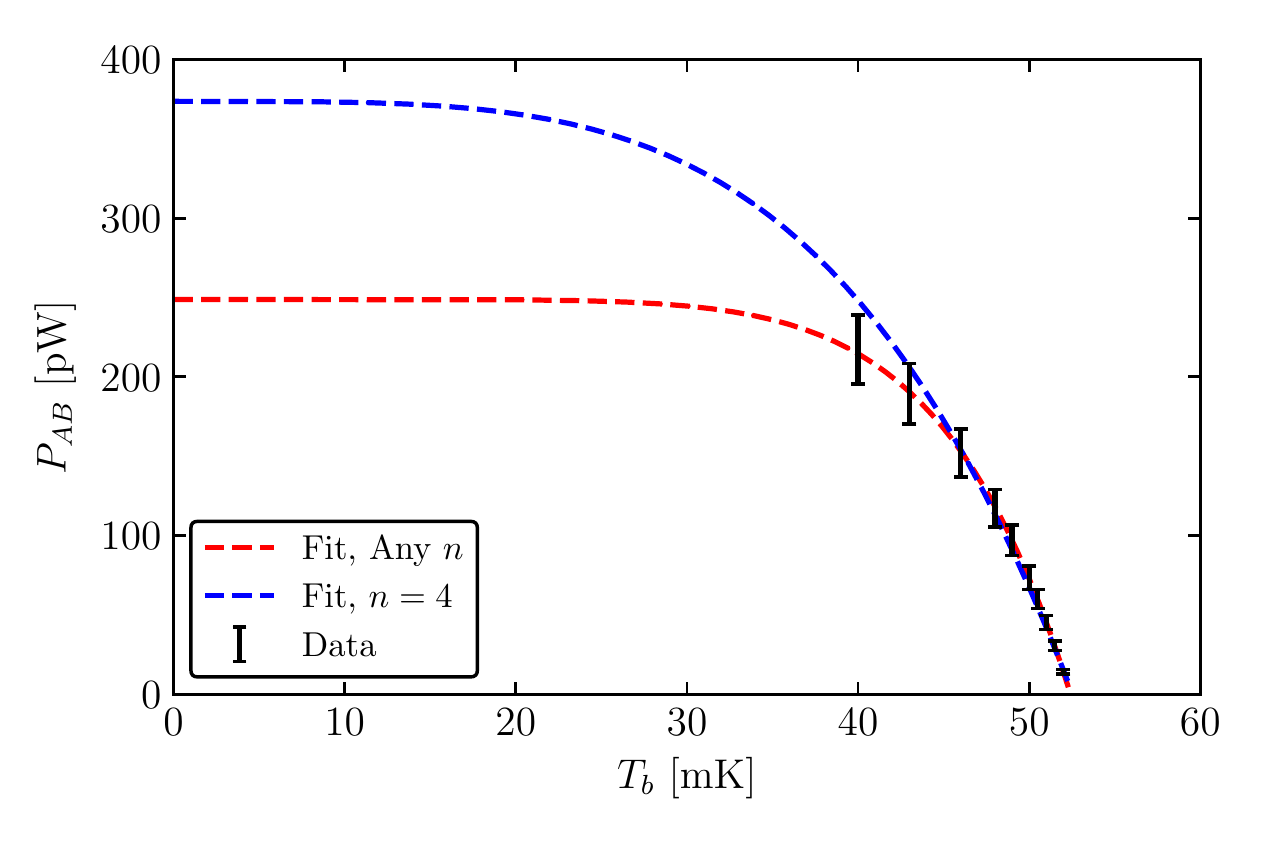}
    \caption{The power flowing from absorber to bath through the rubber cement at various bath temperatures taken for TES200x50. Also shown are the least squares fits with $n$ varying and $n=4$ fixed.}
    \label{fig:gab_rubber}
\end{figure}

\begin{table}
    \centering
    \caption{Values from the least-squares fits to the power flowing from absorber to bath through the rubber cement. Included in the table are the estimated heat capacity via the low-temperature Debye model using Eq.~(\ref{eq:Cv}), as well as the related thermal time constant $\tau$.}
    \begin{tabular}{crrrrr}
    \hline \hline
    \rule{0pt}{10pt} Fit & $G_{AB}$ $\left[\frac{\mathrm{nW}}{\mathrm{K}}\right]$ & $T_a$ [$\mathrm{mK}$] & $n$ & $C^{th}$ $\left[\frac{\mathrm{pJ}}{\mathrm{K}}\right]$ & $\tau$ $[\mathrm{ms}]$ \\ \hline
    Any $n$ & $34.8\pm52.4$ & $52.42\pm0.06$  & $7.3\pm1.5$ & $8.65\pm0.03$ & $0.248 \pm 0.022$  \\
    $n=4$ & $28.4\pm1.2$ & $52.55\pm0.06$  & 4  & $8.72\pm0.03$ & $0.306\pm0.014$ \\ \hline \hline
    \end{tabular}
    \label{tab:gab_rubber}
\end{table}

To measure $G_{AB}$, we use TES200x50 as the heater and TES100x25 as the thermometer. In addition to the $5\, \mathrm{m}\Omega$ shunt resistor, the normal resistance of TES200x50 in this measurement is $423 \, \mathrm{m}\Omega$, with a parasitic resistance of $4.3\, \mathrm{m}\Omega$. The data from this measurement are reported in Table~\ref{tab:pab_data_rubber}. The error in $P_{AB}$ is set to be 10\% of the measured power, based on an estimated 10\% uncertainty in $R_{sh}$. We then fit Eq.~(\ref{eq:Gab}) with $n$ allowed to be any value, and $n=4$ fixed as expected from the phonon mismatch between the absorber and the rubber cement. In Fig.~\ref{fig:gab_rubber}, we show the results of these two fits and compare to the measured data, with the various fit parameters reported in Table~\ref{tab:gab_rubber}.

\subsection{\label{sec:cleangta}Thermal Conductance with an IV Sweep}

For each of the previous sections, we showed examples of the quick $G_{TA}$ measurement method, which is fast, but could be subject to systematic errors as noted in the explanation of the method. In this section, we show examples of clean $G_{TA}$ measurements via IV sweeps at various bath temperatures for TES800x200, TES400x100, and TES200x50.

\subsubsection{$G_{TA}$ Measurement}

\begin{figure}
    \centering
    \begin{subfigure}{.45\textwidth}
        \centering
        \includegraphics[width=1\linewidth]{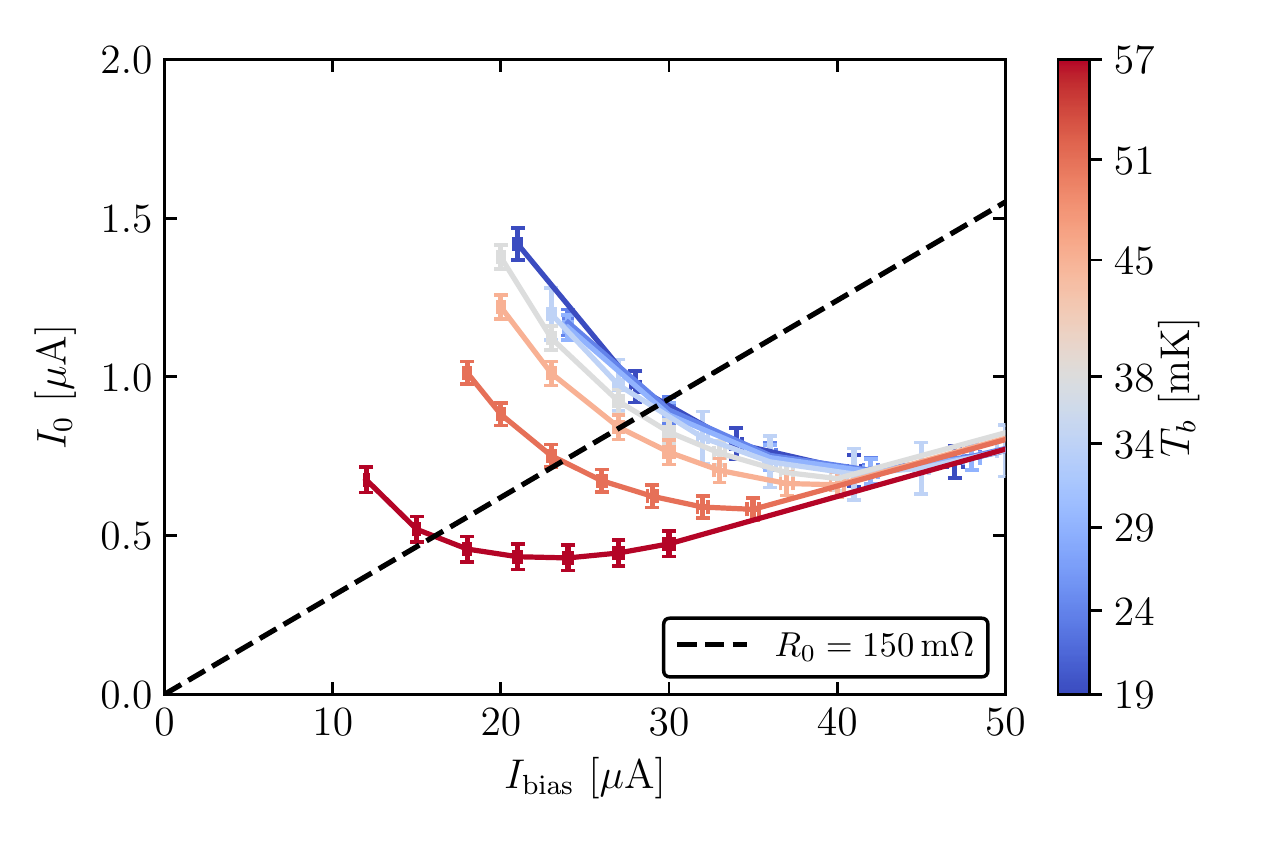}
    \end{subfigure}%
    \begin{subfigure}{.45\textwidth}
        \centering
        \includegraphics[width=1\linewidth]{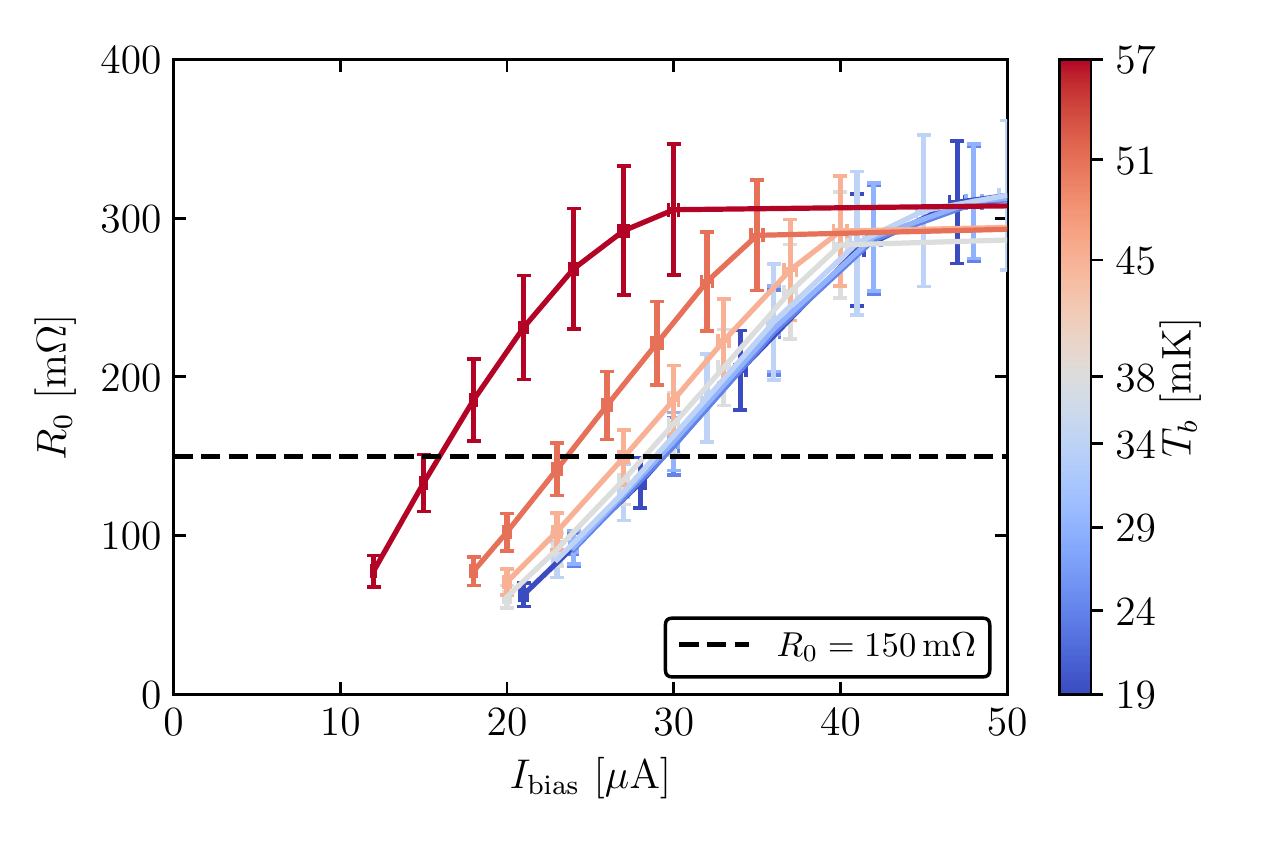}
    \end{subfigure} \\
    \begin{subfigure}{.45\textwidth}
        \centering
        \includegraphics[width=1\linewidth]{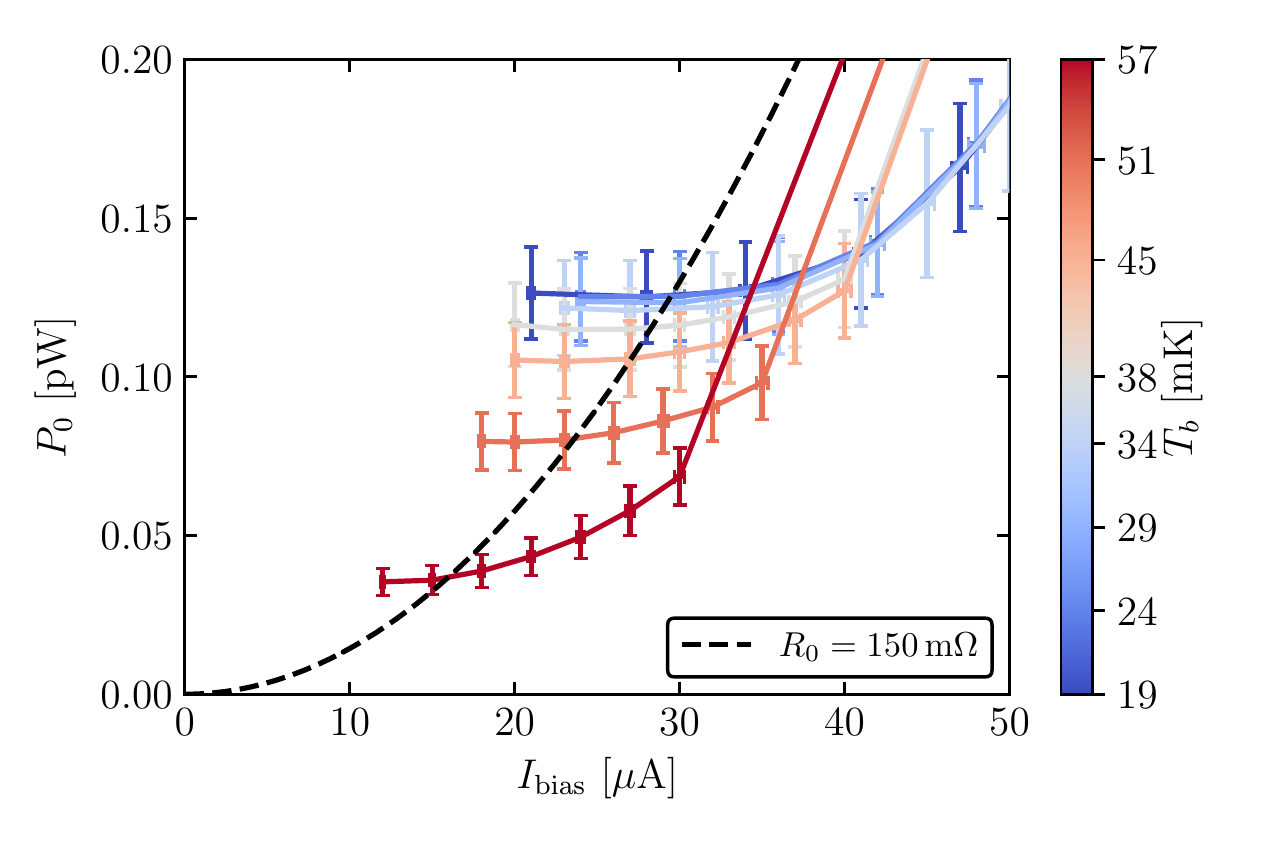}
    \end{subfigure}
    \begin{subfigure}{.45\textwidth}
        \centering
        \includegraphics[width=1\linewidth]{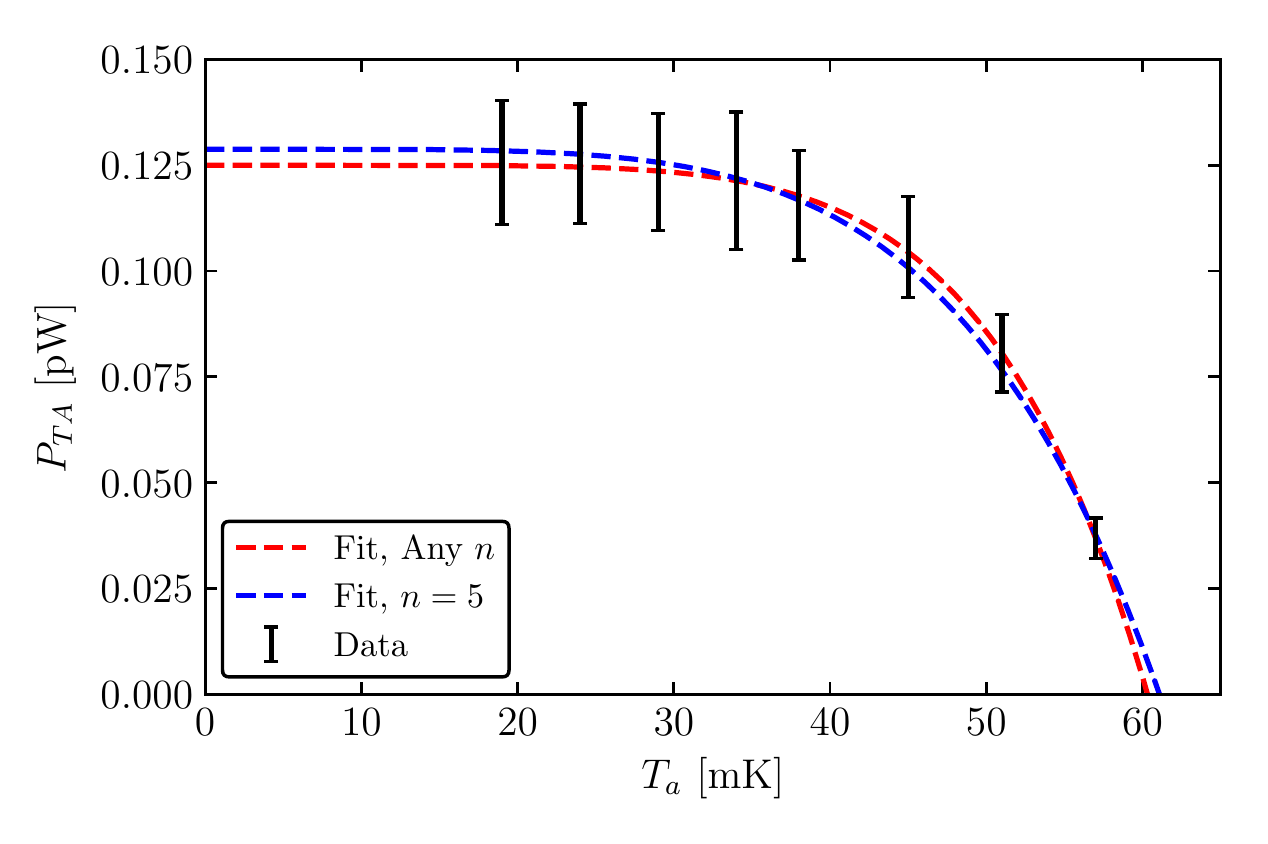}
    \end{subfigure}
    \caption{Example of the clean $G_{TA}$ measurement with TES200x50. (Top left) TES current as a function of bias current. (Top right) TES resistance as a function of bias current. (Bottom left) TES bias power as a function of bias current. For these three plots, the superconducting regions of the curves were not plotted for cosmetic reasons, and the lines connecting data points are simple linear interpolations. (Bottom right) Fit to $G_{TA}$ with $n$ varying and $n=5$ fixed.}
    \label{fig:cleangta_200x50}
\end{figure}

\begin{figure}
    \centering
    \begin{subfigure}{.45\textwidth}
        \centering
        \includegraphics[width=1\linewidth]{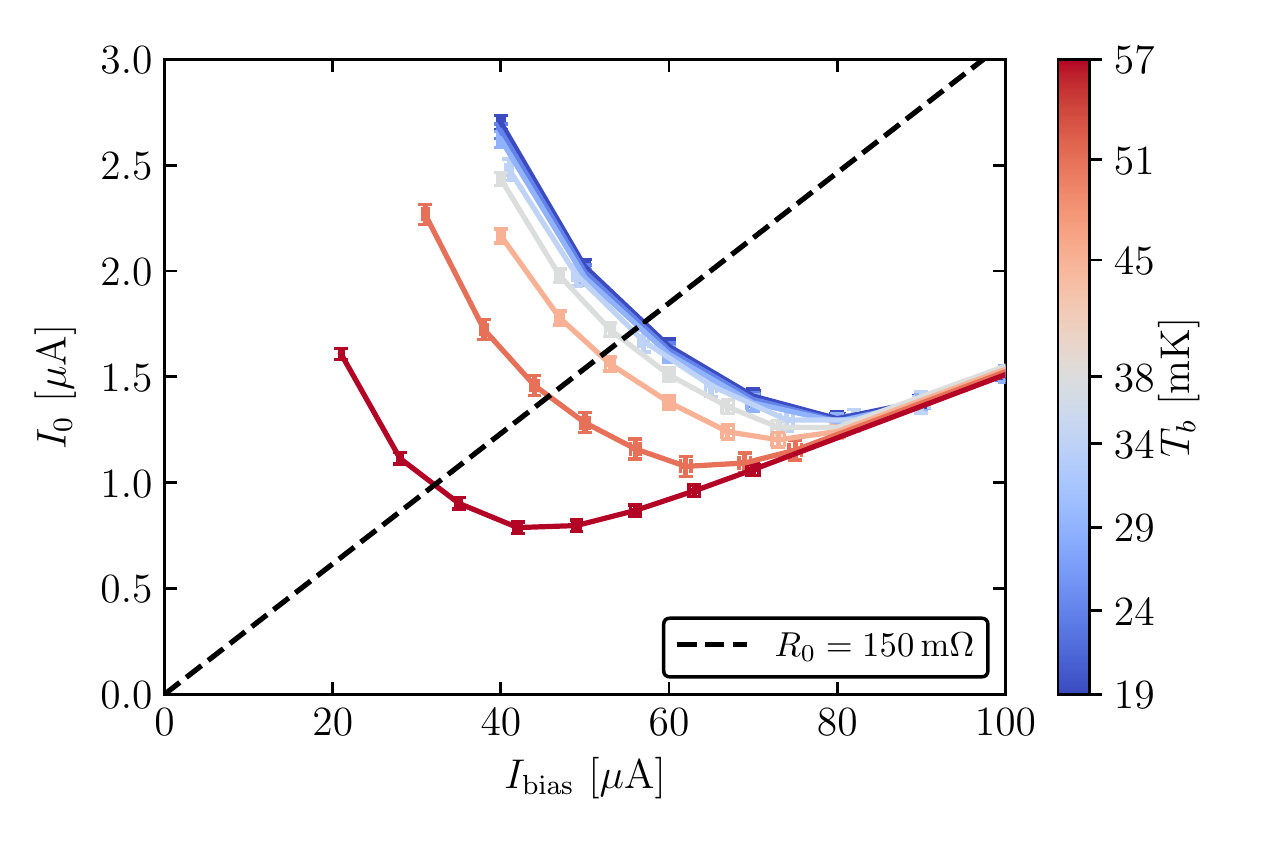}
    \end{subfigure}%
    \begin{subfigure}{.45\textwidth}
        \centering
        \includegraphics[width=1\linewidth]{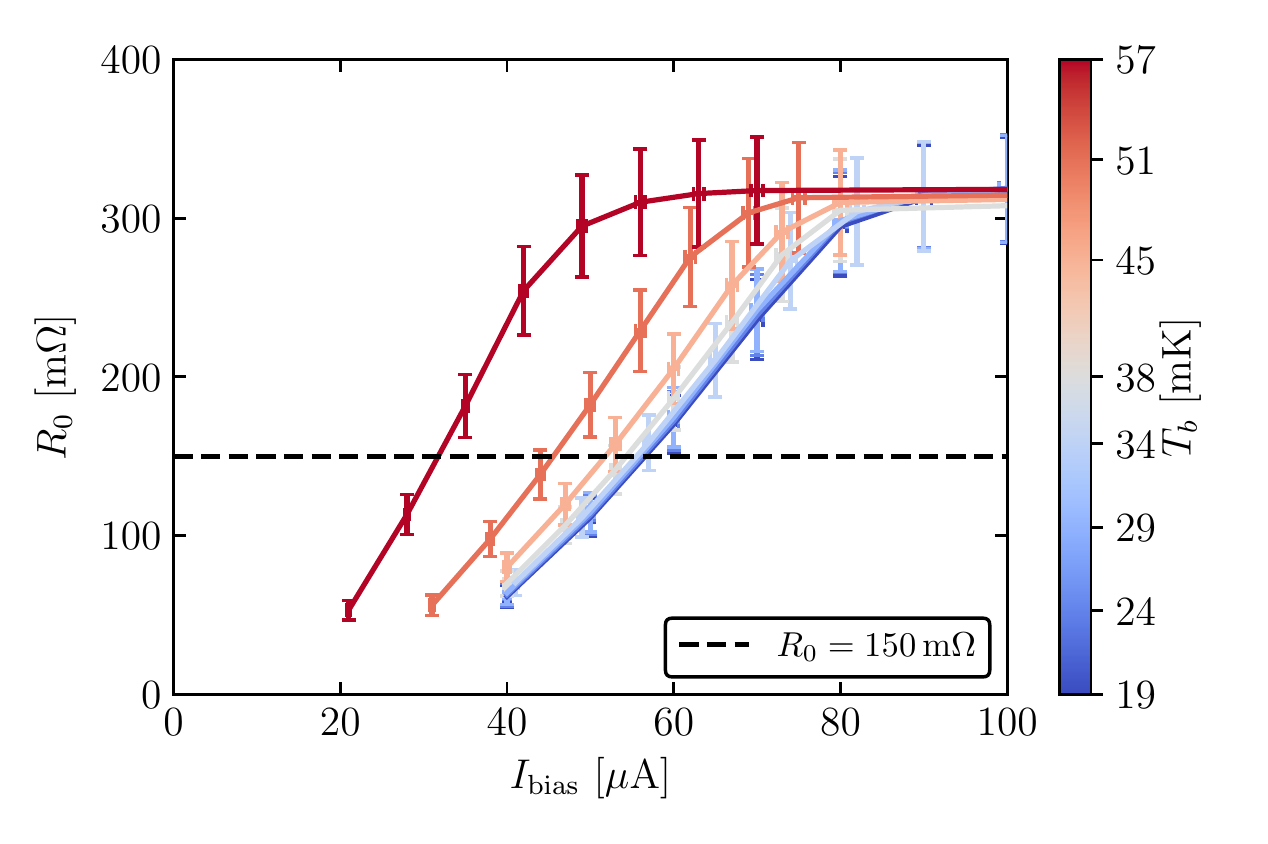}
    \end{subfigure} \\
    \begin{subfigure}{.45\textwidth}
        \centering
        \includegraphics[width=1\linewidth]{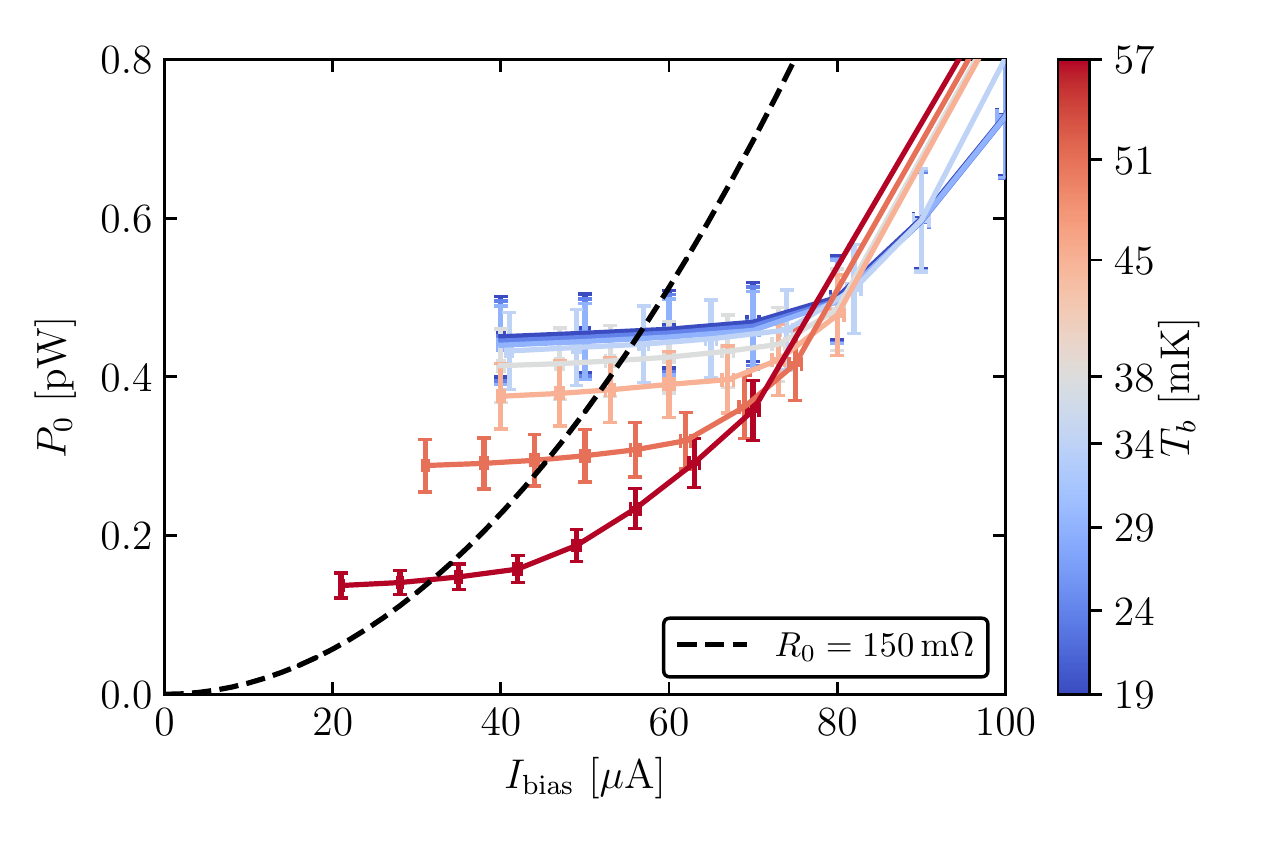}
    \end{subfigure}
    \begin{subfigure}{.45\textwidth}
        \centering
        \includegraphics[width=1\linewidth]{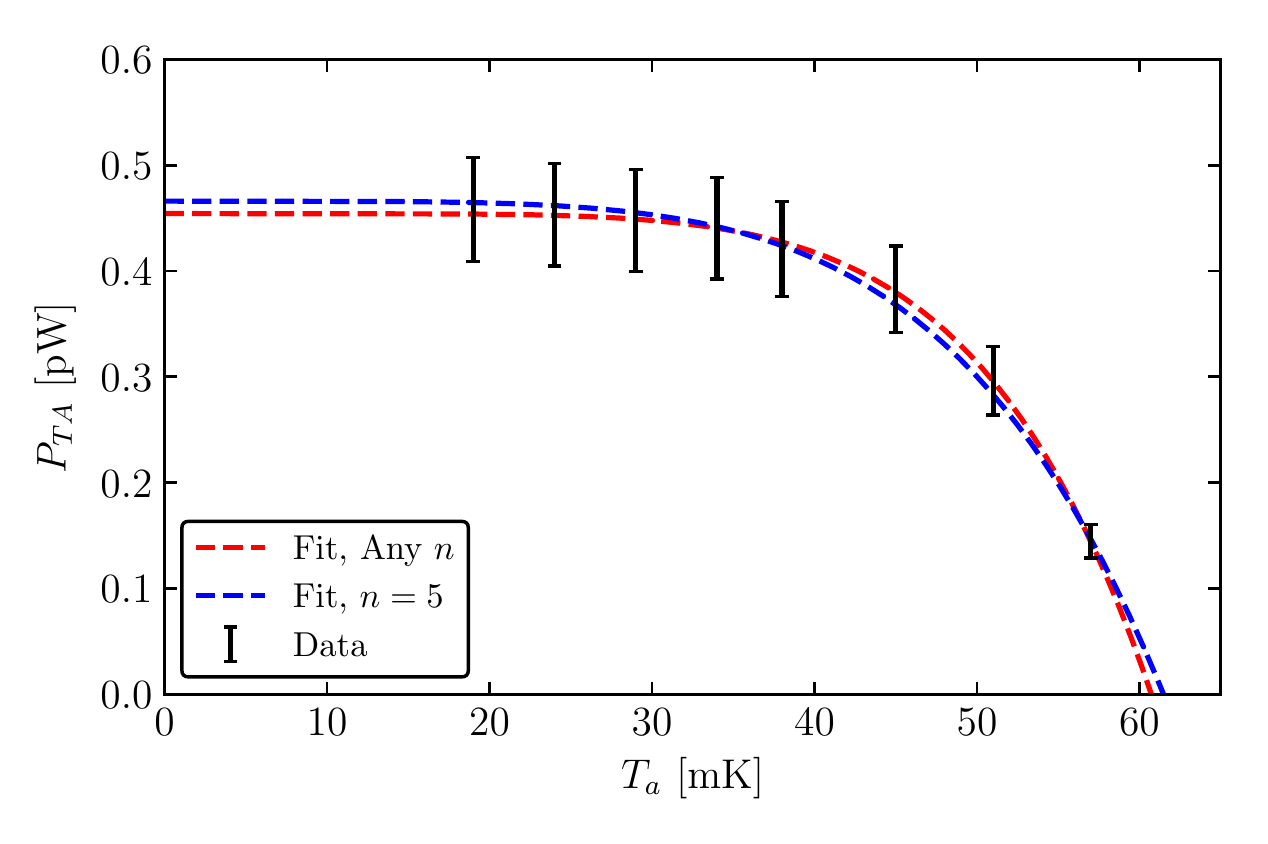}
    \end{subfigure}
    \caption{Example of the clean $G_{TA}$ measurement with TES400x100. (Top left) TES current as a function of bias current. (Top right) TES resistance as a function of bias current. (Bottom left) TES bias power as a function of bias current. For these three plots, the superconducting regions of the curves were not plotted for cosmetic reasons, and the lines connecting data points are simple linear interpolations. (Bottom right) Fit to $G_{TA}$ with $n$ varying and $n=5$ fixed.}
    \label{fig:cleangta_400x100}
\end{figure}

\begin{figure}
    \centering
    \begin{subfigure}{.45\textwidth}
        \centering
        \includegraphics[width=1\linewidth]{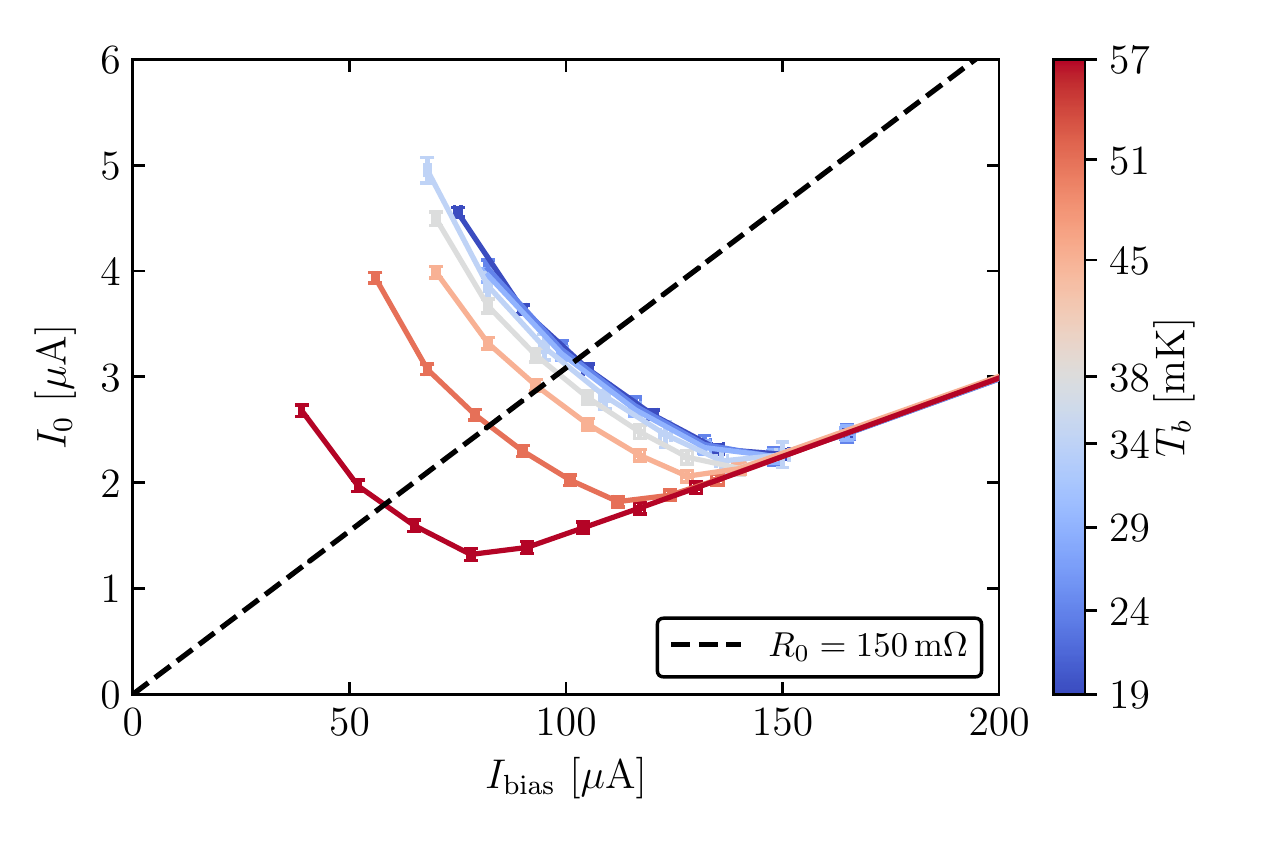}
    \end{subfigure}%
    \begin{subfigure}{.45\textwidth}
        \centering
        \includegraphics[width=1\linewidth]{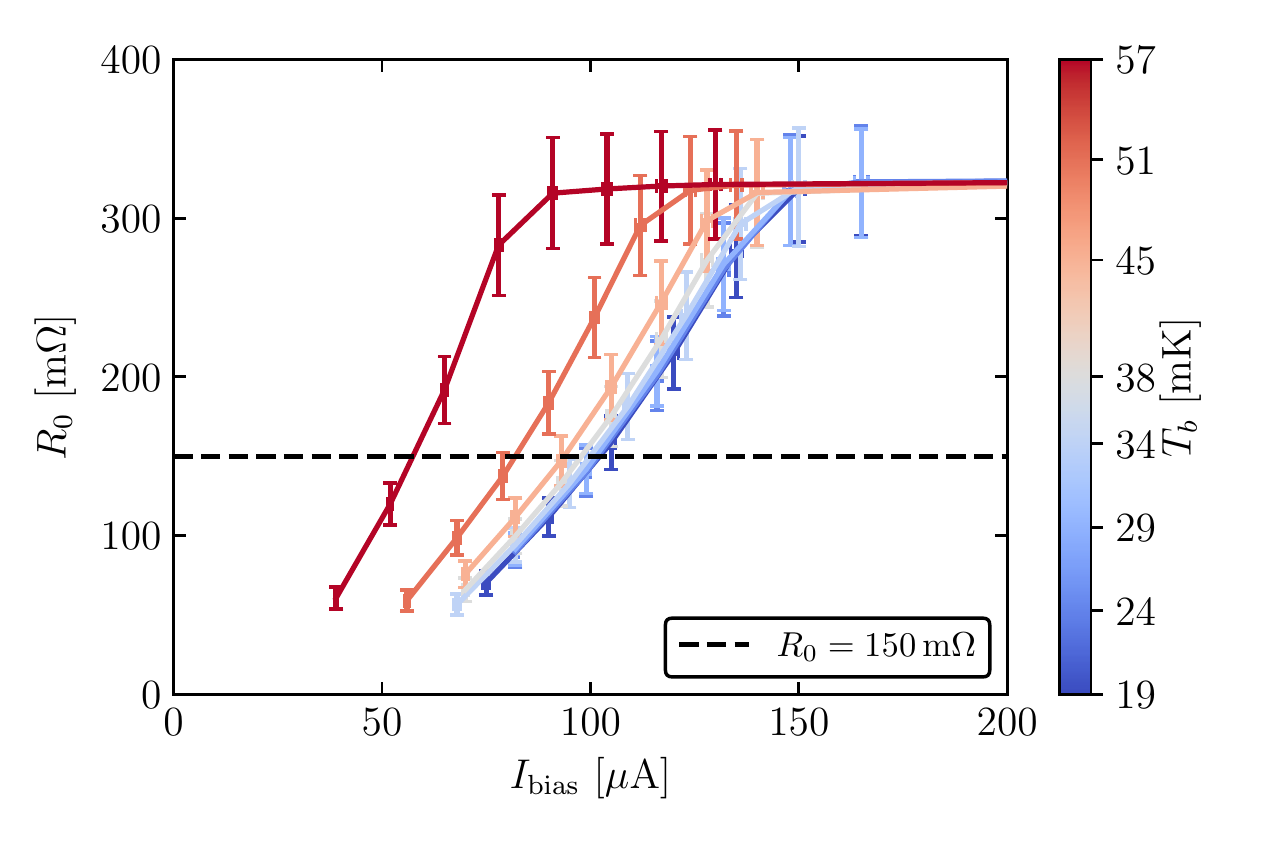}
    \end{subfigure} \\
    \begin{subfigure}{.45\textwidth}
        \centering
        \includegraphics[width=1\linewidth]{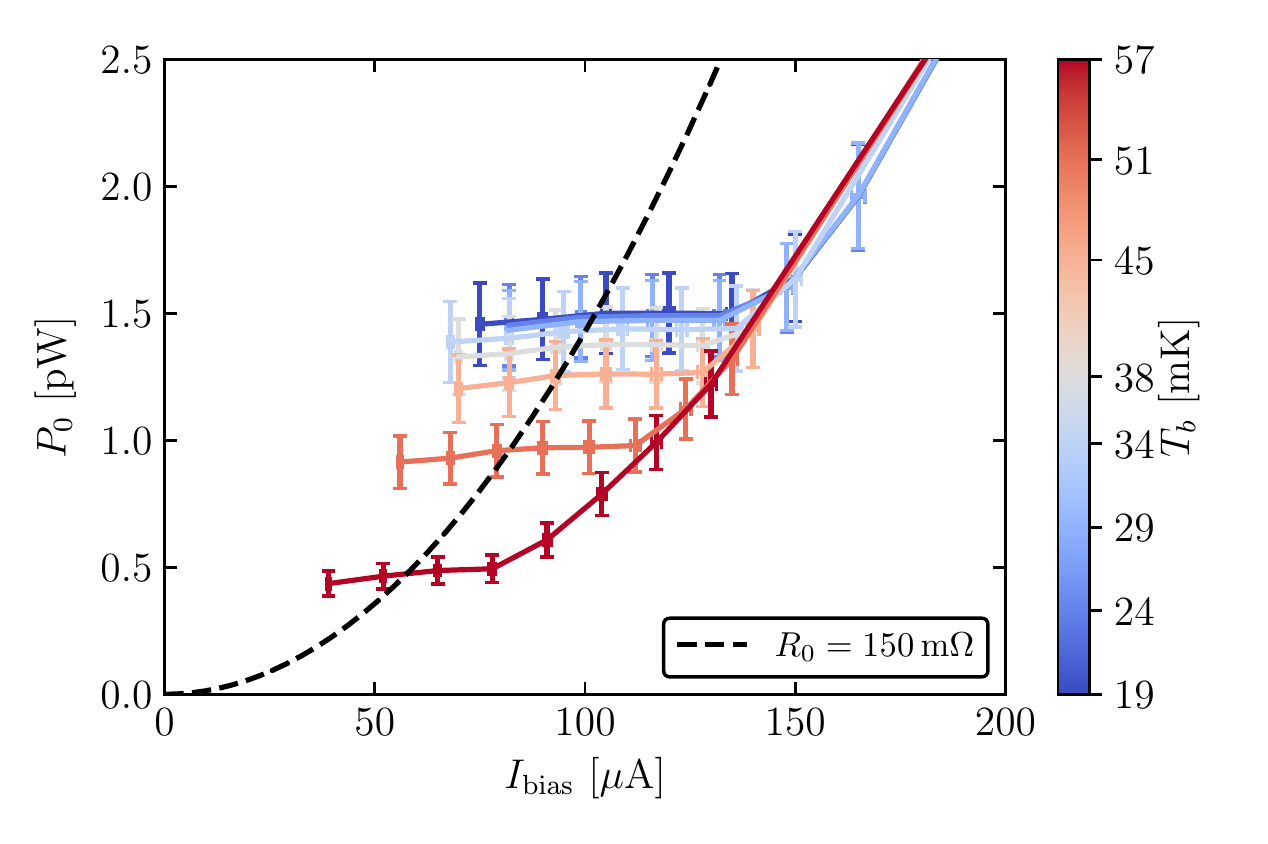}
    \end{subfigure}
    \begin{subfigure}{.45\textwidth}
        \centering
        \includegraphics[width=1\linewidth]{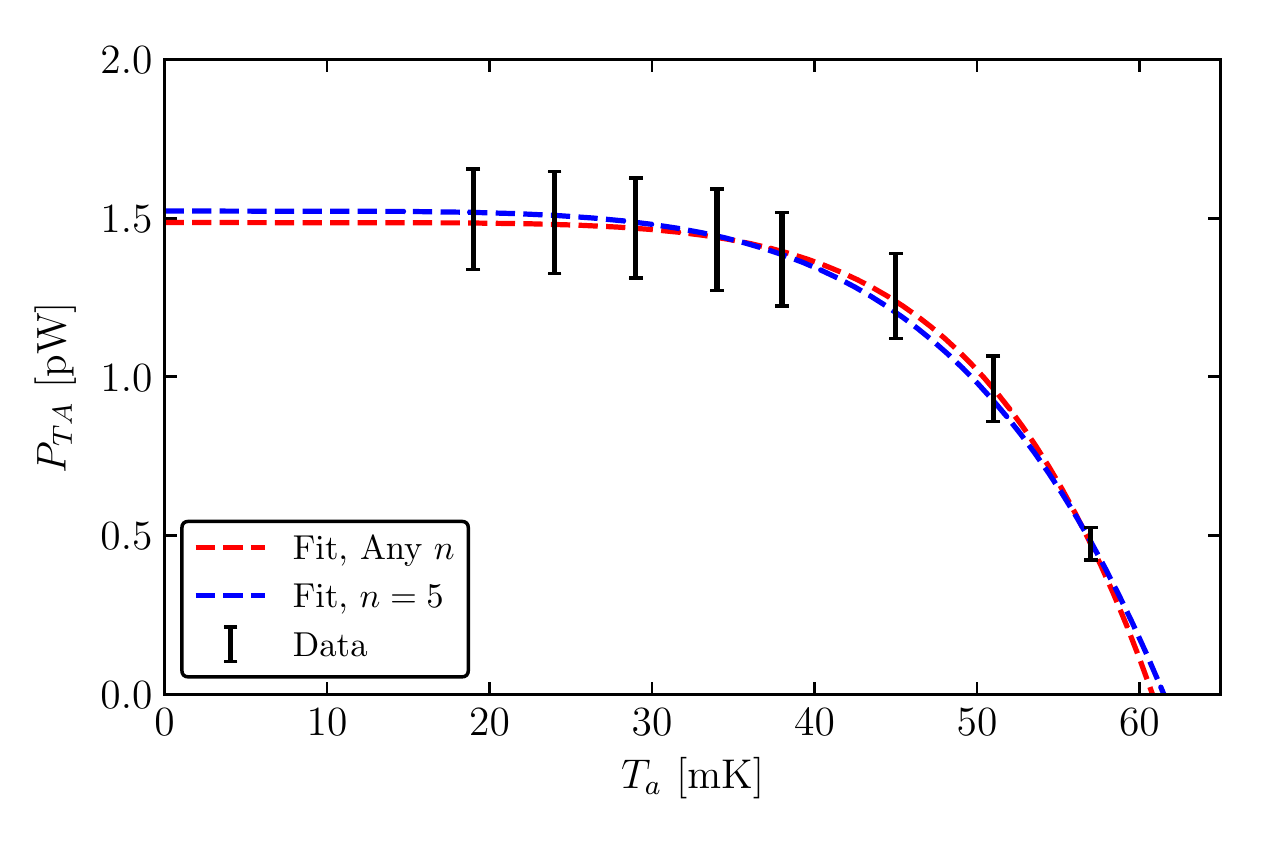}
    \end{subfigure}
    \caption{Example of the clean $G_{TA}$ measurement with TES800x200. (Top left) TES current as a function of bias current. (Top right) TES resistance as a function of bias current. (Bottom left) TES bias power as a function of bias current. For these three plots, the superconducting regions of the curves were not plotted for cosmetic reasons, and the lines connecting data points are simple linear interpolations. (Bottom right) Fit to $G_{TA}$ with $n$ varying and $n=5$ fixed.}
    \label{fig:cleangta_800x200}
\end{figure}

\begin{table}
    \centering
    \caption{The corresponding fitted values from the $G_{TA}$ measurement for TES200x50, TES400x100, and TES800x200 on the GE-7031 varnished chip used as an example of a clean $G_{TA}$ measurement. Included are estimated electron-phonon coupling constants $\Sigma_{ep}$ for these W TESs.}
    \begin{tabular}{ccrrrr}
    \hline \hline
    \rule{0pt}{10pt} & Fit & $G_{TA}$ $\left[\frac{\mathrm{pW}}{\mathrm{K}}\right]$ & $T_c$ [$\mathrm{mK}$] & $n$ &  $\Sigma_{ep}$ $\left[\frac{\mathrm{GW}}{\mathrm{K}^n \mathrm{m}^3}\right]$ \\ \hline
      \multirow{2}{*}{TES200x50} & Any $n$ & $12.8\pm3.7$ & $60.3\pm1.2$  & $6.2\pm1.9$ & \multicolumn{1}{c}{---} \\
    & $n=5$ & $10.5\pm0.6$ & $61.1\pm0.7$  & 5 & $0.379 \pm 0.034$ \\ \hline 
    \multirow{2}{*}{TES400x100} & Any $n$ & $45.1\pm13.1$ & $60.7\pm1.2$  & $6.0\pm1.9$ & \multicolumn{1}{c}{---} \\
    & $n=5$ & $37.9\pm1.9$ & $61.5\pm0.7$  & 5 & $0.331 \pm 0.028$ \\ \hline 

   \multirow{2}{*}{TES800x200} & Any $n$ & $145\pm42$ & $60.8\pm1.3$  & $5.9\pm1.8$ & \multicolumn{1}{c}{---} \\
    & $n=5$ & $123.8\pm6.2$ & $61.5\pm0.7$  & 5 & $0.270 \pm 0.023$ \\ \hline \hline
    \end{tabular}
    \label{tab:cleangta}
\end{table}

In Figs.~\ref{fig:cleangta_200x50}--\ref{fig:cleangta_800x200}, we show the $IV$, $RV$, and $PV$ curves for each of the TESs, as well as the $G_{TA}$ fit. Note that we chose an operating resistance of $R_0=150\,\mathrm{m}\Omega$ and interpolated between data points to ensure estimate the bias power of the TES at this value. We fit Eq.~(\ref{eq:Gta}) with $n$ varying and $n=5$ fixed, and report the various fit parameters in Table~\ref{tab:cleangta}. The power law curves in these measurements follow $n=5$ much more convincingly in these fits, due to the lower amount of systematics as opposed to the quick measurement.

\section{Thermal Conductance Measurements with a SPICE MELANGE Detector}

\begin{figure}
    \centering
    \includegraphics[width=\linewidth]{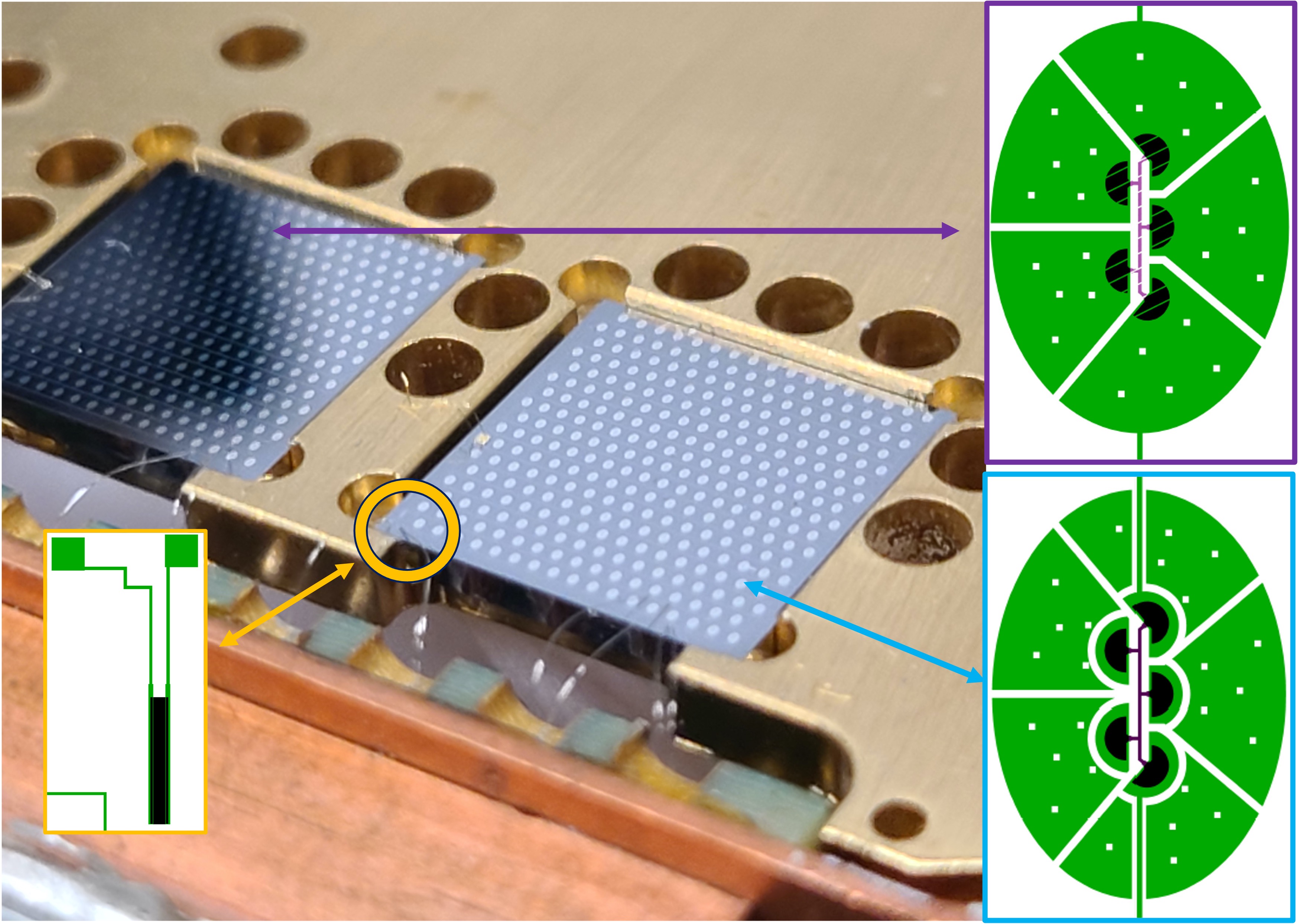}
    \caption{(Figure from Ref.~\cite{finkthesis}) The right $1\, \mathrm{cm}^2 \times 1\, \mathrm{mm}$ chip was used to measure the thermal conductance from absorber to bath for a Au wire bonding pad, which is a 4\% coverage design with ``disconnected`` Al fins (cyan inset). The TES295x40 on the chip is circled in yellow. These QET-based detectors were designed by C. W. Fink and are studied in detail in his thesis~\cite{finkthesis}.}
    \label{fig:run15_melange}
\end{figure}

In Fig.~\ref{fig:run15_melange}, we show the detector setup from one of our runs in the dilution refrigerator at UC Berkeley, where we will use the 4\% detector with disconnected fins to measure the thermal conductance from absorber to bath of the gold pad. Both of these chips are resting on the gold-plated Cu plate, and each has a TES rectangle of dimensions $295 \, \mu \mathrm{m} \times 40 \, \mu\mathrm{m} \times 40 \, \mathrm{nm}$, henceforth referred to as TES295x40. Because the TES rectangle on the connected fin device had a short, we used the disconnected fin device instead.

\subsection{\label{sec:aupad}Au Pad}

On the Si substrate of the detector, a gold pad was deposited with dimensions $300 \, \mu\mathrm{m} \times 300 \, \mu\mathrm{m} \times 750 \, \mathrm{nm}$ for the purpose of thermalization through a gold wire. The sizing of the pad was decided by C. W. Fink, and we will use it as a data point for the thermal conductance of a thin gold film. As noted in Refs.~\cite{maasilta_gold_g, PhysRevB.72.012302}, the power law exponent for a thin gold film can be expected to be either $n=5$ for an ordered metal~\cite{Gantmakher_1974, PhysRevB.49.5942, PhysRevLett.73.2123}, or $n=6$ for a disordered metal~\cite{impure_metal, PhysRevB.61.6041}.

\subsubsection{$G_{AB}$ Measurement}

\begin{table}
    \centering
    \caption{Values used to calculate the change in power flowing from the absorber to bath via the Au wire bond pad at different bath temperatures, where TES295x40 was the absorber heater.}
    \begin{tabular}{rrr}
    \hline \hline
     $T_b$ [mK]     &  $I_\mathrm{bias}$ $[\mu \mathrm{A}]$    & $P_{AB}$ [pW]  \\ \hline
     47            & 133                  & 1.23    \\
     46            &  179                  & 2.23         \\ 
     44            & 232                  &    3.74      \\ 
     42            & 266               &    4.92          \\ 
     39            & 300                   &  6.26          \\ 
     36            & 321                   &  7.17           \\ 
     30            & 347                   & 8.38        \\ 
     22            & 362                   & 9.12        \\ 
     15            & 368                   & 9.42        \\ 
     10            & 370                  &  9.52    \\ \hline \hline
    \end{tabular}
    \label{tab:pab_data_au}
\end{table}

\begin{figure}
    \centering
    \includegraphics{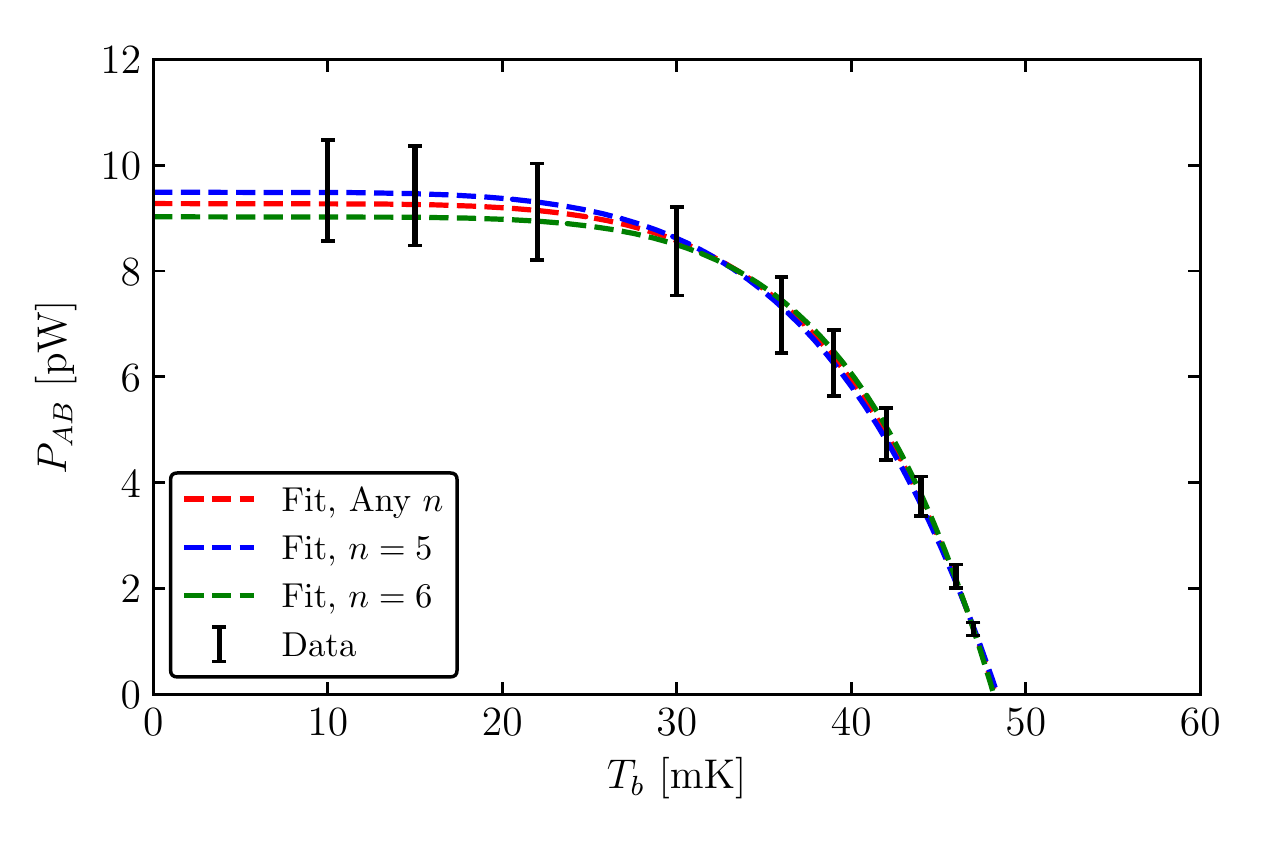}
    \caption{The power flowing from absorber to bath through the Au wire bond pad at various bath temperatures taken for TES295x40. Also shown are the least squares fits with $n$ varying, $n=5$ fixed, $n=6$ fixed.}
    \label{fig:gab_au}
\end{figure}

\begin{table}
    \centering
    \caption{Values from the least-squares fits to the power flowing from absorber to bath through the Au wire bond pad. Included in the table are the estimated heat capacity via the low-temperature Debye model using Eq.~(\ref{eq:Cv}), the related thermal time constant $\tau$, and the electron-phonon coupling constant measured for the Au pad.}
    \begin{tabular}{crrrrrr}
    \hline \hline
    \rule{0pt}{10pt} Fit & $G_{AB}$ $\left[\frac{\mathrm{nW}}{\mathrm{K}}\right]$ & $T_a$ [$\mathrm{mK}$] & $n$ & $C^{th}$ $\left[\frac{\mathrm{pJ}}{\mathrm{K}}\right]$ & $\tau$ $[\mathrm{ms}]$ &  $\Sigma_{ep}$ $\left[\frac{\mathrm{GW}}{\mathrm{K}^n \mathrm{m}^3}\right]$ \\ \hline
    Any $n$ & $1.04\pm0.12$ & $48.28\pm0.22$  & $5.4\pm0.8$ & $6.76\pm0.09$ & $6.45 \pm 0.84$ & \multicolumn{1}{c}{---} \\
    $n=5$ & $0.98\pm0.04$ & $48.38\pm0.15$  & 5 & $6.80\pm0.06$ & $6.93 \pm 0.32$ & $0.531\pm0.026$ \\
    $n=6$ & $1.12\pm0.04$ & $48.7\pm0.13$  & 6  & $6.71\pm0.0$ & $5.97\pm0.26$ &  $10.7\pm0.5$\\ \hline \hline
    \end{tabular}
    \label{tab:gab_au}
\end{table}

To measure $G_{AB}$, we use TES295x40 as the heater and array of disconnected fin QETs as the thermometer.  In addition to the $5\, \mathrm{m}\Omega$ shunt resistor, the normal resistance of TES295x40 in this measurement is $344 \, \mathrm{m}\Omega$, with a parasitic resistance of $2.6\, \mathrm{m}\Omega$. The data from this measurement are reported in Table~\ref{tab:pab_data_au}, where the error in $P_{AB}$ is set to be 10\% of the measured power, based on an estimated 10\% uncertainty in $R_{sh}$. We then fit Eq.~(\ref{eq:Gab}) with $n$ allowed to be any value, and $n=5$ and $n=6$ fixed as expected from ordered and disordered metals, respectively. In Fig.~\ref{fig:gab_au}, we show the results of these fits and compare to the measured data, with the various fit parameters reported in Table~\ref{tab:gab_au}. Both $n=5$ and $n=6$ seem to reasonably describe the power law, so we do not make a claim as to which is correct, but instead report the values for both.  

\section{Summary of Measurements}

In this appendix, we discussed methods for measuring the thermal conductances from TES to absorber and absorber to bath and used the methods for a variety of different measurements, where the substrate holding scheme varied significantly. The various values from that we fit are summarized in Table~\ref{tab:gab_summary} for the $G_{AB}$ fits and Table~\ref{tab:gta_summary} for the $G_{TA}$ fits.

For the $\Sigma_{ep}$ values, previous measurements have seen $\Sigma_{ep, \, \mathrm{Au}}=2.4 \, \mathrm{GW}/(\mathrm{K}^5 \,\mathrm{m}^3)$~\cite{RevModPhys.78.217} and $\Sigma_{ep, \, \mathrm{W}} = 0.3\, \mathrm{GW}/(\mathrm{K}^5 \,\mathrm{m}^3)$~\cite{1439703} for Au and W, respectively. Our measurements for W are completely consistent with that value, while the $n=5$ value for Au is a factor of 5 smaller than previously measured. It is unknown why this is the case. For the $n=6$ power law for Au (i.e. a disordered film), it has been measured as $\Sigma'_{ep, \, \mathrm{Au}}=60 \, \mathrm{GW}/(\mathrm{K}^6 \,\mathrm{m}^3)$~\cite{maasilta_gold_g, PhysRevB.72.012302}, a value which itself depends on the level of disorder. If this were the case, then the $n=6$ value we measured is a factor of 6 lower, which would imply that the level of disorder is different for our film (i.e. less disordered). This could be the case, as our Au film is about a factor of 10 thicker ($750 \, \mathrm{nm}$ vs. $57 \, \mathrm{nm}$), and the boundary scattering effects would be smaller. Another possibility could be that the Au pad itself is not well-connected to the substrate, such that there is an extra interfacial thermal resistance that we cannot measure. More thermal conductance measurements of the Au pad system would be useful to understand the difference.

\begin{table}
    \centering
    \caption{Summary table of the $G_{AB}$ measurements in this appendix. We have added an asterisk on the cirlex clamps measurement due to the poor quality of that dataset.}
    \begin{tabular}{lcr}
    \hline \hline
    \rule{0pt}{10pt} Thermal Bath Connection & Fixed $n$ & $G_{AB}$ $\left[\frac{\mathrm{nW}}{\mathrm{K}}\right]$ \\ \hline
    Cirlex Clamps* & $n=4$   & $34.8\pm1.8$  \\
    GE-7031 Varnish & $n=4$ & $2.83\pm0.11$\\ 
    Al Wire& $n=4$ & $0.054\pm0.002$ \\ 
    Rubber Cement& $n=4$ & $28.4\pm1.2$   \\
    Au Pad & $n=5$ & $0.98\pm0.04$   \\ \hline \hline
    \end{tabular}
    \label{tab:gab_summary}
\end{table}

\begin{table}
    \centering
    \caption{Summary table of the $\Sigma_{ep}$ measurements in this appendix. We have added an asterisk on the G115 measurement due to the poor quality of that dataset.}
    \begin{tabular}{cccr}
    \hline \hline
    \rule{0pt}{10pt} Section & Device & Fixed $n$ & $\Sigma_{ep}$ $\left[\frac{\mathrm{GW}}{\mathrm{K}^n \mathrm{m}^3}\right]$ \\ \hline
   \ref{sec:g115} &  W G115* & $n=5$  & $1.24\pm 0.06$\\ \hline
     \multirow{2}{*}{\ref{sec:gevarn}} & W TES100x25 & $n=5$ & $0.264\pm0.013$ \\  
   & W TES200x50 & $n=5$ & $0.256\pm0.015$ \\  \hline
    \multirow{2}{*}{\ref{sec:albonds}} & W TES100x25 & $n=5$ &  $0.299\pm0.022$ \\ 
    & W TES200x50 & $n=5$ & $0.387 \pm 0.022$ \\  \hline
    \multirow{3}{*}{\ref{sec:cleangta}}& W TES200x50 & $n=5$ &  $0.379 \pm 0.034$ \\ 
    & W TES400x100& $n=5$ &  $0.331 \pm 0.028$ \\ 
    & W TES800x200& $n=5$ & $0.270 \pm 0.023$ \\ \hline
    \multirow{2}{*}{\ref{sec:aupad}}& MELANGE Au Pad & $n=5$ & $0.531\pm0.026$  \\
    & MELANGE Au Pad & $n=6$ & $10.7\pm0.5$   \\ \hline \hline
    \end{tabular}
    \label{tab:gta_summary}
\end{table}

\chapter{\label{chap:vibrations}Reduction of Vibrational Noise}

The decoupling of sensitive detectors or instruments from environmental vibrations has been pursued by many cryogenic experiments, usually using a three-dimensional elastic pendulum~\cite{LYNCH2002345} for passive vibration isolation~\cite{Maisonobe_2018, doi:10.1063/1.5088364, doi:10.1063/1.1149839, Caparrelli2006, PIRRO2000331, PIRRO2006672, doi:10.1063/1.4794767}. These vibrations can provide additional in the experimental, reducing sensitivity of an experiment. This is especially important in rare event searches that work towards the lowest possible noise environments. For the CPD studied in the main body of this thesis, the excess signals observed within the DM ROI could removed with a stress-free holding scheme, which would then be sensitive to vibrations, thus motivating the design and fabrication of an internal vibration decoupler. A common source of vibrations in cryogenic experiments that use a cryogen-free dilution refrigerator is the pulse-tube cryocooler. The pulse-tube cryocooler (PT) is critical to achieve sub-$10 \, \mathrm{mK}$ base temperatures and relies on a oscillating compression system to provide cooling power for the dilution refrigerator. The operating frequency of these oscillations is low, with $1.4 \, \mathrm{Hz}$ as a common frequency. Thus, the goal is to have a system that works as a low-pass filter with a fundamental frequency that is as low as possible. Practically, this will be constrained by the dilution refrigerator dimensions, as we will discuss later in this appendix.

For the system used in the Pyle Lab, we expand upon the designs of two different vibration isolation systems: one developed by R. Maisonobe \emph{et al}.~\cite{Maisonobe_2018} and one developed by L. Gottardi \emph{et al}.~\cite{doi:10.1063/1.5088364}. We will call each of these vibration decouplers the ``Maisonobe system'' and the ``Gottardi system,'' respectively.

\begin{figure}
    \centering
    \includegraphics{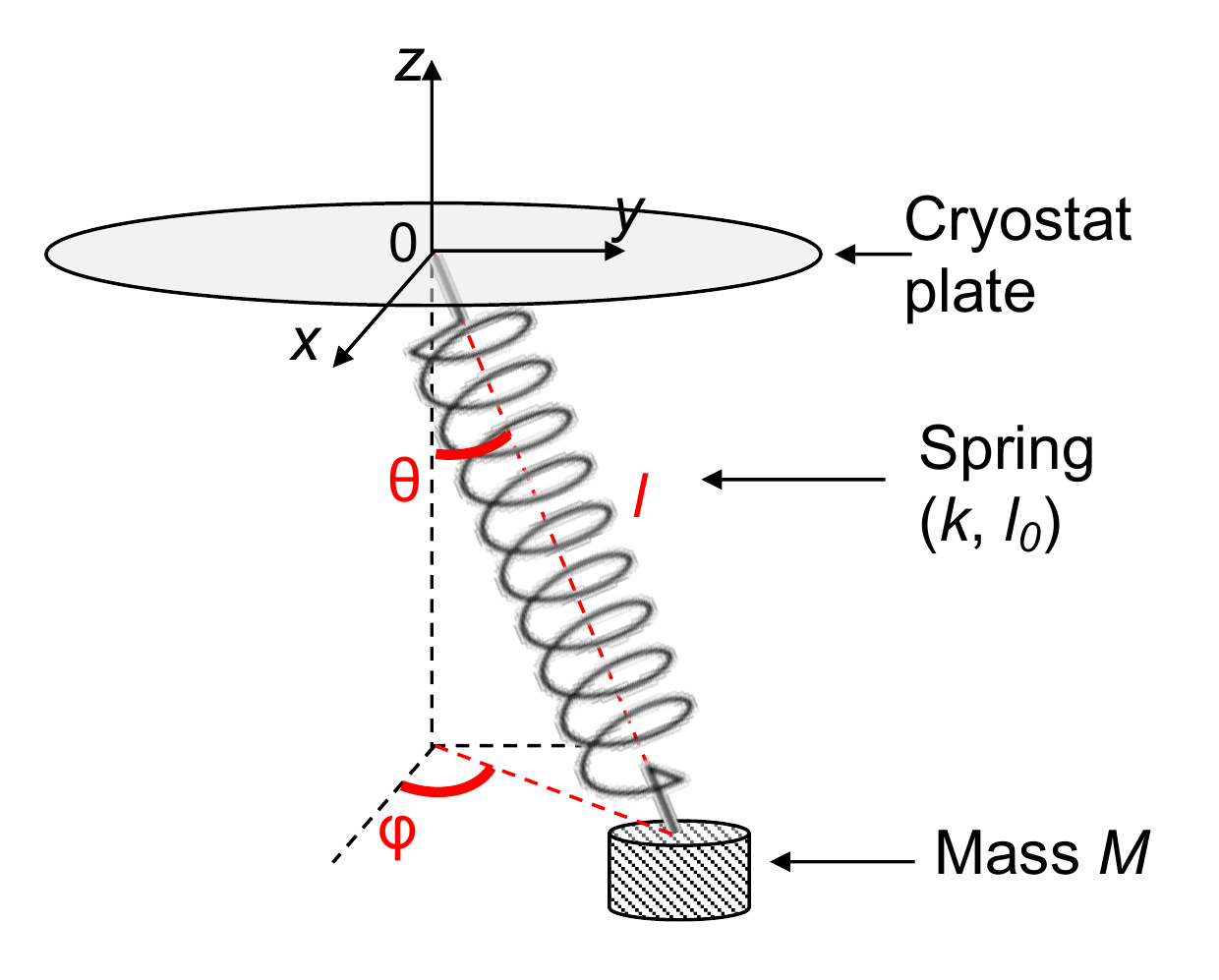}
    \caption{(Figure from Ref.~\cite{Maisonobe_2018}) Design of the Maisonobe system for passive vibration isolation based on an elastic pendulum where motion is driven by perturbations of the plate.}
    \label{fig:edelweisspendulum}
\end{figure}

As shown in Fig.~\ref{fig:edelweisspendulum}, the Maisonobe system is based on a single elastic pendulum placed in the center of the cryostat plate. Their design is simple---they use a single spring pendulum from the $1\,\mathrm{K}$ to $10\,\mathrm{mK}$ stage. The pendulum ultimately giving the best performance is based on a nylon wire and a single stainless steel spring, together giving fundamental frequencies of $1.8 \, \mathrm{Hz}$ in the vertical direction and $0.8 \, \mathrm{Hz}$ in the radial direction. This system allowed the detectors studied to achieve baseline energy resolutions between a factor 2--40 better than without any vibration decoupling.

\begin{figure}
    \centering
    \includegraphics[width=0.8\linewidth]{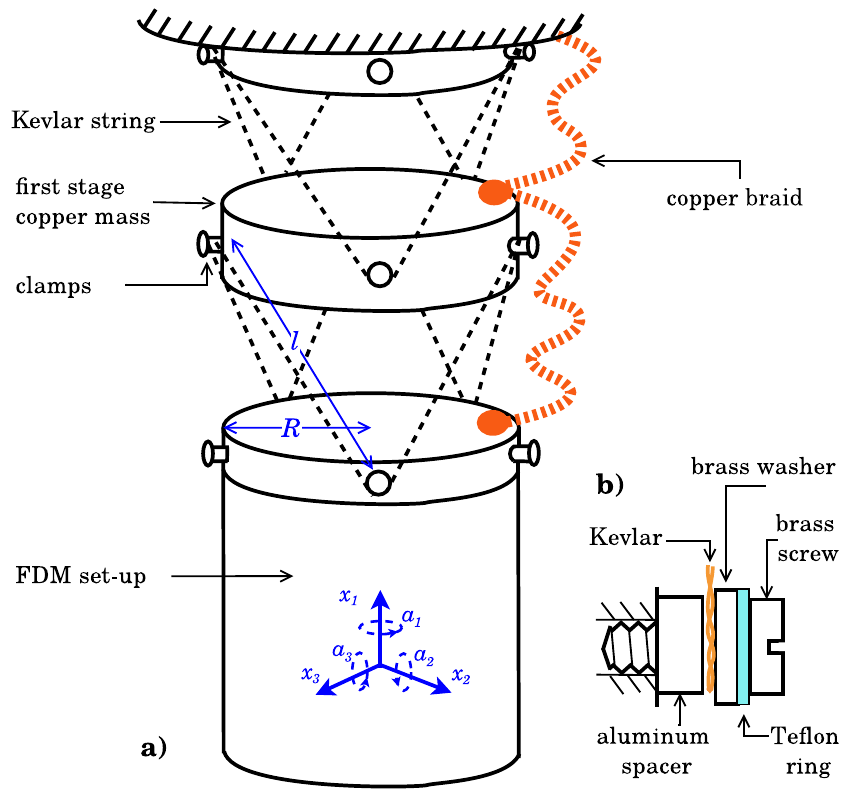}
    \caption{(Figure from Ref.~\cite{doi:10.1063/1.5088364}) (a) Design of the Gottardi system, which uses kevlar wires to act as a elastic double pendulum. (b) The clamping scheme showing how the kevlar wire is installed }
    \label{fig:tvi}
\end{figure}

The Gottardi system is based on the taber vibration isolator proposed in Ref.~\cite{doi:10.1063/1.1149839}. This system uses a kevlar wire which is clamped in a triangular configuration, as shown in Fig.~\ref{fig:tvi}, which effectively acts as three springs symmetrically placed at $120^\circ$ intervals. This allows the design to attenuate vibrations along more degrees of freedom than the Maisonobe system. The design also has two stages, as opposed to one, which increases the attenuation power above the fundamental frequencies with the cost of increasing them by a small amount (as opposed to a one-stage system). For their system, the fundamental frequencies were shown to be on the order of tens of Hz. When adjusting this design to the approximate dimensions of the Pyle Lab fridge it was found that the highest frequencies were still about 10 Hz, thus we decided not to use this design.

\begin{figure}
    \centering
    \includegraphics[width=\linewidth]{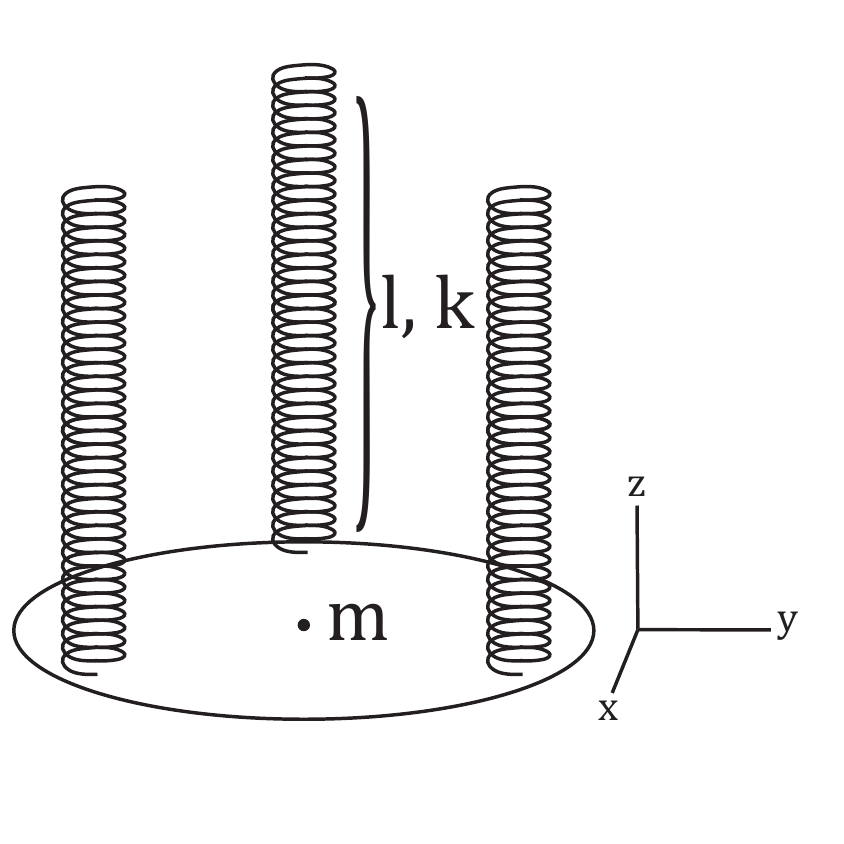}
    \caption{Our proposed vibration decoupling system with three springs spaced equally apart. Each spring has the same equilibrium length $l$ and the same spring constant $k$. Note that the coordinate axis we will use is centered on the disk's center of mass, but is translated to the right to reduce overlapping within the figure.}
    \label{fig:onestage_trans}
\end{figure}

For the Pyle Lab, we considered a model similar to as depicted in Fig.~\ref{fig:onestage_trans}. This system can be thought of as an in-between for both the Maisonobe~\cite{Maisonobe_2018} and Gottardi~\cite{doi:10.1063/1.5088364} systems. With the springs, we should be able to achieve lower fundamental frequencies with the system (as well as in a two-stage version), as well as have attenuation along more degrees of freedom than the Maisonobe system. In order to understand the expected fundamental frequencies in this proposed system, we will derive them using Lagrangian mechanics.

\section{Derivation of Fundamental Frequencies}

In deriving the fundamental frequencies of the vibration decoupler, we will consider to versions of our design: one with a single vibrational stage, and one with two vibrational stages.

\subsection{\label{sec:onestage}One Stage}

The one-stage derivation is as follows. First, we calculate the equilibrium state with the usual sum of forces
\begin{equation}
    -mg + 3 k \left(l_{eq} - l_0\right) = 0\, ,
\end{equation}
where we assume that the system is such that the center of mass is at the disk's center. If we compare to the Maisonobe system, we see that the change in length of each spring will be a third of that in the Maisonobe system, assuming the same spring constants.

In order to describe how the lengths of the springs change under various movements of the disk, we need to first define the coordinate system. Specifically, we will set the origin of the coordinate system as the center of the tops of the three springs. In other words, the center-of-mass of the disk at equilibrium is set to be
\begin{equation}
    \mathbf{x}_{cm} = x\uvec{x} + y \uvec{y} + (z-l_{eq})\uvec{z} \, ,
\end{equation}
where $l_{eq}$ is the equilibrium length of the springs. We also define the $x$-axis as some angle $\theta$ from spring 1, such that spring 2 and spring 3 are placed $\theta + \frac{2 \pi}{3}$ and $\theta + \frac{4 \pi}{3}$ away from the $x$-axis. In other words, we define the horizontal axes to be arbitrary. Using our coordinate system, the locations of the top of the three springs are defined as
\begin{align}
    \mathbf{R}_1 &= R\left[\cos\left(\theta\right)\uvec{x} + \sin\left(\theta\right) \uvec{y}\right]\, ,\\
    \mathbf{R}_2 &= R\left[\cos\left(\theta + \frac{2\pi}{3}\right)\uvec{x} + \sin\left(\theta + \frac{2\pi}{3}\right) \uvec{y}\right]\, ,\\
    \mathbf{R}_3 &= R\left[\cos\left(\theta + \frac{4\pi}{3}\right)\uvec{x} + \sin\left(\theta + \frac{4\pi}{3}\right) \uvec{y}\right]\, .
\end{align}

In order to find the vector of each spring, we can follow the vector path of the top of a spring $\mathbf{R}_i$, to the origin, to the location of the center-of-mass (including its displacement), to the bottom of the spring (with some rotation of the disk). Thus, for spring $i$, its vector is
\begin{equation}
    \mathbf{l}_i = \mathbf{x}_{cm} + \mathbf{R}'_i - \mathbf{R}_i\, ,
\end{equation}
where $\mathbf{R}'_i$ is the rotated $\mathbf{R}_i$.

Because the three springs are placed on the edges of the disk, we expect six degrees of freedom, also seen in the Gottardi system. There will be three translational and three rotational modes. The most general way to calculate the fundamental frequencies of these modes would be to take into account all possible rotations of the disk. However, this quickly becomes a mess of Euler angles and is unwieldy to do by hand.

Thus, to simply our efforts, we will first calculate the modes of the system when we allow rotations of the disk about the horizontal axis, and then calculate the modes of the system when we allow rotations of the disk about the vertical axis. We note that there is an assumption here that the horizontal rotational modes and vertical rotational modes do not couple, which should be a good assumption simply based on intuition of the system.

\emph{Rotations about the horizontal---}To find all of the fundamental frequencies of the system when allowing for rotations about the horizontal, we will begin by defining the kinetic energy, potential energy, and Lagrangian equations as
\begin{align}
    T &= \frac{1}{2}m(\dot{x}^2 + \dot{y}^2 + \dot{z}^2) + \frac{1}{2}I_x \dot{\phi}^2, \\
    V &= \sum_{i=1}^3\left[\frac{1}{2} k (l_i -l_0)^2\right] + mgz,\\
    L &= T - V.
\end{align}
The $l_i$ terms are nonlinear, so the Euler-Lagrange equations that we would get out of this Lagrangian would not be analytically solvable. However, we can Taylor expand the potential energy around the equilibrium point to second order to estimate the natural frequencies for small oscillations. As we are looking at rotations of the disk about the horizontal by some angle $\phi$, we can relate $\mathbf{R}'_i$ to $\mathbf{R}_i$ by
\begin{equation}
    \mathbf{R}'_i = 
    \begin{bmatrix}
        1 & 0 & 0 \\
        0 & \cos{\phi} & -\sin{\phi} \\
        0 & \sin{\phi} & \cos{\phi}
    \end{bmatrix} \mathbf{R}_i.
\end{equation}

For the zeroth order term of the Taylor expansion, the potential energy is simply $V_0 = \frac{3}{2}k(l_{eq} - l_0)^2$, which is the potential energy stored in each spring at equilibrium. For the first order terms, it can be shown that the first derivative of the potential energy with respect to each variable is zero when evaluated at equilibrium (the point at which we are expanding). For the second order terms, it can be shown that the cross-terms all evaluate to zero at equilibrium, while the second derivatives with respect to each variable are nonzero. Specifically, the potential energy can be approximated as
\begin{align}
    V \approx &V_0 + \frac{3}{2} k \left(1 - \frac{l_0}{l_{eq}}\right) x^2 + \frac{3}{2} k \left(1 - \frac{l_0}{l_{eq}}\right) y^2  \nonumber \\
    &+ \frac{3}{2}kz^2 + \frac{3}{4}k R^2 \phi^2.
\end{align}
Using our equilibrium relation, we can simplify this to
\begin{align}
    V \approx &V_0 + \frac{1}{2} \frac{mg}{l_{eq}} x^2 + \frac{1}{2} \frac{mg}{l_{eq}} y^2 + \frac{3}{2}kz^2 + \frac{3}{4}k R^2 \phi^2.
\end{align}

It is clear that the Euler-Lagrange equations will completely decouple, and we do not have any mixing of modes when Taylor expanding up to second order. Applying the Euler-Lagrange equations, we find that the decoupled equations of motion are
\begin{align}
    \ddot{x} + \frac{g}{l_{eq}} x &= 0, \\
    \ddot{y} + \frac{g}{l_{eq}} y &= 0, \\
    \ddot{z} + \frac{3k}{m} z &= 0, \\
    \ddot{\phi} + \frac{3k R^2}{2 I_x} \phi &= 0.
\end{align}

Thus, we have that the fundamental frequencies are
\begin{align}
    f_{x,y} &= \frac{1}{2 \pi} \sqrt{\frac{g}{l_{eq}}}, \\
    f_{z} &= \frac{1}{2 \pi} \sqrt{\frac{3k}{m}}, \\
    f_{x,y,rot.} &= \frac{1}{2 \pi} \sqrt{\frac{3kR^2}{2I_x}}.
\end{align}
Note that the $x$ and $y$ translational modes have the same fundamental frequencies due to the symmetry of system. Because we set up the system to have arbitrary $x$ and $y$ axes, the fundamental frequencies about the $x$ and $y$ axes are identical as well.

\emph{Rotations about the vertical---}To find all of the fundamental frequencies of the system when allowing for rotations about the vertical by some angle $\psi$, we will again define the kinetic energy, potential energy, and Lagrangian equations as
\begin{align}
    T &= \frac{1}{2}m(\dot{x}^2 + \dot{y}^2 + \dot{z}^2) + \frac{1}{2}I_z \dot{\psi}^2, \\
    V &= \sum_{i=1}^3\left[\frac{1}{2} k (l_i -l_0)^2\right] + mgz,\\
    L &= T - V.
\end{align}
We will again Taylor expand the potential energy, where the relation between $\mathbf{R}'_i$ and $\mathbf{R}_i$ is instead
\begin{equation}
    \mathbf{R}'_i = 
    \begin{bmatrix}
        \cos{\psi} & -\sin{\psi} & 0 \\
        \sin{\psi} & \cos{\psi} & 0 \\
        0 & 0 & 1
    \end{bmatrix} \mathbf{R}_i.
\end{equation}

For the zeroth order term of the Taylor expansion, the potential energy is again $V_0 = \frac{3}{2}k(l_{eq} - l_0)^2$, as expected. For the first order terms, it can be shown that the first derivative of the potential energy with respect to each variable is zero when evaluated at equilibrium. For the second order terms, it can be shown that the cross-terms all evaluate to zero at equilibrium, while the second derivatives with respect to each variable are nonzero. Specifically, the potential energy can be approximated as
\begin{align}
    V \approx &V_0 + \frac{3}{2} k \left(1 - \frac{l_0}{l_{eq}}\right) x^2 + \frac{3}{2} k \left(1 - \frac{l_0}{l_{eq}}\right) y^2  \nonumber \\
    &+ \frac{3}{2}kz^2 + \frac{3}{2}k \left( 1- \frac{l_0}{l_{eq}}\right) R^2 \psi^2.
\end{align}
Using our equilibrium relation, we can simplify this to
\begin{align}
    V \approx &V_0 + \frac{1}{2} \frac{mg}{l_{eq}} x^2 + \frac{1}{2} \frac{mg}{l_{eq}} y^2 + \frac{3}{2}kz^2 + \frac{1}{2} \frac{mgR^2}{l_{eq}}\psi^2.
\end{align}

We again have that the Euler-Lagrange equations will completely decouple, and we do not have any mixing of modes when Taylor expanding up to second order. Applying the Euler-Lagrange equations, we find that the decoupled equations of motion are
\begin{align}
    \ddot{x} + \frac{g}{l_{eq}} x &= 0, \\
    \ddot{y} + \frac{g}{l_{eq}} y &= 0, \\
    \ddot{z} + \frac{3k}{m} z &= 0, \\
    \ddot{\psi} + \frac{mg R^2}{I_z l_{eq}} \psi &= 0.
\end{align}

Thus, the translational modes match those that we derived for when studying the system under horizontal rotations. Thus, we can summarize all of the different fundamental frequencies as
\begin{align}
    f_{x,y} &= \frac{1}{2 \pi} \sqrt{\frac{g}{l_{eq}}},\label{eq:fxy_onestage} \\
    f_{z} &= \frac{1}{2 \pi} \sqrt{\frac{3k}{m}}, \\
    f_{x,y,rot.} &= \frac{1}{2 \pi} \sqrt{\frac{3kR^2}{2I_{x,y}}}, \\
    f_{z,rot.} &= \frac{1}{2 \pi} \sqrt{\frac{mgR^2}{I_z l_{eq}}},\label{eq:fzrot_onestage}
\end{align}
where $f_{z,rot.}$ follows from solving the differential equation describing $\psi$.

Thus, we have that, using the Taylor expansion of the potential energy, we can analytically solve the system using the Lagrangian formalism. The next Pyle Lab setup to study will be the two-stage vibration decoupler.

\subsection{\label{sec:twostage}Two Stages}

\begin{figure}
    \centering
    \includegraphics[width=\linewidth]{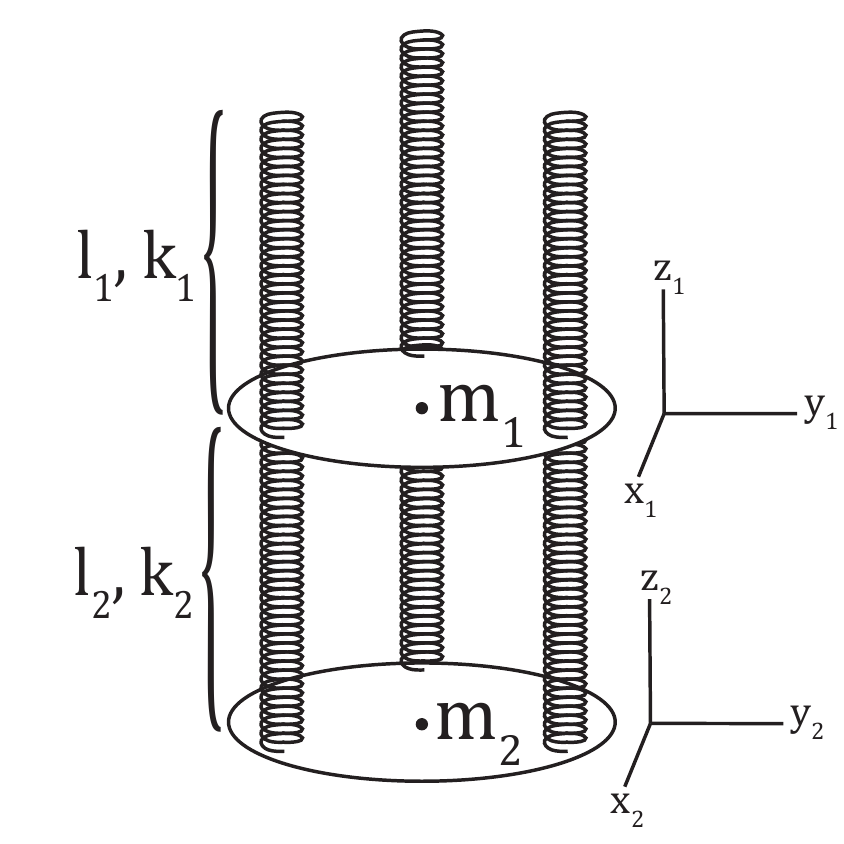}
    \caption{Our proposed vibration decoupling system with three springs spaced equally apart for two different stages. Each spring stage 1 and stage 2 has the same equilibrium length $l_1$, $l_2$ and the same spring constant $k_1$, $k_2$ for those stages, respectively. Note that the coordinate axis we will use is centered on each disk's center of mass, but is translated to the right to reduce overlapping within the figure.}
    \label{fig:twostage_trans}
\end{figure}

As opposed to the derivation of the one-stage system, we will instead go through the various intuitive vibrational modes for the two-stage system case by case. This simplifies the derivation as we are now assuming that the modes decouple, rather than proving this by checking all of the various derivatives. We will also assume that the spring constants for each spring within a stage are the same, but different between the two stages, that the springs for the two stages have different equilibrium lengths, and that the masses of the two stages are also different (but the same disk radius). In Fig.~\ref{fig:twostage_trans}, we show the proposed system, as well as define the coordinate system used.

\emph{Case 1---}Let's start with vertical translations of the two-stage system. The kinetic energy, potential energy, and Lagrangian can be written as:
\begin{align}
    T =& \frac{1}{2}m_1 \dot{z}_1^2 + \frac{1}{2}m_2 \dot{z}_2^2, \\
    V =& \frac{3}{2} k_1 z_1^2 + \frac{3}{2}k_2 (z_2 - z_1)^2 + m_1 g z_1 + m_2 g z_2, \\
    L =& \frac{1}{2}m_1 \dot{z}_1^2 + \frac{1}{2}m_2 \dot{z}_2^2 - \frac{3}{2} k_1 z_1^2 - \frac{3}{2}k_2 (z_2 - z_1)^2\nonumber \\ 
     &- m_1 g z_1 - m_2 g z_2,
\end{align}
where each spring in a stage has identical potential energies (hence the factors of 3). After carrying out the Euler-Lagrange equations and simplifying, we have two coupled differential equations:
\begin{align}
    m_1 \ddot{z}_1 + 3 (k_1 + k_2) z_1 - 3k_2 z_2 &= 0, \label{eq:vert1} \\
    m_z \ddot{z}_2 - 3 k_2 z_1 + 3 k_2 z_2 &= 0,\label{eq:vert2}
\end{align}
where we have again neglected constant values (as they can be removed by a change in coordinate system). After solving this system for the eigenvalues, we have that the natural frequencies are
\begin{align}
    \omega^2 =& \frac{3[(k_1 + k_2)m_2 + k_2 m_1]}{2 m_1 m_2}\left\{ 1 \pm \sqrt{1 - \frac{4 m_1 m_2 k_1 k_2}{[(k_1 + k_2)m_2 + k_2 m_1]^2}}\right\}.
\end{align}

\emph{Case 2---}Next, let's look at horizontal translations of the disks. This can be thought of as the double pendulum modes for the disks' centers of mass. We start by showing the kinetic and potential energy equations in $x$ and $y$ coordinates:
\begin{align}
    T =& \frac{1}{2} m_1 \left(\dot{x}_1 + \dot{y}_1 \right)^2 + \frac{1}{2} m_2 \left(\dot{x}_2 + \dot{y}_2 \right)^2, \\
    V =& m_1 g y_1 + m_2 g y_2.
\end{align}
Since this will be oscillations due to the angles that the disk centers of mass make with the vertical, we need to change our variables to be in terms of these angles for each center of mass ($\alpha_1$ and $\alpha_2$). That is, we have that
\begin{align}
    x_1 &= l_1 \sin \alpha_1, \\
    \dot{x_1} &= l_1 \dot{\alpha}_1 \cos \alpha_1, \\
    y_1 &= -l_1 \cos \alpha_1, \\
    \dot{y}_1 &= l_1 \dot{\alpha}_1 \sin \alpha_1, \\
    x_2 &= l_1 \sin \alpha_1 + l_2 \sin \alpha_2, \\
    \dot{x_2} &= l_1 \dot{\alpha}_1 \cos \alpha_1 + l_2 \dot{\alpha}_2 \cos \alpha_2, \\
    y_2 &= -l_1 \cos \alpha_1 - l_2 \cos \alpha_2, \\
    \dot{y}_2 &= l_1 \dot{\alpha}_1 \sin \alpha_1 + l_2 \dot{\alpha}_2 \sin \alpha_2.
\end{align}
We can plug all of these relations into the kinetic and potential energy equations, which gives us
\begin{align}
    T =& \frac{1}{2} (m_1 + m_2) l_1^2 \dot{\alpha}_1^2 \nonumber\\
    &+ m_2 l_1 l_2 \dot{\alpha}_1 \dot{\alpha}_2 \cos(\alpha_1 - \alpha_2) + \frac{1}{2}m_2l_2 \dot{\alpha}_2^2,\\
    V =& - (m_1 + m_2) g l_1 \cos\alpha_1 - m_2 g l_2 \cos \alpha_2.
\end{align}
Using the Euler-Lagrange equations and the small angle approximation, one can find that this gives us two coupled differential equations:
\begin{align}
    (m_1 + m_2)l_1^2\ddot{\alpha}_1 + m_2 l_1 l_2 \ddot{\alpha}_2 + (m_1 + m_2) g l_1 \alpha_1 &=0,\\
    m_2 l_2^2 \ddot{\alpha}_2 + m_2 l_1 l_2 \ddot{\alpha}_1 + m_2 g l_2 \alpha_2 &=0.
\end{align}
After solving this system of equations for the eigenfrequencies, we have that the natural frequencies are
\begin{align}
    \omega^2 =& \frac{(m_1 + m_2)(l_1 + l_2)g}{2m_1l_1l_2}\left\{ 1 \pm \sqrt{1 - \frac{4m_1l_1l_2}{(m_1 + m_2)(l_1+l_2)^2}}\right\}.
\end{align}

\begin{figure}
    \centering
    \includegraphics[width=\linewidth]{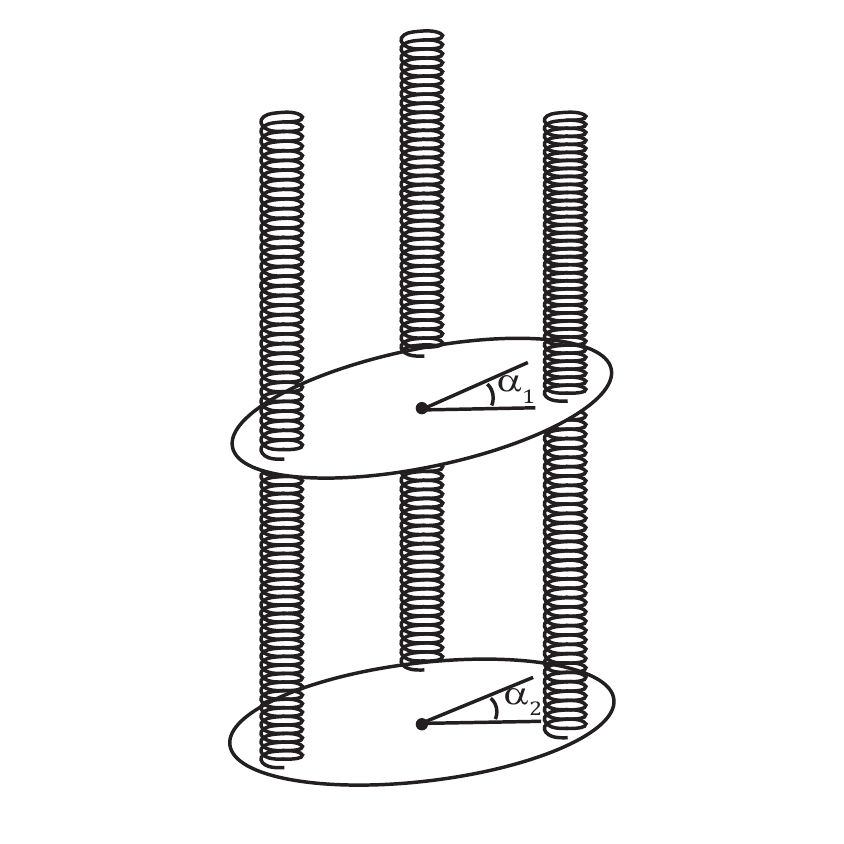}
    \caption{An example of how each stage would pitch under rotations of angle $\alpha_1$ and $\alpha_2$ about an arbitrarily aligned $x$-axis.}
    \label{fig:twostage_rot_horiz}
\end{figure}

\emph{Case 3---}Next, we look at the eigenmodes for rotations about a horizontal axis, where each disk rotates by an angle of $\alpha_1$ and $\alpha_2$, respectively. This is shown in Fig.~\ref{fig:twostage_rot_horiz} In this case, we will need to keep track of the locations of the springs along the disk, remembering that each spring is vertically aligned. In this case, we again remind the reader of the below relations for each of the spring locations, where the same holds true for each stage.
\begin{align}
    \mathbf{R}_1 &= R\left[\cos\left(\theta\right)\hat{\mathbf{x}} + \sin\left(\theta\right) \hat{\mathbf{y}}\right]\, ,\\
    \mathbf{R}_2 &= R\left[\cos\left(\theta + \frac{2\pi}{3}\right)\hat{\mathbf{x}} + \sin\left(\theta + \frac{2\pi}{3}\right) \hat{\mathbf{y}}\right]\, ,\\
    \mathbf{R}_3 &= R\left[\cos\left(\theta + \frac{4\pi}{3}\right)\hat{\mathbf{x}} + \sin\left(\theta + \frac{4\pi}{3}\right) \hat{\mathbf{y}}\right]\, .
\end{align}
Thus, we can write the unsimplified version of the kinetic and potential energy equations:
\begin{align}
    T =& \frac{1}{2}I_1 \dot{\alpha}_1^2 +  \frac{1}{2}I_2 \dot{\alpha}_2^2, \\
    V =& \frac{1}{2}k_1[(R \cos(\theta) \sin(\alpha_1)]^2 \nonumber \\
       &+ \frac{1}{2}k_1\left[R \cos\left(\theta + \frac{2\pi}{3}\right) \sin(\alpha_1)\right]^2  \nonumber \\
       &+ \frac{1}{2}k_1\left[R \cos\left(\theta + \frac{4\pi}{3}\right) \sin(\alpha_1)\right]^2. \nonumber \\
       & +\frac{1}{2}k_2\{R \cos(\theta) [\sin(\alpha_1) - \sin(\alpha_2)]\}^2  \\
       &+ \frac{1}{2}k_2\left\{R \cos\left(\theta + \frac{2\pi}{3}\right) [\sin(\alpha_1) - \sin(\alpha_2)]\right\}^2  \nonumber \\
       &+ \frac{1}{2}k_2\left\{R \cos\left(\theta + \frac{4\pi}{3}\right) [\sin(\alpha_1) - \sin(\alpha_2)]\right\}^2. \nonumber
\end{align}
Using angle-addition formulae and some algebra, we find that the potential energy equation simplifies to:
\begin{align}
    V =& \frac{3}{4} (k_1+k_2) R^2 \sin^2\alpha_1 \nonumber \\
       &- \frac{3}{2}k_2 R^2 \sin \alpha_1 \sin\alpha_2  \\
       &+ \frac{3}{4} k_2 R^2 \sin^2\alpha_2 \nonumber.
\end{align}
We note that, as expected, there is no dependence on the coordinate system used. Now, we can use the Euler-Lagrange equations and the small-angle approximation to find the coupled differential equations.
\begin{align}
    I_1 \ddot{\alpha}_1 + \frac{3}{2} (k_1 + k_2)R^2 \alpha_1 - \frac{3}{2}k_2 R^2 \alpha_2 &=0, \\
    I_2 \ddot{\alpha}_2 - \frac{3}{2} k_2 R^2 \alpha_1 + \frac{3}{2}k_2 R^2 \alpha_2 &=0.
\end{align}
Solving for the eigenfrequencies, we find that
\begin{align}
    \omega^2 =& \frac{3 R^2 [(k_1 + k_2) I_2 + k_2 I_1]}{4 I_1 I_2}\left\{ 1 \pm \sqrt{1 - \frac{4I_1 I_2 k_1 k_2}{[(k_1 + k_2) I_2 + k_2 I_1]^2}}\right\}.
\end{align}

\begin{figure}
    \centering
    \includegraphics[width=\linewidth]{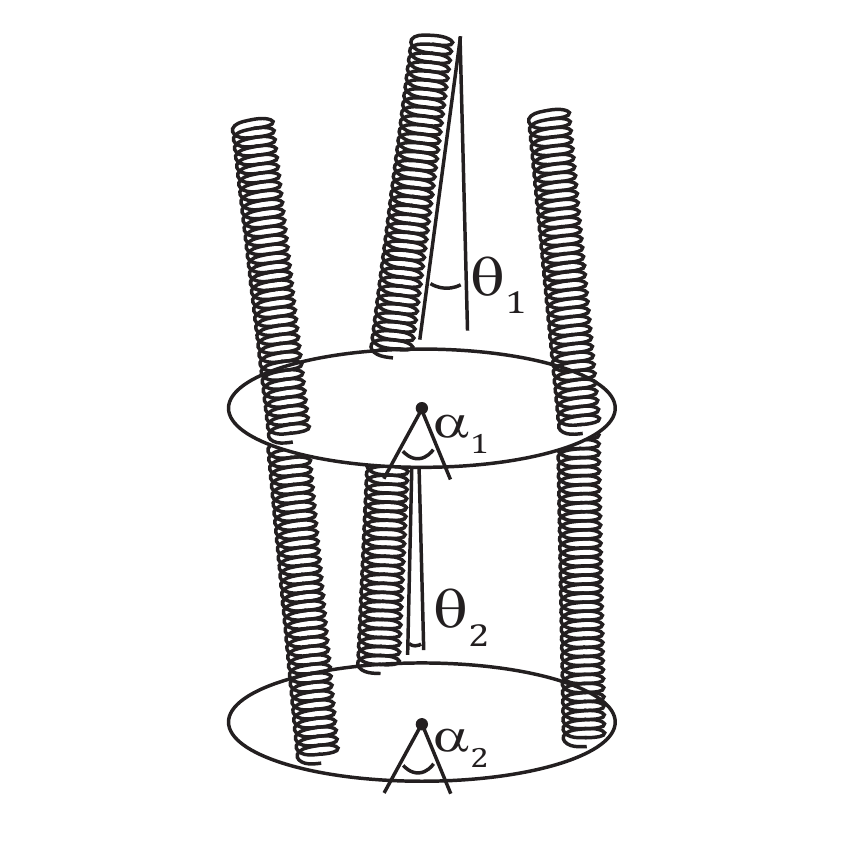}
    \caption{An example of how each stage would twist under rotations of angle $\alpha_1$ and $\alpha_2$ about the vertical.}
    \label{fig:twostage_rot_vert}
\end{figure}

\emph{Case 4---}Lastly, we look at the natural frequencies for rotations of the disks about the vertical axis, i.e. the double torsion pendulum modes. In this case, we will say that each disk rotates about the vertical by angles $\alpha_1$ and $\alpha_2$, respectively. In Fig.~\ref{fig:twostage_rot_vert}, we show an example of this type of rotation. We note that in this case, each spring is rotated by an angle $\theta_1$ and $\theta_2$ with respect to the vertical. Thus, the kinetic energy and potential energy equations are:
\begin{align}
    T &= \frac{1}{2} I_1 \dot{\alpha}_1^2 + \frac{1}{2} I_2 \dot{\alpha}_2^2, \\
    V &= - (m_1 + m_2) g l_1 \cos\theta_1 - m_2 g l_2 \cos \theta_2,
\end{align}
where we have dropped the constants in the potential energy and neglected the kinetic energy due to vertical movement, as this is very small (also the equations are no longer analytically solvable).

Now, we have that the angles $\theta_1$ and $\theta_2$ are related to the $\alpha_1$ and $\alpha_2$ by the following relations:
\begin{align}
    l_1 \theta_1 &= R\alpha_1, \\
    l_1 \theta_1 + l_2\theta_2 &= R\alpha_2.
\end{align}
Next, we carry out the Euler-Lagrange equations, use the small-angle approximation, and use the above relations to arrive at our coupled differential equations:
\begin{align}
    I_1 \ddot{\alpha}_1 + \left( \frac{m_1 + m_2}{l_1} + \frac{m_2}{l_2}\right) g R^2 \alpha_1 - \frac{m_2}{l_2}g R^2 \alpha_2 &=0 , \\
    I_2 \ddot{\alpha}_2 - \frac{m_2}{l_2}g R^2 \alpha_1 + \frac{m_2}{l_2}g R^2 \alpha_2 &=0.
\end{align}
Solving for the eigenfrequencies, we find that

\begin{align}
    \omega^2 =& \frac{\left[ \left( \frac{m_1 + m_2}{l_1} + \frac{m_2}{l_2}\right)I_2 + \frac{m_2}{l_2}I_1\right] g R^2}{2I_1 I_2}\left\{ 1 \pm \sqrt{1 - \frac{4I_1 I_2 \frac{m_1 + m_2}{l_1}\frac{m_2}{l_2}}{\left[ \left( \frac{m_1 + m_2}{l_1} + \frac{m_2}{l_2}\right)I_2 + \frac{m_2}{l_2}I_1\right]^2}}\right\}.
\end{align}

\subsection{Spring Mode Intuition}

We can gain some intuition on the final vibrational system via these fundamental frequencies. First, we note that the frequencies follow two basic forms: pendulum frequencies (i.e. $f \sim \sqrt{g/l}$) and spring frequencies (i.e. $f \sim \sqrt{k/m}$). Thus, we have that the ideal system will have springs that have the smallest spring rates and pendula that are as long as possible. These will be limited by the design constraints of our system, as will be discussed in Section~\ref{sec:designconstraints}.

\begin{figure}
    \centering
    \includegraphics[width=0.8\linewidth]{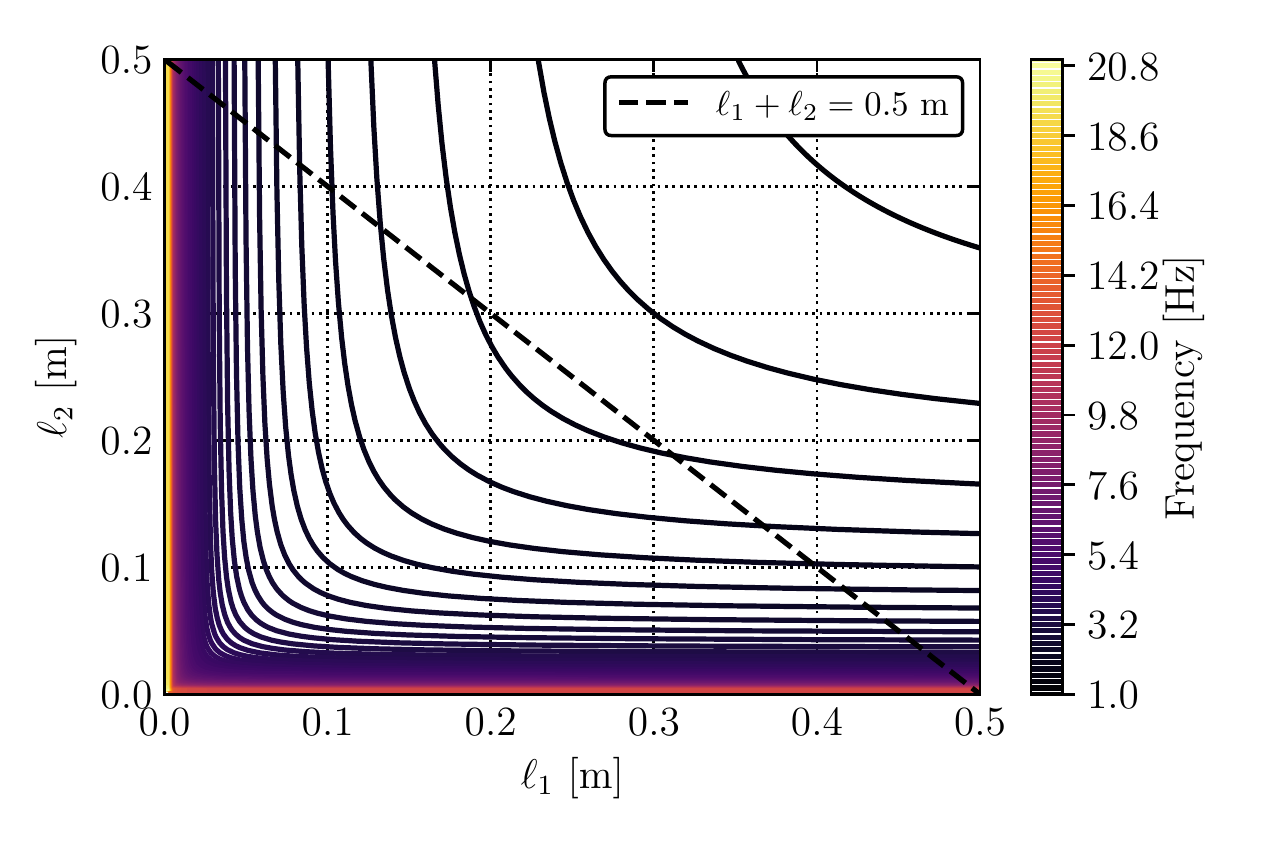}
    \caption{Showing that the lowest frequency is achieved by having the same stretched length, this is for the translational frequencies in the $x,y$ directions. The values shown assume $m_1 = 2 m_2$, but the conclusion is true for any distribution of mass between the two stages.}
    \label{fig:twostage_same_l}
\end{figure}

However, one constraint that we can use to quickly gain intuition on our spring design is that we have a finite vertical volume for our system. For the one-stage system, we should use the full vertical volume. For the two-stage system, this means that the two stretched lengths must add up to some maximum value that is defined by the dilution refrigerator geometry. For our system, we will see that $0.5 \, \mathrm{m}$ is roughly this maximum. If we take this constraint, we can see from Fig.~\ref{fig:twostage_same_l} that a representative pendulum mode for the two-stage system will have its lowest value with the first stage and second stage springs having the same length.

\begin{figure}
    \centering
    \includegraphics[width=0.8\linewidth]{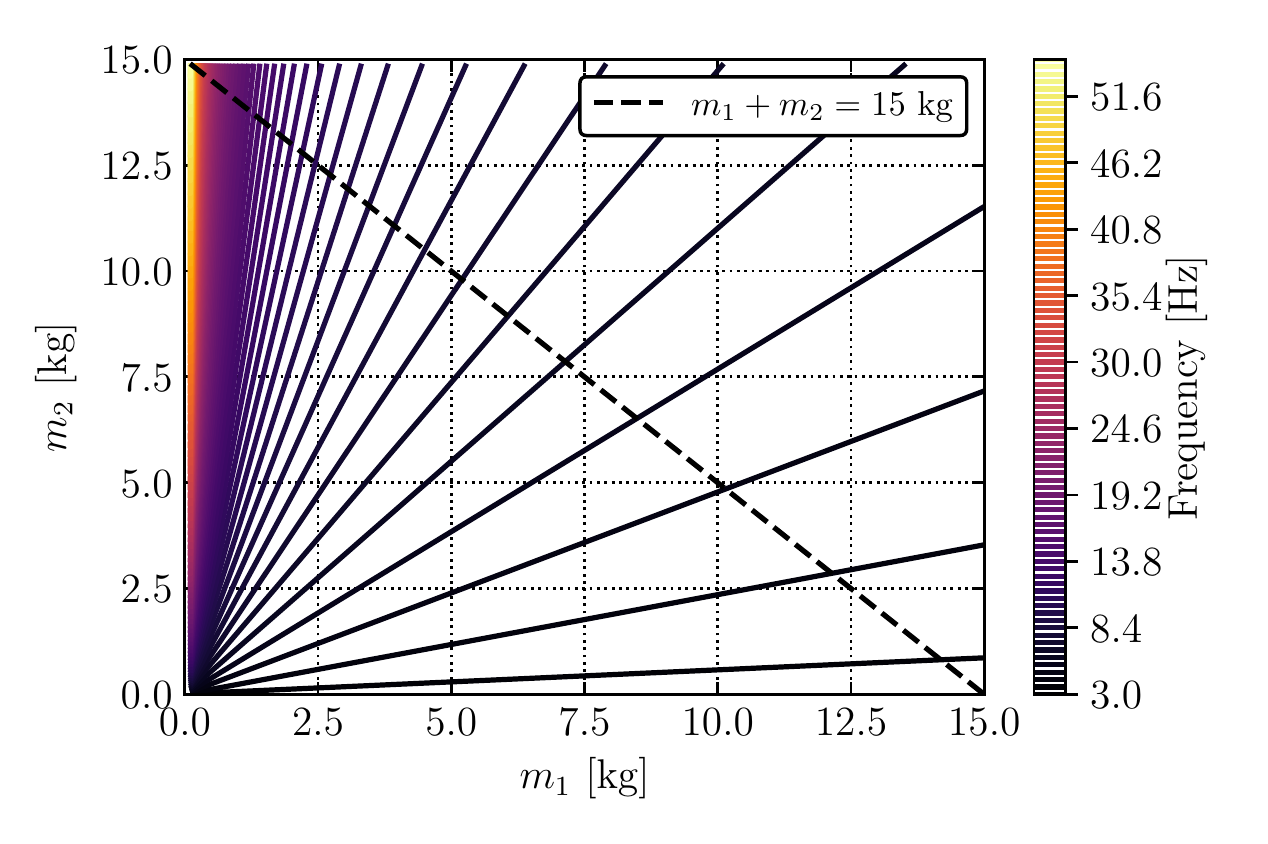}
    \caption{Showing that the lowest frequency is achieved by having the majority of the mass on the upper stage, this is for the translational frequencies in the $x,y$ directions. The values shown assume $\ell_1 = \ell_2$.}
    \label{fig:twostage_mass_pend}
\end{figure}

\begin{figure}
    \centering
    \includegraphics[width=0.8\linewidth]{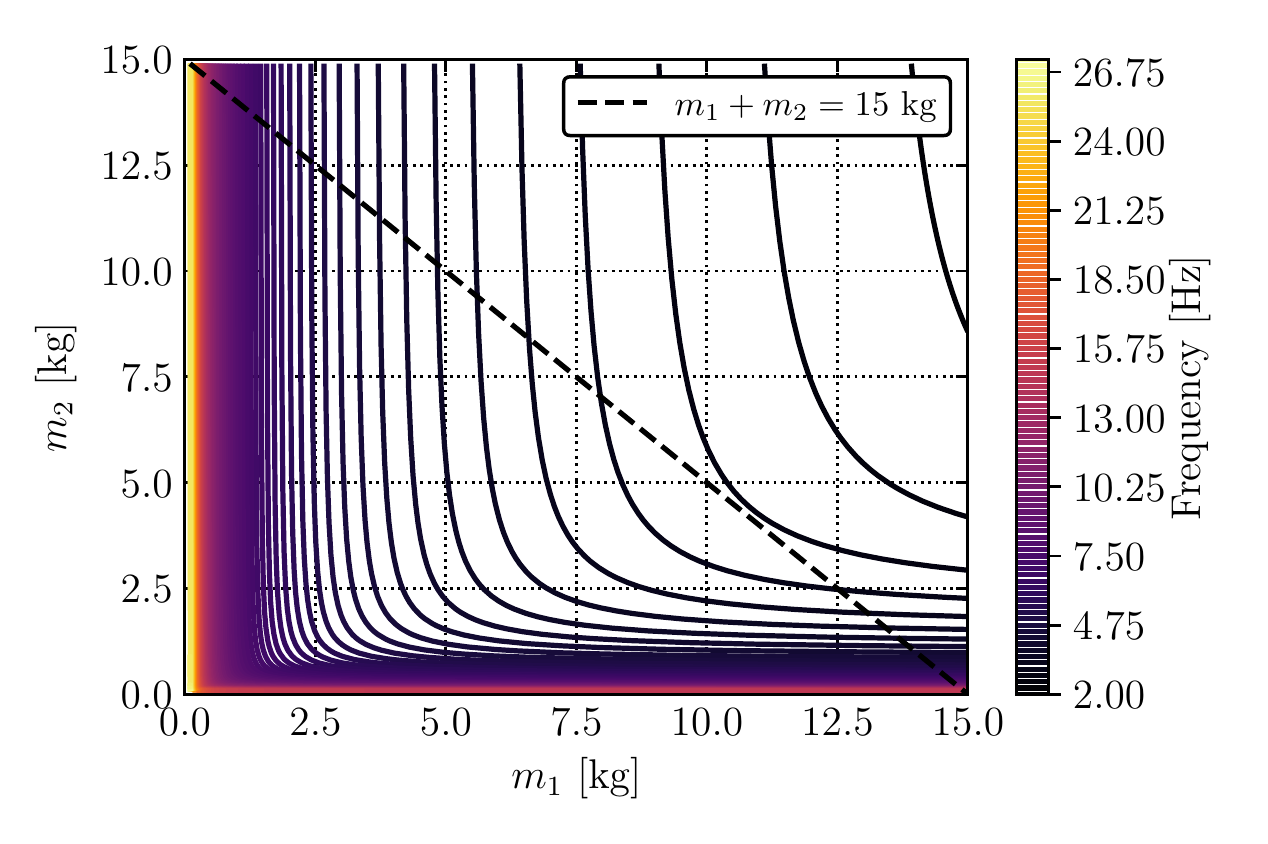}
    \caption{Showing that there may be a raising of the spring mode frequencies if put too much mass on the upper stage, but the $m_1>m_2$ relation still holds. This is for the translational frequencies in the $z$ direction. The values shown assume $\ell_1 = \ell_2$, and $k_1 = 3k_2$. The inequality holds for any pair $k$ values, while the location of frequency degradation will approach $m_1=m_2$ as $k_1$ approaches $k_2$ from below.}
    \label{fig:twostage_mass_spring}
\end{figure}

It is also worth discussing mass distribution between the two stages. For the one-stage system, it is simple, as the spring modes imply to have the single stage be as heavy as possible. For the two-stage system, it will be more complex, as we need to know how to distribute the mass between stages. To do this, we will assume some total mass of $15\, \mathrm{kg}$ as a placeholder value for that of the system. In Fig.~\ref{fig:twostage_mass_pend}, we see that a characteristic pendulum mode implies that as much mass as possible should be placed on the upper stage. On the other hand, if we look at a characteristic spring mode, then we see that there will be an upper bound on this value (depending on the spring constants), as shown in Fig.~\ref{fig:twostage_mass_spring}. Nonetheless, the inequality that $m_1 > m_2$, or that the upper stage should have more mass than the lower stage, will hold. Thus, we will design our system with the intuition that we should put more mass on the upper stage than on the lower stage and make a decision on the distribution based on the spring constants we will be able to use.

\section{Extension Spring Design}

We have our expected fundamental frequencies and a small amount on intuition for some of the two-stage system's values. With this in mind, we need to understand some of the fundamental constraints of our system. That is, we must know what the geometric constraints of the dilution refrigerator are (e.g. the maximum vertical space, the horizontal space), as well as constraints on spring design from an engineering point-of-view.

\subsection{Design Principles}

For spring design, the majority of the equations and rules are well-known in standard mechanical engineering textbooks, such as \emph{Shigley's Mechanical Engineering Design}~\cite{shigley:2011}, which we will follow when designing our springs. 

\subsubsection{Spring Rate}

\begin{figure}
    \centering
    \includegraphics{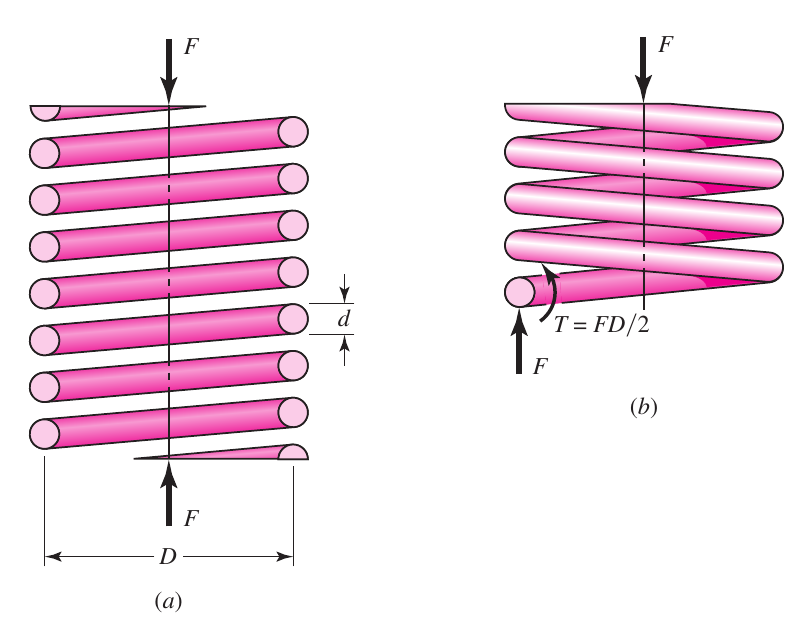}
    \caption{(Figure from Ref.~\cite{shigley:2011}) (a) Diagram of the spring diameter $D$ and the wire diameter $d$, showing the forces being applied vertically. (b) The applied torque on a given segment of wire given the applied force. Note that this diagram is actually for a compression spring, but the definitions are the same for an extension spring (with the force being in the opposite direction).}
    \label{fig:spring_defs}
\end{figure}

We start with the maximum stress in a wire being a superposition of the direct shear stress and the torsional shear stress:
\begin{equation}
    \tau_\mathrm{max} = \frac{F}{A} + \frac{Tr}{J},
    \label{eq:genstress}
\end{equation}
where $F$ is the applied force, $A$ is the cross sectional area of the wire, $T$ is the applied torque, $r$ is the radius of the coiled wire, and $J$ is the polar moment of area. To relate these values directly to the dimensions of the spring, we have that
\begin{align}
    T &= \frac{1}{2}FD, \\
    r &= \frac{d}{2}, \\
    J &= \frac{\pi d^4}{32}, \\
    A &= \frac{\pi d^2}{4},
\end{align}
where $D$ is the average spring diameter, and $d$ is the wire diameter. Each of these values are shown diagrammatically in Fig.~\ref{fig:spring_defs}.

Plugging these values in Eq.~(\ref{eq:genstress}), we have that
\begin{equation}
    \tau_\mathrm{max} = \frac{8FD}{\pi d^3} + \frac{4F\pi}{d^2},
\end{equation}
which can be further simplified with the definition of the spring index $C \equiv D/d$ and the shear stress-correction factor $K_s \equiv \frac{2C + 1}{2C}$, giving
\begin{equation}
    \tau_\mathrm{max} = K_s \frac{8 F D}{\pi d^3}.
\end{equation}
However, this shear stress-correction factor relies on the assumption that the wire is straight, which of course is not true for a spring. Note that a general engineering rule-of-thumb for spring design is that $C$ should fall in the range of $4 < C < 12$, as a spring with index outside of this range will be prone to failure. Generally, in order to be conservative by using the largest empirical stress-correction factor, it is common to replace $K_s$ with the Bergstr{\"a}sser factor
\begin{equation}
    K_B \equiv \frac{4 C + 2}{4C - 3},
\end{equation}
which will be larger for a given $C$, as compared to $K_s$. Thus, this gives maximum stress of
\begin{equation}
    \tau_\mathrm{max} = K_B \frac{8 F D}{\pi d^3}.
    \label{eq:tau_max}
\end{equation}
At this point, it is of importance for us to relate the total deflection of the spring $y$ to the force $F$. As shown again in Ref.~\cite{shigley:2011}, Castigliano's theorem will result in
\begin{equation}
    y = \frac{8 FD^3 N_a}{d^4 G},
\end{equation}
where $G$ is the modulus of rigidity of the spring material, and $N_a$ is the number of active coils (we will define number of active coils in the following paragraphs). Thus, we know that the spring constant is (via Hooke's Law) defined by the ratio of the force to the spring deflection ($k = F/y$), which gives for a spring that
\begin{equation}
    k = \frac{d^4 G}{8D^3N_a}.
\end{equation}

\subsubsection{Adding Hooks}

\begin{figure}
    \centering
    \includegraphics{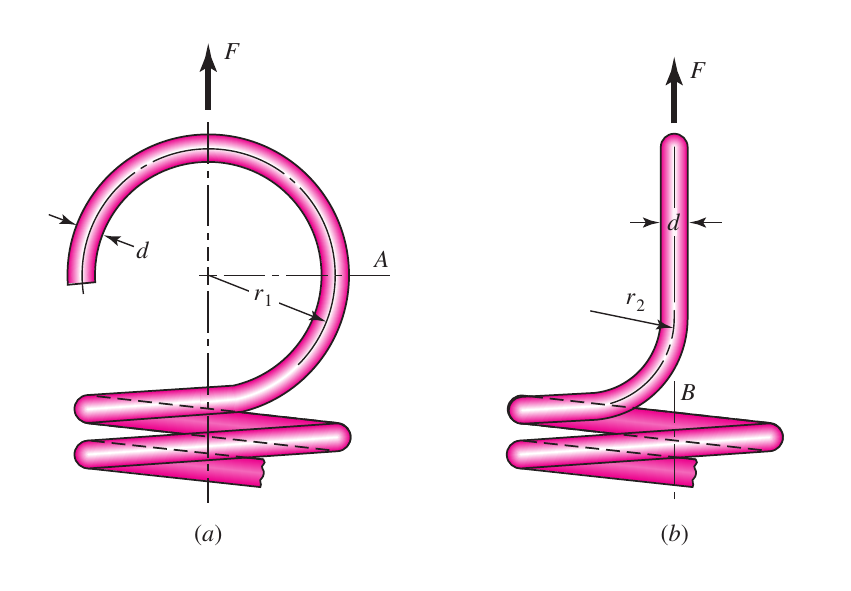}
    \caption{(Figure from Ref.~\cite{shigley:2011}) (a) Dimensions of a hook in relation to point $A$ on the right hand side of the spring hook. (b) Rotating by $90^\circ$, the dimensions of the same hook in relation to point $B$.}
    \label{fig:hooks}
\end{figure}

The maximum tensile stress at $A$ in Fig.~\ref{fig:hooks} is given by
\begin{equation}
    \sigma_A = F \left[(K)_A \frac{16D}{\pi d^3} + \frac{4}{\pi d^2}  \right],
\end{equation}
where $(K)_A$ is a bending stress-correction factor for curvature
\begin{equation}
    (K)_A = \frac{4C_1^2 - C_1 - 1}{4C_1(C_1-1)}
\end{equation}
and
\begin{equation}
    C_1 = \frac{2r_1}{d}.
\end{equation}
The maximum torsional stress at point $B$ is given by
\begin{equation}
    \tau_B = (K)_B \frac{8FD}{\pi d^3},
\end{equation}
where $(K)_B$ is yet another stress-correction factor for curvature
\begin{equation}
    (K)_B = \frac{4C_2 - 1}{4C_2 - 4}
\end{equation}
and
\begin{equation}
    C_2 = \frac{2 r_2}{d}.
\end{equation}
When designing extension springs for loads, both the yield strengths of the coils as well as the hooks must be taken into account to ensure that there is negligible chance of failure.

When making springs, spring manufacturers typically prefer the inclusion of some initial tension such that the free length (i.e. unstretched spring length) can be held more accurately. The deflection of the spring then becomes depending on exceeding the initial tension $F_i$, such that
\begin{equation}
    y = \frac{F - F_i}{k}.
    \label{eq:deflect}
\end{equation}
The free length of a spring including the end hooks is then expressed as
\begin{equation}
    L_0 = 2(D-d) + (N_b + 1)d,
    \label{eq:freel}
\end{equation}
where $D$ is the mean coil diameter, $N_b$ is the number of body coils, and $C$ is the spring index. Note that this assumes that the hook has the same diameter as the coils. For the number of active coils for a spring with ordinary end loops (e.g. Fig.~\ref{fig:hooks}) is given by
\begin{equation}
    N_a = N_b +\frac{G}{E},
\end{equation}
where $G$ is again the modulus of rigidity, and $E$ is the elastic modulus.

Practically, spring manufacturers tend to have a preferred range for what initial tension to lock in for a spring to ensure that the winding process of the wire to create the spring is easily repeatable (i.e. easy to do with the usual equipment available). The preferred range can be expressed in terms of the uncorrected torsional stress
\begin{equation}
    \tau_i = \frac{231.0}{\exp (0.105 C)} \pm 6.895 \left(4 - \frac{C - 3}{6.5} \right) \ \mathrm{MPa}.
\end{equation}
Equivalently, we can plug in the uncorrected torsional stress $\tau = 8 F_i D / (\pi d^3)$ to directly calculate the preferred range of initial tension
\begin{equation}
    F_i = \frac{\pi d^3}{8 D} \tau_i.
    \label{eq:preferredfinit}
\end{equation}

\subsubsection{Safety Factors}

To ensure our system is safe (i.e. the spring system is highly unlikely to fail and drop our detectors inside our dilution refrigerator), we need to keep track of the various safety factors for the various stresses in the system (both the body coils and the hooks). Generally, this is the ratio between the maximum allowable stress of the spring material to the expected stress from the hanging load.

\begin{table}
    \centering
    \caption{Constants $A$ and $m$ for determining the minimum tensile strength of phosphor bronze spring wires. Values are from Ref.~\cite{shigley:2011}, where the measurements are for phosphor bronze wire ASTM B159 and temper CA510.}
    \begin{tabular}{ccc}
         \hline \hline
          $d$ $[\mathrm{mm}]$ & $A$ $\left[\mathrm{MPa}\,\mathrm{mm}\right]$  & $m$ \\ \hline
          0.1--0.6 &  1000 & 0 \\
          0.6--2 & 913 & 0.028 \\
          2--7.5 & 932 & 0.064 \\
          \hline \hline
    \end{tabular}
    \label{tab:tensilestrength}
\end{table}

For various materials, the minimum tensile strength can be represented by
\begin{equation}
    S_{ut} = \frac{A}{d^m}
\end{equation}
where $A$ and $m$ are fitted constants to empirical data. In Table~\ref{tab:tensilestrength}, we show the corresponding values for various wire diameters of phosphor bronze spring wires. For various failure modes, the safety factor will be different, such that we must calculate it for three different cases: failure of the spring body, failure due to bending of the spring hook at point $A$ in Fig.~\ref{fig:hooks}, and failure due to bending of the spring hook at point $B$ in Fig.~\ref{fig:hooks}. For a nonferrous alloy (e.g. phosphor bronze), the maximum allowable stresses in percent of minimum tensile strength are $35\%$, $55\%$, and $30\%$, respectively. Thus, the safety factors $n$ to keep track of when designing extension springs are
\begin{equation}
    n = \begin{cases}
    \frac{0.35S_{ut}}{\tau_\mathrm{max}} & \mathrm{failure\ at\ body}\\
    \frac{0.55S_{ut}}{\sigma_A} & \mathrm{failure \ at \ } A\\
    \frac{0.30S_{ut}}{\tau_B} & \mathrm{failure \ at \ } B\\
    \end{cases}.
    \label{eq:safetyfactors}
\end{equation}

\subsection{\label{sec:designconstraints}Geometric Constraints and Expected Load}

\begin{figure}
    \centering
    \includegraphics[width=0.8\linewidth]{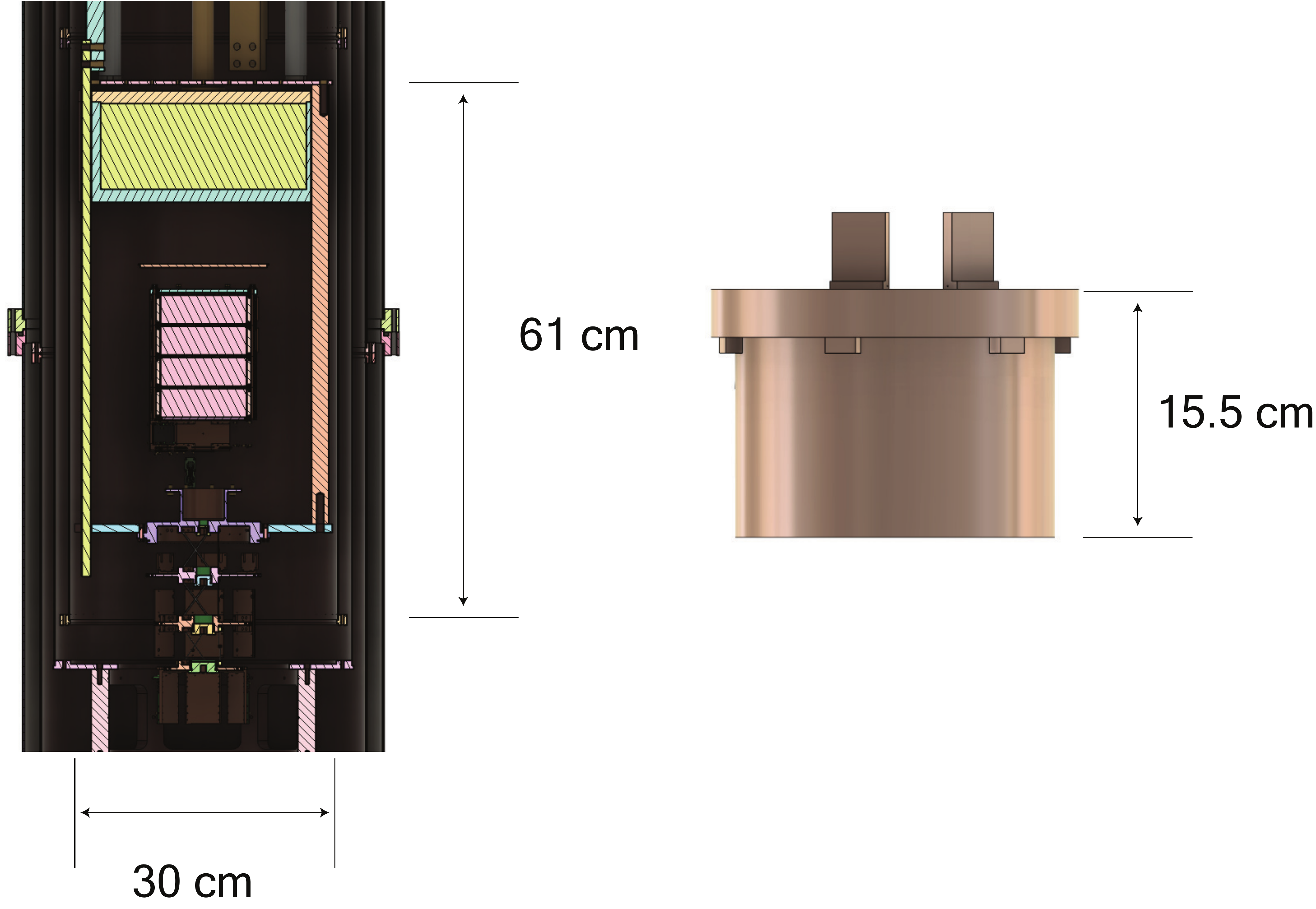}
    \caption{(Left) Cross section of the Pyle Lab dilution refrigerator from the CAD design. The vertical space between the mixing chamber and the bottom of the $1 \, \mathrm{K}$ can is about $61 \, \mathrm{cm}$. Note that the internal payload in this cross section is a placeholder and not what will be installed. (Right) Design of IR-tight detector housing by Yen-Yung Chang and Siddhant Mehrotra. This housing will be attached to the bottom of the lower stage of the vibration decoupler design.}
    \label{fig:fridge}
\end{figure}

Starting with the geometric constraints (Fig.~\ref{fig:fridge}), we have that the fridge has $61 \, \mathrm{cm}$ of vertical space (from mixing chamber to bottom of the $1 \, \mathrm{K}$ can) and a horizontal space of $30 \, \mathrm{cm}$ in diameter, and the detector housing that we expect to install to the bottom of the lower decoupler stage has a height of $15.5\, \mathrm{cm}$. With these dimensions, we are restricted to about $45\, \mathrm{cm}$ of vertical space. Because we do not want the housing to touch the bottom of the $4 \, \mathrm{K}$ can, we will design our springs with a vertical constraint of $35 \, \mathrm{cm}$. This allows some extra space to work with in case we decide to increase the thickness of the various plates, depending on how much vertical space is taken by the spring hooks, and to ensure that there is some vertical space for the system to fall in case a failure (which will then be caught be a safety structure, proposed in Section~\ref{sec:safetystruc}).

The expected mass of the manufactured detector housing is $18 \, \mathrm{kg}$. With detectors, electronics, and other necessities, we will estimate that the total detector payload is about $20 \, \mathrm{kg}$ when considering spring design. Thus, these constraints in geometry and mass will restrict the length of stretched springs and the radius of the stages, and this must be taken into account.

\subsection{Proposed Extension Springs}

With all of these spring design equations to keep track of, we will design our springs using the following workflow for calculating the various spring dimensions and the expected natural frequencies of the system:
\begin{enumerate}
    \item Choose a spring material
    \item Determine the expected total hanging load and per spring
    \item Choose a maximum spring diameter $D$
    \item Solve for the wire $d$ for a given safety factor for failure at the spring body (use 2, as this is generally the least likely failure mode)
    \item Determine the low-end value of the preferred $F_i$ range for spring manufacturers
    \item Solve for the number of body coils $N_b$ needed to achieve the desired stretched length given the geometric constraints, using the sum of the free length and the expected deflection. Round $N_b$ down (to stay within the geometric constraints) to the nearest half-integer to ensure that the hooks are in-plane with each other
    \item Use $N_b$, $D$, and $d$ to calculate the spring rate $k$
    \item Calculate the safety factors for the hook failure modes (at points $A$ and $B$) to ensure they are 1.5 or greater (a reasonable value)
    \item Calculate the expected natural frequencies of the system
\end{enumerate}

\subsubsection{One Stage}

As the one-stage system will be simpler than the two-stage system, we will use it as a relatively in-depth example of the workflow. For our spring material, we will choose phosphor bronze due to its long history of usage in cryogenics for its good thermal conductance. As will be discussed in later sections, this property is important, as the springs themselves should quickly cool down, such that they do not act as a heat source. For the one-stage system, we will assume that the hanging mass is a total of $20 \, \mathrm{kg}$, given by our expected load from the IR-tight housing (Section~\ref{sec:designconstraints}). We next set the maximum spring diameter to $D=25\, \mathrm{mm}$ semi-arbitrarily---given that the stages are up to $27\, \mathrm{cm}$ in diameter, the chosen $D$ is about 10\% of this, and the springs can be installed close to the perimeter of the stage. Furthermore, we should remember that our intuition on springs was such that the spring should have the smallest possible spring rate $k\propto d^4 / D^3$, this implies that we always want the spring diameter $D$ to be as large as possible and the wire diameter $d$ to be as small as possible.

To solve for $d$ given a safety factor of 2 at the spring body, we use Eqs.~(\ref{eq:tau_max}) and (\ref{eq:safetyfactors}), rearranging as
\begin{equation}
    \frac{0.35 S_{ut}}{n} = K_B \frac{8 F_{load}D}{\pi d^3}
\end{equation}
and solving for $d$, where $F_{load} = \frac{1}{3}mg$. In our case of $m = 20\, \mathrm{kg}$ for the one-stage system, we find that $d=3.2\,\mathrm{mm}$ is the smallest that the wire diameter can be (any lower and the safety factor would drop below 2). With both $D$ and $d$ known, we use Eq.~\ref{eq:preferredfinit} to determine the smallest reasonable initial tension $F_i$ for a spring manufacturer to be able to fabricate. In this case, we find that $F_i=40 \, \mathrm{N}$ for each spring. Comparing to the mass per spring of $20/3 \, \mathrm{kg}$ (i.e. $F_{load} = 65 \, \mathrm{N}$ per spring), this initial tension is significant and the deflection of the springs under load will be about 2.6 times less than if there were no initial tension in the spring.

The next step is to solve for the number of body coils $N_b$, which will finish up the spring design parameters. Given our geometric constraints of the fridge, we know that the one-stage system should not have a stretched spring length longer than $l_{max} = 35\, \mathrm{cm}$. Using Eqs.~(\ref{eq:deflect}) and (\ref{eq:freel}), we rearrange the relations to
\begin{equation}
    l_{max} = L_0 + y = 2(D-d) + (N_b + 1)d + \frac{8(F_{load} - F_i)D^3 (N_b + \frac{G}{E})}{G d^4}
    \label{eq:fullpendulumlength}
\end{equation}
and solve for $N_b$. For our one-stage system, we find that the number of body coils needed to fulfill all of our requirements so far is $N_b=77$. Note that this was done assuming that the hook radius about point $A$ was the same as the spring radius (i.e. $r_1 = D/2$), so we will keep this assumption and furthermore set the hook radius about point $B$ to the same value ($r_2 = D/2$).

In the last few steps, we use all of our design parameters to calculate the spring rate $k$ and the remaining safety factors related to hook failure. We find that $k = 440.6 \, \mathrm{N}/\mathrm{m}$ and the safety factors of the hook for failure about point $A$ and point $B$ are 1.63 and 1.82, respectively. Thus, we would expect failure to occur at the hooks before the spring body. Nonetheless, our spring design for the expected load should be safe, as the safety factors are far above 1 (i.e. we would have to overload our system by more than $10 \, \mathrm{kg}$ to cause a failure).

\begin{table}
    \centering
    \caption{Spring design parameters and expected values for the one-stage system.}
    \begin{tabular}{llr}
        \hline \hline
        System & Parameter & Value \\ \hline
         \multirow{17}{*}{One Stage} & Total Hanging Mass $m$ & $20 \,\mathrm{kg}$\\
         & Spring Diameter $D$ & $25 \, \mathrm{mm}$\\
         & Wire Diameter $d$ & $3.2 \, \mathrm{mm}$ \\
         & Hook Bend Radius $r_1$ & $12.5 \, \mathrm{mm}$ \\
         & Hook Bend Radius $r_2$ & $12.5 \, \mathrm{mm}$ \\
         & Spring Index $C$ & $7.8$ \\
         & Initial Tension $F_i$ & $40 \, \mathrm{N}$ \\
         & Number of Body Coils $N_b$ & 77 \\
         & Number of Active Coils $N_a$ & 77.4 \\
         & Spring Rate $k$ & $440.6 \, \frac{\mathrm{N}}{\mathrm{m}}$ \\
         & Free Length $L_0$ & $29.2 \, \mathrm{cm}$\\
         & Expected Deflection $y$ & $5.7 \, \mathrm{cm}$ \\
         & Safety Factors $n$ (body, $A$, $B$)& 2, 1.63, 1.82\\
         & Natural Frequency $f_{z}$ & $1.29\, \mathrm{Hz}$\\
         & Natural Frequency $f_{x,y}$ & $0.84 \, \mathrm{Hz}$\\
         & Natural Frequency $f_{z, rot.}$ & $1.19 \, \mathrm{Hz}$\\
         & Natural Frequency $f_{x,y,rot.}$ & $1.83 \, \mathrm{Hz}$\\ \hline \hline
    \end{tabular}
    \label{tab:onestage_params}
\end{table}

With $k=440.6 \, \mathrm{N}/\mathrm{m}$, $l_{max} = 35 \, \mathrm{cm}$, a hanging mass of $20 \, \mathrm{kg}$, and stage diameter of $27 \, \mathrm{cm}$, we finally return to the natural frequencies that we solved for using the Lagrangian method in Section~\ref{sec:onestage}. Plugging into Eqs.~(\ref{eq:fxy_onestage})--(\ref{eq:fzrot_onestage}), we find that the various natural frequencies are $f_{z} = 1.29 \, \mathrm{Hz}$, $f_{x,y} = 0.84 \, \mathrm{Hz}$, $f_{z, rot.} = 1.19 \, \mathrm{Hz}$, and $f_{x,y, rot.} = 1.83 \, \mathrm{Hz}$. Above these frequencies, all vibrationally-induced noise should be attenuated. Given that the pulse-tube itself oscillates at $1.4 \, \mathrm{Hz}$, these frequencies are not exactly that frequency, and we should be safe from being on a resonance. For this proposed one-stage system, the design and calculated parameters are all stored in Table~\ref{tab:onestage_params}.

\subsubsection{Two Stage}

When determining the spring design parameters for the two-stage system, the workflow is equivalent, but must be done twice. That is, we split up the stages into an upper stage with a total hanging mass of $m_1 + m_2$ (the sum of the masses of both stages) and a lower stage with a total hanging mass of $m_2 = 20\, \mathrm{kg}$ (the mass of the IR-tight housing). Again choosing phosphor bronze, we start with the same spring diameter $D=25 \, \mathrm{mm}$ for the springs on each stage. Although $m_1$ can be any value, we have seen from our intuition that it should be at least as much as $m_2$ if not more. In this design, we will choose $m_1 = 20\, \mathrm{kg}$ as well and will later check that a larger mass would not give significantly greater performance (i.e. significantly lower the natural frequencies of the system).

\begin{table}
    \centering
    \caption{Spring design parameters and expected values for the two-stage system.}
    \begin{tabular}{llr}
        \hline \hline
        System & Parameter & Value \\ \hline
         \multirow{13}{*}{Two Stage, Upper} & Total Hanging Mass $m_1 + m_2$ & $40 \,\mathrm{kg}$\\
         & Spring Diameter $D$ & $25 \, \mathrm{mm}$\\
         & Wire Diameter $d$ & $4.1 \, \mathrm{mm}$ \\
         & Hook Bend Radius $r_{1}$ & $12.5 \, \mathrm{cm}$ \\
         & Hook Bend Radius $r_{2}$ & $12.5 \, \mathrm{cm}$ \\
         & Spring Index $C$ & $6.1$ \\
         & Initial Tension $F_i$ & $105.5 \, \mathrm{N}$ \\
         & Number of Body Coils $N_b$ & 30.5 \\
         & Number of Active Coils $N_a$ & 30.9 \\
         & Spring Rate $k$ & $3028.5 \, \frac{\mathrm{N}}{\mathrm{m}}$ \\
         & Free Length $L_0$ & $17.1 \, \mathrm{cm}$\\
         & Expected Deflection $y$ & $0.8 \, \mathrm{cm}$ \\
         & Safety Factors $n$ (body, $A$, $B$)& 2, 1.64, 1.84\\ \hline
         \multirow{13}{*}{Two Stage, Lower} & Total Hanging Mass $m_2$ & $20 \,\mathrm{kg}$\\
         & Spring Diameter $D$ & $25 \, \mathrm{mm}$\\
         & Wire Diameter $d$ & $3.2 \, \mathrm{mm}$ \\
         & Hook Bend Radius $r_1$ & $12.5 \, \mathrm{mm}$ \\
         & Hook Bend Radius $r_2$ & $12.5 \, \mathrm{mm}$ \\
         & Spring Index $C$ & $7.8$ \\
         & Initial Tension $F_i$ & $40.1 \, \mathrm{N}$ \\
         & Number of Body Coils $N_b$ & 31 \\
         & Number of Active Coils $N_a$ & 31.4 \\
         & Spring Rate $k$ & $1086.0 \, \frac{\mathrm{N}}{\mathrm{m}}$ \\
         & Free Length $L_0$ & $14.6 \, \mathrm{cm}$\\
         & Expected Deflection $y$ & $2.3 \, \mathrm{cm}$ \\
         & Safety Factors $n$ (body, $A$, $B$) & 2, 1.63, 1.82\\ \hline
        \multirow{4}{*}{Two Stage (Full System)} & Natural Frequencies $f_{z}$ & $1.67, 4.11 \, \mathrm{Hz}$\\
         & Natural Frequencies $f_{x,y}$ & $0.91, 2.21 \, \mathrm{Hz}$\\
         & Natural Frequencies $f_{z, rot.}$ & $1.29, 3.12 \, \mathrm{Hz}$\\
         & Natural Frequencies $f_{x,y,rot.}$ & $2.37, 5.82 \, \mathrm{Hz}$\\ \hline \hline
    \end{tabular}
    \label{tab:twostage_params}
\end{table}

In this two-stage system, we also will ensure that the stretched lengths of the upper and lower stage springs are not identical, to ensure that the resonances of each spring are nonoverlapping. To do this, we make another semi-arbitrary choice and choose the upper stage springs to have stretched lengths of $17.9 \, \mathrm{cm}$ and the lower stage springs to have stretched lengths of $16.9 \,\mathrm{cm}$. We then follow our workflow to determine the rest of the values, just as we did for the one-stage system. Using the proposed constraints in the previous paragraph, the parameters for the two-stage system are reported in Table~\ref{tab:twostage_params}.

\begin{figure}
    \centering
    \includegraphics{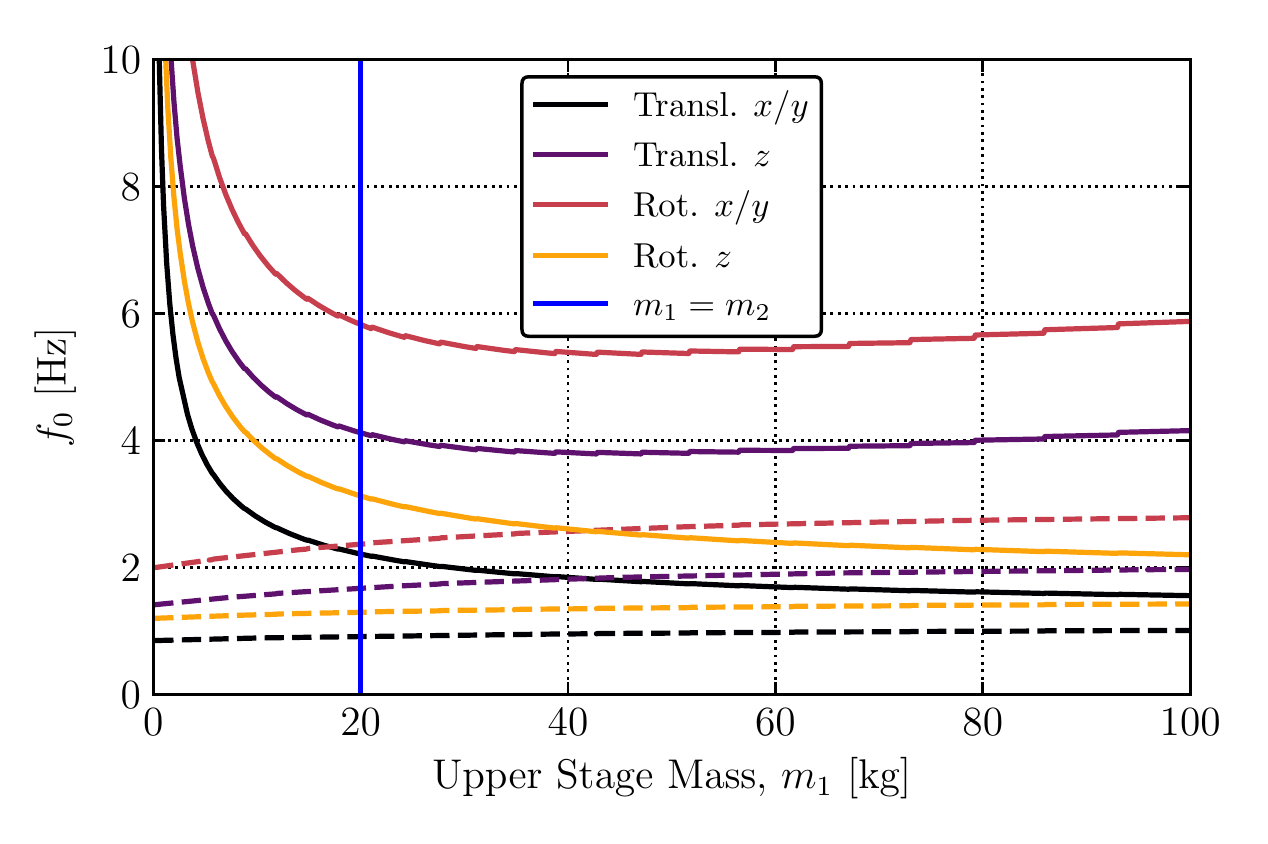}
    \caption{The dependence of the various frequencies on the mass of the upper stage in the two-stage system. The solid lines denote the higher frequency for the given mode, while the dashed lines denote the lower frequency for the given mode. Note that the jumps in the curves corresponding to spring modes are due to rounding the number of body coils $N_b$ to the nearest half-integer---without this constraint, the curves would be smooth.}
    \label{fig:uppermassfreqs}
\end{figure}

We now return to our choice of $m_1 = 20\, \mathrm{kg}$ on the upper stage, such that the upper springs together would be holding up $m_1 + m_2 = 40\, \mathrm{kg}$ in mass. In order to show that this was a reasonable choice, we can show how the various natural frequencies change as a functional of upper stage mass $m_1$. To do this, we simply follow the same workflow for some arbitrary $m_1$ and report the various natural frequencies which would correspond to the optimal design, with the results shown in Fig.~\ref{fig:uppermassfreqs}. In this figure, it is clear that the difference between $m_1 = 20 \, \mathrm{kg}$ and $m_1 = 40\, \mathrm{kg}$ is small, such that are original choice is reasonable. We also note that there is a minimum in both of the highest frequencies such that too large of masses actually begin to increase some of the frequencies. This should not be surprising given what we showed the same behavior in Fig.~\ref{fig:twostage_mass_spring}, from which we would expect worse frequencies for larger upper stage masses after some point for the spring modes.

\begin{figure}
    \centering
    \includegraphics{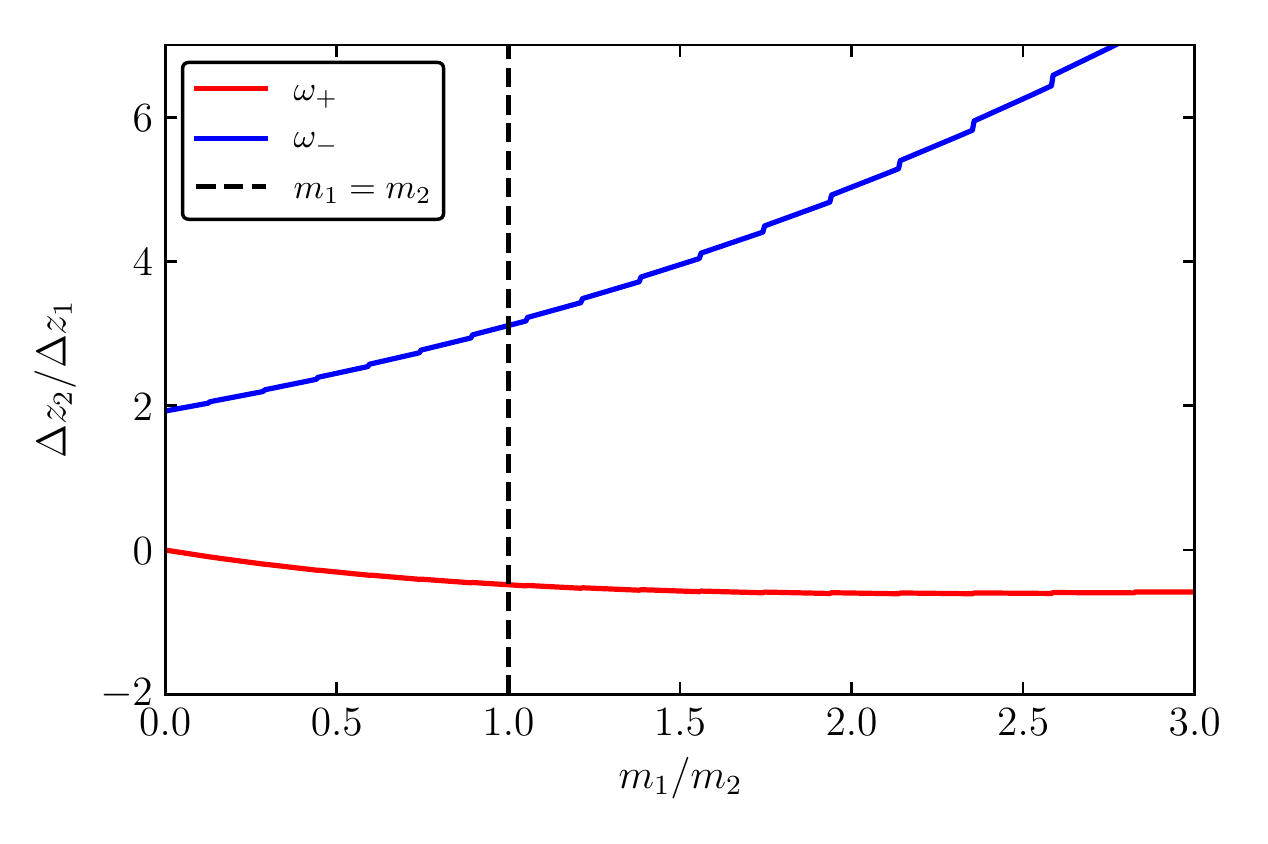}
    \caption{Ratio of the vertical displacement (i.e. eigenvector components) of each stage for the corresponding natural frequencies as the ratio of the upper and lower stage masses varies. Note that the jumps in each of the curves are due to rounding the number of body coils $N_b$ to the nearest half-integer---without this constraint, the curves would be smooth.}
    \label{fig:uppermassvecs}
\end{figure}

We have spent most of our time with the eigenfrequencies of the system, but it is worth also checking how the eigenvectors change as we increase the mass (while keeping our springs within the safety factors that we have chosen). Taking Eqs.~(\ref{eq:vert1}) and (\ref{eq:vert2}), we can solve for the ratio of the eigenvector components of the system for the eigenmodes $\omega_\pm$ corresponding to vertical translations
\begin{equation}
    \frac{\Delta z_2}{\Delta z_1} = \frac{3 k_\mathrm{lower}}{-m_1 \omega_\pm^2 + 3 (k_\mathrm{lower} + k_\mathrm{upper})}.
\end{equation}
This ratio of the vertical displacement of the two stages becomes as shown in Fig.~\ref{fig:uppermassvecs} as the ratio of the upper and lower stage masses varies. As we increase the ratio of the masses in the upper and lower stages, the displacement of the lower stage increases considerably in relation to the the upper stage. As our detectors will be installed on the lower stage (and be sensitive to frictional rubbing), it is preferable to keep the displacement small, while ensuring that we have excellent attenuation. Thus, this brings us back to the choice of $m_1=m_2=20 \, \mathrm{kg}$, where we have this trade-off of low natural frequencies and small relative displacements of the upper and lower stages.

\section{\label{sec:thermal}Thermal Considerations}

We must make sure the heat flow in our vibration decoupler is efficient, such that we do not have any parasitic power sources that will prevent our system from cooling to base temperature (e.g. $8 \, \mathrm{mK}$).

\subsection{\label{sec:thermalcond}Thermal Conductance Between Stages}

In our system, we will be using flexible OFHC (Oxygen-Free High thermal Conductivity) Cu braids for thermal links~\cite{DHULEY201717} and our designed phosphor bronze springs (which themselves will have some thermal conductance). For these metals, the thermal conductivity can be approximated using the empirical Wiedemann--Franz law, which states that the ratio of the electronic contribution to the thermal conductivity $\kappa$ to the electrical conductivity $\sigma$ of a metal is proportional to temperature
\begin{equation}
    \frac{\kappa}{\sigma} = LT,
\end{equation}
where $L$ is the Lorenz number
\begin{equation}
    L \equiv \frac{\pi^2}{3} \left(\frac{k_B}{e} \right)^2 = 2.44\times10^{-8} \frac{\mathrm{V}^2}{\mathrm{K^2}}.
\end{equation}
For the electrical conductance of a metal, one can generally use a measured value at room temperature and a measured value at a few K. Usually this ratio is defined using the electrical resistivity $\rho = 1/\sigma$, with the Residual-resistance ratio (RRR) defined as
\begin{equation}
    \mathrm{RRR} = \frac{\rho_{300 \, \mathrm{K}}}{\rho_{0 \, \mathrm{K}}}.
\end{equation}
Practically, the resistivity of a given sample will be measured at a few K, and the RRR will be determined at this nonzero temperature. Notably, in superconducting metals, the RRR would be defined using the resistivity just above the superconducting transition temperature, as $\rho = 0$ below this. For us, we will use OFHC Cu and CDA 510 phosphor bronze, which each have measured RRRs of 77~\cite{DHULEY201717} and 1.24~\cite{doi:10.1063/1.3402333}, respectively.

For determining the thermal conductance $G$ of a metal, we use the Wiedemann--Franz law, measured RRR values, and that the conductance is related to the conductivity via $G= \kappa A / \ell$, where $A$ is the area through which heat flows (e.g. the cross sectional of a wire) and $\ell$ is the thickness of the metal through which heat is flowing (e.g. the length of a wire). Putting this together, we have that thermal conductance of a normal metal at low temperatures is
\begin{equation}
    G = \frac{L A \cdot \mathrm{RRR}}{\rho_{300 \, \mathrm{K}} \ell} T.
\end{equation}
Note that, in the case of Cu braids, there may be some empty space between the braided wires. Thus, the actual cross sectional area $A$ is smaller than what can be measured from the diameter of the braid. This is parameterized by the porosity $p$, where the true cross section for single braid is given by $A = p \pi r^2$, where $r$ is the radius of the braid. In Ref.~\cite{DHULEY201717}, the porosity of Cu braid was given to be $p=0.645$, which we will use in our design.

\begin{table}
    \centering
    \caption{(Values from Ref.~\cite{ekin:2006}) The thermal conductances of various interfaces at different temperatures and the empirical exponent to be use with Eq.~\ref{eq:thermcontact}.}
    \begin{tabular}{ccccc}
    \hline \hline
    Interface Materials & $G\,[\mathrm{W}/\mathrm{K}]$ at $4.2 \, \mathrm{K}$ & $G\,[\mathrm{W}/\mathrm{K}]$ at $77 \, \mathrm{K}$ & $y$ & Ref. \\ \hline
    Au/Au & 0.2 & --- & 1.3 & \cite{berman:1958} \\
    Cu/Cu & 0.01 & 0.3 & 1.3 & \cite{berman:1958,berman:1956} \\
    Steel/steel & 0.005 & 0.3 & --- & \cite{berman:1956} \\
    \hline \hline
    \end{tabular}
    \label{tab:contactconductance}
\end{table}

To thermally connect our Cu braids and springs to the various stages of the passive vibration isolation system, the means of this will be achieved through solid-solid interfaces. Interestingly, the thermal conductance through a solid-solid interface is not dependent on the area of contact, but instead on the force being applied to the interface. This is because the microscopic roughness of the surfaces means that that there are only points of contact, thus adding force creates a better thermal link by increasing the number of points of contact. As discussed by Ekin~\cite{ekin:2006}, the thermal conductance of a solid-solid interface can be calculated via the empirical expression
\begin{equation}
    G(F,T) = G(F=445\, \mathrm{N}, T = 4.2 \, \mathrm{K}) \frac{F}{445\, \mathrm{N}} \left(\frac{T}{4.2 \, \mathrm{K}} \right)^y,
    \label{eq:thermcontact}
\end{equation}
where $F$ is the force between the two surfaces, $T$ is the temperature of the system, and $y$ is a dimensionless constant with values between 1.3 and 3, depending on the material. For pertinent values for Au/Au, Cu/Cu, and steel/steel interfaces, we have reproduced the table from Appendix A2.3 of Ref.~\cite{ekin:2006} in Table~\ref{tab:contactconductance}, dropping the row on sapphire/sapphire. Further experimentally-measured values have been compiled in Ref.~\cite{DHULEY2019111} for various other interfaces.

To create contact between the Cu braids and the different stages, one generally uses screws to create a large contact force. For our springs, we will be using stud anchors, where the weight of the stages themselves will be shown to be sufficient for a good thermal connection. Because these contact forces are in series with the metals, as well as parallel conductances through the springs and the braids, the thermal system will become more complex. In studying this system, we refer back to the relations between thermal conductances in series and in parallel
\begin{align}
    \frac{1}{G_\mathrm{series}} &= \frac{1}{G_1} + \frac{1}{G_2} + \cdots,\\
    G_\mathrm{par.} &= G_1 + G_2 + \cdots.
\end{align}
When keeping track of the various thermal conductances, we will ensure we follow these relations such that our power loads are modeled correctly.

To calculate steady-state power loads after all the thermal conductances have been estimated, we use the heat flow equation between two bodies of different temperatures
\begin{equation}
    P = \int_{T_\mathrm{low}}^{T_\mathrm{high}} \mathop{dT'} G(T').
\end{equation}

\subsection{Stiffness of Copper Braids}

Because these braids will be a part of the passive vibration isolation system, the stiffness of the braids may have a nonnegligible effect on the spring constants of our springs. The braids will connect between each stage, and thus can be thought of as its stiffness is in parallel in the various spring rates of our phosphor bronze springs. Spring rates in parallel are governed by
\begin{equation}
    k_\mathrm{par.} = k_1 + k_2 + \cdots,
\end{equation}
which implies that the stiffness of the copper braid must be much less than the spring rates of the phosphor bronze springs in order to negligibly affect the natural frequencies.

As discussed in Section 5.1.6.1 of Ref.~\cite{zhao:2019}, the simplest idealization of thermal link stiffness is to model the braid as a fixed-free cantilever beam (i.e. one end of the beam is fixed and one is free). When the braid is bent at with a force $F$ at its free end, the deflection of that end is
\begin{equation}
    x = \frac{F\ell^3}{3 E I},
\end{equation}
where $E$ is the elastic modulus and $I$ is the second moment of area. For a bundle of small wires (e.g. a braid), the second moment of area would be given as
\begin{equation}
    I_\mathrm{braid} = \frac{N \pi d^4}{64}, 
\end{equation}
where $N$ is the number of wires and $d$ is the wire diameter. Thus, the stiffness $k$ in this idealized case can be estimated as
\begin{equation}
    k^\mathrm{ideal}_\mathrm{braid} = \frac{3EN\pi d^4}{64 \ell^3}.
\end{equation}
Unfortunately, this idealized case often does not apply, as the friction between wires has a significant effect on link stiffness. Furthermore, the Cu in the braids will build up oxide layers on its surface over time (when, e.g., the dilution refrigerator is open to ambient air), which can increase friction and create oxide bonds between wire. However, we can still use this for some intuition, such that the stiffness of a braid should decrease with length, less braids should give lower stiffness, and smaller diameter braids should also give lower stiffness. Knowing that the internal friction of a braid will increase the stiffness, a bent braid should also have a large radius (e.g. less friction between internal strands). Lastly, the braids will need to be attached to a small Cu block, which can be screwed into our stages for thermal connection between surfaces, and the braid should have a minimal bend at the Cu block, or otherwise there would be increased stiffness (and a chance for failure). In Section~\ref{sec:geomconstraints_straps}, we will use our geometric constraints of our system to ensure that we design thermal straps that fulfill each of these ideas.

The theoretical studies of metallic strands of wires have been undertaken over the years using Euler-Bernoulli beam theory, e.g. in the textbook \emph{Theory of Wire Rope}~\cite{costello:1997} and more recently in Ref.~\cite{FOTI20161}. However, the complexity of these analyses and models (as well as the dependence on the specific braid design) prevent us from easily calculating the expectations of the stiffness for different thermal links made of braids. Thus, it is better to empirically measure the stiffness of the braid, which a manufacturer of thermal links (e.g. Technology Applications, Inc.~\cite{techapps}) offer as part of the fabrication process (they have also made many thermal links in the past, and thus have many results from previous measurements).

\subsection{\label{sec:geomconstraints_straps}Geometric Constraints}

\begin{figure}
    \centering
    \includegraphics[width=0.4\linewidth]{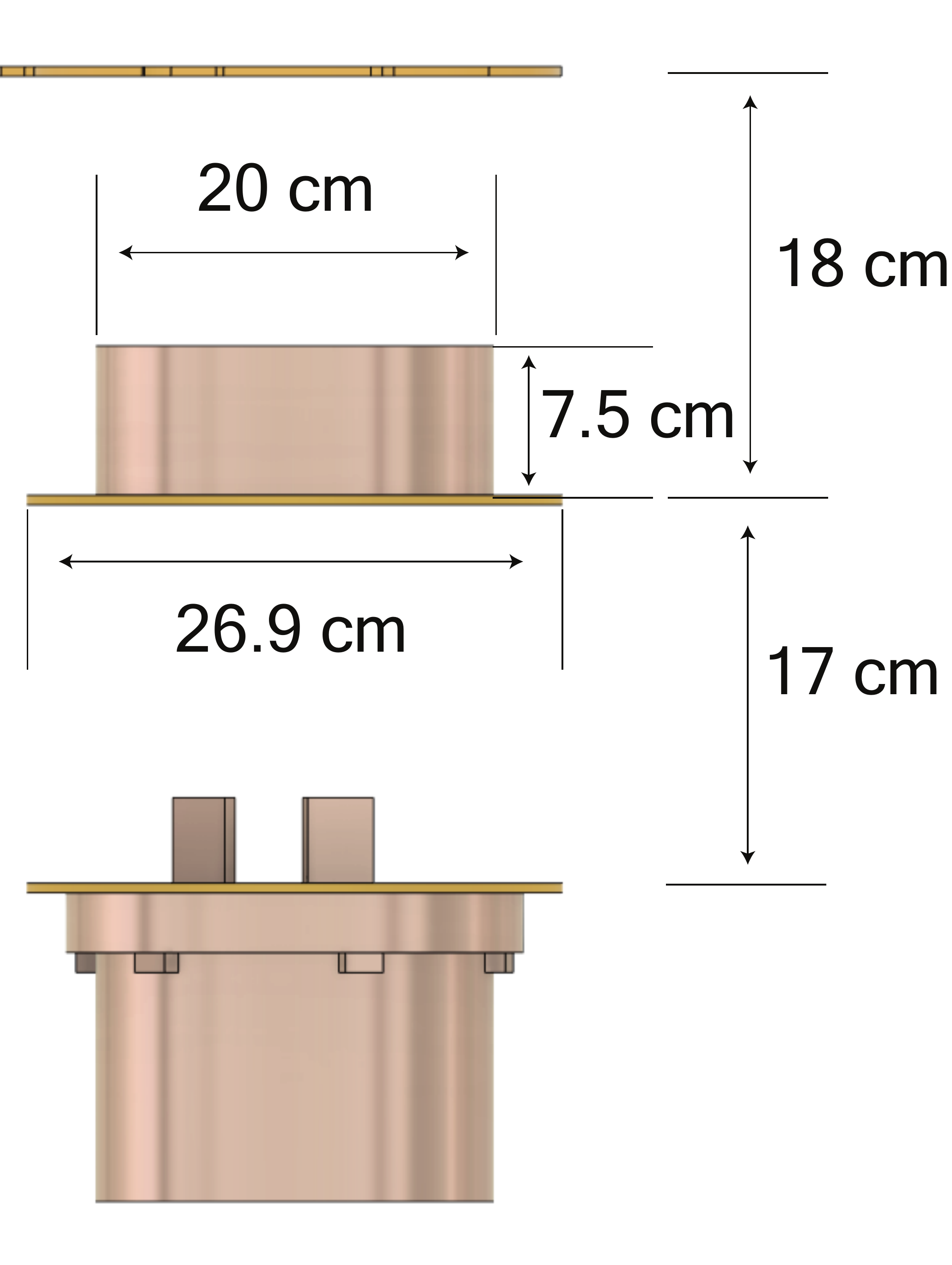}
    \caption{The various pertinent dimensions of a prototype design, where the spacing of the the stages is determined by the spring lengths in Table~\ref{tab:twostage_params}. The top plate is the mixing chamber plate (i.e. what the passive vibration isolation system attaches to). }
    \label{fig:geomcons_straps}
\end{figure}

To understand the allowed lengths of our thermal links, we must understand the various dimensions of our system. As we have estimated spring lengths for the upper and lower stages, we can mock up a prototype design, where we will assume a $20 \, \mathrm{kg}$ Cu weight is being used on the upper stage. This Cu weight is not exactly $20 \, \mathrm{kg}$, but has a mass of $21.1\, \mathrm{kg}$ such that the dimensions are simpler to report. The dimensions of this system are shown in Fig.~\ref{fig:geomcons_straps}. Note that the Cu weight has a smaller radius than the stages, which was done to ensure there was room for the $25 \, \mathrm{mm}$ diameter springs.

\begin{figure}
    \begin{subfigure}{.5\textwidth}
        \centering
        \includegraphics[width=1\linewidth]{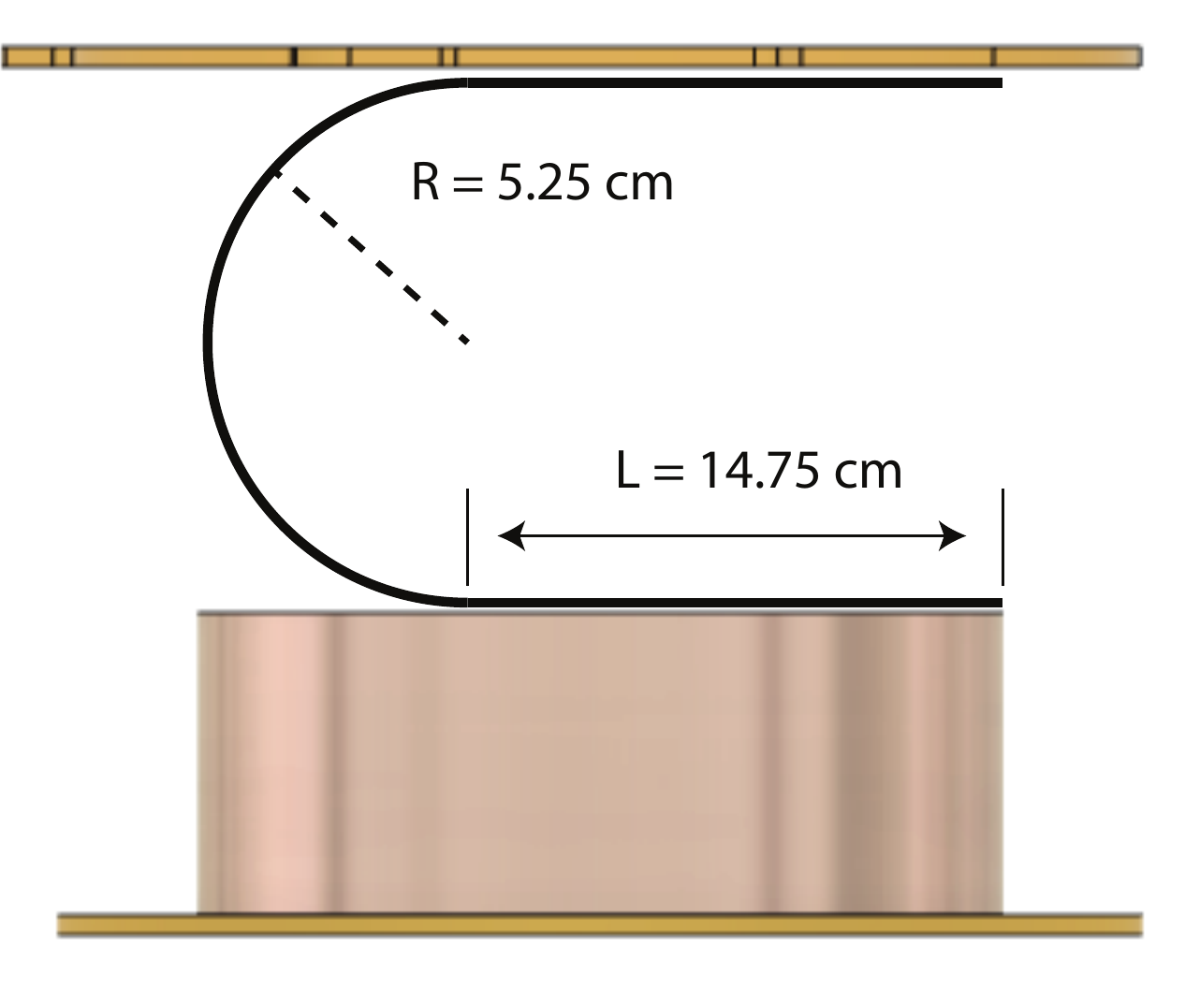}
    \end{subfigure}%
    \begin{subfigure}{.5\textwidth}
        \centering
        \includegraphics[width=1\linewidth]{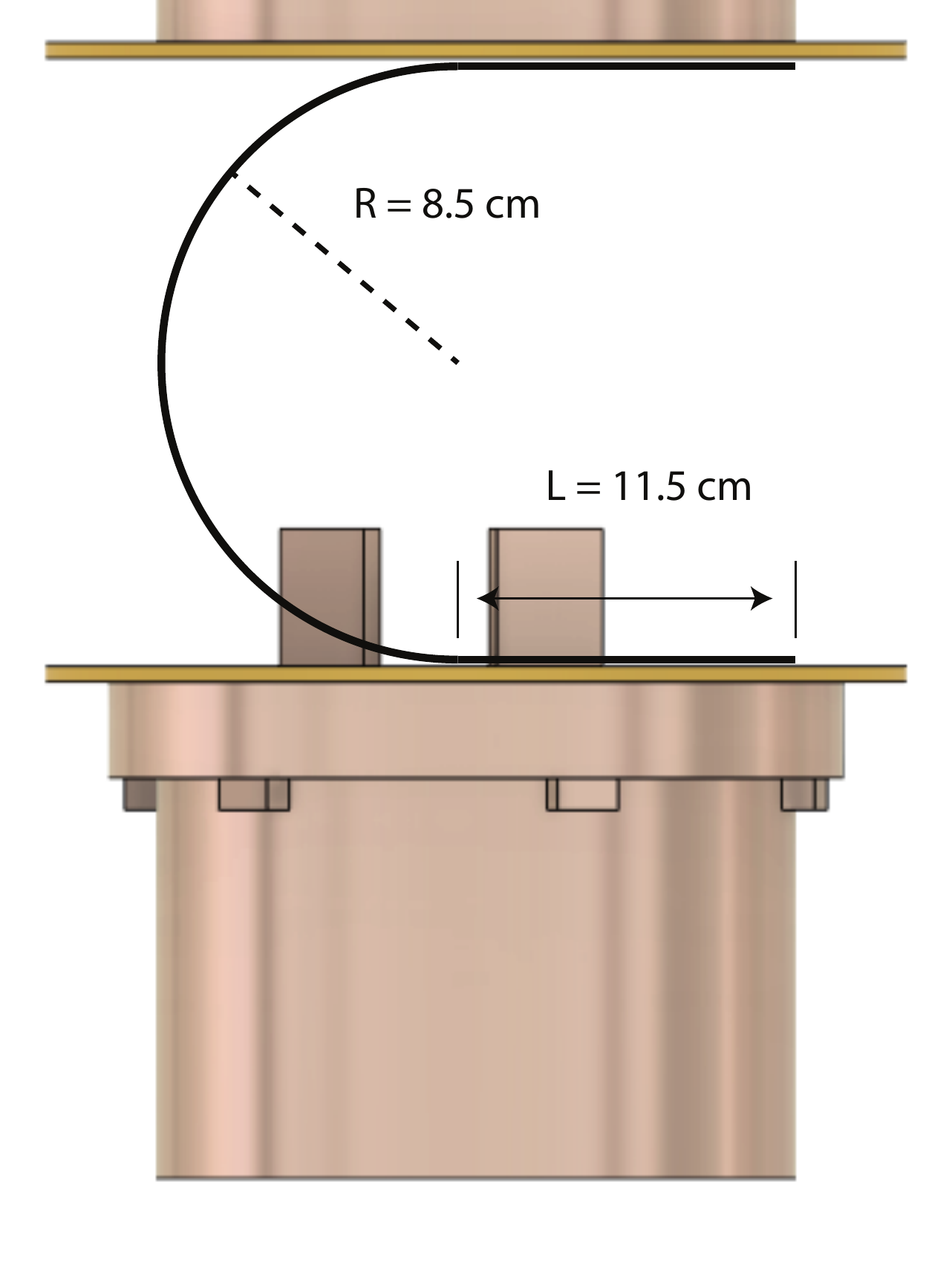}
    \end{subfigure}
    \caption{(Left) The maximum length of a thermal braid which would thermally connect the mixing chamber (top plate) to the first stage (connection via the Cu weight). (Right) The maximum length of a thermal braid which would thermally connect the upper stage (top) to the lower stage (bottom). A prototype design of the IR-tight detector housing is shown connected to the lower stage.}
    \label{fig:thermalstraps_lengths}
\end{figure}

For the Cu braids, we are looking for a tradeoff between excellent thermal conductance and low stiffness. Given the geometric constraints of the system, we can find a range of possible values of the lengths, where we assume that the braids attached between each stage via a semicircle and may have extra length the runs parallel to the stages. To keep the braids from physically touching the springs or the outer cans, we can keep the same horizontal constraints of the Cu weight (cannot go further out that $10\, \mathrm{cm}$ in radius from the center of a stage. In Fig.~\ref{fig:thermalstraps_lengths}, we show the maximum lengths of these braids given the semicircle shape and the maximum extra lengths. The minimum lengths would simply be the semicircle itself, thus giving a range of braid length for the upper stage of 16.5--46.0 cm and for the lower stage of 26.7--49.7 cm. In the next section, we will use these ranges of the lengths to help determine our thermal link design.

\subsection{Proposed Thermal Links}

For a single Cu link, we have that the thermal conductance will given by the conductances of the clamps and the braid added in series
\begin{equation}
    \frac{1}{G_{link}} = \frac{1}{G_{braid}} + \frac{1}{G_{clamp}} + \frac{1}{G_{clamp}},
\end{equation}
which gives the thermal conductance of a single link as
\begin{equation}
    G_{link} = \frac{G_{braid}G_{clamp}}{G_{clamp} + 2 G_{braid}}.
\end{equation}
Thus, we must have $G_{braid} \ll G_{clamp}$ for the thermal conductance of the braid to dominate the total thermal link conductance. Taking the conductance equations from Section~\ref{sec:thermalcond}, we have that for a Cu braid
\begin{align}
    G_{braid} &= \frac{L A \cdot \mathrm{RRR}}{\rho^\mathrm{Cu}_{300 \, \mathrm{K}} \ell} T \nonumber \\
    &= 3.55\times 10^{-4} \frac{n }{\ell} T,
\end{align}
where $n$ is the number of braids and $\ell$ is the braid length. In this, we have used a porosity $p=0.645$, braid radius $r=1.26 \, \mathrm{mm}$, $\mathrm{RRR} = 77$, and $\rho_{300 \, \mathrm{K}}^\mathrm{Cu} = 17.1 \, \mathrm{n}\Omega\mathrm{m}$. Thus, we can only vary the length of the braids and how many there are, as these values are either standard for Cu braids~\cite{DHULEY201717} or physical characteristics of Cu.

For the clamp that will hold the copper, we will use bolted Cu joints, which have been estimated to have a force of $F=3 \, \mathrm{kN}$~\cite{SCHMITT201541}. We will also assume that we have gold-plated the clamps, such that we can use the Au/Au connection values in Table~\ref{tab:contactconductance}. Thus, for one the clamps in the thermal link, we have that
\begin{equation}
    G_{clamp} = 0.2 \cdot \frac{3000}{445} \left(\frac{T}{4.2 \, \mathrm{K}} \right)^{1.3}.
\end{equation}

For the springs, we expect that a single spring would have a thermal conductance that follows
\begin{align}
    G_{spring} &= \frac{L A \cdot \mathrm{RRR}}{\rho^\mathrm{PhBr}_{300 \, \mathrm{K}} \ell} T \nonumber \\
    &= 0.275 \cdot \frac{\pi (d/2)^2}{(N_b +2) \pi D} T,
\end{align}
where $d$ is the wire diameter, $D$ is the spring diameter, and $N_b$ is the number of body coils, each of which can be found in Table~\ref{tab:twostage_params}. We can again assume that we have some type of gold-plating for attaching the springs, such that the interfacial conductance will be determined by the weight of the stages
\begin{equation}
    G_{anchor} = 0.2 \cdot \frac{m_{stage}g/3}{445} \left(\frac{T}{4.2 \, \mathrm{K}} \right)^{1.3}.
    \label{eq:auauconductance}
\end{equation}
For each hanging spring, the conductance in series then similarly becomes
\begin{equation}
    G_{hang} = \frac{G_{spring}G_{anchor}}{G_{anchor} + 2 G_{spring}}.
\end{equation}

Thus, for each of the stages, we have that the effective thermal conductances are given by
\begin{align}
    G_{upper} &= G^{upper}_{link} + 3 G_{hang}^{upper} \\
    G_{lower} &= G^{lower}_{link} + 3 G_{hang}^{lower}.
\end{align}
Equivalently, we can estimate the thermal flow from the mixing chamber to the bottom stage as characterized by these two thermal conductances in series
\begin{equation}
    G_{total} = \frac{G_{upper} G_{lower}}{G_{upper} + G_{lower}},
\end{equation}
where we have neglected the thermal conductance of the Cu weight on the upper stage, as it will be much larger (a factor of 50,000 as compared to the braid due to its larger cross sectional area) and will be negligible when added in series.

At this point, we need to understand the tradeoffs between stiffness and thermal conductance. Ideally, a larger thermal conductance would allow our system to cool down efficiently and quickly. However, we know that a large thermal conductance in a braid would mean that they would be very short, and we would need many of them. Both of these would serve to increase the stiffness, which would degrade the performance of the passive vibration isolation. As the goal of this system is to achieve the lowest possible fundamental frequencies (i.e. lowest stiffness), we should be prioritizing the stiffness of the braids over excellent thermal conductance. In this way, we will check first the least stiff scenario, and study if the cooling power of the system is sufficient.

\begin{figure}
    \begin{subfigure}{.5\textwidth}
        \centering
        \includegraphics[width=1\linewidth]{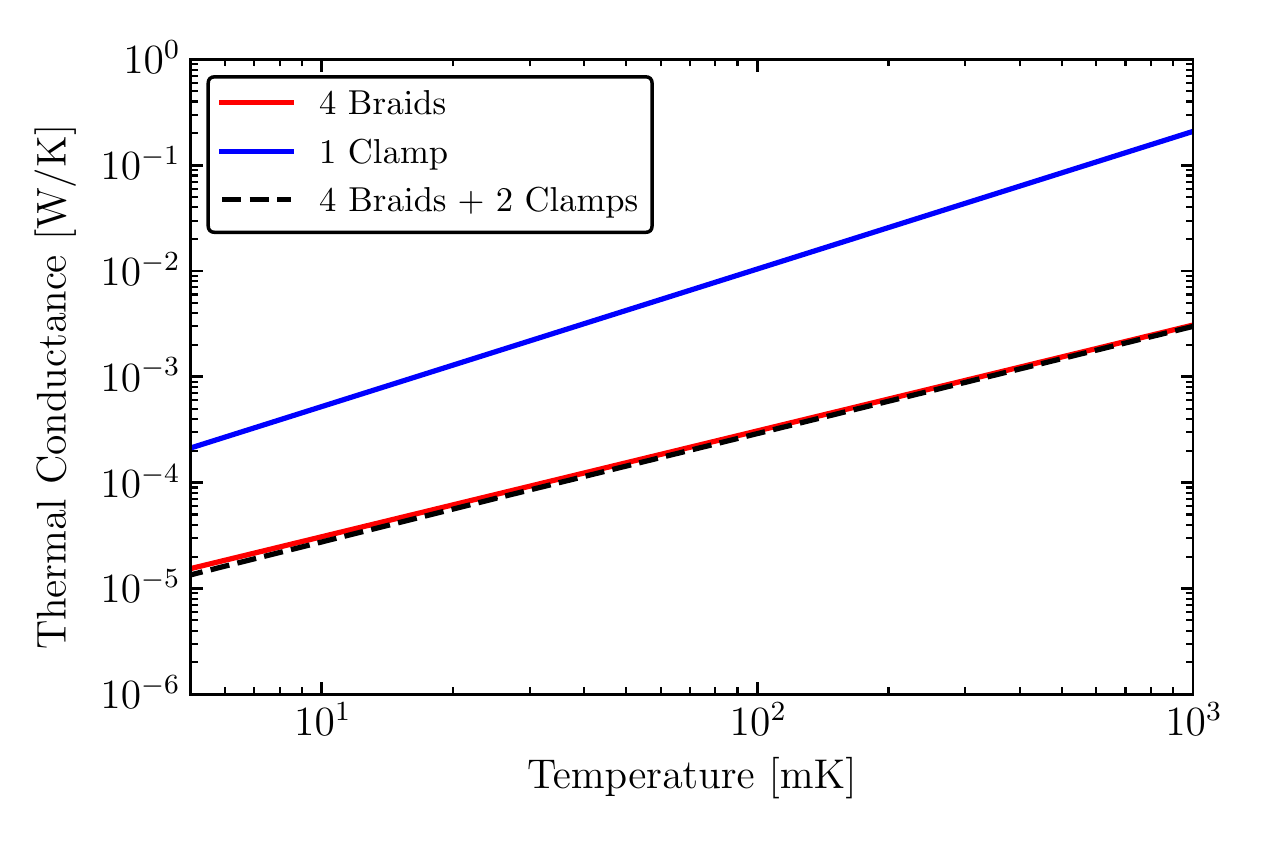}
    \end{subfigure}%
    \begin{subfigure}{.5\textwidth}
        \centering
        \includegraphics[width=1\linewidth]{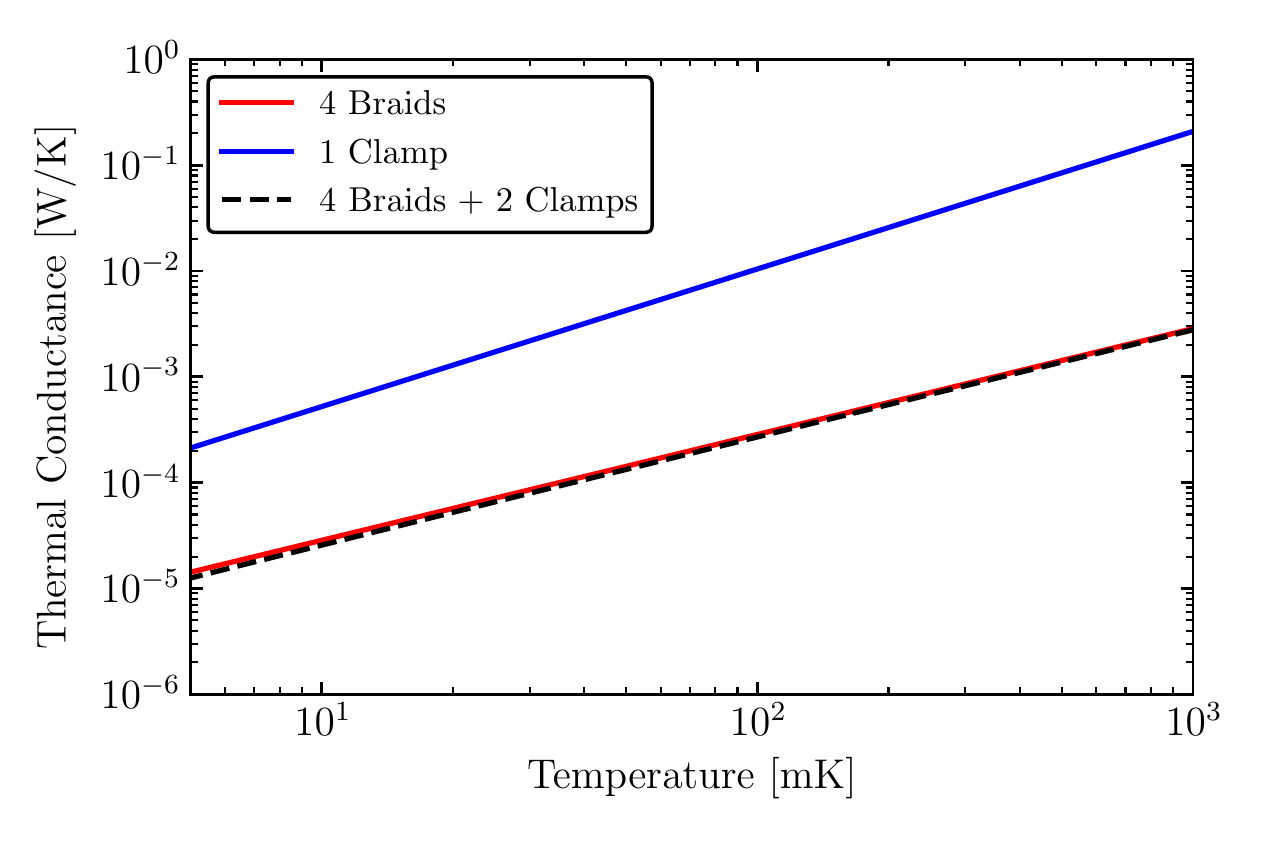}
    \end{subfigure}
    \caption{(Left) For the upper stage, the thermal conductance of the four braids, a clamp, and the braids and clamps in series. The series conductance is solely determined by the braid. (Right)  For the lower stage, the thermal conductance of the four braids, a clamp, and the braids and clamps in series. The series conductance is solely determined by the braid. Because the systems are nearly identical, these two plots are seemingly the same.}
    \label{fig:thermalstrap_conductances}
\end{figure}

The least stiff scenario for the thermal links are to use the longest possible braids and as little as possible. For links from TAI, they appear to manufacture links with four braids as the smallest amount. Thus, we will study the thermal system with the maximum braid lengths ($46.0\, \mathrm{cm}$ for the upper stage and $49.7 \, \mathrm{cm}$ for the lower stage) from the geometric constraints and with four braid. In Figure~\ref{fig:thermalstrap_conductances}, we plot the thermal conductances of the four braids in the link, a comparison to a clamp, and the series conductance of the braids and the two clamps. We see that the braid conductance dominates and the plots for the two stages look identical due to using nearly the same length or braids for both stages.

\begin{figure}
    \centering
    \includegraphics{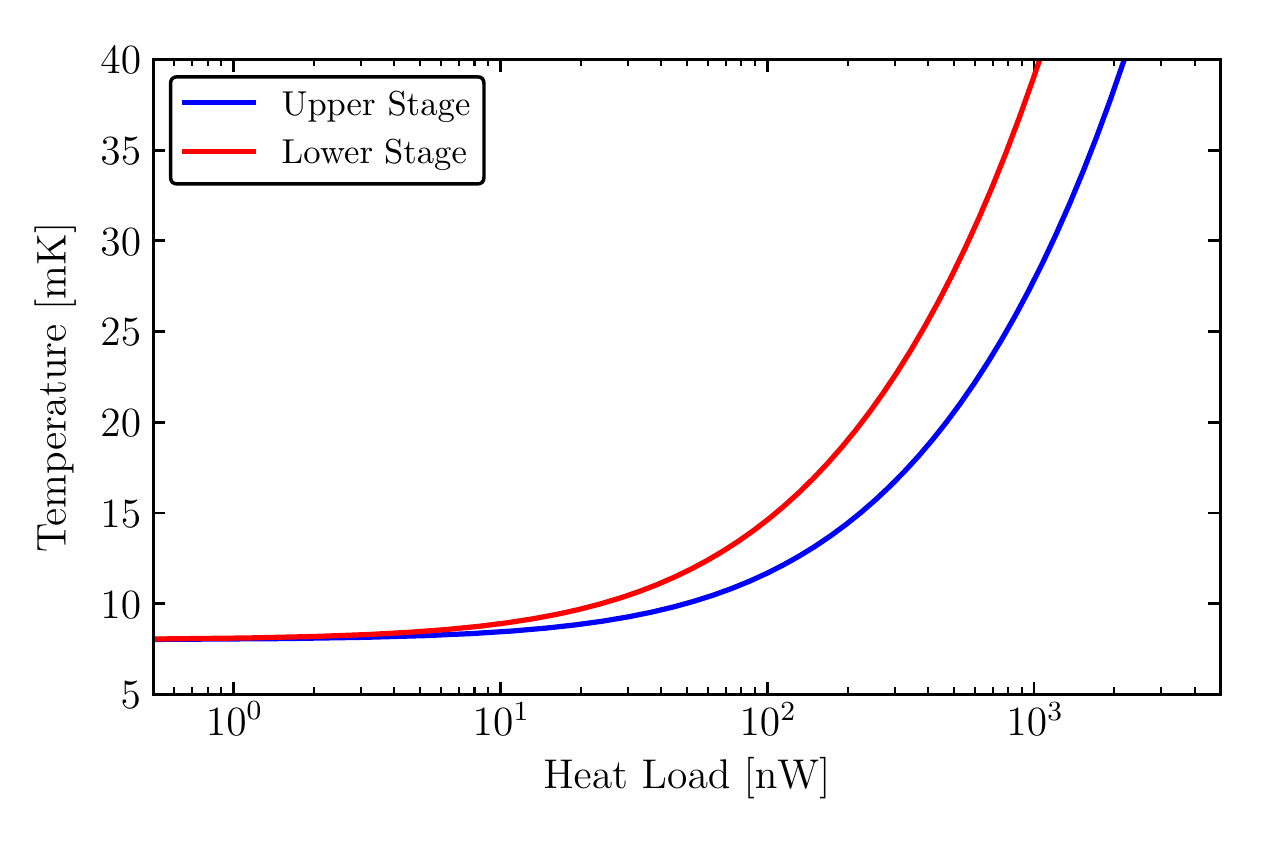}
    \caption{The temperatures of the upper and lower stages for various heat loads given our thermal link design. Note that we would run our detectors on the lower stage.}
    \label{fig:heatload}
\end{figure}

Next, we can check the temperature of each stage under various heat loads for this design. To do this, we assume a mixing chamber temperature of $T_{MC} = 8 \, \mathrm{mK}$, and we can integrate the thermal conductances for each stages to calculate the heat load as a function of temperature. That is, we numerically solve the following equations
\begin{align}
    P(T_1) &= \int_{T_{MC}}^{T_1} \mathop{dT} G_{upper} (T), \\
    P(T_2) &= \int_{T_{MC}}^{T_2} \mathop{dT} G_{total} (T).
\end{align}
In Figure~\ref{fig:heatload}, we plot the temperatures of the stages for various heat loads. Given that our generally have pW-scale or less bias powers, we would not expect heat from the detectors to affect the temperature of the stages, as a temperature change only becomes significant at about $20\, \mathrm{nW}$.

In this discussion, we did not make specific note of the thermal conductance of the springs and anchors. Calculating the thermal conductance of one of the spring assemblies $G_{hang}(T)$, we find that $G_{hang}(10\, \mathrm{mK}) = 8.5 \, \mathrm{nW}/\mathrm{K}$ and $G_{hang}(1\, \mathrm{K}) = 0.85 \, \mu\mathrm{W}/\mathrm{K}$. Comparing to Fig.~\ref{fig:thermalstrap_conductances}, this is about four orders of magnitude less than the thermal conductance of the link $G_{link}$, such that it is negligible when added in parallel for each stage. Nonetheless, the thermal conductances of the springs were included in the calculation of the heat loads.
\begin{figure}
    \centering
    \includegraphics{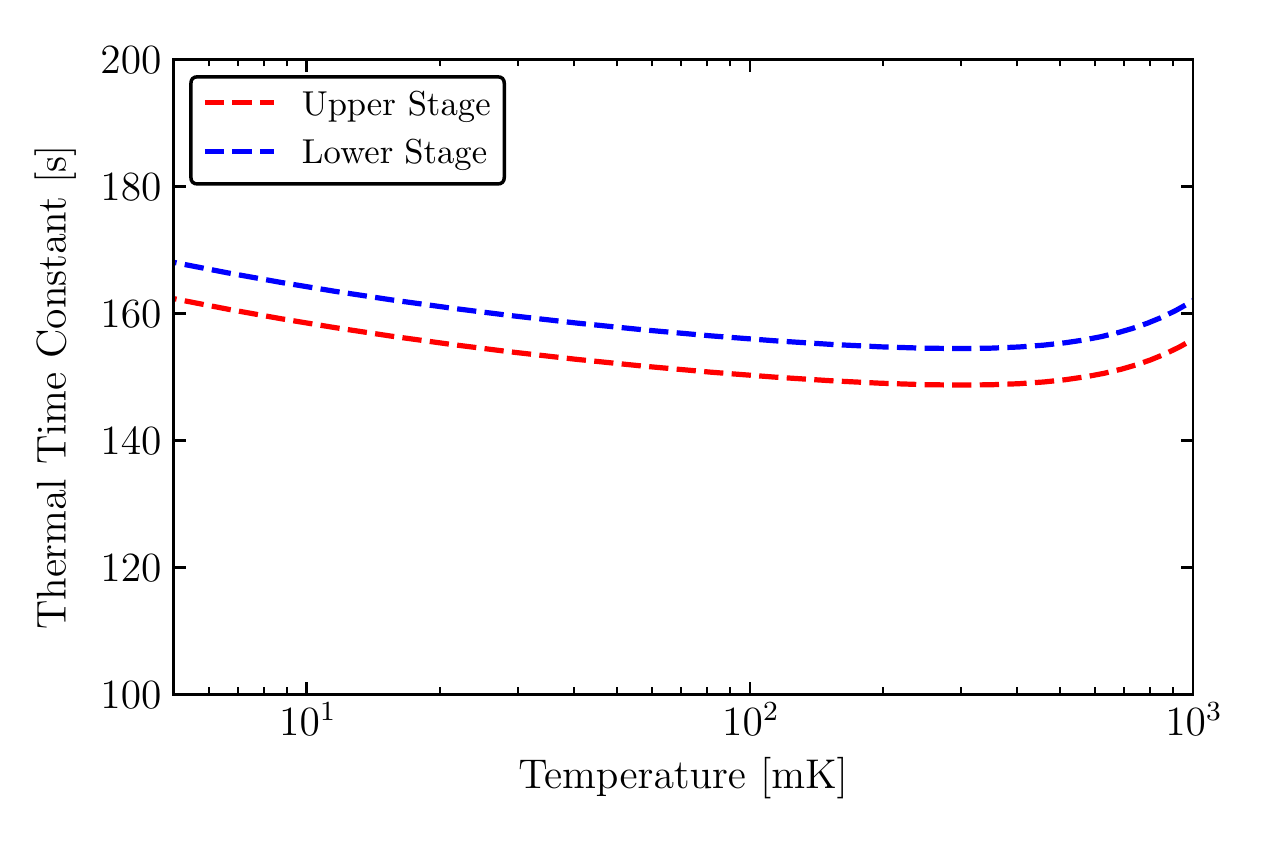}
    \caption{Thermal time of the upper and lower stages as a function of temperature. Both stages have roughly constant time constants on the order of $160 \, \mathrm{s}$.}
    \label{fig:stagetimeconstants}
\end{figure}

We can also check the cooling times of the system to ensure that the stages do not take an unreasonable amount of time to cool down (e.g. a day long cool down time would mean we would have to wait a week for the detectors to fully cool down). The thermal time constant is a good approximation of this, which is given by
\begin{equation}
    \tau(T) = \frac{C(T)}{G(T)},
\end{equation}
where $C$ is the heat capacity and $G$ is the thermal conductance. We can estimate the low-temperature heat capacity of our masses as Cu blocks with
\begin{equation}
    C(T) = \frac{m}{M} \left(\gamma T + A T^3 \right),
\end{equation}
where $m$ is the mass of the body being cooled (e.g. one of the stages), $M$ is the molar mass of the material, $\gamma$ is the electronic heat capacity coefficient, and $A$ is the lattice heat capacity coefficient. Generally, $A$ is parameterized in terms of the Debye temperature $\theta_D$ via $A = 12 \pi^4 R / (5\theta_D^3)$, where $R=8.31\, \mathrm{J}/(\mathrm{mol} \, \mathrm{K})$ is the ideal gas constant. At very low temperatures, the lattice heat capacity term ($T^3$) is usually negligible, but will have some effect at about $1 \, \mathrm{K}$. For Cu, we will use $\gamma = 0.695 \, \mathrm{mJ}/(\mathrm{mol} \, \mathrm{K}^2)$,  $M = 63.55 \, \mathrm{g}/\mathrm{mol}$, and $\theta_D = 344.5 \, \mathrm{K}$~\cite{arblaster2015thermodynamic}. For the masses, we use a $m=40\,\mathrm{kg}$ for total mass hanging from the upper stage and $m=20 \,\mathrm{kg}$ for the total mass hanging from the lower stage. As we have already calculated the thermal conductances for both stages ($G_{upper}$ and $G_{total}$), we can calculate the thermal time constants at low temperatures, as shown in Fig.~\ref{fig:stagetimeconstants}. We see that the time constants are on the order of $160\, \mathrm{s}$, which is fast enough that this thermal linking scheme should not appreciably lengthen the cool down time more than about $15 \, \mathrm{min}$ (whereas we were worried about it being on the order of a day).

The proposed thermal links are such as follows: the upper stage is thermally connected to the mixing chamber by 4 Cu braids which are $46.0 \, \mathrm{cm}$ long, and the lower stage is thermally connected to the upper stage by 4 Cu braids which are $49.7 \, \mathrm{cm}$ long. These designs will keep the thermal links as flexible as possible due to the length and large bend radius, as well as ensure that up to $20 \, \mathrm{nW}$ of heat will not appreciably affect the various stage temperatures, and that the cool down time is not significantly increased.

\subsection{Spring Cross-Checks}

With our spring designs, we have so far neglected a couple of thermal aspects so far. It will be useful to know how long we expect the phosphor bronze springs to remain warm compared to the stages (i.e. the thermal time constant), as well as how much we expect the lengths to contract as we cool to cryogenic temperatures.

\subsubsection{Spring Cool Down Time}

For these springs, it is important that they cool down fast enough such that they are not significantly warmer than the surrounding environment at base temperature. If this were not the case, then the springs could act as a parasitic power source that prevents us from quickly reaching base temperature. To check that this is not the case, we will calculate the thermal time constant $\tau=C/G$ for the upper and lower springs.

\begin{table}
    \centering
    \caption{Volume and masses of the springs for the upper and lower stages of the two-stage system. The masses were determined using a phosphor bronze density of $8800 \, \mathrm{kg}/\mathrm{m}^3$.}
    \begin{tabular}{ccc}
    \hline \hline
    Spring & Parameter & Value \\ \hline
    \multirow{5}{*}{Two Stage, Upper} & $N_b$ & 30.5  \\
     & $D$ & $25\,\mathrm{mm}$  \\
     & $d$ & $4.1\,\mathrm{mm}$  \\
     & $V_{spring}$ & $33.7 \, \mathrm{cm}^3$  \\
     & $m_{spring}$ & $297 \, \mathrm{g}$  \\ \hline
    \multirow{5}{*}{Two Stage, Lower} & $N_b$ & 31  \\
     & $D$ & $25\,\mathrm{mm}$  \\
     & $d$ & $3.2\,\mathrm{mm}$  \\
     & $V_{spring}$ & $20.8 \, \mathrm{cm}^3$  \\
     & $m_{spring}$ & $183 \, \mathrm{g}$  \\
    \hline \hline
    \end{tabular}
    \label{tab:springmasses}
\end{table}

We will use the thermal conductance of the interfacial contact of the spring to the anchor (Eq.~(\ref{eq:auauconductance})) and the heat capacity of the phosphor bronze spring (where we will estimate the heat capacity as that of Cu). The volume of a spring is given by
\begin{equation}
    V_{spring} = (N_b + 2) \pi^2 D \left(\frac{d}{2}\right)^2 
\end{equation}
where $N_b$ is the number of body coils, $D$ is the spring diameter, and $d$ is the wire diameter. For the two-stage system, we report the spring volumes and masses in Table~\ref{tab:springmasses}, which we will use to calculate the heat capacity.

\begin{figure}
    \centering
    \includegraphics{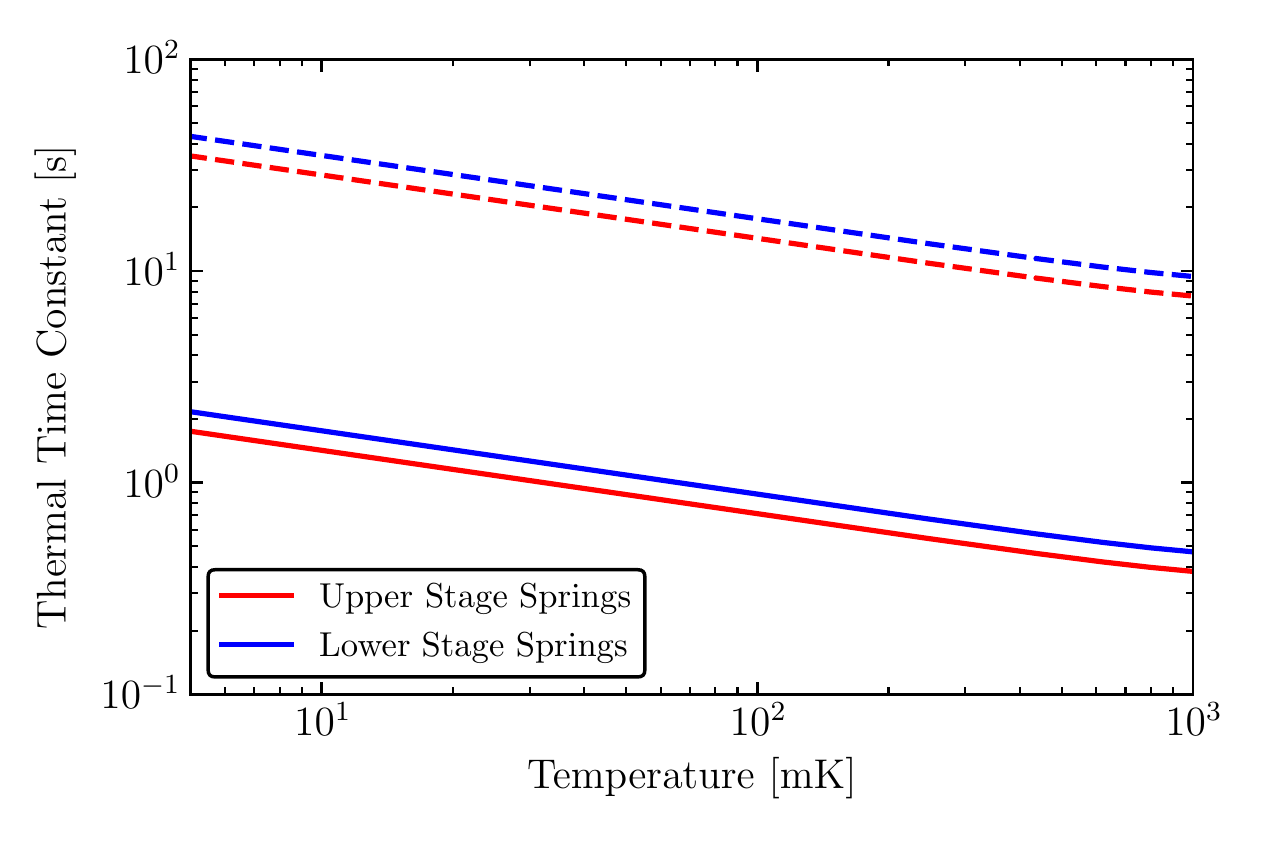}
    \caption{The thermal time constants of the springs in the upper and lower stages of the two-stage passive vibration isolation system. The solid lines represent thermal contact between two Au interfaces, and the dashed lines represent thermal contact between two Cu interfaces. The difference being a factor of 20 is a direct result of the thermal conductance ratios at $4.2\,\mathrm{K}$ and the corresponding empirical formula (see Table~\ref{tab:contactconductance}).}
    \label{fig:springtimeconstants}
\end{figure}

For the thermal conductance, we will use that each spring on the upper stage carries a third of the total mass of the two stages (i.e. the force on each upper spring is $40g/3 \, \mathrm{N}$), and each spring on the lower stage carries a third of the mass of the lower stage (i.e. the force on each lower spring is $20g/3 \, \mathrm{N}$). For the thermal time constant of each spring, we divide the spring heat capacity by the contact thermal conductance of the connection to the upper plate. In Fig.~\ref{fig:springtimeconstants}, the thermal time constants are plotted over the range of $5 \,\mathrm{mK}$ to $1 \, \mathrm{K}$, where the solid lines denote Au/Au contact (gold-plating), and the dashed lines denote Cu/Cu contact (bare Cu). In either case, we see that the cooling down of the springs is less than a minute, with the Au/Au time constant on the order of seconds. As compared to the average dilution refrigerator cool down time over the course of a few days, the springs should not act as parasitic power sources due to these fast time constants.

\subsubsection{Thermal Contraction of Springs}

As the springs cool down to cryogenic temperatures, both the shear modulus $G$ of the springs will change, as well as the coefficient of thermal expansion $\alpha$. We return to the definition of the spring rate $k$
\begin{equation}
    k = \frac{d^4 G}{8D^3N_a},
\end{equation}
where $d$ is the wire diameter, $N_a$ is the number of active coils, and $D$ is the wire diameter. Thus, $D$, $d$, and $G$ will change as we cool down, and so will the spring rate.

For Cu alloys (e.g. phosphor bronze), it has been shown that the elastic modulus $E$ increases by about 9.6\% as the temperatures changes from $300 \, \mathrm{K}$ to $4 \, \mathrm{K}$~\cite{LEDBETTER1982653}. Empirically, the elastic modulus and shear modulus for Cu alloys have been shown to be proportional by $G = 3E/8$~\cite{LEDBETTER1977133}, and we can assume the same percent change is true for the shear modulus.

\begin{figure}
    \centering
    \includegraphics{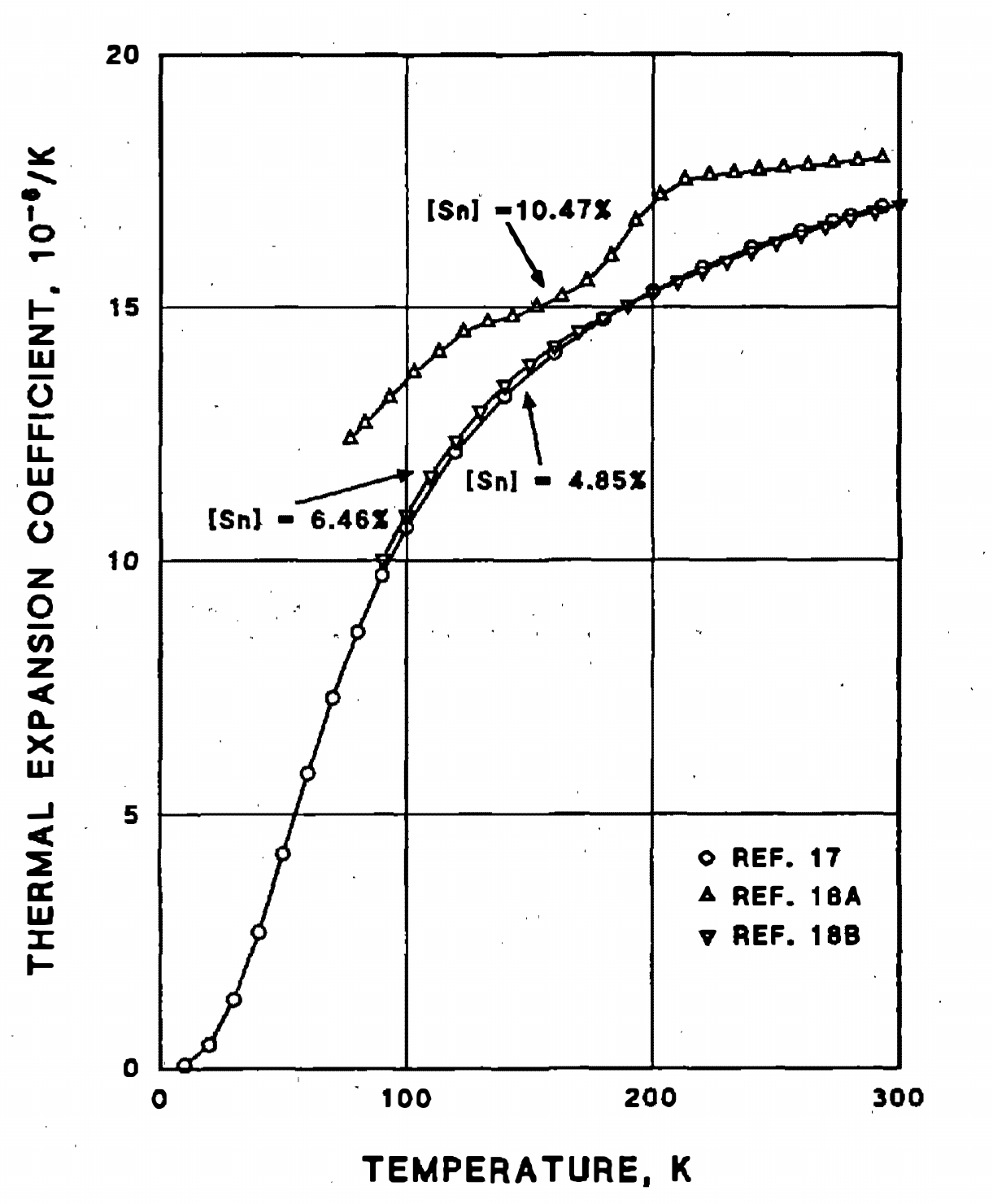}
    \caption{(Figure from Ref.~\cite{osti_5340308}) Coefficient of thermal expansion of phosphor bronze as a function of temperature.}
    \label{fig:cte_phbr}
\end{figure}

To calculate the expected thermal contraction of phosphor bronze after reaching cryogenic temperatures, one can use the coefficient of thermal expansion $\alpha$, which relates an infinitesimal change in temperature $dT$ to an infinitesimal change in length $dL$ via
\begin{equation}
    dL = L_0 \alpha(T) dT,
\end{equation}
where $L_0$ is the initial length of the material. To estimate the percent change in a length dimension for some temperature range, we integrate
\begin{equation}
    \frac{\Delta L}{L_0} = \int_{T_1}^{T_2} \mathop{dT} \alpha(T).
\end{equation}
For phosphor bronze, the coefficient of thermal expansion from $300 \, \mathrm{K}$ to $10 \, \mathrm{K}$ is shown in Fig.~\ref{fig:cte_phbr}. Integrating across this temperature range, we find that phosphor bronze would contract by 0.34\%. Thus, we expect the wire diameter and spring diameters to contract by this same factor. 

Returning to the spring rate and using the expected changes in the various parameters, we have that the spring rate at base temperature can be expected to be
\begin{align}
    k_{base} &\approx \frac{1.096 (0.9966)^4}{(0.9966)^3} k_{300 \, \mathrm{K}} \\
    k_{base} &\approx 1.092 k_{300 \, \mathrm{K}},
\end{align}
such that the spring rate increases by about 9.2\%. As the natural frequencies of our spring modes are proportional to $\sqrt{k/m}$, we expect that these frequencies would increase by about $4.5\%$ (where this increase is the same regardless of the spring design). For example, using the values in Table~\ref{tab:twostage_params}, the translational spring modes in the $z$-axis would increase from $1.67$ and $4.11 \, \mathrm{Hz}$ to $1.75$ and $4.29 \, \mathrm{Hz}$, respectively. For the rotational spring modes about the $x$ and $y$ axes, these frequencies would increase from $2.37$ and $5.82 \, \mathrm{Hz}$ to $2.48$ and $6.08 \, \mathrm{Hz}$, respectively. These increases are small, so we should expect similar vibration isolation performance at cryogenic temperatures in the spring modes, while knowing that the frequencies will increase by small amount.

For the pendulum modes, we have that they are essentially proportional to $\sqrt{g/\ell}$. As the full extended length depends on both the free length and the deflection via Eq.~(\ref{eq:fullpendulumlength}), which we have reproduced below and simplified using the definition of $k$
\begin{equation}
    l_{max} = 2(D - d) + (N_b + 1)d + \frac{F_{load} - F_i}{k}.
\end{equation}
Thus, we can plug in our values from Table~\ref{tab:twostage_params} with and without the expected thermal contractions of the various parameters and calculate the expected change in $l_{max}$ due to thermal contraction of the springs. We expect the stretched lengths to decrease by $0.5\%$ for the upper stage and by $0.9\%$ for the lower stage. Thus, the pendulum natural frequencies will increase by a factor of less than $\sqrt{1/0.99} \approx 1.005$, or an increase of less than 0.5\%. This change is negligible, and we should have effectively the same performance for the pendulum modes.

\section{Intermission}

\begin{figure}
    \centering
    \includegraphics[width=0.33\linewidth]{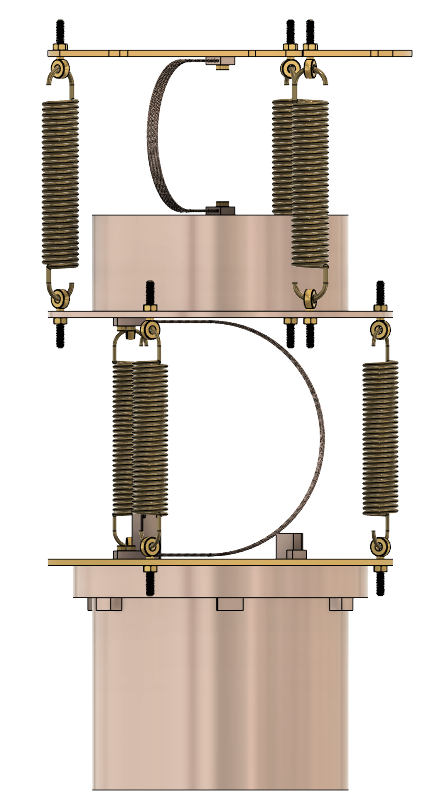}
    \caption{In-progress passive vibration isolation system design as discussed so far in this appendix. Included are the spring designs, the expected plates for each stage, the thermal link designs, and the the various ways of creating connections between interfaces.}
    \label{fig:full_nosafety}
\end{figure}

With our springs and thermal links designed, we can now put together an in-progress design in CAD, the full tentative system being shown in Fig.~\ref{fig:full_nosafety}. Note that the spring hooks in the design make a right angle, which is not the actual proposed design, but was done to save time creating the springs in CAD. To show the proposed connections for the springs and the thermal links, we take a short break from design to help envision what we have actually proposed so far.

\begin{figure}
    \begin{subfigure}{.5\textwidth}
        \centering
        \includegraphics[width=0.8\linewidth]{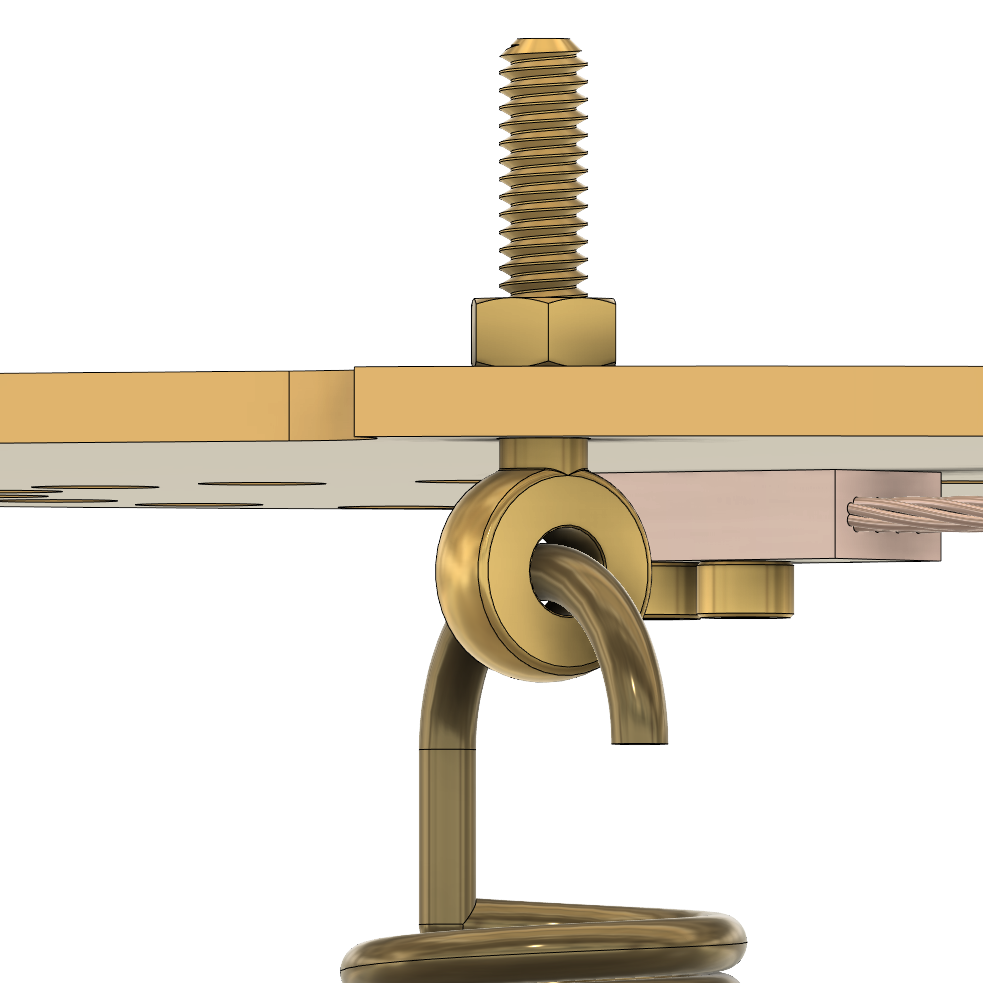}
    \end{subfigure}%
    \begin{subfigure}{.5\textwidth}
        \centering
        \includegraphics[width=0.8\linewidth]{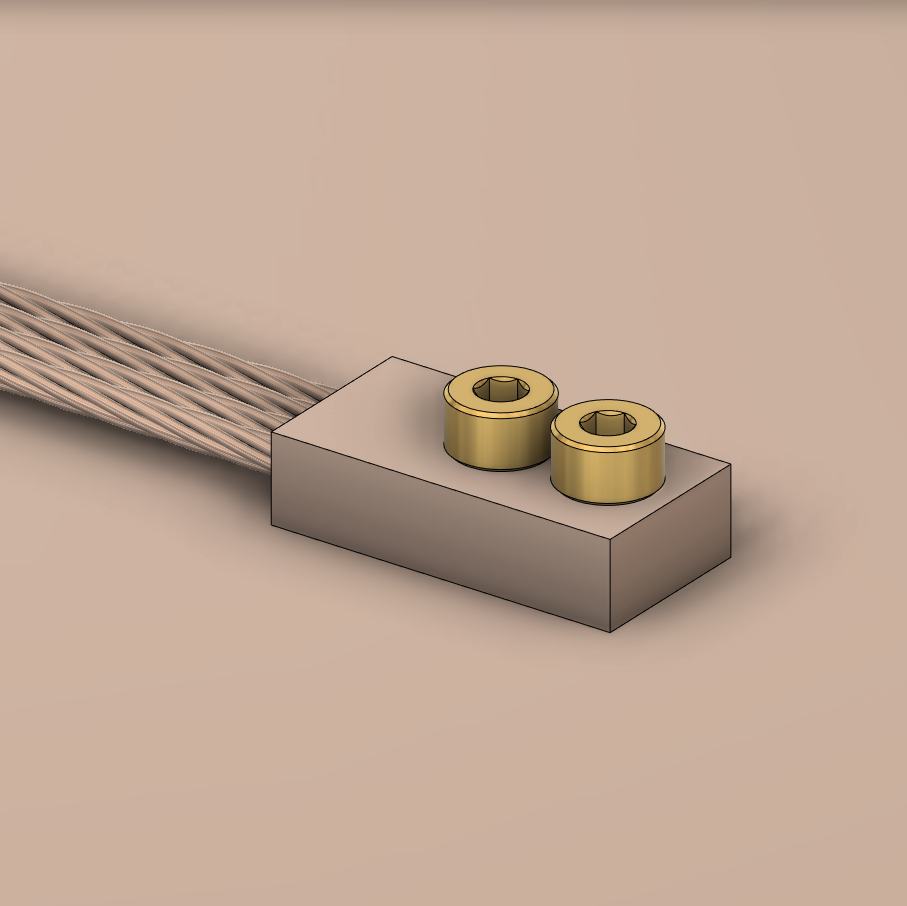}
    \end{subfigure}
    \caption{(Left) Connection of a spring to a plate via a brass eye bolt, which goes through a clearance hole in the plate. A hex nut is threaded onto the bolt to carry the weight of the stages, also allowing angular alignment of the eye bolts. (Right) Simple connection to the first stage, where two M4 brass screws are screwed into the copper blocks of the thermal links and into blind holes in the copper mass. On the left hand side of the subfigure, the expected flexible braids are seen.}
    \label{fig:connections}
\end{figure}

For the various connections, we can use off-the-shelf brass bolts from McMaster-Carr Supply Company~\cite{mcmaster}. In Fig.~\ref{fig:connections}, we show the proposed designs for creating the thermal connections with both the springs and the thermal links we have designed. Note that the design for the spring hooks (based on brass eye bolts) has clearance holes through the plates, such that the weight is held through the brass hex nut, which is in contact with plate through the mass of the stages. The reason for the clearance hole is to allow alignment of the eye bolts on the top and bottom of each stages, such that the springs are not subject to a twisting torque (which would be possibility if the holes in the plates were threaded and the different eye bolts has some manufacturing variability in their threads).

The first stage Cu plate and its Cu weight have been combined into one part, which avoids and interface between the two, and simplifies alignment of the weight to the center of mass. Because the design has a few blind holes at the interfaces of the thermal links, there may be a possibility of virtual leaks (i.e. air pockets in the blind holes could be created under vacuum, which would slowly leave, creating the impression of small leak in the vacuum). To mitigate this, one could drill holes through the center of the screws, which would give a simple path for the air to be pumped out. On the other hand, these air pockets will be small and not within the detector housing, such that it may be negligible for overall detector performance.

There is one important feature left to design for this system, which is a safety structure in case of catastrophic failure at one of the interfaces. That is, if a spring fails, then it would be unwise to allow $40\,\mathrm{kg}$ of Cu to fall to the bottom of a dilution refrigerator (this would cause immense damage). Thus, we will finish off this design appendix with a proposed safety structure to avoid such a catastrophe.

\section{\label{sec:safetystruc}Safety Structure}

The safety structure design for this system needs to be able to catch either the upper or lower stage in case of failure, which can be done by creating a rigid structure below both stages. There are many different possibilities for this type of system, where the principles are less defined than when we were designing the springs and the thermal links. Instead, we will stick to simple principles of ensuring each stage would be caught and would not fall far.

\begin{figure}
    \centering
    \includegraphics[width=0.33\linewidth]{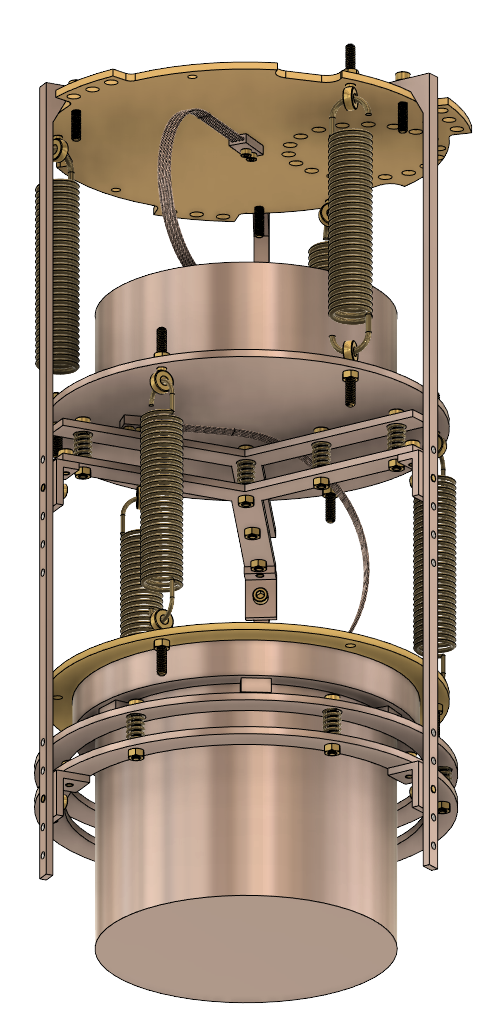}
    \caption{The full design of the passive vibration isolation system with the proposed safety structure.}
    \label{fig:safety_full}
\end{figure}

To keep with the cylindrical symmetry of the design, the safety structure itself will also be cylindrically symmetric. For the upper stage, the springs will block any type of circular design. In contrast, for the lower stage, the IR-tight housing constrains the design radially, such that as circular design will be necessary. In Fig.~\ref{fig:safety_full}, we show such a proposal, for which we will go into more detail in the following paragraphs. The idea follows that we use three Cu pillars to hold the safety structure for each stage, where the upper stage is held by a Y-shaped structure, and the lower stage is held by an O-shaped safety structure.

\begin{figure}
    \centering
    \includegraphics[width=0.8\linewidth]{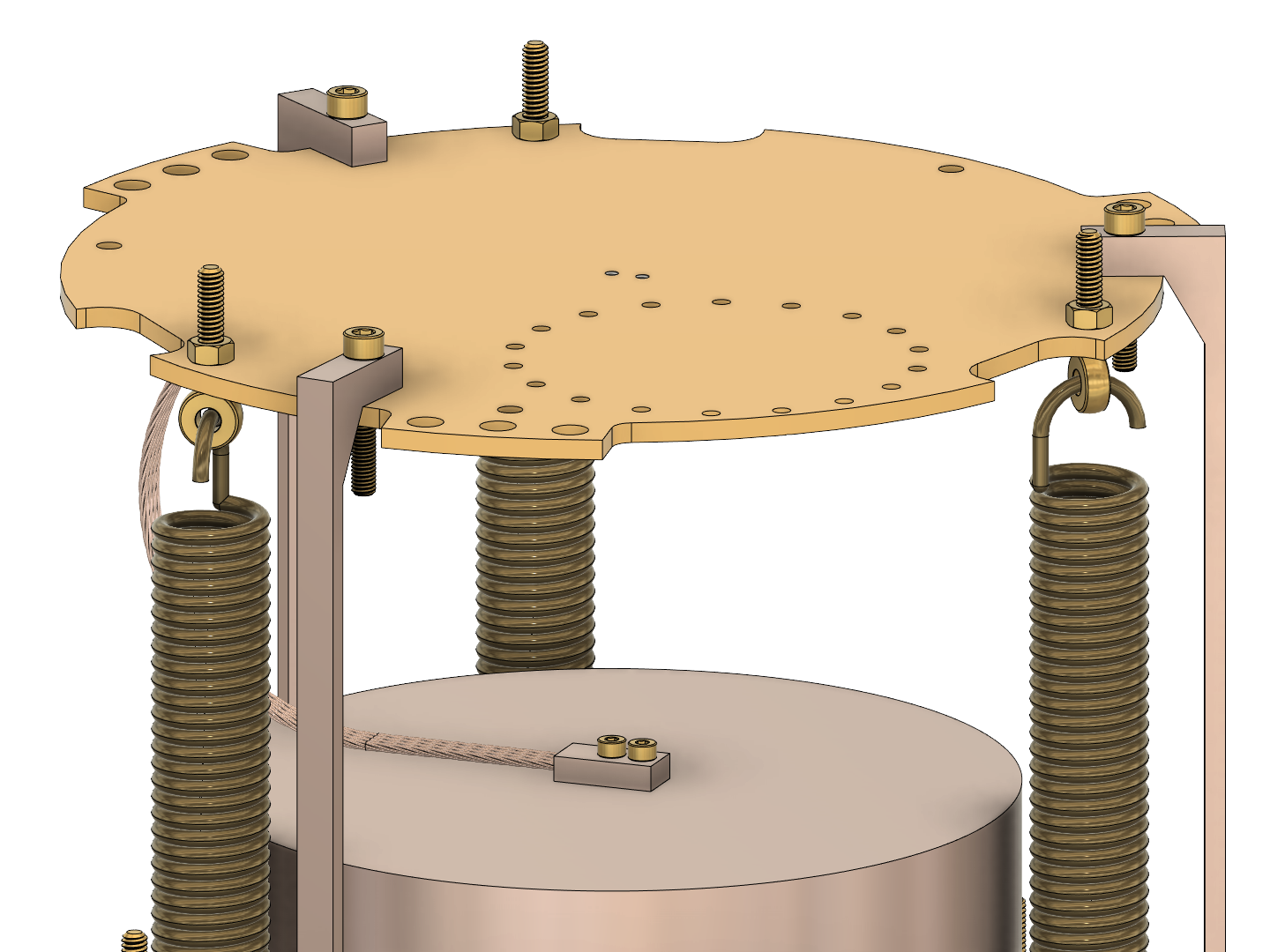}
    \caption{Connection of the safety structure to the mixing chamber plate, such that it is rigidly and thermally connected at three points.}
    \label{fig:safety_top}
\end{figure}

At the mixing chamber plate (Fig.~\ref{fig:safety_top}), the connection of the Cu pillars is given by brass bolts, which are installed through threaded holes in both the pillar connection and the mixing chamber plate. This allows for the equal spacing of $120^\circ$ for each pillar. These pillars go out only as far as the largest radial dimension of the mixing chamber plate, such that the design stays within the radial constraints of the dilution refrigerator. We have also added a chamfer to the corner of the pillars in order to improve the strength at that edge, as opposed to having an abrupt right angle. These pillars then continue downwards to hold the safety structure design for the upper stage.

\begin{figure}
    \centering
    \includegraphics[width=0.8\linewidth]{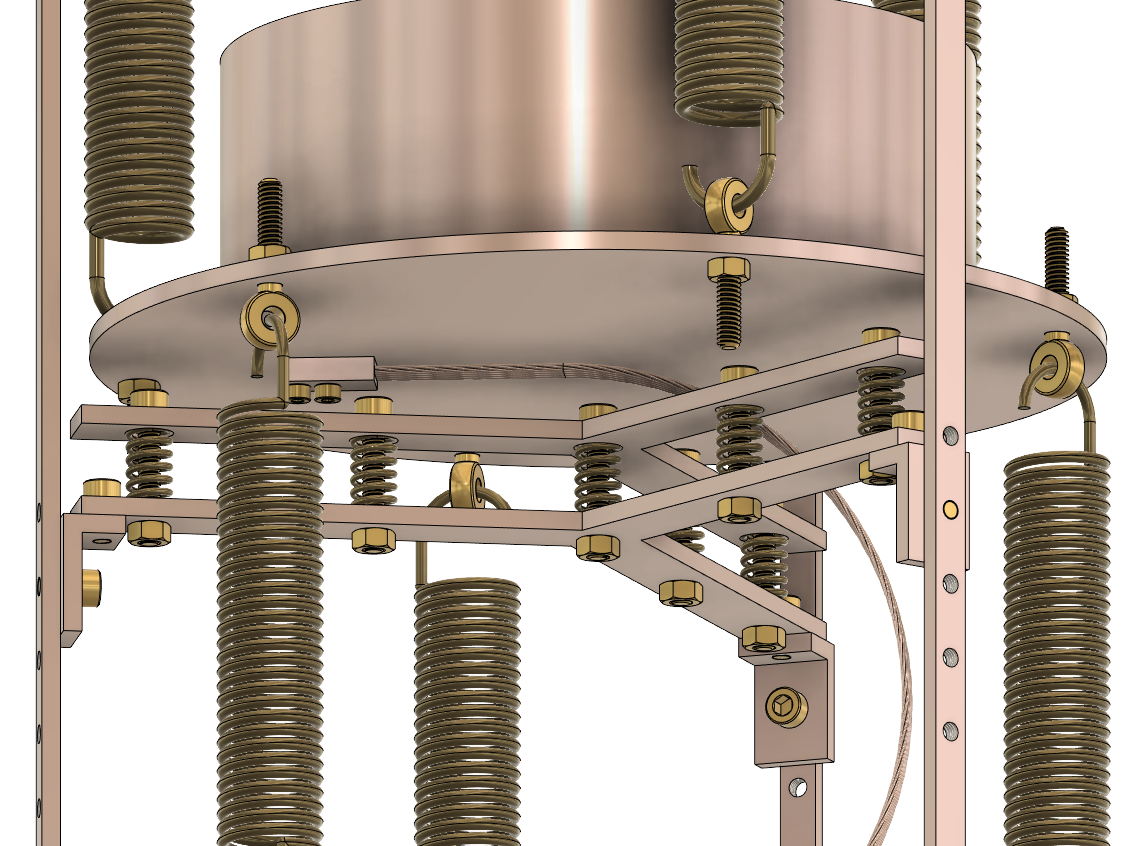}
    \caption{The Y-shaped safety structure to catch the upper stage in case of failure. There are seven phosphor bronze compression springs which will reduce the impulse of the falling mass on the structure.}
    \label{fig:safety_middle}
\end{figure}

To rigidly hold the Y-shaped safety structure at the upper stage, L-shaped brackets are used with tapped holes, as shown in Fig.~\ref{fig:safety_middle}. These brackets can be installed at various heights, depending on any variation in weight of the upper stage, or on how much vertical space is desired between the two. The Y-shaped structure itself is made of Y-shaped Cu plates, which have inset regions for holding seven phosphor bronze compression springs. These springs reduce the impulse on the safety structure, such that there is a more gradual stopping of the weight, rather than being instantaneous. These springs can be bought off-the-shelf from McMaster-Carr, such that they are cheap to replace if they break after a failure. To hold the springs in place and in tension, the top Y-plate is threaded, and the bottom has a clearance hole. The bolts are screwed in to the top plate, travel through the clearance hole, and then each spring can be put under tension by tightening with a nut on the bottom. The main reason to keep these in tension is to create a thermal connection between the spring and the Cu plates, as the interfacial thermal conductance is related to the force being applied at the interface.

\begin{figure}
    \centering
    \includegraphics[width=0.8\linewidth]{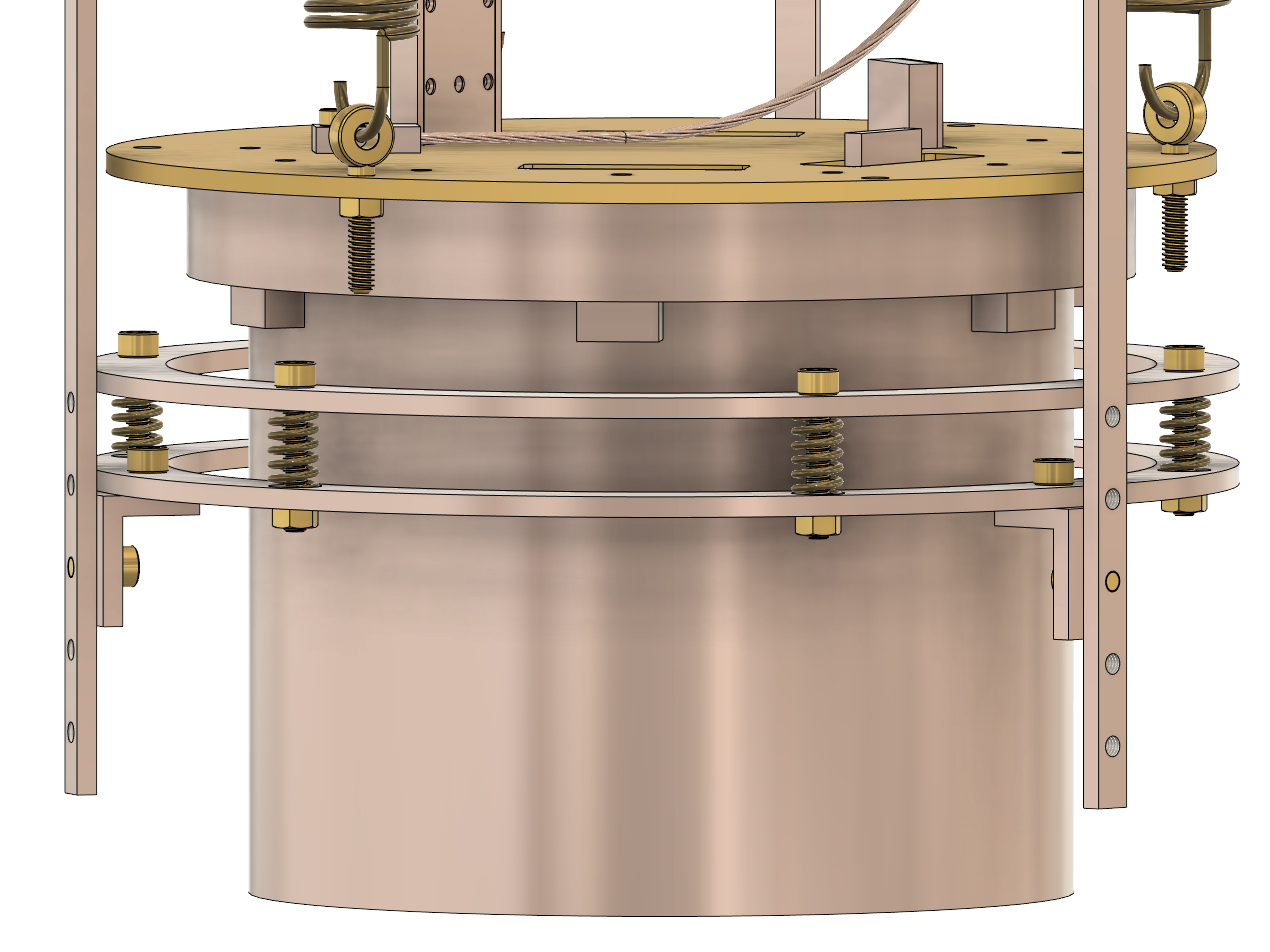}
    \caption{The O-shaped safety structure to catch the lower stage in case of failure. Phosphor bronze springs are again used to reduce the impulse on the structure in case of failure, but in this case six springs are equally spaced in a circle.}
    \label{fig:safety_bottom}
\end{figure}

Lastly, for the lower stage, we use an O-shaped circular disk (i.e. an annulus) to catch the lower stage in case of failure, as shown in Fig.~\ref{fig:safety_bottom}. As with the Y-shaped structure for the upper stage, this structure is attached to the pillars through L-shaped brackets, and there are six equally spaced in-tension springs that will compress if the lower stage falls. The mechanism to keep the springs in-tension and thermally connected is identical to the design of the structure for the upper stage, where we have two plates with inset regions and tightened bolts that allow for movement.

\begin{figure}
    \centering
    \includegraphics{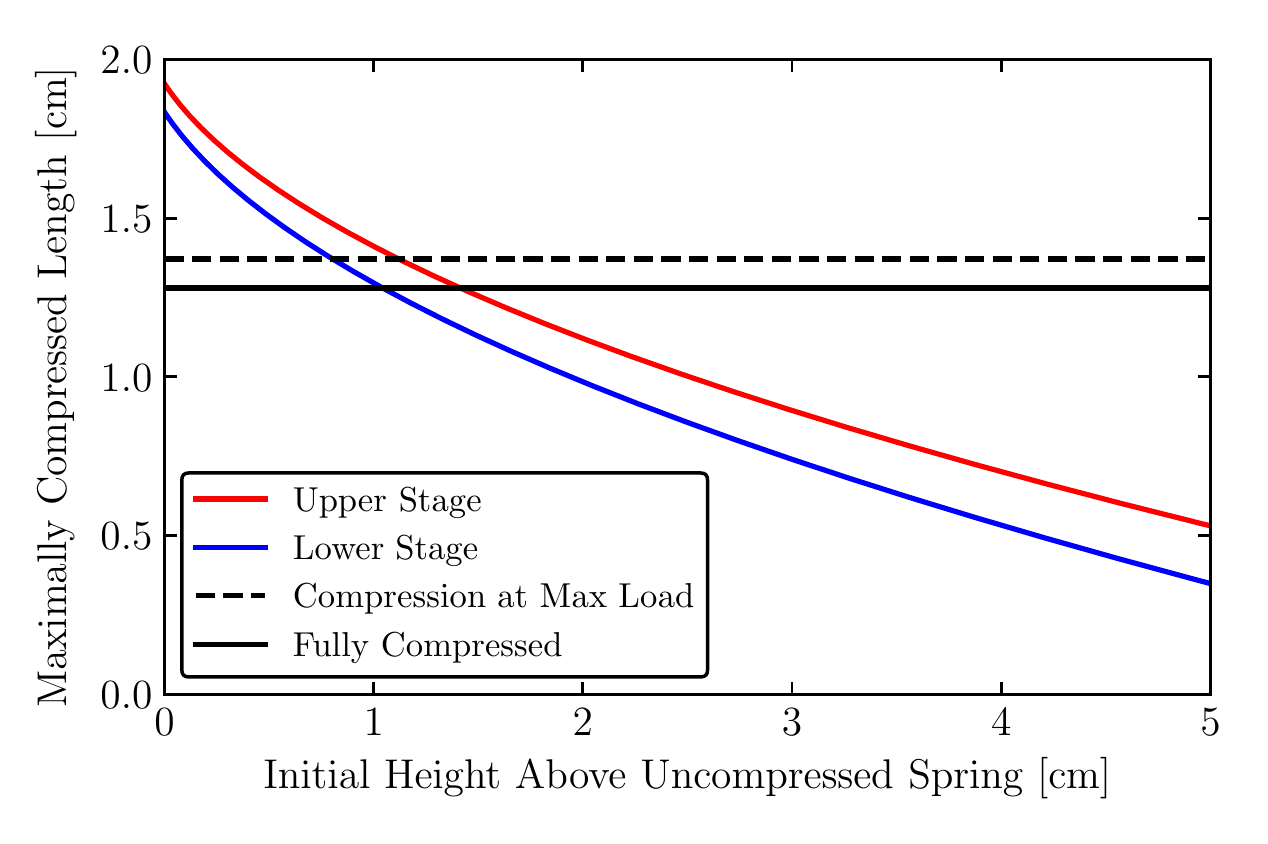}
    \caption{The expected compressed length of the springs as a function of the height above the uncompressed spring, given by Eq.~\ref{eq:maxcompress}. The values that fall below the solid black line are physically unachievable and represent that the spring will have fully compressed and need to be replaced after such a failure of some part of the passive vibration isolation system.}
    \label{fig:safetyspringcompression}
\end{figure}

For the upper and lower stages, the masses are significant, and the expected compression of the springs must be checked against what is physically allowed. For an arbitrary compression spring of spring rate $k$, its maximum compression $x_f$ will be given by some mass $m$ being dropped from some height above the uncompressed spring $h$ (or $h+x_f$ above the maximally compressed spring). Thus, we can use conservation of energy to solve for the expected maximum compression for any mass and any height
\begin{equation}
    mg(h + x_f) = \frac{1}{2}kx_f^2.
\end{equation}
Solving for $x_f$, we have that
\begin{equation}
    x_f = \frac{mg}{k}\left(1 \pm \sqrt{1  + 2\frac{kh}{mg}} \right),
    \label{eq:maxcompress}
\end{equation}
where we throw out the answer corresponding to the negative sign (it is nonphysical as it would mean that the spring had extended). For our system, any stage would drop less than $5\,\mathrm{cm}$ after failure, so we can use Eq.~\ref{eq:maxcompress} to determine the expected maximally compressed lengths as a function of height above the uncompressed spring, as shown in Fig.~\ref{fig:safetyspringcompression}, where it is assumed that each stage is catching $20 \, \mathrm{kg}$ of mass. This also takes into account that there are seven springs in series for the upper stage and six springs in series for the lower spring, where the effective spring constant increases additively (for $n$ identical springs, $k_{eff} = n k$). 

For the specific springs being used from McMaster (part number \texttt{9657K255}, the only phosphor bronze compression spring that they have in inventory at the time of writing), the spring rate is $k=10.4 \, \mathrm{N/mm}$, the initial length is $2.5\, \mathrm{cm}$, its length at maximum load is $1.4\, \mathrm{cm}$, and its fully compressed length is $1.3\, \mathrm{cm}$ (where fully compressed means that each coil has full contact and the spring cannot physically compress any more). Both of these latter two lengths are shown in the figure. Thus, if the spring compresses by more than $1.4\,\mathrm{cm}$, then it should not be reused, as it likely has been compressed passed its yield strength and not longer has the same properties. Fortunately, we know these springs are cheap, so if they can only be used once, that is completely permissible. Furthermore, we designed our system to be safe from failure, so this occurrence should be extremely rare. Nonetheless, we should adjust the height of each stage of the safety structure such that the initial height of above the uncompressed spring is minimized, as is the impulse to the safety structure.

\section{Final Remarks}

At this point, we have fully designed a passive vibration isolation system specifically for the Pyle dilution refrigerator. This design will be implemented in the latter half of 2022, and should do an excellent job of reducing the vibrational sensitivity of our devices. It is my hope that the design principles that I have detailed are applicable beyond the Pyle group lab and can act as a starting point for any group pursuing the reduction of vibrational sensitivity for their detectors. For our group, I am excited to see the final results, as well as the insight it will give us on the excess background signals that have been detailed and discussed in this thesis.

\chapter{\label{chap:of}Optimal Filtering}

The optimal filter (OF) appendix is effectively standard to be included in a CDMS-related thesis, as our energy reconstruction techniques rely on extracting pulse amplitudes from data with known noise. An alternative version of this appendix can be found, e.g., in Thomas Shutt's thesis~\cite{shutt_thesis}, Sunil Golwala's thesis~\cite{golwala}, Arran Phipps' thesis~\cite{phipps_thesis}, Noah Kurinsky's thesis~\cite{Noah_thesis}, and Nicholas Mast's thesis~\cite{mast_thesis}. Nevertheless, I mention the OF throughout this thesis, so it is useful to include the basic principles to ensure this work is self-contained, as well as include the general ideas behind more complex versions of the OF.

\section{\label{sec:offormalism}Optimal Filter Formalism}

The OF formalism can be explained as the least-square minimization of the frequency-weighted chi-square between an expected pulse shape and some measured pulse. Here, we will quickly go through the various common results of a single-pulse OF, where will use the following Fourier transform conventions (previously defined in Section~\ref{sec:tesresponse}), where we define the transform pair (for some function $g$) for the continuous case as
\begin{align}
    \tilde{g}(f) &= \int_{-\infty}^\infty \mathop{dt} g(t) \mathrm{e}^{-i \omega t}, \\
    g(t) &= \int_{-\infty}^\infty \mathop{df} \tilde{g}(f) \mathrm{e}^{i \omega t},
\end{align}
where $\omega \equiv 2 \pi f$ for convenience, and for the discrete case as
\begin{align}
    \tilde{g}_n &= \frac{1}{N} \sum_{k = -N/2}^{N/2 - 1}g_k  \mathrm{e}^{-i \omega_n t_k}, \\
    g_k  &= \sum_{n = -N/2}^{N/2 - 1} \tilde{g}_n \mathrm{e}^{-i \omega_n t_k}.
\end{align}
In this appendix, we will use the continuous approximation for readability, knowing that the calculations are all done in the discrete case.

Before we can extract pulse heights using the OF, we must make a few definitions. We will define $J(f)$ as the two-sided power spectral density (i.e. not summing the positive and negative frequencies) describing the noise of the system (recall the difference between one-sided and two-sided in the discussion at the beginning of Section~\ref{sec:noisederiv}), $s(t)$ as a template (i.e. expected shape) of the signal normalized to a maximum amplitude (pulse height) of 1, $A$ as the amplitude of the signal, and $t_0$ as the start time of the signal. When actually recording data, the noise in the system (which can be represented by $n(t)$ in time domain) will mean that we, in practice, will measure for some event
\begin{equation}
    v(t) = A s(t) + n(t).
\end{equation}
As we have knowledge of $n(t)$ (i.e. $J(f)$) and $s(t)$, we should be able to extract what the best estimate of $A$ is from this noisy signal, which is the purpose of the OF. When Fourier transforming $v(t)$ and $s(t)$, we will denote their frequency space counterparts as $\tilde{v}(f)$ and $\tilde{s}(f)$, respectively.

\subsection{Extracting Pulse Heights}

To extract pulse heights, we use frequency space and calculate the chi-square between the measured signal and the expected signal, weighting the frequencies by the PSD
\begin{equation}
    \chi^2 = \int_{-\infty}^\infty \mathop{df} \frac{\left|\tilde{v}(f) - A \tilde{s}(f) \right|^2}{J(f)},
\end{equation}
where $A$ is as-of-yet undetermined. To find the best estimate of $A$, we minimize chi-square and solve for $A$, which gives the optimal estimator
\begin{equation}
    \hat{A} = \frac{\int_{-\infty}^\infty \mathop{df} \frac{\tilde{s}^*(f) \tilde{v(f)}}{J(f)}}{ \int_{-\infty}^\infty \mathop{df} \frac{\left|\tilde{s}(f) \right|^2}{J(f)}}.
\end{equation}

From this chi-square minimization, we can also find the variance in this best fit value through the general result of $\sigma_A^2 = \left[\frac{1}{2} \frac{\partial^2 \chi^2}{\partial A^2}\right]^{-1}$, which can be shown to be
\begin{equation}
    \sigma_A^2 = \left[ \int_{-\infty}^\infty \mathop{df} \frac{\left|\tilde{s}(f) \right|^2}{J(f)} \right]^{-1}.
    \label{eq:sigmaa2}
\end{equation}
This is the best possible variance achievable given the assumed pulse shape and PSD, and is completely independent on pulse amplitude, as well as the data itself. Thus, given some signal template and known PSD, it is immediately calculable what best possible variance can be expected.

We have not allowed a time degree-of-freedom, so this amplitude estimation is akin to finding the amplitude corresponding to overlapping the expected signal template with the data, and scaling it to have the best fit (lowest chi-square). However, this is only useful if we know the exact location of the event, which will always have some uncertainty (or may be completely unknown). Fortunately, we can add a time degree-of-freedom, which we will do in the following section.

\subsection{\label{sec:timingres}Extracting Pulse Heights and Pulse Times}

With a time degree-of-freedom $t_0$ added to the chi-square, we can take into account the amount of shifting of the template from no shifting (i.e. no delay) to some time of best fit. The inclusion of $t_0$ is added to the template itself in the chi-square, such that we will minimize
\begin{equation}
    \chi^2 = \int_{-\infty}^\infty \mathop{df} \frac{\left|\tilde{v}(f) - A \mathrm{e}^{-i \omega t_0} \tilde{s}(f) \right|^2}{J(f)}.
    \label{eq:chi2of}
\end{equation}
We again minimize chi-square with respect to amplitude, which will result in 
\begin{equation}
    A\left(t_0 \right) = \frac{\int_{-\infty}^\infty \mathop{df} \mathrm{e}^{i \omega t_0} \frac{\tilde{s}^*(f) \tilde{v(f)}}{J(f)}}{ \int_{-\infty}^\infty \mathop{df} \frac{\left|\tilde{s}(f) \right|^2}{J(f)}},
    \label{eq:ampt0}
\end{equation}
with a dependence of the pulse amplitude of the time-shift of the template. At this point, we make the definition of the optimal filter $\phi(f)$
\begin{equation}
    \phi(f) \equiv \frac{s^*(f)}{J(f)},
\end{equation}
such that Eq.~\ref{eq:ampt0} becomes
\begin{equation}
    A\left(t_0 \right) = \frac{\int_{-\infty}^\infty \mathop{df} \mathrm{e}^{i \omega t_0} \phi(f) v(f)}{ \int_{-\infty}^\infty \mathop{df} \frac{\left|\tilde{s}(f) \right|^2}{J(f)}}.
\end{equation}
Note that the numerator is simply the inverse Fourier transform of $\phi(f) v(f)$, meaning that this is computationally simple to numerically calculate for all time-shifts of the signal template via a fast Fourier transform (a fact that we use repeatedly in our OF algorithms in \textsc{QETpy}~\cite{qetpy}).

We can then directly put $A(t_0)$ into the chi-square and simplify the equation to
\begin{equation}
    \chi^2(t_0) = \int_{-\infty}^\infty \mathop{df} \frac{\left| v(f) \right|^2}{J(f)} - A(t_0)^2 \int_{-\infty}^\infty \mathop{df} \phi(f) s(f),
\end{equation}
where it is immediately clear that the lowest chi-square has a one-to-one correspondence to the time shift degree-of-freedom that has the largest pulse amplitude. This is extremely useful in that if we only need to maximize $A(t_0)$ to find the time delay $t_0$, and this will correspond to the minimum chi-square.

On the variance of both $A$ and $t_0$, we have that $\sigma_A^2$ is again defined by Eq.~\ref{eq:sigmaa2} (as the time offset is effectively part of the signal template and drops out when taking its modulus). For the variance of the time offset $t_0$, we again use the general result from a chi-square that $\sigma_{t_0}^2 = \left[\frac{1}{2} \frac{\partial^2 \chi^2}{\partial t_0^2}\right]^{-1}$, and find that
\begin{equation}
    \sigma_{t_0}^2 = \left[A^2 \int_{-\infty}^\infty \mathop{df} \omega^2 \frac{\left|\tilde{s}(f) \right|^2}{J(f)} \right]^{-1}.
\end{equation}
Where, in this case, there is a dependence on the signal amplitude, where $\sigma_{t_0} \propto 1/A$. That is, as the signal becomes larger, it becomes easier to find its location in time, as expected from an increase in the signal-to-noise ratio.

In this appendix, we kept all of the Fourier transforms in the continuous limit, as it is much easier to handle. However, these formulas that we have (quickly) derived can be converted to the discrete limit and applied to discrete data, which is what we do in practice.

\subsection{An Optimal Filter Example}

\begin{figure}
    \begin{subfigure}{.5\textwidth}
        \centering
        \includegraphics[width=1\linewidth]{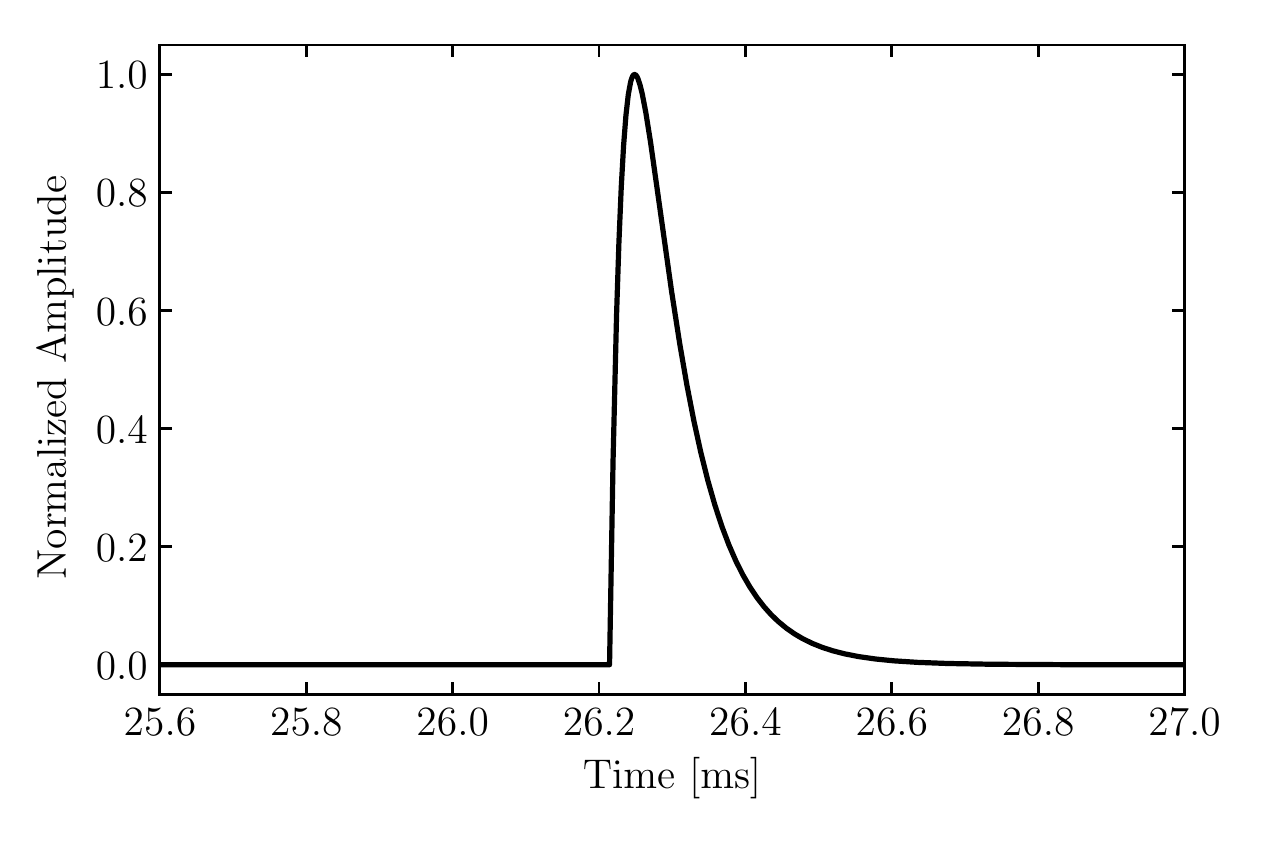}
    \end{subfigure}%
    \begin{subfigure}{.5\textwidth}
        \centering
        \includegraphics[width=1\linewidth]{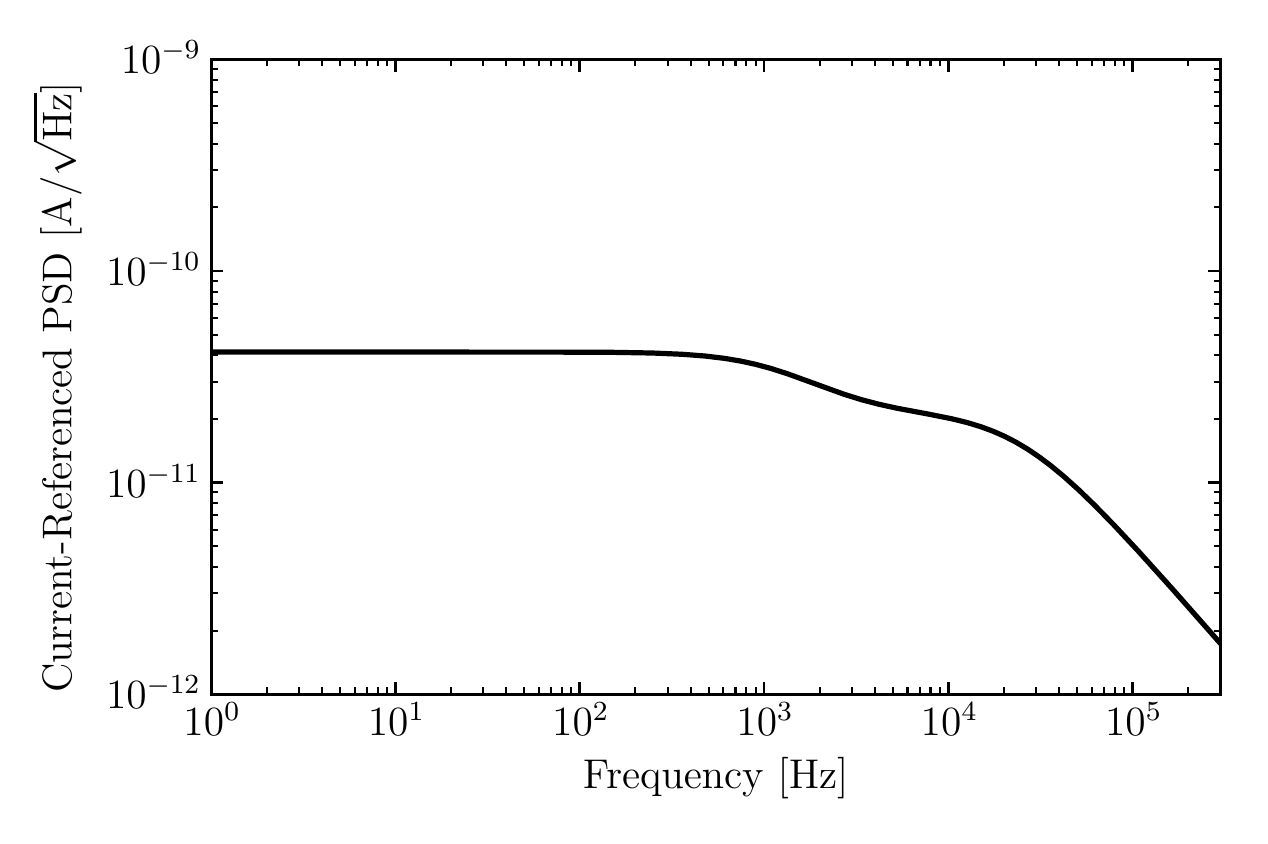}
    \end{subfigure}
    \caption{(Left) The pulse template to be used for this example from simulated noise. (Right) The PSD to simulate noise from and add onto a shifted pulse template.}
    \label{fig:example_templatepsd}
\end{figure}

To see the actual application of the OF to an event, we can use a known pulse template added to noise generated from a PSD. In this case, we will use the pulse template and PSD as shown in Fig.~\ref{fig:example_templatepsd}, where the template has a $20 \, \mu\mathrm{s}$ rise time and a $66 \, \mu\mathrm{s}$ fall time. 

\begin{figure}
    \centering
    \includegraphics{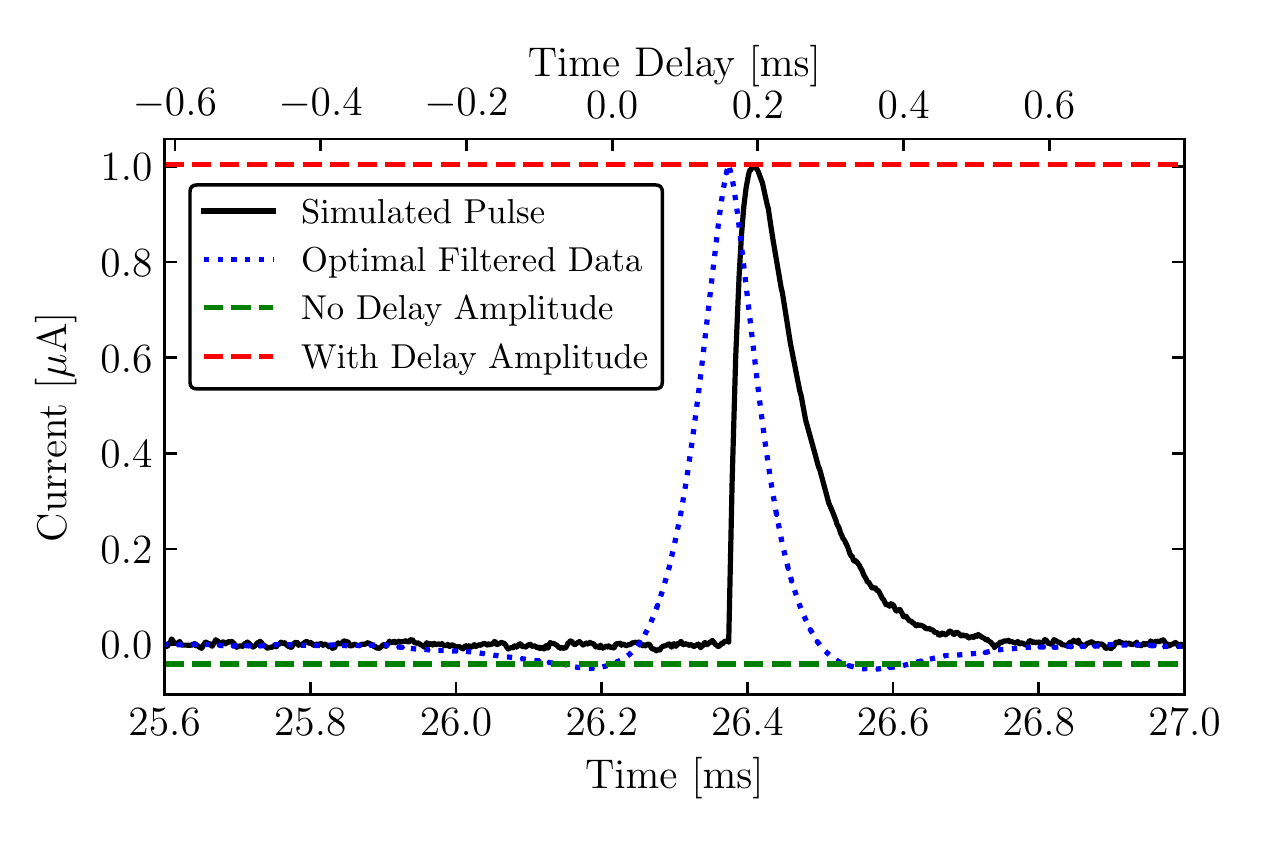}
    \caption{A simulated pulse, where we show the optimal filtered trace $A(t_0)$, where $t_0$ corresponds to a shift defined by the top $x$-axis. The bottom $x$-axis defines the time since the beginning of the trace. We have extracted both OF amplitude with no delay and OF amplitude with delay, showing the importance of allowing the start time degree-of-freedom for accurate determination of pulse heights. For the no delay fit, $\chi^2_0 = 211734$. For the fit with delay, $\chi^2=32767$, matching the expectation of $\chi^2 = 32768$.}
    \label{fig:example_of}
\end{figure}

If we add noise generated from this PSD (using the function \texttt{gen\_noise} in \textsc{QETpy}~\cite{qetpy}) to the pulse template scaled to $1 \, \mu \mathrm{A}$, we can apply the OF and extract amplitudes. We will also shift this pulse to the right (in time) by $160 \, \mu \mathrm{s}$ to show the importance of the time-shift degree of freedom. In Fig.~\ref{fig:example_of}, we show the resulting optimal filtered trace (i.e. $A(t_0)$) as compared to the simulated event, as well the OF amplitudes with no time shifting and the with time shifting.

Thus, it is clear that, unless the pulse start time is known perfectly, the time-shift degree-of-freedom is absolutely necessary for accurate calculation of the pulse height of the event in the OF formalism. We also show in this figure both the time delay and the time elapsed within the trace as the top and bottom $x$-axes, as the two terms can be easily confused. When there is zero time delay in this example, it corresponds to extracting to OF amplitude at a little after $26.2 \, \mathrm{ms}$ (i.e. the time corresponding to the start time of the pulse in the unshifted template).

\section{Two-Pulse Optimal Filters}

Frequently, when the data being taken has a high rate, there may be two events in a single trace (``pileup''). Extracting the pulse amplitudes and pulse heights for each pulse in a single event can be numerically quite expensive. However, there are some different techniques that we can use to be able to speed of the numerical calculation, which we will discuss in the follow sections.

\subsection{Iterative Pileup Optimal Filter}

The iterative pileup OF is based on the assumption that the two pulses are far enough away from one another, such that there is no interference in the optimal filtered trace between the two pulses. Thus, one can simply run the single-pulse OF with the time-shifting degree of freedom, find one of the two pulses, subtract out the fit to that pulse, and then run an OF on the subtracted trace. If there are more pulses, this can be repeated indefinitely. In the OF formalism for two pulses, we would first minimize the chi-square in Eq.~(\ref{eq:chi2of}), and then use those best fit values again to minimize the iterative chi-square
\begin{equation}
    \chi^2_{iter.}(A_1, t_1) = \int_{-\infty}^\infty \mathop{df} \frac{\left|\tilde{v}(f) - \hat{A}_0 \mathrm{e}^{-i \omega \hat{t}_0} - A_1 \mathrm{e}^{-i \omega t_1} \tilde{s}(f) \right|^2}{J(f)}.
\end{equation}
Within the OF formalism as discussed in Section~\ref{sec:offormalism}, the minimization of the chi-square to solve for the best fit values of $A_1$ and $t_1$ is the same, but with the substitution of $\tilde{v}(f) \to \tilde{v}(f) - \hat{A}_0 \mathrm{e}^{-i \omega \hat{t}_0}$.

\begin{figure}
    \begin{subfigure}{.5\textwidth}
        \centering
        \includegraphics[width=1\linewidth]{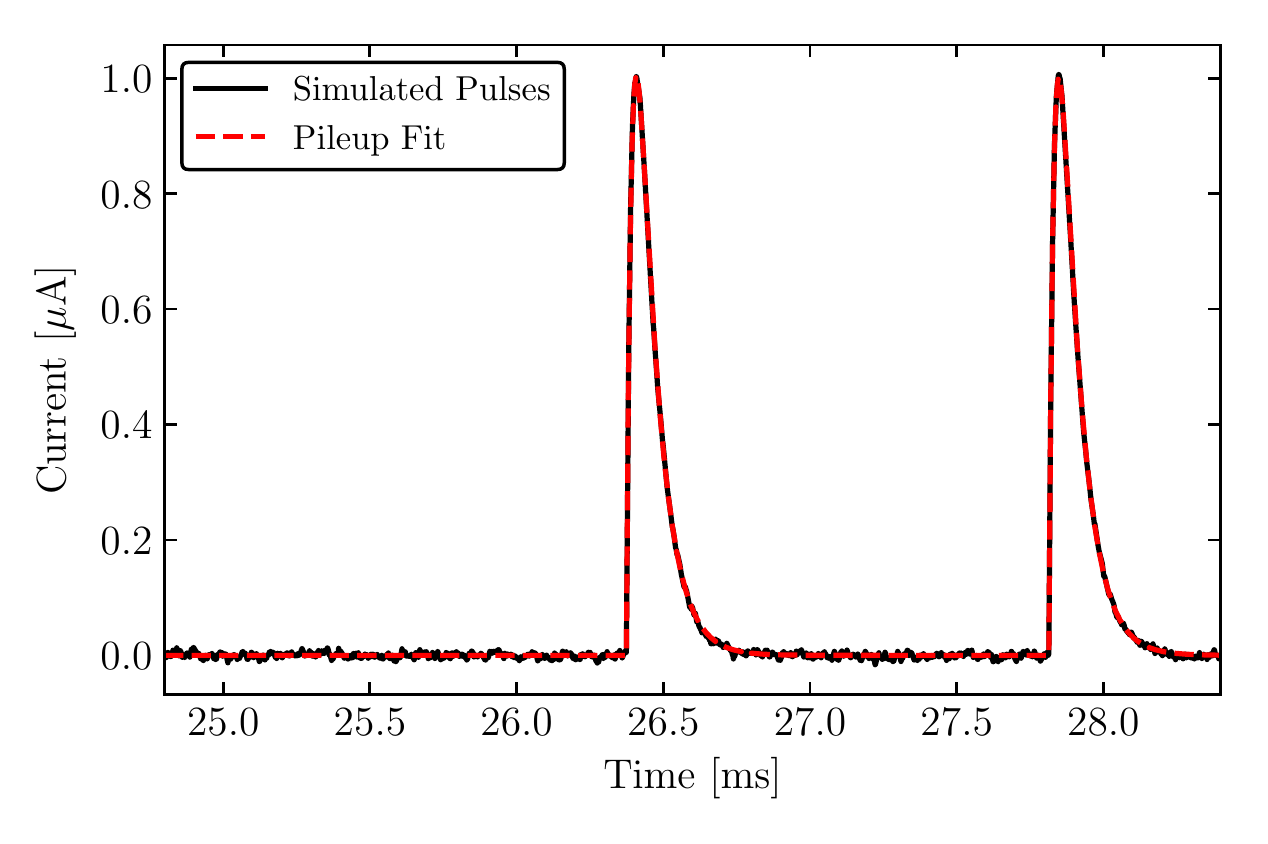}
    \end{subfigure}%
    \begin{subfigure}{.5\textwidth}
        \centering
        \includegraphics[width=1\linewidth]{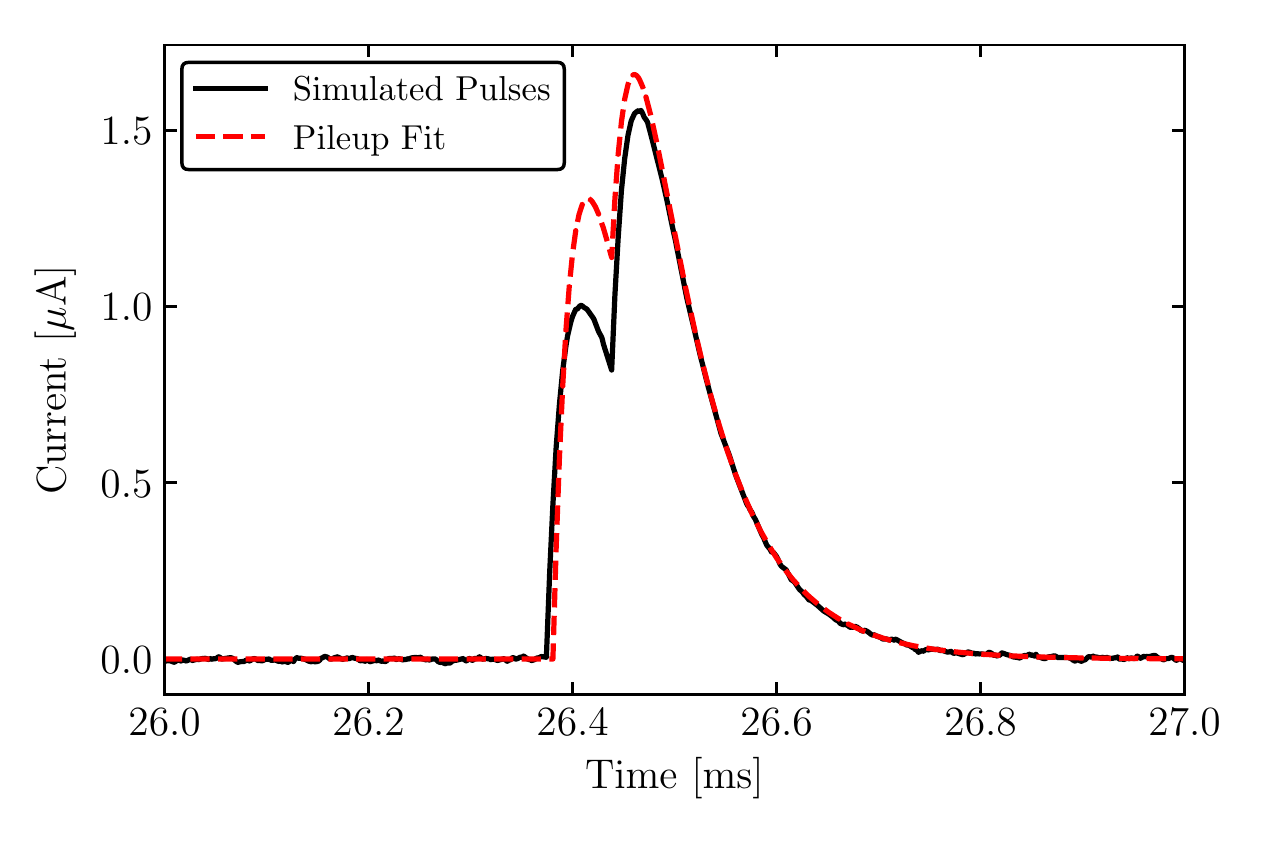}
    \end{subfigure}
    \caption{(Left) Two $1 \, \mu \mathrm{A}$ amplitude pulses added to simulated noise $1.44\, \mathrm{ms}$ apart and fit with an iterative pileup OF. The pulse heights are reconstructed well, as the two pulses are far apart (loosely defined as nonoverlapping). The chi-square for this fit is $\chi^2 = 32767$, matching expectation. (Right)  Two $1 \, \mu \mathrm{A}$ amplitude pulses added to simulated noise $64\, \mu\mathrm{s}$ apart and fit with an iterative pileup OF. The two pulses are overlapping, and the iterative pileup OF does a poor job of reconstructing the amplitudes of the pulses. The chi-square for this fit is $\chi^2 = 64073$, far from the expectation, indicating that this is not a good fit.}
    \label{fig:ipfar_close}
\end{figure}

In Fig.~\ref{fig:ipfar_close}, we show two applications of the iterative pileup OF to events that are spaced far apart and events that are spaced close together. The fit to the closely-spaced pulses poorly reconstructs the pulse amplitudes and demonstrates that a better-performing OF is needed in this regime.

\subsection{\label{sec:simult_of}Simultaneous Pileup Optimal Filter}

To reconstruct amplitudes of overlapping pulses, the OF needs to simultaneously vary the pulse amplitudes and time offsets. Thus, the chi-square to be minimized becomes (where $N$ is the number of pulse amplitudes to reconstruct)
\begin{equation}
    \chi^2_N (\mathbf{A}, \mathbf{t}) = \int_{-\infty}^\infty \mathop{df} \frac{\left|\tilde{v}(f) - \sum_{j=1}^N A_j \mathrm{e}^{-i \omega t_j}\tilde{s}_j(f) \right|^2}{J(f)},
\end{equation}
where $\mathbf{A}=(A_1, A_2,...)$ is a vector containing the pulse amplitudes, and $\mathbf{t} = (t_1, t_2,...)$ is a vector containing the pulse start times.

We start by minimizing this chi-square with respect to an arbitrary amplitude dimension $A_k$, such that
\begin{equation}
    \frac{\partial \chi^2_{N}}{\partial A_k} = \int_{-\infty}^\infty \mathop{df} \left[ -\frac{\tilde{s}_k^*(f)\mathrm{e}^{i \omega t_k} \tilde{v}(f)}{J(f)} + \frac{\tilde{s}_k^*(f)\mathrm{e}^{i \omega t_k} \sum_{j=1}^N A_j \tilde{s}_j(f) \mathrm{e}^{-i \omega t_j}}{J(f)} \right]= 0,
\end{equation}
where we have removed an overall factor of 2. This creates a system of equations relating all of the the amplitudes, which we can define more easily in terms of Einstein notation. That is, we define
\begin{align}
    P_{kj}(t_k, t_j) &\equiv \int_{-\infty}^\infty \mathop{df} \frac{\tilde{s}_k^*(f)  \tilde{s}_j(f) \mathrm{e}^{-i \omega (t_j - t_k)}}{J(f)}, \\
    q_k(t_k) &\equiv  \int_{-\infty}^\infty \mathop{df} \frac{\tilde{s}_k^*(f)\mathrm{e}^{i \omega t_k} \tilde{v}(f)}{J(f)},
\end{align}
and can relate these quantities to an amplitude index of $A_j$ by
\begin{equation}
    P_{kj}(t_k, t_j) A_j = q_k(t_k).
\end{equation}
Note that we have kept the time degrees of freedom explicit, as we have not yet minimized with respect to them. Solving for $A_j$ means that we need to invert the $P$ matrix, where the solution would be
\begin{equation}
    A_j (\mathbf{t}) = \left[P(t_k, t_j) \right]^{-1}_{jk} q_k(t_k).
\end{equation}
Thus, we have that every pulse amplitude is dependent on the start times of itself and the other pulses.

At this point, we will now restrict ourselves to the case of $N=2$ pulses, such that we can directly compare to the iterative pileup OF, as well as set all of the pulse templates to be the same (i.e. $\tilde{s}(f) \equiv \tilde{s}_1(f) = \tilde{s}_2(f)$). In this case, the matrix $P$ is
\begin{equation}
    P_{N=2}(t_1, t_2) = \begin{bmatrix}
        \int_{-\infty}^\infty \mathop{df} \dfrac{\left|\tilde{s}(f)\right|^2}{J(f)} &  \int_{-\infty}^\infty \mathop{df}\dfrac{\left|\tilde{s}(f)\right|^2 \mathrm{e}^{-i \omega (t_1 - t_2)}}{J(f)}\\
        \int_{-\infty}^\infty \mathop{df} \dfrac{\left|\tilde{s}(f)\right|^2 \mathrm{e}^{-i \omega (t_1 - t_2)}}{J(f)} & \int_{-\infty}^\infty \mathop{df} \dfrac{\left|\tilde{s}(f)\right|^2}{J(f)}
    \end{bmatrix}.
\end{equation}
Thus, this matrix must be calculated and then inverted for every combination of $t_1$ and $t_2$ in our simulated trace in order to find all of the possible pulse amplitudes as a function of time, and then find the best chi-square, as we did for the single-pulse version. However, if we blindly did this for the entire trace, we would quickly run into a problem: our computer likely does not have a enough RAM to store this. The traces we are using are 32768 bins long, and we can calculate the number of combinations of pairs of pulse times using the binomial coefficient
\begin{equation}
    \begin{pmatrix}
        n \\
        k 
    \end{pmatrix} = \frac{n!}{k! (n-k)!},
\end{equation}
where we find that
$\big(\begin{smallmatrix}
    32768 \\
    2 
\end{smallmatrix} \big)=536,854,528$
combinations. Thus, to populate this many combinations of a $2\times2$ matrix of ($8\, \mathrm{B}$) floating values would take up $17\,\mathrm{GB}$ of RAM, not to mention that we then have to invert this many matrices. Thus, we must restrict the range of combinations that we will use to create these matrices.

\begin{figure}
    \centering
    \includegraphics[width=0.7\linewidth]{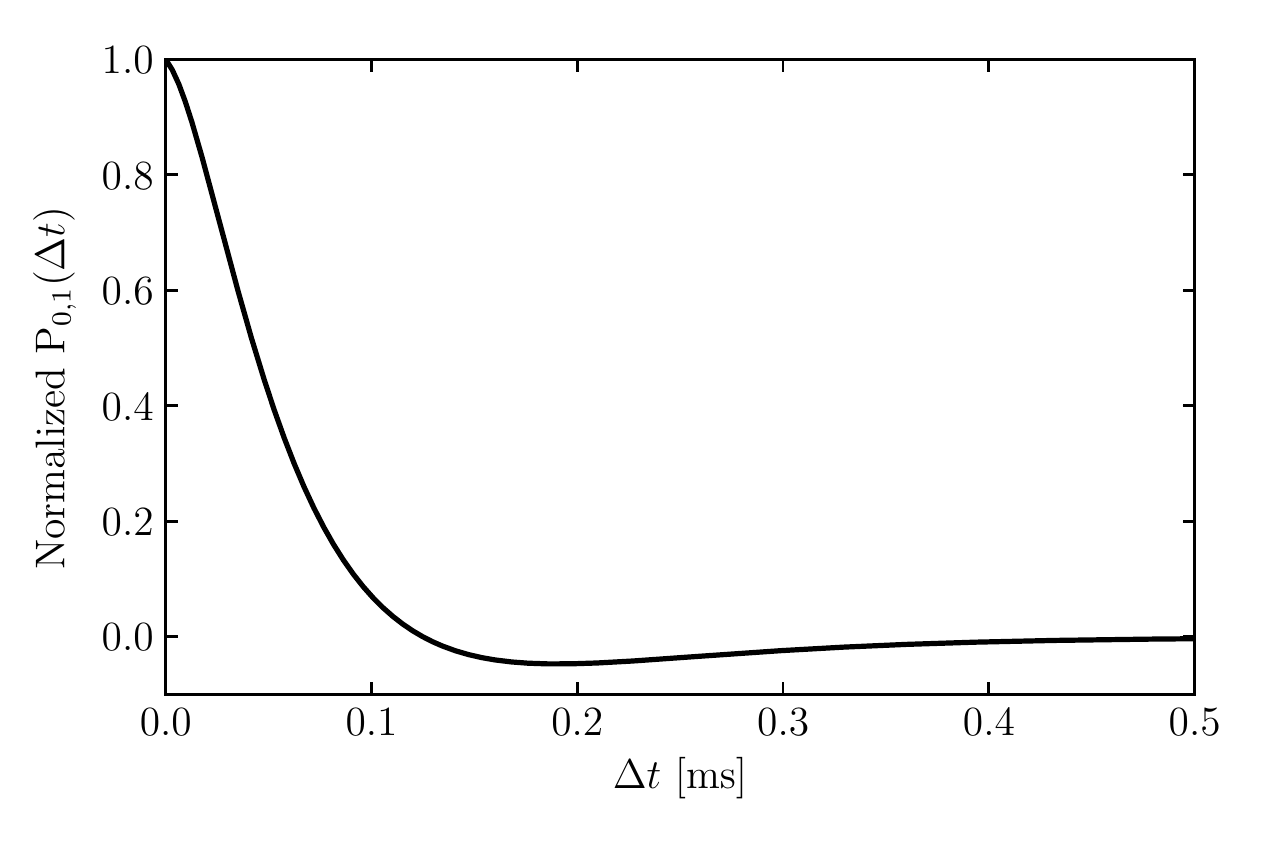}
    \caption{The off-diagonal component of $P_{N=2}$ as a function of $\Delta t$. This gives an estimate of the expected error in the inversion of the matrix as compared to if the matrix were diagonal. For the pulse templates that we are using, the error is less than 7\% for pulses that are spaced greater than $100 \, \mu\mathrm{s}$ apart.}
    \label{fig:poff}
\end{figure}

One way to do this is to go back to our $P$ matrix and estimate a range where the off-diagonal components of the matrix are not small, such that the inversion outside of this range is effectively the iterative pileup OF (these large nonzero off-diagonal components represent the interference between pulses when fitting). If we define $\Delta t = t_2 - t_1$, then we see that the off-diagonal component can be written as
\begin{equation}
    P_{0,1}(\Delta t) = \int_{-\infty}^\infty \mathop{df}\frac{\left|\tilde{s}(f)\right|^2\mathrm{e}^{i \omega \Delta t}}{J(f)},
\end{equation}
which is the inverse Fourier transform of $\left|\tilde{s}(f)\right|^2 / J(f)$, and is thus fast to calculate computationally. Plotting this as a function of $\Delta t$, and normalizing to 1, we have that the behavior of the off-diagonal component is shown in Fig.~\ref{fig:poff}. Because of the symmetry of this matrix, this normalized off-diagonal component provides a good estimate of the error of the matrix if we were to assume that it was simply diagonal~\cite{sprent_1965}. We expect that, if two pulses are spaced further than $100\, \mu\mathrm{s}$ apart, then the iterative and simultaneous versions of the OF should have similar performance, as the error in $P^{-1}$ is estimated to be less than 7\%. Thus, we should restrict the combinations of $t_1$ and $t_2$ to the roughly the range of $|\Delta t| < 100 \, \mu\mathrm{s} \equiv |\Delta t_{max}|$. To know where to calculate this range, it is useful to run a simple OF first for an estimate for where the pileup pulses are in time $t_{start}$. Given this time, we have the range of times to calculate the pileup OF as $[t_{start} - |\Delta t_{max}|, t_{start} + |\Delta t_{max}|]$. Using the 7\% cutoff, this reduces the combinations of times that we need to calculate and store to 7,875, about 5 orders of magnitude as compared to without a time range restriction. With a manageable amount of matrices to store and invert, the simultaneous pileup OF can be run on the closely spaced pulses that the iterative pileup OF could not reconstruct.

\begin{figure}
    \centering
    \includegraphics[width=0.7\linewidth]{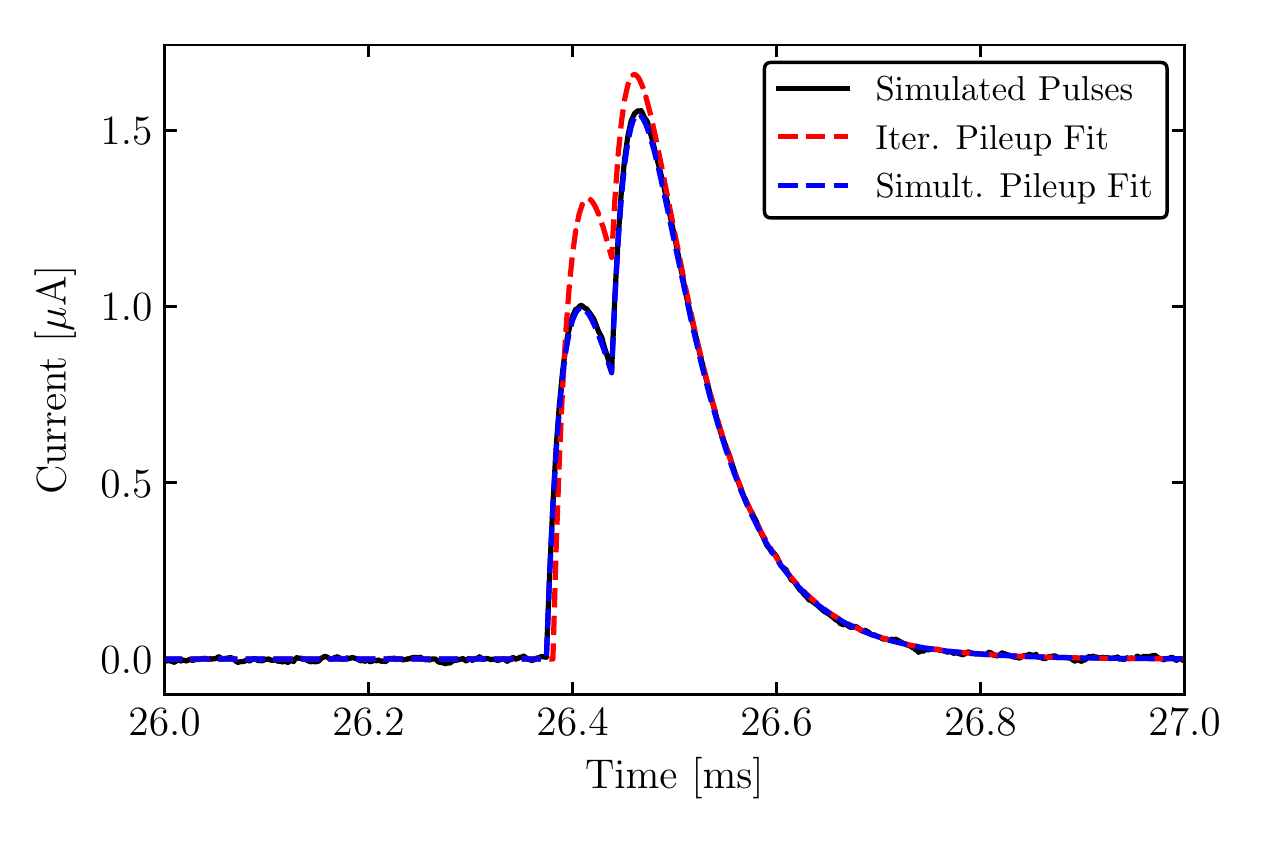}
    \caption{The two $1 \, \mu \mathrm{A}$ amplitude pulses added to simulated noise $64\, \mu\mathrm{s}$ apart and fit with both the iterative pileup OF and the simultaneous pileup OF. The simultaneous pileup OF does an excellent job of reconstructing the pulse heights, with goodness-of-fit $\chi^2 = 32463$.}
    \label{fig:simult_of}
\end{figure}

In Fig.~\ref{fig:simult_of}, we compare the simultaneous and iterative pileup OFs as applied to the closely spaced pileup pulses. The simultaneous pileup OF reconstructs the pulse amplitudes very well, showing the benefit of varying the pulse amplitudes and start times all at once. 

\section{Arbitrary Number of Pulses}

The natural next step is to apply the same OF formalism to an arbitrary number of pileup pulses. However, if we were to follow the same method of restricting ranges, then we would run into computational issues extremely quickly. That is, we would need to take the combinations of $N$ pulses of the time range chosen to calculate. If we stay under 10,000 combinations of the start times of $N=3$ pulses, this corresponds to a time range of 40 bins, or $64\,\mu\mathrm{s}$. This is less than the fall time of a single simulated pulse, which would be too small of a time range to extract three possible pileup pulses.

Thus, we must abandon the method of precomputing these matrices and carrying out computationally efficient robust searches along all time degrees-of-freedom. Because of the nonlinearity of the time degrees-of-freedom, a simple gradient descent algorithm will not reliably minimize the chi-square. As such, a derivative-free optimization algorithm will be needed, with an excellent option being differential evolution, originally introduced by Storn and Price~\cite{diffevolution}. The basic idea of this algorithm is that, given some function minimize, it iteratively improves the candidate solution through a random sampling of the variable space. Though it does not guarantee finding the optimal solution, it can be tuned such that it is quite likely in the case of the $N$ pulse OF.

Returning to the $N$ pulse chi-square, we have that we want to minimize
\begin{equation}
    \chi^2_N (\mathbf{A}, \mathbf{t}) = \int_{-\infty}^\infty \mathop{df} \frac{\left|\tilde{v}(f) - \sum_{j=1}^N A_j \mathrm{e}^{-i \omega t_j}\tilde{s}_j(f) \right|^2}{J(f)},
\end{equation}
for which we have shown the method in Section~\ref{sec:simult_of}. That is, we need to invert the matrix $P$ in order to find the pulse amplitudes at given times. We know that the pulse amplitudes depend on the start times of all pulses, such that we can calculate them for any given pair of times
\begin{equation}
    A_j (\mathbf{t}) = \left[P(t_k, t_j) \right]^{-1}_{jk} q_k(t_k),
\end{equation}
where we again are reproducing from Section~\ref{sec:simult_of}. If we plug this equation into the chi-square, we have
\begin{equation}
    \chi^2_N (\mathbf{t}) = \int_{-\infty}^\infty \mathop{df} \frac{\left|\tilde{v}(f) - \sum_{j=1}^N A_j(\mathbf{t}) \mathrm{e}^{-i \omega t_j} \tilde{s}_j(f) \right|^2}{J(f)},
\end{equation}
where it is now solely a function of the time degrees of freedom. When discussing the simultaneous OF, this is where we set $N=2$ and then precomputed the $P^{-1}$ matrices. Because we cannot do this efficiently for $N>2$, this is where we go to the differential evolution algorithm to minimize the chi-square.

Because differential evolution randomly samples, we can improve the likelihood of finding the global minimum of the chi-square by restricting the range of time values that it considers to where we believe there might be pileup. We can use the same concepts as Section~\ref{sec:simult_of} to set it on the estimated errors of the matrices, where we increase the ranges by factors of the number of pileup pulses to extract, in order to ensure we allow all pulses to be spaced by $|\Delta t_{max}|$.

\begin{figure}
    \centering
    \includegraphics{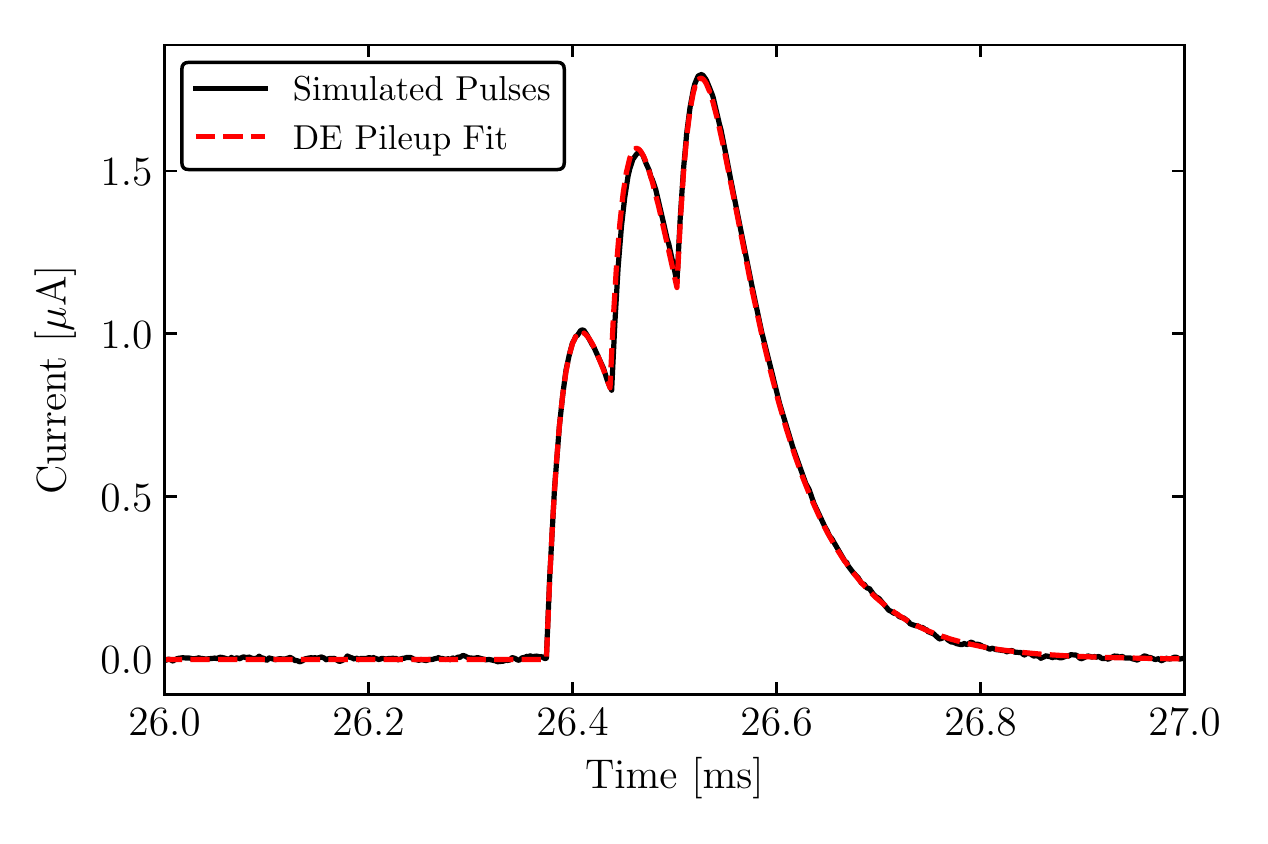}
    \caption{Three $1 \, \mu \mathrm{A}$ amplitude pulses added to simulated noise with $64\, \mu\mathrm{s}$ spacing and fit with differential evolution--based pileup OF. This pileup OF reconstructs the pulse heights and times well, with goodness-of-fit $\chi^2 = 32407$.}
    \label{fig:de_pileup}
\end{figure}

In Fig.~\ref{fig:de_pileup}, we show the result of a three pulse fit using differential evolution (via \textsc{SciPy}~\cite{2020SciPy-NMeth}) to find the minimum chi-square in the time range around the expected location of the pileup. We see that the algorithm found the pulse heights and times quite well, without needing to do a robust grid search. This algorithm can also scale to any arbitrary number of pulses. 

Furthermore, the differential evolution algorithm in \textsc{SciPy} includes an option for creating constraints, based on work by Lampinen~\cite{1004459}. For instance, we might want to constrain the pileup pulses to only have positive amplitudes, if we are in a biasing setup of the TES such that real pulses only go in that direction. Thus, constraints like this or others are easily added, but may significantly slow down the computation time depending on the complexity.

\section{Optimal Filter Code in \textsc{QETpy}}

The various OF codes discussed in this Appendix have been written in \textsc{Python} by me and placed in various locations in our open-source package \textsc{QETpy}~\cite{qetpy}. Although the methods herein are not particularly novel, their existence as usable and publicly available code is an excellent place to start using them without needing to write the algorithms from scratch. Below, I direct the reader to the following classes in \textsc{QETpy}
\begin{itemize}
    \item \texttt{qetpy.OptimumFilter}: contains the single-pulse OF and the iterative pileup OF
    \item \texttt{qetpy.PileupOF}: contains the two-pulse simultaneous pileup OF
    \item \texttt{qetpy.PileupDE}: contains the differential evolution--based pileup OF for an arbitrary number of pulses.
\end{itemize}
For the simultaneous pileup OFs, they do have a built-in assumption that the pulses all have the same template. When analyzing real data, this may not be the case based on the physics involved. This is one possible upgrade to these existing algorithms, which I leave to future \textsc{QETpy} collaborators.


\end{document}